\pdfoutput=1

\newcommand{\thesisTitleFrontmatter}{NEW TOOLS, PROGRAMMING MODELS, AND SYSTEM SUPPORT FOR PROCESSING-IN-MEMORY ARCHITECTURES}

\newcommand{\thesisTitlePlain}{New Tools, Programming Models, and System Support for Processing-in-Memory Architectures}

\newcommand{\thesisAuthor}{Geraldo Francisco de Oliveira Junior}
\newcommand{\thesisUni}{\protect{ETH Z\"urich}}

\newcommand{\thesisYear}{2025}
\newcommand{\thesisVersion}[0]{\theversion.3}

\documentclass[12pt,oneside,a4paper]{ethzthesis}

\usepackage[export]{adjustbox}
\usepackage[en-GB, useregional=numeric]{datetime2}
\usepackage{algorithmic}
\usepackage[acronym,nonumberlist,nowarn]{glossaries}
\usepackage{amsfonts}
\usepackage{amsmath} % for eqautions
\usepackage{balance}
\usepackage{booktabs}
\usepackage{clipboard}
\usepackage[compact]{titlesec}
\usepackage{color, colortbl}
\usepackage{cuted,capt-of}
\usepackage{cleveref}
\usepackage{enumitem}
\usepackage{fancyhdr}
\usepackage{float}
\usepackage{flushend}
\usepackage{footnote}
\usepackage[multiple]{footmisc}
\usepackage{graphbox}
\usepackage{graphicx}
\usepackage{imakeidx}
\usepackage{listings}
\usepackage{longfbox}
\usepackage{ltablex}
\usepackage{makecell}
\usepackage{marginnote}
\usepackage{mathtools}
\usepackage{multirow}
\usepackage{multicol}
\usepackage{pifont}
\usepackage{pifont}
\usepackage{setspace}
\usepackage{setspace}
\usepackage[binary-units=true]{siunitx}
\usepackage{soul}
\usepackage{tocloft} 
\usepackage{subcaption}
\usepackage{textcomp}
\usepackage{tikz}
\usepackage{titlesec}
\usepackage[many]{tcolorbox}
\usepackage[textwidth=20mm, prependcaption,textsize=tiny]{todonotes}
\usepackage{xargs}
\usepackage{xcolor}
\usepackage{xspace}
\usepackage[normalem]{ulem}
\usepackage[labelfont=bf, textfont=bf, skip=2pt]{caption}
\usepackage{wrapfig}

\glsdisablehyper
\newacronym{ADI}{ADI}{Alternating Direction Implicit}
\newacronym{AGU}{AGU}{Address Generation Unit}
\newacronym[longplural={Access History Tables}]{AHT}{AHT}{Access History Table}
\newacronym{AHT-R}{AHT-R}{Access History Table - Read}
\newacronym{AHT-W}{AHT-W}{Access History Table - Write-Back}
\newacronym{DMA}{DMA}{direct memory access}

\newacronym{AIP}{AIP}{Access Interval Predictor}
\newacronym{AL}{AL}{Added Latency for column accesses}
\newacronym{ASIC}{ASIC}{Application Specific Integrated Circuit}
\newacronym{AVX}{AVX}{Advanced Vector Extensions}
\newacronym[longplural={Basic Block Vectors}]{BBV}{BBV}{Basic Block Vector}
\newacronym{BHT}{BHT}{Branch History Table}
\newacronym{HBM}{HBM}{Hybrid Bandwidth Memory}
\newacronym{DDR4}{DDR4}{Double-Data Rate 4}
\newacronym{MIMD}{MIMD}{multiple-instruction multiple-data}
\newacronym{SIMT}{SIMT}{Single Instruction Multiple Threads}
\newacronym{SLP}{SALP}{subarray-level parallelism}
\newacronym{BLP}{BLP}{bank-level parallelism}
\newacronym{ReRAM}{ReRAM}{resistive random access memory}
\newacronym{MRAM}{MRAM}{magnetoresistive random access memory}
\newacronym{FeRAM}{FeRAM}{ferroelectric random access memory}

\newacronym{NN}{NN}{neural network}
\newacronym{BTB}{BTB}{Branch Target Buffer}
\newacronym{CAS}{CAS}{Column Address Strobe}
\newacronym{AMAT}{AMAT}{Average Memory Access Time}
\newacronym{CCD}{CCD}{Column to Column Delay}
\newacronym{AI}{AI}{arithmetic intensity}
\newacronym[longplural={Columnar Database Systems}]{CDBS}{CDBS}{Columnar Database System}
\newacronym[longplural={Chip Multiprocessors}]{CMP}{CMP}{Chip Multiprocessor}
\newacronym{CMT}{CMT}{Chip Multithreading}
\newacronym{SSD}{SSD}{solid-state drive}
\newacronym[longplural={Coarse-Grain Reconfigurable Arrays}]{CGRA}{CGRA}{Coarse-Grain Reconfigurable Array}
\newacronym{C-RAM}{C-RAM}{Computational-RAM}
\newacronym{CWD}{CWD}{Column Write Delay}
\newacronym{DBI}{DBI}{Dynamic Binary Instrumentation}
\newacronym[longplural={Database Management Systems}]{DBMS}{DBMS}{Database Management System}
\newacronym{DDR}{DDR}{Double Data Rate}
\newacronym{DEWP}{DEWP}{Dead Line and Early Write-Back Predictor}
\newacronym{DRAM}{DRAM}{dynamic random access memory}
\newacronym{DSBP}{DSBP}{Dead Sub-Block Predictor}
\newacronym{DSM}{DSM}{Decomposed Storage Manage}
\newacronym{FAW}{FAW}{Four row Activation Window}
\newacronym{FP}{FP}{Floating-Point}
\newacronym{EDP}{EDP}{Energy-Delay Product}
\newacronym{FCFS}{FCFS}{First-Come First-Serve}
\newacronym{FIFO}{FIFO}{Fist In, First Out}
\newacronym{FSM}{FSM}{finite state machine}
\newacronym{FPGA}{FPGA}{field-programmable gate array}
\newacronym{API}{API}{application programming interface}
\newacronym[longplural={Functional Units}]{FU}{FU}{Functional Unit}
\newacronym{GAg}{GAg}{Global Adaptive branch prediction using one Global PHT}
\newacronym{GAs}{GAs}{Global Adaptive branch prediction using per-Set PHT}
\newacronym{GCC}{GCC}{GNU Compiler Collection}
\newacronym{GEMS}{GEMS}{General Execution-driven Multiprocessor Simulator}
\newacronym[longplural={General Purpose Processors}]{GPP}{GPP}{General Purpose Processor}
\newacronym{HMC}{HMC}{Hybrid Memory Cube}
\newacronym{HIVE}{HIVE}{HMC Instruction Vector Extensions}
\newacronym{IATAC}{IATAC}{Inter-Access Time per Access Count}
\newacronym{ILP}{ILP}{Instruction Level Parallelism}
\newacronym{IPC}{IPC}{Instructions per Cycle}
\newacronym{ISA}{ISA}{Instruction Set Architecture}
\newacronym{IRAM}{IRAM}{Intelligent RAM}
\newacronym{KIPS}{KIPS}{Kilo Instructions per Second}
\newacronym{kNN}{kNN}{k-nearest neighbors}
\newacronym{NVM}{NVM}{non-volatile memory}

\newacronym{LDS}{LDS}{Linked Data Structure}
\newacronym{LFSR}{LFSR}{Linear Feedback Shift Register}
\newacronym{LFMR}{LFMR}{Last-to-First Miss-Ratio}
\newacronym{MLP}{MLP}{Memory-Level Parallelism}
\newacronym{ML}{ML}{machine learning}

\newacronym{PCB}{PCB}{printed circuit board}
\newacronym{LLM}{LLM}{large language model}
\newacronym{DLRM}{DLRM}{deep learning recommendation model}
\newacronym{LLC}{LLC}{last-level cache}
\newacronym{LRU}{LRU}{Least Recently Used}
\newacronym{LSU}{LSU}{Load Store Unit}
\newacronym{LTP}{LTP}{Last-Touch Predictor}
\newacronym{LvP}{LvP}{Live-time Predictor}
\newacronym{LWP}{LWP}{Last Write Predictor}
\newacronym{MARSS}{MARSS}{Micro Architectural and System Simulator}
\newacronym{McPAT}{McPAT}{Multi-core Power, Area, and Timing}
\newacronym{PE}{PE}{processing element}
\newacronym{LRDIMM}{LRDIMM}{load-reduced DIMM}
\newacronym{DIMM}{DIMM}{dual in-line memory module}
\newacronym{MIPS}{MIPS}{Microprocessor without Interlocked Pipeline Stages}
\newacronym{MOB}{MOB}{Memory Order Buffer}
\newacronym{MMU}{MMU}{memory management unit}
\newacronym{PuD}{PUD}{processing-using-DRAM}
\newacronym{PUF}{PUF}{physically unclonable function}
\newacronym{COTS}{COTS}{commodity off-the-shelf}
\newacronym{TRNG}{TRNG}{true random number generator}
\newacronym{SpMV}{SpMV}{sparse matrix-vector multiplication}
\newacronym{MPKI}{MPKI}{Misses per Kilo-Instruction}
\newacronym{MSHR}{MSHR}{Miss-Status Handling Registers}
\newacronym{NAS}{NAS}{Numerical Aerodynamic Simulation}
\newacronym{NDP}{NDP}{near-data processing}
\newacronym{NMP}{NMP}{Near Memory Processor}
\newacronym{NoC}{NoC}{Network-on-Chip}
\newacronym{NPB}{NPB}{NAS Parallel Benchmark}
\newacronym{NUCA}{NUCA}{Non-Uniform Cache Architecture}
\newacronym{NUMA}{NUMA}{Non-Uniform Memory Access}
\newacronym{OoO}{OoO}{Out-of-Order}
\newacronym{OpenMP}{OpenMP}{Open Multi-Processing}
\newacronym{OS}{OS}{Operating System}
\newacronym{PAg}{PAg}{Per-address Adaptive branch prediction using one Global PHT}
\newacronym{PAs}{PAs}{Per-address Adaptive branch prediction using per-Set PHT}
\newacronym{PC}{PC}{Program Counter}
\newacronym[longplural={Pointer-Chasing Engines}]{PCE}{PCE}{Pointer-Chasing Engine}
\newacronym{PCM}{PCM}{Performance Counter Monitor}
\newacronym{PIM}{PIM}{processing-in-memory}
\newacronym[longplural={Pattern History Tables}]{PHT}{PHT}{Pattern History Table}
\newacronym{RAPL}{RAPL}{Running Average Power Limit}
\newacronym{GEMM}{GEMM}{general matrix-matrix multiplication}
\newacronym{RAS}{RAS}{Row Address Strobe}
\newacronym{RAT}{RAT}{Registers Alias Table}
\newacronym{RC}{RC}{Row Cycle}
\newacronym{RCD}{RCD}{RAS to CAS Delay}
\newacronym[longplural={Row-based Database Systems}]{RDBS}{RDBS}{Row-based Database System}
\newacronym{ROB}{ROB}{Reorder Buffer}
\newacronym{RRD}{RRD}{Row to Row activation Delay}
\newacronym{DNN}{DNN}{Deep Neural Network}
\newacronym{ANN}{ANN}{Artificial Neural Network}
\newacronym{PnM}{PNM}{processing-near-memory}
\newacronym{PuM}{PUM}{processing-using-memory}
\newacronym{FFD}{FFD}{first-fit-decreasing}
\newacronym{HFF}{HFF}{helper flip-flop}
\newacronym{OBPS}{OBPS}{one-bit per-subarray}
\newacronym{ABOS}{ABOS}{all-bits in one-subarray}
\newacronym{ABPS}{APBS}{all-bits per-subarray}
\newacronym{GEMV}{GEMV}{general matrix-vector product}
\newacronym{RBR}{RBR}{redundant binary representation}
\newacronym{RP}{RP}{Row Precharge}
\newacronym{RTL}{RTL}{Register Transfer Level}
\newacronym{RTP}{RTP}{Read To Precharge}
\newacronym{RVU}{RVU}{Reconfigurable Vector Unit}
\newacronym{SDP}{SDP}{Skewed Dead-Block Predictor}
\newacronym{SESC}{SESC}{Superescalar Simulator}
\newacronym{SSE}{SSE}{Streaming SIMD Extensions}
\newacronym{SFP}{SFP}{Spatial Footprint Predictor}
\newacronym{SiNUCA}{SiNUCA}{Simulator of Non-Uniform Cache Architectures}
\newacronym{SMT}{SMT}{Simultaneous Multi-Threading}
\newacronym{SIMD}{SIMD}{single-instruction multiple-data}
\newacronym{LUT}{LUT}{lookup table}

\newacronym{SPEC}{SPEC}{Standard Performance Evaluation Corporation}
\newacronym{SoC}{SoC}{System-on-Chip}
\newacronym{SPP}{SPP}{Spatial Pattern Predictor}
\newacronym{SRAM}{SRAM}{static random access memory}
\newacronym{SSOR}{SSOR}{Symmetric Successive Over-Relaxation}
\newacronym{SSV}{SSV}{Search Set Vector}
\newacronym{TLB}{TLB}{Translation Look-aside Buffer}
\newacronym{TLP}{TLP}{Thread Level Parallelism}
\newacronym[longplural={Through-Silicon Vias}]{TSV}{TSV}{Through-Silicon Via}
\newacronym{VWQ}{VWQ}{Virtual Write Queue}
\newacronym{WTR}{WTR}{Write Recovery time}
\newacronym{WR}{WR}{Write To Read delay time}
\newacronym{TRA}{TRA}{triple row activation}

\newcommand{\DAMOV}{{DAMOV}\xspace} 
\newcommand{\MIMDRAM}{{MIMDRAM}\xspace} 
\newcommand{\Proteus}{\emph{Proteus}\xspace} 
\newcommand{\DaPPA}{{DaPPA}\xspace} 

\definecolor{amber}{rgb}{1.0, 0.49, 0.0}
\definecolor{awesome}{rgb}{1.0, 0.13, 0.32}
\definecolor{dollarbill}{rgb}{0.52,0.73,0.4}
\definecolor{moegi}{rgb}{0.357, 0.537, 0.188}
\definecolor{burgundy}{rgb}{0.5, 0.0, 0.13}
\definecolor{ballblue}{rgb}{0.13, 0.67, 0.8}
\definecolor{ups-truck}{rgb}{0.53, 0.28, 0.21}
\definecolor{airforceblue}{rgb}{0.36, 0.54, 0.66}
\definecolor{cadmiumgreen}{rgb}{0.0, 0.42, 0.24}
\definecolor{darkcyan}{rgb}{0.0, 0.55, 0.55}
\definecolor{caribbeangreen}{rgb}{0.0, 0.8, 0.6}
\definecolor{flamingopink}{rgb}{0.99, 0.56, 0.67}
\definecolor{jazzberryjam}{rgb}{0.65, 0.04, 0.37}
\definecolor{mediumpersianblue}{rgb}{0.0, 0.4, 0.65}
\definecolor{coolblack}{rgb}{0.0, 0.18, 0.39}
\definecolor{bleudefrance}{rgb}{0.19, 0.55, 0.91}
\definecolor{ao}{rgb}{0.0, 0.0, 1.0}
\definecolor{babyblueeyes}{rgb}{0.63, 0.79, 0.95}
\definecolor{antiquefuchsia}{rgb}{0.57, 0.36, 0.51}
\definecolor{burntorange}{rgb}{0.8, 0.33, 0.0}
\definecolor{blush}{rgb}{0.87, 0.36, 0.51}
\definecolor{MidnightBlue}{rgb}{0.1, 0.1, 0.44}
\definecolor{dodgerblue}{rgb}{0.12, 0.56, 1.0}
\definecolor{brandeisblue}{rgb}{0.0, 0.44, 1.0}
\definecolor{brickred}{rgb}{0.8, 0.25, 0.33}
\definecolor{eggplant}{rgb}{0.38, 0.25, 0.32}
\definecolor{byzantium}{rgb}{0.44, 0.16, 0.39}
\definecolor{ddgreen}{rgb}{0.00, 0.50, 0.00}
\definecolor{mygreen}{rgb}{0,0.6,0}
\definecolor{mygray}{rgb}{0.5,0.5,0.5}
\definecolor{mymauve}{rgb}{0.58,0,0.82}
\definecolor{bluehl}{rgb}{0.8,0.874,1}
\definecolor{pinkhl}{rgb}{0.992156863,0.847058824,1}
\definecolor{macaroniandcheese}{rgb}{1.0, 0.74, 0.53}
\definecolor{mossgreen}{rgb}{0.68, 0.87, 0.68}
\definecolor{greenhl}{rgb}{0.835,0.996,0.939}
\definecolor{yellowhl}{rgb}{0.996,0.957,0.8}
\definecolor{palecerulean}{rgb}{0.61, 0.77, 0.89}
\definecolor{gray(x11gray)}{rgb}{0.75, 0.75, 0.75}
\definecolor{amethyst}{rgb}{0.6, 0.4, 0.8}
\definecolor{cadmiumorange}{rgb}{0.93, 0.53, 0.18}
\definecolor{frenchlilac}{rgb}{0.53, 0.38, 0.56}
\definecolor{heliotrope}{rgb}{0.87, 0.45, 1.0}
\definecolor{peridot}{rgb}{0.9, 0.89, 0.0}
\definecolor{saffron}{rgb}{0.96, 0.77, 0.19}
\definecolor{tuscanred}{rgb}{0.51, 0.21, 0.21}
\definecolor{uscgold}{rgb}{1.0, 0.8, 0.0}
\definecolor{tangerineyellow}{rgb}{1.0, 0.8, 0.0}
\definecolor{rufous}{rgb}{0.66, 0.11, 0.03}
\definecolor{safetyorange}{rgb}{1.0, 0.4, 0.0}
\definecolor{LightGray}{gray}{0.95}

\definecolor{Gray}{gray}{0.9}
\definecolor{darkolivegreen}{rgb}{0.33, 0.42, 0.18}
\definecolor{plum}{rgb}{0.56, 0.27, 0.52}
\definecolor{islamicgreen}{rgb}{0.0, 0.56, 0.0}
\definecolor{lightyellow}{rgb}{0.980, 0.956, 0.623}
\definecolor{dgreen}{rgb}{0.00, 0.80, 0.00}

\newcommand{\gft}[2]{\ifnum#1=\value{version}\textcolor{blue}{#2}\else{#2}\fi}
\newcommand{\gftcomment}[2]{\ifnum#1=\value{version}\todo[size=\scriptsize, linecolor=orange, bordercolor=orange, backgroundcolor=white]{\textcolor{blue}{\textbf{@gy:} #2}}\else{}\fi}

\newcommand{\copiedlabel}[2]{\ifnum#1=\value{version}\todo{\tiny{\textcolor{burgundy}{\textbf{Copied from} #2}}}\else{}\fi}
\newcommand{\omt}[2]{\ifnum#1=\value{version}\textcolor{blue}{#2}\else{#2}\fi}

\newcommand{\circled}[1]{\tikz[baseline=(char.base)]{\node[shape=circle,draw,inner sep=0pt,fill=black, text=white] (char) {#1};}}
\newcommand{\circledgreen}[1]{\tikz[baseline=(char.base)]{\node[shape=circle,draw,inner sep=0pt,fill=dollarbill, text=white] (char) {#1};}}
\newcommand{\circledii}[1]{\tikz[baseline=(char.base)]{\node[shape=circle,draw,inner sep=0pt,fill=gray, text=white] (char) {\itshape#1};}}
\newcommand{\circlediii}[1]{\tikz[baseline=(char.base)]{\node[shape=circle,draw,inner sep=0pt,fill=white, text=black] (char) {\itshape#1};}}

\DeclareSIUnit{\flop}{FLOP}

\newcommand{\tempcommand}[1]{\renewcommand{\arraystretch}{#1}}

\DeclarePairedDelimiter\ceil{\lceil}{\rceil}
\DeclarePairedDelimiter\floor{\lfloor}{\rfloor}

\newcommand\add{\colorbox{BurntOrange}{ADD}}

\newcommand{\paratitle}[1]{\vspace{4pt}\noindent\textbf{#1.}}

\newcommand{\revdel}[1]{}

\newcommand{\li}{(\textit{i})}
\newcommand{\lii}{(\textit{ii})}
\newcommand{\liii}{(\textit{iii})}
\newcommand{\liv}{(\textit{iv})}
\newcommand{\lv}{(\textit{v})}

\crefformat{section}{\S#2#1#3}
\crefformat{subsection}{\S#2#1#3}
\crefformat{subsubsection}{\S#2#1#3}
\crefformat{appendix}{Appendix #2#1#3}

\newcommand{\param}[1]{#1}
\newcommand{\secref}[1]{§\ref{#1}}

\newcommand\aap{\texttt{AAP}/\texttt{AP}\xspace}
\newcommand\aaps{\texttt{AAP}s/\texttt{AP}s\xspace}
\newcommand\uop{\textmu{}Op\xspace}
\newcommand\uops{\textmu{}Ops\xspace}
\newcommand\uprog{\textmu{}Program\xspace}
\newcommand\uprogs{\textmu{}Programs\xspace}
\newcommand\ureg{\textmu{}Register}
\newcommand\upc{\textmu{}PC}
\newcommand\uprogc{\textmu{}Program counter}
\newcommand\bbop{\emph{bbop}\xspace}

\newcommandx{\unsure}[2][1=]{\todo[linecolor=red,backgroundcolor=red!25,bordercolor=red,#1, size=\tiny,fancyline]{#2}}
\newcommandx{\change}[2][1=]{\todo[linecolor=blue,backgroundcolor=blue!25,bordercolor=blue,#1,size=\tiny]{\textbf{#2}}}
\newcommandx{\feedback}[2][1=]{\todo[linecolor=yellow,backgroundcolor=yellow!25,bordercolor=yellow,#1,size=\tiny]{#2}}
\newcommandx{\improvement}[2][1=]{\todo[linecolor=Plum,backgroundcolor=Plum!25,bordercolor=Plum,#1]{#2}}
\newcommandx{\thiswillnotshow}[2][1=]{\todo[disable,#1]{#2}}
\newcommandx{\completedRevision}[2][1=]{\todo[disable,backgroundcolor=red,#1]{#2}}
\newcommandx{\dataSource}[2][1=]{\todo[disable,backgroundcolor=red,#1]{#2}}
\newcommandx{\info}[2][1=]{\todo[linecolor=dollarbill,backgroundcolor=dollarbill!25,bordercolor=dollarbill,#1, size=\tiny,fancyline]{#2}}

\newif\ifthesiscameraready
\thesiscamerareadytrue
\newcounter{version}
\ifthesiscameraready
\else
\fi

\newcommand\membottleneck{\cite{mutlu2013memory, mutlu2015research, dean2013tail, kanev_isca2015, ferdman2012clearing, wang2014bigdatabench, boroumand2018google, boroumand2021google, boroumand2021google_arxiv, mutlu2019enabling, mutlu2019processing, mutlu2020intelligent, ghose.ibmjrd19, alser2020accelerating, cali2020genasm, koppula2019eden, kanellopoulos2019smash, deoliveira2021IEEE, deoliveira2021, oliveira2021.SLS, mutlu.msttalk17, mutlu.gwutalk19, mutlu.isscctalk19, mutlu.glvlsitalk19, mutlu.appttalk19, mutlu.iccdtalk19,narancic2014evaluating,ayers2018memory,jia2016understanding,Manegold_2000,gholami2020ai,stengel2015quantifying,chishti2019memory,burger1995declining,gupta2020architectural,sriraman2019softsku,zhao2022understanding,ruan2019insider,gan2018architectural,sriraman2020accelerometer,yuan2023rambda,delimitrou2018amdahl,hsia2020cross,lottarini2018vbench,wang2022characterizing,richins2020missing,mckee2004reflections}\xspace}

\newcommand\memscalingissue{\cite{kang2014co, mutlu2013memory,
mutlu2015research, mckee2004reflections, wilkes2001memory, kim2014flipping, kim2012case, yoongu-thesis,liu2012raidr,mutlu2017rowhammer,lee2015decoupled,lee-isca2009,yoon2012row,yoon-taco2014,lim-isca09,
wulf1995hitting, chang.sigmetrics2016, Tiered-Latency_LEE, lee2015adaptive,
chang2017understanding, lee2017design, luo2014characterizing,
luo.arxiv17,hassan2017softmc, hassan2016chargecache, chang2017understandingphd, patel2017reach, hassan2019crow, ghose2018your, kim2018solar, kim2020revisiting, mutlu2019rowhammer, wang2018reducing, mutlu2018recent, ghose.sigmetrics20,mutlu2015main,hong2010memory,sites1996,luo2023rowpress,yaglikci2024spatial,olgun2024abacus,bostanci2024comet,yauglikcci2022hira,giray-thesis,bostanci2025understanding,luo2025revisiting,canpolat2025chronus,tuugrul2025understanding}}

\newcommand\pimdef{\cite{ghose.ibmjrd19, mutlu2020modern,deoliveira2021IEEE,pim-book,mutlu2019processing,mutlu2019enabling,mutlu2015research,mutlu2013memory,loh2013processing,Near-Data,stone1970logic,Miss_Mem_Wall_1996,mutlu2023memory, mcciede2024}\xspace}

\newcommand\pnm{\cite{farmahini2015nda,babarinsa2015jafar,devaux2019true,ghiasi2022genstore,gomez2021benchmarkingcut,gomezluna2021benchmarking,gomez2022benchmarking,syncron,singh2020nero,skhynixpim,ke2021near,giannoula2022sparsep,shin2018mcdram,cho2020mcdram,denzler2021casper,asghari2016chameleon,IRAM_Micro_1997,C_RAM_1999,CASES_MVX,Xi_2015,sun2021abc,matam2019graphssd,gokhale1995processing,hall1999mapping,MEMSYS_MVX,lockerman2020livia,ahn2015scalable,nai2017graphpim,boroumand2018google,lazypim, top-pim, gao2016hrl, kim2018grim, drumond2017mondrian, RVU, NIM, PEI, gao2017tetris,Kim2016,gu2016leveraging, boroumand2019conda, hsieh2016transparent, cali2020genasm, NDC_ISPASS_2014,pattnaik2016scheduling,akin2015data,hsieh2016accelerating,lee2015bssync,boroumand2021mitigating,boroumand2021google,boroumand2022polynesia,boroumand2021polynesia,amiraliphd,besta2021sisa,fernandez2020natsa,singh2019napel,kwon202125,lee2021hardware,niu2022184qps,Sparse_MM_LiM,azarkhish2016logic,azarkhish2018neurostream,guo20143d,de2018design,akin2014hamlet,huang2020heterogeneous,dai2018graphh,liu2018processing,tsai:micro:2018:ams,gu2020ipim,DRAMA_CAL_2014,Asghari-Moghaddam_2016,huang2019active,kersey2017lightweight,li2019pims,kim2017grim,boroumand2017lazypim,zhuo2019graphq,zhang2018graphp,lim2017triple,smc_sim,HIVE,jang2019charon,IBM_ActiveCube,hadidi2017cairo,santos2018processing,lenjani2020fulcrum}\xspace}

\newcommand\pim{\cite{farmahini2015nda,babarinsa2015jafar,devaux2019true,ghiasi2022genstore,gomez2021benchmarkingcut,gomezluna2021benchmarking,gomez2022benchmarking,syncron,singh2020nero,skhynixpim,ke2021near,giannoula2022sparsep,shin2018mcdram,cho2020mcdram,denzler2021casper,asghari2016chameleon,IRAM_Micro_1997,C_RAM_1999,CASES_MVX,Xi_2015,sun2021abc,matam2019graphssd,gokhale1995processing,hall1999mapping,MEMSYS_MVX,lockerman2020livia,ahn2015scalable,nai2017graphpim,boroumand2018google,lazypim,top-pim,gao2016hrl,kim2018grim,drumond2017mondrian,RVU,NIM,PEI,gao2017tetris,Kim2016,gu2016leveraging,boroumand2019conda,hsieh2016transparent,cali2020genasm,NDC_ISPASS_2014,pattnaik2016scheduling,akin2015data,hsieh2016accelerating,lee2015bssync,boroumand2021mitigating,boroumand2021google,boroumand2022polynesia,boroumand2021polynesia,amiraliphd,besta2021sisa,fernandez2020natsa,singh2019napel,kwon202125,lee2021hardware,niu2022184qps,Sparse_MM_LiM,azarkhish2016logic,azarkhish2018neurostream,guo20143d,de2018design,akin2014hamlet,huang2020heterogeneous,dai2018graphh,liu2018processing,tsai:micro:2018:ams,gu2020ipim,DRAMA_CAL_2014,Asghari-Moghaddam_2016,huang2019active,kersey2017lightweight,li2019pims,kim2017grim,boroumand2017lazypim,zhuo2019graphq,zhang2018graphp,lim2017triple,smc_sim,HIVE,jang2019charon,IBM_ActiveCube,hadidi2017cairo,santos2018processing,Chi2016,Shafiee2016,seshadri2017ambit,seshadri2019dram,li2017drisa,seshadri2013rowclone,seshadri2016processing,deng2018dracc,xin2020elp2im,song2018graphr,song2017pipelayer,gao2019computedram,eckert2018neural,aga2017compute,dualitycache,seshadri2016buddy,seshadri.bookchapter17,seshadri2018rowclone,seshadri2015fast,li2016pinatubo,ferreira2021pluto,ferreira2022pluto,imani2019floatpim,he2020sparse,flashcosmos,truong2022adapting,truong2021racer,olgun2021quactrng,kim2019d,kim2018dram,bostanci2022dr,olgun2022pidram,ali2019memory,angizi2019graphide,li2018scope,subramaniyan2017parallel,zha2020hyper,fujiki2018memory,orosa2021codic,sharad2013ultra,rezaei2020nom,gao2021parabit,choi2020flash,han2019novel,merrikh2017high,wang2018three,lue2019optimal,kim2021behemoth,wang2022memcore,han2021flash,kang2021s,lee2020neuromorphic,lee20223d,si2019dual,simon2020blade,nag2019gencache,wang2019bit,al2020towards,kang2014energy,kim2021colonnade,jiang2020c3sram,jeloka201628,wang2023infinity,kang2015energy,imani2020dual,chang2016low,hajinazarsimdram,deng2019lacc,sutradhar2021look,sutradhar2020ppim,lenjani2020fulcrum,peng2023chopper,oliveira2022accelerating,singh2021fpga,oliveira2023dappa,oliveira2022methodologies,oliveira2022heterogeneous,shahroodi2023swordfish,chen2023simplepim,gupta2023evaluating,gomez2023evaluating,item2023transpimlib,diab2023framework,mao2022genpip,singh2022accelerating,park2024lpddr,lee2022improving,kim2021aquabolt,samsunghc23,sim2022computational,sudarshan2022critical,chen2025reis,yuksel2025pudhammer,soysal2025mars,canpolat2025easydram,simon2025processing,asquini2025accelerating,mutlu2025memory,simon2025cimba,frouzakis2025pimdal,kabra2025ciphermatch,he2025papi,gu2025pim,bostanci2025revisiting}\xspace}

\newcommand\pimindustry{\cite{devaux2019true,skhynixpim,ke2021near,kwon202125,lee2021hardware,park2024lpddr,lee2022improving,kim2021aquabolt,samsunghc23,sim2022computational,niu2022184qps}\xspace}
\newcommand\pimacademia{\cite{farmahini2015nda,babarinsa2015jafar,gomez2021benchmarkingcut,gomezluna2021benchmarking,gomez2022benchmarking,syncron,ghiasi2022genstore,singh2020nero,giannoula2022sparsep,shin2018mcdram,cho2020mcdram,denzler2021casper,asghari2016chameleon,IRAM_Micro_1997,C_RAM_1999,CASES_MVX,Xi_2015,sun2021abc,matam2019graphssd,gokhale1995processing,hall1999mapping,MEMSYS_MVX,lockerman2020livia,ahn2015scalable,nai2017graphpim,boroumand2018google,lazypim,top-pim,gao2016hrl,kim2018grim,drumond2017mondrian,RVU,NIM,PEI,gao2017tetris,Kim2016,gu2016leveraging,boroumand2019conda,hsieh2016transparent,cali2020genasm,NDC_ISPASS_2014,pattnaik2016scheduling,akin2015data,hsieh2016accelerating,lee2015bssync,boroumand2021mitigating,boroumand2021google,boroumand2022polynesia,boroumand2021polynesia,amiraliphd,besta2021sisa,fernandez2020natsa,singh2019napel,Sparse_MM_LiM,azarkhish2016logic,azarkhish2018neurostream,guo20143d,de2018design,akin2014hamlet,huang2020heterogeneous,dai2018graphh,liu2018processing,tsai:micro:2018:ams,gu2020ipim,DRAMA_CAL_2014,Asghari-Moghaddam_2016,huang2019active,kersey2017lightweight,li2019pims,kim2017grim,boroumand2017lazypim,zhuo2019graphq,zhang2018graphp,lim2017triple,smc_sim,HIVE,jang2019charon,IBM_ActiveCube,hadidi2017cairo,santos2018processing,Chi2016,Shafiee2016,seshadri2017ambit,seshadri2019dram,li2017drisa,seshadri2013rowclone,seshadri2016processing,deng2018dracc,xin2020elp2im,song2018graphr,song2017pipelayer,gao2019computedram,eckert2018neural,aga2017compute,dualitycache,seshadri2016buddy,seshadri.bookchapter17,seshadri2018rowclone,seshadri2015fast,li2016pinatubo,ferreira2021pluto,ferreira2022pluto,imani2019floatpim,he2020sparse,flashcosmos,truong2022adapting,truong2021racer,olgun2021quactrng,kim2019d,kim2018dram,bostanci2022dr,olgun2022pidram,ali2019memory,angizi2019graphide,li2018scope,subramaniyan2017parallel,zha2020hyper,fujiki2018memory,orosa2021codic,sharad2013ultra,rezaei2020nom,gao2021parabit,choi2020flash,han2019novel,merrikh2017high,wang2018three,lue2019optimal,kim2021behemoth,wang2022memcore,han2021flash,kang2021s,lee2020neuromorphic,lee20223d,si2019dual,simon2020blade,nag2019gencache,wang2019bit,al2020towards,kang2014energy,kim2021colonnade,jiang2020c3sram,jeloka201628,wang2023infinity,kang2015energy,imani2020dual,chang2016low,hajinazarsimdram,deng2019lacc,sutradhar2021look,sutradhar2020ppim,lenjani2020fulcrum,peng2023chopper,oliveira2022accelerating,singh2021fpga,oliveira2023dappa,oliveira2022methodologies,oliveira2022heterogeneous,shahroodi2023swordfish,chen2023simplepim,gupta2023evaluating,gomez2023evaluating,item2023transpimlib,diab2023framework,mao2022genpip,singh2022accelerating,cho2020near,shin2025piccolo,seo2025facil,kubo2025mvdram,kubo2025pudtune}\xspace}

\newcommand\pnmthreed{\cite{ahn2015scalable,nai2017graphpim,boroumand2018google,lazypim, top-pim, gao2016hrl, kim2018grim, drumond2017mondrian, RVU, NIM, PEI, gao2017tetris,Kim2016,gu2016leveraging, boroumand2019conda, hsieh2016transparent, cali2020genasm, NDC_ISPASS_2014,pattnaik2016scheduling,akin2015data,hsieh2016accelerating,lee2015bssync,boroumand2021mitigating,boroumand2021google,boroumand2022polynesia,boroumand2021polynesia,amiraliphd,besta2021sisa,fernandez2020natsa,singh2019napel,kwon202125,lee2021hardware,niu2022184qps,Sparse_MM_LiM,azarkhish2016logic,azarkhish2018neurostream,guo20143d,de2018design,akin2014hamlet,huang2020heterogeneous,dai2018graphh,liu2018processing,tsai:micro:2018:ams,gu2020ipim,DRAMA_CAL_2014,Asghari-Moghaddam_2016,huang2019active,kersey2017lightweight,li2019pims,kim2017grim,boroumand2017lazypim,zhuo2019graphq,zhang2018graphp,lim2017triple,smc_sim,HIVE,jang2019charon,IBM_ActiveCube,hadidi2017cairo,santos2018processing}\xspace}

\newcommand\pum{\cite{Chi2016, Shafiee2016, seshadri2017ambit, seshadri2019dram, li2017drisa, seshadri2013rowclone, seshadri2016processing, deng2018dracc, xin2020elp2im, song2018graphr, song2017pipelayer,gao2019computedram, eckert2018neural, aga2017compute,dualitycache,besta2021sisa,seshadri2016buddy,seshadri.bookchapter17,seshadri2018rowclone,seshadri2015fast,li2016pinatubo,ferreira2021pluto,ferreira2022pluto,imani2019floatpim,he2020sparse,flashcosmos,truong2022adapting,truong2021racer,olgun2021quactrng,kim2019d,kim2018dram,bostanci2022dr,olgun2022pidram,ali2019memory,angizi2019graphide,li2018scope,subramaniyan2017parallel,zha2020hyper,fujiki2018memory,orosa2021codic,sharad2013ultra,rezaei2020nom,gao2021parabit,choi2020flash,han2019novel,merrikh2017high,wang2018three,lue2019optimal,kim2021behemoth,wang2022memcore,han2021flash,kang2021s,lee2020neuromorphic,lee20223d,si2019dual,
simon2020blade,nag2019gencache,wang2019bit,al2020towards,kang2014energy,kim2021colonnade,jiang2020c3sram,jeloka201628,wang2023infinity,kang2015energy,imani2020dual, chang2016low,hajinazarsimdram,deng2019lacc,sutradhar2021look,sutradhar2020ppim,peng2023chopper,shahroodi2023swordfish,sudarshan2022fefet,sudarshan2022weighted,sudarshan2022optimization,sudarshan2021novel}\xspace}

\newcommand\drampum{\cite{angizi2019graphide,besta2021sisa,bostanci2022dr,deng2018dracc,ferreira2021pluto,ferreira2022pluto,gao2019computedram,li2017drisa,li2018scope,olgun2021quactrng,olgun2022pidram,seshadri.bookchapter17, seshadri2013rowclone,seshadri2015fast,seshadri2016buddy, seshadri2016processing, seshadri2017ambit,seshadri2018rowclone, seshadri2019dram, xin2020elp2im}\xspace}

\newcommand\srampum{\cite{aga2017compute,eckert2018neural,si2019dual,simon2020blade,nag2019gencache,wang2019bit,al2020towards,kang2014energy,kim2021colonnade,jiang2020c3sram,jeloka201628,wang2023infinity,kang2015energy}\xspace}

\newcommand\nvmpum{\cite{imani2019floatpim,li2016pinatubo,Shafiee2016,song2017pipelayer,song2018graphr,truong2021racer,truong2022adapting,sharad2013ultra,Chi2016,imani2020dual}\xspace}

\newcommand\flashpum{\cite{flashcosmos,gao2021parabit,choi2020flash,han2019novel,merrikh2017high,wang2018three,lue2019optimal,kim2021behemoth,wang2022memcore,han2021flash,kang2021s,lee2020neuromorphic,lee20223d}\xspace}

\newcommand\ambit{\cite{seshadri2017ambit,seshadri2019dram,seshadri2015fast,seshadri.bookchapter17,seshadri2016buddy,seshadri2016processing}\xspace}

\newcommand\drambackground{\cite{hassan2019crow, ghose.sigmetrics20, ghose2018your, kim2016ramulator, seshadri2019dram, kim2012case, zhang2014half, hassan2016chargecache, Tiered-Latency_LEE, seshadri2017ambit, chang2017understanding,
chang2017understandingphd,
chang.sigmetrics2016, chang2014improving, chang2016low, lee2015adaptive, lee2016reducing, lee2016reducingthesis, lee2015decoupled, liu2013experimental, liu2012raidr, seshadri2013rowclone, seshadri2015gather, ipek2008self, lee2016simultaneous, Dennard68field,
keeth2007dram,
mineshphd,
hasanphd,
yauglikcci2022hira,
luo2023rowpress,
keeth2001dram,
frigo2020trrespass,
o2021energy,
cooper2010fine,
jacob_book_2008,
jacob2009memory,
cuppu1999performance,
oconnor2017fine,
kang2014co,
kim1998dram,
giray-thesis,
patel2024rethinkingieee
}\xspace}

\newcommand\pnmshort{\cite{devaux2019true,ghiasi2022genstore,gomez2021benchmarkingcut,gomezluna2021benchmarking,gomez2022benchmarking,syncron,singh2020nero,skhynixpim,ke2021near,giannoula2022sparsep,denzler2021casper,IRAM_Micro_1997,C_RAM_1999,gokhale1995processing,hall1999mapping,ahn2015scalable,boroumand2018google,lazypim, top-pim, kim2018grim, RVU, NIM, PEI,Kim2016, boroumand2019conda, hsieh2016transparent, cali2020genasm,hsieh2016accelerating,boroumand2021mitigating,boroumand2021google,boroumand2022polynesia,boroumand2021polynesia,besta2021sisa,fernandez2020natsa,singh2019napel,lee2021hardware,kim2017grim,boroumand2017lazypim,santos2018processing,lenjani2020fulcrum}\xspace}

\makeatletter
\newcommand\requiredelimiter[2][########]{%
  \ifdefined#2%
    \def\@temp{\def#2#1}%
    \expandafter\@temp\expandafter{#2}%
  \else
    \@latex@error{\noexpand#2undefined}\@ehc
  \fi
}
\@onlypreamble\requiredelimiter
\makeatother

\begin{document}

\frenchspacing
\raggedbottom
\selectlanguage{english}
\pagenumbering{roman}
\pagestyle{plain}

% bibliography concerns
\bstctlcite{IEEEexample:BSTcontrol}
% \bstctlcite[@auxoutS]{IEEEexample:BSTcontrol}
\setbiblabelwidth{1000} % because the custom multibib prefix messes up indents

% IEEE bibliography stuff
\bstctlcite{IEEEexample:BSTcontrol}

% FRONTMATTER

\begin{titlepage}   
    \large
    \begin{center}
        \begingroup
        \MakeUppercase{Diss. ETH No. 31078}
        %\thesisDissNumber{}
        \endgroup
    
        \hfill

        \vfill

        \begingroup
            \textit{\textbf{\thesisTitleFrontmatter}}
        \endgroup

        \vfill

        \begingroup
            A thesis submitted to attain the degree of\\
            \vspace{1.0em}
            \MakeUppercase{Doctor of Sciences}\\
            %\vspace{0.5em}
            (Dr. sc. \thesisUni) \\
            
        \endgroup

        \vfill

        \begingroup
            presented by
            \vfill
            
            \emph{GERALDO FRANCISCO DE OLIVEIRA JUNIOR}

            \vfill
%             M.Sc., Federal University of Rio Grande do Sul\\
%            \vspace{0.5em}
            born on \emph{04.02.1994}\\
        \endgroup

        \vfill

        \begingroup
            accepted on the recommendation of\\
            \vspace{1.0em}
            Prof.\ Dr.\ Onur Mutlu, examiner\\
            \vspace{0.5em}
            Dr. Christian Weis, co-examiner \\
            \vspace{0.5em}
            Dr. Donghyuk Lee, co-examiner \\
            \vspace{0.5em}
            Prof. Dr. Reetuparna Das, co-examiner \\
            \vspace{0.5em}
            Prof. Dr. Tony Nowatzki, co-examiner \\
        \endgroup

        \vfill

        \thesisYear%

        \vfill
    \end{center}
\end{titlepage}

\thispagestyle{empty}

\hfill

\vfill

\noindent\thesisAuthor: \textit{\thesisTitlePlain,}
\textcopyright\ \thesisYear
\setstretch{1.3}
\cleardoublepage
\thispagestyle{empty}

\vspace*{3cm}

\begin{center}
    \textit{To my loving parents, Simone and Geraldo.\\
    (Aos meus amados pais, Simone e Geraldo).}
\end{center}

\medskip

\clearpage

\chapter*{Acknowledgments}
\addcontentsline{toc}{chapter}{Acknowledgments}

This thesis marks the culmination of a transformative journey over the past several years, one made possible through the unwavering support, mentorship, and friendship of many individuals. I take this opportunity to express my deepest gratitude to those who have shaped, supported, and walked beside me along the way.

First and foremost, I thank my advisor, Prof. Dr. Onur Mutlu, for his invaluable guidance and support throughout my PhD. His deep insight, boundless enthusiasm, and high expectations continually pushed me to grow—not only as a researcher but also as a thinker and problem solver. I am immensely grateful for the opportunities he has given me and for believing in my potential, especially at times when I struggled to do so myself.

I owe a great debt of gratitude to my close collaborators and mentors, Juan Gómez-Luna and Saugata Ghose. Their technical mentorship and personal support were instrumental to this dissertation. In particular, I am especially thankful to Juan Gómez-Luna. We began our SAFARI journey at nearly the same time, and over the years, we have grown together—as researchers, as colleagues, and as friends. His dedication, patience, and insight were a constant source of strength throughout this journey.

I also extend my sincere thanks to my PhD committee members: Dr. Christian Weis, Dr. Donghyuk Lee, Prof. Dr. Reetuparna Das, and Prof. Dr. Tony Nowatzki. I am grateful for their time, constructive feedback, and thoughtful questions that helped improve this dissertation and refine my research.

The SAFARI Research Group provided a unique and intellectually stimulating environment that shaped my growth as a researcher. I thank all current and former members for their friendship, support, and countless discussions. I am especially thankful to Nika Mansouri Ghiasi, who welcomed me to Zürich with open arms and a generous heart, easing my transition to a new country and culture. My deepest thanks to Abdullah Giray Yağlıkçı and Can Fırtına—my PhD brothers—who embarked on this path with me and finished alongside me. Their camaraderie, encouragement, and daily support—including our shared hunt for chicken at the university mensa—remain some of the most cherished memories of my PhD. I am deeply proud of what we have accomplished together and forever grateful for their presence. I also thank Minesh Patel for his wise, calm mentorship and his quiet but steady friendship, which always offered clarity and perspective when I needed it most. I am further grateful to Ataberk Olgun, Ismail E. Yüksel, and Nisa Bostancı for their support and for enriching my time in SAFARI.

Beyond the lab, I was fortunate to have met incredible friends who became my chosen family. I thank Vesna, Marcella, Tomer, Nathalia, Barbara, António, Bruna, Larissa, Pedro, Betül, and Çiçek. Each of you left a mark on this journey, and your presence made every challenge more bearable and every joy more meaningful. Vesna, my best friend, has been a pillar of light, love, and unwavering support. Her wisdom and strength have lifted me countless times. Switzerland will never be the same without Marcella, whose companionship and care carried me through both the best and hardest times. I am forever grateful for her generosity and friendship. And Tom--thanks for the memories.

To my coaches and friends at CrossFit Zürich—thank you for keeping me grounded and helping me maintain physical and mental resilience. More than sport, you taught me that consistency breeds strength, and your unwavering positivity gave me the motivation to keep going.

I am also deeply grateful for the early mentorship I received in Brazil, which laid the foundation for this academic journey. I thank Prof. Ricardo dos Santos Ferreira and Prof. Luigi Carro, who first sparked my curiosity about computer architecture and patiently taught me the fundamentals. Their encouragement was critical to my decision to pursue graduate studies abroad. I acknowledge with sincere gratitude the public policies implemented under the administration of President Dilma Rousseff, which enabled access to international educational opportunities for students from all backgrounds. It was through these initiatives that I was able to study abroad in the United States, learn English, and broaden my understanding of the world beyond Brazil—despite coming from a very simple family. This opportunity fundamentally transformed my life and academic path.

Finally, I owe everything to my parents, Simone and Geraldo. Their love, support, and sacrifices form the foundation of everything I have achieved. Their belief in me, even across continents and time zones, gave me the courage to persevere and pursue my dreams.

\clearpage
\chapter*{\vspace{-2em}Abstract\vspace{-1em}}
\addcontentsline{toc}{chapter}{Abstract}

\omt{3}{Continuously} \gft{2}{increasing data intensiveness of modern applications} \omt{3}{has} led to high performance and energy costs for \omt{2}{data movement} in traditional processor-centric computing systems. 
To mitigate these costs, the \gls{PIM} paradigm moves computation closer to where the data resides, reducing (and sometimes eliminating) the need to move data between memory and the processor. 
There are two main approaches to \gls{PIM}: 
\li~\gls{PnM}, where \gls{PIM} logic is added to the same die as memory or to the logic layer of 3D-stacked memory, and 
\lii~\gls{PuM}, which uses the operational principles of memory cells and memory circuitry to perform computation.
Many works from academia and industry have shown the benefits of \gls{PnM} and \gls{PuM} for
a wide range of workloads from different domains. 
However, fully adopting \gls{PIM} in commercial systems is still very challenging due to the lack of tools \omt{2}{as well as programming and} system support for \gls{PIM} architectures across the computer architecture stack, which includes: 
\li~workload characterization methodologies and benchmark suites targeting \gls{PIM} architectures; 
\lii~execution \omt{2}{and programming} models that can take advantage of the available application parallelism to maximize hardware utilization and throughput; 
\liii~compiler support and compiler optimizations targeting \gls{PuM} architectures;  
\liv~data-aware runtime systems that can leverage key characteristics of data to improve overall \gft{2}{\gls{PIM}} performance and energy efficiency. 

Our \emph{goal} in this dissertation is to provide tools\omt{2}{, programming models,} and system support for \gls{PIM} architectures (with a focus on DRAM-based solutions), to ease the adoption of \omt{2}{PIM} in current and future systems.  
To this end, we \omt{2}{make at least \emph{four} new major contributions.} 

\omt{2}{First, we} introduce \DAMOV, the first rigorous methodology to characterize memory-related data movement bottlenecks in modern workloads, and the first data movement benchmark suite.
In \DAMOV, we 
\li~perform the first large-scale characterization of \omt{3}{hundreds of} applications\omt{2}{,}
\omt{2}{\lii~}develop a three-step workload characterization methodology that introduces and evaluates four key metrics to identify the sources of data movement bottlenecks in real applications, and 
\omt{2}{\liii}~study whether \gls{PIM} architectures are a viable data movement mitigation mechanism for different classes of memory-bottlenecked applications.

Second, \omt{2}{we introduce} \MIMDRAM, \omt{2}{a new hardware/software co-designed \gft{2}{substrate} that addresses the major} current \omt{2}{programmability and flexibility} limitations of the bulk bitwise execution model of \gls{PuD} architectures. 
\MIMDRAM \gft{2}{enables the allocation and control of} only the needed computing resources inside DRAM for \gls{PuD} computing. On the hardware side, \MIMDRAM introduces simple modifications to the DRAM architecture that enable the execution of 
\li~different \gls{PuD} operations concurrently inside a single DRAM subarray in a \gls{MIMD} fashion, and 
\lii~native vector reduction computation.
On the software side, \MIMDRAM implements a series of compiler passes that automatically identify and map code regions to the underlying \gls{PuD} substrate, alongside system support for data mapping and allocation of \gls{PuD} memory objects.

Third, \omt{2}{we introduce} \Proteus, \omt{2}{the first hardware framework that addresses} the high \omt{3}{execution} latency \omt{3}{of} bulk bitwise \gls{PuD} operations in state-of-the-art \gls{PuD} architectures by implementing a data-aware runtime engine for \gls{PuD}. 
\Proteus~reduces the latency of \gls{PuD} operations \gft{3}{in three different \omt{3}{ways:}} 
\li~\gft{3}{\Proteus~\emph{concurrently executes}} \gft{2}{independent in-DRAM} primitives \gft{3}{belong to a \emph{single}} \gls{PuD} operation \gft{2}{across DRAM arrays}\gft{3}{.}  
\lii~\gft{3}{\Proteus~\emph{dynamically}} \gft{3}{reduces} the bit-precision \gft{2}{(and consequentially the latency and energy consumption) of} \gls{PuD} operations \omt{3}{by} exploiting narrow values (i.e., values with many leading zeros or ones)\gft{3}{.}
\liii~\gft{3}{\emph{\Proteus~chooses and uses} the most appropriate data representation and arithmetic algorithm implementation for a given \gls{PuD} instruction \emph{transparently} \omt{3}{to} the programmer.}

Fourth, \omt{2}{we introduce} \DaPPA~\gft{2}{(\underline{da}ta-\underline{p}arallel~\underline{p}rocessing-in-memory \underline{a}rchitecture)}, \omt{2}{a new programming framework that eases} programmability for general-purpose \gls{PnM} \gft{3}{architectures} by allowing the programmer to write efficient \gls{PIM}-friendly code without the need to manage hardware resources explicitly. The key idea behind \DaPPA is to remove the responsibility of managing hardware resources from the programmer by providing an intuitive data-parallel pattern-based programming interface that abstracts the hardware components of the \gls{PnM} system. 
Using this key idea, \DaPPA transforms a data-parallel pattern-based application code into the appropriate \gls{PnM}-target code, including the required \glspl{API} for data management and code \omt{3}{partitioning}, which can then be compiled into a \gls{PnM}-based binary transparently from the programmer.

\omt{2}{Overall, our four major contributions demonstrate that we can \emph{effectively} exploit the inherent parallelism of \gls{PIM} architectures and \emph{facilitate} their adoption across a broad spectrum of workloads through end-to-end design of hardware and software support \gft{3}{(i.e., workload characterization methodologies and benchmark suites, execution and programming models, compiler support and programming frameworks, and adaptive data-aware runtime mechanisms)} for PIM, thereby enabling orders of magnitude improvements in performance and energy efficiency \omt{3}{across a \omt{3}{wide} variety of modern workloads}.}

\clearpage
\vspace{-3em}\chapter*{Zusammenfassung}
\addcontentsline{toc}{chapter}{Zusammenfassung}

Die stetig zunehmende Datenintensität moderner Anwendungen hat in herkömmlichen prozessorzentrierten Rechensystemen zu hohen Leistungs- und Energiekosten für die Datenbewegung geführt. 
Zur Minderung dieser Kosten verlagert das \emph{Processing-in-Memory} (PIM)-Paradigma die Berechnung näher an den Ort der Daten, wodurch die Notwendigkeit, Daten zwischen Speicher und Prozessor zu bewegen, verringert (und mitunter ganz eliminiert) wird. 
Es gibt zwei Hauptansätze für PIM: 
\li~\emph{Processing-Near-Memory} (PNM), bei dem PIM-Logik auf demselben Chip wie der Speicher oder in der Logikschicht von 3D-gestapeltem Speicher integriert wird, und 
\lii~\emph{Processing-Using-Memory} (PUM), das die Funktionsprinzipien von Speicherzellen und Speicherschaltungen zur Durchführung von Berechnungen nutzt. 
Zahlreiche Arbeiten aus Wissenschaft und Industrie haben die Vorteile von PNM und PUM für eine breite Palette von Workloads aus unterschiedlichen Domänen aufgezeigt. 
Die vollständige Einführung von PIM in kommerziellen Systemen ist jedoch nach wie vor sehr herausfordernd, da es an Werkzeugen sowie an Programmier- und Systemunterstützung für PIM-Architekturen über den gesamten Computerarchitektur-Stack hinweg fehlt, einschließlich: 
\li~Methodiken zur Workload-Charakterisierung und Benchmark-Suiten, die auf PIM-Architekturen abzielen; 
\lii~Ausführungs- und Programmiermodellen, die die vorhandene Anwendungsparallelität ausnutzen, um Hardwareauslastung und Durchsatz zu maximieren; 
\liii~Compilerunterstützung und Compileroptimierungen für PUM-Architekturen; \liv~datenbewussten Laufzeitsystemen, die zentrale Datenmerkmale ausnutzen, um die PIM-Gesamtleistung und Energieeffizienz zu verbessern.

Ziel dieser Dissertation ist es, Werkzeuge, Programmiermodelle und Systemunterstützung für PIM-Architekturen (mit Fokus auf DRAM-basierte Lösungen) bereitzustellen, um die Einführung von PIM in aktuellen und künftigen Systemen zu erleichtern. Zu diesem Zweck leisten wir mindestens vier wesentliche neue Beiträge.

Erstens stellen wir DAMOV vor, die erste rigorose Methodik zur Charakterisierung speicherbezogener Datenbewegungs-Engpässe in modernen Workloads sowie die erste Benchmark-Suite für Datenbewegung. 
In DAMOV
\li~führen wir die erste groß angelegte Charakterisierung von Hunderten von Anwendungen durch, 
\lii~entwickeln eine dreistufige Workload-Charakterisierungsmethodik, die vier zentrale Metriken einführt und bewertet, um die Ursachen von Datenbewegungs-Engpässen in realen Anwendungen zu identifizieren, und \liii~untersuchen, ob PIM-Architekturen ein tauglicher Mechanismus zur Minderung von Datenbewegung für verschiedene Klassen speichergebundener Anwendungen sind.

Zweitens stellen wir MIMDRAM vor, ein neuartiges Hardware/Software-Codesign-Substrat, das die wesentlichen aktuellen Einschränkungen der Programmierbarkeit und Flexibilität des bitweisen Massenausführungsmodells von \emph{Processing-Using-DRAM} (PUD)-Architekturen adressiert. 
MIMDRAM ermöglicht die Allokation und die Steuerung ausschließlich der jeweils benötigten Rechenressourcen innerhalb von DRAM für PUD-Berechnungen. Auf der Hardware-Seite führt MIMDRAM einfache Modifikationen der DRAM-Architektur ein, die 
\li~die gleichzeitige Ausführung verschiedener PUD-Operationen innerhalb eines einzelnen DRAM-Subarrays in \emph{multiple-instruction multiple-data} (MIMD)-Manier und
\lii~native Vektorreduktionsberechnungen ermöglichen. 
Auf der Software-Seite implementiert MIMDRAM eine Reihe von Compiler-Durchläufe, die Codeabschnitte automatisch identifizieren und auf das zugrunde liegende PUD-Substrat abbilden, zusammen mit Systemunterstützung für Datenabbildung und die Allokation von PUD-Speicherobjekten.

Drittens stellen wir \emph{Proteus} vor, das erste Hardware-Framework, das die hohe Ausführungslatenz von Bulk-bitweisen PUD-Operationen in modernen PUD-Architekturen adressiert, indem es eine datenbewusste Runtime-Engine für PUD implementiert. 
\emph{Proteus} reduziert die Latenz von PUD-Operationen auf drei Arten: 
\li~\emph{Proteus} führt unabhängige \emph{In-DRAM-Primitive}, die zu einer einzelnen PUD-Operation gehören, über DRAM-Arrays hinweg gleichzeitig aus. 
\lii~\emph{Proteus} reduziert die Bitpräzision (und damit Latenz und Energieverbrauch) von PUD-Operationen dynamisch, indem es „narrow values“ ausnutzt (d. h. Werte mit vielen führenden Nullen oder Einsen). 
\liii~\emph{Proteus} wählt und verwendet für eine gegebene PUD-Instruktion die jeweils geeignetste Datenrepräsentation und Implementierung des arithmetischen Algorithmus – für die Programmierenden transparent.

Viertens stellen wir DaPPA (\emph{\underline{da}ta-\underline{p}arallel \underline{p}rocessing-in-memory \underline{a}rchitecture}) vor, ein neues Programmier-Framework, das die Programmierbarkeit für allgemeine PNM-Architekturen erleichtert, indem es Programmierenden erlaubt, effizienten PIM-freundlichen Code zu schreiben, ohne Hardware-Ressourcen explizit verwalten zu müssen. Der zentrale Gedanke von DaPPA ist es, die Verantwortung für das Ressourcenmanagement von den Programmierenden zu nehmen, indem eine intuitive, datenparallele, musterbasierte Programmierschnittstelle bereitgestellt wird, die die Hardwarekomponenten des PNM-Systems abstrahiert. Auf Basis dieses Kerngedankens transformiert DaPPA einen datenparallelen, musterbasierenden Anwendungscode in den geeigneten PNM-Zielcode – einschließlich der notwendigen \emph{Application Programming Interfaces} (APIs) für Datenmanagement und Code-Partitionierung –, der anschließend für eine PNM-basierte Binärdatei kompiliert werden kann, ohne dass die Programmierenden dies explizit wahrnehmen.

Insgesamt zeigen unsere vier Hauptbeiträge, dass sich die inhärente Parallelität von PIM-Architekturen effektiv ausnutzen und ihre Einführung über ein breites Spektrum von Workloads hinweg erleichtern lässt – durch ein ganzheitliches End-to-End-Design von Hardware- und Software-Unterstützung (d. h. Workload-Charakterisierungsmethodiken und Benchmark-Suiten, Ausführungs- und Programmiermodelle, Compiler-Unterstützung und Programmier-Frameworks sowie adaptive, datenbewusste Runtime-Mechanismen) für PIM. Dadurch werden Leistungs- und Energieeffizienzsteigerungen um Größenordnungen über eine Vielzahl moderner Workloads hinweg ermöglicht.

\pagestyle{headings}
\cleardoublepage
\tableofcontents
\newpage
\listoffigures
\newpage
\listoftables

% % MAINMATTER
\cleardoublepage
\pagenumbering{arabic}%

\glsresetall

\chapter{Introduction}

\omt{3}{Modern} computing systems, including servers, cloud platforms, \omt{2}{supercomputers,} mobile/embedded devices, and sensor systems, are designed following a \emph{processor-centric} approach, where the system is divided between computation, communication, and storage/memory elements \omt{2}{and computation is performed only in the computation \omt{3}{engines} (\omt{3}{e.g.,} either a host CPU, GPU, \omt{2}{FPGA}, or accelerator)}. 
%In this model, data processing (i.e., computation) is \omt{3}{performed at} the computation \omt{3}{engines} (\omt{3}{e.g.,} either a host CPU, GPU, \omt{2}{FPGA}, or accelerator).
In contrast, data storage (including main memory) and communication units are treated as unintelligent workers that are incapable of computation. 
As a direct result of this processor-centric design model, data needs to be \emph{constantly} moved back and forth between computation and communication/storage units, so that computation can be performed, \omt{2}{which frequently leads} to \emph{data movement bottlenecks} \omt{2}{that greatly reduce performance and increase energy consumption}.  
Data movement bottlenecks can be observed across a wide variety of computing systems and application domains, including machine learning, databases, graph analytics, genome analysis, high-performance computing, security, data manipulation, and a wide variety of mobile and server-class workloads~\membottleneck, directly \omt{2}{and largely degrading} the performance and energy efficiency of modern processor-centric systems since data movement from storage/memory to computing engines incurs long latency and consumes a \omt{3}{large} amount of energy~\cite{hashemi2016accelerating,hashemi2016continuous, ahn2015scalable, PEI, boroumand2018google, boroumand2021google, boroumand2021google_arxiv, keckler2011gpus}.  

High-performance systems have evolved to include mechanisms that aim to alleviate data movement's impact on system performance and energy consumption, such as \omt{2}{large-capacity and multi-level} cache hierarchies\omt{3}{~\cite{wilkes1965slave,jouppi1990improving,hill1988case,sohi1991high,boggs2004microarchitecture,papworth1996tuning,kalla2010power7,sinharoy2011ibm,sinharoy2015ibm,christie1996developing},} aggressive prefetchers\omt{2}{\omt{3}{~\cite{cooksey2002contentsensitive,zhuang2003ahardwarebased,iacobovici2004effective,jnesbit2004acdc,hur2006memory,bera2021pythia,ishii2009access,fu1992stride,chen1995effective,bakhshalipour2018domino,bera2019dspatch,gonzalez1997speculative,ebrahimi2009coordinated,cao1995study,charney1995,roth1999effective,annavaram2001data,ainsworth2016graph,kaushik2021gretch,nilakant2014prefedge,charneyphd,cooksey2002stateless,joseph1997prefetching,srinath2007feedback,ebrahimi2009techniques,collins2001speculative,sadrosadati2018ltrf,roth1998,LipastiSKR95,KarlssonDS00,CollinsSCT02,DBLP:conf/hpca/HuMK03,DBLP:conf/micro/YuHSD15,smith1978sequential,kroft1981lockupfree,lee1987data,chen1992anefficient,smith1992prefetching,dahlgren1995sequential,pierce1996wrongpath,xia1996instruction,ki1997adaptive,luk1998cooperative,zhang2000hardwareonly,mutlu2003runaheadmicro,mutlu2006efficient,mutlu2005techniques,hashemi2016continuous,mutlu2003runahead}}, sophisticated out-of-order execution techniques\omt{3}{~\cite{patt.critical.micro85,patt.hps.micro85,tomasulo.ibmjrd67,kalla2004ibm,tendler2001power4,yeager1996themips,hinton2001themicroarchitecture,kessler1999thealpha}}, and many levels of multi-threading\omt{3}{~\cite{thornton1964parallel,smith1986pipelined,kongetira2005niagara,smith1982architecture,papadopoulos1991multithreading,tullsen1995simultaneous,akkary1998adynamic,roth2001speculative,spracklen2005chip,baghsorkhi_ppopp2012,van2009mlp,joao2012bottleneck,kumar2004single,joao2013utility,Luk01,kalla2004ibm}}}\gft{2}{\omt{3}{. As a result,} most of the \omt{3}{hardware} real estate within a single compute node (e.g., large caches, memory
controllers, interconnects, communication interfaces and associated circuitry, main memory, and solid-state drives) \omt{3}{is dedicated to} structures \omt{3}{that} to handle data movement and storage~\cite{kumar.isscc2009,howard201048core,jowani2010x8664,gillespie2014steam,singh2017zen,mutlu2023memory}}.
\omt{2}{Unfortunately}, such mechanisms not only come with significant hardware cost and complexity, but also often fail to hide the latency and energy costs of accessing off-chip main memory--which consists mostly of \gls{DRAM}~\omt{2}{\cite{Dennard68field,dennard1974design,markoff2019ibms,electronics2018memory,mutlu2013memory}}, the \emph{de-facto} main memory technology--in many modern and emerging applications~\omt{2}{\cite{boroumand2018google, kanev_isca2015,
jia2016understanding, 
tsai:micro:2018:ams, sites1996,mutlu2013memory}}. 
\omt{2}{One primary reason behind the ineffectiveness of such processor-centric data movement mitigation mechanisms is that modern} applications' memory behavior can differ significantly from more traditional applications \omt{2}{from decades ago (for which many of the aforementioned mechanisms were designed)} since modern applications often have lower memory locality, more irregular access patterns, and \omt{2}{much} larger working sets~\cite{ghose.sigmetrics20, ghose.ibmjrd19,ahn2015scalable,seshadri2015gather,nai2017graphpim,hsieh2016accelerating,ebrahimi2009techniques,mutlu2003runahead, hashemi2016continuous,kim2018grim,cali2020genasm,amiraliphd}.
For example, recent works~\cite{boroumand2021google,boroumand2021google_arxiv,boroumand2018google} show that
\li~more than 62\% of the entire system energy of a mobile device is spent on data movement between the processor and the memory
hierarchy for widely-used mobile workloads~\cite{boroumand2018google}; and
\lii~more than 90\% of the entire system energy is spent on memory when executing large commercial edge neural network models on modern edge machine learning accelerators~\cite{boroumand2021google,boroumand2021google_arxiv}.
At the same time, as \omt{2}{memory manufacturing process} technology \omt{2}{node continues} to shrink, further scaling memory capacity, energy, cost, and performance has become increasingly difficult--especially in the past decade--leading to new reliability and robustness challenges such as DRAM read disturbance~\memscalingissue\omt{3}{, including RowHammer~\cite{kim2014flipping} and RowPress~\cite{luo2023rowpress}, and reduced data retention capabilities~\cite{liu2011flikker,lin2012secret,baek2013refresh,bhati2013coordinated,nair2013acase,arahmati2014refreshing,bhati2014scalable,cui2014dtail,han2014dataaware,jung2014optimized,nair2014refresh,zhang2014cream,bhati2015flexible,jung2015omitting,bhati2016dram_refresh,kotra2017hardwaresoftware,nguyen2018nonblocking,wang2018content,pan2019hiding,pan2019thecolored,choi2020reducing,kim2020chargeaware,qureshi2015avatar,liu2012raidr,yauglikcci2022hira,chou2015reducing,chang2014improving,das2018vrl,mukundan.isca13,ghosh2007smart,takemura2007longretentiontime,kong2008analysis,kim2011characterization,kim2011study,bacchini2014characterization,bacchini2014total,wang2014proactivedram,wang2015radar,weis2015retention}}.
%Thus, the cost of data movement is a \emph{fundamental} issue with the \emph{processor-centric} nature of contemporary computer systems.

One promising solution to fundamentally mitigate \omt{2}{(and even eliminate)} data movement bottlenecks in modern and emerging applications is to move from a \emph{processor-centric} \omt{3}{design paradigm, where computation is performed \emph{only} in processing-using,} to a \emph{memory-centric} design paradigm, where computation is performed where data resides. 
A primary example of \omt{2}{the} memory-centric system \omt2{design paradigm} is 
\emph{\gls{PIM}}~\pim, where the cost of data movement to/from storage/memory is reduced by placing computation capability \omt{2}{inside or} close to memory \omt{2}{structures}. 
In \gls{PIM}, the computational logic \omt{2}{inside or} close to memory \omt{2}{structures} has access to data that resides in memory arrays with significantly higher memory bandwidth, lower latency, and lower energy consumption than the CPU (or \omt{2}{GPU, FPGA, or any other} processor-centric accelerator) has in existing systems.

There are two main approaches to \gls{PIM}~\pimdef.
The first approach, called \emph{\gls{PnM}}~\pnm, 
exploits the ability to implement a wide variety of processing logic (i.e., \omt{2}{computation} capabilities) \omt{2}{\emph{near}} the memory arrays (e.g., in a DRAM chip, next to each memory bank or subarray,  at the logic layer of 3D-stacked
memory~\cite{HMC2, HBM,lee2016simultaneous}, in the memory controllers\omt{2}{, near cache memory arrays, inside a flash memory chip}) and thus leveraging the high internal bandwidth and low latency
available inside the memory chip (e.g., between the logic layer and the memory layers of
3D-stacked memory).
The second approach, called \emph{\gls{PuM}}~\pum, exploits the existing memory \omt{2}{structures} and the analog operational principles of the memory circuitry to enable (bulk) \omt{2}{computation} operations within the memory \omt{2}{arrays}. 
\omt{3}{On the one hand,} \gls{PnM} designs are often a more general approach to \gls{PIM}, where the logic implemented near the memory arrays can be \omt{2}{powerful and} customized\omt{2}{,} and thus can benefit a wide variety of applications\omt{3}{.
On the other hand,} \gls{PuM} designs 
\li~\emph{fundamentally} \emph{eliminate} data movement by performing computation \emph{in situ} and  
\lii~exploit the large internal bandwidth and \omt{2}{large bit-level and array-level} parallelism available \emph{inside} the memory arrays.
Since both approaches offer different \omt{2}{tradeoffs}, they should be viewed as \emph{complementary} to each other and can be combined to exploit the maximum potential of a \gls{PIM} system~\omt{2}{\cite{mcciede2024,mutlu2020modern}}.
Both \gls{PnM} and \gls{PuM} architectures can be implemented using different memory technologies, including \gls{SRAM}~\srampum, \gls{DRAM}~\drampum, emerging non-volatile \omt{2}{memory}~\nvmpum or NAND flash \omt{2}{memory}~\flashpum.

\section{Problem Definition}
\label{sec:intro:problem}

{Many works from {academia}~\pimacademia and {industry}~\pimindustry have shown the benefits of \gls{PnM} and \gls{PuM} for a wide range of workloads from different domains, including  machine learning~\cite{boroumand2021mitigating,he2020newton,kwon202125,lee2015bssync,boroumand2021google_arxiv,oliveira2022heterogeneous,gomez2023evaluating,cho2020mcdram,wang2019bit,kim2021colonnade,boroumand2021google,peemen2013memory,gao2017tetris,Shafiee2016,eckert2018neural,Chi2016,kang2021s,wang2022memcore,lue2019optimal,choi2020flash,shahroodi2023swordfish,oliveira2022accelerating,giannoula2024pygim,imani2019floatpim,liu2018processing,min2019neuralhmc,samsunghc23,park2024lpddr,samsunghc23,zhou2022transpim,deng2018dracc,deng2019lacc,si2019dual,lee20223d,NIM}, genome analytics~\cite{shahroodi2023swordfish,ghiasi2022genstore,cali2020genasm,mao2022genpip,liu20173d,diab2023framework,nag2019gencache}, databases~\cite{oliveira2022heterogeneous,RVU,boroumand2022polynesia,lee2022improving,caminal2022accelerating,Xi_2015,babarinsa2015jafar}, graph analytics~\cite{nai2015instruction,ahn2015scalable,PEI,zhuo2019graphq,angizi2019graphide,matam2019graphssd,angizi2019graphs,zhang2018graphp,song2018graphr,nai2017graphpim,dai2018graphh,huang2020heterogeneous,besta2021sisa}, high-performance computing~\cite{kim2015understanding}, and a wide variety of mobile~\cite{boroumand2018google,boroumand2021mitigating,jo2023nerpim,sun2023efficient} and server-class~\cite{teguia2024vpim,chen2022pimcloud} workloads. 
These past \gls{PIM} approaches improve system performance and energy efficiency by \omt{3}{reducing} costly data movement between memory and compute units while exploiting the massive internal parallelism inherent in modern memory architectures. 
However, adopting \gls{PIM} as a mainstream architecture \emph{holistically}, i.e., in a seamless manner
that does \emph{not} place a heavy burden on the vast majority of programmers, and \emph{efficiently}, i.e., where \gls{PIM} drawbacks \gft{3}{(such as limited computation power, stringent design constraints, and data layout restriction)} can be avoided while its benefits \gft{3}{(such as access to data with low latency, high bandwidth, and low energy cots)} can be enhanced, is still very challenging due to the lack of tools\omt{3}{, programming,} and system support for \gls{PIM} across the computer architecture stack. 

First, there is a lack of \textbf{workload characterization methodologies and benchmark suites} targeting \gls{PIM} architectures, making it difficult to \emph{systematically} identify application-specific data movement bottlenecks that can be potentially mitigated by \gls{PIM}, compare alternative processor-centric and memory-centric data movement mitigation solutions, and quantify performance and energy efficiency gains.
Developing standardized benchmarks and clear metrics is critical to drive adoption in both academic research and industry settings. 
Second, current processor-centric \textbf{execution models} often fail to fully harness the abundant parallelism available in \gls{PIM}-enabled systems.
New, specialized memory-centric execution paradigms need to be designed to map applications effectively onto \gls{PIM} hardware resources, ensuring high \gls{PIM} utilization and throughput. 
Third, \textbf{compiler support} and \textbf{programming frameworks} to aid \gls{PIM} programmability are severely underdeveloped. 
Conventional compiler frameworks are \emph{not} designed to 
\li~generate code that effectively leverages the main characteristics of \gls{PIM} systems, such as abundant \omt{3}{bit-level and array-level} memory and compute parallelism, and 
\lii~schedule and orchestrate \gls{PIM} operations in ways that can maximize data locality and minimize data movement.
Consequently, effective programming of \gls{PIM} systems frequently places significant demands on the programmer, who must acquire an in-depth understanding of the underlying hardware substrate to accurately and efficiently map the target application onto the \gls{PIM} architecture.
Fourth, \textbf{adaptive data-aware runtime} mechanisms are essential to fully harness the benefits of \gls{PIM} architectures.
Besides adopting a \emph{data-centric} design (by co-locating computation near/in memory), an efficient \gls{PIM} system should also be \emph{data-aware}, where the intrinsic properties of the data are used to make informed decisions during execution. 
By dynamically adapting \gls{PIM} resources and execution parameters based on the observed data characteristics at runtime, such data-centric and \omt{3}{a} data-aware \gls{PIM} system can significantly reduce latency and improve overall system performance, throughput, and energy efficiency. 
The dynamic nature of data requires the development of dynamic runtime mechanisms, to adapt \gls{PIM} execution to the characteristics of the target workload (and its dataset).

%Addressing these challenges in a holistic manner--through unified tool-chains, standardized benchmarking, specialized compiler optimizations, and intelligent runtime management--is key to unlocking the full potential of \gls{PIM} architectures. 
%A concerted effort from both academia and industry is required to build a robust ecosystem around \gls{PIM}, fostering innovation and enabling its integration into mainstream computing systems.

\section{Our Goal}
\label{sec:intro:goal}

Our \emph{goal} in this dissertation is to provide tools\omt{3}{, programming,} and system support for \gls{PIM} architectures (with a focus on DRAM-based solutions), to ease the adoption of such architectures in current and future systems.

\section{Thesis Statement}
\label{sec:intro:thesis}

The following thesis statement encompasses our approach:

\begin{center}
\parbox{15cm}{\textit{We can effectively exploit the inherent parallelism of \gls{PIM} architectures and facilitate their adoption across a broad spectrum of workloads through end-to-end design of hardware and software support for PIM\omt{3}{,} including benchmark suites \omt{3}{and workload analysis methodologies}, compiler/programming frameworks, and data-aware runtime systems\omt{3}{,} \omt{3}{thereby enabling large (e.g., factors or orders of magnitude) improvements in performance and energy efficiency.}}}
\end{center}

\section{Our Approach}

In line with our thesis statement, we investigate \gft{3}{four} major directions that enable us to ease
\gls{PIM} adoption and effectiveness in current and future systems: 
\li~identifying and characterizing potential sources of data movement bottlenecks over a broad set of applications,
\lii~providing \gft{3}{programming/compilation and} system support for end-to-end \gft{3}{\gls{PuM}} execution \gft{3}{while maximizing hardware utilization and attained throughput}, 
\gft{3}{\liii~enabling data-aware optimizations for efficient \gls{PuM} execution,} and
\gft{3}{\liv}~designing programming frameworks to aid \gls{PIM} programmability. 
Toward these \gft{3}{four} major directions, we make four key contributions:

\subsection{\gft{3}{A Methodology and Benchmark Suite for Understanding and Mitigating Data Movement Bottlenecks}}

The \emph{goals} of our first \omt{3}{contribution} are twofold. 
First, we aim to methodically identify potential sources of data movement bottlenecks over a broad set of applications and to comprehensively compare traditional processor-centric data movement mitigation techniques (e.g., caching and prefetching) to memory-centric techniques (e.g., \gls{PIM}), thereby developing a rigorous understanding of the best techniques to mitigate each source of data movement.
Second, we aim to develop a benchmark suite for data movement that captures the previously-characterized sources of data movement bottlenecks across many applications.

With these goals in mind, we perform the first large-scale characterization of a wide variety of applications, across a wide range of application domains, to identify \emph{fundamental} program properties that lead to data movement to/from main memory. 
We develop the first systematic methodology to classify applications based on the sources contributing to data movement bottlenecks. 
From our large-scale characterization of 77K functions across 345 applications, we select 144 functions to form the first open-source benchmark suite for main memory data movement studies (called  DAMOV). 
We select a diverse range of functions that \li~represent different types of data movement bottlenecks, and 
\lii~come from a wide range of application domains. 
We identify new insights about the different data movement bottlenecks and use these insights to determine the most suitable data movement mitigation mechanism (from processor-centric to memory-centric solutions) for a particular application.
We show how our DAMOV benchmark suite can aid the study of open research problems for \gls{PIM} architectures, via four case studies, where we evaluate: 
\li~the \omt{3}{performance} impact of load balance and inter-\gls{PIM} communication in \gls{PIM} systems, \lii~the impact of \gls{PIM} accelerators on our memory \gft{3}{workload} analysis, 
\liii~the impact of different core models on \gls{PIM} architecture \omt{3}{performance}, and 
\liv~the potential benefits of identifying  simple \gls{PIM} instructions \gft{3}{for offloading}. 
We conclude that our benchmark suite and methodology can be employed to address many different open research and development questions on data movement mitigation mechanisms, particularly topics related to \gls{PIM} systems and architectures.
We open-source DAMOV and the complete source code for our new characterization methodology at~\url{https://github.com/CMU-SAFARI/DAMOV}.
Chapter~\ref{chap:damov} describes this work in more \omt{3}{detail}.

\subsection{\gft{3}{Enabling Efficient and Programmable MIMD Execution in \omt{4}{Processing-Using-DRAM (PUD)} Architectures}}

Our second \omt{3}{contribution} aims to design a flexible \gls{PuD} system that overcomes the limitations caused by the large and rigid granularity of \gls{PuD} \omt{3}{execution}.
\gls{PuD} is a \gls{PIM} approach that uses a DRAM array's massive internal parallelism to execute very-wide (e.g., 16,384--262,144-bit-wide)} data-parallel operations, in a \gls{SIMD} fashion. 
However, DRAM rows' large and rigid granularity limit the effectiveness and applicability
of \gls{PuD} in three ways. 
First, since applications have varying degrees of \gls{SIMD} parallelism (which is often smaller than the DRAM row granularity), \gls{PuD}  execution often leads
to under-utilization, throughput loss, and energy waste. 
Second, due to the high area cost of implementing interconnects that connect columns in a wide DRAM row, most \gls{PuD}  architectures are limited to the execution of parallel \omt{3}{\emph{map}} operations, where a single operation is performed over equally-sized input and output arrays.
Third, the need to feed the wide DRAM row with tens of thousands of data elements combined with the lack of adequate compiler support for \gls{PuD} systems create a programmability barrier, since programmers need to {manually} extract \gls{SIMD} parallelism from an application and map computation to the  \gls{PuD} hardware.

\omt{3}{To tackle the three limitations caused by the large and rigid granularity of \gls{PuD} execution}, we propose MIMDRAM, a hardware/software co-designed \gls{PuD} system that introduces new mechanisms to allocate and control {only} the necessary resources for a given \gls{PuD} operation. 
The \emph{key idea} of MIMDRAM is to leverage fine-grained DRAM~\omt{3}{\cite{cooper2010fine,udipi2010rethinking,zhang2014half,ha2016improving,lee2017partial,olgun2022sectored,o2021energy,oconnor2017fine,olgun2024sectored}} (i.e., the ability to independently access smaller segments of a large DRAM row) for \gls{PuD} computation. MIMDRAM exploits this key idea to enable a \gls{MIMD} execution model in each DRAM subarray (and \gls{SIMD} execution within each DRAM row segment).
MIMDRAM leverages fine-grained DRAM for \gls{PuD} in hardware and software. 
On the hardware side, MIMDRAM proposes simple modifications to the DRAM subarray and includes new mechanisms \omt{3}{in} the memory controller that 
\li~allow the execution of independent \gls{PuD} operations across the DRAM mats in a single subarray; and 
\lii~enable communication across columns of a DRAM row at varying granularities for the execution of vector-to-scalar reduction in DRAM at low hardware cost. 
On the software side, MIMDRAM implements compiler passes to
\li~automatically vectorize code regions that can benefit from \gls{PuD} execution (called \gls{PuD}-friendly regions);
\lii~for such regions, generate \gls{PuD} operations with the most appropriate \gls{SIMD} granularity; and
\liii~schedule the concurrent execution of independent \gls{PuD} operations in different DRAM \omt{3}{row segments}. 
We discuss how to integrate MIMDRAM in a real system, including how MIMDRAM deals with 
\li~data allocation within a DRAM subarray and 
\lii~mapping of a \gls{PuD}'s operands to guarantee high utilization of the \gls{PuD} substrate.

We evaluate MIMDRAM using twelve real-world applications and 495 multi-programmed application mixes.
\gft{3}{When using 64 DRAM subarrays per bank and 16
banks for PuD computation in a DRAM chip, MIMDRAM provides 
\li~13.2$\times$/0.22$\times$/173$\times$ the performance,
\lii~0.0017$\times$/0.00007$\times$/0.004$\times$ the energy consumption, 
\liii~582.4$\times$/13612$\times$/272$\times$ the performance per Watt of the CPU~\cite{intelskylake}/GPU~\cite{a100}/SIMDRAM~\cite{hajinazarsimdram} (the prior state-of-the-art \gls{PuD} framework) baselines and
\liv~when using a single DRAM subarray, 15.6$\times$ the \gls{SIMD}
utilization of SIMDRAM.} 
%
%Our evaluation shows that MIMDRAM  provides 34$\times$ the performance, 14.3$\times$ the energy efficiency, 1.7$\times$ the throughput, and 1.3$\times$ the fairness of a state-of-the-art \gls{PuD} framework, along with
%30.6$\times$ and 6.8$\times$ the energy efficiency of a high-end CPU and GPU, respectively. 
MIMDRAM adds small area cost to a DRAM chip (1.11\%) and CPU die (0.6\%). 
We open-source MIMDRAM workloads and evaluation framework at \url{https://github.com/CMU-SAFARI/MIMDRAM}.
Chapter~\ref{chap:mimdram} describes this work in more \omt{3}{detail}.

\subsection{\gft{3}{Enabling High-Performance \gls{PuD} Execution via Dynamic Precision \gls{PuD} Arithmetic}}

Our third \omt{3}{contribution} aims to overcome three \omt{3}{major} limitations of \gls{PuD} architectures caused by \omt{3}{the naive} \omt{3}{use of} a bit-serial execution model \omt{3}{in such architectures}.
While \gls{PuD} promises high throughput at low energy and area cost, we uncover three limitations of existing \gls{PuD} approaches that lead to significant \omt{3}{performance and energy} inefficiencies. 
\gft{3}{First, they employ a \textbf{rigid and static data representation}, which is \emph{inefficient} for bit-serial execution.}
\gft{3}{Existing \gls{PuD} engines typically employ a fixed bit-precision, statically defined data representation (commonly two's complement) for all \gls{PuD} operations. 
This rigid and static data format introduces inefficiencies in a bit-serial execution model, where bits of a data word are \emph{individually} and \emph{sequentially} processed. 
Since many applications store data in data representation formats that exceed the necessary precision~\cite{pekhimenko2012base,alameldeen2004adaptive,islam2010characterization,ergin2006exploiting,brooks1999dynamically,ergin2004register,budiu2000bitvalue,wilson1999case} (e.g., 8-bit values stored in a 32-bit integer), a significant fraction of \gls{PuD} computation is wasted on inconsequential bits, such as leading zeros or sign-extended bits, causing significant latency and energy overhead.
Second, \gls{PuD} architectures \emph{only} favor a \textbf{throughput-oriented execution with limited latency tolerance}.
\gls{PuD} operations are composed of high-latency in-DRAM primitives (e.g., in-DRAM row copy~\cite{seshadri2013rowclone} and in-DRAM majority-of-three~\cite{seshadri2017ambit}), making individual \gls{PuD} operations inherently slow. 
To compensate for this latency, \gls{PuD} architectures adopt a \emph{throughput-oriented execution model} that distributes large amounts of data across multiple DRAM subarrays and DRAM banks, enabling a massively parallel execution.
However, this approach is \emph{only} effective when sufficient data-level parallelism is available to amortize the high latency of individual \gls{PuD} primitives. 
In scenarios where data-level parallelism is limited, this throughput-oriented execution model fails to hide the latency of individual in-DRAM primitives, potentially leading to performance degradation. 
Third, bit-serial \gls{PuD} architectures face \textbf{scalability challenges for high-precision operations}. 
\gls{PuD} systems suffer from increased latency as the target bit-precision grows. 
Due to their bit-serial nature, the latency of arithmetic \gls{PuD} operations scales linearly~\cite{hajinazarsimdram} (for operations such as addition/subtraction) or quadratically~\cite{hajinazarsimdram} (for operations such as multiplication/division) with the target bit-precision.
This scaling behavior arises from the inherently serial structure of these operations, which often requires long carry-propagation chains across the individual bits of a data word. As a result, high-precision computations (e.g., beyond 32-bit) become prohibitively slow.}

% \li~static data representation, i.e., 2's complement with fixed bit-precision, leading to \emph{unnecessary computation} over useless (i.e., inconsequential) data; 
% \lii~support for \emph{only} throughput-oriented execution, where the high latency of individual \gls{PuD} operations can \emph{only} be hidden in the presence of bulk data-level parallelism; and 
% \liii~high latency for high-precision (e.g., \omt{3}{greater than} 32-bit) operations. 

To address these issues, we propose \emph{Proteus}, an adaptive \emph{data-representation} and \emph{operation-implementation} framework, which builds on \gft{3}{three} \emph{key ideas}. 
\gft{3}{To solve the \textbf{first limitation} (i.e., rigid and static data representation), \emph{Proteus} \emph{reduces} the bit-precision for \gls{PuD} operations by leveraging \emph{narrow values} (i.e., values with many leading zeros).
\gft{3}{As several works observe~\cite{pekhimenko2012base,alameldeen2004adaptive,islam2010characterization,ergin2006exploiting,brooks1999dynamically,ergin2004register,budiu2000bitvalue,wilson1999case}, programmers often over-provision the bit-precision used to store operands, using large data types (e.g., a 32-bit or 64-bit integer) to store small (i.e., narrow) values. 
Based on this observation, \emph{Proteus} can exploit narrow values to reduce the bit-precision of a \gls{PuD} operation to that of the best-fitting number of bits; thus performing costly in-DRAM operations \emph{only} over consequential bits, which improves overall performance and energy efficiency.}
To solve the \textbf{second limitation} (i.e., throughput-oriented execution with limited latency tolerance), \emph{Proteus} \emph{parallelizes} the execution of \emph{independent} in-DRAM primitives in a \gls{PuD} operation by leveraging DRAM's internal organization combined with \emph{bit-level parallelism}.
\gft{3}{We make the key observation that many in-DRAM primitives that compose a \gls{PuD} operation (e.g., an in-DRAM addition) can be executed \emph{concurrently} across different bits of a data word.
For example, executing an $n$-bit in-DRAM addition (i.e., ${A_{n-1}, \dots, A_0} + {B_{n-1}, \dots, B_0}$) in a bit-serial manner requires performing at least three majority-of-three (\texttt{MAJ3}) operations per bit $i$ to compute the $sum$ and propagate the carry to bit $i{+}1$. However, only one of these operation (i.e., the carry propagation from bit $i$ to bit $i+1$) needs to be serialized across the $n$ bit positions of a data word, while the other two \texttt{MAJ3} operations for bit positions $i$ and $i+1$ can be \emph{concurrently executed}.
To exploit this observation, \emph{Proteus} scatters the $n$ bits of a data word across multiple DRAM subarrays (i.e., $subarray_i \leftarrow  \{A_i,B_i\}$) and employs \gls{SLP}~\cite{kim2012case} to enable each subarray $i$  to \emph{concurrently} execute the in-DRAM primitive associated with its respective bit $i$, thus hiding the high latency of individual in-DRAM primitives in a \gls{PuD} operation over the many bits of the target data word.}
To solve the \textbf{third limitation} (i.e., scalability challenges for high-precision operations), \emph{Proteus} exploit an alternative data representation for high-precision computation. Concretely, we investigate an alternative data representation, i.e., the \gls{RBR}~\cite{guest1980truth,phatak1994hybrid,lapointe1993systematic, olivares2006sad, olivares2004minimum} (where multiple-digit combinations represent the same value), for high-precision computation. 
\gls{PuD} execution can
take advantage of two properties of \gls{RBR}-based arithmetic: 
\li~the operations no longer need to propagate carry bits through the full width of the data (e.g., \gls{RBR}-based addition limits carry propagation to at most two places~\cite{brown2002using}), and 
\liii~the operation latency is \emph{independent} of the bit-precision.}

Based on these two \omt{3}{three key ideas} ideas, we design \emph{Proteus} as a three-\omt{3}{component} hardware framework for high-performance \gls{PuD} computation that \emph{transparently} \omt{3}{(from the user/programmer)} selects 
\li~the most efficient data format (e.g., 2's complement, redundant binary~\cite{guest1980truth,phatak1994hybrid,lapointe1993systematic}),
\lii~the exact bit-precision for a workload, and 
\liii~the fastest \omt{3}{arithmetic} algorithm for latency- or throughput-oriented \gls{PuD} execution. 

We compare \emph{Proteus} to different state-of-the-art computing platforms (CPU, GPU, and the SIMDRAM \gls{PuD} architecture) for {twelve} real-world applications. 
Using \omt{3}{only} a single DRAM bank, \emph{Proteus} provides 
\li~17$\times$, 7.3$\times$, and 10.2$\times$ the performance per mm$^2$; and
\lii~90.3$\times$, 21$\times$, and 8.1$\times$ lower energy consumption than that of the CPU, GPU, and SIMDRAM, respectively, on average across twelve real-world applications.
\emph{Proteus} incurs low area cost on top of a DRAM chip (1.6\%) and CPU die (0.03\%).
Chapter~\ref{chap:proteus} describes this work in more \omt{3}{detail}.

\subsection{\gft{3}{A Data-Parallel Programming Framework for Easing Programmability and Enabling High-Performance \gls{PnM} Execution}}

Our fourth \omt{3}{contribution} aims to ease programmability for the \gft{3}{general-purpose \gls{PnM}} \gft{3}{architectures}, allowing a programmer to write efficient PIM-friendly code \emph{without} the need to manage hardware resources explicitly. 
\gft{3}{\omt{3}{We} use the UPMEM PIM architecture~\omt{3}{\cite{upmem,upmem2018,gomez2021benchmarking,gomez2021benchmarkingcut,gomez2022benchmarking}}, the first publicly-available real-world \gls{PIM} architecture, as a case study for the implementation of our programming framework.}
The UPMEM \gls{PIM} architecture is a many-core system consisting of \emph{UPMEM modules}, which are standard DDR4-2400 DIMMs with multithreaded general-purpose in-order processors (called \omt{3}{DRAM Processing Units, i.e.,} DPUs) coupled together with DRAM banks (called \emph{MRAM}). 
A programmer needs to follow four main steps to implement a given application targeting the UPMEM system. The programmer needs to: 
\li~partition the computation (and input data) across the DPUs in the system, \emph{manually} exposing thread-level parallelism (TLP);
\lii~distribute (copy) the appropriate input data from the CPU's main memory into the DPU's memory space;
\liii~launch the computation kernel that the DPUs will execute; and
\liv~gather (copy) output data from the DPUs to the CPU main memory once the DPUs execute the kernel.
Even though UPMEM's programming model resembles that of widely employed architectures, such as GPUs, it requires the programmer to 
\li~have \omt{3}{detailed} knowledge of the underlying UPMEM hardware \omt{3}{and its idiosyncrasy} and
\lii~\omt{3}{\emph{manually}} manage data movement at a fine granularity \omt{3}{(e.g., guarantee that memory accesses are byte-aligned, orchestrating data reading/writing from DPUs' local scratchpad memory)}. 
Such limitations \omt{3}{and the requirement of introduce knowledge of underlying hardware by the programmer can greatly limit the adoption of \gls{PnM} systems as general-purpose systems.}

To \omt{3}{solve this major adoption and programming problem}, we introduce DaPPA (\underline{da}ta \underline{p}arallel~\underline{p}rocessing-in-memory \underline{a}rchitecture), a programming \omt{3}{and compilation} framework that eases the programmability of \gft{3}{general-purpose \omt{3}{\gls{PnM}}} systems by automatically managing data movement, memory allocation, and workload distribution. 
The \emph{key idea} behind DaPPA is to leverage a high-level data-parallel pattern\omt{3}{-based}~\cite{cole1989algorithmic,cole2004bringing} programming interface to abstract hardware complexities away from the programmer. 
DaPPA comprises three main components:
\li~\emph{data-parallel pattern APIs}, a collection of five primary data-parallel pattern primitives that allows the programmer to express data transformations within an application \omt{3}{at an abstract level};
\lii~\emph{dataflow programming interface}, which allows the programmer to define how data moves across data-parallel patterns; and
\liii~a \emph{dynamic template-based compilation}, which leverages code skeletons and dynamic code transformations to convert data-parallel patterns implemented via the dataflow programming interface into an optimized binary \gft{3}{for the target \gls{PnM} architecture}. 

We evaluate DaPPA using six workloads from the PrIM benchmark suite~\cite{gomezluna2021repo} on a real UPMEM system. Compared to hand-tuned implementations, DaPPA improves end-to-end performance by 2.1$\times$, on average, and reduces programming complexity (measured in lines-of-code) by 94\%. Our results demonstrate that DaPPA is an effective programming framework for efficient and user-friendly programming on UPMEM systems.
\omt{3}{DaPPA is also a general \gls{PnM} programming framework that can be deployed for any \gls{PnM} architecture.}
Chapter~\ref{chap:dappa} describes this work in more \omt{3}{detail}.

\section{Contributions}

\omt{3}{In this} dissertation\omt{3}{, we} make the following contributions:

\begin{itemize}[leftmargin=3mm,itemsep=0mm,parsep=0mm,topsep=0mm]
    \item \omt{3}{We propose} the first methodology to characterize data-intensive workloads based on the \omt{3}{sources} of their data movement bottlenecks. This methodology is driven by insights obtained from a large-scale experimental characterization of 345~applications from 37 different benchmark suites and an evaluation of the performance of memory-bound functions from these applications with three data-movement mitigation mechanisms.
    
    \item \omt{3}{We} \omt{3}{introduce} DAMOV, the first open-source benchmark suite for main memory data movement-related studies, based on our systematic characterization methodology. This suite consists of 144~functions representing different sources of data movement bottlenecks and can be used as a baseline benchmark set for future data-movement mitigation research. \omt{3}{In fact, multiple works have already used DAMOV~\cite{garzon2024fasta,kwon2023mccore,iskandar2023ndp,saglam2024data,asifuzzaman2022evaluating,fernandez2020natsa,hwang2025optimizing,iskandar2023auto,papalekas2022near,glint2022fresh,tian2023novel,olgun2022sectored,olgun2024sectored} since its initial release in 2021.} 
    
    \item \omt{4}{We} \omt{3}{show} how \omt{3}{the} DAMOV benchmark suite \omt{3}{and methodology to identify data movement bottlenecks} can aid the study of open research problems for \gls{PIM} architectures, via four case studies. In particular, we evaluate 
    \li~the \omt{3}{performance} impact of load balance and inter-\omt{3}{\gls{PIM}} communication in \gls{PIM} systems, 
    \lii~the impact of \gls{PIM} accelerators on our \omt{3}{workload} analysis, \liii~the impact of different core models on \gls{PIM} architecture \omt{3}{performance}, and 
    \liv~the potential benefits of identifying  simple \gls{PIM} instructions \omt{3}{for offloading}. We conclude that our benchmark suite and methodology can be employed to address many different open research and development questions on data movement mitigation mechanisms, particularly topics related to \gls{PIM} systems and architectures.

    \item \omt{3}{We} \omt{3}{propose} MIMDRAM, an end-to-end processing-using-DRAM (PUD) system for general-purpose applications, which executes operations in a multiple-instruction multiple-data (MIMD) fashion. MIMDRAM makes low-cost modifications to the DRAM subarray design that enable the \emph{concurrent} execution of multiple independent \gls{PuD} operations in a single DRAM subarray. 
    
    \item \omt{3}{We} \omt{3}{propose} compiler passes that take as input unmodified C/C++ applications and, transparently to the programmer, \li~identify loops that are suitable for \gls{PuD} execution, \lii~transform the source code to use \gls{PuD} operations, and \liii~schedule independent \gls{PuD} operations for concurrent execution in each DRAM subarray, maximizing utilization of the underlying \gls{PuD} architecture.

    \item \omt{3}{We} \omt{3}{propose} \emph{Proteus}, \gft{3}{a novel \gls{PuD} framework that overcomes three major inefficiencies in existing bit-serial \gls{PuD} architectures: static data representation, lack of support for latency-oriented execution, and high-latency for high-precision \gls{PuD} operations. \emph{Proteus} exploits DRAM's internal parallelism to hide the latency of individual in-DRAM primitives (i.e., row copy, majority-of-three), and introduces mechanisms to scatter bits of a data word across multiple DRAM subarrays. This bit-level distribution enables \emph{concurrent} execution of independent in-DRAM primitives, and allows \emph{Proteus} to leverage both bit-serial and bit-parallel algorithms for \gls{PuD} arithmetic.}

    \item \omt{3}{We} \omt{3}{introduce a three-component adaptive data-aware runtime mechanism that transparently (from the user/programmer) selects the most suitable data representation, bit-precision, and arithmetic algorithm (bit-serial or bit-parallel) for each \gls{PuD} instruction, enabling high-performance and energy-efficient \gls{PuD} execution.}

    \item \omt{3}{We} \omt{3}{propose} DaPPA (\underline{da}ta-\underline{p}arallel~\underline{p}rocessing-in-memory \underline{a}rchitecture), a data-parallel pattern-based programming framework that abstracts \emph{both} computation and communication requirements while \emph{automatically} generating code for \gft{3}{general-purpose \gls{PIM} architectures. We equip DaPPA with a series of code optimizations that \omt{3}{greatly} improve the performance of workloads running on \gft{3}{general-purpose \gls{PnM}} systems.} 
    
    \item \gft{3}{We demonstrate that DaPPA's data-parallel pattern-based programming significantly improves both performance and programming efficiency for real-world UPMEM-based \gls{PnM} architectures. The results highlight DaPPA's broad applicability to different workloads and indicate that adopting a data-parallel pattern-based, high-level programming approach can lead to performance improvements while minimizing programmers' effort. }
\end{itemize}

\section{Outline}

This dissertation is organized into eight chapters. 

Chapter~\ref{chap:bg} gives relevant background information about DRAM organization, operation, and the \gls{PIM} architectures we build on top of in this dissertation.
Chapter~\ref{chap:related} provides \omt{3}{a comprehensive} overview of related prior work \omt{3}{in 
\li~\gls{PuM} architectures (\secref{chap:related:pum}), including SRAM-based (\secref{chap:related:pum:sram}), DRAM-based (\secref{chap:related:pum:dram}), \gls{NVM}-based (\secref{chap:related:pum:nvm}), and NAND flash-based (\secref{chap:related:pum:flash}) \gls{PuM} architectures; and 
\lii~system support for \gls{PIM} (\secref{chap:related:pimsys}), including data movement bottleneck characterization (\secref{chap:related:pimsys:workload}), \gls{PIM} suitability (\secref{chap:related:pimsys:suitability}), compiler support for \gls{PIM} (\secref{chap:related:pimsys:compiler}), memory management support for \gls{PIM} (\secref{chap:related:pimsys:memory}), and programming frameworks and high-level \glspl{API} for \gls{PIM} (\secref{chap:related:pimsys:programming}).}
Chapter~\ref{chap:damov} presents \omt{3}{DAMOV}, our workload characterization methodology and benchmark suite for data movement bottlenecks.
Chapter~\ref{chap:mimdram} presents the design of MIMDRAM, our end-of-end hardware/software co-designed \gls{PuD} system. 
Chapter~\ref{chap:proteus} presents \emph{Proteus}, our runtime solution for dynamic precision bit-serial \gls{PuD} computation.
Chapter~\ref{chap:dappa} discusses DaPPA, our programming framework that abstracts hardware components away from the programmer, thus facilitating code development for \gft{3}{general-purpose \gls{PnM} architectures}. 
Chapter~\ref{chap:conc} provides a summary of this dissertation, \omt{3}{along with promising} future research directions and concluding remarks. 
 \chapter{Background}
\label{chap:bg}

This chapter provides an overview of the background
material necessary to understand our discussions, analyses, and contributions.
Section~\ref{sec:background_dramorg} reviews DRAM organization and DRAM operation.
Section~\ref{sec:background:pim} provides background material about \gls{PIM} architectures (both \gls{PnM} and \gls{PuM}) relevant for this dissertation.

\section{DRAM Organization \& Operation}
\label{sec:background_dramorg}

\subsection{DRAM Organization} 

\begin{figure}[ht]
    \centering
    \includegraphics[width=\linewidth]{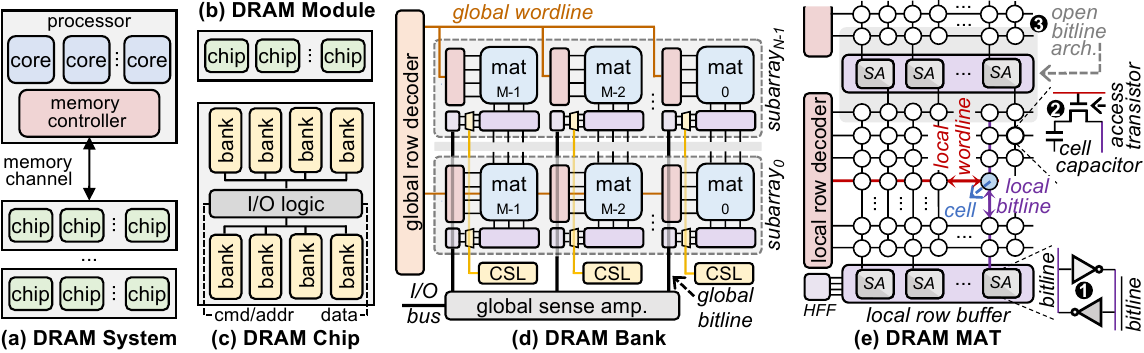}
    \caption{Overview of DRAM organization.}
    \label{fig_subarray_dram}
\end{figure}

A DRAM~\omt{3}{\cite{Dennard68field,dennard1974design}} system \gft{3}{( Figure~\ref{fig_subarray_dram}a)} comprises of a hierarchy of components. 
A \emph{DRAM module} (Figure~\ref{fig_subarray_dram}\gft{3}{b}) has several (e.g., 8--16) DRAM chips. 
A \emph{DRAM chip} (Figure~\ref{fig_subarray_dram}\gft{3}{c}) has multiple DRAM banks (e.g., 8--16). 
A \emph{DRAM bank} (Figure~\ref{fig_subarray_dram}\gft{3}{d}) has multiple (e.g., 64--128) 2D arrays of DRAM cells known as \emph{DRAM mats}. Several DRAM mats (e.g., 8--16) are grouped in a \emph{DRAM subarray}~\omt{3}{\cite{kim2012case,seshadri2013rowclone,Tiered-Latency_LEE,lee2015adaptive,chang2014improving}}. 
In a DRAM bank, there are three global components that are used to access the DRAM mats (as Figure~\ref{fig_subarray_dram}\gft{3}{d} depicts):
\li~a \emph{global row decoder} that selects a row of DRAM cells \emph{across} multiple mats in a subarray,
\lii~a \emph{column select logic} (CSL) that selects portions of the DRAM row based on the column address, and
\liii~a \emph{global sense amplifier} (\omt{3}{also sometimes called} global row buffer) that transfers the selected fraction of the data from the selected DRAM row through the \emph{global bitlines}.

A \emph{DRAM mat} (Figure~\ref{fig_subarray_dram}\gft{3}{e}) consists of a 2D array of DRAM cells organized into multiple \emph{rows} (e.g., 512--1024) and multiple \emph{columns} (e.g., 512--1024)~\cite{kim2018solar, lee2017design, kim2002adaptive}. 
Each mat contains 
\li~a \emph{local row decoder} that drives the local wordlines to the appropriate voltage levels to open (activate) a row, 
\lii~a row of sense amplifiers (also called a \emph{local row buffer}; \gft{3}{\circled{1} in  Figure~\ref{fig_subarray_dram}e}) that \omt{3}{sense} and \omt{3}{latch} data from the activated row, and
\liii~\emph{\glspl{HFF}} that drive a portion (e.g., \SI{4}{\bit}) of the data in the local row buffer to the global bitlines. 
A \emph{DRAM cell} (\gft{3}{\circled{2}}) consists of an \emph{access transistor} and a \emph{storage capacitor}. 
The source nodes of the access transistors of all the DRAM cells in the same column connect the cells' storage capacitors to the same \emph{local bitline}. 
The gate nodes of the access transistors of all the DRAM cells in the same row connect the cells' access transistors to the same \emph{local wordline}. 
To achieve high density, modern DRAM designs employ an \emph{open bitline architecture}~\omt{3}{\cite{lim20121,takahashi2001multigigabit,luo2020clr}}, fitting only enough sense amplifiers in a local row buffer to sense half a row of cells. To sense the entire row of cells, each subarray has local bitlines connecting to two rows of local sense amplifiers--one above and one below the cell array \gft{3}{(\circled{3})}. 
\omt{3}{For more detail on DRAM organization, we refer the reader to prior literature~\drambackground, especially~\cite{seshadri2019dram}.}

\subsection{DRAM Operation} 

\begin{figure}[ht]
    \centering
    \includegraphics[width=\linewidth]{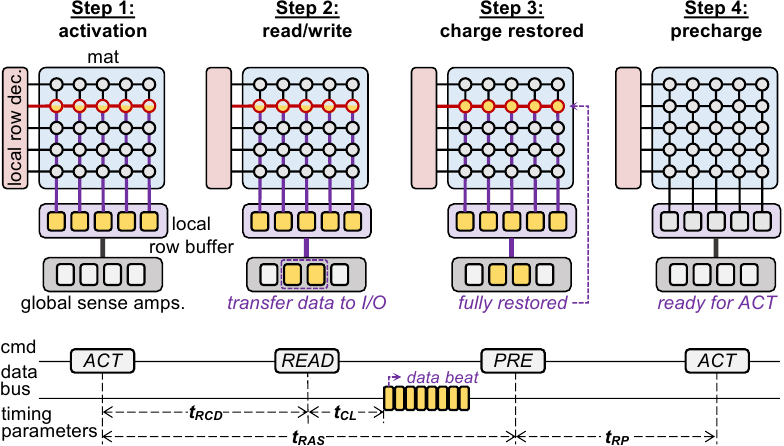}
    \caption{\gft{3}{Overview of DRAM operation. Reproduced from~\cite{chang2017understandingphd}}.}
    \label{fig_subarray_dram_operation}
\end{figure}

The memory controller \gft{3}{(Figure~\ref{fig_subarray_dram}a)} \omt{3}{can issue} three commands to service a DRAM request\gft{3}{, in four main steps: 
\li~\textbf{row activation \& sense amplification}: opening a row to transfer its data to the row buffer,
\lii~\textbf{read/write}: accessing the target column in the row buffer, 
\liii~\textbf{charge restoration:} recharging the capacitor in the DRAM cell to its original voltage level after the stored charge has been partially depleted during the sense amplification, and 
\liv~\textbf{precharge}:
closing the row and the row buffer. 
We use Figure~\ref{fig_subarray_dram_operation} to explain these four steps in detail.
The top part of the figure shows the phase of the cells within the row that is being accessed.
The bottom part shows both the DRAM command and data bus timelines, and demonstrates
the associated timing parameters.}

\paratitle{Initial State} \gft{3}{Initially, the DRAM bank is in the precharged state (\underline{step~4} in Figure~\ref{fig_subarray_dram_operation}), where all of the components are ready for activation, i.e., 
the local bitlines are set at a reference voltage ($\frac{V_{DD}}{2}$), the local wordline is disabled, and the local sense amplifiers is \emph{off} without any data latched in it.}

\paratitle{Row Activation \& Sense Amplification Phases} \gft{3}{To open a row, the memory controller sends an \texttt{ACTIVATE} (\texttt{ACT}) command to raise the local wordline of the corresponding DRAM row, which connects the row to the local bitlines 
(\underline{step~1} in Figure~\ref{fig_subarray_dram_operation}). 
This triggers an activation, where charge starts to flow from the DRAM cell to the local bitline (or the other way around, depending on the initial charge level in the cell) via a process called \emph{charge sharing}. 
This process perturbs the voltage level on the
corresponding local bitline by a small amount. If the DRAM cell is initially charged (which we assume for the rest of this explanation, without loss of generality), the local bitline voltage is perturbed \emph{upwards}. 
Note that this causes the DRAM cell itself to discharge, losing its data temporarily (hence
the partial filling color of the accessed row), but this charge will be restored as we will describe below. 
After the activation phase, the sense amplifier senses and amplifies the voltage perturbation on
the local bitline. 
When the local bitline is amplified to a certain voltage level (e.g., 0.8$V_{DD}$), the sense amplifier latches in the cell's data, which transforms it into binary data, i.e., a logical `1' or `0'. 
At this point in time, the data
can be read from the local sense amplifier. 
The latency of these two phases (activation and sense amplification) is called the \emph{activation latency}, and is defined as $t_{RCD}$ in the standard DDR interface~\cite{jedec2008ddr3,jedec2017jedec,jedec2020jesd795}. 
This activation latency specifies the latency from the time an \texttt{ACT} command is issued to the time the data is ready to be accessed in the local sense amplifiers.}

\paratitle{Read/Write \& Restoration Phases} \gft{3}{Once the local sense amplifiers (row buffer) latches in the data, the memory controller can send a \texttt{READ} (\texttt{RD}) or \texttt{WRITE} (\texttt{WR}) command to access the corresponding column of data within the row buffer (\underline{step 2} in Figure~\ref{fig_subarray_dram_operation}). 
The column access time to read the cache line data is called $t_{CL}$ ($t_{CWL}$ for writes). These parameters define the time
between the column command and the appearance of the first \emph{beat} of data on the data bus, shown at the bottom of Figure~\ref{fig_subarray_dram_operation}. 
A data beat is a 64-bit data transfer from the DRAM to the processor. In a typical DRAM~\cite{jedec2008ddr3,jedec2017jedec,jedec2020jesd795}, a column \texttt{RD} command reads out 8 data beats.}

\gft{3}{After the DRAM bank becomes activated and the local sense amplifier latches in the binary data of a cell, it starts to restore the connected cell's charge back to its original fully-charged state
(\underline{step~3} in Figure~\ref{fig_subarray_dram_operation}). This phase is known as \emph{charge restoration}, and can happen in parallel with column accesses. The timing parameter that defines the restoration latency (from issuing an \texttt{ACT} command to fully restoring a row of
DRAM cells) is called $t_{RAS}$.}

\paratitle{Precharge Phase} \gft{3}{In order to access data from a different DRAM row, the DRAM bank needs to be
re-initialized back to the precharged state (\underline{step~4} in Figure~\ref{fig_subarray_dram_operation}). 
To achieve this, the memory controller
sends a \texttt{PRECHARGE} (\texttt{PRE}) command, which 
\li~disables the local wordline of the corresponding DRAM row, disconnecting the DRAM row from the local sense amplifiers, and 
\lii~resets the voltage level on the bitline
back to the initialized state.
The \omt{3}{timing parameter that defines the} precharge latency is called $t_{RP}$.}

\omt{3}{For more detail on DRAM operation, commands, and latency, we refer the reader to prior literature~\drambackground.}

%The first command, \texttt{ACTIVATE} (\texttt{ACT}), connects DRAM cells in a row to its local bitline,
%and the cell's transistor shares its charge with the bitline to
%shift the bitline voltage higher (or lower) if the cell stores a `1' (`0').
%The local row buffer amplifies the shifts to CMOS-readable values (simultaneously restoring charge to the DRAM cell).
%The \omt{3}{timing parameter that defines the latency} from the start of activation until charge restoration is called $t_{RAS}$.
%The second command, \texttt{READ} (\texttt{RD}), returns a cache line of data from the local row buffer.
%The third command, \texttt{PRECHARGE} (\texttt{PRE}), disconnects DRAM cells from the bitlines, and returns the bitlines to their reference voltage.
%The \omt{3}{timing parameter that defines the} precharge latency is called $t_{RP}$.

\section{Processing-in-Memory Architectures}
\label{sec:background:pim}

\subsection{\gft{3}{A Brief \gls{PIM} Taxonomy}}

\gls{PIM} aims to reduce the impact that data movement between memory and CPU has on the overall system's performance and energy by 
\li~adding \omt{3}{computation} logic to the same die as memory~\cite{farmahini2015nda,babarinsa2015jafar,devaux2019true,ghiasi2022genstore,gomez2021benchmarkingcut,gomezluna2021benchmarking,gomez2022benchmarking,syncron,singh2020nero,skhynixpim,ke2021near,giannoula2022sparsep,shin2018mcdram,cho2020mcdram,denzler2021casper,asghari2016chameleon,IRAM_Micro_1997,C_RAM_1999,CASES_MVX,Xi_2015,sun2021abc,matam2019graphssd,gokhale1995processing,hall1999mapping,MEMSYS_MVX,lockerman2020livia}\omt{3}{, memory controllers~\cite{seshadri2015gather, lee2015decoupled, hashemi2016continuous, hashemi2016accelerating, singh2020nero},} or to the logic layer of 3D-stacked memory~\pnmthreed (i.e., \gls{PnM}); or
\lii~exploiting the analog operational properties of
the memory circuitry (\omt{3}{i.e.,} \gls{PuM}~\pum).

In the literature, the term \gls{PnM} (also processing-in-memory, near-data computing/processing, or near-memory computing/processing) has come to refer to a wide range of architectures.
We categorize these different architectures into three general approaches where \gls{PIM} logic can be integrated into the system. Each of the three approaches has different \omt{3}{trade offs}.

The first approach adds \gls{PIM} logic to the memory controller, enabling the memory controller to perform more complex functions than scheduling reads/writes without the need for CPU involvement. 
For example, several works~\cite{seshadri2015gather, lee2015decoupled, hashemi2016continuous, hashemi2016accelerating, singh2020nero} add computation \omt{3}{logic to} the memory controller to reduce the round-trip cost of data movement, avoiding the latency of \gft{3}{moving data within} the cache hierarchy \gft{3}{until it reaches to the} host \gft{3}{CPU} core. A key advantage of such an approach is that it can work with conventional memory technologies such as DDRx memories. 
Unfortunately, \omt{3}{this approach} does \emph{not} eliminate unnecessary data movement, as the memory controller typically sits on-chip with the CPU cores and still requires data to \omt{3}{move back-and-fourth between} the pin-limited and power-hungry main memory channel \omt{3}{and the memory controller}.

\gft{3}{Beyond reducing data movement overhead, a promising direction enabled by computation-capable memory controllers is the design of \emph{intelligent memory controllers} that actively manage and mitigate reliability and robustness issues in DRAM. 
For example, by integrating specialized logic, the memory controller can detect and respond to phenomena such as read disturbance (e.g., RowHammer~\cite{kim2014flipping} and RowPress~\cite{luo2023rowpress}) and data retention failures more effectively and at lower overhead compared to traditional, reactive system-level techniques.
Several recent works demonstrate that such intelligent mechanisms can proactively identify vulnerable rows~\cite{apple2015about, enterprise2015hpmoonshot,lenovo2015rowhammer,greenfield2012throttling, kim2014flipping, kim2014architectural,  bains2015method, bains2016rowhammer, bains2016distributed, aichinger2015ddrmemory, aweke2016anvil, gomez2016dram_rowhammer, yang2016suppression, son2017making, seyedzadeh2017counterbased, seyedzadeh2017mitigating, seyedzadeh2018mitigating, irazoqui2016mascat, ryu2017overcoming, yang2017scanning, you2019mrloc, lee2019twice, park2020graphene, yaglikci2021security, yauglikcci2021blockhammer, canpolat2024breakhammer, frigo2020trrespass, kang2020cattwo, hassan2021uncovering, qureshi2022hydra, saileshwar2022randomized, brasser2017cant, konoth2018zebram, vanderveen2018guardion, vig2018rapid, hassan2019crow, gautam2019rowhammering, kim2022mithril, lee2021cryoguard, marazzi2023protrr, zhang2022softtrr, joardar2022learning, juffinger2023csirowhammercryptographic, yauglikcci2022hira, saxena2022aqua, manzhosov2022revisiting, ajorpaz2022evax, naseredini2022alarm, joardar2022machine, hassan2022case, zhang2020leveraging,loughlin2021stop, devaux2021method, han2021surround, fakhrzadehgan2022safeguard, saroiu2022theprice, saroiu2022howto, loughlin2022moesiprime, zhou2022ltpim, hong2023dsac, mutlu2023fundamentally, marazzi2022rega, didio2023copyonflip, sharma2022areview, woo2023scalable, park2022rowhammer_reduction, wi2023shadow, kim2023a11v, guderamarao2023defending, guha2022criticality, france2022modeling, france2022reducing, bennett2021panopticon, enomoto2022efficient, arikan2022processor, tomita2022extracting, saxena2023ptguard, zhou2023dnndefender, bostanci2024comet, olgun2024abacus} or dynamically adjust refresh policies~\cite{liu2012raidr,patel2020bit,kim2020revisiting} to maintain data integrity, security, and overall system robustness. Embedding this functionality within a programmable or specialized in-controller PIM substrate allows for fine-grained, real-time memory management that adapts to the dynamic behavior of DRAM devices, thereby improving reliability without incurring significant performance or energy penalties.
Several recent works demonstrate that such intelligent mechanisms can implement different read disturbance mitigation mechanisms~\cite{apple2015about, enterprise2015hpmoonshot,lenovo2015rowhammer,greenfield2012throttling, kim2014flipping, kim2014architectural,  bains2015method, bains2016rowhammer, bains2016distributed, aichinger2015ddrmemory, aweke2016anvil, gomez2016dram_rowhammer, yang2016suppression, son2017making, seyedzadeh2017counterbased, seyedzadeh2017mitigating, seyedzadeh2018mitigating, irazoqui2016mascat, ryu2017overcoming, yang2017scanning, you2019mrloc, lee2019twice, park2020graphene, yaglikci2021security, yauglikcci2021blockhammer, canpolat2024breakhammer, frigo2020trrespass, kang2020cattwo, hassan2021uncovering, qureshi2022hydra, saileshwar2022randomized, brasser2017cant, konoth2018zebram, vanderveen2018guardion, vig2018rapid, hassan2019crow, gautam2019rowhammering, kim2022mithril, lee2021cryoguard, marazzi2023protrr, zhang2022softtrr, joardar2022learning, juffinger2023csirowhammercryptographic, yauglikcci2022hira, saxena2022aqua, manzhosov2022revisiting, ajorpaz2022evax, naseredini2022alarm, joardar2022machine, hassan2022case, zhang2020leveraging,loughlin2021stop, devaux2021method, han2021surround, fakhrzadehgan2022safeguard, saroiu2022theprice, saroiu2022howto, loughlin2022moesiprime, zhou2022ltpim, hong2023dsac, mutlu2023fundamentally, marazzi2022rega, didio2023copyonflip, sharma2022areview, woo2023scalable, park2022rowhammer_reduction, wi2023shadow, kim2023a11v, guderamarao2023defending, guha2022criticality, france2022modeling, france2022reducing, bennett2021panopticon, enomoto2022efficient, arikan2022processor, tomita2022extracting, saxena2023ptguard, zhou2023dnndefender, bostanci2024comet, olgun2024abacus} or perform online profiling of DRAM cell retention times and online
adjustment of refresh rate on a per-row
basis~\cite{liu2012raidr,liu2013experimental,khan2014efficacy,qureshi2015avatar,khan2016parbor,khan.cal16,patel2017reach,khan2017detecting,patel2021harp,patel2020bit,patel2019understanding,patel2024rethinking,patel2024rethinkingieee} to maintain data integrity, security, and overall system robustness.}

\gft{3}{The second approach adds \gls{PIM} logic close physical to (but \emph{not} within) the memory chip, enabling near-data processing without requiring modifications to the memory chip itself. 
Such designs typically exploit one of two integration strategies:
\li~2.5D integration, where \gls{PIM} logic is integrated alongside high-bandwidth memory chips on a silicon interposer, enabling low-latency, high-bandwidth communication between logic and memory dies\cite{zhang2015heterogeneous};
\lii~module-level integration, where \gls{PIM} logic is embedded on the same \gls{PCB} as memory chips. 
One notable example of the latter is the use of the buffer chips in load-reduced DIMMs (LRDIMMs) as hosts for PIM logic~\cite{asghari2016chameleon,sun2021abc,ke2021near,ke2019recnmp}, although similar concepts can extend to PIM chips co-packaged with standard DRAM devices on a DRAM module.
These approaches have the advantage of avoiding data transfers over the off-chip memory channel between the host processor and main memory, thereby reducing memory latency and memory bandwidth pressure. 
However, because data movement still occurs to/from the memory chip to the computation chip(s), and such data movement consumes more energy than data movement that happens \emph{only} inside the memory chip.}

\gft{3}{The third approach adds PIM logic inside the memory chip, allowing the \gls{PIM} logic to access the abundant internal bandwidth inside the memory chip directly at the lowest energy costs. 
Recent advances in memory manufacturing have enabled adding \gls{PIM} logic inside the memory chip in two ways. 
First, \gls{PIM} logic can be added to the \emph{logic layer} of 3D-stacked memories~\pnmthreed. 
In a 3D-stacked memory, multiple layers of memory (typically DRAM in already-existing systems) are stacked on top of each other.
These layers are connected together using vertical through-silicon vias (TSVs)~\cite{loh2008stacked,lee2016simultaneous}. 
Using current manufacturing process technologies, thousands of TSVs can be placed within a single 3D-stacked memory chip. 
As such, the TSVs provide much greater internal memory bandwidth than the narrow memory channel. 
Examples of 3D-stacked DRAM available commercially include \gls{HBM}~\cite{HBM,hbm2,lee2016simultaneous}, 
Wide I/O~\cite{wideio}, 
Wide I/O 2~\cite{wideio2}, 
and the \gls{HMC}~\cite{hmc.spec.2.0}. 
Emerging die-stacking and packaging technologies, like \emph{hybrid bonding}~\cite{kagawa2016novel,niu2022isscc,schmidt1998wafer} and \emph{monolithic 3D integration}~\cite{gopireddy2019m3d,mitra2018vlse,hwang2018cmos,mitra2015nano,rich2020nano,sabry2015abundant,sabry2019n3xt,ghiasi2022revamp3d}, can further amplify the benefits of conventional TSV-based 3D-stacked memory chips by greatly improving internal bandwidth across layers using \emph{high-density} inter-layer vias (ILVs)~\cite{gopireddy2019m3d,sabry2019n3xt} or \emph{direct} wafer-to-wafer connections via Cu--Cu bonding~\cite{lau2023recent,kagawa2016novel,niu2022isscc,schmidt1998wafer}, respectively. 
In addition to the multiple layers of DRAM, a number of prominent 3D-stacked DRAM architectures, including HBM and HMC, incorporate a logic layer inside the chip~\cite{lee2016simultaneous, HBM,hbm2, hmc.spec.2.0,loh2008stacked,ahn2015scalable}. 
The logic layer is typically the bottommost layer of the chip, and is connected to the same TSVs as the memory layers.\footnote{Logic layer(s) in monolithic 3D architectures can also be added \emph{between} memory layers~\cite{ebrahimi2014monolithic,kim2023van}.} 
The logic layer provides area inside the 3D-stacked DRAM system where architects can implement functionality that interacts with both the processor and the DRAM cells. Currently, manufacturers make limited use of the logic layer and there is significant amount of area the logic layer can provide, especially when manufactured with a high-quality logic fabrication process. 
This presents a promising opportunity for architects to implement new and efficient PIM logic in the available area of the logic layer. 
We can potentially add a wide range of computational logic (e.g., general-purpose cores~\cite{boroumand2018google, ahn2015scalable,drumond2017mondrian,boroum2019conda,NDC_ISPASS_2014,singh2019napel,azarkhish2016logic,azarkhish2018neurostream}, 
accelerators~\cite{top-pim,RVU,NIM,gao2017tetris,hsieh2016transparent,cali2020genasm,boroumand2021mitigating,boroumand2021google,boroumand2021polynesia,fernandez2020natsa,LiM_3D_FFT_MM,akin2014hamlet}, 
reconfigurable logic~\cite{gao2016hrl,farmahini2015nda}, special-purpose functional units~\cite{boroumand2018google,nai2017graphpim,kim2018grim,PEI,kwon202125,lee2021hardware}, or a combination of different such types of logic) in the logic layer, as long as the added logic meets area, energy, and thermal dissipation constraints, which are important and potentially limiting constraints in 3D-stacked systems~\cite{boroumand2018google, ahn2015scalable}.
Second, \gls{PIM} logic can be added near memory banks by leveraging DRAM manufacturing processes that support mixing logic and memory~\cite{skhynixpim,kwon202125,lee2021hardware,devaux2019true,gomezluna2021benchmarking,gomez2021benchmarkingcut,gomez2022benchmarking}. 
The main drawbacks of integrating \gls{PIM} logic inside the memory chip are high manufacturing costs, limited area budget inside 3D-stacked memories, and difficulties tight to thermal dissipation.} 

% \begin{figure}[!ht]
%   \medskip
% \centering
% \begin{subfigure}[t]{.33\linewidth}
%   \centering
%   \includegraphics[width=1.1\linewidth]{main/mainmatter/02_background/ddr4_.pdf}
%   \caption{}
%   \label{figure_ddr4}
%   \label{fig:in-controller-PIM}
% \end{subfigure}\hfill
% ~
% \begin{subfigure}[t]{.31\linewidth}
%   \centering
%   \includegraphics[width=1.1\linewidth]{main/mainmatter/02_background/hbm_.pdf}
%   \caption{}
%   \label{figure_hbm}
%   \label{fig:interposer-PIM}
% \end{subfigure}\hfill
% ~
% \begin{subfigure}[t]{.3\linewidth}
%   \centering
%   \includegraphics[width=0.8\linewidth]{main/mainmatter/02_background/hmc_.pdf}
%   \caption{}
%   \label{figure_hmc}
%   \label{fig:in-chip-PIM}
% \end{subfigure}
% \caption{Different approaches to integrating PIM logic in the system architecture.}
% \label{figure_organization}
% \end{figure}

\subsection{\gft{3}{Early \gls{PIM} Proposals}}

\gls{PIM} is not a new idea: the first \gft{3}{works} that foresaw the need to bring computation closer to memory dates from \gft{3}{1969~\cite{Kautz1969} and} 1970~\cite{stone1970logic}. 
\omt{3}{In the 70s and 80s, many works~\cite{raschid1986special,shaw1982von,shaw1985non,takagi1985hardware,emam1979architectural,hsiao1977structure,defiore1973data,lin1975design,su1973retrieval,parker1971logic,minsky1972rotating,jino1978magnetic,doty1980magnetic,bongiovanni1980magnetic,shaw1981non} investigate the implementation of \gls{PIM} architectures using different memory technologies (e.g., hard disks, content-addressable memories) for key applications (e.g., databases, production systems, computer vision, knowledge base management) and kernels (e.g., sorting) at the time.} 
In the 1990s, another wave of works~\gft{3}{\cite{IRAM_Micro_1997,C_RAM_1999,oskin1998active,kang2012flexram,IRAM_WML_1997,Miss_Mem_Wall_1996,hall1999mapping,keeton.1998,gokhale1995processing,kogge1994execube,elliott1992computational}} focused on exploiting the difference between internal and external bandwidth in DDRx memories with \gls{PnM} by integrating computational modules in the same \gls{PCB} as DRAM chips. 
However, these early \gls{PnM} works were not adopted into real systems \omt{3}{in part} due to the challenges of manufacturing logic and memory (particularly DRAM) into a single, high-performance, and high-area-efficient chip~\cite{kim1999assessing,kim1996assessing}.

\subsection{\gft{3}{Major Trends Affecting Main Memory \& \gls{PIM} Resurgence}}

Recently, \gls{PIM} \gft{3}{(particularly \gls{PnM})} has seen a resurgence due to \gft{3}{three} main reasons. 
First, due to the maturation of 3D manufacturing and advances in the CMOS manufacturing process, integrating memory and computation is now possible~\omt{3}{\cite{kgil2006picoserver,ahn2015scalable,HBM,hbm2,lee2016simultaneous,loh2008stacked,hmc.spec.1.1, hmc.spec.2.0, gokhale2015hmc}} \omt{3}{and becoming increasingly effective and cheaper with better packaging, technology nodes, and integration}. 
Second, data movement has become a \omt{3}{very large cause of performance and energy} bottleneck in today's systems\gft{3}{~\membottleneck} \gft{3}{due to a combination of factors.
\li~From a \emph{technology point of view}, the memory subsystem has long been optimized \emph{primarily} for cost per bit and capacity. 
However, memory access latency has remained almost constant over the past two decades (i.e., main memory access latency, measured by the row cycling time~\cite{kim2012case,Tiered-Latency_LEE}, reduced by only 30\%)~\omt{3}{\cite{chang.sigmetrics2016,patel2024rethinkingieee}}, making it a significant performance and energy bottleneck for many modern workloads.
As low-latency computing is becoming ever more important~\cite{mutlu2013memory, mutlu2015research, cali2020genasm, cali2018nano, alser2020accelerating, alser2017gatekeeper, alser2020sneakysnake, alser2019shouji,dean2013tail,boroumand2022polynesia}, e.g., due to the ever-increasing need to process large amounts of data at real time, and predictable performance continues to be a critical concern in the design of modern computing systems~\cite{moscibroda-usenix2007,mutlu2007stall,mutlu2008parallelism,mutlu2015research,lavanya-thesis,lee2015decoupled,subramanian2013mise,usui2016dash,subramanian2015application,kim2010thread,kim2014bounding,kim2016bounding}, it is increasingly important to design low-latency main memory chips.
\lii~From the \emph{application point of view}, the growing data working set sizes of modern applications~\cite{mutlu2013memory, mutlu2015research,dean2013tail, kanev_isca2015, ferdman2012clearing,wang2014bigdatabench, boroumand2018google, boroumand2021google, boroumand2021google_arxiv,pekhimenko2012lcp,pekhimenko2013linearly,churin1988camac,abali2001mxt,friedrich2014power8} impose an ever-increasing demand for higher main memory capacity and performance.
For example, memory capacity requirements of large machine learning models increased by more than 10,000 times in the past five years~\cite{gholami2020ai}. 
Unfortunately, DRAM technology scaling is becoming increasingly challenging: it is increasingly difficult to enlarge DRAM chip capacity at low cost while also maintaining DRAM performance, energy efficiency, and reliability~\cite{liu2013experimental,kim2014flipping,mutlu2017rowhammer, yauglikcci2021blockhammer, hassan2021uncovering, orosa2021deeper,frigo2020trrespass,kang2014co, mutlu2013memory,wilkes2001memory,kim2012case,yoongu-thesis,liu2012raidr,lee2015decoupled,lee-isca2009,yoon2012row,yoon-taco2014,lim-isca09,
wulf1995hitting, chang.sigmetrics2016, Tiered-Latency_LEE, lee2015adaptive,chang2017understanding, lee2017design, luo2014characterizing,luo.arxiv17,hassan2017softmc, hassan2016chargecache, chang2017understandingphd, patel2017reach, hassan2019crow, ghose2018your, kim2018solar, kim2020revisiting, mutlu2019rowhammer, wang2018reducing, mutlu2018recent, ghose2019demystifying,mutlu2015main,hong2010memory,sites1996,luo2023rowpress,yaglikci2024spatial,olgun2024abacus,bostanci2024comet,yauglikcci2022hira, mutlu2015research, dean2013tail, kanev_isca2015, ferdman2012clearing, wang2014bigdatabench, boroumand2018google, boroumand2021google, boroumand2021google_arxiv, mutlu2019enabling, mutlu2019processing, mutlu2020intelligent, ghose.ibmjrd19, alser2020accelerating, cali2020genasm, koppula2019eden, kanellopoulos2019smash, deoliveira2021IEEE, oliveira2021.SLS, mutlu.msttalk17, mutlu.gwutalk19, mutlu.isscctalk19, mutlu.glvlsitalk19, mutlu.appttalk19,mutlu.iccdtalk19,narancic2014evaluating,jia2016understanding,Manegold_2000,gholami2020ai,stengel2015quantifying,chishti2019memory,burger1995declining,gupta2020architectural,sriraman2019softsku,zhao2022understanding,ruan2019insider,gan2018architectural,sriraman2020accelerometer,ayers2018memory,yuan2023rambda,delimitrou2018amdahl,hsia2020cross,lottarini2018vbench,wang2022characterizing,richins2020missing,mckee2004reflections}.}

\gft{3}{Third, main memory (i.e., DRAM) technology scaling to smaller nodes adversely affects DRAM reliability and robustness. 
For a DRAM cell to operate correctly, both the capacitor and the access
transistor (as well as the peripheral circuitry) need to operate
reliably. 
As the size of the DRAM cell reduces, both the capacitor and the access transistor become less reliable, more leaky, and generally more vulnerable to electrical noise and disturbance. 
As a result, reducing the size of the DRAM cell increases the difficulty of
correctly storing and detecting the desired original value in the DRAM, as shown in various recent works that study DRAM reliability by analyzing data retention and other reliability issues of modern DRAM chips
cell~\cite{liu2013experimental,mutlu2013memory,kim2014flipping,mutlu2017rowhammer,khan2014efficacy, khan2016parbor, khan.cal16, khan2017detecting, qureshi2015avatar, patel2017reach, hassan2017softmc, hassan2019crow, liu2012raidr, patel2020bit, patel2021harp, patel2019understanding, kim2020revisiting, yauglikcci2021blockhammer, hassan2021uncovering, orosa2021deeper, mutlu2019rowhammer, frigo2020trrespass, cojocar2020susceptible,luo2023rowpress,yaglikci2024spatial,olgun2024abacus,bostanci2024comet,yauglikcci2022hira}.
Hence, memory technology scaling causes memory errors to appear more frequently, even leading to new robustness-related (including security and safety problems) vulnerabilities.
For instance, many works demonstrated that the \emph{read disturbance} phenomenon in DRAM can be exploited to cause security and safety problems that enable attackers to  take over a system~\cite{aga2017good, bosman.2016, cojocar19exploiting, frigo2020trrespass, gruss2016rowhammerjs, anotherflip, qiao2016new, tatar2018defeating, vanderveen.2016, vanderveen2018guardion, zhang2020pthammer,wang2022research, kurmus2017from,seaborn.2015,seaborn.2016,gruss.2015,glitch-vu,nethammer,throwhammer}, 
read data they do \emph{not} have access to~\cite{ cojocar19exploiting, carre2018openssl, cohen2022hammerscope, frigo2020trrespass, ji2019pinpoint, kwong2020rambleed, qiao2016new, tobah2022spechammer, kaur2022workinprogress, li2023fphammer,vanderveen.2016}, 
break out of virtual machine sandboxes~\cite{razavi.2016,cloudflops}, 
corrupt important data (even rendering machine learning inference useless)~\cite{kim2014flipping,yao2020deephammer,tol2023don,hong2019terminal, cojocar2020susceptible, deridder2021smash, jattke2022blacksmith, glitch-vu, frigo2020trrespass, hassan2021uncovering, kogler2022halfdouble, pessl2016drama, qiao2016new, razavi.2016, zhang2018triggering}, 
steal secret data (e.g., cryptographic keys~\cite{bhattacharya2016curious,bhattacharya2018advanced,carre2018openssl, cohen2022hammerscope, cojocar19exploiting, fahrjr2022when, frigo2020trrespass, ji2019pinpoint, kwong2020rambleed, poddebniak2018attacking, tobah2022spechammer, weissman2020jackhammer, mus2022jolt, fahr2022theeffects, islam2022signature, tomita2022extracting,razavi.2016} and 
steal or alter machine learning model parameters~\cite{hong2019terminal, liu2022generating, rakin2022deepsteal, tol2023don, zheng2022trojvit, cai2022onthe, roohi2022efficient, staudigl2022neurohammer,yang2022sociallyaware,rakin2019bit}).} 

\gft{3}{In summary, the resurgence of \gls{PIM} is fueled by three fundamental and converging trends: 
\li~the increasing feasibility and affordability of memory--compute integration enabled by advances in 3D integration and CMOS scaling, 
\lii~the growing inefficiency and cost of data movement driven by stagnant memory latency and rapidly expanding application data demands, and 
\liii~the declining reliability and security of DRAM due to aggressive technology scaling. 
These trends expose the \emph{fundamental} limitations of traditional processor-centric architectures and motivate a paradigm shift toward architectures that minimize data movement, improve energy efficiency, and enhance system robustness. 
\gls{PIM} is a promising direction to overcome these limitations by enabling computation capability close to or within memory, thereby directly addressing the key bottlenecks in performance, energy, scalability, and reliability faced by modern processor-centric systems.}

\subsection{Real-World \gls{PIM} Architectures} 

In this section, we describe \omt{3}{several} DRAM-based real-world \gls{PIM} architectures that have been introduced by industry in recent years. 
First, we provide details on the general-purpose UPMEM \gls{PIM} architecture, which we use as a baseline \gls{PIM} architecture in Chapter~\ref{chap:dappa}.
Second, we briefly describe \omt{3}{several} application-specific \gls{PIM} architectures \omt{3}{introduced by Samsung, SK hynix, and Alibaba}. 

\subsubsection{\omt{3}{2.2.4.1.~General-Purpose \omt{3}{Real-World} \gls{PIM} \omt{3}{Architecture}: The UPMEM \gls{PIM}}} 

\begin{figure}[ht]
    \centering
    \includegraphics[width=0.85\linewidth]{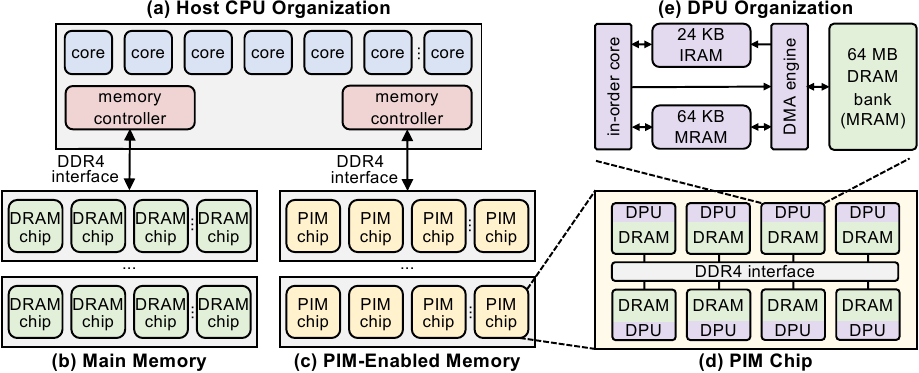}
    \caption{\omt{3}{UPMEM system organization.}}
    \label{fig:architecture}
\end{figure}

In 2019, UPMEM introduced UPMEM \gls{PIM}\omt{3}{~\cite{upmem,upmem2018,gomez2021benchmarking,gomez2022benchmarking,gomez2021benchmarkingcut}}, the first real-world general-purpose \gls{PIM} architecture \omt{3}{that is commercially available}.  Figure~\ref{fig:architecture} illustrates the main components of a UPMEM-enabled system~\cite{upmem,upmem2018,gomez2021benchmarking,gomez2022benchmarking,gomez2021benchmarkingcut}. 
\omt{3}{An} UPMEM-enabled system consists of a host CPU (Figure~\ref{fig:architecture}a) equipped with regular DRAM as its main memory (Figure~\ref{fig:architecture}b) and specialized UPMEM DRAM \glspl{DIMM} (\gls{PIM}-enabled memory in Figure~\ref{fig:architecture}c). 
\omt{3}{An} UPMEM module is a standard DDR4-2400 \gls{DIMM} (module) with 8 (1-rank) or 16 (\omt{3}{2-rank}) \gls{PIM} chips. 
Inside each UPMEM \gls{PIM} chip (Figure~\ref{fig:architecture}d), there are 8 small general-purpose in-order \gft{3}{\emph{\gls{PIM} cores}}, called \emph{DPUs}. 
\omt{3}{Each DPU is placed next to a DRAM bank. Each DPU is} fine-grained multithreaded. 
Each DPU (Figure~\ref{fig:architecture}e) has exclusive access to 
\li~a \SI{64}{\mega\byte} DRAM bank, called \emph{MRAM};
\lii~a \SI{24}{\kilo\byte} instruction memory, called \emph{IRAM}; and 
\liii~a \SI{64}{\kilo\byte} scratchpad memory, called \emph{WRAM}.
The \gls{PIM} cores in \omt{3}{an} UPMEM system operate at \SI{450}{\mega\hertz} and feature \omt{3}{a} 14-stage pipeline, enabling them to execute one integer addition or subtraction per cycle. 
Integer multiplication and division require up to 32 cycles when the pipeline is fully utilized. 
However, floating-point operations incur significantly higher latencies, ranging from tens to 2,000 cycles~\cite{gomez2021benchmarking} \omt{3}{because they are emulated in software due to the lack of native hardware support in DPUs}. 
A standard UPMEM-based \gls{PIM} system is composed of 20 UPMEM \glspl{DIMM}, thus containing up to 2,560 DPUs and \SI{160}{\giga\byte} of PIM-capable memory, achieving a peak compute throughput exceeding 1~TOPS.

\gft{3}{G{\'o}mez-Luna \emph{et al.}~\cite{gomez2021benchmarking,gomez2022benchmarking,gomez2021benchmarkingcut} have conduct the first experimental characterization of the UPMEM PIM system. First, they implement and deploy various microbenchmarks to assess the microarchitectural limitations (such as compute throughput, memory bandwidth) and potential benefits of the UPMEM PIM system. 
All these works demonstrate that the UPMEM PIM system can provide substantial performance improvements compared to processor-centric architectures when executing such workloads.
We highlight three key new insights from their analyses.
\li~The UPMEM PIM architecture is \emph{fundamentally compute bound}. As a result, the most suitable workloads are memory-bound in processor-centric systems (i.e., CPU, GPU).
\lii~The most well-suited workloads for the UPMEM PIM architecture use \emph{no arithmetic operations} or use \emph{only simple operations} (e.g., bitwise operations and integer addition/subtraction). The primary reason is that each DPU in the UPMEM PIM system is a very lightweight processor, featuring native hardware support \emph{only} for 32-bit integer addition/subtraction and 8-bit multiplication while emulating the implementation of complex (e.g., 32-bit integer multiplication/division) and floating-point operations. 
\liii~The most well-suited workloads for the UPMEM PIM architecture require \emph{little or no communication across DPUs}. This is due to the fact that there is \emph{no} direct inter-DPU communication mechanism in the current UPMEM \gls{PIM} system; instead, all communication between \gls{PIM} cores occurs through memory transfers between host main memory and \gls{PIM}-enabled memory.
Second, G{\'o}mez-Luna \emph{et al.} present the \emph{PrIM} benchmarks (\emph{\underline{Pr}ocessing-\underline{I}n-\underline{M}emory benchmarks})~\cite{gomezluna2021repo}, a benchmark suite of 16 workloads from different application domains (e.g., dense/sparse linear algebra, databases, data analytics, graph processing, neural networks, bioinformatics, image processing) targeting the UPMEM PIM system.
Their experimental evaluation results show that the UPMEM PIM systems outperform modern CPUs in terms of performance and energy efficiency on most of PrIM benchmarks and outperform modern GPUs on a majority (10 out of 16) of PrIM benchmarks.
}

\gft{3}{Many recent works~\cite{giannoula2022sparsep, giannoula2022towards, giannoula2022repo,diab2023framework,diab2022high,diab2023repo,gomez2023evaluating,gomezluna2022isvlsi,rhyner2024pimopt,giannoula2024pygim,gogineni2024swiftrl,chen2023uppipe,lavenier2020variant,lavenier2016blast,kang2023pim,baumstark2023accelerating,baumstark2023adaptive,lim2023design,bernhardt2023pimdb,lee2024spid,nider2022bulk,das2022implementation,zarif2023offloading,kim2025accelerating} have explored in depth the suitability of the UPMEM PIM system for important application domains, including bioinformatics workloads (e.g., RNA-seq qualification~\cite{chen2023uppipe}, variant calling~\cite{lavenier2020variant}, BLAST~\cite{lavenier2016blast}), analytics \& databases~\cite{kang2023pim,baumstark2023accelerating,baumstark2023adaptive,lim2023design,bernhardt2023pimdb,lee2024spid}, imagine processing~\cite{nider2022bulk}, and machine learning (ML) workloads (e.g., deep neural networks~\cite{das2022implementation} and recommendation models~\cite{zarif2023offloading}).
To illustrate, we briefly discuss some of such works below. 
First,~\cite{giannoula2022sparsep} introduces \emph{SparseP}~\cite{giannoula2022repo}, the first \gls{SpMV} library for real PIM architectures, which covers a wide variety of sparse matrices with diverse sparsity patterns. 
Second,~\cite{diab2023framework} 
\li~evaluates the suitability of the UPMEM PIM system to accelerate sequence alignment algorithms (such as \emph{Needleman-Wunsch}~\cite{needleman1970general}, \emph{Smith-Waterman-Gotoh}~\cite{gotoh1982improved}, GenASM~\cite{cali2020genasm}, and \emph{wavefront algorithm}~\cite{marco2021fast}) and 
\lii~introduces a framework for PIM-based sequence alignment, called \emph{Alignment-in-Memory (AiM)} ~\cite{diab2023repo}, where the host CPU dispatches sequence pairs across the DPUs available in the UPMEM PIM system.}

\gft{3}{Third, works such as~\cite{gomez2023evaluating,gomezluna2022isvlsi,rhyner2024pimopt,giannoula2024pygim,gogineni2024swiftrl} comprehensively study the potential of the UPMEM PIM architecture to accelerate ML training.  
\li~\cite{gomez2023evaluating,gomezluna2022isvlsi} implements and optimizes several representative classical ML algorithms (namely, linear regression~\cite{legendre1806nouvelles,gauss1809theoria}, logistic regression~\cite{berkson1944application}, decision tree~\cite{morgan1963problems}, k-means clustering~\cite{lloyd1982least,macqueen1967some}) on the UPMEM PIM system.
\lii~\cite{rhyner2024pimopt} proposes \emph{PIM-Opt}, where the authors implement, optimize, and rigorously evaluate 12 representative
distributed stochastic gradient algorithms~\cite{robbins1951stochastic} on a real-world PIM architecture.
\liii~\cite{giannoula2024pygim} proposes \emph{PyGim}, a software library that efficiently maps graph neural networks~\cite{hamilton2017inductive,kipf2016semi,xu2018powerful,zheng2020distdgl} to the UPMEM PIM system.
\liv~\cite{gogineni2024swiftrl} proposes \emph{SwiftRL}, which accelerates reinforcement learning~\cite{sutton2018reinforcement} algorithms and their training phases on the UPMEM PIM system.}

\gft{3}{Some other works~\cite{lee2024pim,hyun2024pathfinding,noh2024pid,kim2023virtual} propose improvements over the UPMEM PIM system, focusing especially on benchmarking and communication mechanisms between processing units.}

\subsubsection{2.2.4.2~Application-Specific \omt{3}{Real-World} \gls{PIM} Architectures}

\paratitle{Samsung Function-In-Memory DRAM (FIMDRAM)} In 2021, Samsung introduced FIMDRAM~\cite{kwon202125, lee2021hardware, kim2021aquabolt} (also known as HBM-PIM~\omt{3}{\cite{lee2021hardware}} or Aquabolt-XL~\omt{3}{\cite{kim2021aquabolt}}), a \gls{PnM} architecture targeted \omt{3}{for accelerating matrix multiplication operations, commonly used during} machine learning inference. 
FIMDRAM embeds one 16-bit floating-point \gls{SIMD} unit with 16 lanes, called \emph{Programmable Compute Unit} (\emph{PCU}), next to two DRAM banks in HBM2 layers~\cite{hbm2}. 
\omt{3}{Each PCU supports} a small instruction set (FP16 add, multiply, multiply-accumulate, multiply-and-add). 
In 2023, Samsung introduced two main extensions of the FIMDRAM architecture~\cite{samsunghc23} targeting generative artificial intelligence (AI)~\cite{goodfellow2014generative,kingma2013auto,vaswani2017attention} applications, e.g., \glspl{LLM} such as GPT-4~\cite{achiam2023gpt}, GPT-J~\cite{mesh-transformer-jax}, T5~\cite{raffel2020exploring}, BERT~\cite{devlin2019bert,devlin2018bert}, and Llama-2~\cite{touvron2023llama}.
The \omt{3}{extended} architectures \omt{3}{are}:
\li~LPDDR-PIM, which adds FIMDRAM's PCU to LPDDRx~\cite{lpddr4} memories and targets on-device generative AI inference, and
\lii~CXL-PNM, which adds \gls{PnM} engines to the compute express link (CXL)~\cite{van2019hoti} memory expander. 

\paratitle{\omt{3}{Samsung CXL-PNM}} Samsung presented two different implementations of their CXL-PNM architecture~\omt{3}{\cite{samsunghc23,park2024lpddr}}. 
The first architecture adds PNM engines to the CXL controller~\omt{3}{\cite{van2019hoti}}, while the second architecture adds PNM engines \omt{3}{(similar to the FIMDRAM architecture)} to the memory chip itself~\cite{samsunghc23}. 
The PNM engines are composed of a \gls{PE} array (targeting the acceleration of general matrix multiplication operations) and an adder tree (targeting the acceleration of general matrix vector multiplication operations). 
The CXL-PNM architecture can provide a capacity of \SI{512}{\giga\byte} and a memory bandwidth of \SI{1.1}{\tera\byte\per\second}.
Samsung provides further details on their CXL-PNM architecture in their HPCA 2024 paper~\cite{park2024lpddr}, where they describe their current realization of the CXL-PNM architecture using LPDDR5X-based CXL memory. 
The paper shows that the proposed architecture (equipped with eight CXL-PNM devices)
achieves 23\% lower latency, 31\% higher throughput, and 2.8$\times$ higher energy efficiency at 30\% lower hardware cost compared to a \omt{3}{modern} GPU-based architecture (with eight A100 NVIDIA GPUs~\cite{a100}) \omt{3}{on} LLM inference \omt{3}{tasks}.

\paratitle{Samsung Acceleration \gls{DIMM} (AxDIMM)} AxDIMM~\cite{ke2021near, kim2021aquabolt, lee2022improving}, also from Samsung, is a \gls{DIMM}-based solution which places an \gls{FPGA} fabric in the buffer chip of \gft{3}{registered DIMMs (RDIMMs).
A RDIMM~\cite{rdimmddr4,rdimmddr5,lrdimmddr3}  is a commonly used DRAM module architecture in server-class systems, where extra buffer chips are added to the DRAM module to decouple I/O signals, improving signal integrity and scalability. This enables scaling the RDIMM DRAM devices to higher memory capacities and frequencies compared to unbuffered DIMMs~\cite{yoon2010virtualized}.
Several prior works~\cite{asghari2016chameleon,ke2021near, kim2021aquabolt, lee2022improving,ke2019recnmp,kwon2019tensordimm,zhou2021gcnear,tian2022g,kim2024sardimm,yi2024gate,yun2023grande,dai2022dimmining,park2021trim,chen2023metanmp,zhou2023dimm,huangfu2019medal,liu2023accelerating,kwon2021tensor,feng2022menda,sun2021abc,tianndpbridge2024,alian2018application,cho2021accelerating,kim2017heterogeneous,xiong2022secndp,liu2021enmc,liu2025make,chen2025asyncdimm,chen2024bridge,chen2025bridge,chen2025mdnmp,lee2024hail}, including AxDIMM, augment the buffer chip with computation in order to implement \gls{PnM} architectures, without modifying the DRAM interface. 
AxDIMM has been tested for \gls{DLRM} inference~\cite{DLRM19, ke2021near} and for database operations~\cite{lee2022improving}. 
In the context of \gls{DLRM} acceleration~\cite{ke2021near} inference, AxDIMM offloads the execution of sparse embedding computations (particularly the element-wise summation of multiple embedding table entries), since embedding operations are a major performance bottleneck in DLRMs~\cite{ke2021near,ke2019recnmp,gupta2019architectural}. 
AxDIMM implements two near-memory accelerators (one per rank) within the FPGA, which operate in parallel to perform the summation, thus reducing data movement and improving inference throughput by up to 1.5$\times$ that of the baseline CPU.}

\paratitle{Accelerator-in-Memory (AiM)} Another major DRAM vendor, SK hynix, introduced \gft{3}{their own}  \gls{PnM} called Accelerator-in-Memory (AiM)~\cite{skhynixpim}, a GDDR6~\cite{jedec2021graphics} \gls{PIM} architecture with specialized units for multiply-and-accumulate and lookup-table-based activation functions for deep learning applications. 
In AIM, \omt{3}{next to} each DRAM bank \omt{3}{in a DRAM chip}, there is a \emph{processing unit} (\emph{PU}) that is composed of an array of 16 16-bit floating-point multipliers, an adder tree, an accumulator, and the necessary logic for activation functions. 
The chip also \omt{3}{adds in the peripheral circuitry of each DRAM bank} a \SI{2}{\kilo\byte} SRAM buffer, called \emph{global buffer} (\emph{GB}), which can store input vectors or serve as an intermediate buffer for copy operations between DRAM banks. 
This inter-bank copy operation is a limited form of a RowClone-like operation~\cite{seshadri2013rowclone} (\gft{3}{which we describe next in} Section~\ref{chap:bgr:pud}).

\paratitle{\gft{3}{SK hynix CXL-Memory Solution (CMS)}} \gft{3}{SK hynix later introduced a CXL~\cite{van2019hoti}-based \gls{PIM} architecture, called} CXL-memory solution (CMS)~\cite{sim2022computational}. 
The CMS architecture~\cite{sim2022computational} comprises of a CXL controller, internal \gft{3}{DDR4 memory chips}, a custom load balancer, and a \gls{PIM} engine \gft{3}{tailored for \gls{kNN} execution}. 
CMS has been prototyped using a Xilinx Alveo U250 \gls{FPGA} board~\cite{AlveoU2590}.
The custom load balancer is designed to, transparently from the host CPU, maximize memory bandwidth utilization by interleaving data across each channel \omt{3}{at multiples} of \gft{3}{the} DDR access granularity (i.e., \SI{64}{\byte}).
The \gls{PIM} engine comprises of three custom logic units (i.e., dot product, \gls{kNN} calculator, and top-K unit) targeting the acceleration of \gls{kNN} application kernels.  
\gft{3}{Their experimental results show that the CMS architecture achieves up to 1.9$\times$ the performance/power of the baseline CPU system.}

\paratitle{Alibaba Logic-to-DRAM Hybrid Bonding with PNM (HB-PNM)}
Alibaba  presented HB-PNM~\cite{niu2022184qps}, a \gls{PnM} system with specialized engines for recommendation systems, which is composed of a DRAM die and a logic die vertically integrated via hybrid bonding (HB)~\cite{lau2023recent}.
The DRAM die contains 36 \SI{1}{\giga\bit} DRAM cores (organized in a $6 \times 6$ 2D-array of DRAM cores) with 8 banks each. 
The logic die contains multiple processing elements called \emph{match} and \emph{neural engines} that perform, respectively, matching and ranking in a recommendation system.
\gft{3}{HB-PNM achieves 9.78$\times$ the throughput and 317.43$\times$ the energy efficiency of the baseline CPU.}

\subsection{DRAM-Based \gls{PuM} Architectures}
\label{chap:bgr:pud}

\paratitle{In-DRAM-Row Copy} RowClone~\cite{seshadri2013rowclone} enables copying a row~$A$ to a row~$B$ in the \emph{same} subarray by issuing two consecutive \texttt{ACT} commands to these two rows, followed by a \texttt{PRE} command. This command sequence is called \texttt{AAP}~\cite{seshadri2017ambit} (\texttt{ACT}-\texttt{ACT}-\texttt{PRE}). The first \texttt{ACT} copies the contents of the source row $A$ into the local row buffer. 
The second \texttt{ACT} connects the DRAM cells in the destination row~$B$ to the local bitlines. Because the sense amplifiers have already sensed and amplified the source data by the time row~$B$ is activated, the data in each cell of row~$B$ is overwritten by the data stored in the row buffer (i.e., row~$A$'s data). LISA~\cite{chang2016low} expands RowClone functionally to enable the execution of in-DRAM row copy operations across DRAM rows in \emph{different} subarrays of a DRAM chip by connecting local row buffers of neighbors subarrays using isolation transistors.
Recent works~\cite{gao2019computedram,olgun2022pidram} experimentally demonstrate the feasibility of executing in-DRAM row copy operations in unmodified off-the-shelf DRAM chips.

\paratitle{In-DRAM Data Relocation} FIGARO~\cite{wang2020figaro} is a mechanism that enables data relocation in DRAM. 
It is built upon the key observation that the local row buffers inside a subarray in a bank are connected to a single shared global row buffer. FIGARO takes advantage of this connectivity to perform column-granularity data relocation across subarray without using the off-chip memory channel. 
To relocate data, FIGARO introduces a new DRAM command called \texttt{RELOC} (\emph{relocate column)}. The \texttt{RELOC} command moves a column $A$ (i.e., \SI{64}{\byte} data) stored in subarray $S_{A}$ to another column $B$ stored in subarray $S_{B}$ by:
\li~activating the DRAM row that contains column $A$ in $S_{A}$, which latches the DRAM row to subarray $S_{A}$'s local row buffer;
\lii~selecting column $A$ from subarray $S_{A}$'s local row buffer, which loads the column into the global row buffer; \liii~simultaneously connecting the global row buffer to subarray $S_{B}$ using subarray's $S_{B}$'s column decoder, which effectively places column $A$ from subarray $S_{A}$ into column $B$ of subarray $S_{B}$;
\liv~activating subarray $S_{B}$, which overwrites column $B$ in the activated row with column $A$, since the global row buffer has a higher drive strength of the local row buffer~\cite{itoh2013vlsi,wang2020figaro}; and
\lv~precharging the entire bank for future accesses.

\paratitle{In-DRAM AND/OR/NOT}
Ambit~\ambit shows that simultaneously activating \emph{three} DRAM rows, via a DRAM operation called \emph{\gls{TRA}}, can perform \emph{in-DRAM} bitwise AND and OR operations.
Upon sensing the perturbation of the three simultaneously activated rows, the sense amplifier amplifies the local bitline voltage to $V_{DD}$ or 0 if at least two of the capacitors of the three DRAM cells are charged or discharged, respectively. 
As such, a \gls{TRA} results in a Boolean majority operation ($MAJ$).
A majority operation $MAJ$ outputs a 1 (0) only if more than half of its inputs are 1 (0). 
In terms of AND ($\cdot$) and OR (+) operations, a 3-input majority operation can be expressed as \texttt{MAJ(A, B, C) = A $\cdot$ B + A $\cdot$ C + B $\cdot$ C.} 
To achieve functional completeness, Ambit implements NOT operations by exploiting the differential design of DRAM sense amplifiers. 
As Section~\ref{sec:background_dramorg} explains, the sense amplifier already generates the complement of the sensed value as part of the activation process.
Therefore, Ambit simply forwards  the complement of the sensed value to a special DRAM row in the subarray that consists of DRAM cells with \emph{two} access transistors, called \emph{dual-contact cells} (DCCs). 
Each access transistor is connected to one side of the sense amplifier and is controlled by a separate wordline (\emph{d-wordline} or \emph{n-wordline}).
By activating either the d-wordline or the n-wordline, the row of DCCs can provide the true or negated value stored in the row's cells, respectively.
Ambit  defines a new command called \texttt{AP} that issues a \gls{TRA} to compute a $MAJ$, followed by a \texttt{PRE} to close all three rows.
\footnote{Although the `\texttt{A}' in \texttt{AP} refers to a \gls{TRA} operation instead of a conventional \texttt{ACT}, we use this terminology to remain consistent with the Ambit paper~\cite{seshadri2017ambit}, since an \texttt{ACT} can be internally translated to a TRA operation by the DRAM chip~\cite{seshadri2017ambit}.} 
Since TRA operations are destructive, Ambit divides DRAM rows into \emph{three groups} for \gls{PuD} computing: 
\li~the \textbf{D}ata group, which contains regular data rows;
\lii~the \textbf{C}ontrol group, which consists of two rows (\texttt{C0} and \texttt{C1})  with all-0 and all-1 values; and 
\liii~the \textbf{B}itwise group, which contains six rows designated for computation (four regular rows, \texttt{T0}, \texttt{T1}, \texttt{T2}, \texttt{T3}; and two rows, \texttt{DCC0} and \texttt{DCC1}, of dual-contact cells for NOT).
The B-group rows are designated to perform bitwise operations. 
They are all connected to a special local row decoder that can simultaneously activate three rows using a single address 
(i.e., perform a \gls{TRA}).

\paratitle{Generalizing In-DRAM Majority} SIMDRAM~\cite{hajinazarsimdram} proposes a three-step framework to implement \gls{PuD} operations. 
In the first step, SIMDRAM converts an AND/OR/NOT-based representation of the desired operation into an equivalent optimized MAJ/NOT-based representation. 
By doing so, SIMDRAM reduces the number of TRA operations required to implement the operation.
In the second step, SIMDRAM generates the required sequence of DRAM commands to execute the desired operation. 
Specifically, this step translates the MAJ/NOT-based implementation of the operation into \texttt{AAPs}/\texttt{APs}.
This step involves 
\li~allocating the designated compute rows in DRAM to the operands and 
\lii~determining the optimized sequence of \texttt{AAPs}/\texttt{APs} that are required to perform the operation. 
This step's output is a \uprog, i.e., the optimized sequence of \texttt{AAPs}/\texttt{APs} that will be used to execute the operation at runtime.
Each \uprog corresponds to a different \emph{bbop} instruction, which is one of the CPU ISA extensions to allow programs to interact with the SIMDRAM framework. 
In the third step, SIMDRAM uses a control unit in the memory controller to execute the \emph{bbop} instruction using the corresponding \uprog.  
SIMDRAM implements 16 \emph{bbop} instructions, including abs, add, bitcount, div, max, min, mult, ReLU, sub, and-/or-/xor-reduction, equal, greater, greater equal, and if-else. 

\begin{figure}[ht]
    \centering
    \includegraphics[width=0.7\linewidth]{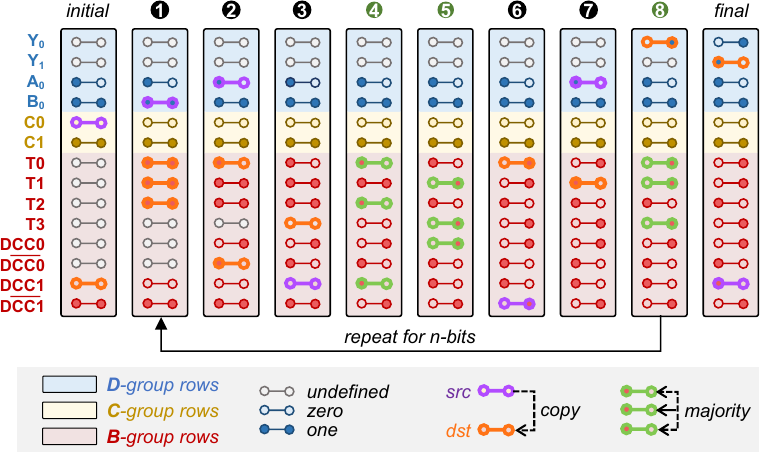}
    \caption{Full adder operation in SIMDRAM.}
    \label{fig:simdram:example}
\end{figure}

Figure~\ref{fig:simdram:example} illustrates how SIMDRAM executes a one-bit full addition operation using the sequence of row copy (\texttt{AAP}) and majority (\texttt{AP}) operations in DRAM. 
The figure shows one iteration of the full adder computation that computes \texttt{Y$_{0}$ = A$_{0}$ + B$_{0}$ + C$_{in}$}.

First, SIMDRAM uses a \emph{vertical} data layout, where all bits of a data element are placed in a single DRAM column when performing \gls{PuD} computation.  
Consequently, SIMDRAM employs a \emph{bulk bit-serial \gls{SIMD}} execution model, where each data element is mapped to a column of a DRAM row. 
This allows a DRAM subarray to operate as a \emph{\gls{PuD} SIMD engine}, where a single bit-serial operation is performed over a large number of independent data elements (i.e., as many data elements as the size of a logical DRAM row, for example, 65,536) at once. 
Second, as shown in the figure, each iteration of the full adder requires five \texttt{AAP}s (\circled{1}, \circled{2}, \circled{3}, \circled{6}, \circled{7}) and three \texttt{AP}s (\circledgreen{4}, \circledgreen{5}, \circledgreen{8}). 
A bit-serial addition of $n$-bit operands needs $n$ iterations, thus $(8 × n + 2)$ \texttt{AP}s and \texttt{AAP}s~\cite{hajinazarsimdram}.

\chapter{Related Work}
\label{chap:related}

This chapter provides an overview of the broader literature related to \gls{PIM} architectures.
First, in \gft{3}{Section~\ref{chap:related:pnm}, we give an overview of the design space of \gls{PnM} architectures.} 
Second, in Section~\ref{chap:related:pum}, we discuss the implementation of \gls{PuM} using various memory technologies, including SRAM (Section~\ref{chap:related:pum:sram}), DRAM (Section~\ref{chap:related:pum:dram}), \gls{NVM} (Section~\ref{chap:related:pum:nvm}), and flash (Section~\ref{chap:related:pum:flash}).
Third, in Section~\ref{chap:related:pimsys}, we discuss closely related works that aim to address system challenges for \gls{PIM} architectures, including
\li~workload characterization \& benchmark suites (Section~\ref{chap:related:pimsys:workload}),
\lii~application suitability (Section~\ref{chap:related:pimsys:suitability}),
\liii~compiler support (Section~\ref{chap:related:pimsys:compiler}),
\liv~memory management (Section~\ref{chap:related:pimsys:memory}), and
\lv~programming frameworks \& high-level APIs (Section~\ref{chap:related:pimsys:memory}).

\section{Processing-Near-Memory (PNM) Architectures}
\label{chap:related:pnm}

\gls{PnM} architectures integrates computation capabilities (via accelerators, simple processing cores, reconfigurable logic, for example) closer to memory (either within the memory chips themselves or in close physical proximity)~\pnm. 
\gft{3}{Despite this physical proximity, \gls{PIM} logic and memory elements in a \gls{PnM} architecture typically maintain a logical and architectural separation, allowing the memory to retain its main role as a storage substrate and the \gls{PIM} logic to be designed independently to match specific performance, area, and energy targets. 
As a result, \gls{PnM} exposes a broad design space in which compute and memory can be co-designed but still optimized separately. 
Hence, the \gls{PnM} design space can be systematically described along three key dimensions: 
\li~\emph{memory technology}, i.e., the memory technology (e.g., SRAM, DRAM, flash, disk) with which the logic is integrated;
\lii~\emph{offloading granularity}, i.e.,  the granularity of computation offloaded from the host processor to near-memory logic (e.g., a single instruction, an application kernel, or an entire application); and 
\liii~\emph{computation type}, i.e., the nature of the PIM logic embedded near or within the memory arrays (e.g., general-purpose, fixed-function unit, or specialized accelerator).
Next, we briefly describe how each one of these dimensions impacts the design of \gls{PnM} architectures.
}

\paratitle{Memory Technology} \gft{3}{\gls{PnM} systems have been explored across a spectrum of memory technologies, including SRAM (e.g., \cite{lockerman2020livia, wang2023affinity, schwedock2024leviathan, schwedock2022tako, wang2022near, wang2023infinity, nori2021reduct, denzler2021casper, denzler2023casper, vieira2021compute,eggermann202316,shacham2009smart,zhang2018minnow}), DRAM (e.g., \cite{ahn2015scalable,akin2014hamlet,akin2015data,Asghari-Moghaddam_2016,asghari2016chameleon,azarkhish2016logic,azarkhish2018neurostream,babarinsa2015jafar,besta2021sisa,boroumand2017lazypim,boroumand2018google, boroumand2019conda,boroumand2021google,boroumand2021mitigating,boroumand2021polynesia,boroumand2022polynesia, cali2020genasm,CASES_MVX,cho2020mcdram,C_RAM_1999,dai2018graphh,de2018design,devaux2019true,DRAMA_CAL_2014, drumond2017mondrian,farmahini2015nda,fernandez2020natsa,gao2016hrl,gao2017tetris,giannoula2022sparsep,gomez2021benchmarkingcut,gomez2022benchmarking,gomezluna2021benchmarking,guo20143d,hsieh2016accelerating,hsieh2016transparent,huang2019active,huang2020heterogeneous,IRAM_Micro_1997,ke2021near,kersey2017lightweight,kim2017grim,kim2018grim,kwon202125,lazypim,lee2021hardware,li2019pims,liu2018processing,nai2017graphpim, NDC_ISPASS_2014, NIM,niu2022184qps,pattnaik2016scheduling, PEI, RVU,shin2018mcdram,singh2019napel,singh2020nero,skhynixpim,Sparse_MM_LiM,sun2021abc,syncron,top-pim,tsai:micro:2018:ams,Xi_2015,zhang2018graphp,zhuo2019graphq,lim2017triple,smc_sim,HIVE,jang2019charon,IBM_ActiveCube,rezaei2020nom,hall1999mapping,hadidi2017cairo,santos2018processing,MEMSYS_MVX}), and non-volatile memories, including NAND flash~\cite{mansouri2022genstore,pei2019registor,jun2018grafboost, do2013query, seshadri2014willow,kim2016storage, gu.isca16, kang2013enabling, wang2019project,jun2015bluedbm, jun2016bluedbm, torabzadehkashi2019catalina, lee2020smartssd, ajdari2019cidr, koo2017summarizer,Cho_2013,jeong2019react, jun2016storage,ghiasimegis2024,tiwari2013active,cho2013active,wang2016ssd,kang2019towards,ruan2019insider,salamat2021nascent,soltaniyeh2022near} and magnetic disks~\cite{riedel2001active,acharya.1998,keeton.1998,riedel.1998}. 
The choice of memory technology for \gls{PnM} is heavily influenced by the characteristics of the target application (e.g., the degree of spatial and temporal locality, data reuse, and memory footprint). 
These characteristics determine not only the suitability of a memory technology in terms of capacity and performance, but also the feasibility and efficiency of performing computation near it. 
Moreover, since \gls{PnM} logic operates within or near the memory hierarchy, it is subject to the latency and energy-per-access constraints of the underlying memory substrate. 
For instance, SRAM provides fast and energy-efficient access, but its low density and high leakage make it less suitable for large-scale data processing. 
In contrast, DRAM and NAND flash offer higher storage densities but exhibit higher access latencies and energy costs, with additional constraints imposed by their respective manufacturing processes.

The feasibility of manufacturing \gls{PnM} logic is closely tied to the fabrication process used for the target memory technology. 
For example, integrating logic near SRAM arrays is relatively straightforward, as both memory and logic can be fabricated using standard CMOS processes, enabling tight integration without requiring process specialization. 
Similarly, NAND flash systems already incorporate logic, such as simple programmable microcontrollers, for purposes such as wear leveling, garbage collection, and address translation~\cite{kim2020evanesco,luo2018improving,luo2018heatwatch,luo2015warm,cai2013error,cai2012error,cai2012flash,yucai.bookchapter18,cai.bookchapter18.arxiv,yucai-thesis,luo.thesis18,cai.procieee17}. 
\gls{PnM} systems targeting NAND flash often repurpose these existing controllers for specialized data processing tasks, such as filtering~\cite{ghiasi2022genstore,jun2015bluedbm,jun2016bluedbm} or aggregation~\cite{zhang2018flashabacus,De_2013}, thus avoiding the need for major fabrication changes.

In contrast, integrating \gls{PIM} logic with DRAM presents significant challenges due to DRAM's specialized manufacturing process targeting high-density and low cost. 
Logic transistors fabricated in a DRAM process suffer from limited performance due to thicker gate oxides, longer channel lengths, and lower drive strengths, making them less energy-efficient and slower than transistors fabricated in standard logic processes~\cite{kim1999assessing}. 
In addition, DRAM chips are built with a limited number of metal layers, which significantly constrains the area available for interconnect and limits the complexity of the logic that can be integrated~\cite{devaux2019true}. Thermal constraints are also critical, as DRAM is sensitive to temperature increases that can affect data retention times and increase refresh frequency~\cite{patel2017reach,liu2013experimental,khan2014efficacy,liu2012raidr,hamamoto1998onthe}, thereby further complicating logic integration. 
These limitations impose tight area and power budgets on any logic embedded directly within DRAM die. 
To address these integration barriers, recent advances in 3D integration technologies, such as \gls{HBM}~\cite{HBM,hbm2} and HMC~\cite{HMC2,hmc.spec.1.1, hmc.spec.2.0}, enable the stacking of logic and memory dies using \glspl{TSV} and hybrid bonding techniques~\cite{lau2023recent}. These approaches allow logic to be fabricated in a more suitable process node, separate from the DRAM die, while maintaining high-bandwidth, low-latency connectivity between logic and memory. By partially decoupling the conflicting fabrication constraints of logic and memory, these integration techniques provide a practical path forward for building more capable \gls{PnM} systems, especially for DRAM-based architectures.
}

\paratitle{Offloading Granularity} \gls{PnM} systems can vary in terms of their offloading granularity, i.e., in the granularity with which computation is offloaded to \gls{PIM} logic. 
At the finest granularity, \gls{PnM} systems that implement \emph{instruction-level offloading} (e.g., ~\cite{PEI,nai2017graphpim,hadidi2017cairo,chen2022general,nai2015instruction,nai2019thermal,sokulski2022sapive,sokulski2022spec,santos2021enabling,gokhale1995processing,nai2018coolpim,li2019pims,ahmed2019compiler}) allow the execution of individual instructions or fine-grained operations directly within memory. 
This fine-grained approach can have significant benefits in terms of potential adoption, since existing processor-centric execution models already operate (i.e., perform computation) at the granularity of individual instructions, and all such machinery can be reused to aid offloading to be as seamless as possible with existing programming models and system mechanisms (such as data coherency). 
For example, the authors of~\cite{PEI} propose \emph{PIM-Enabled Instructions} (PEI), a collection of simple instructions, generated by the compiler or programmer to indicate potentially \gls{PIM}-offloadable operations in the program, that can be executed either on a traditional host CPU (that fetches and decodes them) or on the PIM logic in 3D-stacked memory. 
PEI simplifies system integration by defining PIM instructions that are cache-coherent, virtually addressed by the host processor, and operate on only a single cache block. 
It requires no changes to the sequential execution and programming model, no changes to virtual memory, minimal changes to cache coherence, and no need for special data mapping to take advantage of PIM (because each PEI is restricted to a single memory module due to the single cache block restriction it has). 
Such system-level simplifications allow PEI to implement a \emph{locality-aware execution} runtime mechanism that dynamically decides where to execute a PEI (that is, the host processor or PIM logic) based on simple locality characteristics and simple hardware predictors. 
This runtime mechanism executes the PEI at the location that maximizes performance. 
In summary, PIM-Enabled Instructions provide the illusion that PIM operations are executed as if they were host instructions: the programmer may not even be aware that the code is executing on a PIM-capable system, and the exact same program containing PEIs can be executed on conventional systems that do not implement PIM.
Other works~\cite{nai2017graphpim,hadidi2017cairo,nai2015instruction,nai2019thermal,nai2018coolpim,li2019pims,ahmed2019compiler} built on top of the HMC 2.0 architecture~\cite{HMC2}, which implements a fixed set of atomic-like near-memory instructions that perform simple read-modify-write operations close to memory. 
Despite its simplicity and coherence advantages, instruction-level offloading often incurs increased traffic between the host CPU and PIM logic due to the fine granularity of the offloaded instructions and associated control metadata.

At a coarser level, \gls{PnM} systems that implement \emph{function-level offloading} (e.g., ~\cite{singh2020nero, boroumand2018google, boroumand2021google, kim2017grim, kim2018grim, cali2022segram, cali2020genasm, soysal2025mars, azarkhish2018neurostream,ke2019recnmp,asgari2020mahasim,Kim2018HowMC,liang2019ins,glova2019near, fernandez2020natsa, gu2020dlux,radulovic2015another,boroumand2021mitigating,yavits2021giraf,herruzo2021enabling,asgarifafnir}) move portions of an application (e.g., an application function) to \gls{PIM} logic. 
This design can better partition the compute load, allowing the host CPU to focus on compute-intensive tasks while the PIM logic accelerates data-intensive operations, thereby reducing instruction traffic between the CPU and memory. 
However, function-level offloading introduces two key challenges in system integration. 
First, determining the \emph{appropriate granularity} for function-level offloading is non-trivial, since the semantic definition of a function in software does \emph{not} always align with uniform execution characteristics (i.e., a single software-defined function may interleave memory-bound and compute-bound regions). 
This issue can be addressed in two primary ways.
In the first way, the programmer can \emph{manually} identify memory-bound functions by performing application profiling during development and then selectively transfer them to PIM. 
This methodology is widely used in the literature~\cite{singh2020nero, boroumand2018google, boroumand2021google, kim2017grim, kim2018grim, cali2022segram, cali2020genasm, soysal2025mars, azarkhish2018neurostream,ke2019recnmp,asgari2020mahasim,Kim2018HowMC,liang2019ins,glova2019near, fernandez2020natsa, gu2020dlux,radulovic2015another,boroumand2021mitigating,yavits2021giraf,herruzo2021enabling,asgarifafnir}. 
For example, the authors of~\cite{boroumand2018google} manually profile key consumer workloads (i.e.,  Chrome web browser~\cite{chrome}, TensorFlow Mobile~\cite{mobile-tensorflow}, video playback~\cite{youtube}, and video capturing~\cite{google-hangouts}) and identify memory-bound kernels (such as texture tiling, color blitting, and page compression/decompression in Chrome web browser; quantization and packing in TensorFlow Mobile; and sub-pixel interpolation, deblocking filter, motion
estimation in video playback/capturing) as suitable offloading targets. 
In bioinformatics workloads, prior works such as GRIM-Filter~\cite{kim2017grim, kim2018grim} target memory-intensive bitmap filtering operations in sequence alignment, while GenASM~\cite{cali2020genasm} accelerates the memory-bound dynamic programming steps in genome assembly. 
These works demonstrate that coupled static profiling is effective in identifying offloadable units of computation in various application domains.
Second, an alternative approach is to use dynamic profiling and segmentation, in which the system itself analyzes runtime behavior to automatically partition code into offloadable segments. 

In the second way, compiler and/or hardware runtime mechanisms aim to identify group of instructions belonging to high-level software constructs (e.g., loops, software-defined functions, warps) that can be suitable for \gls{PIM} offloading~\cite{ghiasi2022alp, hsieh2016transparent, kim.sc17, vadivel2020tdo, xu2025identifying, maity2025framework, jiang20243, chen2022general, khan2022cinm,wei2022pimprof}.
For example, TOM (Transparent Offloading and Mapping)~\cite{hsieh2016transparent} introduces new compiler analysis techniques to identify code blocks (in the \emph{wrap} granularity) in GPU kernels that can benefit from offloading to PIM engines. 
The compiler estimates the potential memory bandwidth
savings for each code block. 
To this end, the compiler compares the bandwidth consumption of the code block, when executed on the regular GPU cores, to the bandwidth cost of transmitting/receiving input/output registers, when offloading to the GPU cores in the logic layer of a 3D-stacked memory.  At runtime, a final offloading decision is made based on dynamic system conditions, such as contention for processing resources in the logic layer.   
Another work, ALP~\cite{ghiasi2022alp}, addresses this by introducing the notion of \emph{segments} (a sequence of instructions), and different segments can execute either in CPU cores or PIM cores. 
Such an execution model introduces inter-segment data movement overhead~\cite{suleman2010data,suleman2011data}, i.e., performance and energy overheads due to the cost of moving data between segments of an application that executes in different types of hardware (i.e., CPU cores and PIM cores). 
ALP alleviates the inter-segment data movement overhead by \emph{proactively and accurately} transferring the required data between the segments mapped onto CPU cores and PIM cores. ALP uses a compiler pass to identify these instructions and uses specialized hardware support to transfer data between the CPU cores and PIM cores at runtime. 
Using both the compiler and runtime information, ALP  efficiently maps application segments to either CPU or PIM cores.

Second, maintaining coherence between PIM logic and host CPU during function-level execution is challenging due to the longer-lived and more stateful nature of offloaded functions. Traditional coherence protocols, such as MESI~\cite{goodman.isca83, papamarcos1984low}, often impose high overhead when applied to PIM systems~\cite{lazypim, boroumand2017lazypim, boroum2019conda}. To mitigate this, mechanisms such as CoNDA~\cite{boroum2019conda} and LazyPIM~\cite{lazypim, boroumand2017lazypim} propose lazy coherence approaches that allow the speculative execution of offloaded kernels without immediate coherence enforcement. ConDA defers coherence validation until the end of the execution of the offloaded function, reducing coherence-related delays and synchronization overhead. 
Such designs strike a balance between correctness and performance by ensuring that coherence violations are detected and resolved in a deferred, lightweight manner.

In summary, function-level offloading provides significant performance and energy benefits while amortizing the offloading cost introduced by instruction-level offloading. 
As expected, the benefits are not as high as full offloading and customization of the application level \gls{PnM} system, as we discuss next. This is expected since function-level offloading makes much fewer changes to the system and the programming model than application-level offloading, customization, and rethinking of the system.
As such, function-level offloading is likely easier to adopt in the short-term for a wide variety of workloads and systems.

At the coarsest granularity, \gls{PnM} that implement \emph{application-level offloading} (e.g.,~\cite{ahn2015scalable, mutlu2023tesseractretrospective, boroumand2022icde, fernandez2020natsa, zhuo2019graphq, zhang2018graphp,dai2018graphh}) delegates the execution of entire applications to dedicated PIM cores. 
In such approach, an entire application is re-written to completely execute on the \gls{PnM} substrate, potentially using a specialized programming model and specialized architecture/hardware. 
This approach is especially promising because it can provide the maximum performance and energy benefits achievable from \gls{PnM} acceleration of a given application, since it enables the customization of the \emph{entire} \gls{PnM} system for the application. 
This approach can be especially promising for widely-used data-intensive applications, such as machine learning (including neural networks and large language models), databases, graph analytics, genome analysis, high-performance computing, security, data manipulation, and a wide variety of mobile and server-class workloads.

Tesseract~\cite{ahn2015scalable, mutlu2023tesseractretrospective} exemplifies this model by running entire graph applications directly on memory-side processors, with host CPUs serving primarily as orchestrators. 
Tesseract adopts a message-passing programming model that obviates the need for conventional cache coherence by decoupling PIM cores from the host memory hierarchy.
Tesseract consists of 
\li~new hardware architecture that {effectively} utilizes the
available memory bandwidth in 3D-stacked memory by placing simple in-order processing cores in the logic layer and enabling each core to manipulate data only on the memory partition it is assigned to
control, 
\lii~an efficient method of communication between different
in-order cores within a 3D-stacked memory to enable each core to request computation on data elements that reside in the memory partition controlled by another core, and 
\liii~a message-passing based programming interface, similar to how modern distributed systems are programmed, which enables remote function calls on data that resides in each memory partition. 
The Tesseract design moves functions (i.e., computations and temporary values) to data that is to be updated rather than moving data elements across different memory partitions and cores. It also includes two hardware prefetchers specialized for memory access patterns of graph  processing, which operate based on the hints provided by our programming model. 

Using the Tesseract-based \gls{PnM} approach to accelerate graph processing can lead to more than two orders of magnitude improvements both in performance and in energy efficiency compared to a conventional processor-centric system with high bandwidth memory. 
This demonstrates the potential promise of designing an entire PNM system completely from the ground up for important data-intensive applications.

However, application-level offloading comes with its own challenges, most notably under-utilization of system resources when the workload's memory access patterns or compute phases do not map cleanly to the PIM cores. Moreover, this model often requires re-architecting applications for the PIM execution environment, making it less amenable to incremental adoption in general-purpose systems.

In summary, the choice of offloading granularity in PNM systems influences not only performance but also the complexity of system integration, programming models, and coherence mechanisms. While finer-grained offloading offers greater transparency and compatibility with existing software stacks, coarser-grained approaches provide higher potential performance gains but require deeper software and hardware co-design.

\paratitle{Computation Type} The nature of integrated near-memory compute logic ranges from fixed function accelerators~\cite{RVU, NIM, lee2021hardware,skhynixpim,PEI,kim2021aquabolt,gao2017tetris,yazdanbakhsh2018dram,fernandez2020natsa,ke2019recnmp,ke2021near,niu2022184qps,lee2022improving} designed to support narrow, high-throughput operations (e.g., data copy, bitwise logic, vector operations) to general-purpose processor cores capable of executing arbitrary code~\cite{ghiasi2022alp, ahn2015scalable,upmem, devaux2019true,gomez2022benchmarking, gomez2022machine, giannoula2022sparsep, drumond2017mondrian, item2023transpimlib,top-pim,devic2022pim,zhuo2019graphq,azarkhish2016logic,smc_sim,IRAM_Micro_1997}. 
Fixed-function designs typically offer higher energy efficiency and lower area cost, while general-purpose PNM cores provide programmability and flexibility, often at the cost of increased design and runtime complexity. 
Some designs employ a hybrid model that combines programmable engines with specialized accelerators to strike a balance between efficiency and generality~\cite{boroumand2022polynesia}.

A commonality across this diverse space is the higher data access efficiency that \gls{PnM} logic enjoys relative to conventional off-chip compute units. 
By being co-located with memory, \gls{PnM} logic can exploit higher bandwidth, lower latency, and lower energy per access, which is especially beneficial for data-intensive workloads. However, these benefits come at the cost of reduced memory density and increased system complexity, requiring careful hardware-software co-design.

\section{Processing-Using-Memory (PUM) Architectures}
\label{chap:related:pum}

Many prior works propose to leverage the operational principles of the memory circuitry to enable operations within memory arrays, hence implementing \gls{PuM} architectures. Such works demonstrate that \gls{PuM} is possible using different memory technologies, including
\li~\gls{SRAM}~\cite{aga2017compute,eckert2018neural,dualitycache,kang2014energy},  
\lii~DRAM~\cite{chang2016low,seshadri2017ambit, hajinazarsimdram,seshadri2013rowclone,seshadri2019dram,seshadri2016processing,seshadri.bookchapter17,seshadri2016buddy,seshadri2015fast,angizi2019graphide, ferreira2021pluto,mimdramextended,missingnot,yuksel2024simultaneous,olgun2022pidram}, 
\liii~emerging \gls{NVM} (such as \gls{MRAM}~\cite{angizi2018pima,angizi2018cmp,angizi2019dna}, \gls{ReRAM}~\cite{levy.microelec14,kvatinsky.tcasii14,Shafiee2016,kvatinsky.iccd11,kvatinsky.tvlsi14,gaillardon2016plim,bhattacharjee2017revamp,hamdioui2015memristor,xie2015fast,hamdioui2017myth,yu2018memristive,yavits2021giraf, xi2020memory, zheng2016tcam, truong2021racer, truong2022adapting}, and 
\gls{FeRAM}~\cite{ma20232,slesazeck20192tnc,wang20211t2c}), and 
\liv~NAND flash~\cite{flashcosmos,gao2021parabit,choi2020flash,han2019novel,merrikh2017high,wang2018three,lue2019optimal,kim2021behemoth,wang2022memcore,han2021flash,kang2021s,lee2020neuromorphic,lee20223d}. 
In summary, such \gls{PuM} architectures enable a wide range of different functions, such as bulk as well as finer-grained data copy/initialization~\cite{chang2016low,seshadri2013rowclone,aga2017compute,rezaei2020nom,wang2020figaro,olgun2022pidram,yuksel2024simultaneous}, bulk bitwise operations (e.g., a complete set of Boolean logic operations)~\cite{seshadri2017ambit,li2016pinatubo,angizi2018pima,angizi2018cmp,angizi2019dna,seshadri2015fast,seshadri2016buddy,seshadri2019dram,aga2017compute,li2017drisa,mutlu2019rowhammer,mandelman2002challenges,chang.sigmetrics2016,xin2020elp2im,gao2019computedram, olgun2021pidram, olgun2022pidram, li2018scope}, 
simple arithmetic operations (e.g., addition, multiplication, implication)~\cite{levy.microelec14,kvatinsky.tcasii14,aga2017compute,kang2014energy,li2017drisa,Shafiee2016,eckert2018neural,dualitycache,kvatinsky.iccd11,kvatinsky.tvlsi14,gaillardon2016plim,bhattacharjee2017revamp,hamdioui2015memristor,xie2015fast,hamdioui2017myth,yu2018memristive,deng2018dracc,angizi2019graphide}, 
and lookup table queries~\cite{ferreira2021pluto}.
In this section, we provide an overview of \gls{PuM} such architectures. 

\subsection{SRAM-Based \gls{PuM} Architectures}
\label{chap:related:pum:sram}

SRAM arrays can be used to implement bitwise Boolean operations~\cite{aga2017compute,kang2015energy,jeloka201628,eckert2018neural,dualitycache,simon2020blade,nag2019gencache,1802.shs,si2019dual,wang2019bit,al2020towards,kang2014energy,kim2021colonnade,jiang2020c3sram,wang2023infinity,agrawal2018x}.
Similarly to processing-using-DRAM (described in Section~\ref{chap:bg}), a Boolean operation is realized by simultaneously activating multiple rows, specifically two rows containing the input operands and exploiting the parasitic bitline capacitance produced when sensing the produced bitline voltage level.
If \emph{both} the activated rows store a logic-`1' value, the bitline voltage stays high and thus the sense amplifier senses `1'. 
However, if either one of the activated rows store a logic-`0' value, the bitline voltage goes below the reference sensing voltage $V_{ref}$, and the sense amplifier will sense a `0'. 
A similar process occurs when sensing the complementary bitline, i.e., the sense amplifier connecting the complementary bitline outputs `1' \emph{only} when both activated rows store a logic-`0' value. 
Therefore, a multiple row activation operation in \gls{SRAM} array leads to the computation of logic \texttt{AND} and \texttt{NOR} operations.

One promising direction for \gls{SRAM}-based \gls{PuM} is to leverage the \gls{SRAM} arrays already present in the cache hierarchy of modern computers for \emph{in-situ} computation. 
We illustrate such an approach to \gls{SRAM}-based \gls{PuM} with three examples from recent scientific literature: 
\li~Compute Caches~\cite{aga2017compute}, 
\lii~Neural Cache~\cite{eckert2018neural}, and 
\liii~Duality Cache~\cite{dualitycache}. 
Compute Caches~\cite{aga2017compute} exploits the ability to implement bitwise Boolean operations in \gls{SRAM} by turning the \gls{LLC} of a processor into an in-memory \gls{SIMD} engine, where each bitline of \gls{SRAM} subarray becomes a \gls{SIMD} lane. 
It also extends the functionatilies of \gls{SRAM}-based \gls{PuM} to other operations, including \texttt{NOR}, compare, search, copy, and carryless multiplication operations.
Neural Cache~\cite{eckert2018neural} and Duality Cache~\cite{dualitycache} build on top of Compute Caches by implementing bit-serial arithmetic inside \gls{LLC} slices.
The Neural Cache architecture~\cite{eckert2018neural} targets accelerating \gls{NN} inference by re-purposing \gls{LLC} slices into a \gls{SIMD} engine capable of performing bit-serial arithmetic operations over integer or fixed-point data. 
Similar to SIMDRAM~\cite{hajinazarsimdram}, Neural Cache stores and processes data vertically across the cache rows, with bits from multiple data elements distributed across different wordlines, allowing massive parallelism to be exploited: for example, a \SI{35}{\mega\byte} \gls{LLC} re-purposed as a bit-serial \gls{SIMD} engine can execute an arithmetic operation over up to 1,146,880 1-bit input elements \emph{simultaneously} while incurring only 2\% area overhead when in computation mode~\cite{eckert2018neural}.

Duality Cache~\cite{dualitycache} further enhances Neural Cache by targeting the execution of general-purpose data-parallel program in a \gls{SIMT} execution model.
In Duality Cache's execution model, an \gls{LLC} slice is divided into 
\li~control blocks, each capable of executing 1,024 threads simultaneously;
\lii~each control block is further subdivided into thread blocks;
\liv~each thread block consists of 256 threads; and 
\lv~each thread maps to a bitline in the cache slice.
Each bitline in the cache slice represents a thread lane, with multiple threads executing the same bit-serial instruction \emph{simultaneously}.
Duality Cache further extends the computing capabilities of Compute Caches and Neural Cache by supporting floating-point and transcendental operations using bit-serial computation and CORDIC algorithms~\cite{volder1959cordic,walther1971unified} for functions such as sine and cosine. 

\gls{SRAM}-based \gls{PuM} architectures are quite attractive solutions, since differently from DRAM-based and NVM-based \gls{PuM} , they can be manufactured and tested using standardized and commercially-available CMOS-based electronic design automation tools and library cells. 
However, the much lower density and much higher cost-per-bit of SRAM compared to DRAM and NVM are the major drawbacks of processing-using-SRAM architectures for two main reasons.
First, they limit the applicability of \gls{SRAM}-based \gls{PuM} architectures to workloads with relatively small datasets that can fit within megabyte-sized SRAM arrays. 
Otherwise, the \gls{SRAM}-based \gls{PuM} architecture needs to bring data in/out the memory hierarchy, requiring extra data movement compared to DRAM and NVM-based \gls{PuM} architectures.
Second, they lead to a design that provides lower computation density. 
For example, while a \SI{35}{\mega\byte} \gls{LLC} can be deployed as a processing-using-cache \gls{SIMD} engine with 1,146,880 \gls{SIMD} lanes, an \SI{8}{\giga\byte} DRAM chip deployed as a processing-using-DRAM \gls{SIMD} engine has up to 67,108,864 \gls{SIMD} lanes.\footnote{Assuming that a DRAM row is \SI{8}{\kilo\byte}, and the DRAM module has 16 DRAM banks, each of which with 64 DRAM subarrays.} 

\subsection{DRAM-Based \gls{PuM} Architectures}
\label{chap:related:pum:dram}

\paratitle{Extending \& Improving Bulk Bitwise In-DRAM Operations} Many works have extended Ambit's approach to \gls{PuD} to accelerate various application domains.  
GraphiDe~\cite{angizi2019graphide} and \cite{ali2019memory} leverage quintuple-row activation (QRA) to implement in-DRAM addition operations. 
Even though such implementation is more performant than SIMDRAM's TRA-based implementation for in-DRAM addition, \cite{hajinazarsimdram} shows that TRA is more scalable and variation-tolerant than QRA operations. 
SpDRAM~\cite{kang2024spdram} proposes a framework for the implementation of \gls{SpMV} operations using Ambit-style TRA operations. 
DrAcc~\cite{deng2018dracc} introduces subarray-level modifications to enable in-DRAM computation of ternary-weight neural network operations, particularly bitwise logic and accumulation. It extends Ambit's architecture by repurposing one \texttt{NOT} row into a shift (SHF) row, enabling carry propagation essential for a carry-lookahead adder (CLA).
To support this functionality, DrAcc adds three transistors per column to facilitate signal control and logic operations, and enhances the sense amplifier (SA) with two additional transistors for conditional activation based on intermediate results. A minor wire reconfiguration within the DRAM cell array allows the specialized rows to flexibly support both logic inversion and carry shifting.

Prior works~\cite{li2018scope,afifi2024artemis} use stochastic-based computing to reduce the latency of executing in-DRAM bit-serial multiplication operations. To do so, they approximate  multiplication operations using bitwise \texttt{AND} operations on unary bitstreams (i.e., sequences of bits used to represent numbers probabilistically rather than in traditional binary encoding). In stochastic computing, operations such as multiplication are simplified: to compute $x \times y$, such works generate two unary bitstreams representing and performs a bitwise \texttt{AND} between them. The result is another unary bitstream where the proportion of `1's approximates $x \times y$.

\paratitle{Enabling \gls{PuD} via Enhanced DRAM Cell or Sense Amplifier (SA) Designs} DRISA~\cite{li2017drisa} performs in-situ computing using either 3T1C (3 transistors--1 capacitor) or 1T1C (1 transistor--1 capacitor) designs. In the 3T1C approach, DRISA modifies standard DRAM cells to include two read/write access transistors and an additional transistor that ties the cell onto a bitline in a \texttt{NOR} style, hence allowing the 3T1C design to natively implement a bitwise \texttt{NOR} in a single cycle. In the 1T1C approach, DRISA augments sense amplifiers (SAs) with external logic gates and latches to support operations such as \texttt{AND}, \texttt{OR}, and \texttt{NOT}. 
DRISA uses these bitwise operations (\texttt{NOR}, \texttt{AND}, \texttt{OR}, \texttt{XOR}), alongside with an in-DRAM inter-lane shifting network, to implement complex \gls{PuD} operations such as multiplication, max, and selection.

ReDRAM~\cite{angizi2019redram} enables in-situ DRAM operations using a dual-row activation (DRA) mechanism coupled with a reconfigurable SA design, which integrates additional logic elements (i.e., three inverters with distinct voltage transfer characteristics, a \texttt{NAND} gate, and a multiplexer). These enhancements allow the SA to exploit charge-sharing between two simultaneously activated DRAM rows, enabling the execution of bulk bitwise logic operations, including \texttt{NOT}, \texttt{AND}, \texttt{OR}, and \texttt{XOR}, directly on the bitlines in a single memory cycle. 
Similar to ReDRAM, FlexiDRAM~\cite{zhou2022flexidram} implements in-situ DRAM operations by introducing a reconfigurable SA design that enables native support for complex logic functions, specifically \texttt{XOR2} and \texttt{MAJ3}, using \gls{TRA}. 
The modified SA integrates additional skewed inverters, transistors, and a multiplexer, controlled by four enable signals, allowing it to perform charge-based analog \texttt{NAND}/\texttt{NOR} logic in a single cycle. 
Sieve~\cite{wu2021sieve} also modifies the SAs by including a \texttt{XNOR} gate, an \texttt{AND} gate, and a one-bit latch, which is used to perform k-mer matching.

ROC~\cite{xin2019rocdrambased} implements in-situ DRAM operations by introducing a computing unit (CU) within a DRAM array. 
Different than Ambit and DRISA, which leverage charge-sharing methods for in-DRAM computing, ROC instead utilizes diode-connected transistors between DRAM cells to enable bitwise logic within the DRAM array.
The modification to the DRAM subarray lies in augmenting each CU with two DRAM cells and an additional diode-connected transistor to realize \texttt{AND} and \texttt{OR} operations via sequential data transfers: writing to one cell influences the state of the other based on logical semantics. For instance, to perform an \texttt{OR} operation, data is copied to both cells such that if either operand is `1', the resulting cell becomes `1'. Additionally, a newly added transistor enables the \texttt{NOT} operation. To extend functionality, ROC introduces an enhanced SA equipped with additional transistors to support propagation and shift operations, enabling word-level operations like compare and increment. 

ELP$^2$IM~\cite{xin2020elp2im} executes bulk bitwise in-DRAM operations by introducing a pseudo-precharge-based mechanism that enables in-place logic using modified SA control sequences. 
Instead of relying on TRA (as in Ambit), ELP$^2$IM exploits stable, non-traditional DRAM states: by regulating only one side of the bitline pair, ELP$^2$IM creates a \emph{pseudo-precharge} state that preserves or alters the bitline voltage based on operand values. 
This voltage state is then used to selectively overwrite the destination cell in the following access cycle, thereby realizing basic \texttt{AND} and \texttt{OR} logic operations over two such cycles (an \texttt{APP--AP} sequence). 
ELP$^2$IM proposes modifying the precharge unit to decouple equalizer (EQ) signals and adding a single dual-contact row for \texttt{NOT} operations (similar to Ambit). 

Compared to such approaches, an Ambit-style \gls{PuD} \emph{minimally} modifies the DRAM subarray to efficiently implement in-situ bulk bitwise operations. 
Such design principal provides an edge in adoption Ambit-style \gls{PuD} in future systems, since DRAM's manufacturing process is highly sensitive to the area occupied by the DRAM core array. 

\paratitle{\gls{PuD} for True Random Number Generation}  Intentionally violating DRAM access timing parameters can also be used to generate true random numbers inside DRAM. 
The technique proposed in~\cite{kim2019d} decreases the DRAM row activation latency below the datasheet specifications to induce read errors, or activation failures. 
As a result, some DRAM cells, called \gls{TRNG} cells, fail truly randomly. 
By aggregating the resulting data from multiple such \gls{TRNG} cells, a technique called D-RaNGe, provides a high-throughput and low-latency \gls{TRNG}. Extensive characterization and analysis of the DRAM latency \gls{TRNG} on real \gls{COTS} DRAM chips is provided in~\cite{kim2019d}.

The authors of QUAC-TRNG~\cite{olgun2021quactrng} exploit the  observation that a carefully-engineered sequence of DRAM commands activates four consecutive DRAM rows in rapid succession to generate random numbers. 
This QUadruple ACtivation (QUAC) causes the bitline sense amplifiers to non-deterministically converge to random values when four rows that store conflicting data are activated because the net deviation in bitline voltage fails to meet reliable sensing margins. QUAC-TRNG reads the result of the QUAC operation from the sense amplifiers and performs the SHA-256 cryptographic hash function~\cite{1802.shs} to post-process the result and output random numbers. 

Another recent work proposes DR-STRaNGe~\cite{bostanci2022dr}, an end-to-end system design for DRAM-based \glspl{TRNG} that mitigates three key system integration challenges: 
\li~generating random numbers with DRAM-based \glspl{TRNG} can degrade overall system performance by slowing down concurrently-running applications due to the interference between RNG and regular memory operations in the memory controller (i.e., RNG interference), 
\lii~this RNG interference can degrade system fairness by causing unfair prioritization of applications that intensively use random numbers (i.e., RNG applications), and 
\liii~RNG applications can experience significant slowdown due to the high latency of DRAM-based \glspl{TRNG}. 
DR-STRaNGe proposes an \emph{RNG-aware scheduler} and a \emph{buffering mechanism} in the memory controller to tackle these challenges.

\paratitle{\gls{PuD} for Physically Unclonable Functions}  \Glspl{PUF} are commonly used in cryptography to identify devices based on the uniqueness of their physical microstructures. 
DRAM-based \glspl{PUF} have two key advantages: 
\li~DRAM is present in many modern computing systems, and 
\lii~DRAM has high capacity and thus can provide many unique identifiers. 
However, traditional DRAM \glspl{PUF} exhibit unacceptably high latencies and are not runtime-accessible. 
The authors of~\cite{kim.hpca18}  propose a new class of fast, reliable DRAM \glspl{PUF} that are runtime-accessible, i.e., that can be used during online operation with low latency. 
The key idea is to reduce DRAM read access latency below the reliable datasheet specifications using software-only system calls. 
Doing so results in error patterns
that reflect the compound effects of manufacturing variations
in various DRAM structures (e.g., capacitors, wires, sense amplifiers). 
Some DRAM cells fail always after repeated accesses with violated timing parameters and some others never fail at all.  
A combination of a set of such consistently failing or never failing cells can be used to generate a unique identifier for the device. 
The DRAM latency \gls{PUF} does \emph{not} require any modification to existing DRAM chips -- it only requires an intelligent memory controller that can change timing parameters and identify DRAM regions and cells that can be reliably used as \glspl{PUF}.
An extensive characterization and analysis of the DRAM latency \gls{PUF} on real \gls{COTS} DRAM chips is provided in~\cite{kim.hpca18}.

\paratitle{\gls{PuD} for Lookup Tables (LUT) Operations} LUT-based \gls{PuD}~\cite{ferreira2021pluto,ferreira2022pluto,zhou2022red,sutradhar2023flutpim,sutradhar20233dl,sutradhar2021look,deng2019lacc,gao2016draf} enables efficient implementation of complex operations, such as transcendental functions, that are costly to implement using arithmetic or bulk bitwise Boolean operations in-DRAM. By directly mapping inputs to precomputed outputs, LUT-based \gls{PuD} can improve performance and enhance flexibility of \gls{PuD} system. 
To illustrate how a LUT-based \gls{PuD} operations, we highlight pLUTo~\cite{ferreira2022pluto}, \underline{p}rocessing-using-memory with lookup table (\underline{LUT}) \underline{o}perations, a DRAM-based \gls{PuM} architecture that leverages LUT-based computing via bulk querying of LUTs to perform complex operations beyond the scope of prior \gls{PuD} proposals (i.e., Ambit and SIMDRAM). pLUTo introduces a LUT-querying mechanism, the \emph{pLUTo LUT Query}, which enables the simultaneous querying of multiple LUTs stored in a single DRAM subarray.
In pLUTo, the number of elements stored in each LUT may be as large as the number of rows in each DRAM subarray (e.g., 512{--}1024 rows~\cite{kim2012case,kim2018solar,lee2016reducing}).

pLUTo executes seven main steps to perform a \emph{pLUTo LUT Query}.
First, the \emph{source subarray} stores the \emph{LUT query input vector}, which consists of a set of $N$-bit LUT indices associated with LUT elements.
Second, the \emph{pLUTo Match Logic} comprises a set of comparators that identify matches between 
\li~the row index of the currently activated row in the pLUTo-enabled subarray, and 
\lii~each LUT index in the {LUT query input vector (i.e., the source subarray's row buffer)}.
Third, the \emph{pLUTo-enabled row decoder} enables the successive activation of consecutive DRAM rows in the pLUTo-enabled subarray with a single DRAM command. 
It also outputs the row index of the currently activated row as input to the pLUTo Match Logic.
Fourth, the \emph{pLUTo-enabled subarray} stores multiple vertical copies of a given LUT, which consists of $M$-bit LUT elements.
Fifth, the \emph{pLUTo-enabled row buffer} allows the reading of individual LUT elements from the activated row in the \emph{pLUTo-enabled subarray}. This is possible by extending the DRAM sense amplifier design of the pLUTo-enabled row buffer with switches controlled by the pLUTo Match Logic (using the \emph{matchline} signal).
Sixth, the \emph{flip-flop (FF) buffer} enables pLUTo to temporarily store select LUT elements by copying them from the pLUTo-enabled row buffer, conditioned on the output of the pLUTo Match Logic following each row activation.
Seventh, a LISA~\cite{chang2016low} operation (described in Section~\ref{chap:bg}) copies the entire contents of the FF buffer (i.e., the LUT query output vector) into the destination row buffer, i.e., the row buffer of the \emph{destination subarray}. 

In contrast to the bit-serial paradigm employed by prior \gls{PuD} architectures (e.g., SIMDRAM~\cite{hajinazarsimdram}), pLUTo operates in a \emph{bit-parallel} manner; in other words, the bits that make up each LUT element (e.g., \texttt{A}) are stored \emph{horizontally} (i.e., in adjacent bitlines), and all the copies of each LUT element (i.e., \texttt{\{A,A,...,A\}}) take up \emph{one {whole} row} in the depicted pLUTo-enabled subarray.

\paratitle{\gls{PuD} Using Commodity Off-the-Shelf DRAM} ComputeDRAM~\cite{gao2019computedram} shows that one can mimic the effect of RowClone's back-to-back activation mechanism in off-the-shelf DRAM chips by violating the timing parameters such that two wordlines in a subarray are activated back-to-back as in RowClone. 
This work shows that such a version of RowClone can operate reliably in a variety of \gls{COTS} DRAM chips tested using the SoftMC infrastructure~\cite{hassan2017softmc,softmc.github}. 
More recent works~\cite{missingnot,yuksel2024simultaneous} experimentally demonstrate the \gls{PuM} capabilities of hundred \gls{COTS} DRAM chips using DRAM Bender~\cite{olgun2023drambender,safari-drambender}, an FPGA-based DDR4 testing infrastructure that provides precise control of DDR4 commands issued to a DRAM module.
These works demonstrate that \gls{COTS} DRAM chips are capable of 
\li~performing functionally-complete bulk-bitwise Boolean operations: \texttt{NOT}, \texttt{NAND}, and \texttt{NOR},
\lii~executing up to 16-input \texttt{AND}, \texttt{NAND}, \texttt{OR}, and \texttt{NOR} operations, and \liii~copying the contents of a DRAM row (concurrently) into up to 31 other DRAM rows wit $>$99.98\% reliability.
The authors evaluate the robustness of these operations across data patterns, temperature, and voltage levels, showing that \gls{COTS} DRAM chips can perform these operations at high success rates ($>$94\%).
These fascinating findings demonstrate the fundamental computation capability of DRAM, even when DRAM chips are {\em not} designed for this purpose, and provide a solid foundation for building new and robust \gls{PuD} mechanisms into future DRAM chips and standards.

\subsection{NVM-Based \gls{PuM} Architectures}
\label{chap:related:pum:nvm}

\paratitle{Bitwise Operations in NVMs} Several works~\cite{borghetti2010memristive, linn2012beyond, li2016pinatubo,kvatinsky.tcasii14,kvatinsky.iccd11,kvatinsky.tvlsi14,lehtonen2009stateful,kim2011field,lehtonen2012applications,mahmoudi2013implication,kim2019single,xie2017scouting,gaillardon2016plim} investigate the implementation of \emph{stateful}~\cite{borghetti2010memristive,linn2012beyond,kvatinsky.tcasii14,kvatinsky.iccd11,kvatinsky.tvlsi14,lehtonen2009stateful,kim2011field,lehtonen2012applications,mahmoudi2013implication,kim2019single} and \emph{non-stateful}~\cite{li2016pinatubo,xie2017scouting,gaillardon2016plim} bitwise Boolean operations in emerging NVM technologies. 
When implementing stateful Boolean bitwise operations, all input operands and the produced result are stored in terms of the resistance state variable. 
Hence, many Boolean bitwise operations can be executed back-to-back, at the expense of consuming the limited write cycles (i.e., endurance) of the emerging NVM device. 
In contrast, when implementing non-stateful Boolean bitwise operations, the input operands are stored as resistance state variables while the output value is represented as a voltage level. 
Such implementation avoids write cycles, since the produced output value is not directly written back into the NVM cells, but requires additional sensing circuitry when cascading consecutive operations.
As examples, we highlight the implementation of stateful (e.g., MAGIC~\cite{kvatinsky.tcasii14}) and non-stateful (e.g., Pinatubo~\cite{li2016pinatubo}) NVM-based \gls{PuM} architectures.

In memresistive devices, memory cells use resistance levels to represent logic-`1' (e.g., using a low resistance level $R_{low}$) and logic-`0' (e.g., using a high resistance level $R_{high}$). 
As in DRAM, NVM arrays organize memory cells in rows and columns, and a sense amplifier is used to convert resistance difference of memory cells into voltage levels or current signals based on a reference resistance level $R_{ref}$, determining the result between `\texttt{0}' and `\texttt{1}'. 
MAGIC~\cite{kvatinsky.tcasii14} implements a Boolean \texttt{NOR} operations as follows.
Initially, a \texttt{NOR} logic gate is constructed using two input memristors ($in_{1}$ and $in_{2}$) connected in parallel and an output memristor ($out$) connected in series. 
First, the output memristor $out$ is initialized with a logic-`1' value (i.e., $R_{out}$ = $R_{low}$).
Second, a voltage pulse $V_{0}$ is applied to the gateway of the \texttt{NOR} logic gate. 
Third, $V_{0}$ produces a current that passes through the
circuit. The logical state of the output memristor $out$ switches from the initial logic-`1' value to a logic-`0' value when \emph{either} $in_{1}$ or $in_{2}$ has a low resistance value (i.e., $in_{1}$=$R_{low}$ or $in_{2}$ = $R_{low}$), since the voltage/current will be greater than the memristor threshold voltage/current. 
MAGIC implements other Boolean operations (i.e., \texttt{OR}, \texttt{NAND}, and \texttt{AND}) by changing the connections between $in_{1}$, $in_{2}$, and $out$: 
\li~a Boolean \texttt{OR} operation is implemented by connecting $in_{1}$, $in_{2}$, and $out$ as in a Boolean \texttt{NOR}, but initializing $out$ with a logic-`0' value, i.e., $R_{out}$ = $R_{high}$;
\lii~a Boolean \texttt{NAND} (\texttt{AND}) operation is implemented by connecting $in_{1}$, $in_{2}$, and $out$ in series, and initializing $out$ with a logic-`1' (logic-`0') value, i.e., $R_{out}$ = $R_{low}$ ($R_{out}$ = $R_{high}$).
A Boolean \texttt{NOT} operation is implemented by constructing an inverted logic gate using an input memristor ($in$) and an output memristor ($out$), which is initialized with a logic-`1' value, connected in series.

Pinatubo~\cite{li2016pinatubo} implements bulk bitwise Boolean operations by 
\li~performing \emph{multiple row activation}, similarly to Ambit~\cite{seshadri2017ambit}, and
\lii~modifying the array's sense amplifier by adding new reference resistance levels that are used during \gls{PuM} execution to enable different Boolean operations. 
For example, to compute bulk bitwise \texttt{A} \texttt{OR} \texttt{B} of rows \texttt{A} and \texttt{B}, Pinatubo simultaneously activates the rows containing both input operands. 
The newly-added reference resistance level for a Boolean \texttt{OR} operation (i.e., $R_{ref-or}$) then allows the sense amplifier to output `\texttt{0}' \emph{only} when $R_{A} = R_{B} = R_{ref-or}$. 
Pinatubo follows a similar approach to implement other Boolean operations, such as \texttt{OR}, \texttt{AND}, \texttt{XOR}, and \texttt{INV} operations.

NVM-based \gls{PuM} architectures~\cite{kvatinsky.tcasii14,fernandez2024matsa,lin2022all,parveen2017lowpower,angizi2020exploring,angizi2019dna,angizi2019graphs,leitersdorf2023aritpim} can implement complex arithmetic operations by decomposing such operations into a series of
simple bitwise Boolean operations (e.g., \texttt{AND}, \texttt{NOR}, \texttt{XOR}) that are executed bit-serially over vertically-layout-out input operands.
The authors of~\cite{fernandez2024matsa} use such an approach to accelerate an important data-intensive application, i.e., time series analysis (TSA)~\cite{esling2012time}. 

\paratitle{In-Memory Crossbar Array Matrix-Vector Multiplication (MVM) Operations}
Crossbar array based NVM architectures can natively execute MVM operations based on analog operational principles~\cite{song2018graphr,imani2020dual,challapalle2020gaas,bojnordi2016memristive,feinberg2018enabling,Shafiee2016,Chi2016,ielmini2018memory,wan2022compute,jung2022crossbar,sebastian2020memory,le202364,ankit2019puma,joshi2020accurate,song2017pipelayer,li2019long,valavi201964,imani2019floatpim,le2018mixed,yang2019sparse,jia2020programmable,tang2017binary,marinella2018multiscale,yuan2021forms,wen2020ckfo,ankit2020panther,wen2019memristor,long2018reram,chou2019cascade,feinberg2018making,li2020timely,angizi2019mrima,wang2018snrram,zhu2019configurable,yang2020retransformer,tang2017aepe,xia2016switched,xia2017fault,huang2017highly,cheng2017time,bojnordi2016memristive,chen2018regan,cai2018training,mao2018lergan,nag2018newton}, specifically Kirchhoff's Law~\cite{kirchhoff1859uber,kirchhoff1978verhaltnis,kirchhoff1860relation}. 
A crossbar array based NVM architecture performs a MVM operation between the input vector $V$ and the input matrix $M$, producing an output vector $O$ (i.e., $O = V \times M$) as follows~\cite{Chi2016}.
Before execution, the input matrix $M$ is stored in the crossbar array as resistance levels, where each matrix element $M_{i,j}$ is represented as a resistance level corresponding to the value of the element i.e., $M_{i,j} = \frac{1}{R_{i,j}}$.
Then, the crossbar array performs a \emph{in-situ} MVM operation by applying 
\li~a voltage level $V_i$, which represents the input vector, on the wordlines of the array that stores the matrix $M$ and 
\lii~sensing the output vector $O$ on the bitlines.
Based on Kirchhoff's Law~\cite{kirchhoff1859uber,kirchhoff1978verhaltnis,kirchhoff1860relation}, the current level sensed on the bitlines will be equal to $O_i = \sum\limits_{j} V_i \times M_{i,j} = \sum\limits_{j} V_i \times \frac{1}{R_{i,j}}$. 
Using a crossbar array based NVM, an MVM operation can be performed in nearly a single NVM read cycle (as long as the matrix fits in the crossbar array).

A large number of works leverage crossbar array based NVM's ability to natively perform MVM operations to accelerate different applications~\cite{alibart2012high,angizi2019mrima,ankit2019puma,ankit2020panther,bojnordi2016memristive,cai2018training,challapalle2020gaas,chen2018regan,cheng2017time,Chi2016,chou2019cascade,feinberg2018enabling,feinberg2018making,holmes1993use,hu2012hardware,huang2017highly,ielmini2018memory,imani2019floatpim,imani2020dual,jia2020programmable,joshi2020accurate,jung2022crossbar,le2018mixed,le202364,li2019long,li2020timely,long2018reram,mao2018lergan,marinella2018multiscale,nag2018newton,sebastian2020memory,Shafiee2016,song2017pipelayer,song2018graphr,tang2017aepe,tang2017binary,valavi201964,wan2022compute,wang2014energy,wang2018snrram,wen2019memristor,wen2020ckfo,xia2017fault,yang2019sparse,yang2020retransformer,yuan2021forms,zhu2019configurable,li2013memristor,prezioso2015training,kim2015reconfigurable,chen2015optimized,burr2015large,xia2016switched}, in particular, \gls{NN} inference and training~\cite{Shafiee2016,Chi2016,ielmini2018memory,wan2022compute,jung2022crossbar,sebastian2020memory,le202364,ankit2019puma,joshi2020accurate,song2017pipelayer,li2019long,valavi201964,imani2019floatpim,le2018mixed,yang2019sparse,jia2020programmable,tang2017binary,marinella2018multiscale,yuan2021forms,wen2020ckfo,ankit2020panther,wen2019memristor,long2018reram,chou2019cascade,feinberg2018making,li2020timely,angizi2019mrima,wang2018snrram,zhu2019configurable,yang2020retransformer,tang2017aepe,xia2016switched,xia2017fault,huang2017highly,cheng2017time,bojnordi2016memristive,chen2018regan,cai2018training,mao2018lergan,nag2018newton,prezioso2015training,burr2015large}. 
Even though such proposals are intellectually promising, NVM-based acceleration of MVM operations needs to take into account inherent device non-idealities (e.g., non-ideal analog-to-digital and digital-to-analog converters, write resistance due to imperfect writes, non-ideal sensing circuits), architectural limitations (e.g., limited write endurance), and cost \& latency considerations (e.g., due to analog-to-digital and digital-to-analog conversion and sensing circuitry) that might impact the accuracy and robustness of the target application~\cite{shahroodi2023swordfish,feinberg2018making} as well as the cost and scalability of the system.   

\paratitle{In-Memory Crossbar Array String Matching Operations} Crossbar array based NVM architectures can also perform \emph{in-situ} string matching operations. 
To enable these operations, the crossbar array is operated as a content addressable memory (CAM). 
The CAM array consists of $m \times n$ CAM cells that house \emph{m} reference strings, each of which is \emph{n-bit} long. 
Each CAM cell stores \emph{one} bit and has two programmable resistors, $R_{l}$ and $R_{r}$, and two transistors, $M_{l}$ and $M_{r}$.
To store `\texttt{1}' (or `\texttt{0}') in a CAM cell, $R_{l}$ and $R_{r}$ are programmed to high and low (or low and high) resistance levels, respectively. 
The NVM-based CAM array is able to query the existence of an \emph{n-bit} string in parallel across all \emph{m} rows in four steps.
First, the CAM array precharges the matchline signals to \emph{high} voltage. 
Second, each bit in the input string and its {complement} drive the gate voltages of $M_{l}$ and $M_{r}$ transistors of the CAM cells in the corresponding column, respectively. 
Third, each CAM cell compares its stored bit to the corresponding bit in the input string. If these two bits are different, the pull down network in the CAM cell is turned on and the matchline becomes `\texttt{0}'. Otherwise, the matchline keeps its precharged \emph{high} voltage. 
Fourth, if all bits of the input string match with all corresponding CAM cells in a row, the matchline remains high, indicating an {\emph{exact match}} between the input string and the reference string stored in the CAM array.

The authors of~\cite{mao2022genpip} leverage NVM's ability to perform in-situ MVM and string matching operations to accelerate genome analysis, where a state-of-the-art NVM-based \gls{PuM} array~\cite{lou2020helix} is used to implement basecalling (by performing \gls{NN} inference using the NVM array in-situ MVM operations) and chaining~\cite{chen2020parc} (by performing the dynamic programming algorithm using the NVM array as a CAM for string matching operations).
The authors of~\cite{shahroodi2023swordfish} developed a hardware/software framework to accelerate DNN-based genomic basecallers while taking into account the possible adverse effects of inherent NVM device non-idealities, which can greatly degrade basecalling accuracy.

We conclude that \gls{PuM} architectures with emerging non-volatile memories offer a promising approach for accelerating different classes of data-intensive workloads by minimizing data movement overhead and enabling \emph{in-situ} matrix-vector multiplication and bitwise operations. 
This approach demonstrates significant potential for performance and energy efficiency improvements, yet further work is needed to address the NVM devices' non-idealities, improve overall robustness \& cost efficiency, and investigate applicability to a wider variety of workloads.

\subsection{NAND Flash-Based \gls{PuM} Architectures}
\label{chap:related:pum:flash}

A recent work, ParaBit~\cite{gao2021parabit}, proposes using NAND flash chips for bulk bitwise operations. 
However, ParaBit has two major limitations. 
First, it falls short of maximally exploiting the bit-level parallelism of bulk bitwise operations that could be enabled by leveraging the unique cell-array architecture and operating principles of NAND flash memory.
Second, it is unreliable because it is not designed to take into account the highly error-prone nature of NAND flash memory.

Flash-Cosmos (\underline{Flash} \underline{C}omputation with \underline{O}ne-\underline{S}hot \underline{M}ulti-\underline{O}perand \underline{S}ensing)~\cite{flashcosmos} further enhances NAND flash-based bulk bitwise operations while providing high reliability with two key mechanisms: \li~\underline{M}ulti-\underline{W}ordline \underline{S}ensing (MWS), which enables bulk bitwise operations on a large number of operands (tens of operands) with a \emph{single} sensing operation, and \lii~\underline{E}nhanced \underline{S}LC-mode \underline{P}rogramming (ESP), which enables reliable computation inside NAND flash memory.

MWS leverages the two fundamental cell-array structures of NAND flash memory to perform in-flash bulk bitwise operations on a large number of operands with a \emph{single} sensing operation:
\li~a number of flash cells (e.g., 24--176 cells) are serially connected to form a NAND string (similar to digital \texttt{NAND} logic); 
\lii~thousands of NAND strings are connected to the same bitline (similar to digital \texttt{NOR} logic).
Under these cell-array structures, simultaneously sensing \emph{multiple} wordlines automatically results in 
\li~bitwise \texttt{AND} of \emph{all} the sensed wordlines if they are in the same NAND string or 
\lii~bitwise \texttt{OR} of \emph{all} the wordlines if they are in different NAND strings.
ESP effectively avoids raw bit errors in stored data via more precise programming-voltage control using the NAND flash controller.
A flash cell stores bit data as a function of the level of its threshold-voltage (V$_{\text{TH}}$). 
Reading a cell incurs an error if the cell's V$_{\text{TH}}$ level moves to another V$_{\text{TH}}$ range that corresponds to a different bit value than the stored value, due to various reasons, such as program interference~\cite{cai2013program,cai2017vulnerabilities, park-dac-2016,cai.procieee17}, data retention loss~\cite{cai2015data, cai2012flash, cai2013error, luo2018improving,cai.procieee17}, read disturbance~\cite{cai2015read, ha-ieeetcad-2015,cai.procieee17}, and cell-to-cell interference~\cite{cai2017vulnerabilities,cai.procieee17}.
ESP maximizes the margin between different V$_{\text{TH}}$ ranges by carefully leveraging two existing approaches.
First, to store data for in-flash processing, it uses the single-level cell (SLC)-mode programming scheme~\cite{lee-atc-2009, kim2020evanesco}. 
Doing so guarantees a large V$_{\text{TH}}$ margin by forming only two V$_{\text{TH}}$ ranges (for encoding `\texttt{1}' and `\texttt{0}') within the fixed V$_{\text{TH}}$ window. 
Second, ESP enhances the SLC-mode programming scheme by using 
\li~a higher programming voltage to increase the distance between the two V$_{\text{TH}}$ ranges and 
\lii~more programming steps to narrow the high V$_{\text{TH}}$ range.  
Using ESP makes bulk bitwise computation operations using flash memory completely reliable (i.e., no errors) on tested real NAND flash~\cite{flashcosmos}.

Enabling computation using flash memory to become more capable and programmable is a promising direction future research can take.
Compared to processing-using-DRAM, we believe processing-using-flash is still in its infancy. 
More work is needed to demonstrate the possibility of more complex computations, improve processing-using-flash programmability, and address system integration challenges. 
The good news is that in flash memory, the memory controller can have much more sophisticated capabilities to make computation reliable, as existing flash controllers already use many mechanisms to greatly enhance robustness and avoid errors~\cite{cai.procieee17,cai.bookchapter18.arxiv,yucai.bookchapter18}. For example, existing flash memory controllers perform data randomization, wear leveling, adaptive refresh, block remapping, garbage collection, careful tuning of read reference voltage, and sophisticated error correction~\cite{cai.procieee17,do2013query,lee2020smartssd,kang2013enabling,Cho_2013,liang2019ins,ghiasimegis2024,mansouri2022genstore,pei2019registor,jun2018grafboost, seshadri2014willow,kim2016storage, gu.isca16, wang2019project,jun2015bluedbm, jun2016bluedbm, torabzadehkashi2019catalina, ajdari2019cidr, koo2017summarizer,Cho_2013,jeong2019react, jun2016storage,cai2013error,cai2012error,cai2012flash,cai2013program,cai2013threshold,cai2014neighbor,cai2015data,cai2015read,cai2017vulnerabilities,yucai.bookchapter18,kim2020evanesco,park-dac-2016,nadig2023venice}. 
The ESP (enhanced SLC-mode programming) technique proposed in Flash-Cosmos~\cite{flashcosmos} is a great example of using already  supported mechanisms in flash memory controllers to improve the reliability of processing-using-flash techniques. 
We believe there is significant opportunity for creating novel and robust processing-using-flash techniques by exploiting and enhancing the sophisticated capabilities of flash memory controllers.

\section{System Support for \gls{PIM} Architectures}
\label{chap:related:pimsys}

\subsection{Data Movement Bottlenecks: Identification, Workload Characterization, and Benchmark Suites}
\label{chap:related:pimsys:workload}

\paratitle{Identifying Data Movement Bottlenecks} Many past works investigate how to reduce data movement cost using a range of different compute-centric, e.g., 
prefetchers~\cite{
gonzalez1997speculative,
chen1995effective, 
bera2019dspatch, 
bakhshalipour2018domino, 
fu1992stride, 
ishii2009access, 
hashemi2018learning,
orosa2018avpp,
jog2013orchestrated,
lee2011prefetch,
ebrahimi2009techniques,
srinath2007feedback,
austin1995zero,
ceze2006cava,
kadjo2014b,
kirman2005checkpointed,
cooksey2002stateless,
joseph1997prefetching,
ebrahimi2011prefetch,
ebrahimi2009coordinated,
lee2009improving,
ChangJooLee,charney1995,charneyphd} (by proactively loading data into faster memory closer to the compute units before it is explicitly requested, thus reducing access latency and improving memory bandwidth utilization), speculative execution~\cite{gonzalez1997speculative, mutlu2003runahead, hashemi2016continuous, mutlu2005techniques,mutlu2006efficient, mutlu2005using} (by preemptively executing instructions and fetching required data ahead of time, overlapping computation with memory access to hide latency and improve throughput), value-prediction~\cite{orosa2018avpp, yazdanbakhsh2016rfvp, eickemeyer1993load, endo2017interactions, gonzalez1997speculative, lipasti1996exceeding,lipasti1996value,calder1999selective,wang1997highly,gabbay1996speculative,burtscher1999exploring,fu1998value,goeman2001differential,nakra1999global,sato2002low,sazeides1997predictability,tuck2005multithreaded,tullsen1999storageless,tune2002quantifying} (by speculatively supplying predicted data values, reducing the need to fetch them from memory and thereby decreasing latency and memory bandwidth pressure), 
data compression~\cite{pekhimenko2012base, 
dusser2009zero, 
yang2000frequent,
alameldeen2004adaptive, 
zhang2000frequent,
pekhimenko2016case,pekhimenko2015toggle, pekhimenkoenergy, pekhimenko2015exploiting, pekhimenko2013linearly,chen2009c,hallnor2005unified,hammerstrom1977information,islam2009zero,arelakis2014sc,ekman2005robust,vijaykumar2015case,gaur2016base} (by reducing the volume of data transferred across the memory hierarchy), 
approximate computing~\cite{yazdanbakhsh2016rfvp,koppula2019eden,miguel2015doppelganger,oliveira2018employing}) (by enabling the use of lower-precision or lossy data representations, which reduce data size and consequently the bandwidth and energy required for memory requests), and 
memory-centric techniques~\cite{vijaykumar2018case, yazdanbakhsh2016mitigating, joao2012bottleneck, tsai2018rethinking, sembrant2014direct, PEI,lazypim,MEMSYS_MVX,top-pim,acm,New_PIM_2013,kim2017heterogeneous,pugsley2014comparing,boroumand2018google,azarkhish2016memory,nai2017graphpim,gao2016hrl,de2018design,babarinsa2015jafar,tsai2017jenga,tsai2019compress,vijaykumar2018locality,mutlu2020intelligent,mutlu2020intelligentdate} (by moving computation to where the data resides). 
These works evaluate the impact of data movement on different systems, including mobile systems~\cite{boroumand2018google, pandiyan2014quantifying, narancic2014evaluating, yan2015characterizing,shingari2015characterization}, data centers~\cite{kanev_isca2015, ferdman2012clearing, kozyrakis2010server,yasin2014deep, hashemi2018learning, gupta2019architectural,gan2018architectural,ayers2018memory}, accelerator-based systems~\cite{boroumand2018google, gupta2019architectural, singh2019napel, murphy2001characterization, cali2020genasm, kim2018grim,alser2020accelerating}, and desktop computers~\cite{limaye2018workload, bienia2008parsec, corda2019memory}. 
They use very different profiling frameworks and methodologies to identify the root cause of data movement for a small set of applications. 
Two common approaches to identify data movement bottlenecks in applications is to use profiling tools, such as  the roofline model~\cite{williams2009roofline} for the application--system~\cite{azarkhish2018neurostream,ke2019recnmp,asgari2020mahasim,Kim2018HowMC,liang2019ins,glova2019near, fernandez2020natsa, gu2020dlux, singh2020nero,gomez2022benchmarking,radulovic2015another,boroumand2021mitigating,yavits2021giraf,herruzo2021enabling,asgarifafnir}, or use simple metrics (such as last-level CPU cache \gls{MPKI}) as indication of memory pressure~\cite{hsieh2016accelerating,nai2017graphpim,kim2015understanding,boroumand2018google, nai2015instruction, ghose.ibmjrd19, lim2017triple, tsai:micro:2018:ams,kim2018coda,hong2016accelerating,gries2019performance}. 
However, as we show in Chapter~\ref{chap:damov}, even though both the roofline model and \gls{MPKI} can identify some specific sources of memory bottlenecks, they alone cannot comprehensively identify different possible sources of memory bottlenecks in a system.

\paratitle{Workload Characterization Targeting \gls{PIM}} Many works~\cite{amiraliphd,boroumand2021polynesia,boroumand2022polynesia,boroumand2018google,boroumand2021google,fernandez2020natsa,kim2017grim,gomezluna2021benchmarking,gomez2022experimental,singh2021fpga,diab2022high,he2025papi,fernandez2024matsa,denzler2021casper,lindegger2023scrooge,singh2020nero,li2024pim,giannoula2024pygim,iskandar2023ndp,chen2024pointcim,gogineni2024swiftrl,park2024attacc,ke2019recnmp} perform ad-hoc workload characterization (where they perform in-depth profiling of an application) in order to determine whether an application suffers from memory bottlenecks, and hence would potentially benefit from a \gls{PIM} design. However, such profiling approaches are closely tight to the target application under analysis, making it hard to generalize their application profiling methodology and observed findings to a broader set of applications.

\paratitle{Benchmark Suites Targeting \gls{PIM} Architectures} First, in \cite{murphy2001characterization}, the authors provide the first work that characterizes workloads for \gls{PIM}. 
They analyze five applications (FFT, ray tracing, method of moments, image understanding, data management). 
The \gls{PIM} organization \cite{murphy2001characterization} targets is similar to \cite{IRAM_Micro_1997}, where vector processing compute units are integrated into the DDRx memory modules. 
Even though~\cite{murphy2001characterization} has a similar goal to our work, it understandably does not provide insights into modern data-intensive applications and \gls{PIM} architectures as it dates from 2001. 
Also, \cite{murphy2001characterization} focuses its analysis only on a few workloads, whereas we aim to conduct a broader workload analysis starting from 345 applications. 
Therefore, a new, more comprehensive and rigorous analysis methodology of data movement bottlenecks in modern workloads and modern \gls{PIM} systems is necessary. 
Second, in~\cite{gomez2021benchmarking}, the authors present the \emph{PrIM} benchmark suite (\emph{\underline{Pr}ocessing-\underline{I}n-\underline{M}emory benchmarks})~\cite{gomezluna2021repo}, a benchmark suite of 16 workloads from different application domains (e.g., dense/sparse linear algebra, databases, data analytics, graph processing, neural networks, bioinformatics, image processing) designed for evaluating the UPMEM PIM architecture~\cite{upmem,upmem2018}. Using PrIM, the authors of~\cite{gomez2021benchmarking} conduct an extensive evaluation on two real UPMEM-based PIM systems with 640 and 2556 DPUs, providing new insights about suitability of different workloads to the PIM system, programming recommendations for software designers, and suggestions and hints for hardware and architecture designers of future PIM systems.
Compared to our work, the PrIM benchmark suite is limited to a particular \gls{PIM} system (i.e., the UPMEM \gls{PIM}), while we aim to broadly encompass \gls{PIM} architectures in our analyses.

\subsection{Identification of \gls{PIM} Suitability}
\label{chap:related:pimsys:suitability}

In this section, we discuss works that aim to identify whether an application, kernel, or instruction might benefit from \gls{PIM} execution. 
We call such portions \emph{PIM offloading candidates}.
While PIM offloading candidates can be identified manually by a programmer, the identification would require significant programmer effort along with a detailed understanding of the hardware trade-offs between CPU cores and PIM cores. 
For architects who are adding custom PIM logic (e.g., fixed-function accelerators, which we call PIM accelerators) to memory, the trade-offs between CPU cores and PIM accelerators may not be known before determining which portions of the application are PIM offloading candidates, since the PIM accelerators are tailored for the PIM offloading candidates.
To alleviate the burden of manually identifying PIM offloading candidates, many prior works~\cite{vadivel2020tdo,maity2025framework,xu2025identifying,boroumand2018google,ahmed2019compiler,chen2022general,sokulski2022sapive,PEI,ghiasi2022alp,wei2022pimprof,maity2024coat,baskaran2020decentralized,hadidi2017cairo,jiang20243,maity2023data,hsieh2016transparent,pattnaik2016scheduling} propose systematic approaches for identifying PIM offloading candidates in an application.
We classify \gls{PIM} suitability approaches into three classes:
\li~static-based~\cite{vadivel2020tdo,maity2025framework,xu2025identifying,boroumand2018google,ahmed2019compiler,chen2022general}, where \gls{PIM} suitability is determinate based on characteristics of an application that can be obtained offline (e.g., via previous application profiling or compiler analysis),   
\lii~dynamic-based~\cite{sokulski2022sapive,PEI,ghiasi2022alp}, where \gls{PIM} suitability is determinate by probing the hardware state during application execution,
\liii~hybrid-based~\cite{wei2022pimprof,maity2024coat,baskaran2020decentralized,hadidi2017cairo,jiang20243,maity2023data,hsieh2016transparent,pattnaik2016scheduling}, where \gls{PIM} suitability is determinate using a combination of the static and dynamic solutions.

\paratitle{Static-Based \gls{PIM} Suitability Identification Approaches} 
In~\cite{boroumand2018google}, to identify \gls{PIM} suitability, the authors execute an offline  profiling analysis of the target application. 
Using hardware performance counters and an energy model, they identify that a function is a \gls{PIM} offloading candidate if it meets the following conditions: 
\li~it consumes a significant fraction (e.g., more than 30\%) of the overall workload energy consumption;,
\lii~its data movement consumes a significant fraction (e.g., more than 30\%) of the total workload energy,
\liii~it is memory-intensive (e.g., its last-level cache misses per kilo instruction, or MPKI, is greater than 10~\cite{chou2015reducing,kim2010atlas,muralidhara2011reducing,kim2010thread}),
\liv~data movement is the single largest component of the function’s energy consumption.
In~\cite{ahmed2019compiler}, the authors develop compiler analyses that statically identify candidate instructions for \gls{PIM} offloading based on memory access patterns, data locality, and operation types, leveraging affine analysis and control/data flow properties to ensure correctness and offloading benefit. 
In~\cite{chen2022general}, the authors implement a compiler-based \gls{PIM} suitability solution that decides whether to offload a kernel to PIM using a latency-guided instruction-level offloading model. To avoid expensive per-input profiling, it employs an offline-trained regression predictor to estimate performance benefits based on dataset features, such as expected cache misses.

\paratitle{Dynamic-Based \gls{PIM} Suitability Identification Approaches}
Many works propose runtime mechanisms for \emph{dynamic scheduling} of PIM offloading candidates, i.e.,
mechanisms that decide whether or not to actually offload code that is
marked to be potentially offloaded to PIM engines.
In~\cite{PEI}, the authors develop a locality-aware scheduling
mechanism for PIM-enabled instructions.  
In~\cite{ghiasi2022alp}, the authors add 
\li~an \emph{Offload Management Unit} to the host chip to handle the
offload of code segments to PIM cores and
\lii~a \emph{Monitor Unit} in both the PIM and
CPU cores to collect the necessary runtime information (e.g., L1 cache misses, LLC misses) to feed the  \emph{Offload Management Unit}.
In~\cite{sokulski2022sapive}, the authors propose SAPIVe, a hardware binary translator that dynamically identifies patterns of AVX vector instructions during execution and speculatively converts them into PIM vector instructions. Offloading occurs if enough loop iterations match a recognized pattern, memory accesses are contiguous, and execution can be validated without disrupting the CPU pipeline.

\paratitle{Hybrid-Based \gls{PIM} Suitability Identification Approaches} In~\cite{hadidi2017cairo}, the authors introduce CAIRO, a fine-grained compiler-assisted approach that identifies \gls{PIM}-offloadable instructions using static control flow analysis and memory access disambiguation. CAIRO further refines candidate selection using runtime information to guide offloading decisions dynamically, improving adaptability to program behavior.
In~\cite{jiang20243}, the authors propose A$^3$PIM, a hybrid static-dynamic framework for PIM offloading that analytically estimates the performance impact of offloading candidate code regions. A$^3$PIM leverages offline profiling to extract dynamic information and uses a cost model to predict the trade-offs between PIM execution and CPU execution. 
In~\cite{maity2023data}, the authors first instruments loop regions offline using LLVM to mark them as \gls{PIM} offloadable candidates. Then, at runtime, it decides whether to offload each region by comparing the expected execution time on the host versus the \gls{PIM}, using runtime profiling of data locality (specifically, the number of dirty cache blocks in the \gls{LLC}).
For GPU-based systems~\cite{hsieh2016transparent,pattnaik2016scheduling}, the authors of~\cite{hsieh2016transparent,pattnaik2016scheduling}  explore the combination of compile-time and runtime mechanisms for identification and dynamic scheduling of PIM offloading candidates.

\subsection{Compiler Support for \gls{PIM}}
\label{chap:related:pimsys:compiler}

In this section, we describe previous efforts to implement compiler support for code generation, transformation, and optimization for \gls{PIM} architectures.

Prior works propose programming models for different types of \gls{PuM} architectures, as
\li~CUDA/OpenAcc~\cite{cheng2014professional, OpenACCA1:online} for in-cache computing~\cite{dualitycache};
\lii~tensor dataflow graphs for in-ReRAM computing~\cite{fujiki2018memory}. 
By enabling fine-grained DRAM, we believe such programming models can be now easily ported to \gls{PuD} computing (for example, by assuming that each DRAM mat executes a different CUDA thread block).

CHOPPER~\cite{peng2023chopper} improves SIMDRAM's programming model by leveraging bit-slicing compilers and employing optimizations to reduce the latency of a \uprog. 
Even though CHOPPER simplifies programmability compared to SIMDRAM, it still requires the programmer to re-write applications using the bit-slicing compiler's syntax. Compared to CHOPPER, MIMDRAM has two main advantages. First, MIMDRAM \emph{automatically} generates code for the \gls{PuD} engine without any code refactoring. Second, since CHOPPER maintains the very-wide \gls{SIMD} programming model of SIMDRAM, it also suffers from  \gls{SIMD} underutilization.
Compared to CHOPPER, \emph{Proteus} has two main advantages. 
First, \emph{Proteus} improves a \uprog performance by \emph{fully} leveraging the DRAM parallelism within a single DRAM bank.
Second, although bit-slicing compilers can naturally adapt to different bit-precision values, they require the programmer to specify the target bit-precision manually. In contrast, \emph{Proteus} dynamically identifies the most suitable bit-precision transparently from the programmer.
Some other prior works (e.g.,~\mbox{\cite{caminal2022accelerating, wong2023pumice}}) propose techniques to realize early termination of \gls{PuM} operations for different memory technologies. Compared to these, \emph{Proteus}'s main novelty lies in realizing early termination of bit-serial operation in the context of DRAM/majority-based \gls{PuD} systems.

The authors of~\cite{khan2022cinm} propose CINM, a compiler based on MLIR~\cite{lattner2021mlir} (multi-level intermediate representation) for both UPMEM-like PIM systems and analog-based PIM architectures. 
Its main limitation is the fact that it only supports linear algebra-related kernels.

\subsection{Memory Management Support for \gls{PIM}}
\label{chap:related:pimsys:memory}

When an application needs to access its data inside the main memory, the CPU core must first perform an \emph{address translation}, which
converts the data's virtual address into a \emph{physical} address
within main memory.  
If the translation {metadata} is not
available in the CPU's translation lookaside buffer (TLB), the CPU
must invoke the page table walker in order to perform a long-latency
page table walk that involves multiple \emph{sequential} reads to the
main memory and lowers the application's performance~\cite{kanellopoulos2023victima,kanellopoulos2023utopia,basu2013efficient,karakostas2014performance,kanellopoulos2024virtuoso,li2019framework,mosaic-osr,ausavarungnirun2018mask,ausavarungnirun2017mosaic}. In modern
systems, the virtual memory system also provides access protection mechanisms.

Enabling a unified virtual address space between PIM cores and host cores, where the memory address space is shared across all computing elements in the system, is key for enabling a more flexible programming model (i.e., ``\emph{a point is a pointer'' semantics}~\cite{che2016challenges}), with increased utilization of memory capacity and throughput. 
However, developing a high-performance and scalable unified virtual memory address space for both conventional host systems and PIM architectures can be challenging, since it requires a distributed memory management approach where different components of the system are able to perform address translation while guaranteeing access protection mechanisms for every memory access.

One naive solution to the unified Host--PIM virtual memory address space issue is to make PIM cores reliant on existing CPU-side address translation mechanisms.
However, in this approach, any performance
gains from performing in-/near-memory operations could easily be nullified, as the PIM cores need to send a long-latency translation request to the CPU via the off-chip channel for each memory access.
The translation can sometimes require a page table walk that issues multiple memory requests back to the memory, thereby leading to increased memory traffic on the main memory channels.
  
To improve over such a naive solution and reduce the overhead of page walks, one could utilize
PIM engines to perform page table walks. 
This can be done by duplicating the content of the TLB and {moving} the page walker {to} the PIM processing logic in main memory.  
Unfortunately,
this is either difficult or expensive for three reasons. 
First, coherence {has} to be maintained between the CPU's TLBs and the memory-side TLBs. This introduces extra complexity and off-chip
requests. 
Second, duplicating the TLBs increases the storage and complexity overheads {on the memory side, which should be carefully contained}.  
Third, if main memory is shared across
{CPUs with} different types of architectures, page table structures and the implementation of address translations can be different across {the different} architectures. 
Ensuring
compatibility between the in-memory TLB/page walker and all possible types of {virtual memory} architecture designs can be complicated and often not even practically feasible.

To address these concerns and reduce the overhead of virtual memory, the authors of~\cite{hsieh2016accelerating}  explore a tractable solution for PIM address translation as part of
their in-memory pointer chasing accelerator, IMPICA~\cite{hsieh2016accelerating}.
IMPICA exploits the high bandwidth available within 3D-stacked memory to traverse a chain of virtual memory pointers within DRAM, \emph{without} having to look up virtual-to-physical address
translations in the CPU translation lookaside buffer (TLB) and without using the page walkers within the CPU.  
{IMPICA's key ideas are
\li~to use a region-based page table, which is optimized for PIM acceleration, and 
\lii~to decouple address calculation and memory access with two specialized engines.  

Beyond pointer chasing operations that are tackled by IMPICA~\cite{hsieh2016accelerating}, providing efficient mechanisms for PIM-based
virtual-to-physical address translation (as well as access protection) remains a challenge for the generality of applications,
especially those that access large amounts of virtual
memory~\cite{ausavarungnirun2018mask,ausavarungnirun2017mosaic,mosaic-osr,kanellopoulos2023utopia,kanellopoulos2023victima,basu2013efficient}.

In~\cite{hall2000memory}, the authors present a detailed memory management scheme tailored for \gls{PIM} systems. It introduces a segmented memory model that distinguishes between local and global segments, with \gls{PIM} cores accessing both types based on locality and sharing needs. 
Address translation is handled differently for local and remote accesses, relying on a global naming mechanism and parcel-based communication for remote data retrieval. 
Virtual memory management is extended to support fine-grained allocation and mapping of objects across multiple PIM cores, while a global segment table facilitates object sharing and consistency. The system also supports paging and context switching, enabling robust multi-programming and dynamic memory management across distributed PIM nodes.

In~\cite{picorel2017near}, the authors propose DIPTA (distributed inverted page table) for near-memory address translation tailored for PIM architectures. 
DIPTA enables parallel and independent address translation and data fetch by restricting virtual-to-physical address mapping associativity, which allows \gls{PIM} cores to avoid expensive page walks. The system also supports page faults, TLB coherence, multiprogramming, and unified virtual memory between CPUs and \gls{PIM} cores, with minimal OS changes.

In~\cite{azarkhish2016memory}, the authors implement memory management for PIM hardware by introducing a \emph{zero-copy memory virtualization} scheme that enables direct data sharing between the host processor and the PIM cores, avoiding costly data duplication. 
It employs a \emph{slice-table} mechanism to manage the mapping between virtual and physical memory, supporting fine-grained memory access and enabling merged pages for efficient utilization. 
The system includes TLB support within the PIM to minimize translation overhead and improve efficiency.

In PiDRAM~\cite{olgun2021pidram,olgun2022pidram}, the authors handles memory management for PIM by implementing a custom memory allocation mechanism in its supervisor software to meet the data alignment and allocation constraints of \gls{PuD} operations.
Concretely, it uses a subarray mapping table (SAMT) to ensure allocated pages reside in the same DRAM subarray, and an allocation ID table (AIT) to group allocations with shared constraints. 
For initialization, it maintains an initializer rows table (IRT) to find zero-initialized rows.  These mechanisms are integrated into the PiDRAM framework to enable efficient end-to-end support for commodity DRAM-based \gls{PuM} techniques.

\subsection{Programming Frameworks and High-Level APIs for \gls{PIM}}
\label{chap:related:pimsys:programming}

Several prior works~\cite{chen2023simplepim,item2023transpimlib,giannoula2024pygim} propose different programming frameworks or APIs to ease programmability in UPMEM-based PIM systems. 
First, SimplePIM~\cite{chen2023simplepim} proposes a high-level programming framework that encapsulate management, computation, and communication primitives into software APIs. 
Similarly to DaPPA, SimplePIM uses a MapReduce-like programming model~\cite{jiang2010map}, with support for three execution parallel patterns (i.e., \texttt{map}, \texttt{reduce}, and \texttt{zip}). 
DaPPA builds on top of SimplePIM by 
\li~further extending the programming model to allow for native implementation of more parallel patterns,
\lii~\emph{completely} eliminating the need for the user to handle data communication between CPU and DPUs and across DPUs, and
\liii~allowing for \emph{automatic} cooperative execution between CPU and DPUs for a given kernel. 
By doing so, DaPPA further improves programming productivity compared to SimplePIM.
Second, other works, such as TransPimLib~\cite{item2023transpimlib} and PyGim~\cite{giannoula2024pygim}, provide implementation support for key computation kernels, such as transcendental functions and graph neural networks. 
Even though such works also aid UPMEM programmability, they are limited to a narrow application scope.

\setcounter{version}{99}
% \glsresetall{}
\chapter[DAMOV: A New Methodology and Benchmark Suite for Evaluating Data Movement Bottlenecks]{DAMOV: A New Methodology and Benchmark Suite for Evaluating Data Movement Bottlenecks}
\label{chap:damov}

% Mechanism Name: DAMOV
\newcommand{\bench}{{DAMOV}\xspace} 

% Highlights 
\sethlcolor{lightyellow}
\DeclareRobustCommand{\hlcyan}[1]{{\sethlcolor{cyan}\hl{#1}}}

% Submission flags 
\newif\ifdamovsubmission
\damovsubmissiontrue
%\damovsubmissionfalse
\ifdamovsubmission
    \providecommand\geraldo[1][0]{}
    \providecommand\sg[1][0]{}
    \providecommand\revision[1][0]{}
\else
    \providecommand{\geraldo}[1]{\textcolor{blue}{#1}}
    \providecommand{\sg}[1]{\textcolor{orange}{#1}}
\fi

\newif\ifdamovrevisionri
%\damovrevisionritrue
\damovrevisionrifalse
\ifdamovrevisionri
    \providecommand\geraldorevi[1][0]{}
\else
    \providecommand{\geraldorevi}[1]{\textcolor{black}{#1}} %change to blue
    \providecommand{\geraldorevii}[1]{\textcolor{black}{#1}} %change to orange
    \providecommand{\gfii}[1]{\textcolor{black}{#1}} %change to orange 
    \providecommand{\gfiii}[1]{\textcolor{black}{#1}} %change to orange 
    \providecommand{\gfiv}[1]{\textcolor{black}{#1}} %change to orange 
    \providecommand{\gfv}[1]{\textcolor{black}{#1}} %change to orange 
    \providecommand{\gfvi}[1]{\textcolor{black}{#1}}
    \providecommand{\gfvii}[1]{\textcolor{black}{#1}} %% change to blue

    \providecommand{\gfviii}[1]{\textcolor{black}{#1}} %% change to blue

    \providecommand{\sgv}[1]{\textcolor{black}{#1}} %change to orange 
    \providecommand{\sgrvi}[1]{\textcolor{black}{#1}} % change to blue 
    \providecommand{\juan}[1]{\textcolor{black}{#1}} %chage to blue

    \providecommand{\jgl}[1]{}
    \providecommand{\jgll}[1]{[{\color{ddgreen}JGL: #1}]}
    \providecommand{\juang}[1]{\textcolor{black}{#1}}  %change to orange

    \providecommand{\juangg}[1]{\textcolor{black}{#1}}
    \providecommand{\juanggg}[1]{\textcolor{black}{#1}} 
\fi

\section{Motivation \& Goal}

Many recent works explore how \gls{NDP}\footnote{Particularly in this chapter, we use the term \gls{NDP} to refer to \gls{PnM} architectures.} can benefit various application domains, such as graph processing~\cite{PEI, ahn2015scalable, nai2017graphpim, song2018graphr, lazypim, boroumand2019conda, zhang2018graphp, angizi2019graphs, matam2019graphssd, angizi2019graphide, zhuo2019graphq}, \gfiii{machine learning}~\gfiii{\cite{gao2017tetris, Kim2016,Shafiee2016,Chi2016, boroumand2018google,boroumand2021mitigating,amiraliphd}}, bioinformatics~\cite{kim2018grim,NIM, cali2020genasm}, databases~\gfiii{\cite{drumond2017mondrian, RVU, seshadri2017ambit, lazypim, boroumand2019conda,hsieh2016accelerating,boroumand2021polynesia,amiraliphd}}, security~\gfvii{\cite{gu2016leveraging,kim2019d,kim2018dram}}, data manipulation~\cite{seshadri2017ambit,li2017drisa,seshadri2013rowclone, wang2020figaro, chang2016low, li2016pinatubo, rezaei2020nom, seshadri2015fast}, \geraldorevi{and mobile workloads~\gfiii{\cite{boroumand2018google,amiraliphd}}}. These works demonstrate that simple metrics such as \gfii{last-level CPU cache} \gls{MPKI} and \gls{AI} are useful \gfiii{metrics that serve} as a proxy for the amount of data movement an application experiences. \gfii{These metrics can be used} as a potential guide for choosing when to apply data movement mitigation mechanisms such as \gls{NDP}. However, such metrics \gfiii{(and the corresponding insights)} are often extracted from a small set of applications, with similar \gfii{or not-rigorously-analyzed} data movement characteristics. Therefore, it is difficult to generalize the metrics and insights these works provide to a broader set of applications, making it unclear what different metrics can reveal about a new \gfii{(i.e., previously uncharacterized)} application's data movement behavior (and how to mitigate its associated data movement costs).

We illustrate this issue by highlighting the limitations of two \gfii{different} methodologies commonly used to identify memory bottlenecks and often used as a guide to justify the use of \gls{NDP} architectures \gfii{for an application}: (a)~analyzing a roofline model~\cite{williams2009roofline} of the \gfii{application}, and (b)~using \geraldorevi{last-level CPU cache} \gls{MPKI} as an indicator of \gls{NDP} suitability \gfii{of the application}. The roofline model correlates the computation requirements of an application with its memory requirements under a given system. The model contains two \emph{roofs}: (1)~a diagonal line (\emph{y = Peak Memory Bandwidth $\times$ Arithmetic Intensity}) called the \emph{memory roof}, and (2)~a horizontal line (\emph{y = Peak System Throughput}) called the \emph{compute roof}~\cite{williams2009roofline}. If an application lies under the memory roof, the application is classified as \emph{memory-bound}; if an application lies under the compute roof, it is classified as \emph{compute-bound}. \gfiii{Many prior} works~\gfiii{\cite{azarkhish2018neurostream,ke2019recnmp,asgari2020mahasim,Kim2018HowMC,liang2019ins,glova2019near, fernandez2020natsa, gu2020dlux, singh2020nero,gomez2022benchmarking,radulovic2015another,boroumand2021mitigating,yavits2021giraf,herruzo2021enabling,asgarifafnir}} employ this \gfii{roofline} model to identify memory-bound applications that can benefit from \gls{NDP} architectures. Likewise, many prior works~\gfiii{\cite{hsieh2016accelerating,nai2017graphpim,kim2015understanding,boroumand2018google, nai2015instruction, ghose.ibmjrd19, lim2017triple, tsai:micro:2018:ams,kim2018coda,hong2016accelerating,gries2019performance}} observe that applications with high \geraldorevi{last-level cache} \gls{MPKI}\footnote{Typically, an \gls{MPKI} value greater than 10 is considered \emph{high} by prior works~\cite{hashemi2016accelerating,chou2015reducing,kim2010atlas,kim2010thread,muralidhara2011reducing, subramanian2016bliss,usui2016dash}.} are good candidates for \gls{NDP}.

\begin{figure*}[ht]
    \centering
    \includegraphics[width=0.92\linewidth]{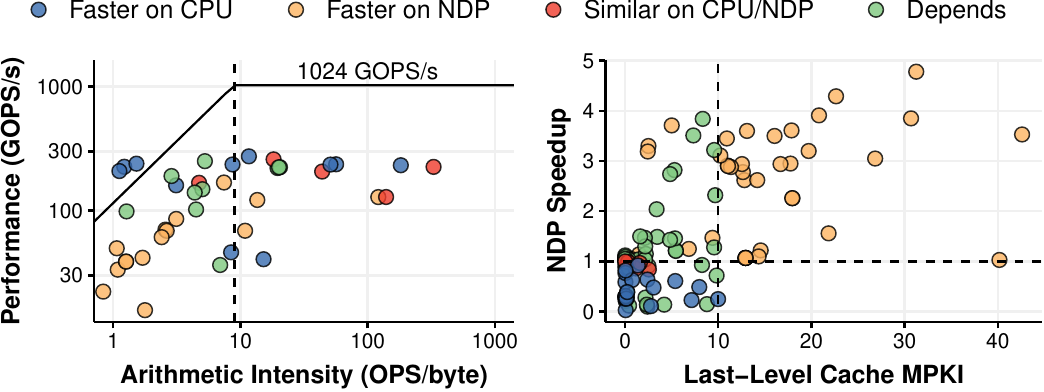}
    \caption{
    Roofline (left) and \gfii{last-level cache} MPKI vs. NDP speedup (right) for \gfii{44} memory-bound applications. Applications are classified into four categories:
    (1)~those that experience performance degradation due to NDP (\gfii{blue; }\geraldorevi{Faster on CPU}), 
    (2)~those that experience performance improvement due to NDP (\gfii{yellow;} \geraldorevi{Faster on NDP}), 
    (\gfii{3})~\gfii{th\gfiii{o}se where the host CPU} and NDP performance are similar (\geraldorevi{\gfii{red;} Similar on CPU/NDP}),
    (\gfii{4})~those that experience either performance degradation or performance improvement due to NDP depending on the microarchitectural configuration (\geraldorevi{\gfii{green;} Depends}).}
    \label{fig_roofline_and_mpki}
\end{figure*}

Figure~\ref{fig_roofline_and_mpki} shows the roofline model (left) and a plot of \gls{MPKI} vs.\ speedup (right) of a system with general-purpose \gls{NDP} support over a baseline system without \gls{NDP} for a diverse set of 44~applications (see Table~\ref{tab:benchmarks}). In the \gls{MPKI} vs. speedup plot, the \gls{MPKI} corresponds to a baseline host \gfii{CPU} system. The speedup represents the performance improvement of a general-purpose \gls{NDP} system over the baseline (see Section~\ref{sec:step3} for our methodology). We make the following observations. First, analyzing the roofline model (Figure~\ref{fig_roofline_and_mpki}, left), we observe that most of the memory-bound applications (yellow dots) benefit from \gls{NDP}, as foreseen by prior works. We later observe (\geraldorevi{in} Section~\ref{sec_scalability_class1a}) that such applications are \gfii{DRAM} bandwidth-bound and are a natural fit for \gls{NDP}.  However, the \gfiii{roofline} model does \gfiii{\emph{not} accurately} account for \gfiii{the \gls{NDP} suitability of} memory-bound applications that 
(i)~benefit from \gls{NDP} \gfii{only} \geraldorevi{under} particular microarchitectural configurations, \gfii{e.g.}, either at low or high core counts (green dots, which are applications that are either bottlenecked by \gfii{DRAM} latency or suffer from \gfii{L3} cache contention; see Sections \mbox{\ref{sec_scalability_class1c}} and \mbox{\ref{sec_scalability_class2a}}); or 
(ii)~experience performance degradation when executed using NDP (blue dots, which are applications that suffer from the lack of a deep cache hierarchy in NDP architectures; see Section~\mbox{\ref{sec_scalability_class2c}}). Second, analyzing the \gls{MPKI} vs.\ speedup plot (Figure~\ref{fig_roofline_and_mpki}, right), we observe that while \geraldorevi{all} applications with high \gls{MPKI} benefit from \gls{NDP} (yellow dots \gfiii{with \gls{MPKI} higher than 10}), some applications with \emph{low} \gls{MPKI} can \geraldorevi{\emph{also}} benefit from \gls{NDP}  \gfiii{in \gfiv{all} of the \gls{NDP} microarchitecture configurations we evaluate (yellow dots with \gls{MPKI} lower than 10) or under specific \gls{NDP} microarchitecture configurations (green dots with \gls{MPKI} lower than 10)}. Thus, even though both the roofline model and \gls{MPKI} can identify \gfiii{some} specific sources of memory bottlenecks and can \gfiii{sometimes} be used as a proxy for \gls{NDP} suitability, \geraldorevi{they \gfiv{alone} cannot definitively determine \gls{NDP} suitability because they cannot} comprehensively identify different \gfiii{possible} sources of memory bottlenecks in a system.

Our \emph{goal} in this work is \geraldorevi{(1)} to understand the major sources of inefficiency that lead to data movement bottlenecks by observing \gfiii{and identifying} relevant metrics and \geraldorevi{(2)} to develop a benchmark suite for data movement that captures each of these sources. To this end, we develop a new three-step methodology to correlate application characteristics with the \emph{primary} sources of data movement bottlenecks and to determine the potential benefits of three example data movement mitigation mechanisms: (1) a deep cache hierarchy, (2) a hardware prefetcher, and (3) a general-purpose \gls{NDP} architecture.\footnote{We focus on these three data movement mitigation mechanisms for two different reasons: (1) deep cache hierarchies and hardware prefetchers are standard mechanisms in \geraldorevi{almost all modern} systems, and (2) \gls{NDP} represents a promising paradigm shift for many modern data-intensive applications.}  \geraldorevi{We use two main profiling strategies to gather key metrics from applications: (i) an architecture-independent profiling tool and (ii) an architecture-dependent profiling tool. The architecture-independent profiling tool provides metrics that characterize the application memory behavior independently of the underlying hardware. In contrast, the architecture-dependent profiling tool evaluates the impact of the system configuration \gfiii{(e.g., cache hierarchy)} on the memory behavior.} Our methodology has three steps. \geraldorevi{In \textit{Step~1}}, we use a hardware profiling tool to identify memory-bound functions \geraldorevi{across} many applications. This step allows for a quick first-level identification of many applications that suffer from memory bottlenecks \gfii{and functions that cause these bottlenecks}. \geraldorevi{In \textit{Step~2}, we use the architecture-independent profiling tool to collect metrics that provide insights about the memory access behavior of the \gfii{memory-bottlenecked} functions. In \textit{Step~3}, we collect architecture-\gfii{dependent} metrics and analyze the performance and energy of each function in an application when each of our three candidate data movement mitigation mechanisms is applied to the system.} By combining the data obtained from all three steps, we can systematically classify the leading causes of data movement bottlenecks \gfii{in an application or function} into different bottleneck classes. 

Using this \geraldorevi{new} methodology, we characterize a large\geraldorevi{,} heterogeneous set of applications (\geraldo{345~applications from \gfiii{37} different workload suites) across a wide range of domains.} \geraldo{\geraldorevi{Within} these applications,} we \gfiv{analyze 77K functions and} find \gfiii{a subset of} \geraldo{144}~functions \gfiii{from 74 different applications} that are memory\gfii{-}bound (and \geraldorevi{that} consume a significant fraction of the overall execution time). \geraldo{We} fully characterize \geraldo{this} set of 144 representative functions to serve as a core set of application kernel benchmarks\gfii{, which we release \gfiii{as the} open\gfiii{-}source \bench \gfiii{(\underline{DA}ta \underline{MOV}ement)} Benchmark Suite~\cite{damov}}. Our analyses reveal \gfii{six} new insights about the sources of memory bottlenecks and their relation \gfiii{to} \gls{NDP}:

\begin{enumerate}[noitemsep, leftmargin=*, topsep=0pt]
    \item \gfii{Applications with high last-level cache \gls{MPKI} and low temporal locality are \emph{DRAM bandwidth-bound}. These applications benefit from the large memory bandwidth available to the NDP system (Section~\ref{sec_scalability_class1a}).} 
    
    \item \gfii{Applications with low last-level cache \gls{MPKI} and low temporal locality are \emph{DRAM latency-bound}. These applications do \emph{not} benefit from L2/L3 caches. The NDP system improves performance and energy efficiency by sending L1 misses directly to DRAM (Section~\ref{sec_scalability_class1b}).} 
    
     \item \gfii{A second group of applications with low LLC MPKI and low temporal locality are \emph{bottlenecked by L1/L2 cache capacity}. These applications benefit from the NDP system at low core counts. However, at high core counts (and thus larger L1/L2 cache space), the caches capture most of the data locality in these applications, decreasing the benefits the NDP system provides  (Section~\ref{sec_scalability_class1c}). We make this observation using a \emph{new} metric that we develop, called \emph{last-to-first miss-ratio (LFMR)}, which we define as the ratio between the number of LLC misses and the total number of L1 cache misses. We find that this metric accurately identifies how efficient the cache hierarchy is in reducing data movement.}
        
    \item \gfii{Applications with high temporal locality and low LLC MPKI are \emph{bottlenecked by L3 cache contention} at high core counts. In such cases, the NDP system provides a cost-effective way to alleviate cache contention over increasing the L3 cache capacity (Section~\ref{sec_scalability_class2a}).}
    
    \item \gfii{Applications with high temporal locality, low LLC MPKI, and low \gls{AI} are bottlenecked by the \emph{L1 cache capacity}. The three candidate data movement mitigation mechanisms achieve similar performance and energy consumption for these applications (Section~\ref{sec_scalability_class2b}).}
    
    \item  \gfii{Applications with high temporal locality, low LLC MPKI, and high AI are \emph{compute-bound}. These applications benefit from a deep cache hierarchy and hardware prefetchers, but the NDP system degrades their performance (Section~\ref{sec_scalability_class2c}).}
    \\
\end{enumerate}

We publicly release our 144~\geraldorevi{\gfii{representative data movement bottlenecked} functions} \gfii{from 74 applications} as \gfii{the first} open-source benchmark suite for data movement\geraldorevi{, called \bench \gfii{Benchmark Suite}}, along with the \gfii{complete} source code for our new characterization methodology~\cite{damov}.

\revdel{This work makes the following key contributions: 
\begin{itemize}[itemsep=0pt, topsep=0pt, leftmargin=*]
    \item We propose the first methodology to characterize data-intensive workloads based on the source of their data movement bottlenecks. This methodology is driven by insights obtained from a large-scale \geraldorevi{experimental} characterization of \geraldo{\sg{345}~applications from \sg{\gfii{37}} different benchmark suites} and an evaluation of \sg{the performance of memory-bound \gfii{functions} from these applications} with three data-movement mitigation mechanisms.
    \item We release \geraldorevi{DAMOV,} the first open-source \gfii{benchmark} suite for main memory data movement-related studies\gfiii{,} based on our systematic characterization methodology. This suite consists of 144~\gfii{functions} representing different sources of data movement bottlenecks and can be used as a baseline benchmark set for future data-movement mitigation research.
    \item \geraldorevi{We show how our \gfiii{\bench} benchmark suite can aid the study of open research problems for \gls{NDP} architectures\gfiii{, via} four case studies. \gfiii{In particular, we} evaluate (i) the impact of load balance and inter-vault communication \gfii{in \gls{NDP} systems}, (ii) the \gfiii{impact of} \gls{NDP} accelerators \gfiii{on our memory bottleneck analysis}, (iii) the \gfiii{impact of different core models on \gls{NDP} architectures}, and (iv) the potential benefits of identifying  simple \gls{NDP} \gfiii{instructions}. \gfii{We conclude that our benchmark suite and methodology can be employed to address many different open research \gfiii{and development} questions on data movement mitigation mechanisms, particularly topics related to \gls{NDP} systems and architectures.}}
\end{itemize}
}
 \glsresetall

\section{Methodology Overview}
\label{sec:methodology}

\sg{We develop a new} workload characterization methodology \sg{to analyze} data movement bottlenecks and the suitability \sg{of} different data movement mitigation mechanisms \sg{for these bottlenecks}, with a focus on \gls{NDP}. \sg{Our} methodology consists of three main steps, \geraldo{as Figure~\mbox{\ref{figure_methodology}} depicts:}
(1)~\sg{\emph{memory-bound function identification} using application profiling};
(2)~\emph{locality-based clustering} to analyze spatial and temporal locality in an architecture-independent manner; and
(3)~\emph{\juan{memory} bottleneck classification} using a scalability analysis to nail down the sources of memory boundedness\geraldorevi{, including architecture-dependent characterization}. \geraldo{Our methodology takes as input an application's source code and its \geraldorevi{input} datasets, \gfiii{and produces as output a classification of} the primary source of memory bottleneck of \gfiii{important functions in an application (i.e., bottleneck class of each key application function).}}
We illustrate the applicability of this methodology with a detailed characterization of 144 \sg{\gfii{functions}} that we select \sg{from} among \gfiv{77K analyzed functions of} \geraldo{345} \sg{characterized} applications. In this section, we give an overview of our workload characterization methodology. 
\sg{We use this methodology to drive the analyses \gfiii{we perform}} \juan{in Section~\ref{sec:characterization}}.

\begin{figure*}[h]
    \centering
    \includegraphics[width=\linewidth]{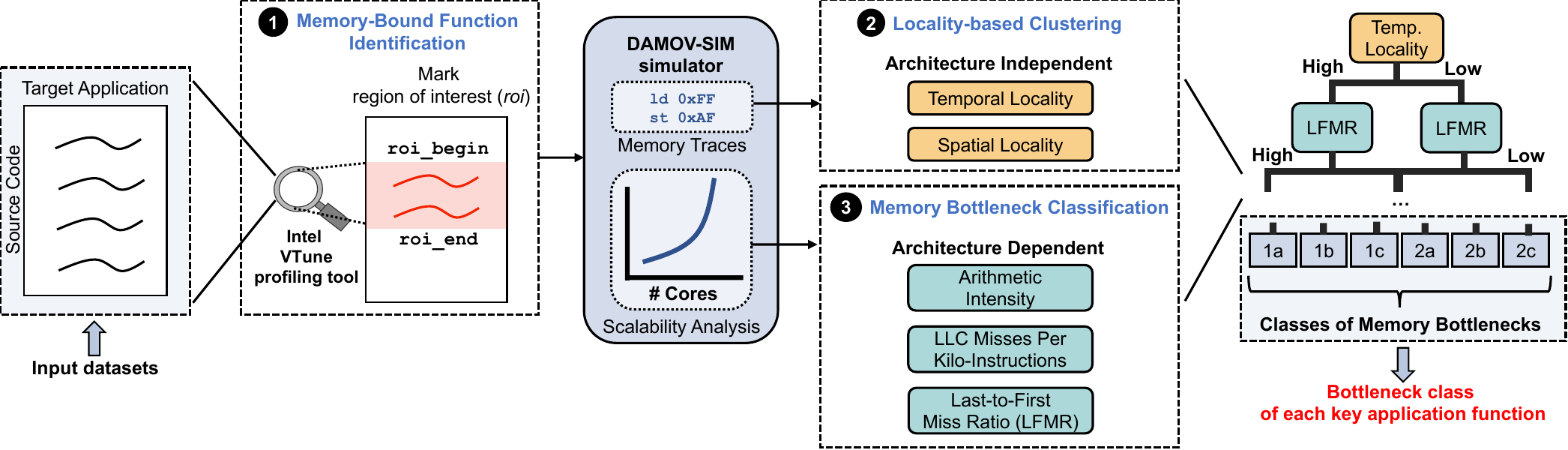}
    \caption{\geraldorevi{Overview of our three-step \gfii{workload characterization} methodology.}}
    \label{figure_methodology}
\end{figure*}

\subsection{Experimental Evaluation Framework}
\label{sec_damovsim}

\sg{As our scalability analysis depends on the \geraldorevi{hardware} architecture, we need \geraldorevi{a hardware} platform that can allow us to replicate and control all of our configuration parameters.  Unfortunately, such an analysis cannot be performed practically using real hardware, as
(1)~there are \geraldorevi{very} few available NDP hardware platforms, and the ones that currently exist do not allow us to comprehensively analyze our general-purpose NDP configuration \gfiii{in a controllable way} (as existing platforms \gfii{are} specialized \gfii{and non-configurable}); and
(2)~the configurations of real CPUs can vary significantly across the range of core counts that we want to analyze\gfiii{, eliminating the possibility of a carefully controlled study}.} \sg{As a result, we must rely on accurate simulation platforms to perform an accurate comparison across different configurations.
To this end, we build a framework that} integrates \sg{the ZSim CPU simulator~\gfiii{\cite{sanchez2013zsim}} with the Ramulator memory simulator~\gfiii{\cite{kim2016ramulator}}} to produce a fast, scalable, and cycle-accurate \gfii{open-source} simulator \geraldorevi{called \bench-SIM~\cite{damov}}. We use ZSim to simulate the core microarchitecture, cache hierarchy, coherence protocol, and prefetchers. \gfiii{W}e use Ramulator to simulate the DRAM \sg{architecture\gfii{, memory controllers,} and memory} accesses. To compute spatial and temporal locality, we modify ZSim to generate a single-thread memory trace for each \gfii{application}, which we use as input for the locality \gfiii{analysis} algorithm \gfiii{described} \sg{in Section~\ref{sec:step2} (which statically computes the temporal and spatial locality at word-level granularity)}.

\subsection{Step 1: Memory-Bound Function Identification}
\label{sec:step1}

The first step (\geraldorevi{labeled \ding{182} in \geraldo{Figure~\ref{figure_methodology})}} aims to identify \sg{\geraldorevi{the} functions of an application \geraldorevi{that} are \emph{memory-bound} (i.e., functions \geraldorevi{that} suffer from data movement bottlenecks)}.
These bottlenecks might be caused at any level of the memory hierarchy. There are various potential sources of memory boundedness, such as cache misses, cache coherence traffic, and long queu\geraldorevi{ing} latencies. Therefore, we need to take \geraldorevi{all such} potential causes into account. This step is optional if the \sg{\geraldorevi{application's \gfii{memory-bound functions (i.e.,} region\gfiii{s} of interest\gfii{, \emph{roi}, in Figure~\ref{figure_methodology})} \gfiii{are}} already known \emph{a priori}}.

Hardware profiling tools, both open-source and proprietary, are available to obtain hardware counters and metrics that characterize the application behavior on a computing system. 
In this work, we use \gfii{the} Intel VTune \gfiii{Profiler}~\cite{vtune}, which %provides a metric 
\juan{implements the well-known \gfiii{\emph{top-down analysis}}~\cite{yasin_ispass2014}\gfii{. \gfiii{T}op-down analysis \gfiii{uses} the available CPU hardware counters to hierarchically identify different sources of CPU system bottlenecks for an application}. \gfii{Among the \gfiii{various} metrics \gfiii{measured} by top-down analysis,} there is a relevant one} called \emph{Memory Bound}~\cite{MemoryBo20} that measures the percentage of CPU pipeline slots that are \gfiii{\emph{not}} utilized \gfii{due to} any issue related to data access. We employ this metric to identify functions that suffer from data movement bottlenecks
\sg{(\gfii{which we define as} functions where \emph{Memory Bound} is greater than 30\%)}.

\subsection{Step 2: Locality-Based Clustering}
\label{sec:meth-locality}
\label{sec:step2}

Two key properties of an \sg{application's memory access pattern are its} inherent spatial locality (i.e., \sg{the likelihood of accessing} nearby memory locations in the near future) and temporal locality (i.e., \sg{the likelihood of} accessing a memory location \gfiii{again} in the near future). \geraldorevi{These properties} are closely related to how well the application can exploit the memory hierarchy in computing systems and how accurate hardware prefetchers can be. Therefore, to understand the sources of memory bottlenecks for an application, we should analyze how much spatial and temporal locality its memory accesses \sg{inherently} exhibit. However, we should isolate these properties from particular configurations of the memory subsystem. Otherwise, it would be unclear if memory bottlenecks are due to the nature of the memory accesses or \sg{due to the characteristics and limitations} of the memory subsystem \gfiii{(e.g., limited cache size, too simple or inaccurate prefetching policies)}. \sg{As a result, in this step (\geraldorevi{labeled} \ding{183} in \geraldo{Figure~\ref{figure_methodology})}, we use \emph{architecture-independent} static} analysis to obtain spatial and temporal locality metrics for the functions selected in the previous step (Section~\ref{sec:step1}). Past works~\gfvii{\cite{Conte91,John98,weinberg2005quantifying, shao2013isa, zhong2009program,gu2009component, beard2017eliminating,lloyd2015memory,conte1991brief,conte1995advances}} propose different ways of analyzing \geraldorevi{spatial} and temporal locality in an architecture-independent manner. In this work, we use the definition of \gfiii{spatial and temporal} metrics presented in \cite{weinberg2005quantifying, shao2013isa}.

The spatial locality metric is calculated for a window of memory references\footnote{\gfiii{We compute both the spatial and temporal locality metrics at the word granularity. In this way, we keep our \gfiv{locality analysis} architecture-independent\gfiv{, using} \emph{only} properties of the application under study.}} of length \emph{W} \sg{using} Equation~\ref{eq:spatial}. First, \sg{for} every \emph{W} memory references, we calculate the minimum distance between any two addresses (\emph{stride}). Second, we create a histogram called \sg{the} \emph{stride profile}, where each bin \emph{i} stores how many times each \emph{stride} appears. Third, to calculate the spatial locality, we divide the \emph{percentage} of times stride $i$ is referenced ($stride\ profile(i)$) by the stride length $i$ \geraldorevi{and sum \gfii{the resulting} value across all \gfii{instances} of $i$}.

\begin{equation}
Spatial\ Locality = \sum_{i=1}^{\#bins } \frac{stride\ profile(i)}{i}
\label{eq:spatial}
\end{equation}

\noindent
A spatial locality \sg{value} close to 0 is caused by \geraldorevi{large} \emph{stride} values (e.g., regular accesses with \geraldorevi{large} strides\gfii{)} or random accesses, while a \sg{value} equal to 1 is caused by a \gfii{completely} sequential access pattern.

The temporal locality metric is calculated by using a histogram of reused addresses. First, we count the number of times each memory address is repeated in a window of \emph{L} memory references. Second, \gfii{we create a histogram called \textit{reuse profile}, where each bin $i$ represents the number of times a memory address is reused, expressed \juan{as} a power of 2.} For each memory address, we increment the bin that represents the corresponding number of repetitions. For example, \textit{reuse profile(0)} represents memory addresses that are reused only once. \textit{reuse profile(1)} represents memory addresses that are reused twice. Thus, if a memory address is reused \gfii{$N$} times, we increment \gfii{\textit{reuse profile($\floor{log_{2} N}$)}} by one. Third, we obtain the temporal locality metric with Equation~\ref{eq:temporal}.

\begin{equation}
Temporal\ Locality = \sum_{i=0}^{\#bins } \frac{ 2^{i} \times reuse\ profile(i)}{total\ memory\ accesses}
\label{eq:temporal}
\end{equation}

A temporal locality \sg{value} \gfii{of} 0 \sg{\gfiii{indicates}} no data reuse, \sg{while a value} close to 1 \sg{indicates \gfii{very} high data} reuse \geraldorevi{(i.e., a value equal to 1 means that the application accesses a \gfii{single} memory address continuously)}.

\sg{To calculate these metrics, we empirically select window lengths \emph{W} and \emph{L} \gfii{to} 32.  \gfii{We find that \gfiii{different values chosen for} \emph{W} and \emph{L} do not significantly change \gfiii{the conclusions of} our analysis. We observe that our \gfiii{conclusions} remain the same when we set both values to 8, 16, 32, 64, and 128.}}

\subsection{Step 3: Bottleneck Classification}
\label{sec:step3}

\sg{While Step~2 allows us to understand inherent application sources for memory boundedness, it is important to understand how \geraldorevi{hardware} architectural features can also result in memory bottlenecks.  As a result, in our third step (\geraldorevi{\ding{184} in} \geraldo{Figure~\ref{figure_methodology}}),} we perform a scalability analysis \gfii{of the functions selected in \emph{Step 1}, where we evaluate performance and energy scaling for three different system configurations. The scalability analysis} \sg{ makes use of} three \sg{\emph{architecture-dependent}} metrics: (1)~\textit{\gls{AI}}, (2)~\textit{\gls{MPKI}}, and (3)~a new metric called \textit{\gls{LFMR}}. We select these metrics for the following reasons. First, \sg{\gls{AI}} can measure the compute intensity of an application. Intuitively, we expect an application with high comput\geraldorevi{e} intensity to not suffer from severe data movement bottlenecks, as demonstrated by prior work~\cite{doerfler2016applying}. Second, \sg{\gls{MPKI} serves} as a proxy for the memory intensity of an application. It can also indicate the memory pressure \gfii{experienced} by the main memory \gfiii{system}~\gfiii{\cite{hashemi2016accelerating, PEI,mutlu2007stall,mutlu2008parallelism,kim2010atlas,hsieh2016transparent,subramanian2016bliss,ebrahimi2010fairness,hashemi2016continuous,ghose.sigmetrics20}}. Third, \sg{\gls{LFMR}, \gfii{a new metric we introduce} \geraldorevi{\gfii{and} is described in detail \juan{later} in this subsection}, indicates how efficient the cache hierarchy is in reducing} data movement. 

\sg{As part of our methodology development, we evaluate other metrics related to data movement, including} raw cache misses, coherenc\geraldorevi{e} traffic, and DRAM row misses/hits/conflicts. We observe that even though such metrics are useful for further characterizing an application (as we do in some of our \sg{later} analyses \geraldorevi{in Section~\ref{sec:scalability}}), they do not necessarily characterize a specific type of data movement \gfii{bottleneck}. We show in Section~\ref{sec_scalability_benchmark_diversity} that the \geraldorevi{three architecture-dependent and two architecture-independent} metrics we select for our classification are enough to \gfii{accurately characterize and} cluster the different types of data movement \gfii{bottlenecks} \gfiii{in a wide variety of applications}.

\subsubsection{Definition \gfiii{of Metrics.}} We define \gfiii{Arithmetic Intensity (\gls{AI})} as the number of \gfiii{arithmetic and logic} operations \gfiii{performed} per \geraldorevi{L1} cache line accessed.\footnote{\geraldorevi{We consider \gls{AI} \gfiii{to be} architecture-dependent since we consider the number of cache lines accessed by the application \gfii{(and hence the hardware cache block size)} to compute the metric. \sgv{This is the same definition of AI used by the hardware profiling tool \gfvi{we employ} in \emph{Step 1} (i.e., the Intel VTune Profiler~\cite{vtune}).}}} This metric indicates how much computation there \sg{is} per memory request. Intuitively, applications with high \gls{AI} are likely to be computationally intensive, while applications with low \gls{AI} tend to be memory intensive. We use \gls{MPKI} \sg{at the \gls{LLC}}, i.e., the number of \sg{\gls{LLC}} misses per one thousand instructions. This metric is considered \sg{to be} a good indicat\gfiii{or} of \gls{NDP} suitability by several \sg{prior} works~\cite{hsieh2016accelerating,nai2017graphpim,kim2015understanding,boroumand2018google, nai2015instruction, ghose.ibmjrd19, lim2017triple, tsai:micro:2018:ams,kim2018coda,hong2016accelerating}. We define the \gls{LFMR} of an application as the ratio between the number of \sg{\gls{LLC}} misses and the total number of L1 cache misses. We find that this metric \geraldorevi{accurately} identifies \geraldorevi{how much} an application benefits from the deep cache hierarchy of a \sg{contemporary} CPU. \sg{An \gls{LFMR} value} close to 0 means that the number of \sg{\gls{LLC}} misses is very small compared to the number of L1 misses, i.e., the L1 misses are likely to hit in \sg{the L2 or L3 caches.} However, \sg{an \gls{LFMR} value} close to 1 means that very few L1 misses \sg{\gfii{hit in} L2 or L3 cache\gfii{s}}, i.e., the application does not benefit \gfiii{much} from the deep cache hierarchy, \sg{and most L1 misses need to be serviced by main memory}.

\subsubsection{Scalability Analysis and System Configuration\gfiii{.}} \gfii{The goal of the scalability analysis \gfiii{we perform} is to nail down the specific sources
of data movement bottlenecks in the application. In \gfiii{this} analysis, we (i) evaluate the performance and energy scaling of an application in three different system configurations; and (ii) collect the key metrics for our bottleneck classification (i.e., \gls{AI}, \gls{MPKI}, and \gls{LFMR}).} \sg{During} scalability analysis, we simulate \gfii{three} \gfii{system} configurations of a general-purpose multicore processor:
\begin{itemize}[itemsep=0pt, topsep=0pt, leftmargin=*]
    \item \gfii{A host CPU with a deep cache hierarchy (i.e., private L1 (32~kB) and L2 (256~kB) caches, and a shared L3 (8~MB) cache with 16 banks). We call this configuration \textit{Host CPU}.}
    \item  \gfii{A host CPU with a deep cache hierarchy (same cache configurations as in \textit{Host CPU}), augmented  with a stream prefetcher~\cite{palacharla1994evaluating}. We call this configuration \textit{Host CPU with prefetcher}.}
    \item \gfii{An \gls{NDP} CPU with a single level of cache (only a private \gfiii{read-only}\footnote{\gfiii{We use read-only L1 caches to simplify the cache coherence model of the NDP system. Enabling efficient synchronization and cache coherence in NDP architectures is an open-research problem, as we discuss in Section~\ref{sec_scalability_limitations}.}} L1 cache (32~kB), as \gfiii{assumed} in many prior \gls{NDP} works~\gfiii{\cite{boroumand2019conda, lazypim, boroumand2018google,smc_sim, tsai:micro:2018:ams,syncron,fernandez2020natsa,singh2019napel,drumond2017mondrian,ahn2015scalable}}) and no hardware prefetcher. We call this configuration \textit{NDP}.}
\end{itemize}

\noindent The remaining components of the processor configuration are \sg{kept} the same (e.g., number of cores, instruction window size, branch predictor) \sg{to isolate the impact of only the caches\gfii{,} prefetcher\geraldorevi{s}\gfii{, and \gls{NDP}}}. This way, we expect that the performance and energy differences between \geraldorevi{the \gfii{three}} configurations \gfii{to} come \emph{exclusively} from the different data movement requirements. \sg{For \gfii{the three} configurations, we sweep the number of CPU cores in our analysis from 1 to 256, as previous works~\cite{drumond2017mondrian,santos2016exploring, ahn2015scalable} show that large core counts are necessary to saturate the bandwidth \geraldorevi{provided by} modern high-bandwidth memories, and because modern CPUs and NDP proposals can have varying core counts.  The core count sweep allows us to} observe \sg{(1)~}how \gfiii{an} \gfii{application's performance} changes when increasing the pressure on the memory subsystem, \sg{(2)~}how much \sg{\gls{MLP}~\gfiii{\cite{glew1998mlp,qureshi2006case,mutlu2008parallelism,mutlu2006efficient,mutlu2005techniques}}} the \gfii{application} has, and \sg{(3)~how much the cores} leverage the cache hierarchy and the available memory bandwidth. \gfii{We proportionally increase the size of the CPU's private L1 and L2 caches  when increasing the number of CPU cores in our analysis (\gfiii{e.g., when} scaling the CPU core \gfiii{count} from 1 to 4, we also scale the aggregated L1/L2 cache size by a factor of 4).} \gfii{We use out-of-order and in-order CPU  cores \gfiii{in} our analysis for all three configurations. In this way, we \geraldorevi{build confidence} that our trends and findings are independent of a specific underlying general-purpose core microarchitecture.} \sg{We simulate a memory architecture similar to the  \gls{HMC}~\cite{HMC2}, where \gfiii{(1)} the host CPU accesses memory through a high-speed off-chip link, and \gfiii{(2)} the \gls{NDP} logic \gfiii{resides} in the logic layer of the memory chip and has direct access to the DRAM banks (thus taking advantage of higher \gfiii{memory} bandwidth and lower \gfiii{memory} latency).} \sg{Table~\ref{table_parameters} lists \geraldorevi{the} \gfiii{parameters} \geraldorevi{of} our \geraldorevi{host CPU, \gfiii{host CPU with prefetcher,} and \gls{NDP}} \geraldorevi{baseline} configuration\geraldorevi{s}.}

\begin{savenotes}
\begin{table*}[!t]
    \tempcommand{1.3}
    \centering
    \caption{Evaluated Host CPU and NDP system configurations.}
    \label{table_parameters}
    \footnotesize
    \resizebox{0.85\textwidth}{!}{
        \begin{tabular}{|c  l|}
        \hline
        \rowcolor{Gray}
        \multicolumn{2}{|c|}{ \textbf{Host CPU Configuration}}\\
        \hline
        \hline 
        \multirow{4}{*}{\shortstack{\textbf{Host CPU}\\ \textbf{Processor}}} &  1, 4, 16, 64, and 256~cores @2.4~GHz, 32~nm; 4-wide out-of-order   \\
                                            & 1, 4, 16, 64, and 256~cores @2.4~GHz, 32~nm; 4-wide in-order   \\
                                            & Buffers: 128-entry ROB; 32-entry LSQ (each)\\
                                            & Branch predictor: Two-level GAs~\cite{yeh1991two}. 2,048~entry BTB; 1~branch per fetch    \\
                                            
        \hline
        \multirow{2}{*}{\shortstack{\textbf{Private}\\ \textbf{L1 Cache}}} & 32~KB, 8-way, 4-cycle; 64~B line; LRU policy \\
                                                   & Energy: 15/33~pJ per hit/miss~\cite{muralimanohar2007optimizing, tsai:micro:2018:ams}                                          \\
            
        \hline
       \multirow{3}{*}{\shortstack{\textbf{Private}\\ \textbf{L2 Cache}}} & 256~KB, 8-way, 7-cycle; 64~B line; LRU policy \\
                                                  & MSHR size: 20-request, 20-write, 10-eviction \\
                                                  & Energy: 46/93~pJ per hit/miss~\cite{muralimanohar2007optimizing, tsai:micro:2018:ams}                                            \\
       \hline
       \multirow{4}{*}{\shortstack{\textbf{Shared}\\ \textbf{L3 Cache}}}  & 8~MB (16-banks), 0.5~MB per bank, 16-way, 27-cycle \\
                                                  & 64~B line; LRU policy; Bi-directional ring~\cite{ausavarungnirun2014design}; Inclusive; MESI protocol~\cite{papamarcos1984low} \\
                                                  & MSHR size: 64-request, 64-write, 64-eviction \\
                                                  & Energy: 945/1904~pJ per hit/miss~\cite{muralimanohar2007optimizing, tsai:micro:2018:ams} \\
       \hline
       \hline 
       \rowcolor{Gray}
       \multicolumn{2}{|c|}{\textbf{Host CPU with Prefetcher Configuration}}\\
       \hline
       \hline 
       \multirow{3}{*}{\shortstack{\textbf{Processor,}\\ \textbf{Private L1 Cache, Private L2 Cache,} \\ \textbf{and Share L3 Cache}}} &  \\
                                                                                                   & Same as in Host CPU Configuration   \\
                                                                                                   & \\ 
       \hline 
       \textbf{L2 Cache Prefetcher}               & Stream prefetcher~\cite{palacharla1994evaluating,srinath2007feedback}: 2-degree; 16 stream buffers; 64 entries \\
       \hline
       \hline 
       \rowcolor{Gray}
       \multicolumn{2}{|c|}{\textbf{NDP Configuration}}\\ 
       \hline
       \hline
       \multirow{4}{*}{\shortstack{\textbf{NDP CPU}\\ \textbf{Processor}}} & 1, 4, 16, 64, and 256~cores @2.4~GHz, 32~nm; 4-wide out-of-order   \\
                                                     & 1, 4, 16, 64, and 256~cores @ 2.4~GHz, 32~nm; 4-wide in-order   \\ 
                                                     & Buffers: 128-entry ROB; 32-entry LSQ (each)\\
                                                    & Branch predictor: Two-level GAs~\cite{yeh1991two}. 2,048~entry BTB; 1~branch per fetch    \\ \hline
        \multirow{2}{*}{\shortstack{\textbf{Private}\\ \textbf{L1 Cache}}}                               &  32~KB, 8-way, 4-cycle; 64~B line; LRU policy; Read-only Data Cache\\ 
                                                     & Energy: 15/33~pJ per hit/miss~\cite{muralimanohar2007optimizing, tsai:micro:2018:ams}  \\
                                                     
       \hline
       \hline 
       \rowcolor{Gray}
       \multicolumn{2}{|c|}{\textbf{Common}}\\ 
       \hline
       \hline 
       \multirow{4}{*}{\textbf{Main Memory}}      & HMC v2.0 Module~\cite{HMC2} 32 vaults, 8 DRAM banks/vault, 256~B row buffer \\
                                                  & 8~GB total size; DRAM@166 MHz; 4-links@8~GHz\\
                                                  & 8~B burst width at 2:1 core-to-bus freq. ratio; 
                                                   Open-page policy; HMC default interleaving~\cite{HMC2, ghose.sigmetrics20}\footnotemark[10]\\
                                                  & Energy: 2~pJ/bit internal, 8~pJ/bit logic layer ~\cite{top-pim,gao2015practical,tsai:micro:2018:ams}, 2~pJ/bit links~\cite{kim2013memory, NDC_ISPASS_2014, tsai:micro:2018:ams}\\
        \hline
        \end{tabular}
    }
\end{table*}

\end{savenotes}

\subsubsection{\gfii{Choosing an NDP Architecture.}}

\geraldo{\sg{We note that across the proposed \gls{NDP} architectures \gfii{in literature}, there is a lack of consensus on whether the architectures should make use of general-purpose \gls{NDP} cores or specialized \gls{NDP} accelerators~\cite{ghose.ibmjrd19, mutlu2020modern}.} In this work, we focus on general-purpose \gls{NDP} cores for \gfiii{two major} reasons. First, many prior works \geraldorevi{(e.g., \gfiii{\cite{tsai:micro:2018:ams,ahn2015scalable,boroumand2018google,lazypim, smc_sim,drumond2017mondrian,NDC_ISPASS_2014, IBM_ActiveCube, gao2015practical,lim2017triple, lockerman2020livia,fernandez2020natsa, syncron,de2018design,singh2019napel}})} suggest that general-purpose cores (especially simple in-order cores) can successfully accelerate memory-bound applications in \gls{NDP} architectures. In fact, UPMEM~\cite{devaux2019true}, \sg{a start-up building some of the first commercial \gfiii{in-DRAM} NDP \gfiii{systems}}, utilizes simple in-order cores \gfii{in} their \gls{NDP} units \gfiii{inside DRAM chips}~\cite{devaux2019true, gomez2022benchmarking}. Therefore, we believe that general-purpose \gls{NDP} cores are a promising candidate for future \gls{NDP} \sg{architectures}. Second, \sg{the goal of our work} is not to perform a design space exploration of \sg{different \gls{NDP} architectures,} but rather to understand the key properties \gfii{of} applications that lead to memory bottlenecks that can be mitigated by a simple \gls{NDP} engine. \sg{While we expect that each application could potentially benefit further from an \gls{NDP}} accelerator tailored to its computational \geraldorevi{and memory} requirements, \sg{such customized architectures open many challenges for a methodical characterization, such as the need for significant code refactoring, change\geraldorevi{s} in data mapping, and code partitioning between \gls{NDP} accelerators and host CPUs.\footnote{\gfiii{We show in Section~\ref{sec:case_study_2} that our \bench benchmark suite is \gfiv{useful} to \gfiv{rigorously} study \gls{NDP} accelerators.}}$^{,}$\footnote{\gfiii{The development of a \gfiv{new} methodology \gfiv{or extension of our methodology} to perform analysis targeting function-specific\gfiv{, customized, or reconfigurable} NDP accelerators is a good \gfiv{direction} for future work.}}}}

\section{Characterizing Memory Bottlenecks}
\label{sec:characterization}

\geraldorevi{In this section, we apply our three-step \gfii{workload characterization methodology} to characterize the sources of memory bottlenecks across a wide range of applications. First, we apply \emph{Step~1} to identify memory-bound functions within an application (Section~\ref{sec:vtune}). 
Second, we apply \emph{Step~2} and cluster \juan{the identified functions using \gfiii{two} architecture-independent metrics (spatial and temporal locality)} 
(Section~\ref{sec:locality}). Third, we apply \emph{Step~3} and combine the architecture-dependent and architecture-independent metrics to classify the different sources of memory bottlenecks we observe (Section~\ref{sec:scalability}).}

\geraldorevi{We \gfii{also} evaluate \gfii{various other} aspects of our three-step \gfii{workload characterization methodology}. We investigate the effect of increasing the last-level cache on our memory bottleneck classification in Section~\ref{sec_scalability_nuca}. We provide a \geraldorevi{validation} of our memory bottleneck classification in Section~\ref{sec_summary}. \gfiii{W}e discuss  the limitations of our proposed methodology in Section~\ref{sec_scalability_limitations}}.

\subsection{\geraldorevi{Step 1: Memory-Bound Function Identification}}
\label{sec:vtune}
\label{sec_selected}

We first \sg{apply Step~1 of our methodology across 345~applications \gfii{(listed in Appendix~\ref{sec:evaluatedapp}}) to} identify functions whose performance is significantly \geraldorevi{affected} by \sg{data movement.  We} use \gfii{the} previously-proposed top-down analysis \gfii{methodology}~\cite{yasin_ispass2014} that has been used by several recent workload characterization \geraldorevi{studies}~\cite{kanev_isca2015, sirin_damon2017, appuswamy_ipdpsw2018}. \sg{As discussed in Section~\ref{sec:step1}, we use \gfii{the} Intel VTune \gfiii{Profiler}~\cite{vtune}, which we run} on an Intel Xeon E3-1240 processor~\gfiii{\cite{E3-1240}} with \sg{four} cores. We disable \sg{hyper-threading for} more accurate profiling results, as recommended by the VTune documentation~\cite{vtune_HT}. For the applications that we \sg{analyze}, we select functions \geraldorevi{(1)} that take at least 3\% of the clock cycles, and \geraldorevi{(2) that have} \geraldorevi{a} Memory Bound percentage \geraldorevi{that is} greater than 30\%.
We choose 30\% as \geraldorevi{the threshold for} this metric because, in preliminary simulation experiments, we \sg{do not observe significant performance improvement or energy savings with data movement mitigation mechanisms for functions whose Memory Bound \geraldorevi{percentage} is} less than 30\%.

\sg{The applications we analyze come from a \gfii{variety of} sources, such as} popular \gfiii{workload} suites (Chai~\cite{gomezluna_ispass2017}, CORAL~\cite{coral}, Parboil~\cite{stratton2012parboil}, PARSEC~\cite{bienia2008parsec}, Rodinia~\cite{che_iiswc2009}, SD-VBS~\cite{Venkata_iiswc2009}, SPLASH-2~\cite{woo_isca1995}), benchmarking (STREAM~\cite{mccalpin_stream1995}, HPCC~\cite{luszczek_hpcc2006}, HPCG~\cite{dongarra_hpcg2015}), 
bioinformatics~\cite{ahmed2016comparison}, databases~\cite{balkesen_TKDE2015, gomezluna_icpp2015}, graph processing frameworks (GraphMat~\cite{sundaram_vldbendow2015}, Ligra~\cite{shun_ppopp2013}), a map-reduce framework (Phoenix~\cite{yoo_iiswc2009}), \sg{and} neural networks (AlexNet \cite{alexnet2012}, Darknet~\cite{redmon_darknet2013}). We explore different input \gfiii{dataset} sizes for the applications and choose real-world input datasets that impose high pressure on the memory subsystem (as we expect that such \gfiii{real-world} inputs are best suited for stressing the memory hierarchy). We also use different inputs for \gfii{applications} whose performance is tightly related to the input \gfiii{data}set properties. For example, we use two different graphs with varying connectivity degrees (rMat~\cite{rMat} and USA~\cite{dimacs}) to evaluate graph processing applications and two different read sequences to evaluate read alignment algorithms\gfiii{\cite{cali2020genasm,alser2017gatekeeper,alser2020accelerating}}. \addtocounter{footnote}{1}\footnotetext{\geraldorevi{The default HMC interleaving scheme (Row:Column:Bank:Vault~\cite{HMC2}) interleaves consecutive cache lines across vaults, and then across banks~\cite{hadidi2017demystifying}.}}

\begin{figure*}[!t]
    \centering
    \includegraphics[width=0.9\linewidth]{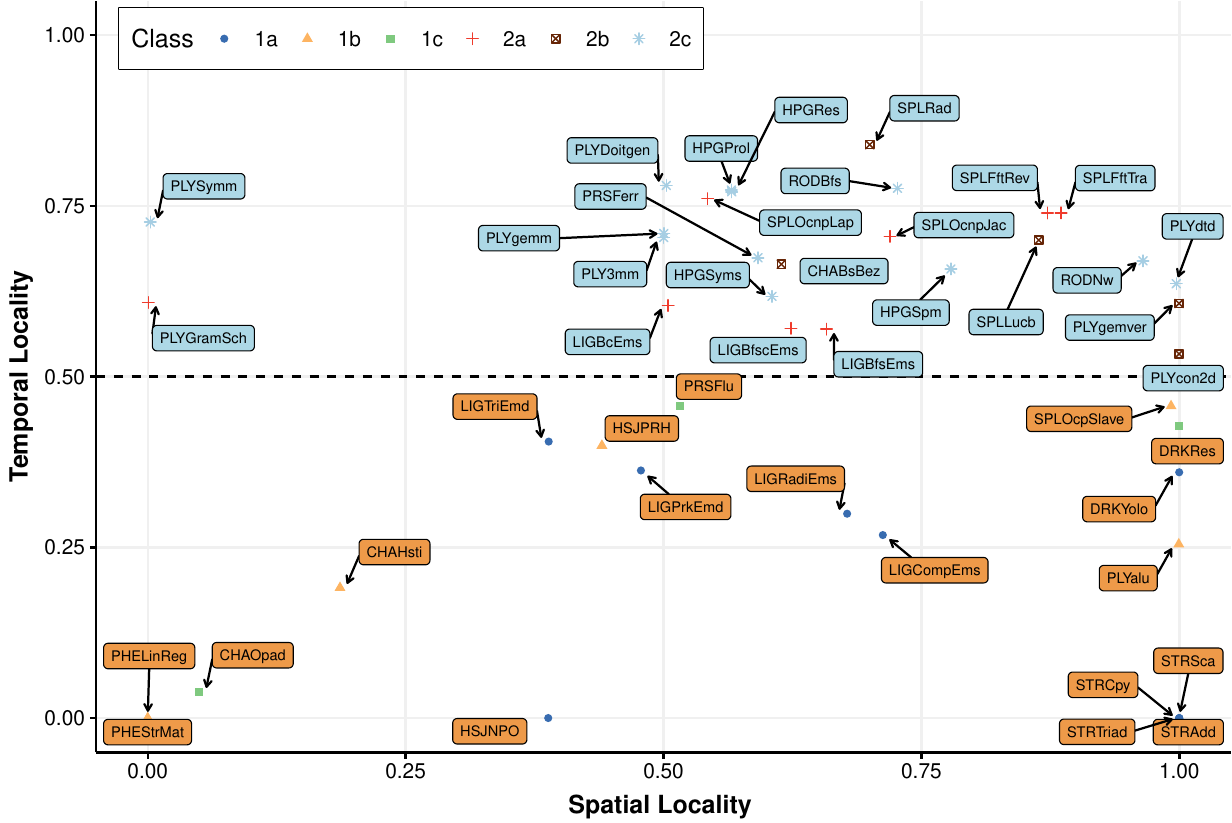}
     \caption{Locality-based clustering \gfii{of 44 representative functions}.}
  \label{fig:locality_chart}
\end{figure*}

{In total, our application analysis covers more than 77K functions.} To date, this is the most extensive analysis of data movement bottlenecks in real-world applications. We find a set of \geraldo{144} functions that take at least 3\% of the total clock cycles and have a value of the Memory Bound metric greater or equal to 30\%, \sg{which forms the basis of \gfiii{\bench,} our new data movement benchmark suite. \geraldorevi{We provide a} list of all 144 functions selected \geraldorevi{based on our analysis} \gfii{and their major characteristics} in Appendix~\ref{sec:benchlist}.} 

After identifying memory-bound functions over a wide range of applications, we apply Steps 2 and 3 of our methodology to classify the primary sources of memory bottlenecks for \gfii{our selected functions}. We evaluate a total of 144 functions out of the \gfiii{77K functions we analyze} in Step 1. \geraldorevi{These functions} \sg{span} across 74 different applications, belonging to 16 different widely-used benchmark suites \sg{or frameworks}.

From the 144~\sg{functions that} we analyze \gfiii{further}, we select a subset of 44 \sg{representative functions to explore in-depth in \gfiii{Sections~\ref{sec:locality} and \ref{sec:scalability}} \geraldorevi{and to drive our bottleneck classification analysis.} \gfiii{We use the 44 representative functions to ease our explanations and make figures more easily readable.} Table~\ref{tab:benchmarks} in Appendix~\ref{sec:benchlist} lists the \geraldorevi{44} representative functions that we select.} The table includes one column that indicates the class \sg{of data movement bottleneck experienced by each function (we discuss the classes \geraldorevi{in Section~\ref{sec:scalability}})}, and another column representing the percentage of clock cycles of the selected function in the whole application. \geraldorevi{We} select \gfiii{representative} \gfii{functions} that belong to a variety of domains: benchmarking, bioinformatics, data analytics, databases, data mining, data reorganization, graph processing, neural networks, physics, and signal processing. \geraldorevi{In Section~\ref{sec_summary}, we validate our classification using the remaining 100 functions \gfiv{and provide a summary of the results of our methodology when applied to all 144 functions}.}

\subsection{Step 2: Locality-Based Clustering}
\label{sec:locality}

We cluster the \gfii{44 representative functions} across both spatial and temporal locality using \gfii{the} K-means \gfii{clustering algorithm}~\cite{hartigan1979algorithm}. \geraldorevi{Figure~\ref{fig:locality_chart} shows how each \geraldorevi{\gfii{function}} is grouped.} \geraldorevi{We} find that two groups emerge from the clustering:
(1)~low temporal locality \gfii{functions} (orange \geraldorevi{boxes in Figure~\ref{fig:locality_chart}}), and
(2)~high temporal locality \gfii{functions} (blue \geraldorevi{boxes in Figure~\ref{fig:locality_chart}}). Intuitively, the closer a \geraldorevi{\gfii{function}} is to the bottom-left corner of the figure, the less likely it is to take advantage of a multi-level cache hierarchy. These \geraldorevi{\gfii{functions}} are more likely to be good candidates for NDP.  However, as we see \geraldorevi{in Section~\ref{sec:scalability}}, the \gfii{\gls{NDP}} suitability of a \gfii{function} also depends on a number of other factors.

\begin{figure*}[!t]
    \centering
  \includegraphics[width=\linewidth]{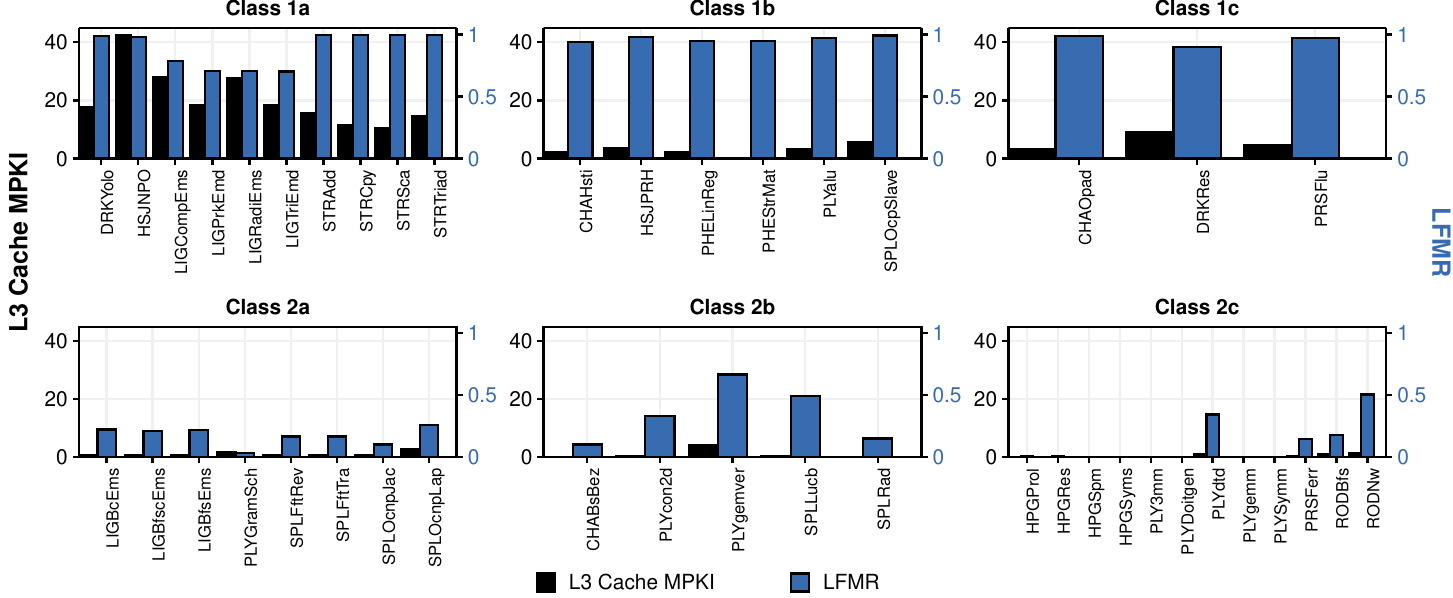}
  \caption{\gfii{L3 Cache \gls{MPKI} and} Last-to-First Miss Ratio (LFMR) \geraldorevi{for 44 representative \gfii{functions}. 
  }}
  \label{fig:lfmr}
\end{figure*}

\subsection{Step 3: Bottleneck Classification}
\label{sec:scalability}

Within the two groups of \gfii{functions} identified in Section~\ref{sec:locality}, we use three key metrics (\gls{AI}, \gls{MPKI}, and \gls{LFMR}) to classify the memory bottlenecks.
We observe that the \gls{AI} of the analyzed low temporal locality \gfii{functions} is low (i.e., \gfiii{always} less than 2.2 op\geraldorevi{s}/cache line\gfiii{, with an average of 1.3 ops/cache line}). Among the high temporal locality \gfii{functions}, there are some with low \gls{AI} (minimum of 0.3 op\geraldorevi{s}/cache line) and others with high \gls{AI} (maximum of 44 op\geraldorevi{s}/cache line). \gls{LFMR} indicates whether a \gfii{function} benefits from a deeper cache hierarchy.  When \gls{LFMR} is low (i.e., less than 0.1), then a \gfii{function} benefits significantly from a deeper cache hierarchy, as most misses from the L1 cache hit in either the L2 or L3 caches.  When \gls{LFMR} is high (i.e., greater than 0.7), then most L1 misses are not serviced by the the L2 or L3 caches, and must go to memory.  A medium \gls{LFMR} (0.1--0.7) indicates that a deeper cache hierarchy can mitigate some, but not \gfiii{a very large fraction of} L1 cache misses. \gls{MPKI} indicates the memory intensity of a \gfii{function} (i.e., the rate at which requests are issued to DRAM). We say that a \gfii{function} is memory-intensive (i.e., it has a high \gls{MPKI}) when the \gls{MPKI} is greater than 10, which is the same threshold used by prior works~\cite{hashemi2016accelerating,chou2015reducing,kim2010atlas,kim2010thread,muralidhara2011reducing, subramanian2016bliss,usui2016dash}.

We find that six classes of \gfii{functions} emerge, based on their temporal locality, \gls{AI}, \gls{MPKI}, and \gls{LFMR} values\geraldorevi{, as we observe from Figures~\ref{fig:locality_chart} and \ref{fig:lfmr}}. We observe that spatial locality is not a key metric for our classification (i.e., it does not define a bottleneck class) \geraldorevi{because} the L1 cache, \juan{which is \gfii{present} in both host \gfii{CPU} and NDP \gfii{system} configurations,} can capture most of the spatial locality for a \gfii{function}. Figure~\ref{fig:lfmr} shows the \gls{LFMR} and \gls{MPKI} values for each class.  Note that we do not have classes of \gfii{functions} for all possible combinations of metrics. In our analysis, we obtain the \geraldorevi{temporal locality, \gls{AI}, \gls{MPKI}, and \gls{LFMR} values} and their combinations empirically. \gfii{F}undamentally, \geraldorevi{not all value combinations of different metrics are possible.} We list some of the combinations we do \gfiii{\emph{not}} observe in our analysis \geraldorevi{of 144 \gfii{functions}}:
\begin{itemize}[noitemsep, leftmargin=*, topsep=0pt]
    \item A \gfii{function} with high \gfii{LLC} \gls{MPKI} does \emph{not} display low \gls{LFMR}\geraldorevi{.} \gfii{This is because} \gfii{a low LFMR happens when most L1 misses hit the L2/L3 caches. Thus,} \geraldorevi{it becomes highly unlikely for the L3 cache to suffer many misses when the L2/L3 caches do a good job \juan{in} fulfilling L1 cache misses.}
    
    \item  A \gfii{function} with high temporal locality does \emph{not} display \geraldorevi{both} high \gls{LFMR} and high \gls{MPKI}. \gfii{This is because} \geraldorevi{a \gfii{function} with high temporal locality will likely issue repeated memory requests to few memory addresses, which will likely \gfiii{be} serviced by the cache hierarchy.}
    \item A \gfii{function} with low temporal locality does \emph{not} display low \gls{LFMR} since there is little data locality to be captured by the cache hierarchy. 
\end{itemize}

We discuss each class in detail below, identifying the memory bottlenecks for each class and whether \gfii{the} NDP \gfii{system} can alleviate these bottlenecks. To simplify our explanations, we focus on a \gfiii{smaller} set of \geraldorevi{12} representative \gfii{functions} \gfii{(out of the 44 representative functions)} for this part of the analysis. \geraldo{Figure~\ref{figure_performance} shows how each \gfiii{of the 12} \gfii{function\gfiii{s}} scales in terms of performance for the \geraldorevi{\textit{host \gfii{CPU}}, \textit{host \gfii{CPU} with prefetcher}, and \textit{NDP}} \gfii{system} configurations.}

\begin{figure*}[h]
    \centering
    \includegraphics[width=\linewidth]{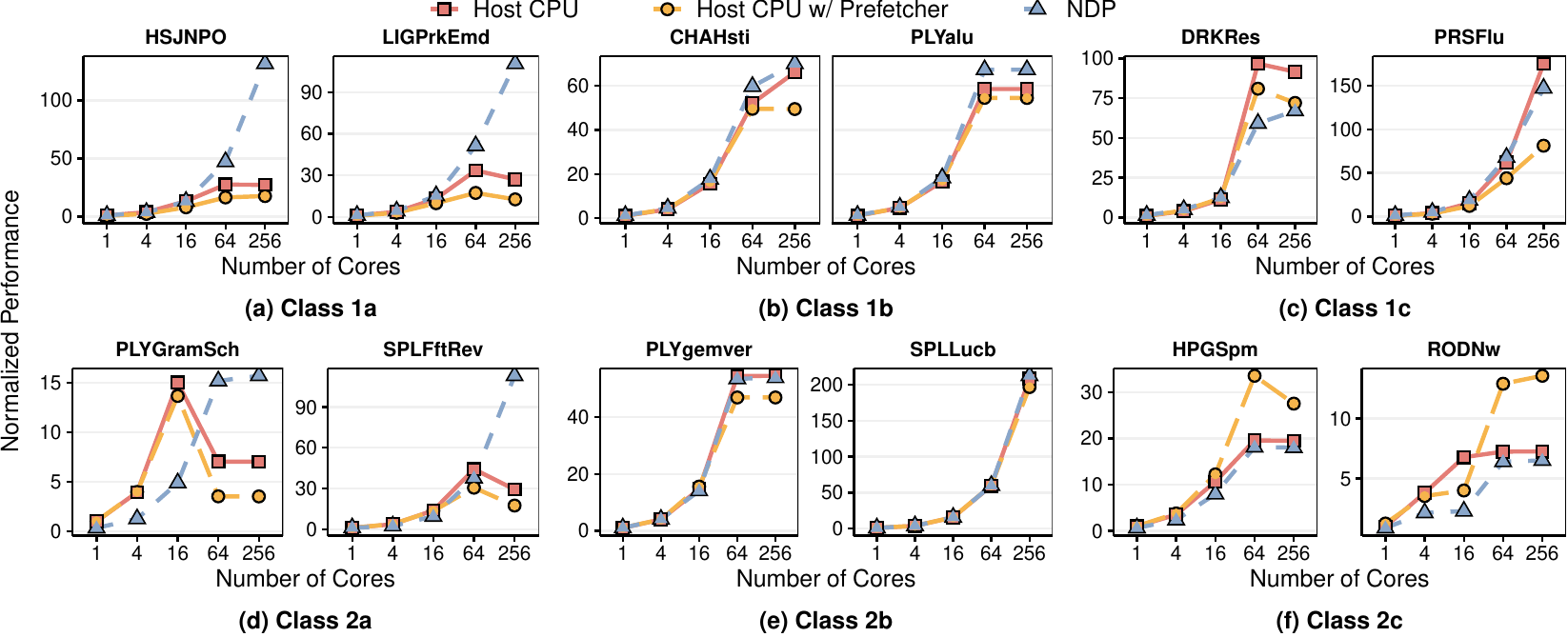}
    \caption{Performance of \gfiii{12} representative  \gfii{functions} on \geraldorevi{three systems:} host \gfii{CPU}\geraldorevi{, host \gfii{CPU} with prefetcher,} and NDP, normalized to one host \gfii{CPU} core.}
    \label{figure_performance}
\end{figure*}

\subsubsection{Class~1a: Low Temporal Locality, Low \gls{AI}, High \gls{LFMR}, and High \gls{MPKI} \gfii{(\textit{DRAM Bandwidth-Bound Functions})}}
\label{sec_scalability_class1a}

\gfii{Functions} in this class exert high \geraldorevi{main} memory pressure \geraldorevi{since they are} highly memory intensive \geraldorevi{and have low data reuse}.  To understand how this affects \geraldorevi{a} \gfii{function}\geraldorevi{'s} suitability for \gls{NDP}, we study how performance scales as we increase the number of cores available to a \gfii{function}, for \geraldorevi{the host \gfii{CPU}, host \gfii{CPU} with prefetcher, and NDP \gfii{system} configurations}.  Figure~\ref{figure_performance}(a) \geraldorevi{depicts} performance\footnote{\geraldorevi{P}erformance \geraldorevi{is} the inverse of \geraldorevi{application} execution time.} as we increase the core count, normalized to the performance of one host \gfii{CPU} core, for two representative \gfii{functions} from \geraldorevi{Class~1a} (\texttt{HSJNPO} and \texttt{LIGPrkEmd}; we see similar trends for all \gfii{functions} in the class).

We make \geraldorevi{three} observations from the figure. First, as the number of host \gfii{CPU} cores increases, performance eventually stops increasing significantly.  For \texttt{HSJNPO}, \gfii{host CPU} performance increases by \geraldorevi{27.5$\times$ going from 1 to 64 host \gfii{CPU} cores but} only 27\% going from 64 host \gfii{CPU} cores to 256 host \gfii{CPU} cores\geraldorevi{. Fo}r \texttt{LIGPrkEmd}, \gfii{host CPU} performance \geraldorevi{increases by 33$\times$ going from 1 to 64 host \gfii{CPU} cores but} \emph{decreases} by 20\% \geraldorevi{going from 64 to 256 host \gfii{CPU} cores}.
We find that the lack of performance improvement \geraldorevi{at large host \gfii{CPU} core counts} is due to \gfii{main} memory bandwidth saturation, as shown in Figure~\ref{fig:bw_group_1}. 
Given the limited \gfii{DRAM} bandwidth available across the off-chip memory channel, we find that Class~1a \gfii{functions} saturate the \gfii{DRAM} bandwidth once enough \gfii{host CPU} cores (e.g., 64) are used,
\gfiii{and thus} these \gfii{functions} \gfiii{are} \emph{bottlenecked by the DRAM bandwidth}.
\geraldorevi{Second, the host \gfii{CPU system} with prefetcher slows down the execution of the \gfiii{\texttt{HSJNPO} (\texttt{LIGPrkEmd})}  \gfii{function} compared with the host \gfii{CPU system without prefetcher} by 43\% (38\%)\gfii{, on average across all core counts}. The prefetcher is \gfii{ineffective} since these \gfii{functions} have low temporal and spatial locality.} \geraldorevi{Third}, when running on \gfii{the} \gls{NDP} \geraldorevi{\gfii{system}}, the \gfii{functions} see \geraldorevi{continued performance improvements} as the number of \gfii{NDP} cores increase\gfii{s}.  By providing the \gfii{functions} with access to the much higher bandwidth available inside memory, the \gls{NDP} \gfii{system} can \geraldorevi{greatly} outperform the host \gfii{CPU system} at a high enough core count. \geraldorevi{For example,} at \geraldorevi{64/}256~cores, \gfii{the} \gls{NDP} \gfii{system} outperform\gfiii{s} \gfii{the} host \gfii{CPU system} by \geraldorevi{1.7$\times$/}4.8$\times$ for \texttt{HSJNPO}, and by \geraldorevi{1.5$\times$/}4.\gfii{1}$\times$ for \texttt{LIGPrkEmd}.

% \begin{figure}[ht]
% \centering
%   \centering
%    \includegraphics[width=\linewidth]{mainmatter/04_damov/Plots/bw-class1a-crop.pdf}
%    \caption{Host \gfii{CPU} \geraldorevi{system} IPC vs. \gfiii{utilized} \geraldorevi{\gfii{DRAM}} Bandwidth for \geraldorevi{representative} Class~1a \geraldorevi{\gfii{functions}}.}
%    \label{fig:bw_group_1}
% \end{figure}

% \begin{figure}[ht]
%   \centering
%    \includegraphics[width=\linewidth]{mainmatter/04_damov/Plots/eg_group_1-crop.pdf}
%   \caption{Cache and DRAM energy \geraldorevi{breakdown} for \geraldorevi{representative} Class~1a \geraldorevi{\gfii{functions}} \gfii{at 1, 4, 16, 64, and 256 cores}.}
%   \label{fig:energy_group_1}
% \end{figure}

\begin{figure}[h]
%  \vspace{-5pt}
\centering
\begin{minipage}[t]{.47\textwidth}
  \centering
   \includegraphics[width=\linewidth]{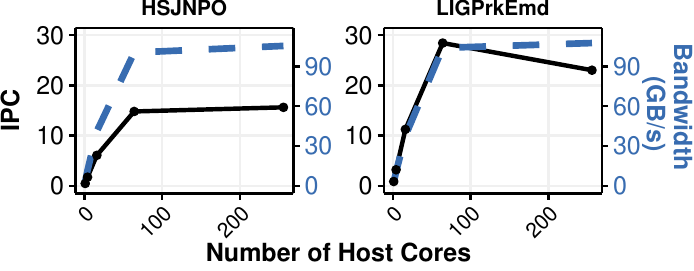}
   \caption{Host \gfii{CPU} \geraldorevi{system} IPC vs. \gfiii{utilized} \geraldorevi{\gfii{DRAM}} Bandwidth for \geraldorevi{representative} Class~1a \geraldorevi{\gfii{functions}}.}
   \label{fig:bw_group_1}
\end{minipage}%
\qquad
\begin{minipage}[t]{.47\textwidth}
  \centering
   \includegraphics[width=\linewidth]{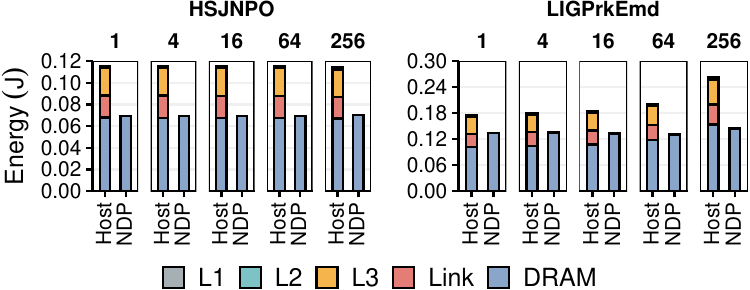}
  \caption{Cache and DRAM energy \geraldorevi{breakdown} for \geraldorevi{representative} Class~1a \geraldorevi{\gfii{functions}} \gfii{at 1, 4, 16, 64, and 256 cores}.}
  \label{fig:energy_group_1}
\end{minipage}
 % \vspace{-5pt}
\end{figure}

Figure~\ref{fig:energy_group_1} depicts the energy breakdown for our two representative \gfii{functions}. We make two observations from the figure. First, for \texttt{HSJNPO}, the energy spent on DRAM for both host \gfii{CPU} \gfii{system} and \gls{NDP} \gfii{system} are similar. This is due to the \gfii{function}'s poor locality, as 98\% of its memory requests miss in the L1 cache.  \geraldorevi{Since} \gls{LFMR} \geraldorevi{is} near 1, \gfii{L1 miss requests} almost always miss in the L2 and L3 caches and go to DRAM \geraldorevi{in the host \gfii{CPU} system} \gfiii{for all core counts we evaluate}, which requires significant energy to query the large cache\geraldorevi{s} and then to \geraldorevi{perform} off-chip \geraldorevi{data transfer\gfii{s}}. 
\geraldorevi{The} NDP \geraldorevi{system} does not \geraldorevi{access} L2, L3, and off-chip link\geraldorevi{s}, \geraldorevi{leading to large system} energy reduction. Second, for \texttt{\texttt{LIGPrkEmd}}, the DRAM energy is higher in \gfii{the} \gls{NDP} \gfii{system} than in the host \gfii{CPU system}. Since the \gfii{function\gfiii{'s}} \gls{LFMR} is 0.7, some memory requests that would be cache hits in the host \gfii{CPU}'s L2 \geraldorevi{and} L3 caches are instead sent directly to DRAM \geraldorevi{in the} \gls{NDP} \geraldorevi{system}. However, the total energy consumption on the host \gfii{CPU} \geraldorevi{system} is still larger than \gfii{that} on \geraldorevi{the} \gls{NDP} \geraldorevi{system}, again because \gfii{the} \gls{NDP} \gfii{system} eliminates the L2\geraldorevi{,} L3 and off-chip link energy.

\geraldorevi{\gfii{DRAM}} bandwidth-bound applications such as those in Class~1a have been the primary focus of a large number of proposed NDP architectures (e.g., \cite{IBM_ActiveCube,azarkhish2018neurostream,ahn2015scalable,azarkhish2016memory,gao2017tetris, nai2017graphpim, boroumand2018google,ke2019recnmp, NDC_ISPASS_2014, pugsley2014comparing}), as they benefit from increased \gfii{main memory} bandwidth and do not have high \gls{AI} (and, thus, do not benefit from complex cores \geraldorevi{on} the host \gfii{CPU} \geraldorevi{system}).  %However, while 
An NDP architecture for \geraldorevi{a} \gfii{function} in Class~1a  \gfii{needs to} extract enough \gls{MLP}~\gfiv{\cite{glew1998mlp, mutlu2003runahead, qureshi2006case, mutlu2008parallelism,mutlu2006efficient,mutlu2005techniques,chou2004microarchitecture,tuck2006scalable,phadke2011mlp,van2009mlp,everman2007memory,patsilaras2012efficiently}} to maximize the usage of the available internal \geraldorevi{memory} bandwidth. \gfii{However,} prior work has shown that this can be challenging due to the area and power constraints in the logic layer of a 3D-stacked DRAM~\cite{boroumand2018google, ahn2015scalable}. To exploit the high \geraldorevi{memory} bandwidth while \gfii{satisfying} these \gfii{area and power} constraints, the NDP architecture should leverage application \gfii{memory} access patterns to efficiently maximize \gfii{main memory bandwidth} utilization. 

We find that there are two dominant types of memory access patterns among our Class~1a \gfii{functions}. First, \gfii{functions} with regular access patterns (\texttt{DRKYolo}, \texttt{STRAdd}, \texttt{STRCpy}, \texttt{STRSca}, \texttt{STRTriad}) can take advantage of specialized accelerators or \gls{SIMD} architectures~\cite{drumond2017mondrian, boroumand2018google}, which can exploit the regular access patterns to issue many memory requests concurrently. Such accelerators or \gls{SIMD} architectures have \geraldorevi{hardware} area and thermal dissipation that fall \gfii{well} within the constraints of 3D-stacked DRAM~\cite{top-pim,eckert2014, boroumand2018google, ahn2015scalable}. Second, \gfii{functions} with irregular access patterns (\texttt{HSJNPO}, \texttt{LIGCompEms}, \texttt{LIGPrkEmd}, \texttt{LIGRadiEms}) require techniques to extract \gls{MLP} while still fitting within the design constraints.  This requires techniques that cater to the irregular memory access patterns, such as prefetching algorithms designed for graph processing~\gfiv{\cite{ahn2015scalable,nilakant2014prefedge,kaushik2021gretch,ainsworth2016graph,basak2019analysis,yan2019alleviating}}, \gfii{pre-execution of difficult access patterns~\gfiv{\cite{mutlu2003runahead, hashemi2016continuous, srinivasan2004continual, mutlu2006efficient, mutlu2005techniques,hashemi2016accelerating,annavaram2001data,collins2001dynamic,dundas1997improving,mutlu2003runaheadmicro,ramirez2008runahead,ramirez2010efficient,zhang2007accelerating}}} or \gfiii{hardware accelerators for} pointer chasing~\gfiv{\cite{hsieh2016accelerating, ebrahimi2009techniques, cooksey2002stateless,roth1999effective,santos2018processing,lockerman2020livia,hong2016accelerating}}. 

\subsubsection{Class~1b: Low Temporal Locality, Low \gls{AI}, High \gls{LFMR}, and Low \gls{MPKI} \gfii{(\textit{DRAM Latency-Bound Functions})}}
\label{sec_scalability_class1b}

While \gfii{functions} in this class do not effectively use the host \gfii{CPU} caches, they do \emph{not} exert high pressure on the \gfii{main }memory due to their low \gls{MPKI}. Across \geraldorevi{all Class~1b} \gfii{functions}, the average \geraldorevi{\gfii{DRAM}} bandwidth consumption is only 0.5 GB/s.  However, all the \gfii{functions} have very high \gls{LFMR} values (the minimum is 0.94 for \texttt{CHAHsti}), indicating that the \geraldorevi{host} \gfii{CPU} L2 and L3 caches are ineffective. Because the \gfii{functions} cannot exploit significant \gls{MLP} but still incur long-latency requests to DRAM, the DRAM requests fall on the critical path of execution and stall forward progress~\gfiv{\cite{mutlu2003runahead, ghose.isca13, mutlu2008parallelism,hashemi2016accelerating,hashemi2016continuous}}. Thus, Class~1b \gfii{functions} are \emph{bottlenecked by \gfii{DRAM} latency}. Figure~\ref{figure_performance}(b) shows performance of both \gfiii{the} host \gfii{CPU system} and \gfiii{the} NDP \gfii{system} for two representative \gfii{functions} from Class~1b (\texttt{CHAHsti} and  \texttt{PLYalu}). We observe that while \gfii{performance of} both \gfiii{the} host \gfii{CPU} \gfii{system} and \geraldorevi{the} NDP \geraldorevi{system} scale well as the core count increases, NDP \geraldorevi{system} performance is always higher than the host \gfii{CPU system} performance for the same core count.  The \geraldorevi{maximum (average)} speedup with NDP over host \gfii{CPU} at the same core count is \geraldorevi{1.15$\times$ (1.12$\times$)} for \texttt{CHAHsti} and \geraldorevi{1.23$\times$ (1.13$\times$)} for \texttt{PLYalu}.

We find that \gfii{the} NDP \gfii{system}'s improved performance is due to a reduction in the \gls{AMAT}~\cite{amatref}.
Figure~\ref{figure_amat_group_2} shows the \gls{AMAT} for our two representative \gfii{functions}.  Memory accesses take significantly longer in the host \gfii{CPU} \geraldorevi{system} than in \geraldorevi{the} NDP \geraldorevi{system} due to the additional latency of looking up requests in the L2 and L3 caches, \geraldorevi{even though} data \geraldorevi{is rarely present in} those caches\gfiii{, and going through the off-chip links}.

% \begin{figure}[ht]
%     \centering
%      \includegraphics[width=1.0\linewidth]{mainmatter/04_damov/Plots/lt_group_2-crop.pdf}%
%       \caption{\gfii{Average Memory Access Time (\gls{AMAT})} for \geraldorevi{representative} Class~1b \geraldorevi{\gfii{functions}}. }
%       \label{figure_amat_group_2}
% \end{figure}

Figure~\ref{fig:energy_group_2} shows the energy breakdown for Class~1b \gfii{representative functions}. Similar to Class~1a, we observe that the L2/L3 caches and off-chip links are a large source of energy usage in the host \gfii{CPU} \geraldorevi{system}. While DRAM energy increases in \gfii{the} NDP \gfii{system}, as L2/L3 hits in the host \gfii{CPU system} become DRAM lookups \gfii{with} NDP, the overall energy consumption \geraldorevi{in the NDP system} is \geraldorevi{greatly smaller (by 69\% {maximum} and 39\% on average)} due to the lack of L2 and L3 caches.

% \begin{figure}[ht]
%   \centering
%    \includegraphics[width=1.0\linewidth]{mainmatter/04_damov/Plots/eg_group_2-crop.pdf}%
%    \caption{Energy breakdown for \geraldorevi{representative} Class~1b \geraldorevi{\gfii{functions}}.}
% \label{fig:energy_group_2}
% \end{figure}

\begin{figure}[ht]
\centering
\begin{minipage}[t]{.47\textwidth}
  \centering
 \includegraphics[width=1.0\linewidth]{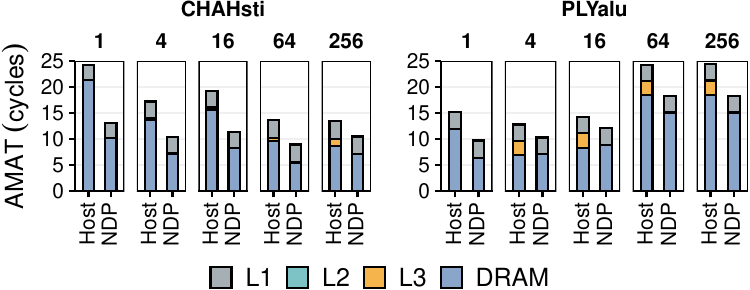}%
  \caption{\gfii{Average Memory Access Time (\gls{AMAT})} for \geraldorevi{representative} Class~1b \geraldorevi{\gfii{functions}}. }
  \label{figure_amat_group_2}
\end{minipage}%
\qquad
\begin{minipage}[t]{.47\textwidth}
  \centering
   \includegraphics[width=1.0\linewidth]{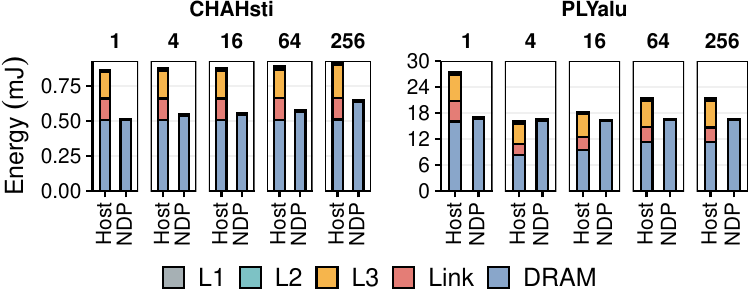}%
   \caption{Energy breakdown for \geraldorevi{representative} Class~1b \geraldorevi{\gfii{functions}}.}
\label{fig:energy_group_2}
\end{minipage}
\vspace{-5pt}
\end{figure}

Class~1b \gfii{functions} benefit from \gfii{the} NDP \gfii{system}, but primarily because of the lower memory access latency \geraldorevi{(and energy)} that \gfii{the} NDP \gfii{system} provides for memory requests that need to be serviced by DRAM.  These \gfii{functions} could benefit from other latency \geraldorevi{and energy} reduction techniques, such as L2/L3 cache bypassing~\gfvi{\cite{tsai:micro:2018:ams, johnson1999run, sembrant2014direct,tsai2017jenga, seshadri2014dirty,seshadri2015mitigating,johnson1997run,tyson1995modified,memik2003just,kharbutli2008counter,gupta2013adaptive,li2012optimal,sridharan2016discrete}}, \gfvii{low-latency DRAM~\cite{Tiered-Latency_LEE, lee2015adaptive,lee2017design,chang2017understanding,chang.sigmetrics2016,hassan2016chargecache,hassan2019crow,seshadri2013rowclone,wang2018reducing,kim2018solar,das2018vrl,chang2016low,chang2014improving,chang2017understandingphd,hassan2017softmc,lee2016reducing,luo2020clr,choi2015multiple,son2013reducing,liu2012raidr,rldram,sato1998fast,orosa2021codic,seongil2014row}, and better memory access scheduling~\gfvii{\cite{mutlu2008parallelism, subramanian2016bliss, kim2010atlas,kim2010thread, mutlu2007stall,ausavarungnirun2012staged, usui2016dash, subramanian2014blacklisting, muralidhara2011reducing,subramanian2015application,subramanian2013mise,rixner2004memory,rixner2000memory,zuravleff1997controller,moscibroda2008distributed,ebrahimi2011parallel,ipek2008self,hur2004adaptive,ebrahimi2010fairness,ghose.isca13,xie2014improving,yuan2009complexity,wang2014memory}}}.  However, they generally do \geraldorevi{\emph{not}} benefit significantly from prefetching (as seen in Figure~\ref{figure_performance}(b)), since infrequent memory requests make it difficult for the prefetcher to successfully train on an access pattern. 

\subsubsection{Class~1c: Low Temporal Locality, Low AI, Decreasing LFMR with Core Count, and Low MPKI \gfii{(\textit{L1/L2 Cache Capacity Bottlenecked Functions})}}
\label{sec_scalability_class1c}

We find that the behavior of \gfii{functions} in this class depends on the number of cores they are using. Figure~\ref{figure_performance}(c) shows the host \gfii{CPU system} and \geraldorevi{the} NDP \geraldorevi{system} performance as we increase the core count for two representative \gfii{functions} (\texttt{DRKRes} and \texttt{PRSFlu}). We make two observations from the figure. First, at low core counts, \geraldorevi{the} NDP \geraldorevi{system} outperforms the host \gfii{CPU} \geraldorevi{system}. \gfii{With} a low number of cores, the \gfii{functions} have medium to high LFMR (0.5 for \texttt{DRKRes} \gfii{at 1 and 4 host CPU cores; 0.97 at 1 host CPU core and \gfii{0.91} at 4 host CPU cores for \texttt{PRSFlu})}, and behave like Class~1b \gfii{functions}, where they are \gfiii{DRAM} latency-sensitive. Second, as the core count increases, \geraldorevi{the} host \gfii{CPU} \geraldorevi{system} \gfii{begins} to outperform \geraldorevi{the} NDP \geraldorevi{system}.  For example, beyond 16 \geraldorevi{(64)} cores, the host \gfii{CPU system}  outperforms \gfii{the} NDP \gfii{system} for \texttt{DRKRes} \geraldorevi{(\texttt{PRSFlu})}.  This is because as the core count increases, the \gfii{aggregate} L1 and L2 cache size available at the host \gfii{CPU system} grow\gfii{s}, \geraldorevi{which} \geraldorevi{reduces} the \gfii{miss rates of both L2 and L3} caches. As a result, the LFMR decreases significantly (e.g., at 256 cores, \gfii{LFMR is} 0.09 for \texttt{DRKRes} and 0.35 for \texttt{PRSFlu}). This indicates that the \emph{available \gfiii{L1/L2} cache capacity} bottlenecks Class 1c \gfii{functions}.

Figure~\ref{fig:energy_group_3} shows the energy breakdown for Class~1c \gfii{functions}. We make three observations from the figure. First, for \gfii{functions} with larger LFMR values (\texttt{PRSFlu}), \gfii{the} NDP \gfii{system} provides energy savings over the host \gfii{CPU system} at lower core counts, \gfii{since} \gfii{the} NDP \gfii{system} \gfii{eliminates the energy consumed due to L3 and off-chip link accesses}. Second, for \gfii{functions} with smaller LFMR \gfii{values} (\texttt{DRKRes}), \gfii{the} NDP \gfii{system} does not provide energy savings even for \gfii{low} core counts. Due to the medium LFMR, \geraldorevi{enough} requests still hit in the host \gfii{CPU system} L2/L3 caches, and these \geraldorevi{cache} hits become DRAM accesses in \gfii{the} NDP \gfii{system}, which consume more energy than the cache hits. Third, at high\geraldorevi{-enough} core counts, \gfii{the} NDP \gfii{system} consum\gfii{es} more energy than the host \gfii{CPU system} for all Class~1c \gfii{functions}. 
\gfii{As the LFMR decreases, the \gfii{functions} effectively utilize the caches in the host CPU system, reducing the off-chip traffic and, consequently, the energy Class~1c functions spend on accessing DRAM.} \gfiii{The \gls{NDP} system, which does not have L2 and L3 caches, pays the larger energy cost of a DRAM access for all L2/L3 hits in the host CPU system.}

\begin{figure}[ht]
    \centering
    \includegraphics[width=0.7\linewidth]{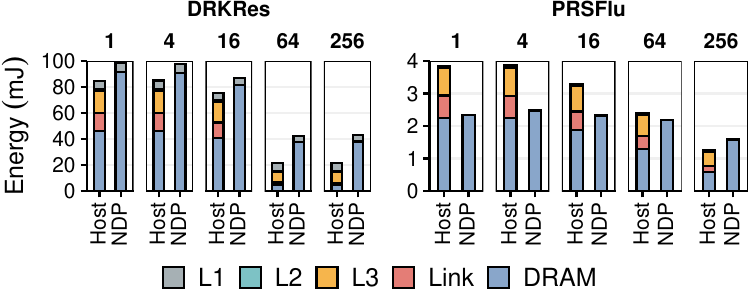}%
    \caption{Energy breakdown for \geraldorevi{representative} Class~1c \geraldorevi{\gfii{functions}}.}
    \label{fig:energy_group_3}
\end{figure}

We find that \gfii{the primary source of the memory bottleneck in Class~1c functions is limited \gfiii{L1/L2} cache capacity. Therefore,} while \gfii{the} NDP \gfii{system} \geraldorevi{improve\gfii{s} performance and energy of \gfii{some} Class~1c \gfii{functions} at low core counts \gfii{(with lower associated \gfiii{L1/L2} cache capacity)}}, \gfii{the} NDP \gfii{system} does not provide \geraldorevi{performance and energy} benefits across all core counts for Class~1c \gfii{functions}.

\subsubsection{Class~2a: High Temporal Locality, Low AI, Increasing LFMR with Core Count, and Low MPKI \gfii{(\textit{L3 Cache Contention Bottlenecked Functions})}}
\label{sec_scalability_class2a}

Like Class~1c \gfii{functions}, the behavior of \gfii{the} \gfii{functions} in this class depends on the number of cores that they \gfii{use}.  Figure~\ref{figure_performance}(d) shows the host \gfii{CPU system} and \geraldorevi{the} NDP \geraldorevi{system} performance as we increase the core count for two representative \gfii{functions} (\texttt{PLYGramSch} and \texttt{SPLFftRev}).  We make two observations from the figure. First, at low core counts, the \gfii{functions} do \emph{not} benefit from \gfii{the} NDP \gfii{system}. In fact, for a single core \geraldorevi{(16 cores)}, \texttt{PLYGramSch} \emph{slows down} by 67\% \geraldorevi{(3$\times$)} when running on \geraldorevi{the} NDP \geraldorevi{system}, compared to running on the host \gfii{CPU} \geraldorevi{system}. This is because, at low core counts, the\gfiii{se} \gfii{functions} make reasonably good use of the cache hierarchy, with LFMR values of 0.03 for \texttt{PLYGramSch} and \gfii{lower than} 0.44 for \texttt{SPLFftRev} \gfii{until 16 host CPU cores}.
We confirm this in Figure~\ref{fig:latency_class_4}, where we see that very few memory requests for \texttt{PLYGramSch} and \texttt{SPLFftRev} go to DRAM (5\% for \texttt{PLYGramSch}, and at most 13\% for \texttt{SPLFftRev}) \geraldorevi{at \gfiii{core counts} low\gfii{er than 16}}.
Second, at high core counts (\gfii{i.e.,} 64 for \texttt{PLYGramSch} and 256 for \texttt{SPLFftRev}), the host \gfii{CPU} \geraldorevi{system} performance starts to \emph{decrease}.  This is because Class~2a \gfii{functions} are \emph{bottlenecked by cache contention}.  At 256 cores, this contention undermines the \gfii{cache effectiveness} and causes the LFMR to increase to 0.97 for \texttt{PLYGramSch} and 0.93 for \texttt{SPLFftRev}.  With the last-level cache rendered  \geraldorevi{essentially} ineffective, the NDP \geraldorevi{system greatly improves} performance over the host \gfii{CPU} \geraldorevi{system:} by 2.23$\times$ for \texttt{PLYGramSch} and 3.85$\times$ for \texttt{SPLFftRev} \gfii{at 256 cores}.

% \begin{figure}[h]
%   \centering
%  \includegraphics[width=1.0\linewidth]{mainmatter/04_damov/Plots/mb_group_4-crop.pdf}%
%  \caption{Memory request breakdown for \geraldorevi{representative} Class~2a \geraldorevi{\gfii{functions}}.}
%  \label{fig:latency_class_4}
% \end{figure}

One impact of the increased cache contention is that it \geraldorevi{converts} these high-\geraldorevi{temporal-}locality \gfii{functions} into \gfii{memory} latency-bound \gfii{functions}.  We find that with the increased number of requests going to DRAM \gfii{due to cache contention}, the \gls{AMAT} increases significantly, in large part due to queuing at the memory controller.  At 256~cores, the queuing becomes so severe that a large fraction of requests (\gfii{24\% for \texttt{PLYGramSch} and 67\% for \texttt{SPLFftRev}}) must be reissued because the \gfiii{memory controller} queues are full.
The increased \gfii{main memory} bandwidth available \gfii{to the NDP cores allows} the NDP \geraldorevi{system} to issue many more requests concurrently, which reduces the average length of the queue and, thus, \gfii{the main memory latency}. \geraldorevi{\gfii{T}he NDP system \gfii{also reduces} memory access latency \gfii{by getting} rid of \gfiii{L2/L3} cache lookup and interconnect latencies.}

Figure~\ref{fig:energy_group_4} shows the energy \gfii{breakdown} for \geraldorevi{the two representative} Class~2a \geraldorevi{\gfii{functions}}. We make two observations. First, the host \geraldorevi{\gfii{CPU} system} is more energy-efficient than the NDP \geraldorevi{system} \gfiii{at} low core count\geraldorevi{s}, as most of the memory requests are \geraldorevi{\gfii{served} by} on-chip \geraldorevi{caches in the host \gfii{CPU} system}. Second, \geraldorevi{the} NDP \geraldorevi{system} provides \gfii{large} energy savings over the host \geraldorevi{\gfii {CPU} system} at high core counts.  This is due to the increased cache contention, which increases the number of off-chip requests that the host  \gfii{CPU system} must make, increasing the L3 and off-chip link energy.

% \begin{figure}[h]
%   \centering
%   \includegraphics[width=1.0\linewidth]{mainmatter/04_damov/Plots/eg_group_4-crop.pdf}%
%     \caption{Energy breakdown for \geraldorevi{representative} Class~2a \geraldorevi{\gfii{functions}}.}
%     \label{fig:energy_group_4}
% \end{figure}

\begin{figure}[h]
\centering
\begin{minipage}[t]{.47\textwidth}
  \centering
 \includegraphics[width=1.0\linewidth]{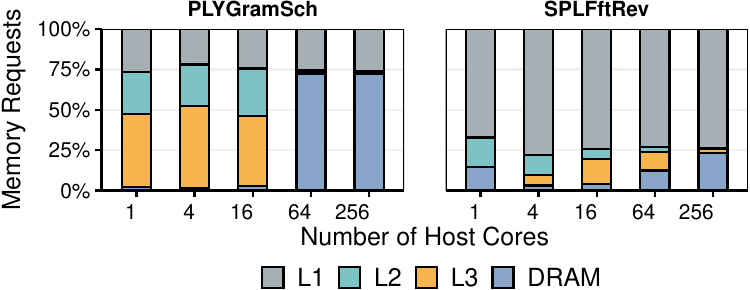}%
 \caption{Memory request breakdown for \geraldorevi{representative} Class~2a \geraldorevi{\gfii{functions}}.}
 \label{fig:latency_class_4}
\end{minipage}%
\qquad
\begin{minipage}[t]{.47\textwidth}
  \centering
  \includegraphics[width=1.0\linewidth]{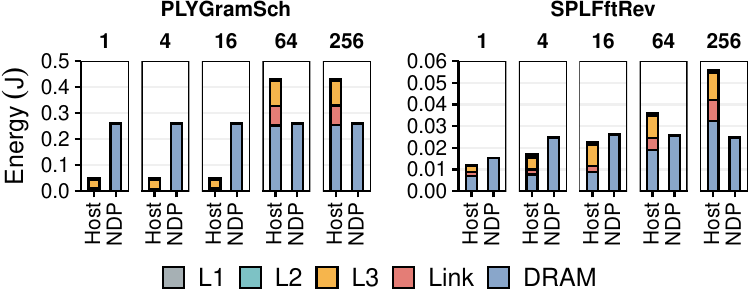}%
    \caption{Energy breakdown for \geraldorevi{representative} Class~2a \geraldorevi{\gfii{functions}}.}
    %\vspace{-3mm}
    \label{fig:energy_group_4}
\end{minipage}
\vspace{-5pt}
\end{figure}

We conclude that cache contention is the primary \geraldorevi{scalability} bottleneck \geraldorevi{for Class~2a \gfii{functions}}, \gfii{and}
\gfii{the} NDP \gfii{system} can provide an effective way of mitigating \gfii{this} cache contention \gfii{bottleneck} without incurring the high area and energy overheads of providing additional cache capacity in the host \gfii{CPU system}, \gfii{thereby improving the scalability of these applications to high core counts.}

\subsubsection{Class~2b: High Temporal Locality, Low AI, Low/Medium LFMR, and Low MPKI \gfii{(\textit{L1 Cache Capacity Bottlenecked Functions})}}
\label{sec_scalability_class2b}

Figure~\ref{figure_performance}(e) shows the host \gfii{CPU system} and \gfii{the} NDP \geraldorevi{system} performance for \texttt{PLYgemver} and \texttt{SPLLucb}. We make two observations from the figure. First, \geraldorevi{as} the number of cores \geraldorevi{increases}, performance \geraldorevi{of} the host \gfii{CPU system} and \geraldorevi{the} NDP \geraldorevi{system} scale in a \geraldorevi{very} similar fashion. The NDP \geraldorevi{system and the host \gfii{CPU} system perform essentially on par with \gfiii{(\gfiii{i.e., }within 1\% of)} each other at all core counts.} Second, even though \gfii{the} NDP \gfii{system} does not \geraldorevi{provide} any performance improvement for \geraldorevi{Class~2b} \gfii{functions}, it also does not hurt performance. Figure~\ref{fig_lt_group_5} shows the \gls{AMAT} for our two representative \gfii{functions}. When \texttt{PLYgemver} executes on the host \gfii{CPU} \geraldorevi{system}, up to 77\% of the memory latency \gfii{comes from} accessing L3 and DRAM, which can be explained by the \gfii{function}\geraldorevi{'s} medium LFMR \gfii{(0.5)}. For \texttt{SPLLucb}, even though up to 73\% of memory latency \gfii{comes from} L1 accesses, some requests still hit in the L3 cache (\geraldorevi{its} LFMR \geraldorevi{is} 0.2), translating to around 10\% of the memory latency. However, the latency \gfii{that comes from} L3 + DRAM for the host \gfii{CPU system} is similar to the latency to access DRAM \gfii{in} the NDP \gfii{system}, resulting in similar performance between the host \gfii{CPU system} and \geraldorevi{the} NDP \geraldorevi{system}.

% \begin{figure}[ht]
%   \centering
%   \includegraphics[width=1.0\linewidth]{mainmatter/04_damov/Plots/lt_group_5-crop.pdf}%
%   \caption{\gls{AMAT} for \geraldorevi{representative} Class~2b \geraldorevi{\gfii{functions}}.}
%    \label{fig_lt_group_5}
% \end{figure}

We make a similar observation for the energy consumption for the host \gfii{CPU system} and \geraldorevi{the} NDP \geraldorevi{system} (Figure~\ref{fig:energy_group_5}). Even though \geraldorevi{a small number of} memory requests hit \gfii{in} L3, the total energy consumption for both \gfii{the} host \gfii{CPU system} and \gfii{the} NDP \gfii{system} is \gfii{similar} due to L3 and off-chip link energy. \gfii{For some functions in Class~2b, we observe that the NDP system slightly reduces energy consumption compared to the host CPU system. For example, the NDP system provides an \gfiii{12\%} average reduction in energy consumption, across all core counts, compared to the host CPU system for \texttt{PLYgemver}}.

% \begin{figure}[ht]
%  \centering
%     \includegraphics[width=1.0\linewidth]{mainmatter/04_damov/Plots/eg_group_5-crop.pdf}%
%     \caption{Energy breakdown for \geraldorevi{representative} Class~2b \geraldorevi{\gfii{functions}}.}
%     \label{fig:energy_group_5}
% \end{figure}

\begin{figure}[ht]
\vspace{-5pt}
\centering
\begin{minipage}[t]{.47\textwidth}
  \centering
  \includegraphics[width=1.0\linewidth]{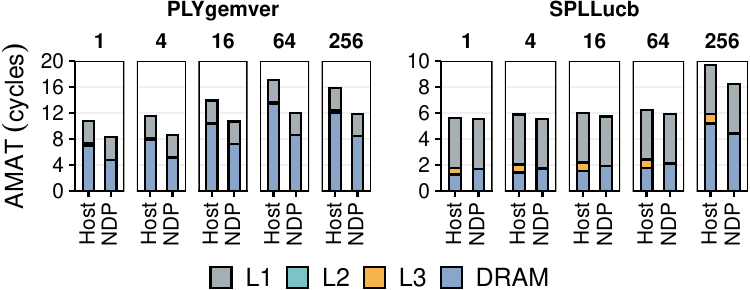}%
  \caption{\gls{AMAT} for \geraldorevi{representative} Class~2b \geraldorevi{\gfii{functions}}.}
   \label{fig_lt_group_5}
\end{minipage}%
\qquad
\begin{minipage}[t]{.47\textwidth}
  \centering
    \includegraphics[width=1.0\linewidth]{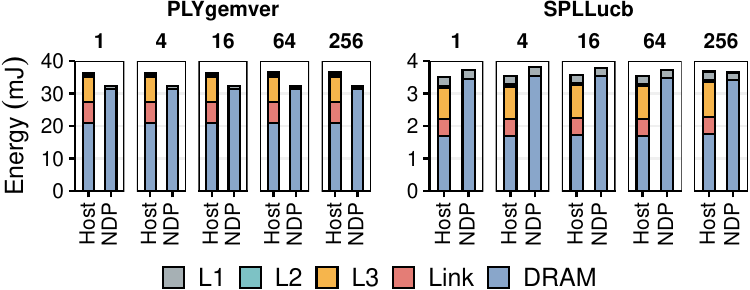}%
    \caption{Energy breakdown for \geraldorevi{representative} Class~2b \geraldorevi{\gfii{functions}}.}
    \label{fig:energy_group_5}
\end{minipage}
\vspace{-5pt}
\end{figure}

We conclude that while \gfii{the} NDP \gfii{system} does not solve any memory bottlenecks for Class~2b \gfii{functions},
it can be used to reduce the overall SRAM area in the system without \gfii{any} performance or energy penalty \gfii{(and sometimes with energy savings)}.

\subsubsection{Class~2c: High Temporal Locality, High AI, Low LFMR, and Low MPKI \gfii{(\textit{Compute-Bound Functions}).}}
\label{sec_scalability_class2c}

Aside from one exception (\texttt{PLYSymm}), all of the \gfii{11} \gfii{functions} in this class exhibit high temporal locality.  When combined with the high AI and low memory intensity, we find that these characteristics significantly impact how \geraldorevi{the} NDP \geraldorevi{system performance} scales for this class. Figure~\ref{figure_performance}(f) shows the host \gfii{CPU system} and \geraldorevi{the} NDP \geraldorevi{system} performance for \texttt{HPGSpm} and \texttt{RODNw}, two representative \gfii{functions} from the class.  We make two observations from the figure. First, the host \geraldorevi{\gfii{CPU} system} performance is \emph{always} greater than the NDP \geraldorevi{system} performance \geraldorevi{(by 44\% for \texttt{HPGSpm} and 54\% for \texttt{RODNw}, on average)}. The high AI (more than 12~ops per cache line), combined with the high temporal locality \geraldorevi{and low MPKI}, \geraldorevi{enables} these \gfii{functions}  \geraldorevi{to make} excellent use of the host \geraldorevi{\gfii{CPU} system} resources. Second, \geraldorevi{both of the} \gfii{functions} benefit \gfii{greatly} from prefetching in the host \gfii{CPU system}. This is a direct result of these \gfii{functions}’ high spatial locality, which allows the prefetcher to be highly accurate \geraldorevi{and effective} in predicting which lines to retrieve \gfii{from main memory}.

\geraldorevi{Figure~\ref{fig:energy_group_6} shows the energy \gfii{breakdown} consumption for the two representative Class~2c \gfii{functions}. We make two observations. First, the host \gfii{CPU} system is 77\% more energy-efficient than the NDP system \gfii{for \texttt{HPGSpm}}, on average \gfii{across all core counts}. Second, the NDP system provides energy savings over the host \gfii{CPU} system at high core counts for \texttt{RODNw} (up to 65\% \gfii{at 256 cores}). \sgv{When the core count increases, the aggregate L1 cache capacity across all cores increases as well, which in turn decreases the number of L1 cache misses. Compared to executing on a single core, executing on 256~cores decreases the L1 cache miss count by 43\%, reducing the memory subsystem energy consumption by 40\%.} \gfv{However, due to \texttt{RODNw}'s medium LFMR of 0.5, the host CPU system \gfiv{still suffers from L2 and L3 cache misses at high core counts}, which \gfiii{require} the \gfiii{large} L3 and off-chip link energy. \gfiv{In contrast, the NDP system eliminates the energy of accessing the L3 cache and the off-chip link energy by directly sending L1 cache misses to DRAM, which\gfv{, at high core counts, leads to lower} energy consumption \gfv{than} the host CPU system.} }}

\begin{figure}[ht]
    \centering
    \includegraphics[width=0.7\linewidth]{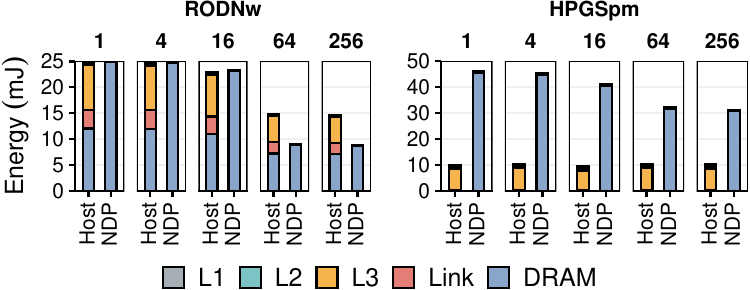}%
    \caption{Energy breakdown for \geraldorevi{representative} Class~2c \geraldorevi{\gfii{functions}}.}
    \label{fig:energy_group_6}
\end{figure}

We conclude that Class~2c \gfii{functions} do not experience \gfiii{large} memory bottlenecks and are not a good fit for \gfii{the} NDP \gfii{system} \geraldorevi{in terms of performance. However, \gfii{the} NDP \gfii{system} can \gfii{sometimes} provide energy savings for \gfii{functions} that experience medium LFMR. }

\subsection{\gfvii{Effect of the Last-Level Cache Size}}
\label{sec_scalability_nuca}
 
\geraldorevi{The bottleneck classification we present in Section~\ref{sec:scalability} depends %of 
\juan{on} two key architecture-dependent metrics \gfii{(\gls{LFMR} and \gls{MPKI})} that are directly \gfii{affected} by the \gfiii{parameters} and the organization of the cache hierarchy. \gfii{O}ur analysis in Section~\ref{sec:scalability} partially evaluate\gfii{s} \gfii{the effect of caching} by scaling the aggregated size of the private \gfiii{(L1/L2)} caches with the number of cores in the system while maintaining the size of the L3 cache fixed at 8~MB for the host \gfii{CPU} system. However, we \gfiii{also} need to understand the impact of the L3 cache size on our bottleneck classification analysis. To this end, t}his section evaluates the effects on our \geraldorevi{bottleneck classification analysis} of using an alternative cache hierarchy configuration, \geraldorevi{where we employ} a \gls{NUCA}~\cite{kim2002adaptive} model \geraldorevi{to scale the size of the L3 cache \gfii{with the number of cores} in the host \gfii{CPU} system.}

 \begin{figure*}[!t]
    \centering
    \includegraphics[width=\linewidth]{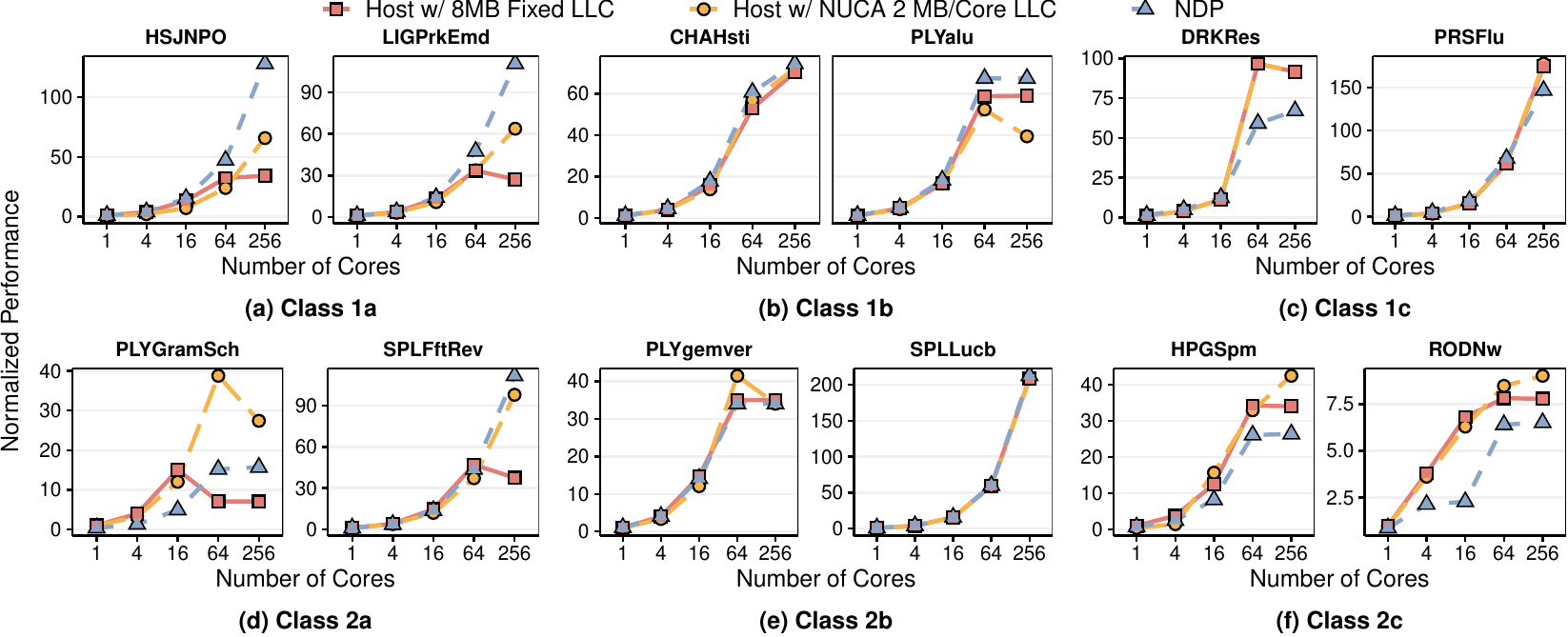}%
    \caption{Performance of \geraldorevi{the} host and \geraldorevi{the} NDP \geraldorevi{system} \geraldorevi{as we vary the} LLC size, normalized to one host core with \geraldorevi{a} fixed 8MB LLC size.}
    \label{fig_nuca_analysis}%  
\end{figure*}

In this configuration, we maintain the sizes of the private L1 and L2 caches (32~kB and 256~kB \juan{per core}, respectively) while increasing the shared L3 cache size with the core count (we use 2~MB/core) \gfii{in the host CPU system}. The cores, shared L3 caches\geraldo{, and DRAM memory controller} are interconnected using a 2D-mesh \gls{NoC}~\gfiv{\cite{moscibroda2009case,das2009application,das2010aergia,nychis2012chip,besta2018slim,fallin2011chipper,grot2011kilo,benini2002networks}} of size $(n+1) \times (n+1)$ \geraldorevi{\gfii{(an extra interconnection dimension is added to place the DRAM memory controller\gfiii{s})}}. To faithfully simulate the \gls{NUCA} model (e.g., including network contention in our simulations), we integrate the M/D/1 network model proposed by ZSim++~\cite{zsimplusplus} in our \geraldorevi{\bench} simulator~\cite{damov}. We use a latency of 3 cycles per hop in our analysis, as suggested by prior work~\cite{zhang2018minnow}. \gfii{We adapt our energy model to account for the energy consumption of the \gls{NoC} in the \gls{NUCA} system. We consider router energy consumption of 63 pJ per request and energy consumed per link traversal of 71 pJ, same as previous work~\cite{tsai2017jenga}.}

Figure~\ref{fig_nuca_analysis} shows the \gfii{performance} scalability curves for representative \geraldorevi{\gfii{functions}} from each one of our bottleneck classes \geraldorevi{presented in Section~\ref{sec:scalability}} \gfii{for the baseline host CPU system (\emph{Host with 8MB Fixed LLC}), the host CPU NUCA system (\emph{Host with NUCA 2MB/Core LLC}), and the NDP system}. We make \geraldorevi{two} observations. First, \geraldorevi{the observations we make for our bottleneck classification \juangg{(Section~\ref{sec:scalability})}} \geraldorevi{are} \gfiii{\emph{not}} affected by increasing the L3 cache size for \geraldorevi{C}lasses 1a, 1b, 1c, 2b, and 2c. \geraldorevi{We observe that Class~1a \gfii{functions} benefit from a large L3 cache size (by up to \juangg{1.9$\times$/2.3$\times$} for \juangg{\texttt{HSJNPO}/\texttt{LIGPrkEmd}}
at 256 cores). However, the NDP system still provides performance benefits compared to the \gfii{host CPU} NUCA system. We observe that increasing the L3 size reduces some of the pressure on main memory but cannot fully reduce the \gfii{DRAM} bandwidth bottleneck for \gfii{Class~1a functions}. \gfii{Functions} in Class~1b do \emph{not} benefit from extra L3 capacity} (we do not observe a decrease in \gls{LFMR} %and 
\juangg{or} \gls{MPKI}). \gfii{Functions} in Class 1c do \gfiii{\emph{not}} benefit from extra \gfiii{L3} cache \geraldorevi{capacity}. \gfii{W}e observe that the private L1 and L2 caches capture most of their data locality, \gfiii{as mentioned \geraldorevi{in Section~\ref{sec_scalability_class1c}}}, \geraldorevi{and} thus, \geraldorevi{these \gfii{functions} do \emph{not}} benefit \geraldorevi{from} \gfii{increasing the} L3 size. \gfiii{Functions in Class~2b do \emph{not} benefit from extra L3 cache capacity, which can even lead to a decrease in performance at high core counts for the host CPU NUCA system in some Class~2b functions} \gfiv{due to long NUCA L3 access latencies}. \gfii{For example, we observe that  \texttt{PLYgemver}'s performance drops 18\% \juangg{when increasing the core count} from 64 to 256 in the host CPU NUCA system. We do \emph{not} observe \juangg{such} a performance drop for the host CPU system with fixed LLC size. \juangg{The} performance drop in the host CPU NUCA system is due to the increase in the number of hops that L3 requests need to travel in the NoC at high core counts, which increase the function's AMAT.} \gfii{Class~2c \gfii{functions} benefit from a \juangg{larger} last-level cache. We observe that \juangg{their} performance improves by 1.3$\times$/1.2$\times$ for \texttt{HPGSpm}/\texttt{RODNw} compared to the host CPU system with 8MB fixed LLC at 256 cores.}

\geraldorevi{Second}, we observe two different \geraldorevi{types of} behavior for \gfii{functions} in Class 2a. Since cache conflicts \geraldorevi{are the major bottleneck for} \gfii{functions} in this class, we observe that increasing the L3 cache size can mitigate \geraldorevi{this} bottleneck. In \gfiii{Figure~\ref{fig_nuca_analysis}}, we observe that for both \texttt{PLYGramSch} and \texttt{SPLFftRev}, the \gfii{host system with} \geraldorevi{NUCA} 2MB/Core LLC provides better performance than the \gfii{host system with 8MB fixed LLC}. However, \gfii{the} \gls{NDP} \gfii{system} can still provide performance benefits in case of contention \gfiii{on the L3 \gls{NoC}} \gfii{(e.g., in \texttt{SPLFftRev})}. \geraldorevi{For \gfii{example}, the NDP system provides 14\% performance improvement \gfii{for \texttt{SPLFftRev}} compared to the NUCA system \gfiii{(with 512~MB L3 cache)} for 256 cores.}

\geraldorevi{In summary, we conclude that the key takeaways and observations we present \gfii{in} our bottleneck classification in Section~\ref{sec:scalability} are also valid for a host system with a shared last-level cache \gfii{whose size scales with core count.}} \gfiii{In particular, different workload classes get affected by an increase in L3 cache size as expected by their characteristics distilled by our classification.}

\gfii{Figure~\ref{fig_nuca_analysis_energy} shows the \gfii{energy} consumption for representative \gfii{functions} from each one of our bottleneck classes  presented in Section~\ref{sec:scalability}. \gfiii{We observe that the NDP system can provide \gfiii{substantial} energy savings for functions in different bottleneck classes, even compared against a system with \gfiii{very} large \gfiii{(e.g., 512~MB)} cache sizes.} We make the following observations for each bottleneck class:}

 \begin{figure*}[h]
    \centering
    \includegraphics[width=\linewidth]{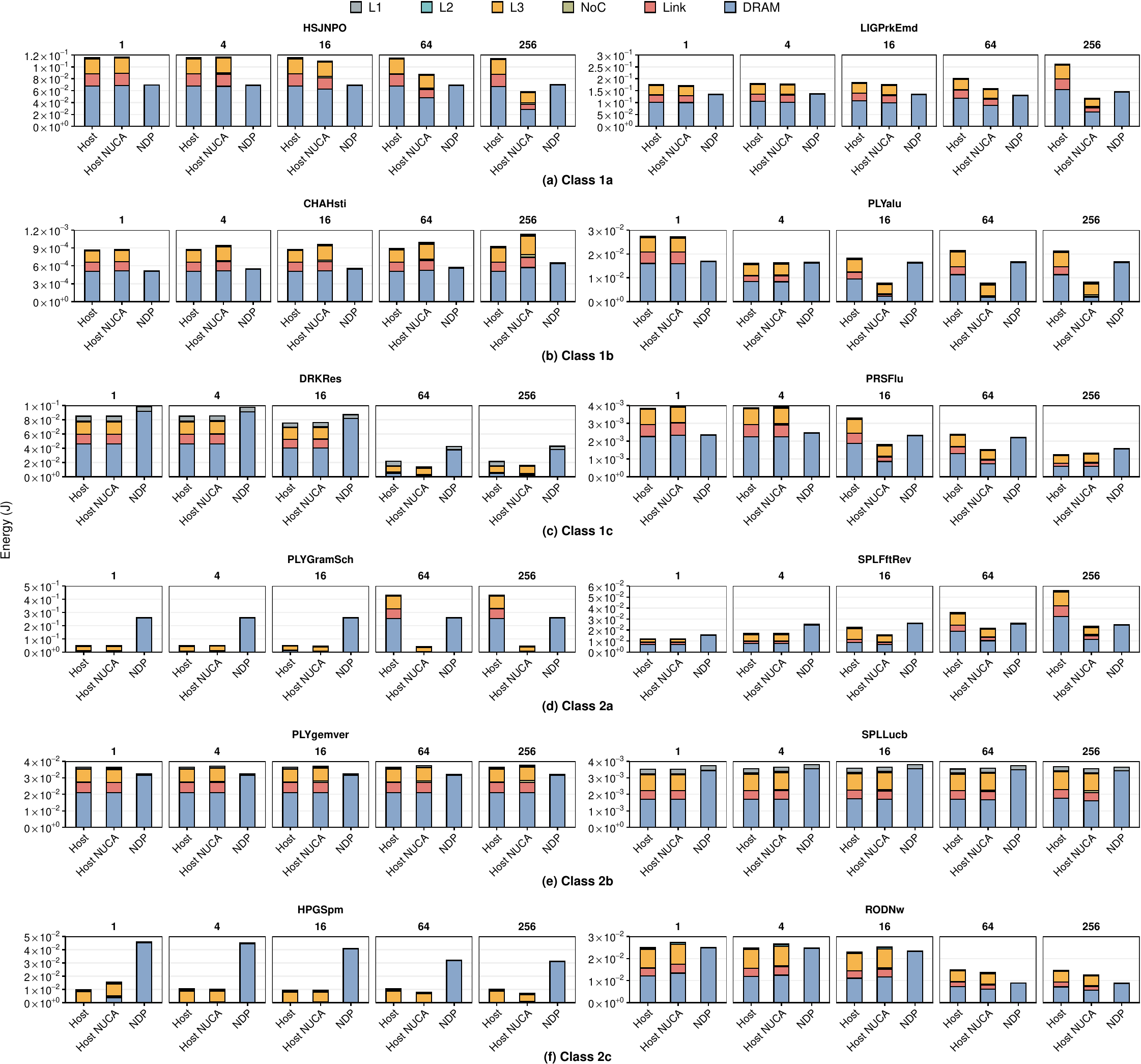}%
    \caption{\gfii{Energy of the host and the NDP system as we vary the LLC size. \emph{Host} refers to the host system with a fixed 8MB LLC size; \emph{Host NUCA} refers to the host system with 2MB/Core LLC.}}
    \label{fig_nuca_analysis_energy}%    
\end{figure*}

\begin{itemize}[noitemsep, leftmargin=*, topsep=0pt]
    \item \gfii{\emph{Class~1a}: First, for both representative functions in this bottleneck class, the host CPU NUCA system and the NDP system reduce energy consumption compared to the baseline host CPU system. However, we observe that the NDP system provides \juangg{larger} energy savings than the host CPU NUCA system. On average, across all core counts, the NDP system and the host CPU NUCA system reduce energy consumption compared to the host CPU system for \texttt{HSJNPO}/\texttt{LIGPrkEmd} by  46\%/65\% and 25\%/22\%, respectively. Second, at 256 cores, the host CPU NUCA system provides larger energy savings than the NDP system for both representative functions. This happens because at 256 cores, the large L3 cache \gfiii{(i.e., 512~MB)} capture\gfiii{s a large} portion of the dataset for these functions, reducing costly DRAM traffic. The host CPU NUCA system reduces energy consumption compared to the host CPU system for \texttt{HSJNPO}/\texttt{LIGPrkEmd} \gfiii{at 256 cores} by 2.0$\times$/2.2$\times$ while the NDP system reduces energy consumption by 1.6$\times$/1.8$\times$. \gfiii{The L3 cache capacity needed to make the host CPU NUCA system more energy efficient than the NDP system is \gfv{\emph{very}} large (512~MB SRAM), which is likely not cost-effective.}}
    
    \item \gfii{\emph{Class~1b}: First, for \texttt{CHAHsti}, the host CPU NUCA system \emph{increases} energy consumption compared to the host CPU system by 9\%\gfiii{, on average across all core counts}. In contrast, the NDP system \emph{reduces} energy consumption by 57\%. Due to its low spatial and temporal locality (Figure~\ref{fig:locality_chart}), this function does not benefit from a deep cache hierarchy. In the host CPU NUCA system, the extra energy from the \gfiii{large amount of} \gls{NoC} traffic further increases the cache hierarchy's overall energy consumption. Second, for \texttt{PLYalu}, the host CPU NUCA system and the NDP system reduce energy consumption compared to the host CPU system by 76\% and 23\%, on average across all core counts. Even though the increase in LLC size does not translate to performance improvements, the large LLC sizes in the host CPU NUCA system aid to reduce DRAM traffic, \gfiii{thereby} providing energy savings compared to the baseline host CPU system.} 
    
    \item \gfii{\emph{Class~1c}: First, for \texttt{DRKRes}, the host CPU NUCA system reduces energy consumption compared to the host CPU system by 15\%\gfiii{, on average across all core counts}. In contrast, the NDP system increases energy consumption by 30\%, which is due to the function's medium \gls{LFMR} (Section~\ref{sec_scalability_class1c}). Second, for \texttt{PRSFlu}, we observe that the NDP system provides large energy savings than the host CPU NUCA system. The host CPU NUCA system reduces energy consumption compared to the host CPU system by 21\%, while the NDP system reduces energy consumption by 25\%, on average across all core counts. However, the energy savings of both host CPU NUCA and NDP systems compared to the host CPU system reduces \gfiii{at high-enough core counts} (\gfiii{the energy consumption of the host CPU NUCA system (NDP system) is 0.6$\times$ (0.9$\times$) that of the host CPU system at 64 cores and 1.1$\times$ (1.3$\times$) that of the host CPU system at 256 cores).} This result is expected for Class~1c functions since \gfiii{the functions in this class have decreasing \gls{LFMR}, i.e., the function\gfiii{s} effectively utilize the private L1/L2 caches in the host CPU system at high-enough core counts}.} 
    
    \item \gfii{\emph{Class~2a}: First, for \texttt{PLYGramSch}, \gfiii{compared to the host CPU system} the host CPU NUCA system reduces energy consumption  by 2.53$\times$ and the NDP system increases energy consumption by 55\%\gfiii{, on average across all core counts}. \gfiii{Even though at high core counts (64 and 256 cores) the host CPU NUCA system provides larger energy savings than the NDP system compared to the host CPU system (the host CPU NUCA system and the NDP system reduce energy consumption compare to the host CPU system by 9$\times$ and 65\% respectively, \gfiii{averaged across 64 and 256 cores}), such large energy savings come at the cost of very large (e.g., 512~MB) cache sizes}. Second, for \texttt{SPLFftRev}, the host CPU NUCA system and the NDP system reduce energy consumption compared to the host CPU system by 42\% and 7\%, on average across all core counts. The NDP system increases energy consumption compared to the host CPU system at low core counts (an increase of 33\%, averaged across 1, 4, and 16 cores). However, it provides similar energy savings as the host CPU NUCA system for large core counts (99\% and 75\% energy \gfiii{reduction compare to the host CPU system} for the host CPU NUCA system and the NDP system, respectively, averaged across 64 and 256 cores counts). Since the function suffers from high network contention, the increase in core count increases \gls{NoC} traffic, which \gfiii{in turn} increases energy consumption for the host CPU NUCA system. \gfiii{We conclude that the NDP system provides energy savings for Class~2a applications compared to the host CPU system at lower cost than the host CPU NUCA system.}} 
    
    \item \gfii{\emph{Class~2b}: First, for \texttt{PLYgemver}, the host CPU NUCA system increases energy consumption compared to the host CPU system by 2\%\gfiii{, on average across all core counts}. In contrast, the NDP system reduces energy consumption by 13\%. This function does not benefit from large L3 cache sizes since Class~2b functions are bottlenecked by L1 capacity. Thus, the \gls{NoC} only adds extra \gfiii{static and dynamic} energy consumption. Second, for \texttt{SPLLucb}, the host CPU NUCA system consumes the same energy as the host CPU system while the NDP system increases energy consumption by 5\%, averaged across all core counts.}
    
    \item \gfii{\emph{Class~2c}: For both representative functions in this class, the host CPU NUCA system reduces energy consumption compared to the host CPU system while the NDP system increases energy consumption. For \texttt{HPGSpm}/\texttt{RODNw}, the host CPU NUCA system reduces energy consumption by 6\%/9\% while the NDP system increases energy consumption by 74\%/22\%, averaged across all core counts. This result is expected since Class~2c functions are compute-bound and highly benefit from a deep cache hierarchy.}
\end{itemize}

\gfii{In conclusion, the NDP system can provide \gfiii{substantial} energy savings for functions in different bottleneck classes, even compared against a system with \gfiii{very} large \gfiii{(e.g., 512~MB)} cache sizes.}

\subsection{ \geraldorevi{Validation} \gfiii{and Summary} of Our \geraldorevi{Workload} Characterization \juanggg{Methodology}}
\label{sec_summary}

\gfiii{In this section, we present \gfiv{the} validation \gfiii{and a summary} of our \gfiv{new} workload characterization \gfiv{methodology}. \gfiii{First, we use the remaining 100 memory-bound functions we obtain from \emph{Step~1} \gfiv{(see Section~\ref{sec:vtune})} to \juanggg{validate our workload characterization methodology. To do so, we calculate the accuracy of our workload classification by using the remaining 100 memory-bound functions\gfiv{, which were not used to identify the six classes we found and described in Section~\ref{sec:scalability}}.}
\gfiii{Second}, we present a summary of the key metrics we obtain for all 144 memory-bound functions, including our analysis of the host CPU system and the NDP system using \gfv{two types of cores (in-order and out-of-order)}.}}

\subsubsection{Validation of Our Workload Characterization  \juanggg{Methodology}}
\label{sec::sub::validation}

\gfiii{Our goal is to evaluate the accuracy of our workload characterization \gfv{methodically} on a large set of \gfiv{functions}. To this end, we apply \emph{Step 2} and \emph{Step 3} of our memory bottleneck classification methodology \gfiv{(as described in Sections~\ref{sec:step2} and \ref{sec:step3})} to the remaining 100 memory-bound functions we obtain from \emph{Step~1} (in Section~\ref{sec:vtune}). Then, we \gfiv{perform} a two-phase validation to calculate the accuracy of our workload characterization.}

\gfiii{In \emph{phase 1} of our validation, we calculate the threshold values that define the low/high boundaries of each of the four metrics we use to cluster the initial 44 functions in the six memory bottleneck \gfiv{classes \gfv{in}} Section~\ref{sec:scalability} (i.e., temporal locality, \gls{LFMR}, LLC \gls{MPKI}, and \gls{AI}). We also include the LFMR curve slope to indicate when \gfii{the LFMR} increases, decreases or stays constant \gfiv{as we} scal\gfiv{e} the core count. We calculate the threshold values for a metric \texttt{M} by computing the middle point between (i) the average value of \texttt{M} across the memory bottleneck \gfiv{classes} with \emph{low} values of \texttt{M} and (ii) the average value of \texttt{M}  across the memory bottleneck \gfiv{classes} with \emph{high} values of \texttt{M} values out of the 44 functions. In \emph{phase 2} of our validation, we calculate the accuracy of our workload characterization by classifying the remaining 100 memory-bound functions using the threshold values obtained from \emph{phase 1} and the \gls{LFMR} curve slope. \gfiv{After \emph{phase 2}\gfv{,} a \gfv{function} is \gfv{considered to be \emph{accurately}} classified into a correct memory bottleneck \gfiv{class} if \gfv{and only if} it \gfv{(1)} fits the definition of the assigned class using the threshold values obtained from \emph{phase 1} \gfiv{and \gfv{(2)} follows the expected performance trends \gfv{of the assigned class} when the \gfv{function} is executed in the host CPU system and the NDP system}. For example, a \gfv{function} is correctly classified into Class~1a \emph{if and only if} it \gfv{(1)} displays low temporal locality, low \gls{AI}, high \gls{LFMR}, high \gls{MPKI} \gfiv{and \gfv{(2)} the NDP system outperforms the host CPU system \gfv{as we scale} the core count when executing the \gfv{function}}.} The final \gfv{\emph{accuracy of our workload characterization methodology}} is calculated by \gfiv{computing} the percentage of the functions that are \gfv{\emph{accurately}} classified into one of the six memory bottleneck \gfiv{classes}.
}

\begin{figure*}[!h]
\vspace{-8pt}
\begin{subfigure}{\textwidth}
  \centering
 \includegraphics[width=\textwidth]{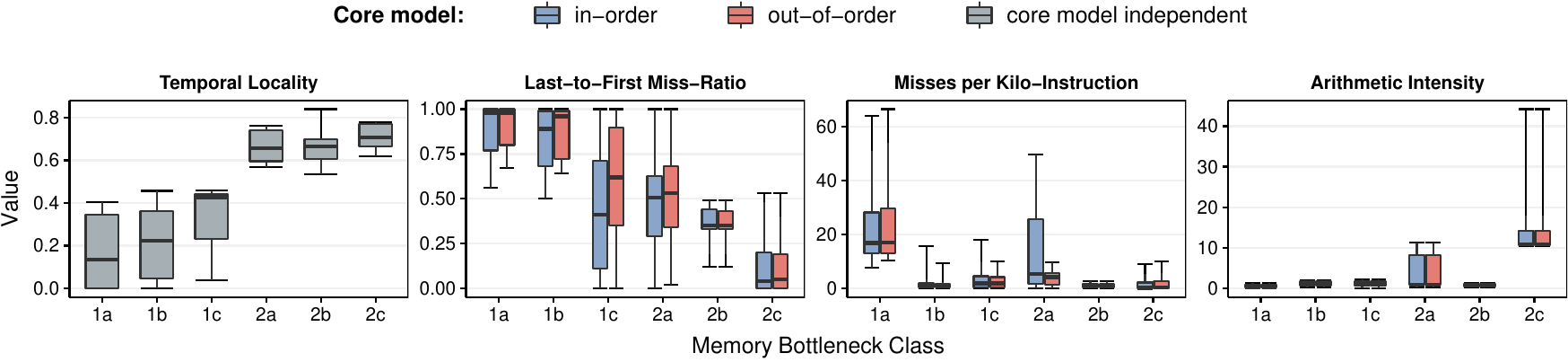}%
 \vspace{-3pt}
  \caption{\gfiv{Summary of the key metrics for each memory bottleneck class. }}
  \label{fig:summary:metrics}
\end{subfigure}
\par\bigskip
\begin{subfigure}{\textwidth}
  \centering
 \includegraphics[width=0.99\textwidth]{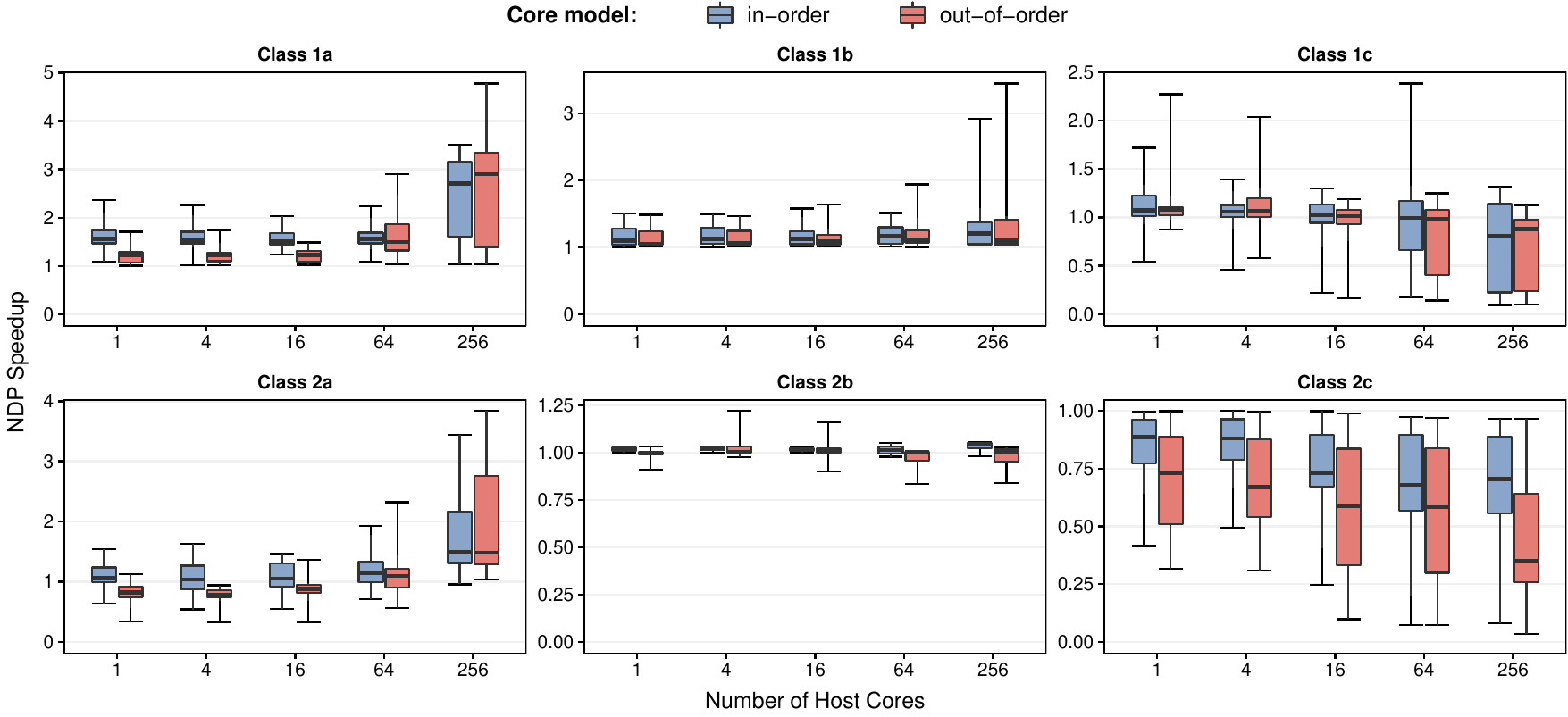}%
 \vspace{-3pt}
  \caption{\gfiv{Summary of NDP speedup for each memory bottleneck class at 1, 4, 16, 64, and 256 cores.}}
  \label{fig:summary:speedup}
\end{subfigure}
     \caption{Summary of our characterization for \gfiv{all} 144 \gfv{memory-bound} functions. \gfv{Each box is lower-bounded by the first quartile and upper-bounded by the third quartile. The median falls within the box. The inter-quartile range (IQR) is the distance between the first and third quartiles (i.e., box size). Whiskers extend to the minimum and maximum data point values on either sides of the box.}}
     \label{fig_summary}
\end{figure*}

\gfiii{First, by applying \emph{phase 1} of our two-phase validation, we obtain that the threshold values are: 0.48 for \emph{temporal locality}, 0.56 for \emph{\gls{LFMR}}, 11.0 for \emph{\gls{MPKI}}, and 8.5 for \emph{\gls{AI}}. Second, by applying \emph{phase 2} of our two-phase validation, we \gfv{find that we can accurately} classify \textit{97\% of the 100 memory-bound functions} into one of our six memory bottleneck \gfiv{classes} \gfiv{(i.e., the accuracy of our workload characterization methodology is 97\%)}. We observe that three functions (\textit{Ligra:ConnectedComponents:compute:rMat}, \textit{Ligra:MaximalIndependentSet:edgeMapDense:USA}, and \textit{SPLASH-2:Oceanncp:relax}) could not be \gfv{accurately} classified into their \gfv{correct} memory bottleneck \gfiv{class} (Class~1a). We observe that these functions have LLC \gls{MPKI} values \emph{lower} than the \gls{MPKI} threshold expected for Class~1a functions. We expect that the accuracy of our \gfv{methodology} can be further improved by incorporating more workloads into our workload suite and fine-tuning each metric to encompass an even large\gfiv{r} set of applications.}

\gfiii{\gfiv{W}e conclude that our workload characterization methodology can accurately classify \gfiv{a given new} application\gfiv{/function} into \gfiv{its} appropriate memory bottleneck \gfiv{class}.}

\subsubsection{Summary of \gfiv{O}ur \gfiv{W}orkload \gfiv{C}haracterization \gfiv{Results}.} 
\label{sec::sub::summary}

\gfiii{Figure~\ref{fig:summary:metrics} summarizes the metrics we collect for all 144 functions across all core counts (i.e., from 1 to 256 cores) and different core microarchitectures \gfiv{(i.e., out-of-order and in-order cores)}. The figure shows the distribution of the key metrics we use during our workload characterization for each memory bottleneck class in Section~\ref{sec:scalability}, including architecture-independent metrics (i.e., temporal locality) and architecture-dependent metrics (i.e., \gls{AI}, \gls{LFMR}, and LLC \gls{MPKI}). We report the architecture-dependent metrics for two core models: (i) in-order and (ii) out-of-order cores.\footnote{\gfiv{In Section~\ref{sec:scalability}, we collect and report the values of the architecture-independent metrics and architecture-dependent metrics for a subset of 44 representative functions out of the 144 memory-bound functions we identify in \emph{Step 1} of our workload characterization methodology. In Section~\ref{sec::sub::summary}, we report values for the \emph{complete set} of 144 memory-bound functions.}} Together with the out-of-order core model that we use in Section~\ref{sec:scalability}, we incorporate an in-order core model to \gfiv{our} analysis, \gfiv{so as} to show that our memory bottleneck classification methodology focuses on data movement requirements and \gfiv{works} independent\gfiv{ly} of the core microarchitecture. \gfiv{Figure~\ref{fig:summary:speedup}} shows the distribution of speedups we observe for when \gfiv {we} offload the function to our general-purpose \gls{NDP} cores, \gfiv{while} employing the same core type as the host CPU system.}

\gfiii{We make two key observations from \gfv{Figure~\ref{fig_summary}}. First, \gfiv{we observe similar values for each \gfv{architecture-dependent} key metric \gfv{(i.e., \gls{LFMR}, \gls{MPKI}, \gls{AI})} regardless of core type for all 144 functions \gfv{(in Figure~\ref{fig:summary:metrics})}.}} \gfiii{Second, we observe that the \gls{NDP} system achieves similar speedup\gfv{s over the host CPU system, when using both} in-order and out-of-order core configurations \gfv{(in Figure~\ref{fig:summary:speedup})}. The speedup provided by the NDP system compared to the host CPU system when both systems use out-of-order (in-order) cores for Classes 1a, 1b, 1c, 2a, 2b, and 2c is 1.59 (1.77), 1.22 (1.15), 0.96 (0.95), 1.04 (1.22), 0.94 (1.01), and 0.56 (0.76), respectively, on average across all core counts and functions within a \gfiv{memory} bottleneck class. \gfiv{The NDP system greatly outperforms the host CPU system across \emph{all core counts} for Class~1a and 1b functions, with a maximum speedup for the out-of-order (in-order) core model of 4.8 (3.5) and 3.4 (2.9), respectively. The NDP system greatly outperforms the host CPU system at \emph{low core counts} for Class~1c functions and at \emph{high core counts} for Class~2a functions, with a maximum speedup for the out-of-order (in-order) core model of 2.3 (2.4) and 3.8 (3.4), respectively. The NDP system provides a modest speedup compared to the host CPU system across \emph{all core counts} for Class~2b functions and slowdown for Class~2c functions, with a maximum speedup for the out-of-order (in-order) core model of 1.2 (1.1) and \gfv{1.0 (1.0)}, respectively.} We observe that\gfv{, averaged across all classes and core types,} the average speedup provided by the NDP system using in-order cores is \gfiv{11\%} higher than the average speedup offered by the NDP system using out-of-order cores. This \gfiv{is} because the host CPU system \gfiv{with} out-of-order cores can hide the \gfiv{performance} impact of memory access latency to some degree (e.g., using \gfiv{dynamic} instruction \gfiv{scheduling})~\gfv{\cite{mutlu2006efficient,mutlu2005techniques,mutlu2003runahead,hashemi2016continuous,oooexec,mutlu2003runaheadmicro}}. On the other hand, the host CPU system using in-order cores has \gfv{little tolerance} to hide memory access latency~\gfv{\cite{mutlu2006efficient,mutlu2005techniques,mutlu2003runahead,hashemi2016continuous,oooexec,mutlu2003runaheadmicro}}.}

\gfiii{We conclude that our methodology to classify memory bottlenecks \gfv{of} application\gfv{s} is \emph{robust} \gfv{and \emph{effective}} \gfiv{since we observe similar trends for the six memory bottleneck classes across a large range of \gfv{(144)} functions and two \gfv{very} different core models.}}

\subsection{Limitations of Our Methodology}
\label{sec_scalability_limitations}

\geraldorevi{We identify three limitations to our workload characterization methodology. We discuss each limitation next.}

\noindent
\textbf{NDP Architecture \geraldorevi{Design Space}.} Our methodology uses the same type and number of cores in \gfii{the} host \gfii{CPU} and \gfiii{the} NDP \gfii{system} configurations \gfiii{for our} scalability analysis (Section~\ref{sec:scalability}) because our main goal is to highlight the performance and energy differences between \gfii{the} host \gfii{CPU system} and \gfii{the} NDP \gfii{system that are} caused by data movement. We do not consider practical limitations related to area or thermal dissipation that could affect the type and the maximum number of cores \gfii{in} the NDP \gfii{system}, because our goal is \textbf{not} \geraldorevi{to propose} NDP architectures but \geraldorevi{to} characteriz\geraldorevi{e data movement \gfiii{and understand the different data movement bottlenecks} in modern} workloads. Proposing NDP architectures for the \geraldorevi{workload classes} that our methodology identifies as \gfiii{suitable for} NDP is a \geraldorevi{\gfii{promising}} topic for future work.

\noindent
\textbf{Function-level Analysis.} We \geraldorevi{choose} to conduct our analysis at a function granularity rather than at the application \geraldorevi{granularity} for two major reasons. First, general-purpose NDP architectures are typically leveraged as accelerator\geraldorevi{s} to which only \emph{parts} of the application or specific \gfiii{functions} are offloaded~\gfv{\cite{boroumand2018google, PEI, hsieh2016transparent, ke2019recnmp, top-pim, kim2018grim, seshadri2017ambit, nai2017graphpim,hadidi2017cairo,hajinazarsimdram,gao2016hrl,lockerman2020livia,boroumand2021polynesia,dualitycache,xin2020elp2im,devaux2019true,IBM_ActiveCube,Asghari-Moghaddam_2016,alves2016large, seshadri2013rowclone,azarkhish2016design,asghari2016chameleon}}, rather than the entire application. Functions typically form natural boundaries for \geraldorevi{parts} of algorithms/applications that can potentially be offloaded. Second, it is well-known that applications go through distinct phases during execution. Each phase may have different characteristics (e.g., a phase might be more compute-bound, while another one might be \gfiii{more} memory-bound) \geraldorevi{and thus fall into different classes in our analysis}.  A fine-grained analysis at the function level enables us to identify each of those phases and hence, identify more fine-grained opportunities for NDP offloading. \geraldorevi{However, the main drawback of function-level analysis is that \gfii{it does not take into account} data movement across function boundaries\gfii{, which affect\gfiii{s}} the performance and energy benefits the NDP system provides \gfiii{over} the host \gfii{CPU} system. For example, the NDP system might hurt overall system performance and energy consumption when a large amount of data needs to be continuously moved between a function executing on the NDP \gfii{cores} and another executing on the host \gfii{CPU cores}~\cite{boroumand2019conda, lazypim}.}

\noindent
\textbf{Overestimating NDP Potential.} Offloading kernels to NDP cores incur\gfiv{s} overheads that \gfii{our analysis do\gfiii{es}} not account for \gfiii{(e.g., maintaining coherence between the host CPU and the NDP cores~\gfiv{\cite{lazypim,boroumand2019conda}}, efficiently synchronizing computation across NDP cores~\gfiv{\cite{syncron,gomez2022benchmarking}}, providing virtual memory support for the NDP system~\gfiv{\cite{hsieh2016accelerating,PEI,picorel2017near}}, and dynamic offloading support for NDP-friendly functions~\gfiv{\cite{hsieh2016transparent}})}. \gfiii{Such} overheads \geraldorevi{can} impact the performance benefits NDP can provide when considering the end-to-end application. However, deciding \geraldorevi{how to and} whether or not to offload computation to NDP is an open research topic, which involves several architecture-dependent components in the system\gfiii{, such as the following two examples}. \gfiii{First}, maintaining coherence between \gfii{the} host \gfii{CPU} and \geraldorevi{the} NDP cores is a challenging task that \geraldorevi{recent} works tackle~\cite{lazypim,boroumand2019conda}. \gfiii{Second, enabling efficient synchronization across NDP cores is challenging due to the lack of shared caches and hardware cache coherence protocols in NDP systems. Recent works, such as \cite{syncron,liu2017concurrent}, provide solutions to the NDP synchronization problem.}  \gfiii{Therefore, to focus our analysis on the data movement characteristics of workloads and the broad benefits of NDP, we minimize our assumptions about our target NDP architecture, making our evaluation as broad\gfiv{ly} \gfiv{applicable} as possible.}

\geraldorevi{\section{\bench: The Data Movement Benchmark Suite}}
\label{sec:benchmark}

\geraldorevi{In this section, we present \bench, the DAta MOVement Benchmark Suite. \bench is the collection of the 144 functions we use to drive our memory bottleneck classification in Section~\ref{sec:characterization}. The benchmark suite is divided into each one of the six classes of memory bottlenecks presented in Section~\ref{sec:characterization}. \bench is the first benchmark suite that encompasses \emph{real} applications from a diverse set of application domains tailored to stress different memory bottlenecks in a system. We present the complete description of the functions in \bench in Appendix~A. We highlight the benchmark diversity of the function\gfii{s} in \bench in Section~\ref{sec_scalability_benchmark_diversity}.} \jgl{To me, it does not make much sense to have a section with a single subsection. We can just say that in this section we study benchmark diversity.} \gfiii{We open source DAMOV~\cite{damov} to facilitate further rigorous research in mitigating data movement bottlenecks, including in near data processing.}

\subsection{Benchmark Diversity}
\label{sec_scalability_benchmark_diversity}

\geraldorevi{We perform a hierarchical clustering algorithm with the 44 \gfii{representative} functions we employ in Section~\ref{sec:scalability}.\footnote{\gfii{\gfiii{In Section~\ref{sec_scalability_benchmark_diversity}, w}e use the same 44 representative functions that we use during our bottleneck classification instead of the \gfiii{entire set of} 144 functions in DAMOV, \gfiii{in order} to visualize better the clustering produced by the hierarchical clustering algorithm.}} \gfii{Our goal is to showcase our benchmark suite's diversity and observe whether a clustering algorithm produces a noticeable difference from the application clustering presented Section~\ref{sec:characterization}}. \gfii{The} hierarchical clustering \gfii{algorithm}~\cite{friedman2001elements} \gfii{takes as input a dataset containing features that define each object in the dataset. The algorithm works} by incrementally grouping \gfii{objects} in \gfii{the dataset} \gfii{that are} similar to each other in terms of some distance metric (called \juan{\emph{linkage distance}}), \gfii{which is calculated based on the features' values}. \gfiii{Two} \gfii{objects} with a short linkage distance have more affinity \gfii{to} each other than \gfiii{two} \gfii{objects} with \gfiii{a} large linkage distance. \gfii{To apply the hierarchical clustering algorithm, we create a dataset where each object is one of the 44 representative functions from \bench. We use as features} the same metrics we use for our analysis, i.e., temporal locality, MPKI, LFMR, and AI. We also include the LFMR curve slope to indicate when \gfii{the LFMR} increases, decreases or stays constant when scaling the core count. We use Euclidean distance~\cite{friedman2001elements} to calculate the linkage distance across \gfii{features} in our dataset. We evaluate other linkage distance metrics (\gfiii{such} as Manhattan distance~\cite{friedman2001elements}), and we observe similar clustering results.}

\geraldorevi{Figure~\ref{fig_dendo_validation} shows the dendrogram \juan{that} the hierarchical clustering algorithm produces for our 44 representative functions. \gfii{We indicate in the figure }%Each color represents 
the application class each \gfii{function} belongs to, according to our classification. 
We make \gfii{three} observations from the figure.} 

\begin{figure}[ht]
    \centering
 \includegraphics[width=0.95\linewidth]{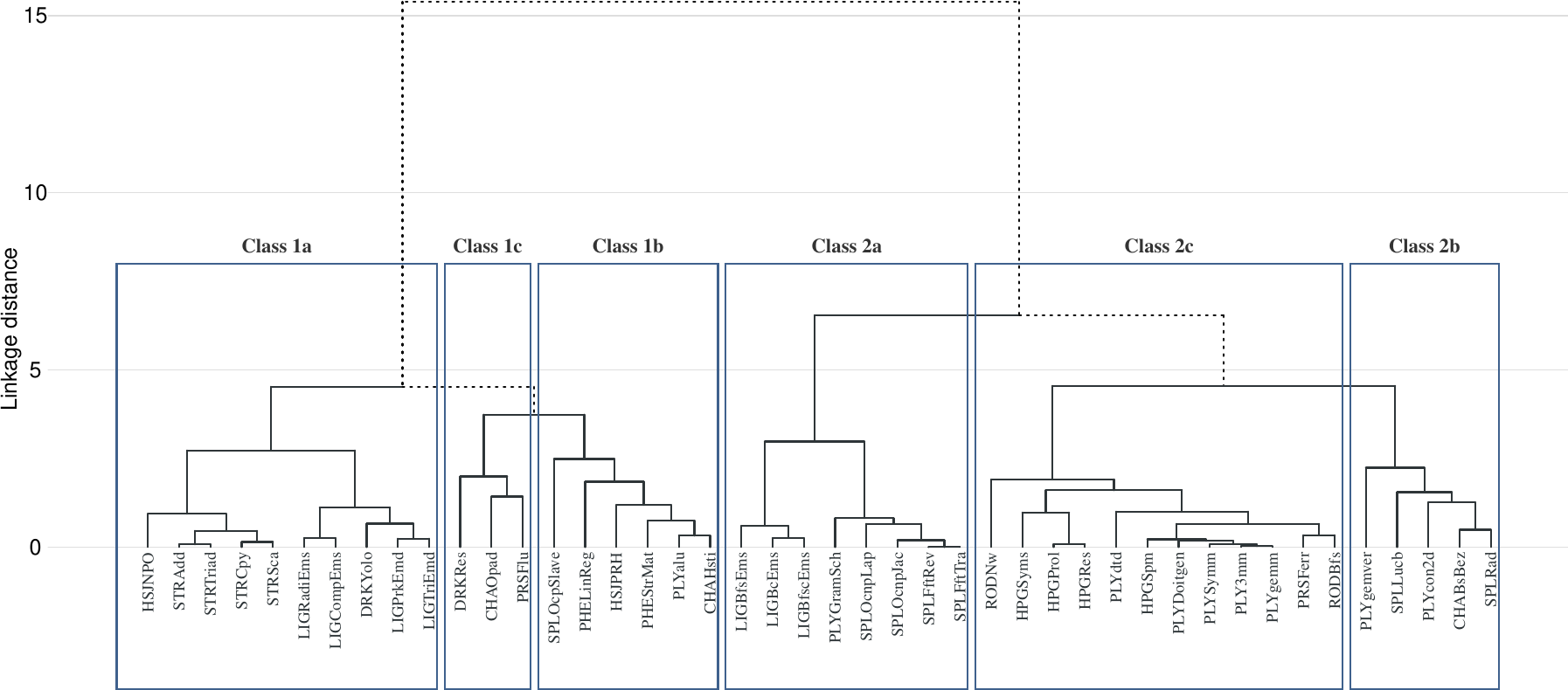}%
    \caption{\geraldorevi{Hierarchical clustering of 44 \gfii{representative} \gfii{functions}.}}
    \label{fig_dendo_validation}
\end{figure}

\geraldorevi{First, our benchmarks exhibit a wide range of behavior diversity, even among those belonging to the same class. For example, we observe that the functions from Class~1a are divided into two groups, with a linkage distance of 3. Intuitively, \gfii{functions} in the first group (\texttt{HSJNPO}, \texttt{STRAdd}, \texttt{STRCpy}, \texttt{STRSca}, \texttt{STRTriad}) have regular access patterns while \gfii{functions} in the second group (\texttt{DRKYolo}, \texttt{LIGCompEms}, \texttt{LIGPrkEmd}, \texttt{LIGRadiEms}) have  irregular access patterns. We observe a similar clustering in Section~\ref{sec_scalability_class1a}}. 

\geraldorevi{Second, we observe that our application clustering \gfii{(Section~\ref{sec:scalability})} matches the clustering \juan{that the} \gfii{hierarchical clustering} algorithm provides \gfii{(Figure~\ref{fig_dendo_validation})}. From the dendrogram root, we observe that the \juan{right} part of the dendrogram consists of \gfii{functions} with high temporal locality (from \geraldorevi{C}lasses 2a, 2b, and 2c). \juan{Conversely}, the \juan{left} part of the dendrogram consists of \gfii{functions} with low temporal locality (from \geraldorevi{C}lasses 1a, 1b, and 1c). \gfii{The \gfii{functions} in the right and left part of the dendrogram} have a high linkage distance (\gfiii{higher than} 15), which implies that the metrics we use for our clustering are significantly different from each other for these \gfii{functions}. \gfii{Third}, we observe that \gfii{functions} within the same class are clustered into groups with a linkage distance \gfiii{lower than} 5. This grouping matches the six classes of data movement bottlenecks present in \bench. Therefore, we conclude that our methodology can successfully cluster \gfii{functions} into distinct classes, each one representing a different memory bottleneck.}

\geraldorevi{We conclude that (i) \bench provides a heterogeneous \gfiv{and diverse} set of functions \gfiii{to study data movement bottlenecks} and (ii) our memory bottleneck clustering methodology matches the clustering provided by \gfii{a} hierarchical clustering algorithm \gfii{(this section; Figure~\ref{fig_dendo_validation})}. }

\section{Case Studies}

In this section, we \gfii{demonstrate} how our benchmark suite is useful to study open questions \geraldorevi{related to} NDP \geraldorevi{system} designs. We \gfii{provide} four case studies. The first study analyzes the impact of load balance and communication on NDP execution. The second study \juan{assesses} \geraldorevi{the impact of tailored NDP accelerators on our memory bottleneck analysis}. \geraldorevi{T}he third study evaluates the \geraldorevi{effect} \juan{of} \geraldorevi{different core \gfii{designs} on NDP \gfii{system performance}}. The fourth study \juan{analyzes} \geraldorevi{the impact of \gfiv{fine-grained} offloading \gfiv{(i.e., offloading} \gfiii{\gfv{small blocks of} instructions to NDP cores}\gfiv{)} on performance.}

\subsection{Case Study 1: Impact of Load Balance and Inter-Vault Communication on NDP Systems} 
\label{sec:case_study_1}

Communication between NDP cores is one of the key challenges for future NDP \geraldorevi{system} designs, especially for NDP architectures based on 3D-stacked memories, where accessing a remote vault incurs extra latency overhead due to network traffic~\cite{ESMC_DATE_2015, ahn2015scalable,syncron}. This case study aims to evaluate the load \geraldorevi{im}balance and inter-vault communication that the NDP cores \geraldorevi{experience} when executing \gfii{functions} from \gfiii{the \bench} benchmark suite. We statically map a \gfii{function} to an NDP core, and  we assume that NDP cores are connected using a 6x6 2D-mesh \geraldorevi{Network-on-Chip (NoC)}, similar to previous works~\gfiii{\cite{drumond2017mondrian,Kim2016,hadidi2018performance,min2019neuralhmc,dai2018graphh}}. \geraldo{Figure~\ref{fig_interconnection_overhead} shows the \geraldorevi{performance} overhead that the \geraldorevi{interconnection network} imposes to NDP cores when running several \gfii{functions} from our benchmark suite.} \geraldorevi{We report performance overheads of \gfii{functions} from different bottleneck classes (i.e., from \geraldorevi{C}lasses 1a, 1b, 2a, and 2b) that experience at least 5\% of performance overhead due to the interconnection network.} We calculate the interconnection \geraldorevi{network performance} overhead by comparing performance \gfii{with} the 2D-mesh versus \gfii{that with} \geraldorevi{an ideal} zero-latency interconnection network. We observe that the interconnection \geraldorevi{network performance} overhead varies across \gfii{functions}, with a minimum overhead of 5\% for \texttt{SPLOcpSlave} and a maximum overhead of 26\% for \texttt{SPLLucb}. 

% \begin{figure}[ht]
%  \centering
%   \includegraphics[width=0.9\linewidth]{mainmatter/04_damov/Plots/interconnection_overhead-crop.pdf}%
%   \caption{\geraldorevi{Interconnection network performance overhead \gfii{in our NDP system}.}}
%   \label{fig_interconnection_overhead}
% \end{figure}

% \begin{figure}[ht]
%  \centering
%   \includegraphics[width=0.9\linewidth]{mainmatter/04_damov/Plots/interconnection_traffic-crop.pdf}%
%   \caption{\geraldorevi{Distribution of NoC hops traveled per memory request.}}
%   \label{fig_traffic_distribution}
% \end{figure}

\begin{figure}[h]
\centering
\begin{minipage}[t]{.47\textwidth}
  \centering
  \includegraphics[width=0.9\linewidth]{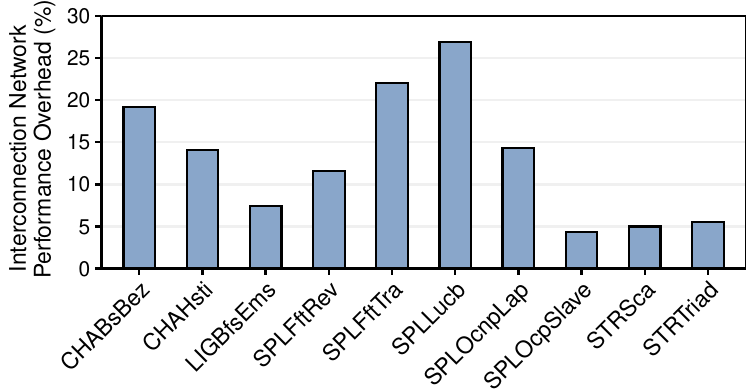}%
  \caption{\geraldorevi{Interconnection network performance overhead \gfii{in our NDP system}.}}
  \label{fig_interconnection_overhead}
\end{minipage}%
\qquad
\begin{minipage}[t]{.47\textwidth}
  \centering
  \includegraphics[width=\linewidth]{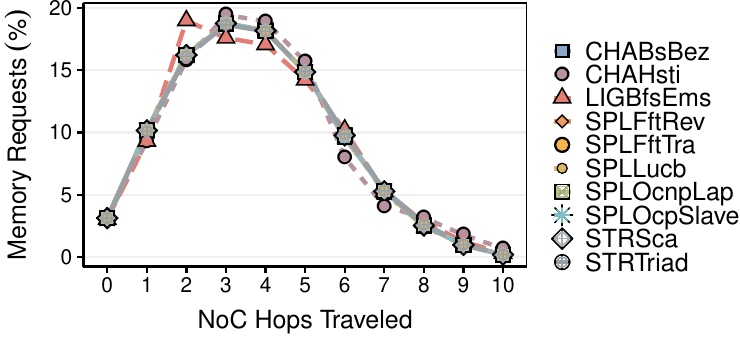}%
  \caption{\geraldorevi{Distribution of NoC hops traveled per memory request.}}
  \label{fig_traffic_distribution}
\end{minipage}
\end{figure}

\geraldo{We further characterize the \geraldorevi{traffic of memory requests injected into the interconnection network} for these \gfii{functions},  aiming to understand the communication pattern\geraldorevi{s} across NDP cores. Figure~\ref{fig_traffic_distribution} shows \gfii{the distribution of all memory requests (y-axis) in terms of} how many hops \gfii{they} need to travel \geraldorevi{in the NoC between NDP cores \gfiii{(x-axis)} for each \gfii{function}}. We make the following observations.} First, we observe that, on average, 40\% of all memory requests need to travel 3 to 4 hops in the \gls{NoC}, and less than 5\% of all requests are issued to a local vault (0 hops). Even though the \gfii{functions} follow different memory access patterns, they all inject similar network traffic \gfiii{into} the \gls{NoC}.\footnote{\geraldorevi{We use the default HMC data interleaving scheme in our experiments (Table~\ref{table_parameters}).}} Therefore, we conclude that the NDP design can be further optimized by (i) employing \geraldorevi{more intelligent} data mapping and scheduling mechanisms that can efficiently allocate data nearby the NDP core \geraldorevi{that access\juan{es} the data} (\geraldorevi{thereby} reducing inter-vault communication \geraldorevi{and improving data locality}) and (ii) designing interconnection networks that can better fit the traffic \geraldorevi{patterns} that NDP workloads produce. \gfiii{The \bench} \geraldorevi{benchmark suite} can be used to develop \geraldorevi{new ideas as well as evaluate existing ideas in} both directions.

\subsection{\geraldorevi{Case Study 2: Impact of NDP Accelerators on Our Memory Bottleneck Analysis}}
\label{sec:case_study_2}

In \geraldorevi{our second} case study, we \gfii{aim to leverage our memory bottleneck classification to} evaluate the benefits \geraldorevi{an} NDP accelerator provides compared to the same accelerator accessing memory externally. We use the Aladdin accelerator simulator\gfiii{~\cite{shao2014aladdin}} to tailor an accelerator \geraldorevi{for an} application \gfii{function}. \geraldorevi{Aladdin works by estimating the performance of a custom accelerator based on the data-flow graph of the application.} The main difference between \gfii{an} NDP accelerator and \gfii{a} regular accelerator \geraldorevi{(i.e., \gfiii{compute-centric} accelerator)} is that the former \geraldorevi{is placed in the logic layer of a 3D-stacked memory device and thus} can leverage larger memory bandwidth\geraldorevi{,} shorter memory access latency\geraldorevi{, and lower memory access energy, compared to the \gfiii{compute-centric} accelerator \gfii{that is exemplary of existing compute-centric accelerator designs}}.

\gfii{To evaluate the benefits of NDP accelerators, we select \gfii{three} functions from our benchmark suite for this case study: \texttt{DRKYolo} (from Class~1a), \texttt{PLYalu} (from Class~1b), and \texttt{PLY3mm} (from Class~2c). \gfii{We select these functions and memory bottleneck classes \gfii{because we expect them}  \juan{\gfii{to} benefit the most (\gfii{or to show no benefit}) from the near-memory placement of an accelerator.} \gfii{According to our memory bottleneck analysis, we expect that the functions we select to } (i) benefit from NDP due to its high DRAM bandwidth (Class~1a), (ii) benefit from NDP due to its shorter DRAM access latency (Class~1b), or (iii) do \emph{not} benefit from NDP in any way \gfiii{ (Class~2c)}.}}

Figure~\ref{fig:accelerator} shows the speedup \juan{that} \geraldorevi{\gfii{the} NDP accelerator provides for the different \gfii{functions} compared to \gfii{the} \gfiii{compute-centric} accelerator}. We make \gfii{four} observations. First, \gfiii{as expected based on our classification,} the NDP accelerator provides performance benefits \geraldorevi{compared to the \gfiii{compute-centric} accelerator} for \gfii{functions} in \geraldorevi{C}lasses 1a and 1b. It does not provide performance improvement for the \gfii{function} in Class 2c. \gfii{Second,} the NDP accelerator for \texttt{DRKYolo} shows the largest performance benefits (1.9$\times$ \geraldorevi{performance improvement compared to the \gfiii{compute-centric} accelerator}). Since this \gfii{function} is \gfii{DRAM} bandwidth-bound \geraldorevi{(Class~1a, Section~\ref{sec_scalability_class1a})}, the \geraldorevi{NDP} accelerator can leverage the larger memory bandwidth available \geraldorevi{in the logic layer of the 3D-stacked memory device}. \gfii{Third,} we observe that the NDP accelerator also provides speedup (1.25$\times$) for the \texttt{PLYalu} \gfii{function} \geraldorevi{compared to the \gfiii{compute-centric}  accelerator,} since \geraldorevi{the NDP accelerator provides shorter memory access latency to the \gfii{function}, which is latency-bound (Class~1b, Section~\ref{sec_scalability_class1b}).} \gfii{Fourth,} the NDP accelerator does not provide performance improvement for the \texttt{PLY3mm} \gfii{function} since this \gfii{function} is compute-bound \geraldorevi{(Class~2\gfiii{c}, Section~\ref{sec_scalability_class2c})}. 

\begin{figure}[h]
     \centering
     \includegraphics[width=0.55\linewidth]{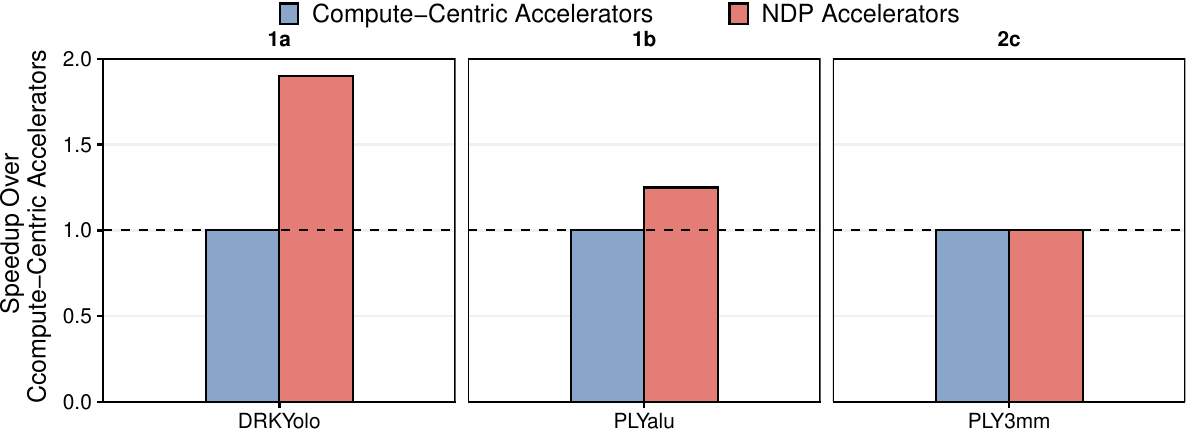}%
     \caption{Speedup \gfii{of the NDP Accelerators over the \gfiii{Compute-Centric} Accelerators for three functions from Classes 1a, 1b, and \gfiii{2}c.}}
     \label{fig:accelerator}
 \end{figure}

\gfiii{In conclusion, our observations for the performance of NDP accelerators are in line with the characteristics of the three memory bottleneck \gfiv{classes} we evaluate in this case study.} Therefore, our memory bottleneck classification can be applied to study other types of system configurations, e.g., the accelerators used in this section. \gfiii{However, since NDP accelerators are often employed under restricted area and power constraints (e.g., limited area available in the logic layer of a 3D-stacked memory~\cite{lazypim, boroumand2019conda}), the core model of the compute-centric and NDP accelerators cannot always be the same}. We leave a thorough analysis that takes area and power constraints \gfiii{in the study of NDP accelerators} into consideration \gfiv{for future research}.

\subsection{\geraldorevi{Case Study 3: Impact of Different Core Models on NDP Architectures}} 
\label{sec:case_study_3}

This case study aims to \gfii{analyze} when a workload can benefit from different core models and numbers of cores while respecting the area and power envelope of the logic layer of a 3D-stacked memory. \gfii{Many prior works employ 3D-stacked memories as the substrate to implement NDP architectures~\gfv{\cite{ahn2015scalable, nai2017graphpim, boroumand2018google, lazypim, top-pim, gao2016hrl, kim2018grim, drumond2017mondrian, RVU, NIM, PEI, gao2017tetris, boroumand2019conda,Kim2016, hsieh2016transparent, cali2020genasm, NDC_ISPASS_2014, farmahini2015nda,pattnaik2016scheduling, hsieh2016accelerating,fernandez2020natsa,syncron,boroumand2021polynesia,boroumand2021mitigating,amiraliphd, IBM_ActiveCube,akin2015data,jang2019charon,nai2015instruction,alves2016large,xie2017processing,lenjani2020fulcrum,kersey2017lightweight,huang2019active,hassan2015near,guo20143d,liu20173d,de2018design,LiM_3D_FFT_MM,Sparse_MM_LiM,glova2019near}}. However, 3D-stacked memories impose severe area and power restrictions on NDP architectures. For example, the area and power budget of the logic layer of a single HMC vault %is
\juang{are} 4.4~$mm^2$ and 312~$mW$,
respectively~\cite{lazypim, boroumand2018google}.} 

\gfii{In the case study, we perform an iso-area and iso-power performance evaluation of three functions from our benchmark suite. We configure the host CPU system and the NDP system to guarantee an iso-area and iso-power evaluation, considering the area and power budget for a 32-vault HMC device~\cite{lazypim, boroumand2018google}. \gfiii{W}e use four out-of-order cores with a deep cache hierarchy for the host system configuration and} two \gfiii{different} NDP configurations\gfiii{:} (1) \gfiii{one} using six out-of-order \gfiii{NDP} cores (\textit{NDP+out-of-order}) and (2) using 128 in-order \gfiii{NDP} cores (\textit{NDP+in-order}), without a deep cache hierarchy.  We \geraldorevi{choose} \gfii{functions} from \geraldorevi{C}lasses 1a, 1b, \juan{and 2b} for this case study since the major \gfiii{effects} distinct microarchitectures have on the memory system are: (a) how much DRAM bandwidth they can \juan{sustain,} and (b) how much DRAM latency they can hide. \geraldorevi{\juan{Classes 1a, 1b, and 2b are the most} affected by memory bandwidth and access latency (as shown in Section~\ref{sec:characterization})}. We \geraldorevi{choose} two \gfiii{representative} \gfii{functions} from each of these classes. 

\geraldorevi{Figure~\ref{fig:inorder} shows the speedup provided by \gfiii{the two} NDP system configurations compared \juan{to} the baseline host system. We make two observations. \gfii{First, in all cases, the \emph{NDP+in-order} system provides higher speedup than the \emph{NDP+out-of-order} system, both compared to the host system. On average across all six functions, the \emph{NDP+in-order} system provides 4$\times$ the speedup of the \emph{NDP+out-of-order} system. The larger speedup the \emph{NDP+in-order} system provides is due to the high number of NDP cores in the \emph{NDP+in-order} system\gfiv{. W}e can fit 128 in-order cores in the logic layer of the 3D-stacked memory \gfiv{as opposed to} \gfiii{only} six out-of-order cores in the same area/power budget.  Second, we observe that the speedup the  \emph{NDP+in-order} system provides compared to the \emph{NDP+out-of-order} system does not scale with the number of cores. For example, the \emph{NDP+in-order} system provides \emph{only} 2$\times$ the performance of the \emph{NDP+out-of-order} system for \texttt{DRKYolo} and \texttt{PLYalu}, even though the \emph{NDP+in-order} system has 21$\times$ the number of NDP cores of the \emph{NDP+out-of-order} system. This implies that even though the \gfii{functions} benefit from \gfii{a large number of NDP cores available in the \emph{NDP+in-order} system}, static instruction scheduling limits performance on the \emph{NDP+in-order} system.}} 

 \begin{figure}[ht]
  \vspace{-5pt}
     \centering
     \includegraphics[width=0.55\linewidth]{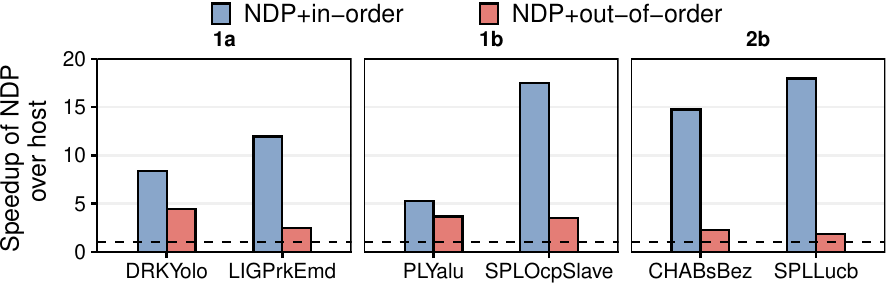}%
    \caption{\geraldorevi{Speedup of NDP architectures \gfiii{over} 4 out-of-order host \gfiii{CPU} cores for two NDP configurations: using \geraldorevi{128 in-order \gfiii{NDP cores} (\emph{NDP+in-order}) and 6 out-of-order \gfiii{NDP cores} (\emph{NDP+out-of-order}) for \gfiii{representative functions} from \geraldorevi{C}lasses 1a, 1b, and \juan{2b.}}}}
     \label{fig:inorder}
 \end{figure}

\geraldo{We believe, and \juang{our previous observations suggest,} that a\gfii{n} \gfii{efficient} NDP architecture can be achieved by leveraging \gfii{mechanisms} that can exploit \juang{both} \gfii{dynamic instruction scheduling} and \gfii{many-core design while fitting in the area and power budget \gfiii{of} 3D-stacked memories.} For example, \gfiii{past works~\gfv{\cite{padmanabha2017mirage, fallin2014heterogeneous,villavieja2014yoga,suleman2012morphcore,ipek2007core,suleman2010data,petrica2013flicker,kim2007composable,lukefahr2012composite,mutlu2006efficient,mutlu2005techniques,mutlu2003runahead,suleman2009accelerating,joao2012bottleneck,chaudhry2009simultaneous, hashemi2016continuous, chou2004microarchitecture,joao2013utility,tarjan2008federation,alipour2020delay,kumar2019freeway,alipour2019fiforder,kumar2003single,kumar2004single}} propose techniques that enable the benefits of simple and complex cores at the same time\gfv{, via heterogeneous \gfvi{or adaptive} architectures}. These \gfiv{ideas} can be examined to \gfiv{enable} better core \gfv{and system} design\gfv{s} for NDP systems, and \bench can facilitate their proper \gfv{design, exploration, and} evaluation.} 
}

\subsection{\geraldorevi{Case Study 4: Impact of \gfiv{Fine-Grained} \gfiii{Offloading to NDP} on Performance}}
\label{sec:case_study_4}

\gfv{Several p}rior works on \gls{NDP} \gfv{(e.g., \cite{PEI,nai2017graphpim,nai2015instruction,hadidi2017cairo,azarkhish2016design,hajinazarsimdram, seshadri2017ambit, seshadri2013rowclone,ahmed2019compiler,baskaran2020decentralized,li2019pims})} propose to identify and offload to the \geraldorevi{NDP system} simple primitives (e.g., \geraldorevi{instructions}, atomic operations). \geraldorevi{We refer to this \gls{NDP} offloading scheme as a \emph{fine-grained \gls{NDP} offloading}, in contrast to a \emph{coarse-grain\geraldorevi{ed} \gls{NDP} offloading scheme} that offloads whole functions and applications to \gls{NDP} systems}. \gfiii{A fine-grained \gls{NDP} offloading scheme provides two main benefits compared to a coarse-grained \gls{NDP} offloading scheme. First, a fine-grained \gls{NDP} offloading scheme allows for a reduction in} the complexity of the processing elements used as \gls{NDP} logic\geraldorevi{, since the \gls{NDP} logic can consist of simple processing elements (e.g., arithmetic units, fixed function units) instead of \gfiii{entire} in-order or out-of-order core\gfiii{s often utilized when employing a coarse-grained \gls{NDP} offloading scheme}. \gfiii{Second, a fine-grained \gls{NDP} offloading scheme} can help developing simple} coherence mechanism needed to allow shared host and \gls{NDP} execution~\cite{PEI}. However, identifying arbitrary \gls{NDP} \gfiii{instructions} can be a daunting task since there is no comprehensive methodology that indicates \geraldorevi{what types} of instructions are good offloading candidate\geraldorevi{s}. 

As the first step in this direction, we exploit the key insight provided by~\cite{ayers2020classifying, hashemi2016accelerating} to identify potential regions of code that can be candidates for fine-grain\geraldorevi{ed} \gls{NDP} offloading. \cite{ayers2020classifying, collins2001speculative, hashemi2016accelerating} show that few instructions are responsible for generating most of the cache misses during program execution in memory-intensive applications. Thus, these instructions are naturally good candidates for fine-grain\geraldorevi{ed} \gls{NDP} offloading. \geraldorevi{Figure~\ref{fig_bbc_dist} shows the distribution of unique basic blocks (x-axis) and the percentage of last-level cache misses (y-axis) the basic block produces for three representative functions from our benchmark suite. We select functions from Classes 1a (\texttt{LIGKcrEms}), 1b (\texttt{HSJPRH}), and 1c (\texttt{DRKRes}) since \gfii{functions} in these classes have higher L3 MPKI than \gfii{functions} in Classes 2a, 2b, and 2c. We observe from the figure that 1\% to 10\% of the basic blocks in each function are responsible for up to 95.3\% of the LLC misses. \geraldorevi{We call these basic blocks the} \emph{hot\geraldorevi{test}} basic block\gfiii{s}.}\footnote{We observe for \geraldorevi{the} 44 \gfii{functions} \geraldorevi{we evaluate in Section~\ref{sec:characterization}} that in many cases (for 65\% of the evaluated workloads), a single basic block is responsible for 90\% to 100\% of the \gls{LLC} misses during the \gfii{function}'s execution.} \geraldorevi{W}e investigate the data-flow of each basic block and observe that \geraldorevi{these basic blocks} often execute simple read-modify-write operations, with few arithmetic operations. Therefore, we believe that such basic blocks are good candidates for fine-grain\juan{ed} offloading. \geraldorevi{Figure~\ref{fig_bbc_speedup} shows the speedup obtained by offloading \gfii{(i)} the hottest basic block we identified for the three representative functions \gfii{and (ii) the entire function} to the \gls{NDP} system, compared to the host system.} Our initial evaluations show that offloading the hot\geraldorevi{test} basic block \geraldorevi{\gfii{of} each function} to \geraldorevi{the} \gls{NDP} \geraldorevi{system} can provide up to 1.25$\times$ speedup compared to the host \gfiii{CPU}\gfii{, which is \gfiii{half of} the 1.5$\times$ speedup achieved when offloading the entire function}. \geraldorevi{Therefore, we believe that \gfiii{methodically} identifying simple NDP \gfiii{instructions} can be a promising research direction for future NDP system designs, which \gfii{our} DAMOV \gfii{Benchmark Suite} can help with.} 

% \begin{figure}[ht]
%  \centering
%   \includegraphics[width=\linewidth]{mainmatter/04_damov/Plots/bb_dist.pdf}%
%   \caption{\geraldorevi{Distribution of unique basic blocks (x-axis) and the percentage of last-level cache misses they produce (y-axis) for three representative functions from Classes 1a (\texttt{LIGKcrEms}), 1b (\texttt{HSJPRH}), and 1c (\texttt{DRKRes}).}}
%   \label{fig_bbc_dist}
% \end{figure}

% \begin{figure}[ht]
%   \centering
%   \includegraphics[width=0.9\linewidth]{mainmatter/04_damov/Plots/bbc_speedup_v2-crop.pdf}%
%   \caption{\geraldorevi{Speedup of offloading to NDP the \emph{hottest} basic block in each function \gfiii{versus} the entire function.}}
%   \label{fig_bbc_speedup}
%  \end{figure}

\begin{figure}[!h]
\centering
\begin{minipage}[t]{.64\textwidth}
  \centering
  \includegraphics[width=\linewidth]{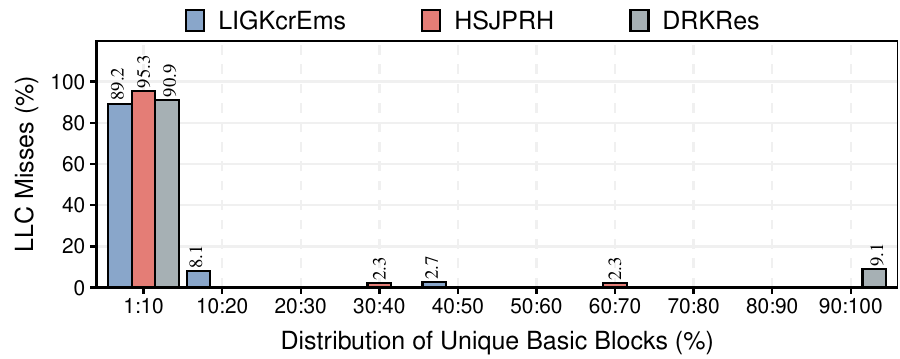}%
  \caption{\geraldorevi{Distribution of unique basic blocks (x-axis) and the percentage of last-level cache misses they produce (y-axis) for three representative functions from Classes 1a (\texttt{LIGKcrEms}), 1b (\texttt{HSJPRH}), and 1c (\texttt{DRKRes}).}}
  \label{fig_bbc_dist}
\end{minipage}%
\qquad
\begin{minipage}[t]{.305\textwidth}
  \centering
  \includegraphics[width=\linewidth]{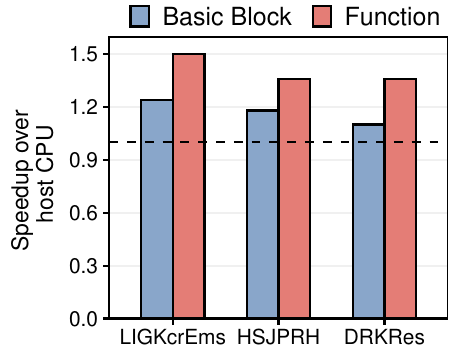}%
  \caption{\geraldorevi{Speedup of offloading to NDP the \emph{hottest} basic block in each function \gfiii{versus} the entire function.}}
  \label{fig_bbc_speedup}
\end{minipage}
\end{figure}

\section{Key Takeaways}

\gfiii{W}e summarize the key takeaways \geraldorevi{from} our \geraldorevi{extensive characterization of 144 functions using our new three-step methodology to identify data movement bottlenecks.} We also highlight when NDP is a good architectural choice to mitigate a particular memory bottleneck. 

\gfii{Figure~\ref{fig:decision} pictorially represents the key takeaways we obtain from our memory bottleneck classification. Based on four key metrics, we classify workloads into six classes of memory bottlenecks. We provide the following key takeaways:}

\begin{figure*}[h]
  \centering
  \includegraphics[width=\textwidth]{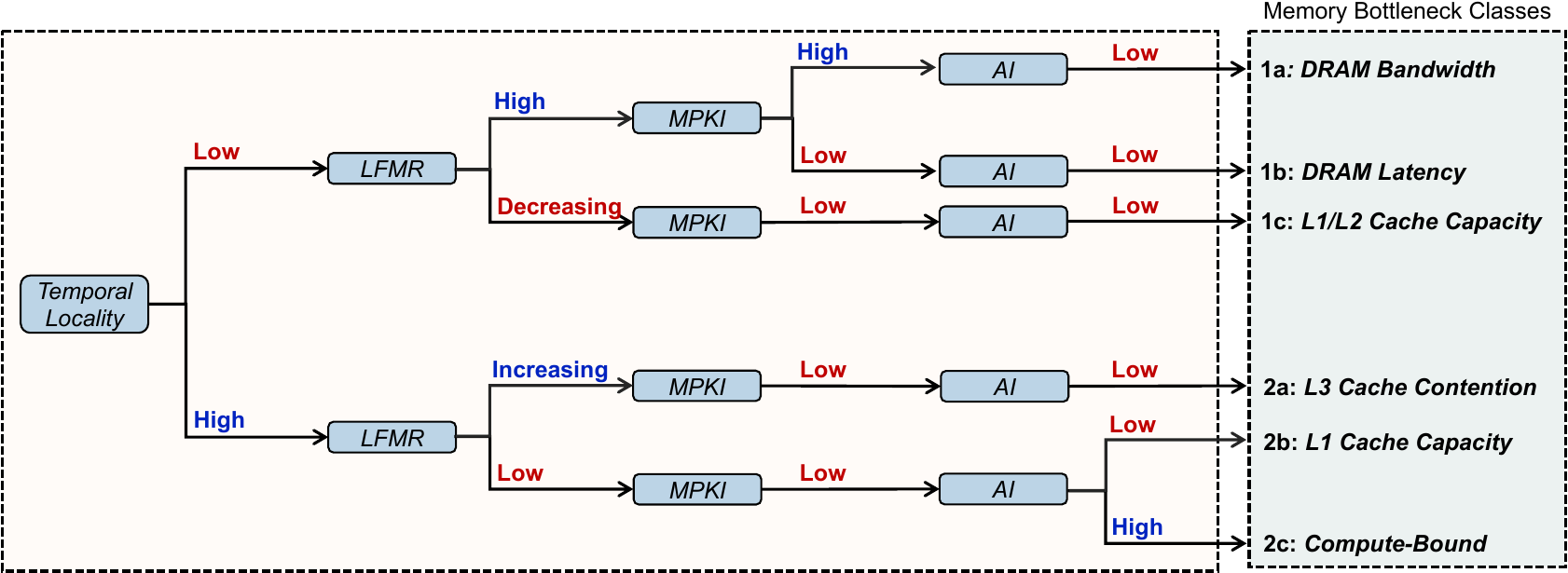}
  \caption{\gfii{Summary of our memory bottleneck classification}.} 
  \label{fig:decision}
\end{figure*}

\begin{enumerate}
\item Applications with low temporal locality, high LFMR, high MPKI, \gfii{and low AI} are \geraldorevi{\gfii{DRAM}} \emph{bandwidth-bound} \geraldorevi{(Class~1a, Section~\ref{sec_scalability_class1a})}. They are bottlenecked by the limited off-chip memory bandwidth as they exert high pressure on main memory. \gfii{We make three observations for Class~1a applications. First,} \geraldorevi{these} applications do benefit from prefetching since they display \geraldorevi{a low} degree of spatial locality.  \gfii{Second,} these applications highly benefit from NDP architectures because they take advantage of the high memory bandwidth available within the memory device. \geraldorevi{\gfii{Third}, NDP architectures significantly improve energy for these applications since they eliminate the off-chip I/O traffic between \juan{the} CPU and \juan{the} main memory.}

\item Applications with low temporal locality, high LFMR, low MPKI, and low AI are \geraldorevi{DRAM} \emph{latency-bound} \geraldorevi{(Class~1b, Section~\ref{sec_scalability_class1b})}. \gfii{We make three observations for Class~1b applications. First,} these applications do not significantly benefit from prefetching since infrequent memory requests make it difficult for the prefetcher to train \geraldorevi{successfully} on an access pattern. \gfii{Second, t}hese applications benefit from NDP architectures since they take advantage of NDP's lower memory access latency \geraldorevi{and the elimination of deep \gfiii{L2/L3} cache hierarchies\gfiii{, which fail to capture data locality} for these workloads}. \geraldorevi{\gfii{Third}, NDP architectures significantly improve energy for these applications since they eliminate costly (and unnecessary) L3 cache look-ups and the off-chip I/O traffic between \juan{the} CPU and \juan{the} main memory.}

\item Applications with low temporal locality, decreasing LFMR with core count, low MPKI, and low AI are \emph{bottlenecked by the available \gfii{L1/L2} cache capacity} \geraldorevi{(Class~1c, Section~\ref{sec_scalability_class1c})}. \gfii{We make three observations for Class 1c applications. First, these} applications are \gfii{DRAM} latency-bound at low core counts, thus taking advantage of NDP architectures\geraldorevi{, both \gfiii{in terms of} performance improvement and energy \gfiii{reduction}}. \gfii{Second,} NDP's benefits reduce when core count \geraldorevi{becomes larger}, which consequently allows the working set\juan{s} of such applications \gfii{to} fit inside the cache hierarchy at high core counts. \juan{\gfii{Third}, NDP architectures can be a good design choice for such workloads in systems with limited area budget \gfiii{since NDP architectures do not require large L2/L3 caches to \gfiv{outperform or perform similarly to} the host CPU \gfiv{(in terms of both \gfv{system throughput} and energy)} for these workloads}.}

\item Applications with high temporal locality, increasing LFMR with core count, low MPKI, and low AI  are \emph{bottlenecked by \gfii{L3} cache contention} \geraldorevi{(Class~2a, Section~\ref{sec_scalability_class2a})}. \gfii{We make three observations for Class~2a applications. First, these} applications benefit from a deep cache hierarchy and do not take advantage of NDP architectures at low core counts. \gfii{Second}, the number of cache conflicts increases when the number of cores in the system \geraldorevi{increases}, \geraldorevi{leading to more} pressure on main memory. We observe that NDP can effectively mitigate such cache contention for these applications without incurring the high area and energy overheads of providing additional cache capacity in the host. \geraldorevi{\gfii{Third,} NDP can improve energy for these workloads at high core counts, since it eliminates the costly data movement between the last-level cache and \juan{the} main memory.}
 
\item Applications with high temporal locality, low LFMR, low MPKI, and low AI are \gfii{bottlenecked by} \emph{L1 cache capacity} (Class~2b, Section~\ref{sec_scalability_class2b}). \gfii{We make two observation for Class~2b applications. First,} NDP can provide similar performance \geraldorevi{and energy consumption than the host system by leveraging lower memory access latency and avoiding off-chip energy consumption for these applications}. \geraldorevi{\gfii{Second}, NDP can be used to reduce the overall SRAM area \gfiii{(by eliminating L2/L3 cache\gfiv{s})} in the system without a performance or energy penalty}. 

\item Applications with high temporal locality, low LFMR, low MPKI, and high AI are \emph{compute-bound} (Class~2c, Section~\ref{sec_scalability_class2c}). \gfii{We make three observation\gfiii{s} for Class~2c applications. First, these applications} suffer performance and energy penalties due to the lack of a deep \gfiii{L2/L3} cache hierarchy when executed on the NDP architecture. \geraldorevi{Second, these applications highly benefit from prefetching due to their high temporal \gfiii{and spatial} locality.} \geraldorevi{\gfii{Third}, these applications are not good candidate\gfiii{s} to execute on NDP architectures.}

\end{enumerate}

\subsection{Shaping Future Research with \bench}

\gfii{A key contribution of our work is \bench, the first benchmark suite for \gfiii{main memory} data movement studies. \bench is the collection of 144 functions from 74 different applications, belonging to 16 different benchmark suites or frameworks, classified into six different classes of data movement bottlenecks.}

\geraldorevi{We believe that \bench can be used to explore a wide range of research directions on the study of data movement bottlenecks,  appropriate mitigation mechanisms, \gfii{and open research topics on NDP architectures.}
\gfii{We highlight \bench's usability \gfiii{and potential benefits} with four \gfiii{brief} case studies, which we summarize below:}
}

\begin{itemize}
\item \gfii{In the first case study (Section~\ref{sec:case_study_1}), we use \bench to evaluate the interconnection network overheads that NDP cores placed in different vaults of a 3D-stacked memory suffer \gfiii{from}. We observe that a large portion of the memory requests \gfiii{an} NDP core issues go to %remove 
\juang{remote} vaults, which increases the memory access latency for the NDP core. We believe that \bench can be employed to study better data mapping \gfiii{techniques} and interconnection network \gfiii{designs} that \gfiii{aim} to minimize (i) the number of remote memory accesses the NDP cores execute and (ii) the interconnection \gfiii{network} latency overheads.}
    
\item \gfii{In the second case study (Section~\ref{sec:case_study_2}), we evaluate the benefits that NDP accelerators can provide for three applications \juang{from} our benchmark suite. We compare the performance improvements an NDP accelerator provides against the \gfiii{compute-centric version of the} same accelerator. We observe that the NDP accelerator provides significant performance benefits compared to the \gfiii{compute-centric} accelerator for applications in Classes 1a and 1b. At the same time, it does not improve performance for an application in Class 2c. We believe that \bench can \gfiii{aid the design of} NDP accelerators that target different memory bottlenecks in the system.} 
    
\item \gfii{In the third case study (Section~\ref{sec:case_study_3}), we perform an iso-area/-power performance evaluation to compare NDP systems using in-order and out-of-order cores. We observe that the in-order cores' performance benefits for some applications are limited by the cores' static instruction scheduling mechanism. We believe that better NDP systems can be built by leveraging techniques that enable dynamic instruction scheduling without incurring the large area and power overheads of out-of-order cores. \bench can help in the analysis and development of such NDP architectures. }
    
\item \gfii{In the fourth case study (Section~\ref{sec:case_study_4}), we evaluate the benefits of offloading small portions of code (i.e., a basic block) to NDP, which simplifies the design of NDP systems. We observe that for many applications, a small percentage of basic blocks is responsible for most of the last-level cache misses. By offloading these basic blocks to an NDP core, we observe a performance improvement of up to 1.25$\times$. We believe that \bench can be used to identify simple NDP \gfiii{instructions} that enable building efficient NDP systems in the future.} 
     
\end{itemize}

\section{Summary}

\geraldorevi{We introduce the first \gfii{rigorous} methodology to characterize \gfii{memory-related} data movement bottlenecks \juan{in modern workloads} and the first data movement benchmark suite, called \bench. We perform the first large-scale characterization of applications to develop a \gfii{three-step workload characterization methodology that introduces and evaluates four key metrics} 
\gfii{to identify} the sources of data movement bottlenecks \gfiii{in real applications}. We \gfii{use our new methodology to} classify \gfii{the primary sources of memory bottlenecks of a broad range of applications into six different classes of memory bottlenecks.} We highlight the benefits of our benchmark suite with four case studies, which showcase how \gfii{representative} \gfii{workloads} in \bench can be used to explore open-research topics on NDP systems \gfii{and reach architectural as well as workload-level \gfiii{insights and} conclusions}. We open-source our benchmark suite and our bottleneck analysis toolchain~\cite{damov}\gfii{. \gfiii{We hope that our work enables} further studies \gfii{and research} \gfiii{on} \gfii{hardware and software solutions for} data movement \gfii{bottlenecks}\gfiv{, including near-data processing}.}}

% % \glsresetall{}
\chapter[MIMDRAM: An End-to-End Processing-Using-DRAM System for High-Throughput, Energy-Efficient, and Programmer-Transparent Multiple-Instruction Multiple-Data Computing]{MIMDRAM: An End-to-End Processing-Using-DRAM System for High-Throughput, Energy-Efficient, and Programmer-Transparent Multiple-Instruction Multiple-Data Computing}
\label{chap:mimdram}

% Results Macros
\newcommand\efficiencysimdram{14.3$\times$\xspace}
\newcommand\efficiencygpu{6.8$\times$\xspace}
\newcommand\efficiencycpu{30.6$\times$\xspace}

% Mechanism Name
\newcommand{\prop}{MIMDRAM\xspace}

% PUD-related macros
\providecommand\aap{\texttt{AAP}/\texttt{AP}\xspace}
\providecommand\aaps{\texttt{AAP}s/\texttt{AP}s\xspace}
\providecommand\matrange{\texttt{\lbrack}\emph{mat begin}\texttt{,}\emph{mat end}\texttt{\rbrack}\xspace}
\providecommand\uop{\textmu{}Op\xspace}
\providecommand\uops{\textmu{}Ops\xspace}
\providecommand\uprog{\textmu{}Program\xspace}
\providecommand\uprogs{\textmu{}Programs\xspace}
\providecommand\ureg{\textmu{}Register}
\providecommand\upc{\textmu{}PC}
\providecommand\uprogc{\textmu{}Program counter}
\providecommand\bbop{\emph{bbop}\xspace}
\providecommand\underutilization{underutilization\xspace}
\providecommand\propA{\emph{\prop-AOp}\xspace}
\providecommand\propT{\emph{\prop-TOp}\xspace}
 
% Paper Submission Macros 
\newif\ifmimdramrevision
%\mimdramrevisiontrue
\mimdramrevisionfalse
\ifmimdramrevision 
    \providecommand{\revA}[1]{\textcolor{red}{#1}}
    \providecommand{\revB}[1]{\textcolor{purple}{#1}}
    \providecommand{\revC}[1]{\textcolor{blush}{#1}}
    \providecommand{\revD}[1]{\textcolor{burntorange}{#1}}
    \providecommand{\revE}[1]{\textcolor{ao}{#1}}
    \providecommand{\revCommon}[1]{\textcolor{blue}{#1}}

    \let\marginpar\marginnote
    \setlength{\marginparwidth}{0.4in}

    \newcommandx{\changeCM}[2][1=]{\todo[linecolor=blue,backgroundcolor=blue!25,bordercolor=blue,#1,size=\scriptsize]{\revCommon{\textbf{#2}}}}
    
    \newcommandx{\changeA}[2][1=]{\todo[linecolor=red,backgroundcolor=red!25,bordercolor=red,#1,size=\scriptsize]{\revA{\textbf{#2}}}}
    
    \newcommandx{\changeB}[2][1=]{\todo[linecolor=purple,backgroundcolor=purple!25,bordercolor=purple,#1,size=\scriptsize]{\revB{\textbf{#2}}}}
    
    \newcommandx{\changeC}[2][1=]{\todo[linecolor=blush,backgroundcolor=blush!25,bordercolor=blush,#1,size=\scriptsize]{\revC{\textbf{#2}}}}
    
    \newcommandx{\changeD}[2][1=]{\todo[linecolor=orange,backgroundcolor=orange!25,bordercolor=orange,#1,size=\scriptsize]{\revD{\textbf{#2}}}}
    
    \newcommandx{\changeE}[2][1=]{\todo[linecolor=ao,backgroundcolor=ao!25,bordercolor=ao,#1,size=\scriptsize]{\revE{\textbf{#2}}}}
\else
    \providecommand{\revA}[1]{#1}
    \providecommand{\revB}[1]{#1}
    \providecommand{\revC}[1]{#1}
    \providecommand{\revD}[1]{#1}
    \providecommand{\revE}[1]{#1}
    \providecommand{\revCommon}[1]{#1}

    \newcommandx{\changeCM}[2][1=]{\todo[disable,#1]{#2}}
    \newcommandx{\changeA}[2][1=]{\todo[disable,#1]{#2}}
    \newcommandx{\changeB}[2][1=]{\todo[disable,#1]{#2}}
    \newcommandx{\changeC}[2][1=]{\todo[disable,#1]{#2}}
    \newcommandx{\changeD}[2][1=]{\todo[disable,#1]{#2}}
    \newcommandx{\changeE}[2][1=]{\todo[disable,#1]{#2}}
\fi

\newif\ifmimdramsubmissionmicro
\mimdramsubmissionmicrotrue
%\mimdramsubmissionmicrofalse
\ifmimdramsubmissionmicro
    \providecommand{\juan}[1]{#1}
    \providecommand{\juani}[1]{#1}
    \providecommand{\sg}[1]{#1}
    \providecommand\jgl[1][0]{}
    \providecommand\gfbox[1][0]{}
    \providecommand{\gf}[1]{#1}
    \providecommand{\gfi}[1]{#1}
    \providecommand{\atb}[1]{#1}
    \providecommand{\agy}[1]{#1}
    \providecommand{\agycomment}[1]{}
\else
    \providecommand{\juan}[1]{#1}
    \providecommand{\juani}[1]{\textcolor{dollarbill}{#1}}
    \providecommand\jgl[1]{\noindent{\color{dollarbill} {\bf \fbox{JGL}} {\it#1}}}
    \providecommand{\sg}[1]{\textcolor{purple}{#1}}
    \providecommand{\gf}[1]{\textcolor{blue}{#1}}
    \providecommand{\gfi}[1]{\textcolor{orange}{#1}}
    \providecommand\gfbox[1]{\noindent{\color{red} {\bf \fbox{TODO:}} {\it#1}}}
    \providecommand{\atb}[1]{\textcolor{ao}{#1}}
    \providecommand{\agy}[1]{\textcolor{orange}{#1}}
    \providecommand{\agycomment}[1]{\textcolor{orange}{\textbf{[agy:} #1\textbf{]}}}
\fi 

\newif\ifmimdramsubmissionhpca
\mimdramsubmissionhpcatrue
%\mimdramsubmissionhpcafalse
\ifmimdramsubmissionhpca
    \providecommand{\gfhpca}[1]{#1}
\else
    \providecommand{\gfhpca}[1]{\textcolor{blue}{#1}}  
\fi 

\newif\ifmimdramrevisionhpca
%\mimdramrevisionhpcatrue
\mimdramrevisionhpcafalse
\ifmimdramrevisionhpca 
    \providecommand{\rA}[1]{\textcolor{red}{#1}}
    \providecommand{\rB}[1]{\textcolor{purple}{#1}}
    \providecommand{\rC}[1]{\textcolor{blush}{#1}}
    \providecommand{\rD}[1]{\textcolor{burntorange}{#1}}
    \providecommand{\rE}[1]{\textcolor{ao}{#1}}
    \providecommand{\rCommon}[1]{\textcolor{blue}{#1}}

    \let\marginpar\marginnote
    \setlength{\marginparwidth}{0.4in}
    
    \newcommandx{\changerCM}[2][1=]{\todo[linecolor=blue,backgroundcolor=blue!25,bordercolor=blue,#1,size=\scriptsize]{\revCommon{\textbf{#2}}}}
    
    \newcommandx{\changerA}[2][1=]{\todo[linecolor=red,backgroundcolor=red!25,bordercolor=red,#1,size=\scriptsize]{\revA{\textbf{#2}}}}
    
    \newcommandx{\changerB}[2][1=]{\todo[linecolor=purple,backgroundcolor=purple!25,bordercolor=purple,#1,size=\scriptsize]{\revB{\textbf{#2}}}}
    
    \newcommandx{\changerC}[2][1=]{\todo[linecolor=blush,backgroundcolor=blush!25,bordercolor=blush,#1,size=\scriptsize]{\revC{\textbf{#2}}}}
    
    \newcommandx{\changerD}[2][1=]{\todo[linecolor=orange,backgroundcolor=orange!25,bordercolor=orange,#1,size=\scriptsize]{\revD{\textbf{#2}}}}
    
    \newcommandx{\changerE}[2][1=]{\todo[linecolor=ao,backgroundcolor=ao!25,bordercolor=ao,#1,size=\scriptsize]{\revE{\textbf{#2}}}}
\else
    \providecommand{\rA}[1]{#1}
    \providecommand{\rB}[1]{#1}
    \providecommand{\rC}[1]{#1}
    \providecommand{\rD}[1]{#1}
    \providecommand{\rE}[1]{#1}
    \providecommand{\rCommon}[1]{#1}
    \newcommandx{\changerCM}[2][1=]{\todo[disable,#1]{#2}}
    \newcommandx{\changerA}[2][1=]{\todo[disable,#1]{#2}}
    \newcommandx{\changerB}[2][1=]{\todo[disable,#1]{#2}}
    \newcommandx{\changerC}[2][1=]{\todo[disable,#1]{#2}}
    \newcommandx{\changerD}[2][1=]{\todo[disable,#1]{#2}}
    \newcommandx{\changerE}[2][1=]{\todo[disable,#1]{#2}}

\fi

\newif\ifmimdramcameraready
\mimdramcamerareadytrue
%\mimdramcamerareadyfalse
\ifmimdramcameraready
    \providecommand{\gfcr}[1]{#1}
    \providecommand{\gfcri}[1]{#1}
    \providecommand{\sgcri}[1]{#1}
    \providecommand{\gfcrii}[1]{#1}
    \providecommand{\gfcriii}[1]{#1}
    \providecommand{\sgcriii}[1]{#1}
    \providecommand{\gfcriv}[1]{#1}
    \providecommand{\om}[1]{#1}
    \providecommand{\omi}[1]{#1}
    \providecommand{\omii}[1]{#1}
    \providecommand{\omiii}[1]{#1}
\else 
    \providecommand{\gfcr}[1]{#1}
    \providecommand{\gfcri}[1]{#1}
    \providecommand{\gfcrii}[1]{#1}
    \providecommand{\gfcriii}[1]{\textcolor{red}{#1}} 
    \providecommand{\gfcriv}[1]{\textcolor{blue}{#1}}
    \providecommand{\sgcri}[1]{#1}
    \providecommand{\sgcriii}[1]{#1}
    \providecommand{\om}[1]{#1}
    \providecommand{\omi}[1]{#1}
    \providecommand{\omii}[1]{\textcolor{red}{#1}} 
    \providecommand{\omiii}[1]{\textcolor{blue}{#1}} 

    \let\marginpar\marginnote
    \setlength{\marginparwidth}{0.4in}
\fi

% Problematic citation blocks

\providecommand\ambit{\cite{seshadri2017ambit,seshadri2019dram,seshadri2015fast,seshadri.bookchapter17,seshadri2016buddy,seshadri2016processing}\xspace}

\providecommand\pnmshort{\cite{devaux2019true,ghiasi2022genstore,gomez2021benchmarkingcut,gomezluna2021benchmarking,gomez2022benchmarking,syncron,singh2020nero,skhynixpim,ke2021near,giannoula2022sparsep,denzler2021casper,IRAM_Micro_1997,C_RAM_1999,gokhale1995processing,hall1999mapping,ahn2015scalable,boroumand2018google,lazypim, top-pim, kim2018grim, RVU, NIM, PEI,Kim2016, boroumand2019conda, hsieh2016transparent, cali2020genasm,hsieh2016accelerating,boroumand2021mitigating,boroumand2021google,boroumand2022polynesia,boroumand2021polynesia,besta2021sisa,fernandez2020natsa,singh2019napel,lee2021hardware,kim2017grim,boroumand2017lazypim,santos2018processing,lenjani2020fulcrum}\xspace}

\section{Motivation}
\label{sec:motivation}

\gf{The efficiency of \juani{state-of-the-art} \gfi{\gls{PuD}} \juani{substrates} can be subpar \juani{when the\revdel{amount of} \gls{SIMD} parallelism that exists in an application\revdel{'s code} is smaller than} or not a multiple of the size of a DRAM row.} 
To quantify the\revdel{ amount of} \gf{SIMD} parallelism \omi{some} real-world applications inherently \gfcrii{possess}, we \omi{profile} the \emph{maximum vectorization factor} of \gfcrii{twelve real-world} applications. The vectorization factor\revdel{ (or vectorization width)} of a\revdel{ given} loop is the number of scalar operands that fit into a \gls{SIMD} register~\cite{pohl2018cost,trifunovic2009polyhedral}. % (e.g., a common vectorization factor for a loop that operates over \SI{4}{\byte} integers in a system with Intel AVX-512~\cite{avx512doc} \gls{SIMD} support is 16, or $\frac{512}{4\times8}$). 
%
% Since the loops we are interested in for processing-using-DRAM offloading are embarrassingly parallel, there is no data dependency across loop iterations.
%
We calculate the maximum vectorization factor by multiplying the vectorization factor of a single loop iteration and the loop's trip count~\cite{sokulski2022spec}.
%Since manually extracting \gls{SIMD} parallelism from an application can be daunting, 
We leverage modern compilers' loop auto-vectorization engines, which allows us to have an initial understanding of the \gls{SIMD} parallelism that a large number of applications \gfcrii{possesses}. 

\gfcrii{For our analysis, 
we use 
\li~the LLVM compiler toolchain~\cite{lattner2008llvm} (version 12.0.0) to \emph{automatically} vectorize loops in the application, and 
\lii~an LLVM pass~\cite{sarda2015llvm, lopes2014getting,writingpass} that instruments each application's loop to, during execution, gather  \emph{dynamic} information \omii{about} each vectorized loop, i.e., the loop trip count, execution count, execution time, and instruction breakdown~\cite{llvmprofiler}.
We compile each application using the clang compiler~\cite{lattner2008llvm}, using the appropriate flags to enable the loop auto-vectorization engine and its loop vectorization report (i.e., \texttt{-O3 -Rpass-analysis=loop-vectorize -Rpass=loop-vectorize}).\footnote{\gfcrii{See \cref{sec:methodology} for the description of our applications and their dataset.}}} 
\gfi{We assume SIMDRAM~\cite{hajinazarsimdram} as the target \gls{PuD} architecture.}

%To understand the in-DRAM \gls{SIMD} utilization, performance, and energy consumption when offloading such \gls{SIMD}-prone operations to the processing-using-DRAM architecture, we perform a \emph{loop offloading analysis}, where we offload the automatically vectorized loops in an application to the SIMDRAM architecture. In this analysis, the vectorization factor we use for all vectorized loops is equal to the size of a logical DRAM row (i.e., 65'536 operands). In case the maximum vectorization factor of a loop is smaller than 65'536, we pad the remaining columns in the target DRAM row with zeros (which represents \underutilization). 

%\paratitle{Vectorization Factor Analysis} 
\gfi{Figure}~\ref{fig_max_utilization} shows the \gfi{distribution of} maximum vectorization factors (y-axis) \omi{of} all the vectorizable loops in an application (x-axis). \changeB{\#B3}\revB{We indicate different amounts of SIMD parallelism with horizontal dashed lines for reference.} We make two observations. 
First, the maximum vectorization factor varies \omi{both} within \omi{an} application and across different applications. Our analysis shows maximum vectorization factors as low as \gf{8} and as high as \gf{134,217,729}. 
Second, \omi{only} a small \omi{fraction} of vectorized loops have enough maximum vectorization factor \changeB{\#B3}\revB{(i.e., values above the green horizontal dashed line)} to fully exploit the \gls{SIMD} parallelism of SIMDRAM. On average, only \gf{0.11}\% of all vectorized loops have a maximum vectorization factor equal to or \juani{greater} than a DRAM row \gfcrii{(i.e., \juani{greater} than 65,536 data elements)}. 
We conclude that 
\li~real-world applications have \juani{varying} degrees of \gls{SIMD} parallelism\revdel{ that can be exploited for \gfi{PuD} computation}; and
\lii~\juani{these varying degrees of \gls{SIMD} parallelism rarely take \omi{full} advantage of the \gfcrii{very}-wide \gls{SIMD} width of state-of-the-art \gls{PuD} substrates.}

\begin{figure}[ht]
    \centering
    \includegraphics[width=0.85\linewidth]{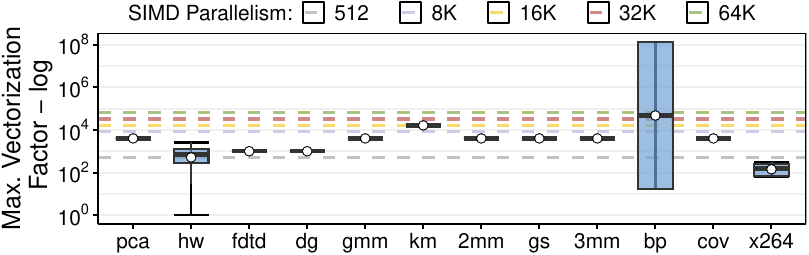}
    \caption{\gfi{Distribution of} maximum vectorization factor \omi{across all vectorized loops}. Whiskers extend to the minimum and maximum data points on either side of the box. \omi{Bubbles depict} average values.}
    \label{fig_max_utilization}
\end{figure}

\paratitle{Problem \& Goal} \gf{We observe that the rigid granularity of \gfi{\gls{PuD}} architectures limits their efficiency \omi{(and thus their \emph{effective} applicability)} for many applications.} Such applications \juani{would benefit from} a variable-size \gls{SIMD} substrate that can \gfcriii{dynamically} adapt to the \omii{varying levels of \gls{SIMD} parallelism (i.e., different vectorization factors) an application exhibits during its execution}. 
Therefore, our \textit{goal} is to design a flexible \gfi{\gls{PuD}} substrate that
\li~adapts to the \omi{varying} \gfcriii{levels of} \gls{SIMD} parallelism \omiii{present in} an application, and 
\lii~maximizes the utilization of the \omi{very-wide} \gfi{\gls{PuD}} engine \changeD{\#D1}\revD{by \omi{concurrently \omii{exploiting}} parallelism across \omi{\emph{different}} \gls{PuD} \gfcrii{operations} \omii{(potentially from different applications)}.}

%
%
% We draw three conclusions from our analysis. 
% First, the rigid granularity of processing-using-DRAM architectures limits their applicability and efficiency for many applications.
% Such applications require a variable-size \gls{SIMD} substrate that can adapt to the various-size vectorization factor the application presents.
% Second, naively offloading all vector operations to be executed by the processing-using-DRAM architecture leads to subpar performance or even slowdowns.
% Third, enabling communication across DRAM bitlines can allow the execution of a broader set of vector operations in DRAM (i.e., vector reduction operations).

% overcomes the two limitations caused by the large and rigid granularity of in-DRAM execution. in this work is to enhance state-of-the-art processing-using-DRAM substrates with solutions that can
% \li~flexibility and efficiently extract vectorization opportunities from general-purpose application targeting in-DRAM computing while
% \lii~maximizing \gls{SIMD} utilization of the underlying processing-using-DRAM substrate. 

\section{\prop: A MIMD \gfi{\gls{PuD}} Architecture}
\label{sec:idea}

\prop is a hardware/software co-designed \gfi{\gls{PuD} system} that enables fine-grained \gfi{\gls{PuD}} computation at low cost and low programming effort. The \emph{key idea} of \prop is to leverage fine-grained DRAM activation for \gfi{\gls{PuD}}, \gfi{which} provides three benefits. First, it enables \prop to allocate \emph{only} the appropriate computation resources (based on the maximum vectorization factor \juani{of a} loop) for a target loop, \omi{thereby} reducing \underutilization and energy waste. 
Second, \prop can currently execute \omi{multiple} independent operations inside a single DRAM subarray \omi{\emph{independently} in separate} \gfi{DRAM mats}. This allows \prop to \juani{operate} as a \gls{MIMD} \gfi{\gls{PuD}} substrate, increasing overall throughput. Third, \prop implements low-cost \gfcri{interconnects} that enable moving data across DRAM columns \gfcrii{\emph{across} and \emph{within} DRAM mats} by combining fine-grained DRAM activation with simple modifications to the DRAM I/O circuitry. This enables \prop to implement reduction operations in DRAM without \juani{any intervention of the host CPU cores.}

\subsection{\prop: Hardware Overview}
%\label{sec:idea:hardware}

%\subsection{Fine-Grained Processing-using-DRAM Execution}

\gfi{Figure}~\ref{fig_subarray_matdram} shows an overview of the DRAM organization of \prop. Compared to the baseline Ambit subarray organization, \prop adds four new components (\juani{colored} in green) to a DRAM subarray and DRAM bank, which \juani{enable} 
\li~fine-grained \gfi{\gls{PuD}} execution;
\lii~global I/O data movement; and
\liii~local  I/O data movement. 
%In this section, we first introduce each component at a high-level, highlighting their goal and operation.
% In the following sections, we describe their detailed hardware implementation and timing analysis. 

\paratitle{Fine-Grained \gfi{\gls{PuD}} Execution} To enable fine-grained \gfi{\gls{PuD}} execution, \prop modifies Ambit's subarray and the DRAM bank with three new hardware structures: the \emph{mat isolation transistor}, the \emph{row decoder latch}, and the \emph{mat selector}. 
\gfcrii{At a high level, 
the \emph{mat isolation transistor} allows for the  independent access and operation of DRAM mats within a subarray while 
the \emph{row decoder latch} enables the execution of a \gls{PuD} operation in a range of DRAM mats that the \emph{mat selector} defines.}

First, the \emph{mat isolation transistor} \gfcrii{(\circlediii{i} in Figure~\ref{fig_subarray_matdram})} segments the global wordline connected to the local row decoder in \emph{each} DRAM mat \gfi{in} a subarray. 
Second, the \emph{row decoder latch} \gfcrii{(\circlediii{ii})} stores the bits from the global wordline used to address the local row decoder. 
Third, the \emph{mat selector} \gfcrii{(\circlediii{iii})}, shared across all DRAM mats \gfi{in} a subarray, asserts one or more mat isolation transistors. \gfcrii{The \emph{mat selector} enables the connection between the global wordline and  the \emph{row decoder latches} belonging to a range of DRAM mats. When issuing \gls{PuD} operations, the memory controller specifies the \emph{logical} address of the \emph{first} and \emph{last} DRAM mats that the \gls{PuD} operation targets (called \emph{logical mat range}). 
Internally, each DRAM chip \li~identifies whether \emph{any} of its DRAM mats belong to the logical mat range and 
\lii~translates the logical mat range into the appropriate \emph{physical mat range}, which is used as input for the \emph{mat selector}. 
With these structures, \omi{\emph{different} \gls{PuD}  operations can execute in \emph{different} ranges of DRAM mats.}}  

For example, to execute a \gfi{\gls{TRA}} in \juani{only} $mat_0$, \prop performs \gfi{four} steps:
\li~when issuing a \gfi{\gls{TRA}}, the memory controller sends, alongside the row address information, the \emph{logical mat range} \matrange = \texttt{[\#0,\#0]} to address  $mat_0$ (\circled{1} in \gfi{Figure}~\ref{fig_subarray_matdram});  
\lii~the \emph{mat selector} (\circled{2}) receives the logical mat range, translates it to the appropriate \emph{physical mat range}, and raises the \emph{matline} \juani{corresponding} to $mat_0$, which asserts $mat_0$'s \emph{mat isolation transistor} (\circled{3}) and connects the global wordline to $mat_0$'s row decoder latch;
\liii~the bits of the global wordline used to drive $mat_0$'s local row decoder are stored \juani{in} $mat_0$'s \emph{row decoder latch} (\circled{4});
\liv~finally, $mat_0$'s local row decoder drives the appropriate rows in $mat_0$'s DRAM array based on the value stored \omi{in} the row decoder latch. From here, the DRAM row activation (and thus, \gfi{\gls{PuD}} computation) proceeds as described in \gfi{\cref{sec:background_dramorg}}, only \juani{involving} the DRAM rows in $mat_0$. Since the \omi{per-mat} row decoder latch stores the local row address for a given row activation \omi{in a mat}, the memory controller can issue a \gfi{\gls{TRA}} to another DRAM mat while $mat_0$ is being activated (\circled{5}).
%Therefore, enabling concurrent execution of different triple-row activation commands across different DRAM mats in a \gls{MIMD} fashion.  

\begin{figure}[ht]
    \centering
    \includegraphics[width=0.85\linewidth]{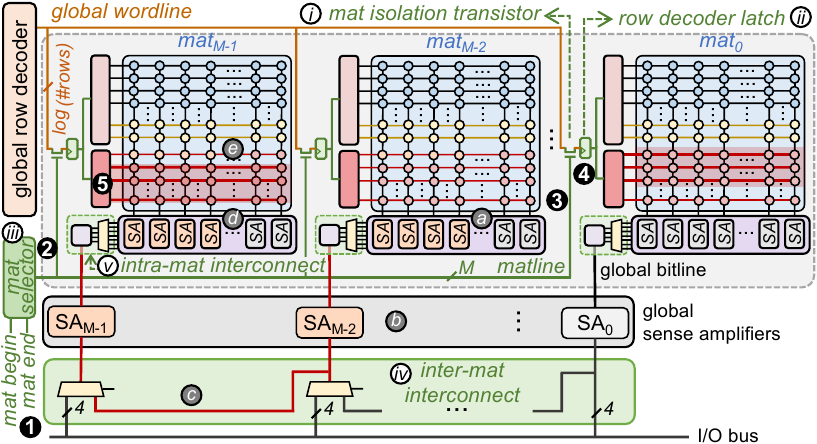}
    \caption{\omiii{\prop subarray and bank organization. {Green-colored boxes represent newly added hardware components.}}}
    \label{fig_subarray_matdram}
\end{figure}

\paratitle{Global I/O Data Movement} To enable data movement across different mats, \prop implements an  \emph{inter-mat \gfcri{interconnect}} by slightly modifying the connection between the I/O bus and the global sense amplifiers \gfcrii{(\circlediii{iv} in Figure~\ref{fig_subarray_matdram})}. The inter-mat \gfcri{interconnect} relies on the observation that the global sense amplifiers \juani{have \emph{higher} drive} than the sense amplifiers in the local row buffer~\cite{keeth2007dram, wang2020figaro}, allowing to directly drive data from the global sense amplifiers into the local row buffer.\footnote{Prior work~\cite{wang2020figaro} leverages the same observation to \omi{copy} DRAM columns from one subarray to another.} \juani{To leverage this observation}, \prop adds a 2:1 multiplexer to the input/output port of each \gfcrii{set of \emph{four} 1-bit  global sense amplifiers}. The multiplexer selects whether the data that is written to the sense amplifier \omi{set} $SA_i$  comes from the I/O bus or from the neighbor global sense amplifier \omi{set} $SA_{i-1}$. 
%Since each sense amplifier $SA_M$ is connected to $mat_M$, the inter-mat network creates a path from $mat_{i-1}$ $\rightarrow$ $SA_{i-1}$ $\rightarrow$  $SA_{i}$ $\rightarrow$ $mat_i$. 

To manage inter-mat data movement, \prop exposes a new DRAM command to the memory controller called \texttt{GB-MOV} (\underline{g}lo\underline{b}al I/O \underline{mo}ve). 
%
%\paratitle{The \underline{G}lo\underline{b}al I/O \underline{Mo}ve Command} 
The \texttt{GB-MOV} command takes as input: 
\li~the logical mat range \matrange, row address, and column address of the \emph{source} DRAM row \gfcrii{and column}; and
\lii~the logical mat range \matrange, row address, and column address of the \emph{destination} DRAM row \gfcrii{and column}. 
With the inter-mat \gfcri{interconnect} and new DRAM command, \prop can move \emph{four} bits\footnote{\gfi{The number of bits the inter-mat \gfcri{interconnect} can move at once depends on the number of \glspl{HFF} \gfcrii{\emph{already present}} in a DRAM mat. We assume that each mat has four \glspl{HFF}, as prior works suggest~\cite{zhang2014half,ha2016improving,oconnor2017fine}.}} of data from a source row \gfcriii{and column (\texttt{row$_{src}$}, \texttt{column$_{src}$}) in $mat_{M-2}$ to a destination row \gfcrii{and column (}\texttt{row$_{dst}$}, \texttt{column$_{dst}$}) in $mat_{M-1}$\gfcrii{, in a DRAM subarray with $M$ DRAM mats}.} Once the memory controller receives a \texttt{GB-MOV} command, it \juani{performs} three steps. 
First, the memory controller issues an \texttt{ACT} to \gfcrii{the source} row in $mat_{M-2}$, which loads the target DRAM \gfcriii{\texttt{row}$_{src}$} to $mat_{M-2}$'s local sense \omi{amplifiers} (\circledii{a} in \gfi{Figure}~\ref{fig_subarray_matdram}). Concurrently, the memory controller issues an \texttt{ACT} to \gfcrii{the destination} row  in $mat_{M-1}$, which connects \gfcriii{\texttt{row}$_{dst}$} to $mat_{M-1}$'s local sense \omi{amplifiers}. 
Second, the memory controller issues a \texttt{RD} with the address of the four-bit \emph{source} column to $mat_{M-2}$. The column select command loads the four-bit \gfcriii{\texttt{column}$_{src}$} from $mat_{M-2}$'s local sense \omi{amplifiers} to its \glspl{HFF}, and $mat_{M-2}$'s \gfcrii{\glspl{HFF} drive} the corresponding \gfcrii{set of four one-bit global sense amplifiers} $SA_{M-2}$ (\circledii{b}).
Third, the memory controller issues a \texttt{WR} \gfcrii{with the address of the four-bit \emph{destination} column to $mat_{M-1}$}. Since the \texttt{WR} corresponds to a \texttt{GB-MOV} command, the multiplexer that connects $mat_{M-1}$'s \glspl{HFF} to the global sense amplifiers takes as input the added datapath coming from $SA_{M-2}$ instead of the conventional datapath coming from the I/O bus  (\circledii{c}). \juani{As a result}, the data stored in $SA_{M-2}$ \omii{is} loaded into $SA_{M-1}$, which in turn drives $mat_{M-1}$'s \glspl{HFF} and local sense amplifiers (\circledii{d}). 
\gfcrii{Once the four-bit column coming from \gfcrii{\texttt{row}$_{src}$} is written into $mat_{M-1}$'s local sense amplifiers, the local sense amplifiers finish the \texttt{WR}  by restoring the local bitlines in $mat_{M-1}$ to \texttt{VDD} or \texttt{GND}, thereby storing the four-bit column coming from \gfcriii{\texttt{column}$_{src}$} as a column of \gfcriii{\texttt{row}$_{dst}$} (\circledii{e}).}

\gfcrii{The \emph{conservative worst-case latency} of a \texttt{GB-MOV} command (i.e., where the addresses of the source and the destination rows differ) is equal to $t_{RAS} + t_{RELOC} + t_{WR} + t_{RP}$; where 
$t_{RAS}$ is latency from the start of row activation until the completion of the DRAM cell's charge restoration,
$t_{RELOC}$~\cite{wang2020figaro} is the latency of turning on the connection between the source and destination local sense amplifiers; 
$t_{WR}$ is the minimum time interval between a \texttt{WR} and a \texttt{PRE} command, which allows the sense amplifiers to restore the data to the DRAM cells;
$t_{RP}$ is the latency between issuing a \texttt{PRE} and when the DRAM bank is ready for a new row activation.}

 %In Section~\ref{sec:design:indramexec}, we discuss the execution latency of a \texttt{GB-MOV} command and how \prop employs it to implement in-DRAM vector reduction. 

\paratitle{Local I/O Data Movement} To enable data movement \omi{across columns} \emph{within} a DRAM mat, \prop implements an  \emph{intra-mat \gfcri{interconnect}} \gfcrii{(\circlediii{v} in Figure~\ref{fig_subarray_matdram})}, which does \emph{not} require any hardware modifications. 
Instead, it modifies the sequence of steps DRAM executes during a column access operation. There are two \emph{key observations} \juani{that enable} the intra-mat \gfcri{interconnect}. 
First, we observe that the local bitlines \gfcrii{of a DRAM mat} \emph{already} share an interconnection path via the \glspl{HFF} and column select logic \gfcrii{(as Figure~\ref{fig_intra_mat} illustrates)}. 
Second, the \glspl{HFF} in a DRAM mat can latch and \emph{amplify} the local row buffer's data~\cite{keeth2007dram,o2021energy}. 
%This happens because the local sense amplifiers have limited drive capability and cannot quickly drive the global bitlines. %Therefore, modern DRAM architectures employ several layers of data amplification to improve signal integrity and DRAM performance. \Based on the two key observations, the intra-mat network operates by latching and amplifying the source data into the mat's \glspl{HFF} and then allowing the \glspl{HFF} to drive the latched data into the destination column. 

\begin{figure}[ht]
    \centering
    \includegraphics[width=0.85\linewidth]{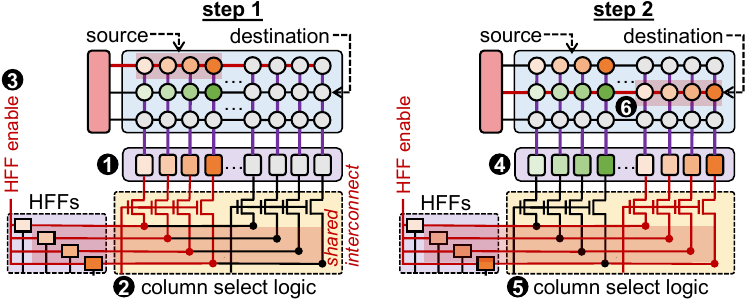}
    \caption{\prop intra-mat \gfcri{interconnect}.}
    \label{fig_intra_mat}
\end{figure}

To manage intra-mat data movement, \prop exposes a new DRAM command to the memory controller called \texttt{LC-MOV} (\underline{l}o\underline{c}al I/O \underline{mo}ve). 
The \texttt{LC-MOV} command takes as input: 
\li~the logical mat range \matrange of the target row, 
\lii~the row and column \gfcrii{addresses} of the \emph{source} DRAM row and column; and
\liii~the row and column \gfcrii{addresses} of the \emph{destination} DRAM row and column. 
With the intra-mat \gfcri{interconnect} and new DRAM command, \prop can move \emph{four} bits of data from a source row \gfcrii{and column (\texttt{row$_{src}$}, \texttt{column$_{src}$}) to a destination row \gfcrii{and column (}\texttt{row$_{dst}$}, \texttt{column$_{dst}$}) in $mat_{M}$.} 
Once the memory controller receives \omi{an} \texttt{LC-MOV} command, it \omi{performs} two steps, which \gfi{Figure}~\ref{fig_intra_mat} illustrates. 
In the \emph{first step}, the memory controller performs an \texttt{ACT}--\texttt{RD}--\texttt{PRE}\revdel{ command sequence} targeting \gfcrii{\texttt{row$_{src}$}, \texttt{column$_{src}$}} in $mat_{M}$. The \texttt{ACT} loads \gfcrii{\texttt{row$_{src}$}} to $mat_{M}$'s local sense amplifier (\circled{1} in \gfi{Figure}~\ref{fig_intra_mat}).
The \texttt{RD} moves \gfcrii{four bits from  \texttt{row$_{src}$}, as indexed by \texttt{column$_{src}$},} into the mat's \glspl{HFF} by \omi{enabling} the appropriate transistors in the column select logic~(\circled{2}). 
The \glspl{HFF} are then enabled by transitioning the \emph{\gls{HFF} enable} signal from low to high. This allows the \glspl{HFF} to \emph{latch} and \emph{amplify} the selected \gfcrii{four-bit} data column from the local sense amplifier~(\circled{3}).
The \texttt{PRE} closes \gfcrii{\texttt{row$_{src}$}}. 
Until here, the \texttt{LC-MOV} command operates exactly as a regular  \texttt{ACT}--\texttt{RD}--\texttt{PRE} command sequence. 
However, differently from a regular \texttt{ACT}--\texttt{RD}--\texttt{PRE}, the \texttt{LC-MOV} command does \emph{not} lower the \emph{\gls{HFF} enable} signal when the \texttt{RD} finishes. This allows the \gfcrii{four-bit} data \gfcrii{from \texttt{column$_{src}$}} to reside in the mat's \glspl{HFF}. 
In the \emph{second step}, the memory controller performs an \texttt{ACT}--\texttt{WR}--\texttt{PRE} targeting \gfcrii{\texttt{row$_{dst}$}, \texttt{column$_{dst}$}} in $mat_{M}$. 
The \texttt{ACT} loads \texttt{row$_{dst}$} into the mat's local row buffer (\circled{4}), and the \texttt{WR} asserts the column select logic to \gfcrii{\texttt{column$_{dst}$}}, creating a path between the \glspl{HFF} and the local row buffer (\circled{5}). Since the \emph{\gls{HFF} enable} signal is kept high, the \glspl{HFF} will \emph{not} sense and latch the data from \gfcrii{\texttt{column$_{dst}$}}. Instead, \gfcrii{the \glspl{HFF} overwrite} the data stored in the local sense amplifier with the previously \gfcrii{four-bit}  data latched from \gfcrii{\gfcrii{\texttt{column$_{src}$}}}. The new data stored in the mat's local sense amplifier propagates through the local bitlines and is written to the destination DRAM cells (\circled{6}). 

\gfcrii{The \emph{conservative worst-case latency} of an \texttt{LC-MOV} command (i.e., where the addresses of the source and the destination rows differ) is equal to $2 \times (t_{RAS} + t_{RP}) + t_{RELOC} + t_{WR}$.}
% Finally, the memory controller issues a \texttt{PRECHARGE} command to prepare the subarray to subsequent requests. 
%In case the \texttt{LC-MOV} command targets columns in the \emph{same} DRAM row \juani{of the same mat}, the memory controller does \emph{not} need to issue the first \texttt{PRE}, reducing the data movement's latency. 

%Note that the \texttt{LC-MOV} command operates by following regular DRAM \texttt{RD} and \texttt{WR} commands. The primary difference is that the \texttt{LC-MOV} command manipulates the behavior of the \glspl{HFF} to enable data to be moved across DRAM columns. 

\subsubsection{\gfi{\gls{PuD}} Vector Reduction} We describe how \prop uses the inter-\omi{mat} and intra-mat \gfcri{interconnects} to implement \gfi{\gls{PuD}} vector reduction. To do so, we use a simple example, where \prop executes a vector addition followed by a vector reduction, i.e., \texttt{out+=(A[i]+B[i])}. We assume that DRAM has \omi{only} \gfcrii{two} mats, and the \gfcrii{data elements of the} input arrays \texttt{A} and \texttt{B} \omi{are evenly distributed} across \gfcrii{the two DRAM mats, as Figure~\ref{fig_vector_reduction} illustrates. \prop executes a vector reduction in three steps.}
%
% \prop executes the \gfi{\gls{PuD}} vector reduction operation in three main steps: 
% \li~map operation (i.e., \texttt{C[i] = A[i] + B[i]});
% \lii~512-element vector reduction operation (i.e., \texttt{tmp[j] += C[i]}); and
% \liii~4-element vector reduction operation (i.e., \texttt{out[3:0] += tmp[j]}). To execute the vector reduction operations, \prop uses the  \texttt{GB-MOV} and \texttt{LC-MOV} to implement an adder tree in DRAM in seven main steps. 
%
\gfcrii{In the first step,} \prop executes a \gls{PuD} addition operation over the data in  \gfcrii{the two DRAM mats (\circled{1})}, storing the temporary output data \texttt{C} into the same mats \omi{where} the computation takes place (i.e., \omi{\texttt{C} = \texttt{\{}\texttt{C[9:5]}$_{mat1}$, \texttt{\omii{C[4:0]}}$_{mat0}$\texttt{\}}}). 
\gfcrii{In the second step}, \prop issues a \texttt{GB-MOV} to move \omii{part} of the temporary output \texttt{C[4:0]} stored in \gfcrii{$mat_{0}$ to a temporary row \texttt{tmp} in $mat_{1}$} (\omi{\texttt{tmp[9:5]}$_{mat1}$$\leftarrow$\texttt{C[4:0]}$_{mat0}$}) \gfcrii{via the inter-mat interconnect (\circled{2}--\circled{3}})\gfcrii{,  \omii{four bits} (i.e., four data elements) \omii{at a} time. \prop \emph{iteratively} executes step 2 until \emph{all} data elements of \texttt{C[4:0]} are copied to $mat_{1}$}.   
% Third, %once the \texttt{GB-MOV} command finishes executing, 
% \prop issues a second \texttt{GB-MOV} to move the portion of the temporary output stored in $mat_{0}$ to $mat_{1}$ ($mat_{1}$  $\leftarrow$ \texttt{C[0]}$_{mat0}$). 
% Fourth, \prop concurrently executes addition operations in $mat_3$ and $mat_1$, which computes \texttt{C[3] + C[2]} in $mat_3$ and \texttt{C[1] + C[0]} in  $mat_1$.   
% Fifth, \prop issues a \texttt{GB-MOV} to move the produced temporary output of \texttt{C[1] + C[0]}, from $mat_1$ to $mat_3$.
\gfcrii{In the third step, once the \texttt{GB-MOV} finishes, \prop executes the final addition operation, i.e. \texttt{tmp} + \texttt{C[1]}, in $mat_1$. The \gfcrii{final} output of the \gfcrii{vector reduction operation} is stored in the destination row \texttt{out} in $mat_1$ (\circled{4}).} 
Once the vector reduction operation finishes, the temporary output array stored in \gfcrii{$mat_1$} holds as many \gfcrii{data} elements as the number of DRAM columns in a mat (\gfcrii{e.g.}, 512 \gfcrii{data} elements). 
\prop allows reducing the temporary output vector further to an output vector with \gfcrii{\emph{four} \gfcrii{data} elements} using the intra-mat \gfcri{interconnect}. The process is analogous to that employed during the 512\omii{-element} vector reduction: \prop uses the intra-mat \gfcri{interconnect} and the \texttt{LC-MOV} command to implement an adder tree inside a single DRAM mat.\footnote{\gfi{The number of \texttt{GB-MOV} and \texttt{LC-MOV} commands \omii{issued depends on} the bit-precision of the input operands~\omii{\cite{hajinazarsimdram}}.}}  

 \begin{figure}[ht]
     \centering
     \includegraphics[width=\linewidth]{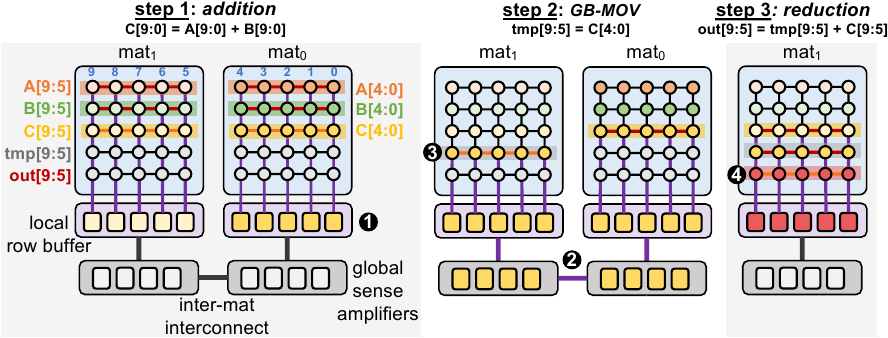}
    \caption{\gfcriii{An example of a \gls{PuD} vector reduction in \prop.}}
     \label{fig_vector_reduction}
 \end{figure}

%\subsubsection{Timing Analysis}

%\subsection{Intra-MAT Communication}

%\subsubsection{Timing Analysis}

% \textcolor{red}{\noindent\rule{8.5cm}{2pt}}
% \begin{center}
%     \vspace{-12pt}
%     \textcolor{red}{\textbf{STALE TEXT BEGIN.}}
% %    \vspace{-15pt}
% \end{center}

\subsection{\prop: \gfcrii{Control \& Execution}}

\gfhpca{To \gfcrii{enable} \prop to execute in a \gls{MIMD} fashion, we need to efficiently 
\li~\gfcrii{\emph{encode} and \omi{\emph{communicate}} information regarding the target DRAM mats (i.e., the target mat range) in a \emph{timely} manner (i.e., respecting DRAM timing parameters) while
\lii~\emph{orchestrating} the execution of independent \gls{PuD} operations across the DRAM mats of a DRAM subarray.
To do so, we take a \emph{conservative} design approach: we aim to integrate \prop in commodity DRAM chips by providing an implementation 
\li~\emph{compatible} with existing DRAM standards and 
\lii~\gfcriii{that does \emph{not} add new pins to a DRAM chips.}}}

\paratitle{Encoding MAT Information} \gfhpca{\prop needs a compact way to encode the target mat information, since a DRAM module often contains many DRAM mats. To solve this issue, \prop \omi{only allows} a \gls{PuD} \gfcrii{operation} \omi{to be executed in a \emph{physically contiguous} set of DRAM mats}.\footnote{In \cref{sec:operating:system}, we describe how we enforce \gfcrii{physically contiguous} mat allocation.} 
In this way, \gfcrii{when executing the DRAM commands (i.e., \texttt{ACT}s and \texttt{PRE}s) that realize a \gls{PuD} operation,} the memory controller only needs to \omi{provide} the \emph{first} and \emph{last} (\emph{logical}) mats an \texttt{ACT} target. 
Then, \prop internally decides which (\emph{physical}) mats fit into the \omi{provided} mat range. 
To do so, \prop implements a simple \emph{chip select logic} and \emph{mat identifier logic} inside the I/O circuitry of each DRAM chip. The \emph{chip select logic} and \emph{mat identifier logic} take as input the \emph{logical mat range} and \omi{output} 
\li~if DRAM mats placed in a chip belong to the mat range, and 
\lii~the physical mat range. 
\gfcrii{\omiii{In case a DRAM mat placed in a chip belongs to the mat range,} the DRAM chip queues the physical mat range in the \emph{mat queue} (which we describe later in this section). 
The \emph{physical mat range} is used as input for the \emph{mat selector} (see Figure~\ref{fig_subarray_matdram}).}
Since there are up to 128 DRAM mats in a DDR4 module~\cite{lee2021greendimm}, \prop uses \omii{14~bits} to encode the logical mat range \gfcrii{(\omii{7~bits each} for \emph{mat begin} and \emph{mat end}, each)}\gfcrii{, from which 
\li~the three most significant bits are used to identify the target DRAM chip and 
\lii~the four \omii{least} significant bits are used to identify individual mats.} The \emph{chip select logic} and \emph{mat identifier logic} comprise simple hardware elements: four \gf{comparators}, two \omi{2-input} AND gates, two 2:1 multiplexers, and a \omii{3-bit} \emph{chip id register} in each DRAM chip. }

\paratitle{\omi{Communicating} MAT Information} 
\prop needs to \omi{communicate} to the DRAM \gfcrii{chip} information regarding the target mats during a \gfi{\gls{PuD}} operation. 
However, it is challenging to \omi{communicate} the mat information alongside an \texttt{ACT} due to the narrow DRAM command/address (C/A) bus interface, since the memory controller uses most of the available pins during a row activation for row address and command \omi{communication}.\footnote{There are 27 C/A pins in a DDR4 chip~\cite{jedec2017jedec}, from which only three pins are \emph{not} used during an \texttt{ACT} command.} 
Our \emph{key idea} to solve this issue is to overlap the latency of \omi{communicating} the mat information to DRAM with the latency of DRAM commands in a \uprog in two ways: 
\li~\texttt{ACT}--\texttt{ACT} overlap, and 
\lii~\texttt{PRE}--\texttt{ACT} overlap.
The first case (\texttt{ACT}--\texttt{ACT} overlap) happens when issuing a row copy operation (\texttt{AAP}). 
In this case, the mat information required by the second \texttt{ACT} is transmitted immediately \emph{after} issuing the first \texttt{ACT}, exploiting the delay between \omi{two} activations. The mat information is buffered once it reaches DRAM.
The second case (\texttt{PRE}--\texttt{ACT} overlap) happens when issuing the first \texttt{ACT} in a row copy operation or the \texttt{ACT} in a \gls{TRA}. 
We notice that
\li~\gfcrii{the first \texttt{ACT}  command in an \texttt{AAP}/\texttt{AP}} is \emph{always} preceded by a \texttt{PRE} (due to a previous \texttt{AAP}/\texttt{AP}, or due to a previous DRAM request), and 
\lii~a \texttt{PRE} does \emph{not} use the row address pins, since it targets a DRAM bank (not a DRAM row). \gfcrii{Thus, \prop uses the row address pins during a \texttt{PRE} \gfcrii{that immediately precedes the first \texttt{ACT} in an \texttt{AAP}/\texttt{AP} command sequence to communicate the mat information}.\footnote{\rA{\changerA{\rA{\#A2}}If there are insufficient pins in the DDRx interface to communicate mat information (e.g., as in DDR5~\cite{jedec2020jesd795}), \prop utilizes multiple DRAM C/A cycles to propagate the mat information. For example, in DDR5, \prop still performs \texttt{PRE-ACT} overlap, communicating the mat information in two cycles. Note that \omi{an} extra cycle does \emph{not} impact \prop's performance, since in a \texttt{PRE-ACT} command sequence, the \texttt{PRE} still needs to wait for the completion of the \texttt{ACT} for more than two DRAM C/A cycles.}}} 

\paratitle{Timing \omi{of} MAT Information} \gfhpca{\prop needs to \gfcrii{communicate} the mat information \emph{before} a respective \texttt{ACT} in a \uprog{}. \gfcrii{Communicating} the mat information immediately \emph{after} the memory controller issues the \texttt{ACT} would open the \emph{entire} DRAM row (instead of only the relevant portion of the DRAM row). To solve this issue, we devise a simple \gfi{queuing}-based mechanism for partial row activation. 
Our mechanism relies on the fact that the \gfcrii{execution} order \gfcrii{of} \texttt{ACT}s and \texttt{PRE}s \gfcrii{in} a \uprog is \emph{deterministic}.\footnote{To realize a \gls{PuD} \omii{operation}, the memory controller \emph{must} respect the order in which \texttt{ACT} and \texttt{PRE} commands are specified in the \uprog{}. Therefore, during \gls{PuD} execution, \texttt{ACT}s and \texttt{PRE}s in a \uprog{} cannot be reordered, and the behavior of the \uprog{} is thus deterministic. If the memory controller is \omi{performing} maintenance operation to a DRAM bank, the \texttt{AAP}/\texttt{AP} commands of a \gls{PuD} operation wait until the maintenance operation finishes.} Thus, we can add to each DRAM command in \gfcrii{an \texttt{AAP}/\texttt{AP}} the information about when the  DRAM circuitry should propagate the mat information. 
\revdel{The memory controller enforces that \glspl{TRA} from different \uprogs \emph{are} overlapped, but \glspl{TRA} from the \emph{same} \uprog \emph{are serialized}. }\prop leverages \gfcrii{this} key idea by \gfcrii{adding} a \emph{mat \gfi{queue}} to the I/O logic of each DRAM chip and adding extra \omi{functionality} to the existing \texttt{ACT} and \texttt{PRE} commands to control the mat queue:
\li~\texttt{ACT-enqueue} issues an \texttt{ACT} to \texttt{row\_addr} in the first DRAM clock cycle and enqueues [\texttt{mat\_begin,mat\_end}] in the second DRAM clock cycle; 
\lii~\texttt{PRE-enqueue} issues a \texttt{PRE} to \texttt{bank\_\omi{id}} and enqueues [\texttt{mat\_begin,mat\_end}];
\liii~\texttt{ACT-dequeue} issues an \texttt{ACT} to \texttt{row\_addr} and dequeues from the mat queue.  }

%\vspace{-15pt}
\paratitle{Orchestrating MAT Information} \prop needs to execute different \gfi{\gls{PuD}} \gfcrii{operations} concurrently. To this end, we implement a control unit inside the memory controller \gfcrii{on the CPU die}, which Figure~\ref{fig_control_unit} illustrates. 
\prop leverages SIMDRAM control unit to translate \omi{each} \emph{bbop} \omi{instruction} into \omi{its corresponding} \uprog{} and adds extra circuitry to 
\li~schedule \omi{each \uprog{}} based on \omi{its} target \omi{mats} and
\lii~maintain multiple \uprog contexts. 
\gfcrii{\prop control unit consists of \gf{four} main components.
First, \emph{bbop buffer}, which stores \emph{bbops} dispatched by the host CPU.
Second,  \emph{mat scheduler}, which schedules the most appropriate \emph{bbop} to execute depending on the \emph{bbop}'s mat range and current mat utilization. 
Third, \emph{mat scoreboard}, which tracks whether a given mat is being \omii{used} by a \emph{bbop} instruction. The \emph{mat scoreboard} stores \omii{an} $M$-bit \emph{mat bitmap} \omii{that keeps track of which mats are \omiii{currently} in use}, where $M$ is the number of mats in the DRAM module. 
The \emph{mat scoreboard} can index a range of positions in the \emph{mat bitmap} using a \emph{mat index}.
Fourth, several \gfi{(\omii{e.g.}, eight)} \emph{\gf{\uprog} processing engines}, \omii{each of} which \omii{translates} a \emph{bbop} into its respective \gf{\uprog} and \omiii{controls} \omii{the \emph{bbop}'s} execution. \revdel{A \gf{\uprog} processing engine is equivalent to \omii{a} single SIMDRAM control unit.}
}

\begin{figure}[!ht]
    \centering
    \includegraphics[width=0.55\linewidth]{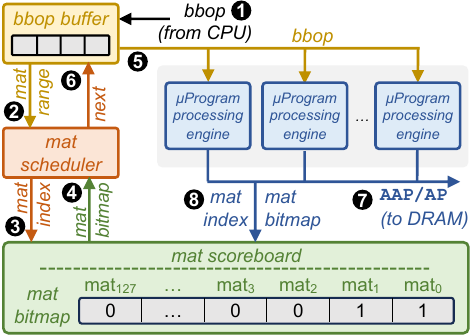}
    \caption{\prop control unit \omi{in the memory controller}.}
    \label{fig_control_unit}
\end{figure}

\gfcrii{\prop control unit works in four steps.} In the first step, \prop control unit enqueues an incoming \emph{bbop} instruction dispatched by the host CPU (\circled{1} in Figure~\ref{fig_control_unit}) in the \emph{bbop} buffer. 
In the second step, the mat scheduler scans the \emph{bbop} buffer from the oldest to the newest element. Then, the mat scheduler employs an online first fit algorithm~\cite{garey1972worst} to select a \emph{bbop} to be executed. For each \emph{bbop} in the \emph{bbop} buffer, the algorithm:
\li~extracts the mat range information encoded in the \emph{bbop} (\circled{2}), which is used to index the \emph{mat scoreboard} (\circled{3});
\lii~reads the mat bitmap to identify whether the mats belonging to the \emph{bbop}'s mat range are currently free or busy (\circled{4});
\liii~in case the mats are free, the mat scheduler writes a new mat bitmap to the mat scoreboard, indicating that the given mat range is now busy, selects the current \emph{bbop} to be executed by allocating \gfcrii{and copying the \bbop to} a free \uprog processing engine \gfcrii{(\circled{5})}, and \omi{removes} the current \emph{bbop} from the \emph{bbop} buffer (\circled{6});
\liv~in case the mats \gfcrii{belonging to the \bbop's range} are busy, the mat scheduler reads the next available \emph{bbop} from the \emph{bbop} buffer and repeats \li--\liii.
In the third step, one or multiple \uprog processing engines execute their allocated \emph{bbop}, issuing \texttt{AAP}s/\texttt{AP}s to \gfcrii{the DRAM chips (\circled{7})}\revdel{, one \uprog processing engine per cycle}. \rA{The \uprog processing engine is \changerA{\rA{\#A4}}responsible for maintaining the timing of \texttt{AAP}/\texttt{AP}  commands\revdel{(i.e., the timing delay between \texttt{ACT} and \texttt{PRE})}. In our design, we avoid the need to maintain state for \emph{all} DRAM mats in a DRAM module \emph{individually} by: 
\li~only allowing a \gls{PuD} \gfcrii{operation} to address a contiguous range of DRAM mats, \omi{which share state as} they execute the same sequence of \texttt{ACT-PRE} \omii{commands} and 
\lii~limiting the number of concurrent \gls{PuD} \gfcrii{operations} to the number of \uprog processing engines available in the control unit.} In the fourth step, when a \uprog processing engine finishes executing, it frees \omi{its allocated} mats by \omi{correspondingly updating the mat bitmap in the} mat scoreboard (\circled{8}) \omi{and notifies the CPU that the execution of the \bbop instruction is done}.

\section{\prop: Software Support}
\label{sec:idea:software}

%\prop provides a hardware substrate that exploits data-level and instruction-level parallelism from applications. However, manually identifying, transforming, and scheduling computation to \prop can be challenging. Thereby, 
To ease \prop's programmability, we provide compiler support to transparently map \gls{SIMD} operations to \prop. 
\gfi{Figure}~\ref{fig_compilation_flow} illustrates \gf{\prop's} compilation flow, which we implement using LLVM~\cite{lattner2008llvm}\omi{: we take} a C/C++ application's source code as input, 
\omi{perform} \gf{three} transformations passes, and  \omi{output} a binary with a mix of CPU and \gfcrii{\gfi{\gls{PuD}} instructions.} \revdel{\rC{\changerC{\rC{\#C2}}We implement {two} transformations and analysis passes at the middle-end, and one at the back-end of the LLVM's compiler toolchain.}}  

\begin{figure}[ht]
    \centering
    \includegraphics[width=\linewidth]{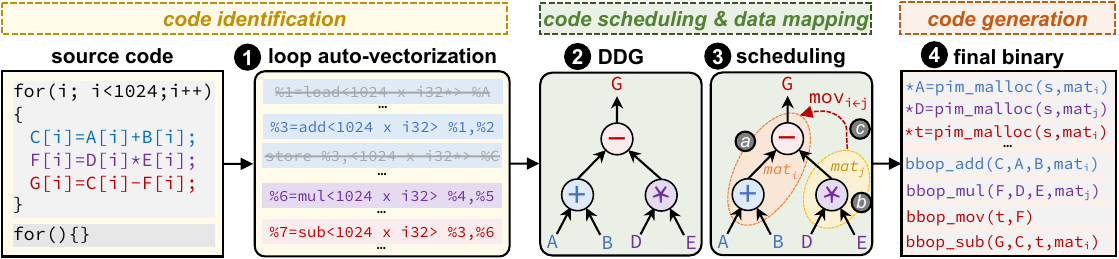}
    \caption{\revA{\prop's compilation flow.}}
    \label{fig_compilation_flow}
\end{figure}

\paratitle{Pass 1: Code Identification} The first pass is responsible for \emph{code identification}. Its goal is to identify 
\li~loops that can be successfully auto-vectorized and 
\lii~the appropriate vectorization factor of a given vectorized loop. The code identification pass takes as input the application's LLVM intermediate representation (IR) generated by the compiler's front-end. It produces as output an optimized IR containing \gls{SIMD} instructions that will be translated to \emph{bbop} instructions. 
We leverage the native LLVM's loop auto-vectorization pass~\cite{AutoVect65:online} to identify and transform loops into their vectorized form \revA{(\circled{1} in \gfi{Figure}~\ref{fig_compilation_flow})}.\footnote{Prior works~\cite{ahmed2019compiler,devic2022pim} also \omi{leverage} modern \omi{compilers'} loop auto-vectorization engines to generate instructions to \gfcrii{processing-near-memory (\gls{PnM}) architectures equipped with \gls{SIMD} engines.}} We apply two modifications to LLVM's loop auto-vectorization pass. 
First, \revdel{we modify how the loop auto-vectorization pass selects the best-performing vectorization factor for a  loop. I}instead of using a cost model to choose the vectorization factor that leads to the highest performance improvement compared to a scalar version of the same loop, we \emph{always} select the \emph{maximum} vectorization factor for the loop. \rC{\changerC{\rC{\#C2}}This is important because the native cost model takes into account the hardware characteristics of \omi{a} target CPU \gls{SIMD} engine (i.e., number of available vector registers, SIMD width of the target execution engine, the latency of different \gls{SIMD} instructions), which are not representative of \omi{our \prop} engine with a variable \gls{SIMD} width.}
Second, \rC{we modify the code generation routine for a given vectorized loop. Concretely, for a given vectorized loop, }we identify and remove memory instructions related to {each} arithmetic \gls{SIMD} operation (i.e., load/store instructions that manipulate vector registers)\revdel{. This optimization is \juani{necessary} } since \gfi{\gls{PuD}} operations directly manipulate the data stored in DRAM\rC{; \changerC{\rC{\#C2}}thus, there is no need to \emph{explicitly} move data into/out \gls{SIMD} registers}. %

\paratitle{Pass 2: Code Scheduling \& Data Mapping} The \gf{second} pass is responsible for \emph{code scheduling and data mapping}. Its goal is to improve overall SIMD utilization by allowing the distribution of independent \gfi{\gls{PuD}} instructions across DRAM mats. \revdel{The key idea behind our code scheduling pass is to distribute \emph{bbop} instructions across DRAM mats based on the \emph{data dependency} of operands.}
Since \gls{PuD} instructions operate \changerC{\rC{\#C2}}directly on the data stored in DRAM, the DRAM mat where the data is allocated determines the efficiency and utilization of the \gls{PuD} SIMD engine. If operands of \omi{independent} instructions are distributed across different DRAM mats, such instructions can be executed concurrently. Likewise, operands of dependent instructions are mapped to the same DRAM mat. 
In that case, intermediate data that one instruction produces and the next instruction consumes do \emph{not} need to be moved across different DRAM mats, improving energy efficiency. \omi{Leveraging these observations,} the code scheduling pass takes as input all \emph{bbop} instructions the code identification pass generates and outputs new \emph{bbop} instructions containing metadata regarding \gf{their} mat location (i.e., \emph{mat label}). The code scheduling pass works in two steps.

In the first step, the code scheduling pass creates a data-dependency graph (DDG) \revA{of the vectorized instructions} \revA{(\circled{2})}. \rC{\changerC{\rC{\#C2}}Each node represents a \emph{bbop} instruction, incoming edges represent input, and outgoing edges represent output of the \emph{bbop}.} In the second step, the code scheduling pass
takes as input the DDG and employs a data scheduling algorithm to distribute \emph{bbop} instructions across DRAM mats \revA{(\circled{3})}. The data scheduling algorithm traverses the DDG \gfcrii{in \emph{topological order} to respect dependencies between \emph{bbop} instructions} using a depth-first search (DFS) kernel\gfcrii{, which is a common algorithm for topological ordering~\cite{cormen2022introduction,tarjan1976edge},} and performs \gf{three} operations. 
First, the algorithm traverses the \emph{left} nodes in the DDG, assigning a single \emph{mat label i} to nodes in the \emph{left} path \revA{(\circled{3}-\circledii{a})}. \revdel{This guarantees that instructions with data dependency reside in the same DRAM mat.}
Second, when the algorithm reaches a leaf node, it traverses the \emph{right} sub-tree in the DDG. In this case, the algorithm assigns a new \emph{mat label j} to the nodes in the \emph{right} path in the sub-tree \revA{(\circled{3}-\circledii{b})}.
Third, once the algorithm visits all the nodes in the \emph{right} sub-tree, it returns to the parent node of the sub-tree. Since the parent node has already been visited when descending into the left path, the left and right sub-tree nodes will be assigned to different mats while having data dependencies across them (as indicated by the parent node). In this case, the algorithm creates a data movement  \emph{bbop} instruction (see \gfi{\cref{sec:design:isa}}) to move the output produced by the right sub-tree from \emph{mat label j} to \emph{mat label i} \revA{(\circled{3}-\circledii{c})}. 
\rC{This process \changerC{\rC{\#C2}}repeats until the algorithm visits all nodes in the DDG.}

% backtrand
% \lii~different mat IDs to all edges and nodes of different connect components, which guarantees that independent in-DRAM \emph{bbop} vector instructions are executed concurrently in different DRAM mats.  
% -> malloc with virtual mat id malloc (size, virtual\_mat\_id, vectorization\_size).
% -> size comes from the data structure 
% -> vectorization\_size = total number of mats that will be required, comes from step 1.
% -> virtual\_mat\_id = indicates grouping. Produced by the data scheduling algorithms.
% -> algorithm has three steps.
% 1: identify connected components 
% 2: nodes within the same connect component gets assign the same  virtual mat id 
% 3: nodes across different connect components get assigned different virtual mat ids. 
% * Pim malloc informs the operating system about the allocation need. The OS then:
% 1. allocates the data accordling (we discuss how this can be implemnted later)
% 2. register the virtual mat id and the assgined mat range to the memory controller (see x). 

\paratitle{Pass \gf{3}: Code Generation}  The \gf{third} pass is responsible for 
\li~\emph{data allocation} and \lii~\emph{code generation}. It takes as input the LLVM IR containing both CPU and \emph{bbop} instructions \omi{(with metadata)} and produces a binary to the target ISA \revA{(\circled{4})}. 
To implement data allocation, the code generation pass first identifies calls for memory allocation routines (e.g., \texttt{malloc}) associated with  operands of \emph{bbop}s and replaces such memory allocation routines with a specialized \gls{PIM} memory allocation routine (i.e., \texttt{pim\_malloc}, see \gfi{\cref{sec:operating:system}}). \texttt{pim\_malloc} receives as input the \emph{mat label} assigned to its associated \emph{bbop} instruction. 
Second, the pass inserts \omi{a} \texttt{bbop\_trsp\_init} \omi{instruction} right after \omi{each} \texttt{pim\_malloc} \omi{call} for each memory object that is an input/output of \gfcrii{a} \emph{bbop} instruction. This instruction registers the memory object in \prop's transposition unit (\gfi{\cref{sec:exc:control}}). Similar to the \texttt{pim\_malloc} call, the \texttt{bbop\_trsp\_init} instruction receives as input the \emph{mat label} assigned to its associated \emph{bbop} instruction. To implement code generation, we modify LLVM's X86 back-end to identify \emph{bbop} instructions and generate the appropriate \gfcrii{assembly code}. \gfhpca{In case the application uses parallel primitives (e.g., OpenMP pragmas~\cite{dagum1998openmp}) to parallelize outermost loops, the code generation pass interacts with the underlying runtime system to statically distribute \emph{bbop} instructions from innermost loops across the  available DRAM mats  in a subarray, i.e., mats \gfcrii{with unassigned \emph{mat labels}}. 
\rC{\changerC{\rC{\#C2}}This allows \prop to execute in a \gfcrii{\gls{SIMT}}~\omi{\cite{lindholm2008nvidia,nvidia2009nvidia}} fashion for manually parallelized applications.} }

% \subsection{Putting All Together: MAT-Based Computing}
% \label{sec:idea:hardware}

% - MAT-based computing enables a fine-grained multiple-instruction multiple-data processing-using-DRAM substrate
% - Each MAT becomes an independent SIMD engine 
% -> Enables different operations to be mapped to a single DRAM subarray 
% -> Enables fine-grained SIMD parallelism inside DRAM
% - Enables new operations: vector reduction 

\section{System Support \omi{for \prop}}
\label{sec:system:integration}

We envision \prop as a tightly-coupled accelerator for the host processor. As such, \prop relies on the host processor for \omi{its system integration}, \omi{which includes} ISA support \gfcrii{(\cref{sec:design:isa})}, 
\gfcrii{instruction execution \& data transposition} \gfcrii{(\cref{sec:exc:control})}, and 
\gfcrii{operating system support for address translation and data allocation \& alignment} \gfcrii{(\cref{sec:operating:system}).}

\subsection{Instruction-Set Architecture}
\label{sec:design:isa}

Table~\ref{table_isa_format} shows the CPU ISA extensions that \prop exposes to the compiler.\footnote{\changeE{\#E3}\revE{\prop ISA extensions are vector-oriented by design. We did \emph{not} use \omii{an} \omi{existing} ISA because we needed to define new fields for \prop that do \emph{not} exist in current vector ISAs (e.g., mat label information). \rD{\changerD{\rD{\#D6}}Instead, we propose to extend the baseline CPU ISA with \prop instructions since there is \omi{usually} more than enough unused opcode space to support the extra opcodes that \prop requires~\cite{lopes2013isa, lopes2015shrink}. Extending the CPU ISA to interface with accelerators is a common approach~\gfcrii{\cite{hajinazarsimdram, PEI, seshadri2017ambit, doblas2023gmx,razdan1994high}}}.}} There are five types of instructions: 
\li~object initialization instructions, 
\lii~1-input arithmetic instructions, 
\liii~2-input arithmetic instructions, 
\liv~predication instructions, and
\lv~data movement instructions. 
The first three types of \prop instructions are inherited from \omi{the} SIMDRAM ISA~\omi{\cite{hajinazarsimdram}}. 
\gfcrii{These instructions can be further divided into two categories:
\li~operations with one input operand (e.g., bitcount, ReLU), 
and 
\lii~operations with two input operands (e.g., addition, division, equal, maximum). 
To enable predication, \prop uses the \texttt{bbop\_if\_else} instruction \gfcriii{that SIMDRAM introduces}, \omii{which takes as input three operands: two input arrays (\texttt{src$_1$} and \texttt{src$_2$}) and \omiii{one} predicate array (\texttt{select}).}} We modify such instructions by including two new fields:
\li~\emph{mat label} (ML), which identifies groups of instructions that must execute inside the same DRAM mat, and
\lii~\emph{vectorization factor} (VF), which dictates how many scalar operands are packed within the vector instruction. These two new fields are automatically generated by \prop's compiler passes (\gfi{\cref{sec:idea:software}}).

\begin{table}[ht]
\centering
\tempcommand{0.8}
\vspace{5pt}
\caption{\prop ISA extensions.}
\vspace{-5pt}
\label{table_isa_format}
\resizebox{0.7\linewidth}{!}{
    \begin{tabular}{@{}ll@{}}
    \toprule
    \textbf{Type}                & \multicolumn{1}{c}{\textbf{ISA Format}}    \\ \midrule
    Initialization & \texttt{bbop\_trsp\_init addr, size, n, ML}                         \\
    1-Input Arith.           & \texttt{bbop\_op dst, src, size, n, ML, VF}                                \\
    2-Input Arith.           & \texttt{bbop\_op dst, src$_1$, src$_2$ size, n, ML, VF}               \\
    Predication                  & \texttt{bbop\_if\_else dst, src$_1$, src$_2$, sel, size, n, ML, VF} \\ 
    Data Move & \texttt{bbop\_mov dst, dst\_idx, src, src\_idx, size, n} \\ 
    \bottomrule
    \end{tabular}
}
%\vspace{-5pt}
\end{table}

% In such instructions, \texttt{bbop\_op} represents the opcode of the \prop operation,
% \texttt{src} and \texttt{dst} represent source and destination \emph{arrays}; \texttt{size} represents the number of elements in the source and destination arrays; \texttt{n} represents the number of bits in each array element; and \texttt{sel} represents the
% predicate array.

%\prop supports all 16 operations SIMDRAM proposes.
Data movement instructions allow the compiler to trigger inter-\omi{mat} and intra-mat data movement operations. In a data movement instruction, \texttt{dst} and \texttt{src} represent the source and destination \emph{arrays}; \texttt{dst\_idx} and \texttt{src\_idx} represent the first position of the first element inside the source and destination arrays to be moved; \texttt{size} represents the number of elements to move from source to the destination array; \texttt{n} represents the number of bits in each array element. \prop control unit automatically identifies the \emph{mat range} the data movement instruction targets by calculating the distance between the source and destination arrays, taking into account the indexes and number of elements to move. In case the source and destination mats are the same, \prop control unit translates the data movement instruction into \omi{an} \texttt{LC-MOV} command\omi{:} otherwise, \omi{a \texttt{GB-MOV}} command.

\subsection{Execution \& Data Transposition}
\label{sec:exc:control}

\paratitle{Instruction Fetch and Dispatch}
\prop relies on the host CPU to offload \emph{bbop} instructions to DRAM since they are part of the CPU ISA. Assuming that the host CPU consists of one or more out-of-order cores, \prop leverages the host processor's front-end to 
\li~identify and 
\lii~dispatch to \prop control unit \emph{only} \omi{independent} \emph{bbop}s. This simplifies the design of \prop control unit since no in-flight \emph{bbop} instructions will have data dependencies. As a result, \prop control unit can freely schedule \uprogs to the \gfi{\gls{PuD}} \gls{SIMD} engine as they arrive.  

\paratitle{\gfcrii{Data Coherence}} \gfcrii{Input arrays to \prop may be generated or modified by the CPU, and the data updates may reside only in the cache (e.g., because the updates have not yet been written back to DRAM). To ensure that \prop does not operate on stale data, programmers are responsible for flushing cache lines~\cite{guide2016intel, manual2010arm} modified by the CPU. \prop can leverage coherence optimizations tailored to \gls{PIM} to improve overall performance~\cite{lazypim,boroumand2019conda}.}

\paratitle{\prop Transposition Unit}
\prop transposition unit shares the same hardware components and functionalities as \omi{the} SIMDRAM transposition unit~\omi{\cite{hajinazarsimdram}}, which includes:
\li~\emph{object tracker}, a small cache that keeps track of the memory objects used by \emph{bbop} instructions;
\lii~an \emph{horizontal to vertical transpose} unit, which converts cache lines of memory objects stored in the object tracker from a horizontal to vertical data layout during \omi{a} \gls{LLC} writeback; 
\liii~a \emph{vertical to horizontal transpose} unit, which converts cache lines of memory objects stored in the object tracker from a vertical to horizontal data layout  during an LLC read request;
\liv~\emph{store} and \emph{fetch} units, which generate memory read/write requests using the transpose units' output data. 
One main limitation of \omi{the} SIMDRAM transposition unit is that it needs to fill \emph{at least} an entire DRAM row with vertically-\omi{laid out} data before the execution of a \emph{bbop}. \revdel{In the case number of operands the \emph{bbop} uses is lower than the DRAM row size, SIMDRAM transposition unit fills the remaining space in the DRAM row with `0's. This issue worsens when considering that a vertically laid-out $n$-bit operand spans $n$ different cache lines in DRAM (with each cache line in a different DRAM row). However, since \prop enables fine-grained DRAM activation, \prop transposition unit does \emph{not} need to fill the entire DRAM row with vertically-layout data.} Instead, \prop transposes \omi{only} as much data as required to fill the segment of the DRAM row that the \emph{bbop} instruction operates over. To do so, \omi{the} \prop transposition unit adds information regarding the mat range a memory object operates to the object tracker. 

\subsection{Operating System Support}
\label{sec:operating:system}

\paratitle{Address Translation} 
 As SIMDRAM, \prop operates directly on physical addresses. When the CPU issues a \emph{bbop} instruction, the instruction's virtual memory addresses are translated into their corresponding physical addresses using the same translation lookaside buffer (TLB) lookup mechanisms used by regular load/store operations. 
 
\paratitle{Data Allocation \& Alignment}
\revCommon{\changeCM{\#CQ1}\prop (as other \omi{\gls{PuD}} architectures~\gfcrii{\cite{seshadri2013rowclone, seshadri2018rowclone,
ferreira2022pluto,
seshadri2017ambit,seshadri2019dram,seshadri2015fast,seshadri.bookchapter17,seshadri2016buddy,seshadri2016processing,olgun2022pidram}}) requires OS support to guarantee that data is properly mapped and aligned within the boundaries of the \omi{bank/subarray/mat} that will perform computation. Particularly,} since \gfi{\gls{PuD}} \gfcrii{operations} are executed in-situ, it is essential to enforce that memory objects belonging to the same \emph{bbop} (and their \omi{dependent instructions}) are placed together in the same DRAM mats.
%Otherwise, \prop must execute costly data copy operations to store all memory objects of a given \emph{bbop} inside the same set of DRAM mats before execution. 
%
To achieve this functionality, we propose the implementation of a new data allocation API called \texttt{pim\_malloc}. The main idea is to allow the compiler to inform the OS memory allocator about the memory objects that must be allocated inside the same set of DRAM mats. The \texttt{pim\_malloc} API takes as input the \emph{size} of the memory region to allocate (as a regular \texttt{malloc} instruction) and the \emph{mat label} that the compiler generates (\gfi{\cref{sec:idea:software}}). Then, it  ensures that \emph{all} memory objects with the same \emph{mat label} are \omi{placed together} within \gfcrii{a set of} DRAM mats \gfcrii{that satisfies the target memory allocation size}. \revdel{In case there is no available space within the target mat due to previously allocated data that is not associated with a \emph{bbop} instruction, the \texttt{pim\_malloc} triggers a page fault operation to move one or more pages \revCommon{last-recently used} out of the target mat, freeing space in the target mat for the \texttt{pim\_malloc} to complete.\footnote{\revCommon{\changeCM{\#CQ1}Our analysis assumes that the mat is free whenever a mat is allocated.}} }
%The \texttt{pim\_malloc} also guarantees that memory objects are contiguous in physical memory.

\gfi{To allow the \texttt{pim\_malloc} API to influence the OS memory allocator and ensure that memory objects are placed within specific DRAM mats,\revdel{However, this is challenging due to two main reasons. Firstly, most systems have no knowledge about how the memory controller scrambles data across DRAM module/chip/banks/subarray/mats. Secondly, the \texttt{pim\_malloc} API requires modifications to the OS memory allocation policies to guarantee the allocation of contiguous physical frames for a given memory object. To address these challenges, the OS can obtain information about the DRAM interleaving scheme (e.g., by reverse engineering the bit locations of memory addresses~\cite{kim2020revisiting, orosa2021deeper,yauglikcci2022understanding}), and the \texttt{pim\_malloc} API can create pools of \revCommon{contiguously} allocated pages using huge pages~\cite{santos2022aggressive}.}}
\revCommon{\changeCM{\#CQ1}we propose a new \emph{lazy data allocation routine} (in the kernel) for \texttt{pim\_malloc} objects, which Figure~\ref{fig:mimdram:allocation} illustrates. This routine has three main components: 
\li~information \omi{about} the DRAM organization (e.g., row, column, and mat sizes), 
\lii~the DRAM interleaving scheme, which the memory controller provides via an open firmware device tree~\cite{devicethree};\footnote{The DRAM interleaving scheme can be obtained by reverse engineering the bit locations of memory addresses~\gfcrii{\cite{kim2020revisiting, orosa2021deeper,yauglikcci2022understanding,loughlin2023siloz}}. \rA{\changerA{\rA{\#A3}}Even though typical DRAM interleaving does \emph{not} take mats into account, it is relatively straightforward to reverse-engineer how a memory address is distributed across the DRAM mats in a DRAM module\gfcrii{, since the mat interleaving is a function of the DRAM chip's organization}. \omi{For example, in} a DDR4 module with 8 chips, 16 mats per chip, and 4 HFFs per mat, a 64~B cache line is evenly distributed across all 128 total mats; i.e., the four least-significant bits of the cache line are placed in mat 0, chip 0, and the four most-significant bits of the cache line are placed in mat 15, chip 7. Our \texttt{pim\_malloc} API takes into account such mat interleaving.}} and 
\liii~a huge \omi{page} pool for \texttt{pim\_malloc} objects (configured during boot time), which guarantees that virtual addresses assigned to a \texttt{pim\_malloc} object are contiguous in the physical address space and \gfhpca{that DRAM mats are free whenever a \texttt{pim\_malloc} object is allocated.}  
The allocation routine uses the DRAM address mapping knowledge to split the huge pages into different memory regions. Then, when an application calls the \texttt{pim\_malloc} API, the allocation routine selects the appropriate memory region that satisfies \texttt{pim\_malloc}.  \rD{\changerD{\rD{\#D7}}Internally, the \texttt{pim\_malloc} API operates using three main sub-tasks, depending on the order of the data allocation: 
\li~\texttt{pim\_preallocate}, for data pre-allocation;
\lii~\texttt{pim\_alloc}, for the first data allocation; and
\liii~\texttt{pim\_alloc\_align}, for subsequent aligned allocations.}}

\begin{figure}[ht]
    \centering
    \includegraphics[width=0.95\linewidth]{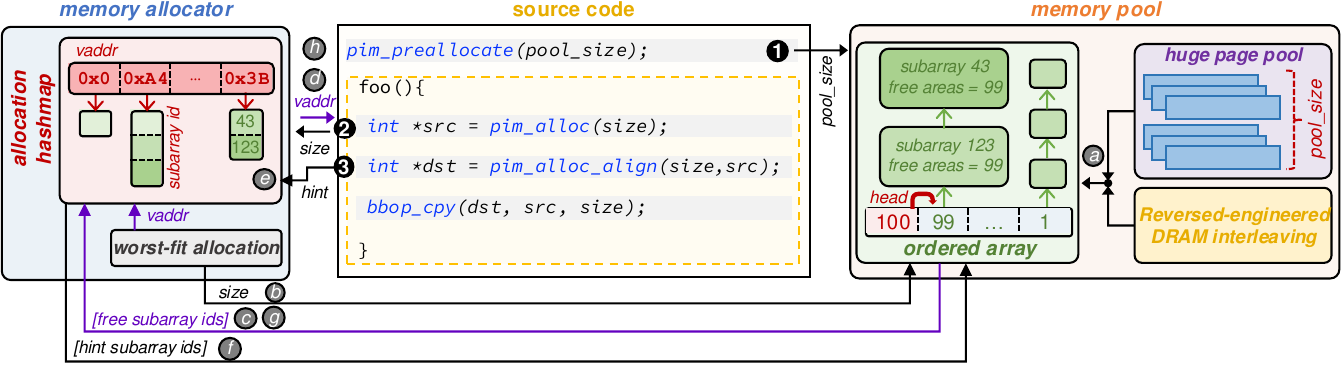}
    \caption{Overview of the MIMDRAM's memory allocation routine.}
    \label{fig:mimdram:allocation}
\end{figure}

\rD{\changerD{\rD{\#D7}}\li~\underline{Pre-Allocation.} The first sub-task's role is to indicate the number of huge pages available for \gls{PuD} allocations (\circled{1} in Figure~\ref{fig:mimdram:allocation}). We \omii{leave it to the user} to provide the number of huge pages used for \gls{PuD} \gfcrii{operations} (\circledii{a} in Figure~\ref{fig:mimdram:allocation}) because huge pages are scarce in the system.}

\rD{\changerD{\rD{\#D7}}\lii~\underline{First Allocation.} The second sub-task uses the \emph{worst-fit allocation scheme}~\cite{johnson1973near} to manage the allocation of memory regions in the huge page pool. The \emph{key idea} behind this placement strategy is to optimize the remaining \omi{memory} space \omi{after} allocations to increase the chances of accommodating another process in the remaining space. For the first \gls{PuD} memory allocation, the \texttt{pim\_alloc} sub-task (\circled{2}) simply scans an \emph{ordered array} data structure (similar to the one used in the Linux Kernel buddy allocator algorithm~\cite{knowlton1966programmer}, where each entry represents the number of memory regions in a single subarray) to select the subarray with the \emph{largest} amount of memory regions available (\circledii{b}). If the requested memory allocation requires more than one memory region, \prop \gfcrii{iteratively} scans the  \emph{ordered array}, searching for the next largest memory region until the memory allocation is fully satisfied. Once enough space is allocated (\circledii{c}), \texttt{pim\_alloc} sub-task creates a new allocation object and inserts it in an \emph{allocation hashmap}, which is indexed by the allocation's virtual address (\circledii{d}). The sub-task needs to keep track of allocations since it might need to find a memory region from the \emph{same} subarray/mat when performing future \omi{\emph{aligned allocations}}.} 

\rD{\changerD{\rD{\#D7}}\liii~\underline{Aligned Allocation.} After allocating \gfcrii{a} memory \gfcrii{region} for the first operand in a \gls{PuD} \gfcrii{operation}, the user can use this memory region as a regular memory object. However, when allocating the remaining operands for a \gls{PuD} \gfcrii{operation}, the \texttt{pim\_malloc} API needs to guarantee data alignment for all memory objects within the same DRAM subarray/mat. To this end, the third sub-task (\texttt{pim\_malloc\_align}; \circled{3}) identifies \gfcrii{a} previously allocated memory region to which the current memory allocation must be aligned (based on the compiler-generated \emph{mat labels}). The \texttt{pim\_malloc\_align} sub-task works in five main steps. 
First, it searches the \emph{allocation hashmap} using the \emph{mat label} as a \emph{hint} for a match with previously allocated memory regions (\circledii{e}). If a match is not found, the allocation fails.
Second, if a match is found, the \texttt{pim\_malloc\_align} sub-task iterates through the identified previously-allocated memory regions (\circledii{f}). 
Third, for each memory region, the sub-task identifies its source subarray/mat address and tries to allocate another memory region \omi{in} the same subarray/mat for the new allocation (\circledii{g}).
Fourth, if the subarray/mat of a given memory region has no free region, the sub-task allocates a new memory region from another subarray/mat following the worst-fit allocation scheme (\circledii{h}). Since we use a worst-fit allocation scheme \gfcrii{that always selects the \emph{largest} \omii{number} of memory regions available during memory allocation for the \emph{first} operand of a \gls{PuD} operation}, we have a good chance of having a single subarray/mat holding memory regions for \gfcrii{the remaining operands of a \gls{PuD} operation}.
Fifth, since memory regions might come from different huge pages, we must perform \texttt{re-mmap} to map such memory regions into contiguous virtual addresses.}

% This work assumes that the \texttt{pim\_malloc} API can influence the OS memory allocator and the system's \gls{MMU} to ensure that memory objects are continuous in physical space and placed within particular DRAM mats. However, enabling such behavior in practice is challenging for two main reasons. 
% First, in most systems, the OS and the \gls{MMU} have \emph{no} knowledge about how the memory controller scrambles data across DRAM module/chip/banks/subarray/mats (i.e., the DRAM interleaving scheme the memory controller employs is oblivious to the OS). This means that even if the OS tries to allocate continuous physical frames for a given memory object, there are no guarantees that such physical frames will be stored continuously within a DRAM module. 
% Second, the \texttt{pim\_malloc} API requires non-trivial modifications to the OS memory allocation policies to ensure the allocation of continuous physical frames for a given memory object.  
% These challenges can be addressed as follows. 
% First, the OS can obtain information regarding the DRAM interleaving scheme either by cooperating with memory manufacturers or by reverse engineering the bit locations of memory addresses, as done by prior works~\cite{kim2020revisiting, orosa2021deeper,yauglikcci2022understanding}. 
% Second, the \texttt{pim\_malloc} API can create pools of continuously allocated pages using huge pages~\cite{santos2022aggressive}. We leave the concrete implementation of both solutions to future work.

\paratitle{Mat Label Translation} To keep track of the \omi{mapping} between \emph{mat label}s and allocated \emph{mat ranges}, \prop adds a small \emph{mat translation table} alongside the page table. The table is indexed by hashing the \emph{mat label} with the \emph{process ID}. It stores in each entry the associated \emph{mat range} that the memory allocator assigned to that particular  \emph{mat label}. When the CPU dispatches a \emph{bbop}, the CPU  \li~accesses the \emph{mat translation table} to obtain the \emph{mat range} assigned to the given \emph{bbop}, and
\lii~replaces the \emph{mat label} with the \emph{mat range}.   
% \subsection{Programming \& Execution Model}
% \label{sec:design:programming}

%\paratitle{Vertical Data Layout for Bit-Serial Computation.}

%\paratitle{SIMD + VLIW}

% \subsection{Compiler Design}
% \label{sec:design:compiler}

% \subsubsection{Identifying Loop Candidates}

% \subsubsection{Offload Cost Model}

% \subsubsection{Static Instruction Scheduling}

% \subsubsection{Binary Generation}

% \subsection{In-DRAM Execution}
% \label{sec:design:indramexec}

%\subsection{Limitations \& Next Steps}
%\label{sec:design:limitations}

 \section{Methodology}
\label{sec:methodology}

 We implement \prop using the gem5 simulator~\cite{gem5} and compare it to a real multicore CPU (Intel Skylake~\cite{intelskylake})\revC{, a real high-end GPU (NVIDIA A100~\cite{a100}),} and a state-of-the-art \gfi{\gls{PuD} framework}  (SIMDRAM~\cite{hajinazarsimdram}). \rD{\changerD{\rD{\#D1}}In all our evaluations, the CPU code is optimized to leverage AVX-512 instructions~\cite{firasta2008intel}.} Table~\ref{table_parameters} shows the system parameters we use\revdel{ in our evaluations}.\revdel{ To measure CPU performance, we implement a set of timers in \texttt{sys/time.h}~\cite{systime}.} To measure CPU energy consumption, we use Intel RAPL~\cite{hahnel2012measuring}. \revC{We capture GPU kernel execution time that excludes data initialization/transfer time. To measure GPU energy consumption, we use the \texttt{nvml} API~\cite{NVIDIAMa14}.}
We implement SIMDRAM on gem5\gfcrii{, taking into account that the latency of executing the back-to-back \texttt{ACT}s is only \omii{1.1}$\times$ the latency of $t_{RAS}$~\ambit,} and validate our implementation rigorously with the results reported in \cite{hajinazarsimdram}. \gf{We use CACTI~\cite{cacti} to evaluate \prop and SIMDRAM energy consumption, where we take into account that each additional simultaneous row activation increases energy consumption by 22\%~\cite{seshadri2017ambit, hajinazarsimdram}. 
Our simulation accounts for the additional latency imposes by \prop's mat isolation transistors and row decoder latches\revdel{ upon DRAM operations} (i.e., \gfcrii{measured (using CACTI~\cite{cacti, muralimanohar2007optimizing}) to incur} less than 0.5\% extra latency for an \texttt{ACT}).
\gfcrii{We open-source our simulation infrastructure  at \url{https://github.com/CMUSAFARI/MIMDRAM}}.} 

% We evaluate \emph{two} \prop's implementations:
% \li~\emph{\prop-AOp}, an area-optimized implementation that implements fine-grained DRAM activation \emph{only} to the DRAM rows in the B-group's portion of the DRAM subarray (as described in Section~\ref{}); and 
% \lii~\emph{\prop-TOp}, a throughput-optimized implementation that implements fine-grained DRAM activation  in the \emph{entire} DRAM subarray (i.e., for both D-group and B-group portions of the DRAM subarray). 
% While \emph{\prop-AOp} minimally modifies the design of a DRAM subarray, its peak throughput is limited by the fact that row copy operations must be serialized (Section~\ref{}). In contrast, \emph{\prop-TOp} relaxes this limitation and allows both row copy and triple-row activation operations to be concurrently executed across different DRAM mats, at the cost of a higher area cost. 

%\omii{We \omiii{use} \omiii{the same} vertical data layout in our Ambit \omiii{and SIMDRAM implementations}, \omiii{which} enables us to (1) evaluate all 16 SIMDRAM operations in Ambit using their equivalent AND/OR/NOT-based implementation\omiii{s}, and (2) highlight the benefits of Step 1 in the \mech framework (i.e., using an optimized MAJ/NOT-based implementation of the operations).} %We \omi{modify Ambit to operate on vertically-laid-out data, to illustrate} the benefits of Steps 1 and 2 of our methodology. 
%\omvuii{Our synthetic throughput analysis \omviii{(\cref{sec_performance})} uses 64M-element input arrays.}

\begin{table}[ht]
\centering
  \vspace{5pt}
   \caption{Evaluated system configurations.}
   \vspace{-5pt}

   \centering
   \footnotesize
   \tempcommand{1.3}
   \renewcommand{\arraystretch}{0.7}
   \resizebox{0.7\columnwidth}{!}{
   \begin{tabular}{@{} c l @{}}
   \toprule
   \multirow{5}{*}{\shortstack{\textbf{\omi{Real} Intel}\\ \textbf{Skylake CPU~\cite{intelskylake}}}} & x86~\cite{guide2016intel}, 16~cores, 8-wide, out-of-order, \SI{4}{\giga\hertz};  \\
                                                                           & \emph{L1 Data + Inst. Private Cache:} \SI{256}{\kilo\byte}, 8-way, \SI{64}{\byte} line; \\
                                                                           & \emph{L2 Private Cache:} \SI{2}{\kilo\byte}, 4-way, \SI{64}{\byte} line; \\
                                                                           & \emph{L3 Shared Cache:} \SI{16}{\mega\byte}, 16-way, \SI{64}{\byte} line; \\
                                                                           & \emph{Main Memory:} \SI{64}{\giga\byte} DDR4-2133, 4~channels, 4~ranks \\
   \midrule
      \multirow{3}{*}{\shortstack{\textbf{\omi{Real} \revC{NVIDIA}}\\ \textbf{\revC{A100 GPU~\mbox{\cite{a100}}}}}} &  \revC{\SI{7}{\nano\meter} technology node; 6912 CUDA Cores;}\\ 
                                                                            & \revC{108 streaming multiprocessors, \SI{1.4}{\giga\hertz} base clock;} \\
                                                                            & \revC{\emph{L2 Cache:} \SI{40}{\mega\byte} L2 Cache; \emph{Main Memory:} \SI{40}{\giga\byte} HBM2~\mbox{\cite{HBM,lee2016simultaneous}}} \\
   \midrule

   \multirow{8}{*}{\shortstack{\omi{\textbf{Simulated}} \\ \textbf{SIMDRAM~\cite{hajinazarsimdram}}\\ \textbf{\& \prop}}} &  gem5 system emulation;  x86~\cite{guide2016intel}, 1-core, out-of-order, \SI{4}{\giga\hertz};\\
                                                                             & \emph{L1 Data + Inst. Cache:} \SI{32}{\kilo\byte}, 8-way, \SI{64}{\byte} line;\\
                                                                             & \emph{L2 Cache:} \SI{256}{\kilo\byte}, 4-way, \SI{64}{\byte} line; \\
                                                                             & \emph{Memory Controller:}  \SI{8}{\kilo\byte} row size, FR-FCFS~\cite{mutlu2007stall,zuravleff1997controller}\\
                                                                             & \emph{Main Memory:}  DDR4-2400, 1~channel, 8~chips, 4~rank \\ &
                                        16~banks/rank, 16~mats/chip, 1~K rows/mat, 512~columns/mat\\      &
                                        
                                       \emph{\prop's Setup:} 8~entries mat queue, \SI{2}{\kilo\byte}~\emph{bbop} buffer \\ &
                                       8~\textit{\uprog{} processing engines},   \SI{2}{\kilo\byte}~\emph{mat translation table} \\ 
                                      % \midrule
%   \multirow{5}{*}{\textbf{\prop}} &  gem5 system emulation;  x86~\cite{guide2016intel}, 1-core, out-of-order, \SI{4}{\giga\hertz};\\
%                                                                              & \emph{L1 Data + Inst. Cache:} \SI{32}{\kilo\byte}, 8-way, \SI{64}{\byte} line;\\
%                                                                              & \emph{L2 Cache:} \SI{256}{\kilo\byte}, 4-way, \SI{64}{\byte} line; \\
%                                                                              & \emph{Memory Controller:}  \SI{8}{\kilo\byte} row size, FR-FCFS~\cite{mutlu2007stall,zuravleff1997controller} scheduling\\
%                                                                              & \emph{Main Memory:}  DDR4-2400, 1~channel, 1~rank, 16~banks \\
  
   \bottomrule
   \end{tabular}
   }
   \label{table_parameters}
\end{table}

%\gfbox{Copy paste table 2 ``DRAM'' from sectored dram please myself}

\paratitle{Real-World Applications} We analyze \gf{117 applications from \omi{seven} benchmark suites (SPEC 2017~\cite{spec2017}, SPEC 2006~\cite{spec2006}, Parboil~\cite{stratton2012parboil}, Phoenix~\cite{yoo_iiswc2009}, Polybench~\cite{pouchet2012polybench}, Rodinia~\cite{che_iiswc2009}, and SPLASH-2~\cite{woo_isca1995}) to select applications that} 
\li~are memory-bound, and 
\lii~the most \juani{\omi{time}-consuming} loop can be auto-vectorized. 
From this analysis, we collect \gf{12} multi-threaded \revC{CPU} applications \rD{(\changerD{\rD{\#D3}}as Table~\ref{table:workload:properties} describes)} from different domains (i.e., video compression, data mining, pattern recognition, medical imaging, stencil computation)\revC{, and their respective GPU implementations, when available}\omi{. Our evaluated applications are:} 
\li~525.x264\_r (\texttt{x264}) from SPEC 2017;
\lii~heartwall (\texttt{hw}), kmeans (\texttt{km}), and backprop (\texttt{bs}) from Rodinia;
\liii~\texttt{pca} from Phoenix; and
\liv~\texttt{2mm}, \texttt{3mm}, covariance (\texttt{cov}), doitgen (\texttt{dg}), fdtd-apml (\texttt{fdtd}), gemm (\texttt{gmm}), and gramschmidt (\texttt{gm}) from Polybench.\footnote{\rE{\omi{Several} prior works~\cite{damov,devic2022pim,dualitycache,fujiki2018memory,vadivel2020tdo,iskandar2023ndp,pattnaik2016scheduling} \omi{show} \changerE{\rE{\#E1}} that our selected \gfcrii{twelve} \omi{workloads} can benefit from different types of \gls{PIM} architectures. }}
\gf{Since our base \gfi{\gls{PuD}} substrate (SIMDRAM) does \emph{not} support floating-point, we manually modify the selected floating-point-heavy auto-vectorized loops to operate \juani{on} fixed-point data arrays.\footnote{\changerD{\rD{\#D10}}\rD{We only modify the three applications from the Rodinia benchmark suite to use fixed-point operations. \omi{Prior works~\cite{fujiki2018memory,yazdanbakhsh2016axbench,ho2017efficient} also \omi{employ} fixed-point for the same three Rodinia applications.} The applications from Polybench can be configured to use integers; the auto-vectorized loops in 525.x264\_r use \texttt{uint8\_t}; pca uses integers. }We do \emph{not} observe an output quality degradation when employing fixed-point for the selected loops.} We use the largest input dataset available \revE{and execute each application \emph{end-to-end}} in our evaluations.}\changeE{\#E1}

\begin{table}[ht]
   \caption{Evaluated applications and their characteristics.}
   \tempcommand{1}
   \resizebox{\columnwidth}{!}{%
    \begin{tabular}{|c|c||c|c|c|c|}
\hline
\textbf{\begin{tabular}[c]{@{}c@{}}Benchmark\\  Suite\end{tabular}} & \textbf{\begin{tabular}[c]{@{}c@{}}Application\\ (Short Name)\end{tabular}} & \textbf{\begin{tabular}[c]{@{}c@{}}Dataset \\ Size\end{tabular}} & \textbf{\begin{tabular}[c]{@{}c@{}}\# Vector \\ Loops\end{tabular}} & \textbf{\begin{tabular}[c]{@{}c@{}}VF\\  \{min, max\}\end{tabular}} & \textbf{\begin{tabular}[c]{@{}c@{}}PUD \\ Ops.{$^\dag$}\end{tabular}} \\ \hline \hline
\begin{tabular}[c]{@{}c@{}}Phoenix~\cite{yoo_iiswc2009}\end{tabular} & $^\ddag$pca (\texttt{pca}) & reference & 2 & \{4000, 4000\} & D, S, M, R \\ \hline
\multirow{7}{*}{\begin{tabular}[c]{@{}c@{}}Polybench\\ \cite{pouchet2012polybench}\end{tabular}} & 2mm (\texttt{2mm}) & \begin{tabular}[c]{@{}c@{}}NI = NJ = NK = NL = 4000\end{tabular} & 6 & \{4000, 4000\} & M, R \\ \cline{2-6} 
 & $^\ddag$3mm (\texttt{3mm}) & \begin{tabular}[c]{@{}c@{}}NI = NJ = NK = NL = NM = 4000\end{tabular} & 7 & \{4000, 4000\} & M, R \\ \cline{2-6} 
 & covariance (\texttt{cov}) & \begin{tabular}[c]{@{}c@{}}N = M = 4000\end{tabular} & 2 & \{4000, 4000\} & D, S, R \\ \cline{2-6} 
 & doitgen (\texttt{dg}) & \begin{tabular}[c]{@{}c@{}}NQ = NR = NP = 1000\end{tabular} & 5 & \{1000, 1000\} & M, C, R \\ \cline{2-6} 
 & $^\ddag$fdtd-apml (\texttt{fdtd}) & \begin{tabular}[c]{@{}c@{}}CZ = CYM = CXM = 1000\end{tabular} & 3 & \{1000, 1000\} & D, M, S, A  \\ \cline{2-6} 
 & gemm (\texttt{gmm}) & \begin{tabular}[c]{@{}c@{}}NI = NJ = NK = 4000\end{tabular} & 4 & \{4000, 4000\} & M, R  \\ \cline{2-6} 
 & gramschmidt (\texttt{gs}) & \begin{tabular}[c]{@{}c@{}}NI = NJ = 4000\end{tabular} & 5 & \{4000, 4000\} & M, D, R  \\ \hline
\multirow{3}{*}{\begin{tabular}[c]{@{}c@{}}Rodinia\\ \cite{che_iiswc2009}\end{tabular}} & backprop (\texttt{bs}) & 134217729 input elm.  & 1 & \{17, 134217729\} & M, R  \\ \cline{2-6} 
 & heartwall (\texttt{hw}) &  reference & 4 & \{1, 2601\} & M, R  \\ \cline{2-6} 
 & kmeans (\texttt{km}) & 16384 data points & 2 & \{16384, 16384\} & S, M, R  \\ \hline
\begin{tabular}[c]{@{}c@{}}SPEC 2017\\\cite{spec2017}\end{tabular} & 525.x64\_r (\texttt{x264}) & \begin{tabular}[c]{@{}c@{}}reference input\end{tabular} & 2 & \{64, 320\} & A  \\ \hline
\end{tabular}%
}
\scriptsize{$^\dag$: D = division, S = subtraction, M = multiplication, A = addition, R = reduction, C = copy
}
\newline
\scriptsize{$^\ddag$\gfcrii{: application with independent \gls{PuD} operations}
}
\label{table:workload:properties}
\end{table}

\paratitle{\gfcrii{Multi-Programmed Application Mixes}} \gfcrii{To measure system throughput and fairness, we \emph{manually} create 495 application mixes by randomly selecting eight applications (from our group of 12 applications) for execution co-location. We classify each application mix into \agy{one of} three categories: 
\emph{low}, \emph{medium}, and \emph{high} vectorization factor (VF) mixes based on \gfi{Figure}~\ref{fig_max_utilization}.
In the \omii{\emph{low}} mix, the maximum VF of \emph{all} eight applications is lower than 16K; 
in the \omii{\emph{medium}} mix, \emph{at least} one application has a maximum vectorization factor between 16K (inclusive) and 64K; and
in the \omii{\emph{large}} mix, \emph{at least} one application has a maximum VF larger than 64K (inclusive). }

\paratitle{\gfcrii{\omii{Comparison to State-of-the-Art} \gls{PIM} Architectures}} \gfcrii{We compare \prop to two other state-of-the-art \gls{PIM} architectures: DRISA~\cite{li2017drisa} and Fulcrum~\cite{lenjani2020fulcrum}. 
DRISA is a \omi{combined} \gls{PuM} \omi{and} \gls{PnM} architecture that \emph{significantly} modifies the DRAM \omii{microarchitecture} and organization to enable bulk in-DRAM computation (\omi{e.g.}, by using 3T1C DRAM cells to execute in-situ bitwise NOR operations \omi{and} by adding logic gates \emph{near} the subarray's sense amplifiers). \revdel{DRISA employs a fine-grained interconnection network to shift data across DRAM columns, thus executing operations in a \emph{bit-parallel} mode (in contrast with SIMDRAM and \prop, which execute operations in a \emph{bit-serial} mode). 
In this analysis, we employ DRISA's 3T1C implementation to contrast both bit-serial and bit-parallel \gls{PuD} execution models.}
Fulcrum is a \gls{PnM} architecture that adds \omi{computation} logic \emph{near} subarrays. 
Fulcrum's primary components are a series of shift registers (called walkers) that latch input/output DRAM rows and a narrow scalar ALU \omi{that executes arithmetic and logic operations}. We model \omi{the} DRISA 3T1C implementation and Fulcrum  \li~\omii{using a DRAM module of equal} \gfcrii{dimensions (i.e., number of DRAM ranks, chips, banks, mats, rows, and columns) as the} baseline DDR4 \gfcrii{DRAM} we use for SIMDRAM and \prop (see Table~\ref{table_parameters}) \gfcriii{and} 
\lii~including all the changes that the DRISA 3T1C and Fulcrum architectures propose to the DRAM cell array and DRAM subarray.}

\section{Evaluation}
\label{sec:eval}

We demonstrate the advantages of \prop 
by evaluating 
\li~\gls{SIMD} utilization and energy efficiency (i.e., performance per Watt) for single applications \gfcrii{(\cref{sec:eval:singleapp})}; 
\lii~\gf{system throughput (in terms of weighted speedup~\cite{snavely2000symbiotic, eyerman2008systemlevel, michaud2012demystifying}), job turnaround time (in terms of harmonic speedup~\cite{luo2001balancing,eyerman2008systemlevel}), and fairness (in terms of maximum slowdown~\cite{kim2010thread, kim2010atlas,subramanian2014blacklisting,subramanian2016bliss, subramanian2013mise, mutlu2007stall, subramanian2015application, ebrahimi2010fairness, ebrahimi2011prefetch, das2009application, das2013application})} for multi-\gfcrii{programmed application mixes} in comparison to the baseline CPU, GPU, and a state-of-the-art \gls{PuD} architecture, i.e., SIMDRAM~\cite{hajinazarsimdram} \gfcrii{(\cref{sec:eval:multiapp})}; 
\lii~\gfcrii{area-normalized performance analysis for single applications  and throughput analysis for multi-programmed application mixes in comparison to state-of-the-art \gls{PIM} architectures, i.e., DRISA~\cite{li2017drisa} and Fulcrum~\cite{lenjani2020fulcrum}   (\cref{sec:eval:otherpims}).}
\gfcriii{\omiii{In most of} our analyses (\cref{sec:eval:singleapp}--\cref{sec:eval:multiapp}), \omiii{to keep our analyses pure,} we \omiii{very} \emph{conservatively} allow \prop to use \omiii{only} a \emph{single} DRAM subarray in a \emph{single} DRAM bank for \gls{PuD} computation. 
In \cref{sec:eval:scalability}, we perform a scalability analysis to evaluate \prop's performance when enabling \emph{multiple} DRAM subarrays and banks for \gls{PuD} computation\omiii{, which reflects a more accurate evaluation of the true benefits of \prop and \gls{PuD}}.} 
Finally, we evaluate \prop's DRAM and CPU area cost \gfcrii{(\cref{sec:eval:area})}.

\subsection{Single-Application \omii{Results}}
\label{sec:eval:singleapp}

\gfi{Figure}~\ref{fig_single_app_analysis} shows \prop's \gls{SIMD} utilization \gf{and normalized energy efficiency (in performance per Watt)} for all \gf{12} applications. Values are normalized to the baseline CPU. 
%\agycomment{Isn't it better to split this figure into two separate figures and have SIMD Utilization and Energy Efficiency as separate subsections of evaluation?}

\begin{figure}[ht]
\centering
\begin{subfigure}{0.95\linewidth}
  \centering
  \includegraphics[width=\linewidth]{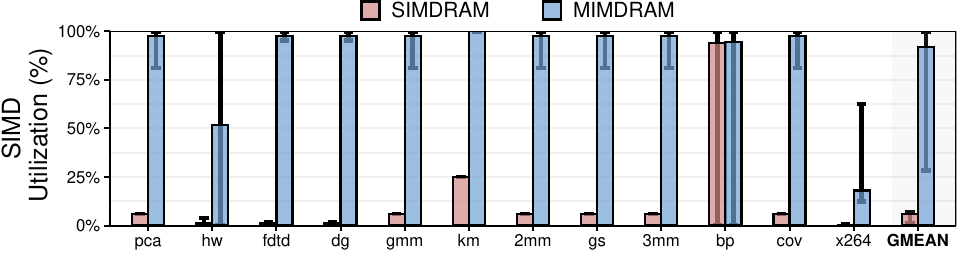}  
  \caption{\omi{\gls{SIMD} utilization. \gf{Whiskers extend to the minimum and maximum \omi{observed} data point values.}}}
  \label{fig:sub-first}
\end{subfigure}
~
\begin{subfigure}{0.95\linewidth}
\centering\includegraphics[width=\linewidth]{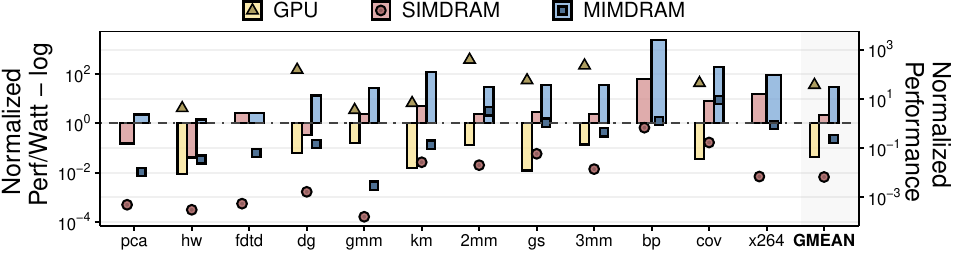}  
  \caption{CPU-normalized performance per Watt (left y-axis; bars) and \revC{performance (right y-axis; dots)}.}
  \label{fig:sub-second}
\end{subfigure}
\caption{Single-application \omii{results} \gfcrii{for processor-centric (i.e., CPU and GPU) and memory-centric (i.e., SIMDRAM and \prop) architectures executing twelve real-world applications}. }
\label{fig_single_app_analysis}
\end{figure}

% \begin{figure}[ht]
%     \vspace{-8pt}
%     \centering
%     \includegraphics[width=\linewidth]{mainmatter/05_mimdram/plots/utilization_perf_watt-crop.pdf}
%     \caption{Single-application analysis. \gf{Whiskers extend to the minimum and maximum data point values.}}
%     \label{fig_single_app_analysis}
%     \vspace{-8pt}
% \end{figure}

\paratitle{SIMD Utilization} 
We make two observations from \gfi{Figure}~\ref{fig_single_app_analysis}a. First, \prop \emph{significantly} improves \gls{SIMD} utilization \omi{over} SIMDRAM. On average across all applications, \prop provides 15.6$\times$ the \gls{SIMD} utilization of SIMDRAM. This is because \prop matches the available \gls{SIMD} parallelism in an application with the underlying \gfi{\gls{PuD}} resources (i.e., \gfi{\gls{PuD}} \gls{SIMD} lanes) by \omi{using} \emph{only} as many DRAM mats as the maximum vectorization factor of a given application's loop. In contrast, SIMDRAM always occupies \omi{\emph{all}} available \gfi{\gls{PuD}} \gls{SIMD} lanes \omi{(i.e., entire subarrays)} for a given operation, resulting in low \gls{SIMD} utilization for applications without a \omi{very}-wide vectorization factor. 
Second, we observe that \gls{SIMD} utilization can vary considerably within an application. For example, \prop's \gls{SIMD} utilization for \texttt{hw} and \texttt{bp} goes from as low as 0.2\% to as high as 100\%. This happens because the \gls{SIMD} parallelism for each vectorized loop in these applications changes at different execution phases. \prop can \gfi{better} adjust to the variation in \gls{SIMD} parallelism \omi{(than SIMDRAM)} due to its flexible design. 
We conclude that \prop \omi{greatly improves} overall \gls{SIMD} utilization for many applications. 

\paratitle{\rC{Performance \&} Energy Efficiency} We make \gfhpca{three} observations from \gfi{Figure}~\ref{fig_single_app_analysis}b. 
First, \prop \emph{significantly} improves energy efficiency \rC{and performance} \omi{over} SIMDRAM. On average across all applications, \prop provides \efficiencysimdram the energy efficiency \rC{and 34$\times$ the performance} of SIMDRAM. \gf{\prop's higher energy efficiency is due to three main reasons.}
\li~\prop parallelizes the computation of \omi{independent} \emph{bbops} in a single application loop across different mats, improving overall performance. \prop reduces execution time \changerC{\rC{\#C1}}by 2.8$\times$ compared with SIMDRAM, on average across applications with \omi{independent} \emph{bbops} \gfcrii{(i.e., \texttt{pca}, \texttt{3mm}, and \texttt{fdtd}).}
\lii~\prop implements in-situ \gfi{\gls{PuD}} vector reduction operations, while SIMDRAM requires the assistance of the CPU \gfcrii{for} vector reduction, increasing latency and energy consumption. \rC{\changerC{\rC{\#C1}}\prop reduces execution time and energy consumption by 1.6$\times$ and 266$\times$ \omi{over} SIMDRAM, on average across the applications with vector reduction \gfcrii{operations} \gfcrii{(from our twelve applications, only \texttt{fdtd} and \texttt{x264} do \emph{not} require vector reduction operations)}.}
\liii~\prop activates \omi{only} the \omi{necessary} {\gls{PuD}} \gls{SIMD} lanes during an application loop's execution, significantly saving energy  when the application has low \gls{SIMD} utilization. \rC{\changerC{\rC{\#C1}}\prop reduces energy consumption by 325$\times$ \omi{over} SIMDRAM, on average across applications with a maximum vectorization factor lower than 65,536 \gfcrii{(from our twelve applications, only \texttt{bs} exhibits a vectorization factor \emph{higher} than 65,536)}.}
\rD{\changerD{\rD{\#D2}}Second, \prop provides \efficiencycpu/\efficiencygpu the energy efficiency of CPU{/GPU baselines}. \prop's higher energy efficiency is due to its inherent ability to avoid costly data movement operations for memory-bound applications. 
{Third, even though \prop improves performance (by 3.1$\times$, 8.6$\times$, 1.1$\times$, and 1.3$\times$) compared to the baseline CPU for some applications (i.e., \texttt{2mm}, \texttt{cov}, \texttt{gs}, and \texttt{bp}), it \omi{leads to} performance loss compared to the baseline CPU and GPU on average across all applications. This is because, for some applications, the bulk parallelism available inside a \emph{single} DRAM \omiii{subarray and bank} is insufficient to hide the latency of costly bit-serial operations (e.g., multiplication). We observe that enabling \prop in \gfcriii{16} DRAM banks \gfcriii{and 64 subarrays (per bank) allows \prop to provide performance gains compared to} the CPU and the GPU (see \cref{sec:eval:scalability}).}}
%\revC{However, such an increase in energy efficiency comes at the cost of lower performance compared to the baseline CPU/GPU. This is because while the CPU and GPU take advantage of multi-threaded execution, \prop's only source of parallelism comes from the data parallelism available in a vectorized loop, which penalizes performance. This issue can be overcome by allowing \prop to leverage other forms of parallelism for computation (e.g., application-level parallelism, as we evaluate in \cref{sec:eval:multiapp}).} 
We conclude that \prop is an energy-efficient \omiii{and high-performance} \gfi{\gls{PuD}} system.

\subsection{Multi-\gfcrii{Programmed} \omii{Workload Results}}
\label{sec:eval:multiapp} 
\gf{We evaluate SIMDRAM and \prop's impact on system throughput (in terms of weighted speedup~\cite{snavely2000symbiotic, eyerman2008systemlevel, michaud2012demystifying}), job turnaround time (in terms of harmonic speedup~\cite{luo2001balancing,eyerman2008systemlevel}), and fairness (in terms of maximum slowdown~\cite{kim2010thread, kim2010atlas,subramanian2014blacklisting,subramanian2016bliss, subramanian2013mise, mutlu2007stall, subramanian2015application, ebrahimi2010fairness, ebrahimi2011prefetch, das2009application, das2013application}) when executing applications concurrently. 
\revdel{\rD{\changerD{\rD{\#D5}}To do so, we \emph{manually} create 495 application mixes by randomly selecting eight applications (from our group of 12 applications) for execution co-location.} We classify each application mix into \agy{one of} three categories: \emph{low}, \emph{medium}, and \emph{high} vectorization factor (VF) mixes based on \gfi{Figure}~\ref{fig_max_utilization}.
In the low VF application mix, the maximum VF of \emph{all} eight applications is lower than 16K; in the medium VF application mix, \emph{at least} one application has a maximum vectorization factor between 16K (inclusive) and 64K; and
in the large VF application mix, \emph{at least} one application has a maximum VF larger than 64K (inclusive).} 
To provide a fair comparison, we introduce \gls{MIMD} parallelism in SIMDRAM with bank-level parallelism (BLP)~\cite{mutlu2008parallelism,kim2010thread,kim2012case,kim2016ramulator,lee2009improving}, where each SIMDRAM-capable DRAM bank can independently run an application. We evaluate four configurations of SIMDRAM where 1 (\emph{SIMDRAM:1}), 2 (\emph{SIMDRAM:2}), 4 (\emph{SIMDRAM:4}), and  8 (\emph{SIMDRAM:8}) banks have SIMDRAM computation capability.} \gf{\gfi{Figure}~\ref{fig_mult_app_analysis} shows the system throughput, job turnaround time (which measures a balance of fairness and throughput), and fairness that SIMDRAM and \prop provide on average across all application mixes. Values are normalized to \emph{SIMDRAM:1}. We make \omiii{three} observations\revdel{ from the figure}.}

\begin{figure}[ht]
    \centering
    \includegraphics[width=\linewidth]{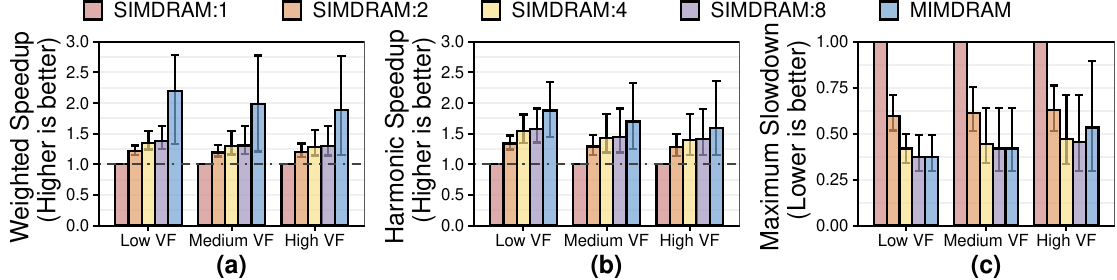}
    \caption{\gfcriii{Multi-\gfcrii{programmed} \omii{workload results} \gf{for three types of application mixes:} 
    \gfcrii{(a) low VF, 
    (b) medium VF, and 
    (c) high VF.
    \emph{VF} stands for vectorization factor.
    \emph{SIMDRAM:X} uses $X$ DRAM banks for computation}. Values are normalized to \emph{SIMDRAM:1}. Whiskers extend to the minimum and maximum observed data point values.}}
    \label{fig_mult_app_analysis}
\end{figure}

\gf{First, \prop \emph{significantly} improves system throughput, job turnaround time, and fairness compared with SIMDRAM. On average, across all application groups, \prop achieves:
\li~1.68$\times$ (min. 1.52$\times$, max. 2.02$\times$) \emph{higher} weighted speedup,
\lii~1.33$\times$ (min. 1.17$\times$, max. 1.72$\times$) \emph{higher} harmonic speedup, and
\liii~1.32$\times$ (min. 0.95$\times$, max. 2.29$\times$) \emph{lower} maximum slowdown than SIMDRAM (averaged across all four configurations). 
\omiii{Second, \prop using a single subarray and single bank for computation, provides 1.68$\times$, 1.54$\times$, and 1.52$\times$ the system throughput of SIMDRAM using 2, 4, and 8 banks for computation, respectively.}
%\agycomment{why not best performing?}
This happens because \prop
\li~utilizes idle resources \agy{\omi{at} DRAM mat granularity} to execute computation as soon as \agy{a mat is}
% they are 
available, thus reducing queuing time \omi{and improving parallelism}; and  
%(in contrast, for SIMDRAM, applications
% need to wait in the execution queue when\agy{are queued as long as} the number of \agy{idle} SIMDRAM-capable DRAM banks is lower than eight);\agycomment{is this true? we schedule resources in a DRAM bank granularity. applications should not wait if there is a bank}
\lii~reduces execution latency of a single application due to its concurrent execution of \omi{independent} \emph{bbop} instructions and support for \gfi{\gls{PuD}} vector reduction.
\omiii{Third}, even though \prop achieves similar fairness compared with \emph{SIMDRAM:4} and \emph{SIMDRAM:8} for application mixes with low and medium VF, \prop's maximum slowdown is 15\% (12\%) higher than \emph{SIMDRAM:8} (\emph{SIMDRAM:4}) for application mixes with high VF. 
This is because in \prop 
\li~applications share the DRAM mats available inside a single DRAM bank and
\lii~\emph{bbops} are dispatched to execution using an online \agy{\emph{first fit}} \agy{algorithm}. In this way, an application in a mix with \emph{high} occupancy and execution latency penalizes an application with \emph{low} occupancy and execution time, \agy{negatively} impacting fairness. In contrast, such interference does \emph{not} happen in \emph{SIMDRAM:8} since each application is assigned \agy{to} a different DRAM bank to execute \agy{at the cost of occupying eight banks instead of one}. \prop's fairness can be further improved by
\li~employing better scheduling algorithms that target quality-of-service~\cite{mutlu2008parallelism, luo2001balancing, xie2014improving,subramanian2013mise,subramanian2014blacklisting,ebrahimi2010fairness} or
\lii~using \omiii{\gls{SLP}~\cite{kim2012case} and} BLP~\cite{mutlu2008parallelism,kim2010thread,kim2012case,kim2016ramulator,lee2009improving}  in MIMDRAM\omiii{, i.e., exploiting multiple subarrays and multiple banks for \prop computation (\cref{sec:eval:scalability}}).\footnote{\omiii{In our extended version~\cite{mimdramextended}, we provide multi-programmed workload results while exploiting \gls{SLP} and BLP for \prop computation.}} We conclude that \prop is an efficient \omi{and high-performance} \gfi{\gls{PuD}} substrate when the system \agy{concurrently} executes several applications. }

\paratitle{\changeC{\#C1}\revC{CPU Multi-\omii{Programmed Workload Results}}} \revC{We evaluate how \prop performance compares to that of \omi{a state-of-the-art} CPU when executing multiple applications. To do so, we randomly \omi{generate} ten different application mixes, each containing eight applications out of our 12 applications. Then, we run each application mix in our baseline CPU (using multi-threading) and in \prop and compute the achieved system throughput for each system (using weighted speedup). Figure~\ref{fig_mult_app_analysis_cpu} shows the system throughput \prop achieves compared to the baseline CPU. We observe that \prop improves overall throughput by 19\%. This is because \prop can parallelize the execution of the applications in each application mix across the DRAM mats in a subarray. In contrast, when executing each application mix, the baseline CPU often suffers from contention in its shared resources (e.g., shared cache and DRAM bus). We conclude that \prop is an efficient substrate for highly-parallel environments.}  

\revdel{
\begin{table}[ht]
   \resizebox{0.8\columnwidth}{!}{
\begin{tabular}{|c||c|}
\hline
\textbf{Application Mix} & \textbf{Applications}                 \\ \hline \hline
Mix1                     & 2mm, 3mm, cov, dg, fdtd, gmm, hw, km  \\ \hline
Mix2                     & 2mm, 3mm, cov, fdtd, bp, hw, km, pca  \\ \hline
Mix3                     & 3mm, cov, dg, gmm, bp, hw, km, pca    \\ \hline
Mix4                     & 2mm, 3mm, dg, fdtd, gs, hw, km, pca   \\ \hline
Mix5                     & 2mm, cov, dg, fdtd, gs, bp, hw, pca   \\ \hline
Mix6                     & 2mm, cov, dg, fdtd, gs, hw, km, pca   \\ \hline
Mix7                     & 2mm, 3mm, cov, dg, gmm, bp, hw, pca   \\ \hline
Mix8                     & 2mm, 3mm, cov, dg, gmm, gs, hw, km    \\ \hline
Mix9                     & 2mm, fdtd, gmm, gs, bp, hw, km, pca   \\ \hline
Mix 10                   & 2mm, 3mm, cov, dg, fdtd, gmm, bp, pca \\ \hline
\end{tabular}
}
\end{table}
}

\begin{figure}[ht]
    \centering
    \includegraphics[width=0.85\linewidth]{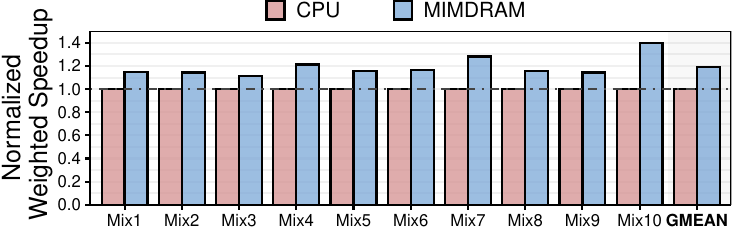}
    \caption{\revC{Multi-\gfcrii{programmed} \omii{workload results} for ten application mixes. Values are normalized to the baseline CPU.}}
    \label{fig_mult_app_analysis_cpu}
\end{figure}

\subsection{\revCommon{Comparison to Other PIM Architectures}}
\label{sec:eval:otherpims}

\revdel{\changeCM{\#CQ2}\revCommon{We compare \prop to two other state-of-the-art \gls{PIM} architectures: DRISA~\cite{li2017drisa} and Fulcrum~\cite{lenjani2020fulcrum}. 
DRISA is a \gls{PuM}/\gls{PnM} architecture that \emph{significantly} modifies the DRAM micro-architecture and organization to enable bulk in-DRAM computation (i.e., by using 3T1C DRAM cells to execute in-situ bitwise NOR operations or by adding logic gates \emph{near} the subarray's sense amplifiers). \revdel{DRISA employs a fine-grained interconnection network to shift data across DRAM columns, thus executing operations in a \emph{bit-parallel} mode (in contrast with SIMDRAM and \prop, which execute operations in a \emph{bit-serial} mode). 
In this analysis, we employ DRISA's 3T1C implementation to contrast both bit-serial and bit-parallel \gls{PuD} execution models.}
Fulcrum is a \gls{PnM} architecture that adds logic \emph{near} subarrays. 
Fulcrum's primary components are a series of shift registers (called walkers) that latch input/output DRAM rows and a narrow scalar ALU.\revdel{By using a single narrow scalar ALU for operations, Fulcrum provides a more flexible execution model than row-wide bitwise \gls{PuD} architectures. 
We compare \prop's and Fulcrum since both works have a similar goal.} We model DRISA 3T1C implementation and Fulcrum using the same baseline DDR4 device we use for SIMDRAM/\prop (see Table~\ref{table_parameters}).}}

\paratitle{\revCommon{Single-Application \omii{Results}}} \revCommon{We compare the performance of each \gls{PIM} architecture and \prop. Since DRISA and Fulcrum \omi{use large additional} area \omi{(i.e., 21\% and 82\% DRAM area overhead, respectively, over our baseline DDR4 DRAM chip)} to implement \gls{PIM} operations, we report area-normalized results (i.e., performance per area) for a fair comparison. We use the area values reported in both DRISA and Fulcrum's papers, scaled to the baseline DDR4 DRAM device we employ. We allow each mechanism to leverage the data parallelism available in each application by dividing the work evenly across DRISA's \gls{PIM}-capable DRAM banks and Fulcrum's \gls{PIM}-capable subarrays. Figure~\ref{fig_single_app_analysis_pim} shows the normalized performance per area for all 12 applications. Values are normalized to \prop. We make two observations. 
First, \prop achieves the highest performance per area compared to DRISA and Fulcrum. On average across the 12 applications, \prop performance per area is \mbox{1.18$\times$/1.92$\times$} that of DRISA and Fulcrum. This is because although DRISA and Fulcrum achieve higher absolute performance than \prop (\mbox{7.5$\times$} and \mbox{3.0$\times$}, respectively), such performance benefits come at the expense of \omi{very large} area overheads. \rCommon{\changerCM{\rCommon{\#CQ1}}While MIMDRAM incurs \omi{small} area cost on top of a DRAM array \changerC{\rC{\#C1}}(1.11\% DRAM area overhead, see~\cref{sec:eval:area}), DRISA and Fulcrum incur significantly larger area costs\revdel{ (21\% and 82\% DRAM area overhead, respectively)}.}
Second, for some applications (namely \texttt{hw}, \texttt{dg}, \texttt{km}, and \texttt{x264}), DRISA and Fulcrum achieve higher performance per area than \prop. We observe that such applications are dominated by multiplication operations, which are costly to implement using \prop's bit-serial approach.  
We conclude \prop \gfcrii{is an area-efficient \gls{PIM} architecture, which provides performance benefits compared to state-of-the-art \gls{PIM} architectures for a fixed area budget.}}

\begin{figure}[ht]
    \centering
    \includegraphics[width=0.8\linewidth]{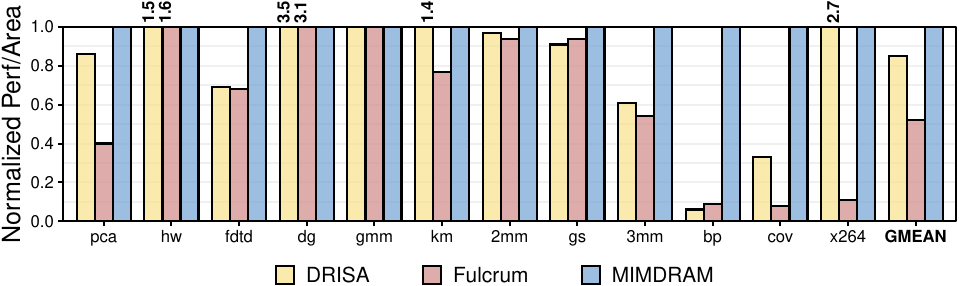}
    \caption{\revCommon{Single-application \omii{results} for different state-of-the-art \gls{PIM} architectures.}}
    \label{fig_single_app_analysis_pim}
\end{figure}

\paratitle{\revCommon{Multi-\omii{Programmed Workload Results}}}
Figure~\ref{fig_multi_app_analysis_pim} shows the system throughput, job turnaround time, and fairness that DRISA, Fulcrum, and \prop provide on average across all application mixes. \revdel{Since each architecture represents a different execution model (i.e., bit-parallel \gls{PuD}, scalar-based \gls{PnM}, and bit-serial \gls{PuD} computing), we consider the performance of \emph{each} architecture when computing the \emph{performance alone} component of the weighted speedup, harmonic speedup, and maximum slowdown metrics.} We employ BLP \omiii{in DRISA and \prop, and} \gls{SLP}~\cite{kim2012case} \omiii{in Fulcrum} to enable \gls{MIMD} execution. \changerCM{\rCommon{\#CQ1}}We make two observations. First, \omiii{all three \gls{PIM} architectures achieve similar system throughput. On average across all application mixes and configurations, DRISA, Fulcrum, and \prop achieve 1.20$\times$,  1.17$\times$, and  1.19$\times$ the system throughput of \emph{DRISA:1}, respectively.
Second, when considering a \emph{single} DRAM subarray for computation, \prop achieves 8\%  and 11\% \emph{higher} fairness than DRISA and Fulcrum, respectively. }

\begin{figure}[ht]
    \centering \includegraphics[width=0.8\linewidth]{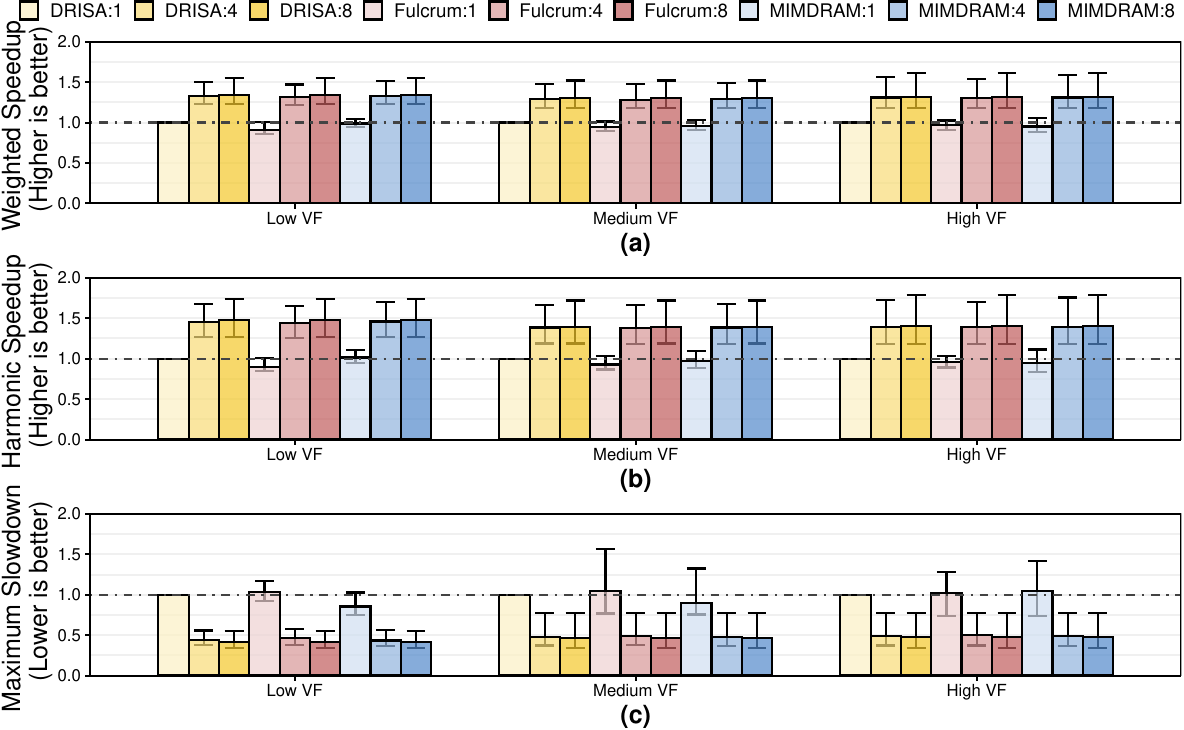}
    \caption{\revCommon{Multi-\omii{programmed workload results} for different \gls{PIM} architectures and three types of application mixes. \emph{VF} stands for vectorization factor. \emph{DRISA:X}\omiii{/\emph{MIMDRAM:X}} (\emph{Fulcrum:X}) uses \emph{X} DRAM banks (subarrays) for computation. \omiii{Values are normalized to \emph{DRISA:1}.} Whiskers extend to the minimum and maximum \gfcrii{observed} data points.}}
    \label{fig_multi_app_analysis_pim}
\end{figure}

\subsection{\gfcriii{\prop with \gls{SLP} \& BLP}}
\label{sec:eval:scalability}

One of the main advantages of \gls{PuD} architectures is the ability to exploit the \omiii{large} internal DRAM parallelism for computation.
A \gls{PuD} substrate can leverage \gls{SLP}~\cite{kim2012case} and BLP~\cite{mutlu2008parallelism,kim2010thread,kim2012case,kim2016ramulator,lee2009improving} techniques to operate \emph{simultaneously} \omiii{exploit} the many DRAM subarrays (e.g., 8--64 \omiii{per bank}) and banks (e.g., 8--16 \omiii{per rank}) in a DRAM chip for \gls{PuD} computation. To this end, we perform a sensitivity analysis of  SIMDRAM and \prop's performance for our twelve applications when using multiple DRAM subarrays (1--64 \omiii{per bank}) and DRAM banks (1--16 \omiii{per rank}) for \gls{PuD} computation, as Figure~\ref{fig_scaling} depicts.
We make two observations from the figure. 

\begin{figure}[ht]
    \centering
    \includegraphics[width=0.7\linewidth]{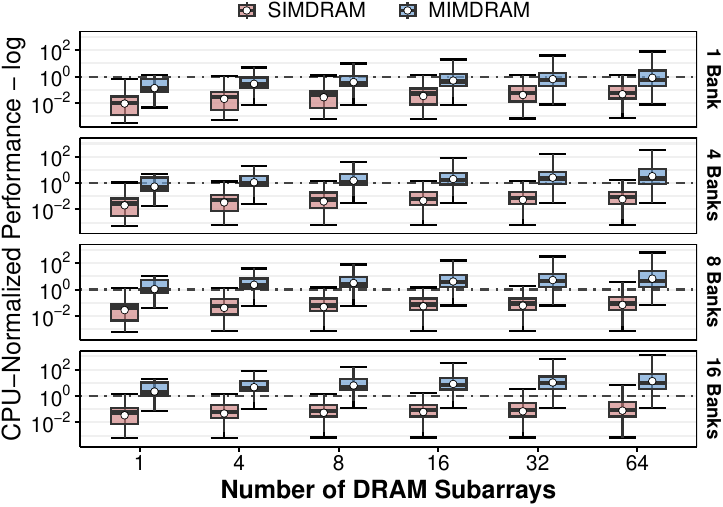}
    \caption{\gfcriii{Distribution of single-application performance across all twelve applications when varying the number of DRAM subarrays and banks for SIMDRAM and \prop. Values are normalized to the baseline CPU. Whiskers extend to the minimum and maximum observed data points on either side of the box. Bubbles depict average values.}}
    \label{fig_scaling}
\end{figure}

First, by \emph{fully} leveraging the internal DRAM parallelism in a DRAM chip, \prop can provide \emph{significant} performance gains compared to the baseline CPU. On average across all twelve applications, \prop (using 64 DRAM subarrays \omiii{per bank} and 16 banks for \gls{PuD} computation) achieves 13.2$\times$ the performance of the CPU. 
Second, in contrast, SIMDRAM \emph{fails} to outperform the baseline CPU, even when fully utilizing the internal DRAM for computation (0.08$\times$ the performance of the CPU when using 64 DRAM subarrays \omiii{per bank} and 16 banks). This is because:
\li~\prop unlocks further parallelism by leveraging idle DRAM mats for computation and
\lii~\prop reduces the latency of costly vector reduction operations.
Third, we observe that \prop \omiii{can lead} to performance loss compared to the baseline CPU \omiii{for some workloads}, even when using all available DRAM subarrays and banks for computation, for two main reasons:
\li~quadratically-scaling \gls{PuD} operations (i.e., multiplication and divisions) or
\lii~\gls{PuD} vector reduction operations dominate \prop's execution time of the application. 
In the first case, \prop's performance could be further improved by leveraging lower-latency algorithms for costly \gls{PuD} operations (e.g., bit-parallel multiplication and division algorithms~\cite{leitersdorf2023aritpim}) \omiii{or performing such complex operations near memory (close to DRAM)~\pnmshort.}
In the second case, \prop would benefit from the assistance of \gls{PnM} architectures to perform faster vector reduction operations in DRAM, at the cost of an increase in area cost.
We conclude that \prop highly benefits from exploiting \gls{SLP} and BLP for \gls{PuD} computation.

\paratitle{Performance \& Energy Efficiency} Figure~\ref{fig_scaling_perf_energy} shows the CPU-normalized performance (Figure~\ref{fig_scaling_perf_energy}a), CPU-normalized energy (Figure~\ref{fig_scaling_perf_energy}b), and CPU-normalized performance per Watt (Figure~\ref{fig_scaling_perf_energy}c) when allowing SIMDRAM and MIMDRAM to \omi{simultaneously} issue in-DRAM operations across \omii{all} 16 banks and 64 subarrays \omi{per bank} in a DRAM rank. We make two key observations.

\begin{figure}[ht]
    \centering
    \includegraphics[width=0.7\linewidth]{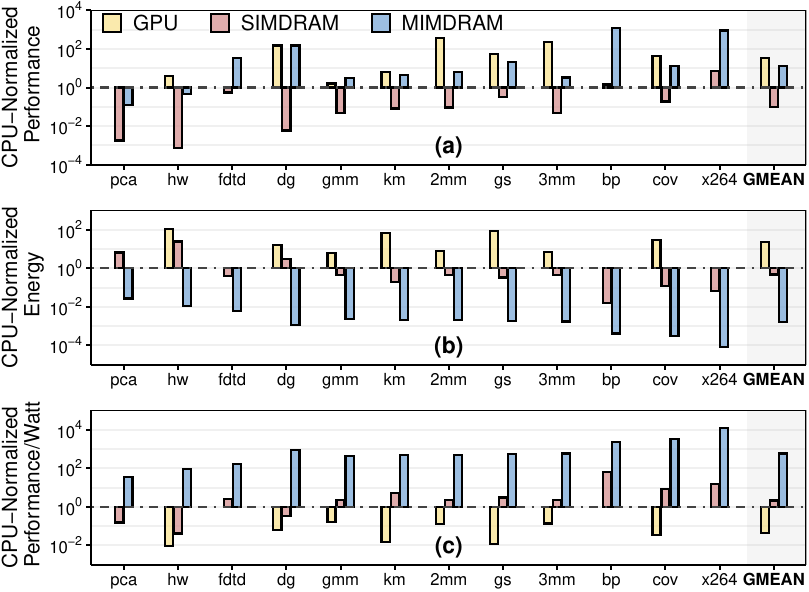}
    \caption{\gfcriii{GPU, SIMDRAM, and MIMDRAM performance (a), energy (b), and performance per Watt (c) for twelve real-world applications when using 16 DRAM banks and 64 DRAM subarrays per bank for in-DRAM computing. Values are normalized to the baseline CPU. }}
    \label{fig_scaling_perf_energy}
\end{figure}

First, MIMDRAM \omii{provides} \li~13.2$\times$/0.22$\times$/173$\times$ the performance,  \lii~0.0017$\times$/0.00007$\times$/0.004$\times$ the energy consumption, and \liii~582.4$\times$/13612$\times$/272$\times$ the performance per Watt of the CPU/GPU/SIMDRAM baseline.
\gf{Second, we observe that MIMDRAM's end-to-end performance gains are limited by the throughput of the inter- and intra-mat interconnects, which are utilized during in-DRAM reduction operations. 
\omi{If we} consider only MIMDRAM's arithmetic throughput \omi{(i.e., no reduction operations)}, we observe that MIMDRAM \omii{provides} 272$\times$ and 11$\times$ the performance of the CPU and GPU baselines, respectively. 
We believe that combining \gls{PuM} and \gls{PnM} holistically, where auxiliary logic placed within the logic layer of 3D-stacked memories is used for high-throughput in-DRAM reduction and DRAM cells are used for high-throughput in-DRAM bulk arithmetic, would be beneficial to improve MIMDRAM's end-to-end performance.}
We conclude that MIMDRAM \omi{enables effective exploitation of DRAM} bank\omi{-level} and subarray-level parallelism \omi{for massively-parallel bulk bitwise execution.}

\subsection{\omt{2}{Hardware} Area Analysis}
\label{sec:eval:area}

\atb{We use CACTI~\cite{cacti,muralimanohar2007optimizing} to model the area of a DRAM chip (Table~\ref{table_parameters}) using a \SI{22}{\nano\meter} technology node. We implement \prop{}'s chip select and mat identifier logic in Verilog HDL and synthesize the HDL using the Synopsys Design Compiler~\cite{synopsysdc} with a \SI{65}{\nano\meter} process technology node.}\footnote{\rB{\changerB{\rB{\#B1}}We use a \SI{65}{\nano\meter} technology node since that is the best CMOS standard cell library we have \omi{access} \omii{to} in \omi{our} environment. We scaled our design to a \SI{22}{\nano\meter} technology node following prior works' methodology~\cite{stillmaker2017Scaling,ghiasi2022genstore,dai2022dimmining,zhang2021sara}.}} 

\paratitle{DRAM Bank Area} 
%\atb{We model the baseline DRAM chip's area as \SI{22.8}{\milli\meter\squared}. 
We evaluate the \agy{area} overhead of
\li~mat isolation transistors, 
\lii~row decoder latches, 
\liii~mat selectors,  
\liv~the wiring to propagate mat selector output to mat isolation transistors (matlines), and
\lv~\omi{multiplexers} and wiring of the inter-mat \gfcri{interconnect}. \prop{} incurs 1.15\% area overhead over the baseline DRAM bank.

% 116.3

\paratitle{DRAM Chip I/O Area} \atb{The total  area overhead for \prop{}'s chip select and mat identifier is \emph{only}
\SI{825.7}{\micro\meter\squared} at a \SI{65}{\nano\meter} technology node. We estimate the equivalent area overhead at a \SI{22}{\nano\meter}  technology node to be \SI{116.3}{\micro\meter\squared}~\cite{stillmaker2017Scaling}.}

% 1.00111210

Overall, \prop{} increases the area of the evaluated DRAM chip (16 banks and I/O) by only 1.11\%.

\paratitle{\omi{\prop} Control Unit \& Transposition Unit Area} \gf{The main components in \omi{the} \prop control unit are the
\li~\emph{bbop} buffer, 
\lii~mat scoreboard, and
\liii~\uprog{} processing engines. We set the size of the \emph{bbop} buffer to \SI{2}{\kilo\byte}, which accommodates up to 1024 \emph{bbop}s. 
%We empirically find that this size is sufficient for our real-world applications.
%\agycomment{how would it scale if real-workload application demands more?} 
\gf{The mat scoreboard requires \omi{128~bits} of storage, one bit per DRAM mat \gfcrii{per subarray}. 
A single \uprog{} processing engine has an area of \SI{0.03}{\milli\meter\squared}. We \gfcrii{empirically} include eight \uprog{} processing engines in our design. \omi{W}e estimate\gfcrii{, using CACTI,} that \prop control unit area is \SI{0.253}{\milli\meter\squared}.} \prop transposition unit has an \omi{area equal to}  \omi{the} SIMDRAM transposition unit (of \SI{0.06}{\milli\meter\squared}).\revdel{\footnote{\gfcrii{We slightly modify the fields of the \emph{object tracker} to fit the \omi{14~bits} mat range information for a memory object in \prop. We  
\li~occupy the unused \omi{7~bits} and 
\lii~reduce the bits used \gfcrii{to store the size information for a memory object from \omi{32~bits} to \omi{25~bits}.}}}} Considering the area of the control and transposition units, \prop has a low area overhead of 0.6\% \omi{over} the die area of a \omi{state-of-the-art} Intel Xeon E5-2697 v3 CPU~\cite{dualitycache}.}

%\agycomment{shouldn't we at least compare its area cost to area cost of SIMDRAM?}

%%% total area is 0.313 mm2 

 %\input{mainmatter/05_mimdram/sections/07_related_work}
%\glsresetall

\section{Summary}
\label{conclusion}

We introduce \prop, a hardware/software co-designed \gfcrii{processing-using-DRAM (\gls{PuD})} substrate that can allocate and control only the needed computing resources inside DRAM for \gfi{\gls{PuD}} computing. 
On the hardware side, \prop introduces simple modifications to the DRAM architecture that enables the execution of
\li~different \gfi{\gls{PuD}} \gfcrii{operations} concurrently inside a single DRAM \gfcrii{subarray} in a \gfcrii{multiple-instruction multiple-data (\gls{MIMD})} fashion, and
\lii~native vector reduction computation.
On the software side, \prop implements a series of compiler passes that automatically identify and map code regions to the underlying \gfi{\gls{PuD}} substrate. 
We experimentally demonstrate that \prop provides significant benefits over state-of-the-art CPU, GPU, and \gfcrii{processing-using-memory (\gls{PuM}) and processing-near-memory (\gls{PnM})} systems. \gfcrii{We hope and believe that our work can inspire more efficient and easy-to-program \gls{PuD} systems. The source code of \prop is freely available at \url{https://github.com/CMU-SAFARI/MIMDRAM}.}
%We hope that future work builds on \prop to further ease the adoption of \gfi{\gls{PuD}} architectures.

% On the hardware side, \prop introduces simple modifications to the DRAM architecture that enables the execution of
% \li~different \gfi{\gls{PuD}} instructions concurrently inside a single DRAM array in a \gls{MIMD}-fashion, and
% \lii~native vector reduction computation.
% On the software side, \prop implements a series of compiler passes that automatically identify and map code regions to the underlying \gfi{\gls{PuD}} substrate. 
% We experimentally demonstrate that \prop provides significant benefits over state-of-the-art CPU, GPU, and \gfi{\gls{PuD}} systems. 
% %We hope that future work builds on \prop to further ease the adoption of \gfi{\gls{PuD}} architectures.

% % \glsresetall{}
\chapter[Proteus: Achieving High-Performance Processing-Using-DRAM with
Dynamic Bit-Precision, Adaptive Data Representation, and Flexible Arithmetic]{\emph{Proteus}: Achieving High-Performance Processing-Using-DRAM with Dynamic Bit-Precision, Adaptive Data Representation, and Flexible Arithmetic}
\label{chap:proteus}

\newif\ifcameraready
\camerareadytrue
%\camerareadyfalse

\ifcameraready
    \providecommand{\gfcr}[1]{\textcolor{black}{#1}}
    \providecommand{\gfcri}[1]{\textcolor{black}{#1}} 
    \providecommand{\gfcrii}[1]{\textcolor{black}{#1}} 
    \providecommand{\gfcriii}[1]{\textcolor{black}{#1}} 
    \providecommand{\gfcriv}[1]{\textcolor{black}{#1}} 

    \providecommand{\omcri}[1]{\textcolor{black}{#1}}
    \providecommand{\omcrii}[1]{\textcolor{black}{#1}}
    \providecommand{\omcriii}[1]{\textcolor{black}{#1}}
    \providecommand{\omcriv}[1]{\textcolor{black}{#1}}

\else 
    \providecommand{\gfcr}[1]{\textcolor{black}{#1}} 
    \providecommand{\gfcri}[1]{\textcolor{black}{#1}} 
    \providecommand{\gfcrii}[1]{\textcolor{black}{#1}} 
    \providecommand{\gfcriii}[1]{\textcolor{black}{#1}} 
    \providecommand{\gfcriv}[1]{\textcolor{blue}{#1}} 
    
    \providecommand{\omcri}[1]{\textcolor{black}{#1}}
    \providecommand{\omcrii}[1]{\textcolor{black}{#1}}
    \providecommand{\omcriii}[1]{\textcolor{black}{#1}}
    \providecommand{\omcriv}[1]{\textcolor{blue}{#1}}

\fi

\renewcommand{\prop}{\textit{Proteus}\xspace}

\providecommand\add{\red{\textbf{X}}}
\providecommand\bbop{\emph{bbop}\xspace}

\providecommand\uproglib{\emph{Parallelism-Aware \uprog Library}\xspace}
\providecommand\dynengine{\emph{Dynamic Bit-Precision Engine}\xspace}
\providecommand\uprogunit{\emph{\uprog Select Unit}\xspace}
\providecommand\bitprec{\emph{Bit-Precision Calculator Unit}\xspace}
\providecommand\prelut{\emph{Pre-Loaded Cost Model \glspl{LUT}}\xspace}

\providecommand\objtrack{\emph{Object Tracker}\xspace}
\providecommand\costmodel{\emph{Cost Model Logic}\xspace}
\providecommand\gbidx{\emph{\textmu{}Program\_addr}\xspace}
\providecommand\idx{\emph{\textmu{}Program\_id}\xspace}

\providecommand\aap{\texttt{AAP}/\texttt{AP}\xspace}
\providecommand\aaps{\texttt{AAP}s/\texttt{AP}s\xspace}

\providecommand\uop{\textmu{}Op\xspace}
\providecommand\uops{\textmu{}Ops\xspace}

\providecommand\uprog{\textmu{}Program\xspace}
\providecommand\uprogs{\textmu{}Programs\xspace}

\providecommand{\paratitle}[1]{\vspace{4pt}\noindent\textbf{#1.}}

%
%%%%%%%%%% Text editing macros %%%%%%%%%%
\providecommand{\tempcommand}[1]{\renewcommand{\arraystretch}{#1}}

\providecommand\ignore[1]{ }
\providecommand{\revdel}[1]{}
\providecommand{\sgdel}[1]{}

%\setdeletedmarkup{\textcolor{red}{\sout{#1}}}

\providecommand{\prtagA}[1]{\lfbox[padding=1pt, border-color=red, background-color=red!20]{\revA{\textbf{\scriptsize #1}}}}
\providecommand{\prtagB}[1]{\lfbox[padding=1pt, border-color=mygreen, background-color=mygreen!20]{\revB{\textbf{\scriptsize #1}}}}
\providecommand{\prtagC}[1]{\lfbox[padding=1pt, border-color=safetyorange, background-color=safetyorange!20]{\revC{\textbf{\scriptsize #1}}}}
\providecommand{\prtagD}[1]{\lfbox[padding=1pt, border-color=heliotrope, background-color=heliotrope!20]{\revD{\textbf{\scriptsize #1}}}}
\providecommand{\prtagE}[1]{\lfbox[padding=1pt, border-color=rufous, background-color=rufous!20]{\revE{\textbf{\scriptsize #1}}}}
\providecommand{\prtagCM}[1]{\lfbox[padding=1pt, border-color=blue, background-color=blue!20]{\revCommon{\textbf{\scriptsize #1}}}}

\newif\ifasplosrevisionsubmission
\asplosrevisionsubmissiontrue
%\asplosrevisionsubmissionfalse

\ifasplosrevisionsubmission
    \providecommand{\omrev}[1]{#1}
\else
    \providecommand{\omrev}[1]{\textcolor{orange}{#1}}
\fi

\newif\ifasplosrevision
%\asplosrevisiontrue
\asplosrevisionfalse
\ifasplosrevision 
    \providecommand{\asplosrev}[1]{\textcolor{blue}{#1}}
\else
    \providecommand{\asplosrev}[1]{\textcolor{black}{#1}}
    \renewcommand{\hl}[1]{#1}
\fi

\newif\ifasplossubmission
\asplossubmissiontrue
%\asplossubmissionfalse
\ifasplossubmission 
    \providecommand{\gfasplos}[1]{\textcolor{black}{#1}}
    \providecommand{\agyasplos}[1]{\textcolor{black}{#1}}
    \providecommand{\agyasploscomment}[1]{}
\else
    \providecommand{\gfasplos}[1]{\textcolor{blue}{#1}}
    \providecommand{\agyasplos}[1]{\textcolor{red}{#1}}
    \providecommand{\agyasploscomment}[1]{\textcolor{red}{\textbf{!!!~Giray:} #1}}
    
\fi

\newif\ifmicrosubmission
\microsubmissiontrue
%\microsubmissionfalse
\ifmicrosubmission 
    \providecommand{\gfmicro}[1]{\textcolor{black}{#1}}
    \providecommand{\agymicro}[1]{\textcolor{black}{#1}}
    \providecommand{\param}[1]{#1}
    \providecommand{\agymicrocomment}[1]{}
\else
    \providecommand{\gfmicro}[1]{\textcolor{blue}{#1}}
    \providecommand{\param}[1]{\textcolor{red}{#1}}
    \providecommand{\agymicro}[1]{\textcolor{red}{#1}}
    \providecommand{\agymicrocomment}[1]{\textcolor{red}{\textbf{!!!~Giray:} #1}}
    
\fi

\newif\ifiscarevision
%\iscarevisiontrue
\iscarevisionfalse
\ifiscarevision 
    \providecommand{\revA}[1]{\textcolor{red}{#1}}
    \providecommand{\revB}[1]{\textcolor{mygreen}{#1}}
    \providecommand{\revC}[1]{\textcolor{safetyorange}{#1}}
    \providecommand{\revD}[1]{\textcolor{heliotrope}{#1}}
    \providecommand{\revE}[1]{\textcolor{rufous}{#1}}
    \providecommand{\revCommon}[1]{\textcolor{blue}{#1}}

\else
    \providecommand{\revA}[1]{\textcolor{black}{#1}}
    \providecommand{\revB}[1]{\textcolor{black}{#1}}
    \providecommand{\revC}[1]{\textcolor{black}{#1}}
    \providecommand{\revD}[1]{\textcolor{black}{#1}}
    \providecommand{\revE}[1]{\textcolor{black}{#1}}
    \providecommand{\revCommon}[1]{\textcolor{black}{#1}}
\fi

\newif\ifcut
%\cuttrue
\cutfalse

\ifcut
   \providecommand{\gfcut}[1]{} 
\else
    \providecommand{\gfcut}[1]{\textcolor{red}{\sout{#1}}}
\fi

\newif\ifiscasubmission
\iscasubmissiontrue
%\iscasubmissionfalse

\ifiscasubmission
    \providecommand{\gfisca}[1]{#1}
    \providecommand{\gfbisca}[1]{}
    \providecommand{\sg}[1]{#1}
    \providecommand{\sgi}[1]{#1}
    \providecommand{\om}[1]{#1}
\else
    \providecommand{\gfbisca}[1]{\textcolor{blue}{\textit{GF: #1}}}
    \providecommand{\gfisca}[1]{\textcolor{blue}{#1}}

    \providecommand{\sg}[1]{\textcolor{red}{#1}}
    \providecommand{\sgi}[1]{\textcolor{brickred}{#1}}

    \providecommand{\om}[1]{\textcolor{orange}{#1}}
\fi

\newif\ifsubmission
\submissiontrue
%\submissionfalse

\ifsubmission
    \providecommand{\kangqi}[1]{#1}
    \providecommand{\juan}[1]{#1}
    \providecommand{\gf}[1]{#1}
    \providecommand{\gfii}[1]{#1}
    
    \providecommand{\jgl}[1]{}

    \providecommand{\gfb}[1]{}
    \providecommand{\mayank}[1]{}
    \providecommand{\ms}[1]{#1}
    \providecommand{\agy}[1]{#1}
    \providecommand{\agycomment}[1]{}
\else
    \providecommand{\jgl}[1]{\textcolor{brickred}{\textit{JGL: #1}}}
    \providecommand{\gfb}[1]{\textcolor{blue}{\textit{GF: #1}}}
    \providecommand{\todo}[1]{\textcolor{red}{\textbf{TODO: #1}}}
    \providecommand{\juan}[1]{\textcolor{brickred}{#1}}
    \providecommand{\mayank}[1]{\textcolor{green}{\textit{Mayank: #1}}}
    \providecommand{\kangqi}[1]{\textcolor{byzantium}{#1}}
    \providecommand{\gf}[1]{\textcolor{blue}{#1}}

    \providecommand{\gfii}[1]{\textcolor{red}{#1}}
    
    \providecommand{\ms}[1]{\textcolor{cyan}{#1}}
    \providecommand{\agy}[1]{\textcolor{orange}{#1}}
    \providecommand{\agycomment}[1]{\agy{\textbf{[@gy:} #1\textbf{]}}}

\fi

\section{Motivation \& Goal}
\label{sec_motivation}

\gfcr{We discuss three \omcrii{major shortcomings} of prior \gls{PuD} architectures: \emph{static data representation}, \emph{support for only throughput-oriented execution}, and \emph{high latency for high-precision operands.}
}

\paratitle{\omcrii{\emph{Limitation 1}:}~\gf{Static Data Representation}} 
\sgi{\Gls{PuD} architectures naively \omcrii{utilize} conventional data formats (e.g., two's complement) and fixed operand bit-precision (e.g., 32-bit integers) to implement bit-serial computation. However, because bit-serial latency \omcrii{directly increases} with bit-precision, these architectures experience subpar performance since \asplosrev{\hl{\omcrii{an} application's data with small dynamic range (i.e., narrow values) are often stored in large data formats}}~\cite{pekhimenko2012base,alameldeen2004adaptive,islam2010characterization,ergin2006exploiting,brooks1999dynamically,ergin2004register,budiu2000bitvalue,wilson1999case} that waste most of the bit-precision.} \asplosrev{\hl{Note that data values \omcrii{often} become narrow dynamically at runtime.}}
\gf{\sgdel{Prior bit-serial \gls{PuD} architectures \emph{undermine} the full benefits of the underlying bit-serial execution model by using a \emph{static data representation} for \gls{PuD} computation. This is because prior works naively utilize conventional data formats (e.g., two's complement) and operands' bit-precision (e.g., 32-bit integers) for \gls{PuD} computation, which leads to subpar performance. 
In a bit-serial execution model, the number of operations required to perform a given computation is tight to the bit-precision of the computation. \gfisca{Thus, using fewer bits to represent operands results in lower latency. 
One way to minimize the bit-precision for bit-serial \gls{PuD} computation is to exploit \emph{narrow values}. As several works observe~\cite{pekhimenko2012base,alameldeen2004adaptive,islam2010characterization,ergin2006exploiting,brooks1999dynamically,ergin2004register,budiu2000bitvalue,wilson1999case}, programmers often over-provision the bit-\gfisca{precision} used to store operands, using large data types (e.g., a 32-bit \sg{or 64-bit} integer) to store small \gfisca{(i.e., narrow}) values \sg{(e.g., an up to 255-pixel value in an RGB image)}.}}%
%For example, prior works~\cite{jang2022encore} show that 67.8\% of the weights of some layers of quantized neural networks (NNs) are narrow values. 
Narrow values have been exploited in many scenarios, \sgi{e.g.,} cache compression~\gfcrii{\cite{pekhimenko2012base,alameldeen2004adaptive,wilson1999case,islam2010characterization,duan2014exploiting,molina2003non,kong2012exploiting,pekhimenko2015toggle,pekhimenko2016case,pekhimenko2015exploiting,gena-thesis}}, 
register files~\gfcrii{\cite{ergin2004register,wang2017gpu,hu2006register,wang2009exploiting,ergin2006exploiting,ozsoy2010dynamic,mittal2017design,ergin2006exploitingpatmos}}, 
logic synthesis \gfcrii{\& circuit optimizations}~\cite{canesche2022polynomial,canis2013software,pilato2013bambu,onur2009exploiting,osmanlioglu2009reducing}\gfcrii{, 
neural network quantization~\cite{jang2022encore, albericio2017bit},
error tolerance~\cite{ergin2006exploiting,karsli2012enhanced,ergin2008reducing}}.}

\noindent \gf{\omcrii{\textbf{\emph{Opportunity 1:} Narrow Values for \gls{PuD} Computation.}}\revdel{In the context of \gls{PuD},} \gfmicro{N}arrow values \gfisca{can} be exploited to reduce the bit-precision of a \gls{PuD} operation to that of the best-fitting number of bits\omcrii{, thereby}, improving overall performance. 
\gfisca{We quantify the required bit-precision in \gls{PuD}-friendly real-world applications in Figure~\ref{fig:narrow_values}. 
We define as \emph{required bit-precision} the minimum number of bits required to represent the input operands of the \gls{PuD} operation.\footnote{For example, if the input operand is an integer storing the value `\texttt{2}', the required bit-precision for such an input operand would be \sg{3} bits (\sg{two bits to represent the data \omcrii{value} and one bit to represent the sign}).} 
\revD{\changeD{D2}We collect the required bit-precision dynamically in three main steps: we 
\li~instrument \asplosrev{\hl{loops in applications that can be auto-vectorized using LLVM's loop auto-vectorization pass~\mbox{\cite{ lattner2008llvm,sarda2015llvm, lopes2014getting,writingpass}} (since prior work~\mbox{\cite{mimdramextended}} shows that such loops are \omcrii{well-suited} for \mbox{\gls{PuD}} execution)}} to output the data \omcrii{values} such \omcrii{loops use} \gfcrii{(i.e., we collect the data values of \emph{each} data array that is used as input/output of an auto-vectorized arithmetic instruction across the auto-vectorized loops in an application)}, 
\lii~execute the application to completion, \liii~post-process the output file containing the loop information data to calculate the required bit-precision.}}} 

\begin{figure}[ht]
    \centering
    \includegraphics[width=0.95\linewidth]{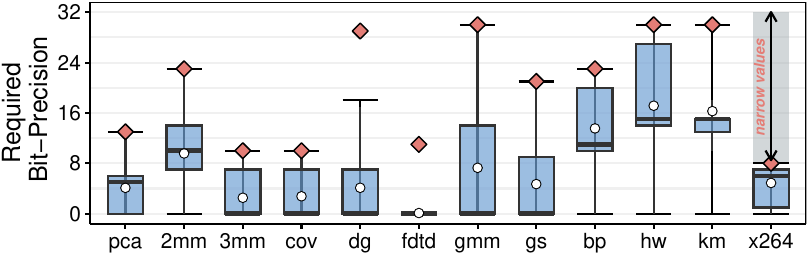}
    \caption{\revA{\gfisca{\sgi{\omcrii{Required b}it-precision distribution\revdel{ (i.e., minimum number of bits needed for an operand)} for \gfcrii{input/output data arrays of auto-vectorized arithmetic instructions in loops across} 12 applications.
    The box represents the 25th to 75th percentiles, with whiskers extending to the smallest/largest precision (with a diamond at the largest precision and a bubble at the mean precision).}
    \sgdel{Bit-precision distribution of twelve real-world applications.
    The y-axis indicates the \emph{required bit-precision} for a given \gls{PuD} operation, defined as the minimum number of bits required to represent input operands. Whiskers extend to the minimum and maximum data points on either side of the box. A bubble depicts average values. A diamond indicates the \emph{minimum} bit-precision that covers 100\% of the input operands.\prtagA{Fix.}}}}}
    \label{fig:narrow_values}
\end{figure}

{{We make two observations.
First, \gfisca{all our real-world applications display a \emph{significant} amount of narrow values. 
In such applications, the input bit-precision can be reduced from the native 32-bit to 20-bit (min. of 8-bit, max. of 30-bit)} \sg{on average across \emph{all} applications}. 
By doing so, the performance of the underlying \gls{PuD} architecture can improve by 1.6$\times$, \sg{in case the application utilizes linearly-scaling \gls{PuD} operations (such as addition~\omcrii{\cite{hajinazarsimdram}}),} or 2.6$\times$, \sg{in case the application utilizes quadratically-scaling \gls{PuD} operations (such as multiplication~\omcrii{\cite{hajinazarsimdram}})}. 
Second, the bit-precision significantly varies \omcrii{across data arrays within} a given application. This indicates the need for a mechanism that can \emph{dynamically} identify the target bit-precision for a given \gls{PuD} operation \gfmicro{(similar to prior works that leverage narrow values for tasks other than \gls{PuD}~\cite{pekhimenko2012base,alameldeen2004adaptive,islam2010characterization,ergin2006exploiting,brooks1999dynamically,ergin2004register,budiu2000bitvalue,wilson1999case,hu2006register,wang2017gpu,lipasti2004physical,loh2002exploiting})}.}}
\revD{\label{rd.2}\changeD{D2}As prior work points out~\cite{brooks1999dynamically}, 
static compiler analyses \emph{cannot} identify the bit-precision of dynamically allocated and initialized data arrays. \revD{We investigated several prior compiler works~\cite{canesche2022polynomial,stephenson2000bidwidth,rodrigues2013fast,campos2012speed,cong2005bitwidth} that perform bit-width identification. However,  such works are limited to identifying the bit-precision of statically\omcrii{-}allocated variables.}} 
 
\paratitle{\gf{\omcrii{\emph{Limitation 2:}}~Throughput-Oriented Execution}} \gf{\omcrii{Existing} \gls{PuD} architectures favor throughput-oriented execution as DRAM parallelism can \sgi{partially} hide the activation latency in a \uprog. To further improve throughput, prior works~\omcrii{\cite{hajinazarsimdram, peng2023chopper,mimdramextended}} use DRAM's \gls{BLP} to 
\li~distribute \sgi{independent \uprogs} across DRAM banks~\cite{hajinazarsimdram}, or
\lii~parallelize data writing and \gls{PuD} computation of \emph{different} \sgi{\uprogs} targeting \emph{different} banks~\cite{peng2023chopper}. However, such approaches cannot reduce the latency of a \emph{single} \sgi{\uprog}.}

\noindent \gf{\omcrii{\textbf{\emph{Opportunity 2:} DRAM Parallelism for Latency-Oriented Execution.}} 
\sgi{\revdel{We aim to leverage the internal parallelism of a DRAM bank to reduce the latency of a \emph{single} \uprog.}
We make the \emph{key observation} that several \gfasplos{primitives in} a \uprog \gfcrii{(\omcriii{i.e.,} \aap primitives that execute in-DRAM row copy or in-DRAM \texttt{MAJ3}/\texttt{NOT} operations)} can be executed concurrently, as they are independent of one another.
Figure~\ref{fig:uprogexample} shows this opportunity for a two-bit addition.
\gfcrii{In conventional bit-serial execution (Figure~\ref{fig:uprogexample}a), \emph{all} bits of the input arrays $A$ and $B$ are placed in a \emph{single} DRAM subarray.
Because of that, \emph{all} in-DRAM primitives in a \uprog are \emph{serialized}, enabling the execution of \omcriii{only} a single bit-position at a time.
In our example,  
\li~DRAM cycles \circled{1}--\circled{3} execute \texttt{MAJ3}/\texttt{NOT} operations over the \emph{least-significant bits} (LSBs) of the input arrays $A$ and $B$, i.e., $A_0$ and $B_0$; and afterwards
\lii~DRAM cycles \circled{4}--\circled{6} execute \texttt{MAJ3}/\texttt{NOT} operations over the \emph{most-significant bits} (MSBs) of the input arrays $A$ and $B$, i.e., $A_1$ and $B_1$.\footnote{\omcriii{For simplicity,} \gfcrii{we do \emph{not} depict DRAM cycles that perform in-DRAM row copy operations in Figure~\ref{fig:uprogexample}.}}  
However, the only \emph{inter-bit} dependency in the \uprog is the \emph{carry propagation} (\circlediii{i} in Figure~\ref{fig:uprogexample}a). 
In contrast, we can leverage \emph{bit-level parallelism} to \emph{concurrently} execute bit-independent in-DRAM primitives \omcriii{\emph{across multiple}} DRAM subarrays. 
In our example, we can reduce the overall latency of the bit-serial \gls{PuD} addition operation by 
\li~\emph{distributing} the individual bits of data-elements from arrays $A$ and $B$ across two DRAM subarrays (i.e., $subarray_0 \leftarrow  \{A_0,B_0\}; subarray_{1} \leftarrow  \{A_{1},B_{1}\}$), 
\lii~executing the required in-DRAM row copies (not shown) and \texttt{MAJ3}/\texttt{NOT} operations for the LSBs (DRAM cycles \circled{1}--\circled{3} in Figure~\ref{fig:uprogexample}b) and MSBs (DRAM cycles \circled{2}--\circled{4} in Figure~\ref{fig:uprogexample}b) \emph{concurrently}, and
\liii~serializing \emph{only} the carry generated from the LSBs to the MSBs of the input arrays (\circlediii{ii} in Figure~\ref{fig:uprogexample}b).}}

\revdel{and how we could concurrently execute primitives for the addition if its operands were distributed across two different subarrays\revdel{ (compared to the serialization of all primitives in a single subarray, as existing \gls{PuD} architectures do currently)}\gfasplos{
, since} the only \omcrii{dependence} between the two bits of the addition is the \emph{carry propagation} (\circled{1} in Figure~\ref{fig:uprogexample}a).
\gfasplos{By placing} each bit of the operand into a different subarray, we can parallelize independent primitives and serialize \emph{only} for the carry propagation (\circled{2} in Figure~\ref{fig:uprogexample}b).}
\sgdel{We aim to leverage DRAM's internal parallelism to reduce the latency of a single \gls{PuD} operation. We make the \emph{key observation} that several row copies and TRAs in a \uprog can be executed in parallel. Figure~\ref{fig:uprogexample} illustrates such observation.
The figure depicts the graph representation of the \uprog for a two-bit bit-serial addition operation and the execution time diagram of such \uprog using one (Figure~\ref{fig:uprogexample}a) and two (Figure~\ref{fig:uprogexample}b) DRAM subarrays. 
In the conventional sequential execution, all row copies and TRAs in the \uprog are serialized inside a single DRAM subarray, executing the required operations for the first-bit of the input operands first and then the second-bit of the input operand. However, the only data dependency between the operations for the first-bit and second-bit in the \uprog is the \emph{carry propagation} (\circled{1} in Figure~\ref{fig:uprogexample}a). Therefore, by using two DRAM subarrays for computation, we can parallelize independent operations and \emph{only} serialize the carry propagation (\circled{2} in Figure~\ref{fig:uprogexample}b), reducing the overall latency of the \gls{PuD} operation.} }

\begin{figure}[ht]
    \centering
    \includegraphics[width=\linewidth]{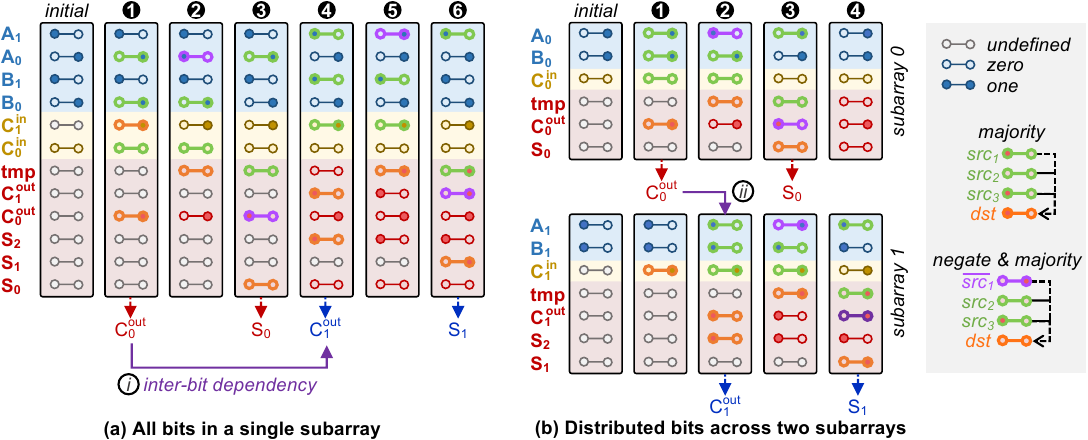}
    \caption{\gfcrii{Simplified bit-serial \gls{PuD} addition of two input arrays $A$ and $B$, each of which with two-bit data elements using (a)~one and (b)~two DRAM subarrays.}}
    \label{fig:uprogexample}
\end{figure}

%\agymicrocomment{Too much whitespace in figure above}
%\jgl{In fact, as part of Opportunity 2, you can also mention that bit-parallel algorithms can also be used if more than one subarray can be activated concurrently.}

\gfcrii{Besides reducing the latency of bit-serial \gls{PuD} operations, carefully distributing individual bit positions across different DRAM subarrays within a DRAM bank enables the efficient realization of latency-friendly \emph{bit-parallel} \gls{PuD} arithmetic operations.
By mapping each bit position of a data element to a distinct subarray, our \gls{PuD} substrate can \emph{concurrently} perform \omcriii{bitwise} operations across all bits of the operand, thereby fully exploiting the parallelism inherent to bit-parallel arithmetic algorithms.}

\paratitle{\gf{\omcrii{\emph{Limitation 3:}}~High-Precision Computation}} 
\sgi{\gls{PuD} suffers from high latency for high bit-precision operations.}
\gf{\sgdel{\gls{PuD} suffers from \gfii{high latency that undermines the potential benefits of \gls{PuD} when implementing} key operations using high bit-precision.}%
For example, even when employing multiple (i.e., 16) \sgi{parallel} DRAM banks, SIMDRAM's throughput for \sg{32-bit and} 64-bit  division is 0.8$\times$ and 0.5$\times$ that of a \sg{16-core} CPU system~\cite{hajinazarsimdram}. \sgi{This is because} the latency of bit-serial multiplication and division scales \emph{quadratically} with the bit-precision. \sgdel{Such high latency directly impacts the performance of the \gls{PuD} substrate when executing multiplication/division-heavy workloads, leading to performance loss in such cases~\cite{hajinazarsimdram}.}}

\noindent \gf{\omcrii{\textbf{\emph{Opportunity 3:} Alternative Data Representation for High-Precision Computation.}} The high latency associated with high-precision computation is an \emph{inherent} property of coupling the binary numeral system with bit-serial computation. \omcrii{We} investigate \gfasplos{an} alternative data representation\gfasplos{, i.e.  the \emph{redundant binary representation (RBR})~\gfcrii{\cite{guest1980truth,phatak1994hybrid,lapointe1993systematic, olivares2006sad, olivares2004minimum}},} for high-precision computation.
\sgi{\gls{RBR} is a positional number system where each \gfcrii{bit-position}~$i$, which encodes $2^i$, is represented by two bits that can take on a value $v \in \{-1, 0, 1\}$, such that the magnitude of \gfcrii{bit-position}~$i$ is $v \times 2^i$.
For example, the 4-\gfcrii{bit-position} number \texttt{<0,1,0,-1>} represents $2^2 - 2^0 = 3$.
\gls{PuD} execution can take advantage of two \omcrii{key} properties of \gls{RBR}-based arithmetic:
\li~the operations no longer need to propagate carry bits through the full width of the data
(e.g., \gls{RBR}-based addition limits carry propagation to \emph{at most} two places~\sg{\cite{brown2002using}}), and
\lii~the operation latency is \emph{independent} of the bit-precision.}
\sgdel{, a positional numeral system that uses more bits than needed (e.g., two bits) to represent a single binary number.
In a \gls{RBR}-based system, each digit can take on any value from the set \texttt{\{-1, 0, 1\}}. Because there are three possible digit values, two bits are required to encode each digit. In conventional unsigned binary format, the $i^{th}$ digit represents $2^i$ multiplied by 0 or 1. In \gls{RBR}, the $i^{th}$ digit represents $2^i$ multiplied by \texttt{-1}, \texttt{0}, or \texttt{1}. An $n$-digit \gls{RBR} number $X = x_{n-1},x_{n-2}, ..., x_0$, where $x_i \in $  \texttt{\{-1, 0, 1\}} represents the value $\sum_{i=0}^{n-1} x_{i}2^{i}$. For example, the 4-digit number \texttt{<0,1,0,-1>} represents $2^2 - 2^0 = 3$. A \gls{RBR} number $X$ is often represented by two sets of bits, $X^{+}$ and $X^{-}$, representing the positive and negative powers of 2, respectively, that are added to the computed value. In the previous example, $X^{+} = $ \texttt{<0,1,0,0>} and $X^{-} = $ \texttt{<0,0,0,1>}. The two's complement of $X$ can be computed by subtracting $X^{-}$ from $X^{+}$ in the two's complement number system.
There are two main properties of RBR arithmetic that can benefit \gls{PuD} execution. 
First, using RBR for arithmetic operations (e.g., addition~\omcrii{\cite{wang2004new,jose2006delay}}) is almost \emph{carry-free}, i.e., there is no need to propagate carry bits through the full width of the addition operation.\footnote{\gf{\gls{RBR}-based addition limits carry propagation to \emph{at most} two digits~\sg{\cite{brown2002using}}.}} Second, the latency of an arithmetic operation is \emph{independent}. 
%However, using RBR-based arithmetic can incur higher latency than using the traditional binary representation for low precision operations~\cite{hossen2012design}.
}}

\paratitle{Goal} \gf{Our \emph{goal} in this work is to mitigate the three limitations of \gls{PuD} architectures \omcriii{that arise due to the naive use of} a bit-serial execution model. 
To do so, we aim to \emph{fully} exploit the opportunities that DRAM's internal parallelism and dynamic bit-precision can provide to \gfcrii{reduce} the latency \gfcrii{and energy} of \gls{PuD} operations\gfcriii{. 
Concretely, we aim to} \gfcrii{\gfcriii{enable} 
\li~adaptive data-representation formats (two's complement and \gls{RBR}) for \gls{PuD} operands and 
\lii~flexible execution of different arithmetic algorithm implementations} (bit-serial and bit-parallel) \gfcrii{for \gls{PuD} instructions}.}  

\section{\prop Overview}
\label{sec:overview}

{\gf{\prop is a \omcrii{new data-aware runtime} \gls{PuD} framework \gfcrii{that \emph{dynamically} adjust the bit-precision and, based on that, \emph{chooses} and \emph{uses} the most appropriate data representation and arithmetic algorithm implementation for a given \gls{PuD} operation.}
The \emph{key ideas} behind \prop are: 
\li~to \gfcrii{\emph{reduce} the bit-precision for \gls{PuD} operations by dynamically leveraging \emph{narrow values};
\lii~to \emph{parallelize} the execution of \emph{independent} in-DRAM primitives (i.e., \aaps in a \uprog) by leveraging \gls{SLP}~\cite{kim2012case} combined with \emph{bit-level parallelism}; and
\liii~to \emph{\omcriii{use}} alternative data representations (i.e., \gls{RBR}~\cite{guest1980truth,phatak1994hybrid,lapointe1993systematic, olivares2006sad, olivares2004minimum}) for high-precision computation.}

\gf{Figure~\ref{fig:high-level} provides a high-level overview of \prop' framework, its main components, and execution flow. \prop is composed of three main components: 
\omcrii{
\li~\uproglib, 
\lii~\dynengine, and 
\liii~\uprogunit}.  
These components are implemented in hardware \omcrii{inside the} DRAM memory controller. The \uproglib and \uprogunit are part of \emph{\prop \gfcrii{C}ontrol \gfcrii{U}nit}. }
%With these three components, \prop can adapt the data representation and \gls{PuD} operation implementation depending on the target bit-precision.
}
\gfcrii{In this section, we provide a high-level overview of \prop' main components (\cref{sec:overview:components}) and its execution flow (\cref{sec:overview:execution}). 
Detailed information on the implementation of each component is provided in~\cref{sec:implementation}. }

\begin{figure}[ht]
    \centering
    \includegraphics[width=\linewidth]{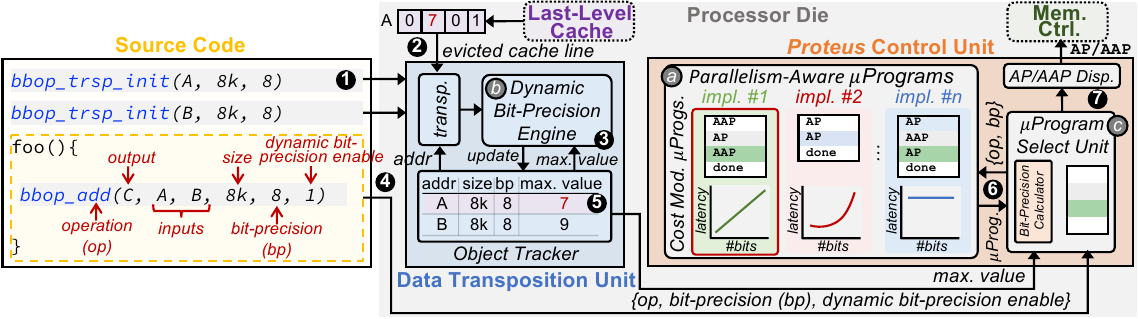}
    \caption{\gfcrii{\hl{Overview of the \mbox{\prop} framework.}}} 
     \label{fig:high-level}
\end{figure}
% \begin{figure*}
% \centering
% \begin{minipage}[t]{.71\textwidth}
%   \centering
%     \adjincludegraphics[width=\linewidth,valign=C]{figures/daftpum_overview-crop.pdf}
%   \caption{\asplosrev{\hl{Overview of the \mbox{\prop} framework.}} \prtagA{Fix}}
%     \label{fig:high-level}
% \end{minipage}\hfill
% \begin{minipage}[t]{.27\textwidth}
%   \centering
%     \adjincludegraphics[width=0.75\linewidth,valign=C]{figures/subarray_org-crop.pdf}
%   \caption{Subarray organization.\prtagA{Fix}}
%   \label{fig:implementation:subarray}
% \end{minipage}
% \end{figure*}

% \begin{figure*}
% \centering
% \begin{minipage}[b]{.9\textwidth}
%   \centering
%     \adjincludegraphics[width=\linewidth,valign=C]{figures/proteus_overview-crop.pdf}
%   \caption{\gfcrii{\hl{Overview of the \mbox{\prop} framework.}}}
%     \label{fig:high-level}
% \end{minipage}
% \end{figure*}

\subsection{Main Components of \prop}
\label{sec:overview:components}

\paratitle{\gf{Parallelism-Aware \uprog Library} \gfcrii{(\cref{sec:implementation:library})}}
\sgi{\prop incorporates a \uproglib  (\gfcrii{\circledii{a}} in Figure~\ref{fig:high-level}) that consists of
\li~hand-optimized implementations of different \uprogs for key \gls{PuD} operations (each with different performance and bit-precision trade-offs), and
\lii~\costmodel.
For each operation, we implement multiple \uprogs (\cref{sec:implementation:library:optimization}) that use different
\li~bit-serial or bit-parallel algorithms and
\lii~data representation formats (i.e., two's complement or \gls{RBR}).
Each \uprog uses a novel data mapping that enables the \emph{concurrent} execution of multiple independent primitives across bits (\cref{sec:implementation:library:obps}).
The performance of each \uprog depends on the bit-precision, and
the \costmodel selects the best-performing \uprog for a given operation and target bit-precision.
}
\gf{\sgdel{\prop' \uproglib (\gfcrii{\circledd{a}} in Figure~\ref{fig:high-level})  provides a range of optimized \uprogs for a given \gls{PuD} operation, each of which with different trade-offs in terms of performance and bit-precision. 
The \uproglib consists of 
\li~hand-tuned implementations of different \uprogs targeting key arithmetic operations, and
\lii~\costmodel.
First, to build a robust set of \gls{PuD} operations, we manually implement \uprogs using different
\li~bit-serial (e.g., ripple-carry adder) and bit-parallel 
\sgdel{(e.g., carry-select adder~\cite{bedrij1962carry}, Kogge–Stone adder~\cite{kogge1973parallel}, Ladner-Fischer adder~\cite{ladner1980parallel})}%
algorithms and 
\lii~data representation formats (i.e., two's complement and \gls{RBR}).  
We optimize each \uprog by 
\li~distributing input operands across different DRAM subarrays using an \gls{OBPS} data mapping, where bits from an input operand are scattered across DRAM subarrays; and 
\lii~parallelizing the execution of data-independent operations in a \uprog across different DRAM subarrays using \gls{SLP}~\cite{kim2012case} (to concurrent\juan{ly{}} execute \aaps in several subarrays) and LISA~\cite{chang2016low} (to transfer partial results across subarrays).
In the end, the \uproglib contains a collection of possible implementations of a desired \gls{PuD} operation, each with different time complexities that depend on the \gls{PuD} bit-precision. 
%As SIMDRAM, \prop reserves a region of DRAM for storing all \uprogs and uses an on-chip scratchpad memory to store the most commonly use \uprogs.
%Second, each \uprog implementation has an associated \texttt{\uprog\_index}, which the \uproglib uses to fetch a given \uprog from main memory or the on-chip scratchpad.
Second, the \costmodel aims to select the best-performing \uprog for a given operation and target bit-precision.}%
The \costmodel comprises of \prelut, which \sgi{list the most-suitable \uprog for each} bit-precision, and \emph{Select Logic} to identify the target \gls{LUT} for a \emph{bbop} instruction. 
\sgi{We \gfcrii{\emph{empirically} measure the throughput and energy efficiency of \uprogs in \uproglib while scaling the target bit-precision to} populate the \prelut.}
}

\paratitle{\gf{Dynamic Bit-Precision Engine} \gfcrii{(\cref{sec:implementation:dynamic-precision})}} \gf{\prop' \dynengine (\gfcrii{\circledii{b}} in Figure~\ref{fig:high-level}) aims to identify the dynamic range of \emph{memory objects} associated with a \gls{PuD} operation.  
To do so, we dynamically identify \gfcrii{(in hardware)} the \emph{largest} input operand a \gls{PuD}'s memory object stores. In state-of-the-art \gls{PuD} architectures~\cite{hajinazarsimdram, peng2023chopper,gao2019computedram,ali2019memory,angizi2019graphide}, cache lines belonging to a \gls{PuD}'s memory object need to be transposed from the traditional horizontal data layout to a vertical data layout \emph{prior} to the execution of a \gls{PuD} operation. To efficiently perform such data transformation, SIMDRAM~\cite{hajinazarsimdram} implements a \emph{Data Transposition Unit}, which hides the data transposition latency by overlapping cache line evictions and data layout transformation. 
The \omcrii{\emph{Data Transposition Unit}} consists of an \emph{Object Tracker} table (a small cache that keeps track of memory objects that are used by \gls{PuD} operations) and \emph{Data \omcrii{Transposition} Engines}. The user/compiler informs the \emph{Object Tracker} of \gls{PuD}'s memory objects (both inputs and outputs) using a specialized instruction called \texttt{bbop\_trsp\_init}.
\prop leverages such \omcrii{a} \emph{Data Transposition Unit} to \omcrii{dynamically} identify \omcrii{in hardware} the largest value in a \gls{PuD}'s memory object by adding:
\li~a \gfcrii{\emph{new field}} in the \emph{Data Transposition Unit} called \emph{maximum value}, which stores the largest value in a given memory object\revdel{ \sg{evicted from a cache line}}; and
\lii~a \dynengine, which \gfmicro{scans} \omcrii{the} data elements of evicted cache lines, identifies the largest data value across all data elements and updates the stored \emph{maximum value} entry in the \emph{Data Transposition Unit}.\revdel{when necessary.}} 

\paratitle{\gf{\uprog Select Unit} \gfcrii{(\cref{sec:implementation:control-unit})}}  \gf{\prop' \uprogunit (\gfcrii{\circledii{c}} in Figure~\ref{fig:high-level}) identifies the appropriate bit-precision for a \gls{PuD} operation based on \omcrii{the operation's} input data.
The \uprogunit has of a \li~\bitprec, which evaluates the target bit-precision based on the input operands of the \gls{PuD} operation and their associated maximum values, and
\lii~buffers to store the selected \uprog. 
}

\subsection{Execution Flow}
\label{sec:overview:execution}

\gf{\prop works in five} \gf{main steps.
In the first step (\omcrii{\circled{1}} in Figure~\ref{fig:high-level}), the programmer/compiler utilizes specialized instructions (i.e., \omcriii{\texttt{bbop\_trsp\_init}}) to 
\li~register \omcrii{in} the \emph{Object Tracker} the address, size, and initial bit-precision \gfcrii{for each memory object used as an input, output, or temporary operand in a \gls{PuD} operation}; and
\lii~execute \gfcrii{a} \gls{PuD} \gfcriii{operation} over previously-registered memory objects. 
When issuing \gfcriii{an arithmetic} \bbop instruction, the programmer/compiler indicates whether or not dynamic bit-precision is enabled or disabled \omcrii{for that \bbop instruction}. \revD{\label{rd.2}\changeD{D2}When dynamic bit-precision is disabled, \prop' \dynengine is turned off, and the \uprogunit utilizes the user-provided bit-precision for \gfcrii{the} upcoming \gls{PuD} \gfcrii{operation related to the issued \bbop instruction}.\revdel{ This allows \prop to avoid performing redundant computations if the user has domain expertise and knows the dynamic range of \gls{PuD} operations beforehand.}}
In the second step, \sgi{if} the \dynengine is enabled, it intercepts evicted cache lines belonging to previously registered memory objects (\gfcrii{\circled{2}}) and identifies the largest value stored in the cache line. \sgi{If} the identified value is \emph{larger than} the current maximum \omcrii{value} stored in the \emph{Object Tracker}, the \dynengine updates the \emph{Object Tracker} with the up-to-date value (\omcrii{\circled{3}}). 
\gfcrii{The second step is repeated for all cache lines belonging \omcriii{to the} memory objects \omcriii{registered} in the \emph{Object Tracker}}. 
As in SIMDRAM~\cite{hajinazarsimdram}, our system employs lazy allocation \gfcrii{and maintains data coherence for \gls{PuD} memory objects through cache line flushing, using the \texttt{clflush} instruction~\cite{guide2016intel}.}
%\footnote{\label{ft.rcm.1}\revCommon{The \texttt{clflush} instruction evicts the cache line containing the target memory address from \emph{all} levels of the cache hierarchy~\cite{guide2016intel}. Thus, even \gfmicro{in case of} non-inclusive \gfmicro{caches}, the target cache line will be flushed to \gfmicro{DRAM}.}} 
Thus, all memory objects initially reside within the CPU caches, and prior to PUD execution, all cache lines belonging to a \gls{PuD} operation \omcrii{are} evicted to DRAM, \gfmicro{which allows} \prop' \dynengine \gfmicro{to} access \emph{all} data elements of a \gls{PuD} operation prior to computation.\footnote{\gfasplos{\omcrii{Some real}-world \gls{PnM} architectures~\omcrii{\cite{upmem,devaux2019true,gomez2022benchmarking,gomez2021benchmarkingcut, gomezluna2021benchmarking,gomez2023evaluating}}} employ a similar execution model, where \emph{all} inputs need to be copied to \gls{PnM} cores \emph{prior} to \gls{PnM} execution.}} 
In the third step, the host CPU dispatches \gfcriii{the} \gfcrii{arithmetic} \bbop \gfcriii{instruction} (\omcrii{\circled{4}}) to \prop' Control Unit.
In the fourth step, \prop' \emph{Control Unit} receives the \bbop instruction from the CPU and the maximum values from the \dynengine (\omcrii{\circled{5}}), which are used as \omcrii{inputs} to the \uprogunit. Based on \omcrii{this} information, the \bitprec computes the target bit-precision and probes the \uproglib (\omcrii{\circled{6}}), which returns the best-performing \uprog and data format representation for the target \gfcrii{\gls{PuD}} operation \omcrii{and} \gfcrii{its associated} bit-precision.
In the fifth step, the \uprogunit dispatches the sequence of \aaps in the selected \uprog to DRAM (\omcrii{\circled{7}}).
When the host CPU reads back \gls{PuD} memory objects (not shown in the \juan{f}igure), 
\prop 
\li~performs the necessary data format conversions either 
from the reduced bit-precision to the user's specified bit-precision or from \gls{RBR} to two's complement \gfasplos{(thus maintaining system compatibility), and
\lii~\gfasplos{prepares the \dynengine for future accesses by resetting the current maximum data value stored in the \emph{Object Tracker}}}.  
} 
\section{\prop Implementation}
\label{sec:implementation}

\revdel{In this section, we describe the implementation details of \prop' main components. We first describe \prop' underlying subarray organization (Section~\ref{sec:implementation:subarray}). Then, we discuss the design of \prop'
\li~\uproglib (Section~\ref{sec:implementation:library}),
 \lii~\dynengine (Section~\ref{sec:implementation:dynamic-precision}), and
 \liii~\uprogunit (Section~\ref{sec:implementation:control-unit}).
}
\subsection{Subarray Organization}
\label{sec:implementation:subarray}

\gfcrii{To efficiently perform \gls{PuD} operations, \prop uses a subarray organization that incorporates additional functionality to perform 
\li~in-DRAM logic primitives (i.e., \aaps), 
\lii~inter-subarray data copy, and 
\liii~subarray-level parallelism (\gls{SLP})~\cite{kim2012case}. 
Figure~\ref{fig:implementation:subarray} illustrates the internal organization of a subarray in \prop, which is replicated across \emph{all} subarrays in a \prop-enabled DRAM bank.}

\paratitle{Performing Logic Primitives with Ambit}
\gf{%
\sgi{\prop reuses the subarray organization of Ambit~\cite{seshadri2017ambit} and SIMDRAM~\cite{hajinazarsimdram} \omcrii{(shown in Figure~\ref{fig:implementation:subarray})} to enable logic primitive execution with only \omcrii{small} subarray modifications\revdel{ (a small row decoder that can activate up to three rows simultaneously)}. DRAM rows are divided into three groups:}
\sgdel{To enable the execution of logic primitives, \prop follows the same subarray organization of Ambit~\cite{seshadri2017ambit} and SIMDRAM~\cite{hajinazarsimdram}, requiring only minimal modifications to the DRAM subarray (namely, a small row decoder that can activate three rows simultaneously).
Like Ambit~\cite{seshadri2017ambit}, \prop divides DRAM rows into \emph{three groups}:}
\li~the \textbf{D}ata group (D-group), containing regular rows that store program data;
\lii~the \textbf{C}ontrol group (C-group), containing two rows pre-initialized with all-`0' and all-`1' values\revdel{\footnote{\gf{Ambit uses \texttt{AP} command sequences to implement Boolean AND and OR operations by simply setting one of the inputs (e.g., $C$) to 1 or 0. The AND operation is computed by setting $C$ to 0 (i.e., \texttt{MAJ(A, B, 0) = A AND B}). The OR operation is computed by setting $C$ to 1 (i.e., \texttt{MAJ(A, B, 1) = A OR B}).}}}; and 
\lii~the \textbf{B}itwise group (B-group), containing six rows (called \emph{compute rows}) to perform bitwise operations. 
The B-group rows are all connected to a special row decoder that can simultaneously activate \label{rd.5}\changeD{D5}\revD{three rows when performing an \texttt{AP} and two when performing an \texttt{AAP}} (\circled{1} in Figure~\ref{fig:implementation:subarray}). }

\begin{figure}[!htp]
    \centering
    \includegraphics[width=0.45\linewidth]{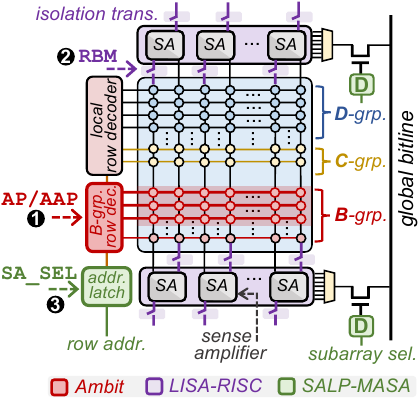}
    \caption{\gfcrii{\prop' subarray organization.}}
    \label{fig:implementation:subarray}
\end{figure}

% \begin{figure}[ht]
%   \centering
%     \adjincludegraphics[width=0.65\linewidth]{figures/subarray_org-crop.pdf}
%   \caption{Subarray organization.}
%   \label{fig:implementation:subarray}
% \end{figure}

\paratitle{Inter-Subarray Data Copy with LISA}
\sgi{\prop leverages LISA-RISC~\cite{chang2016low},
which dynamically connects adjacent subarrays using isolation transistors\revdel{ to copy an entire DRAM row}, to propagate intermediate data\revdel{ (e.g., carry)} \gfmicro{across subarrays}.
LISA-RISC \gfmicro{works} in four steps:
\li~activate the source row in the source subarray (latency: $t_{RAS}$);
\lii~use the LISA \emph{row buffer movement} command \omcrii{(\texttt{RBM}, \circled{2} in Figure~\ref{fig:implementation:subarray})} to turn on isolation transistors, which copies data from the source subarray's local row buffer (LRB) to the destination subarray's LRB (latency: $t_{RBM}$, \SI{5}{\nano\second}\revdel{ in SPICE simulations}~\cite{chang2016low});
\liii~activate the destination row, to save the contents of the destination LRB into the \gfcrii{destination} row  (latency: $t_{RAS}$); and
\liv~precharge the bank (latency: $t_{RP}$).
\gfmicro{D}ue to DRAM's open bitline architecture~\cite{lim20121,takahashi2001multigigabit},
each LRB stores half of the row, so we must perform steps \lii--\liv\ twice to copy both halves of the row.}
\sgdel{\gf{To enable fast propagation of intermediate data (e.g., carry-propagation) across DRAM subarrays, \prop leverages LISA~\cite{chang2016low}. LISA creates a datapath between adjacent subarrays by connecting their
local row buffers using isolation transistors, which are controlled via a new DRAM operation called \emph{row buffer movement} (\texttt{RBM}, \circled{2} in Figure~\ref{fig:implementation:subarray}). 
Such newly added datapath allows LISA to implement \underline{r}apid \underline{i}nter-\underline{s}ubarray bulk data {c}opy (i.e., LISA-RISC) using \sg{four} main steps \sg{(and associated DRAM timing parameters)}:
\li~activate \sg{($t_{RAS}$)} a source row in a subarray;
\lii~rapidly transfer the data in the activated source row buffers to the destination subarray's row buffer via the link created by LISA's isolation transistors \sg{($t_{RBM}$)};
\liii~activate \sg{($t_{RAS}$)} the destination row, which enables the contents of the destination row buffer to be latched into the destination row;
\liv~\sg{precharge ($t_{RP}$) the DRAM bank for subsequent memory requests. Since during an \texttt{ACT} the sense amplifier stores \emph{half} the data for a DRAM row (due to the open bitline DRAM architecture~\cite{lim20121,takahashi2001multigigabit}), to copy an entire DRAM row, the LISA-RISC operation needs to perform a second sequence of \texttt{ACT}-\texttt{RBM}-\texttt{PRE} commands. Thus,} \gfisca{the latency of a LISA-RISC operation is of $3t_{RAS}\ +\ 2t_{RP}\ +\ 2t_{RBM}$.\footnote{\gfisca{According to \cite{chang2016low} SPICE simulation analysis, the latency of a \texttt{RBM} command (i.e., $t_{RBM}$) is of 5~ns.}}}}}

\paratitle{Enabling Subarray-Level Parallelism with SALP} \gf{To enable the concurrent execution of \gfcrii{bit-}independent \gfcrii{primitives} in a \uprog, \prop leverages \gls{SLP}~\cite{kim2012case}.
% 's \emph{multitude of activated subarrays} (MASA) design. 
SALP-MASA \sgi{(\emph{\omcrii{M}ultitude of \omcrii{A}ctivated \omcrii{S}ubarrays})} allows multiple subarrays in a bank to be activated concurrently by 
\li~pushing the global row-address latch to individual subarrays, 
\lii~adding a \sgi{designated}-bit latch (\textbf{D} in Figure~\ref{fig:implementation:subarray}) to each subarray to ensure that only a single subarray's row buffer is connected to the global bitline, and
\liii~routing a new global wire (called \emph{subarray select}), controlled by a new DRAM command (\texttt{SA\_SEL}, \circled{3} in Figure~\ref{fig:implementation:subarray}), allowing the memory controller to set/clear \sgi{each designated-bit latch}.}

% \paratitle{\prop's Near-Bank Shifter} \gf{To enable the efficient execution of applications with non-trivial access patterns (e.g., strided access in a \gls{GEMV} kernel), \prop implements a low-cost shift unit called \emph{near-bank shifter} (NBS). 
% The NBS (\circled{4} in Figure~\ref{fig:implementation:subarray}) supports bit-shifting for up to 128-bit data elements deployed with horizontal data layout and bulk data shifting across up to 128 adjacent bitlines of the vertical data layout. To do so, the NBS reads and stores up to two DRAM columns of an activated DRAM row via the 64-bit global row buffer. This allows for the efficient implementation of pre-defined shifting patterns (i.e., shift-by-1, shift-by-2, and stride-by-2) while maintaining a low area cost and minimizing complex modifications to the DRAM array design. }
%\jgl{I don't get how this NBS works.}

\subsection{Parallelism-Aware \uprog Library}
\label{sec:implementation:library}

\subsubsection{\gfisca{One-Bit Per-Subarray (OBPS) Data Mapping}} 
\label{sec:implementation:library:obps}
To reduce the latency of \gls{PuD} operations (\cref{sec_motivation}), \prop employs a specialized data mapping called \emph{\gls{OBPS}}.
Bit-serial \gls{PuD} architectures can employ three data mappings, as Figure~\ref{fig:obps} illustrates: 
\omcrii{\li~}\gls{ABOS}, 
\omcrii{\lii~}\gls{ABPS}, and 
\omcrii{\liii~}\gls{OBPS}. 
\gfcrii{Assume an example DRAM bank with four subarrays, a DRAM row size of three and an input array $A$ with six two-bit data elements.}

\begin{figure}[ht]
\centering
  \centering
 \includegraphics[width=0.85\linewidth]{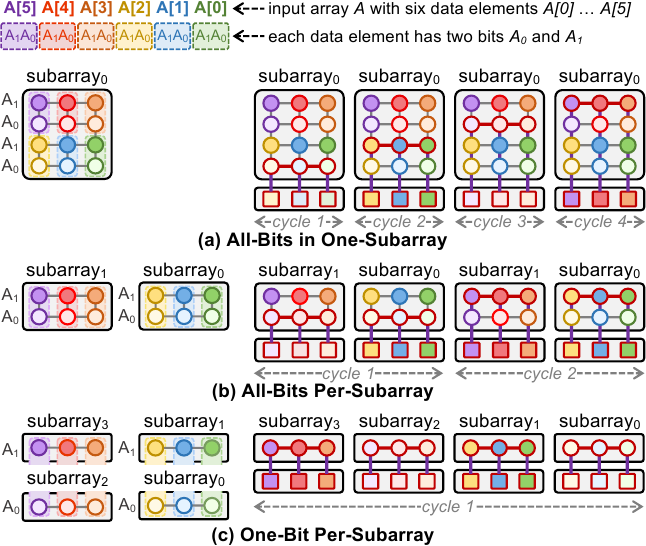}
  \caption{\gfcrii{Three data mappings for bit-serial computing.}}
  \label{fig:obps}
\end{figure}

\gfcrii{First, the \emph{\gls{ABOS}} data mapping stores \emph{all} six two-bit data elements in \emph{one} DRAM subarray (Figure~\ref{fig:obps}a). 
This data mapping limits the parallelism available for \gls{PuD} execution to that of a \emph{single} DRAM subarray, i.e., the number of DRAM columns simultaneously activated by a single \gls{PuD} primitive  (e.g., 65,536 DRAM columns per cycle in DDR4 memory \omcriv{chips}~\cite{standard2012jesd79}). 
In our example, the latency of executing a single \gls{PuD} primitive over \emph{all} data elements of the input array $A$ is four \gls{PuD} cycles \omcriii{(as shown in Figure~\ref{fig:obps}a)}.\footnote{\gfcrii{We refer to a \emph{\gls{PuD} cycle} as the end-to-end latency required to execute a single \aap in-DRAM primitive}.}
Second, the \emph{\gls{ABPS}} data mapping distributes \emph{all} bits of multiple sets of the input array across \emph{multiple} DRAM subarrays (Figure~\ref{fig:obps}b), allowing a \gls{PuD} primitive to execute concurrently on different portions of the input data stored in each subarray \omcriii{by} exploiting \emph{data-level parallelism}.
In our example, the latency of executing a single \gls{PuD} primitive over \emph{all} data elements of the input array $A$ while employing the \gls{ABPS} data mapping is two \gls{PuD} cycles \omcriii{(as shown in Figure~\ref{fig:obps}b)}.
This is because, although execution across data elements can be parallelized by distributing them across multiple DRAM subarrays, the \gls{PuD} system must still serialize the execution of \gls{PuD} primitives across different bit positions of each data element, since \omcriii{\emph{all bits of a given data element}} are co-located within a single DRAM subarray under \gls{ABPS}.
Third, the \gls{OBPS} data mapping distributes \omcriii{each of} the $m$ individual bits of \omcriii{a given} data element of the input array to $m$ DRAM subarrays (Figure~\ref{fig:obps}c), i.e., $subarray_0 \leftarrow  \{A_0\}, \dots$, $subarray_{m-1} \leftarrow  \{A_{m-1}\}$, allowing a \gls{PuD} primitive to execute concurrently on different bits of the input array stored in each subarray \omcriii{by} exploiting \emph{bit-level parallelism}.\footnote{If the number of subarrays is \omcriii{smaller} than the target bit-precision, \gls{OBPS} \emph{evenly} distributes the bits of input operands across the available subarrays.}  
In our example, the latency of executing a single \gls{PuD} primitive over \emph{all} data elements of the input array $A$ while employing the \gls{OBPS} data mapping is \omcriii{\emph{only}} a single \gls{PuD} cycle \omcriii{( as shown in Figure~\ref{fig:obps}c)}.}

% \begin{figure}[!t]
% \centering
% \begin{minipage}[t]{0.5\linewidth}
%   \centering
%  \includegraphics[width=\linewidth]{figures/data_mappings_example-crop.pdf}
%   \caption{Three data mappings for bit-serial computing.\prtagA{GF: REDO}}
%   \label{fig:obps}
% \end{minipage}
%   \hfill %%
% \begin{minipage}[t]{0.44\linewidth}
%   \centering
%     \includegraphics[width=\linewidth]{figures/cost_model_unit-crop.pdf}
%  \caption{\prop's cost model logic.}
%      \label{fig:cost-model}
% \end{minipage}
% \end{figure}

\subsubsection{\uprog Library Implementation}
\label{sec:implementation:library:optimization}

\gf{\prop leverages the subarray organization illustrated in Figure~\ref{fig:implementation:subarray} and our \gls{OBPS} data mapping \omcrii{(Figure~\ref{fig:obps})} to implement parallelism-aware \uprogs for key arithmetic operations \omcrii{(e.g., addition, multiplication)}.
We implement three classes of algorithms for arithmetic \gls{PuD} computations: \emph{bit-serial}, \emph{bit-parallel}, and \emph{RBR-based algorithms}. In \prop, each \uprog implementation 
\li~has an associated \gfcrii{\gbidx}, and 
\lii~is stored in a reserved memory space in DRAM (i.e., \emph{\uprog Memory}).} 
%\jgl{Besides OBPS, we had a fancy wrap-around mapping, which would be really necessary if we don't have up to 64 subarrays. Isn't it useful?}

\paratitle{Bit-Serial Algorithms} \gf{We optimize \uprogs for bit-serial arithmetic operations (i.e., addition, subtraction, division, and multiplication)
\sgdel{in SIMDRAM}%
by concurrent\juan{ly} executing independent \aaps across different DRAM subarrays. Figure~\ref{fig:uprogexample}b illustrates such a process for addition \gfcriv{using the \gls{OBPS} data mapping} (the process is analogous %to 
\juan{for} other arithmetic operations). \sgdel{In this example,}%
\prop implements a ripple-carry adder using majority gates in two main steps. 
\sgi{First},  \prop utilizes SALP-MASA to concurrent\juan{ly} execute the appropriate row copies and majority operations across $N$ different subarrays. 
\sgi{Second}, \prop utilizes LISA-RISC \sgdel{(and its associated \texttt{RBM} DRAM command)}%
to pipeline the carry propagation process (\gfcrii{\circledii{ii} in Figure~\ref{fig:uprogexample}b)} from $subarray_i$ (e.g., $C_{out}^{0}$) to $subarray_{i+1}$ (e.g., $C_{in}^{1}$). This process repeats for all $N$ bits in the input operand. \prop reduces the latency of executing \omcriv{an $N$-bit} bit-serial addition from $8N + 1$ \aap cycles~\omcrii{\cite{hajinazarsimdram}} to $2N +7$ \aap cycles + $2(N-1)$ \texttt{RBM} cycles.\footnote{\label{ft.rc.7}\revC{To compute the number of \aap and \texttt{RBM} cycles in a \uprog, we implement each \uprog's algorithm using our cycle-\omcrii{level} data-accurate simulator (see~\cref{sec:methodology}\omcrii{; open sourced at~\cite{proteusgit}}). 
\revdel{Our simulator allows us to 
\li~compose a \uprog at the granularity of \aap and \texttt{RBM} operations and 
\lii~have individual control of the bits stored in a DRAM array while taking data modeling into consideration.}\revdel{ This means that, when simulating a TRA, our simulator actually computes the result of the majority operation for the data stored in the target DRAM rows.}  
We verified the correctness of a \uprog by testing it against several randomly generated data set combinations.}}  }

\paratitle{Bit-Parallel Algorithms}
\gf{We implement \sgi{bit-parallel variants of our \uprogs that}
leverage \emph{carry-lookahead logic} to decouple the calculation of the carry bits and arithmetic logic (e.g., addition). 
\sgi{Carry-lookahead logic can identify if any arithmetic on a bit will \emph{generate} a carry (e.g., both operands bits are `1' for an addition),
or if it will \emph{propagate} the carry value (e.g., only one operand bit for an addition is a `1').}
\sgdel{Carry-lookahead logic uses the concepts of \emph{generating} and \emph{propagating} carries. For example, in a binary addition, two input bits 
\li~\emph{generate} a carry only if \emph{both} bits are `1', and 
\lii~\emph{propagate} a carry only if there is an input carry.}%
\sgi{For $N$-bit operands, this} reduces time complexity compared to ripple-carry logic from $\mathcal{O}(N)$ to $\mathcal{O}(\log{}N)$, \gfcrii{where $N$ is the number of bits in the input operands.} We implement several carry-lookahead algorithms in \prop, including the carry-select~\cite{bedrij1962carry}, Kogge--Stone~\cite{kogge1973parallel}, Ladner--Fischer~\cite{ladner1980parallel}, and Brent--Kung~\cite{brent1982regular} adders, 
as building blocks to implement subtraction, multiplication, and division. 
Figure~\ref{fig:paralleladder} \sgi{shows an example \prop implementation of a Kogge--Stone adder.}
\sgdel{illustrates \prop' implementation of an exemplary bit-parallel adder (i.e., the Kogge–Stone adder).}%

% \begin{figure}[ht]
%     \centering
%     \includegraphics[width=0.45\linewidth]{figures/bit_parallel_example-crop.pdf}
%     \caption{\gf{\prop's implementation of a 4-bit Kogge–Stone adder. Bits $A_i$ and $B_i$ are stored \emph{vertically} in the same DRAM bitline. \gf{$P$ and $G$ refer to \emph{propagate} and \emph{generate}. } }}
%     \label{fig:paralleladder}
% \end{figure}

\begin{figure}[ht]
\begin{subfigure}{0.46\linewidth}
  \centering
 \includegraphics[width=\linewidth]{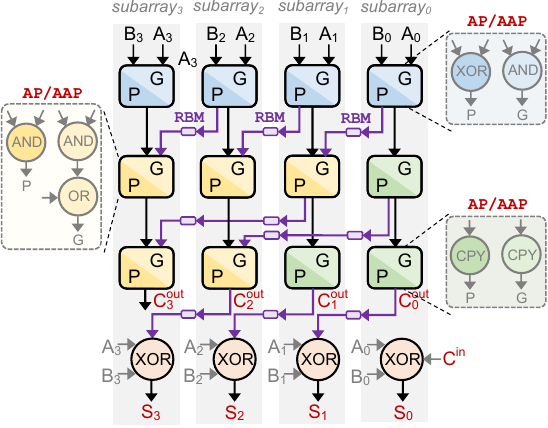} 
  \caption{4-bit Kogge–Stone adder}
  \label{fig:paralleladder}
\end{subfigure}
  \hfill %%
\begin{subfigure}{0.46\linewidth}
  \centering
    \includegraphics[width=\linewidth]{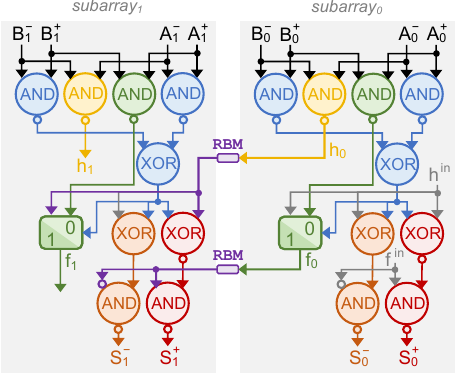}
 \caption{2-bit RBR adder}
  \label{fig:rbr_exemple}
\end{subfigure}
\caption{\prop' implementation of different adders. Bits $A_i$ and $B_i$ are stored \emph{vertically} in the same DRAM bitline \gfcrii{of subarray $i$ using the \gls{OBPS} data mapping}.}
\end{figure}

\revdel{The execution is divided into two main steps.}
In the first step, \prop performs \gfcrii{$2N +4$} inter-subarray data copies (using LISA-RISC) to copy the \emph{generate} and \emph{propagate} bits from $subarray_i$ to $subarray_{i+1}$.
In the second step, \prop performs a series of Boolean operations (using \aaps) to compute the next generate and propagate bits in parallel (using SALP-MASA) across \emph{all} DRAM subarrays.
These two steps repeat for $log(N)$ iterations. 
\gfcriv{The latency of executing \omcriv{an $N$-bit bit-parallel addition} using \prop is} $3log_2 N + 13$ \aap cycles + $2N + 4$ \texttt{RBM} cycles.
\gf{Even though the bit-parallel algorithms have a lower time complexity than the bit-serial algorithms, the \gfcrii{former} can require more inter-subarray copies\gfcrii{, i.e., $2N + 4$ \texttt{RBM} cycles for bit-parallel algorithms versus $2(N - 1)$  \texttt{RBM} cycles for bit-serial algorithms}.}  
}

\paratitle{RBR-Based Algorithms} \gfcrii{\gf{Figure~\ref{fig:rbr_exemple} illustrates \prop' implementation of a two-bit \gls{RBR}-based adder~\cite{makino19968}. 
The adder operates in three steps. 
First, each digit $i$ generates an intermediate value $h_i$, computed \emph{only} from the corresponding input digit $i$. Second, the output value $f_i$ at digit $i$ is computed as a function of both the current digit and the preceding intermediate value $h_{i-1}$. 
Third, the $sum$ at digit $i$ depends on the current digit, $h_{i-1}$, and $f_{i-1}$. 
To propagate intermediate results between digits, \prop uses \texttt{RBM} commands to transfer the values of $h_i$ and $f_i$ from $subarray_i$ to $subarray_{i+1}$. 
The \gls{RBR}-based addition executes with a constant latency of 34 \aaps cycles and 8 \texttt{RBM} cycles. Beyond addition, \prop reuses the same \gls{RBR}-based adder design to support additional arithmetic operations, including subtraction and multiplication in the \gls{RBR} format.}}

% \begin{figure}[!t]
%     \centering
%     \includegraphics[width=0.45\linewidth]{figures/rbr_example-crop.pdf}
%     \caption{\gf{\prop's implementation of a 2-bit \gls{RBR} adder. Bits $A_i$ and $B_i$ are stored \emph{vertically} in the same DRAM bitline. }}
%     \label{fig:rbr_exemple}
% \end{figure}

\subsubsection{Cost Model Logic Implementation}
\label{sec:implementation:cost}

\gf{Figure~\ref{fig:cost-model} depicts the hardware design of the \costmodel. 
The \costmodel has two main components: 
\li~one \gls{LUT} per \gls{PuD} operation, and
\lii~\emph{Select Logic}. 
Each \sgi{LUT row} represents a different bit-precision, and stores the \sgi{index} of the best-performing \uprog in the \sgi{library} for that \sgi{operation\omcrii{--}precision \omcrii{combination}}. 
\sgi{\gfcrii{We \emph{empirically} sized each} \gls{LUT} \gfcrii{to} contain 64 eight-bit rows (i.e., supporting up to 64-bit computation, and indexing up to 64 different \uprog implementations per \gls{PuD} operation)}. 
\revB{\label{rb.2}\changeB{B2}The \costmodel works in four main CPU cycles. It receives as input the \emph{bit-precision} (6 bits) and the \emph{bbop\_op} opcode (4 bits) of the target \gls{PuD} operation. 
In the first cycle, the \emph{bit-precision} indexes all the \glspl{LUT} in parallel (\circled{1} in Figure~\ref{fig:cost-model}), selecting the best-performing \idx for the given bit-precision for all implemented \gls{PuD} operations (\circled{2}).
The \costmodel can \emph{quickly} query the \glspl{LUT} since they consist of a few (i.e., 16) small (i.e.,  64~B in size\revdel{; 64 rows, each holding eight-bit \idx}) SRAM arrays indexed in parallel. 
In the second cycle, based on the 4-bit \emph{bbop\_op} opcode, the \emph{Select Logic} chooses the appropriate \idx (\circled{3}).
In the third cycle, the \idx is concatenated with the \emph{bbop\_op} opcode to form the \gfcrii{\gbidx} (\circled{4}).
In the fourth \gfmicro{cycle}, the \gfcrii{\gbidx} indexes and fetches the best-performing \uprog from the \emph{\uprog Scratchpad} (\circled{5}). If the target \uprog is not loaded in the \emph{\uprog Scratchpad}, the \costmodel fetches it from the \emph{\uprog Memory} (not shown). }} 
\gfmicro{We estimate, using CACTI~\cite{cacti}, that the access latency and energy per access of the \SI{64}{\byte} SRAM array (used in our \costmodel)  is of  \SI{0.07}{\nano\second} (i.e., less than 1 CPU cycle) and \SI{0.00004}{\nano\joule}.\revdel{As a simple comparison point, the energy it takes to execute a 32-bit \gls{PuD} addition operation is 210.6~nJ~\cite{hajinazarsimdram}.}}
%\agymicrocomment{\circled{4} and \circled{5} can go to the right side to save space.}

% \begin{figure}[ht]
%     \centering
%     \includegraphics[width=0.45\linewidth]{figures/cost_model_unit-crop.pdf}
%     \caption{\prop's cost model logic implementation.}
%     \label{fig:cost-model}
% \end{figure}

\begin{figure}[ht]
\centering
  \centering
 \includegraphics[width=0.45\linewidth]{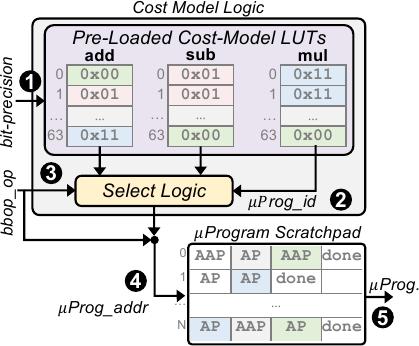}
  \caption{\omcrii{\prop' Cost Model Logic}.}
  \label{fig:cost-model}
\end{figure}

\subsubsection{\gfisca{Pareto Analysis}} 
\label{sec:implementation:pareto}
\gfisca{~We conduct a performance and energy Pareto \omcrii{analysis} to populate the \prelut. 
We model each \uprog using an analytical cost model that takes as input the target bit-precision, the number of elements used during computation, and the number of DRAM subarrays available. 
The analytical cost model outputs the throughput (in GOPs/s) and energy efficiency (in throughput/Watt) for each \uprog in the \uproglib.
We highlight our analyses for two main operations (i.e., addition and multiplication) since they represent linearly and quadratically-scaling \gls{PuD} operations, respectively. The analyses for subtraction and division follow similar observations.
In our analyses, we evaluate a SIMDRAM-like \gls{PuD} architecture using the three data mapping schemes described in Figure~\ref{fig:obps}.
We assume a DRAM bank with 64 \omcrii{\gls{PuD}-capable} DRAM subarrays and a subarray with 65,536 columns.
We vary the number of input elements as multiples of the number of DRAM columns per subarray (from 1 DRAM subarray with \omcriv{64K} input elements to 64 DRAM subarrays with \omcriv{4M} input elements) for our measurements. 
} 

\paratitle{\gfisca{Linearly\omcrii{-}Scaling \gls{PuD} Operations}} \gfisca{Figure~\ref{fig:pareto_addition} shows the throughput (\gfcriv{$y$-axis}; top) and energy efficiency (\gfcriv{$y$-axis}; bottom) of six \uprog implementations for a linearly\omcrii{-}scaling \gls{PuD} operation (i.e., \omcrii{integer} addition) \omcriv{for different bit-precision values ($x$-axis)}.
\gfcriv{Each subplot depicts the different input data sizes we use in our analysis.}
\gfcrii{For this analysis, we implement the following addition algorithms: 
ripple-carry adder (RCA), 
carry-select adder (CSA)~\cite{bedrij1962carry},
Brent-Kung adder~\cite{brent1982regular},
Kogge–Stone adder~\cite{kogge1973parallel}, Ladner-Fischer adder~\cite{ladner1980parallel}, using 
\gfcrii{\li~both} two's complement and \gls{RBR} \gfcrii{data format representations; and
\lii~\gls{ABOS}, \gls{ABPS}, and \gls{OBPS} data mappings}. 
Note that the bit-parallel adder can \emph{only} be implemented using the \gls{OBPS} data mappings.}
We make two observations. 
First, \omcrii{in terms of} throughput, the best-performing adder implementation varies depending on the target bit-precision and number of input elements.
\gfcrii{The achievable throughput ultimately depends on a combination of the number of \aaps that can be concurrently executed across DRAM subarrays and the number of inter-DRAM subarray operations required to implement the adder. 
In general, we empirically observe that as the input data size increases \gfcriv{(see subplots' titles)}, the number of inter-DRAM subarray operations also increases and eventually dominates the overall execution time.}
For \emph{small bit-precision} and \emph{small input size} (i.e., bit-precision smaller than 8, and \omcriv{fewer} than \omcriv{256K} input elements), the bit-serial RCA using the \gls{OBPS} data mapping provides 
\gfcrii{the highest throughput, while for \emph{large bit-precision} and \emph{small input size} (i.e., bit-precision larger than 8, and \omcriv{fewer} than \omcriv{256K} input elements), the \gls{RBR} adder using the \gls{OBPS} data mapping provides the highest throughput.}
%\gfcrii{up to} 2.7$\times$ (1.4$\times$) the throughput of \gfcrii{the second-best performing adder (i.e., \omcrii{the same RCA) using the} \gls{ABOS} (\gls{ABPS}) \gfcrii{data mapping}. 
%For large bit-precision and small input size (i.e., bit-precision larger than 8, and less than 256~K input elements), the \gls{RBR} adder using the \gls{OBPS} data mapping provides 7.7$\times$ (3.9$\times$) the throughput of \omcrii{the same adder using} \gls{ABOS} (\gls{ABPS}) \omcrii{data mapping}.
For large-enough input sizes \gfcrii{(i.e., more than \omcriv{1M} input elements)}, employing the \gls{ABPS} data mapping leads to the highest throughput, independent of the bit-precision. 
This is because when more DRAM subarrays are involved in the execution of the target \gls{PuD} operation, the inter-subarray data transfers dominate overall execution time in the \gls{OBPS} implementations. 
Second, \omcrii{in terms of} energy efficiency, the bit-serial implementation of RCA provides the best throughput/Watt for \gls{ABOS}, \gls{ABPS}, and \gls{OBPS}, independent of the bit-precision and input size. 
This is because 
\li~the number of \aaps performed to execute RCA is the same \emph{independent} of the data mapping, and
\lii~the energy the bit-parallel algorithms consume is dominated by inter-subarray operations\gfcrii{, which is \emph{not} present in bit-serial implementations}.}

\begin{figure}[ht]
    \centering
    \includegraphics[width=\linewidth]{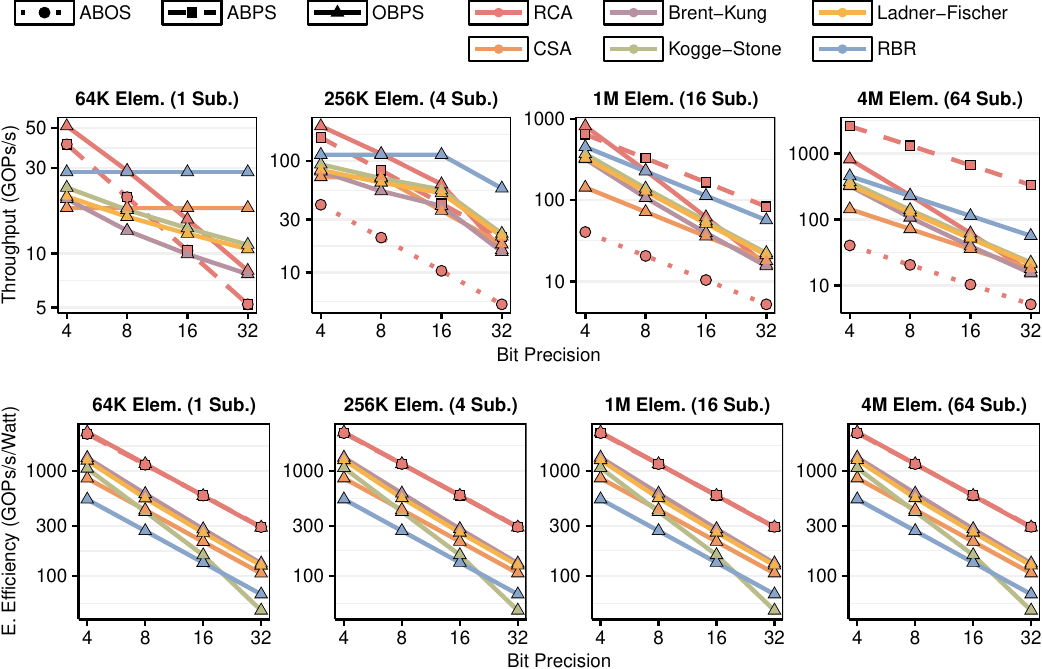}
    \caption{\omcrii{Pareto analysis for throughput (top) and energy efficiency (bottom) for \gls{PuD} addition operations. 
    Dotted lines represent \gls{ABOS};
    dashed lines represent \gls{ABPS}; straight lines represent \gls{OBPS} data mapping. }}
    \label{fig:pareto_addition}
\end{figure}

\paratitle{\gfisca{Quadratically\gfcrii{-}Scaling \gls{PuD} Operations}} \gfisca{Figure~\ref{fig:pareto_multiplication} shows the throughput (top) and energy efficiency (bottom) of six \uprog implementations for a quadratically\omcrii{-}scaling \gls{PuD} operation (i.e., \omcrii{integer} multiplication).
We implement \gls{PuD} multiplication \gfcrii{operations as a triplet composed of \li~the multiplication method (i.e., \gfcrii{Booth's multiplication algorithm~\cite{booth1951signed}} or the divide-and-conquer Karatsuba~\cite{karatsuba1962multiplication} multiplication);
\lii~different methods for addition (i.e., bit-serial RCA, bit-parallel Ladner-Fischer~\cite{ladner1980parallel}, and \gls{RBR}-based adder); and 
\liii~data mappings (i.e., \gls{ABOS}, \gls{ABPS}, and \gls{OBPS})}.
\gfcrii{Note that \gls{PuD} multiplication operations that use bit-parallel and RBR-based adders can \emph{only} be implemented using the \gls{OBPS} data mapping.}
%For this analysis, we implement the following addition algorithms: 
%ripple-carry adder (RCA), 
%carry-select adder (CSA)~\cite{bedrij1962carry},
%Brent-Kung adder~\cite{brent1982regular},
%Kogge–Stone adder~\cite{kogge1973parallel}, Ladner-Fischer adder~\cite{ladner1980parallel}, using 2's complement; 
%and the adder design in  Figure~\ref{fig:rbr_exemple} using RBR.
We make two observations.
First, \omcrii{in terms of} throughput, the best-performing multiplier implementation varies depending on the bit-precision and number of input elements. 
For small bit-precision and small input size (i.e., bit-precision smaller than 8, and \omcriv{fewer} than \omcriv{64K} input elements), \gfcrii{Booth's} bit-serial multiplication \omcrii{with \gls{ABOS} data mapping} provides the highest throughput\gfcrii{, while for medium bit-precision and small input size (i.e., bit-precision from 8 to 16 and \omcriv{fewer} than \omcriv{64K} input elements), \gfcrii{Booth's} bit-parallel multiplication \omcrii{with the \gls{OBPS} data mapping} provides the highest throughput.}
%, i.e., 1.5$\times$ (1.5$\times$) the throughput of \gls{ABOS} (\gls{ABPS}) multiplication implementation.
For high bit-precision and small-to-medium input size (i.e., bit-precision larger than 32 and \omcriv{fewer} than \omcriv{256K} input elements), \gfcrii{RBR-based multiplication using \gls{OBPS} data mapping} provides the highest throughput.
\omcrii{For} large-enough input sizes (i.e., larger than \omcriv{1M} input elements), employing \gfcrii{Booth's bit-serial RCA-based multiplication using} \gls{ABPS} data mapping leads to the highest throughput, independent of the bit-precision. 
Second, \omcrii{in terms of} energy efficiency, \gfcrii{Booth's bit-serial RCA-based} multiplication implementation provides the best throughput/Watt for \gls{ABOS}, \gls{ABPS}, and \gls{OBPS}, independent of the bit-precision and input size\gfcrii{, since
\li~the number of \aaps required to execute the addition step is the same regardless of the data mapping and
\lii~the energy of the bit-parallel-based algorithms is dominated by the large number of inter-subarray operations they require.
}
}

\begin{figure}[ht]
    \centering
    \includegraphics[width=\linewidth]{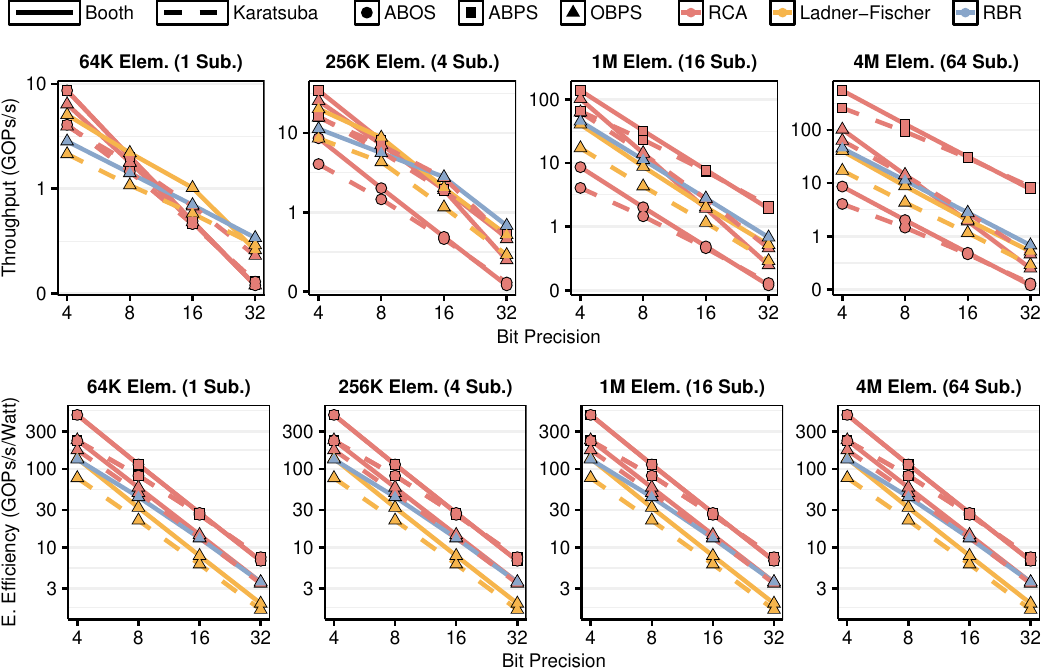}
    \caption{\omcrii{Pareto analysis for throughput (top) and energy efficiency (bottom) for multiplication. Straight lines represent the \gfcrii{Booth's} multiplication method~\cite{booth1951signed}; dashed lines represent the Karatsuba~\cite{karatsuba1962multiplication} multiplication method.  }}
    \label{fig:pareto_multiplication}
\end{figure}

\subsubsection{\gfisca{Non-Arithmetic \gls{PuD} Operations.}} 

\gfisca{We also equip \prop' \uproglib with SIMDRAM's implementations of non-arithmetic \gls{PuD} operations~\omcrii{\cite{hajinazarsimdram}}, including
\li~$N$-bit logic operations (i.e., \texttt{AND}/\texttt{OR}/\texttt{XOR} of more than two input bits), \lii~relational operations (i.e., equality/inequality check, greater than, maximum, minimum), 
\liii~predication, and \liv~bitcount and ReLU~\cite{goodfellow2016deep}.}

\subsection{Dynamic Bit-Precision Engine}
\label{sec:implementation:dynamic-precision}

\gf{The \dynengine comprises a simple reconfigurable $n$-bit comparator and a \gls{FSM}. For each evicted cache line, the \gls{FSM} probes the \emph{Object Tracker} and identifies if the \omcrii{incoming  evicted} cache line belongs to a \gls{PuD}'s memory object. If it does, the \gls{FSM} executes \gfmicro{four} operations.
First, it reads the bit-precision value (specified by the \texttt{bbop\_trsp\_init} instruction) and the current maximum value stored in the \emph{Object Tracker} for the given memory object.
Second, it uses the bit-precision value to configure the $n$-bit comparator.
Third, it inputs to the $n$-bit comparator all $n$-bit values in the \omcrii{incoming} cache line (one \omcrii{at a} time) and the current maximum value. 
Fourth, after all the $n$-bit values are processed, if any value in the \omcrii{incoming} cache line is larger than the current maximum value, the \gls{FSM} sends an update signal to the \emph{Object Tracker} alongside the new maximum value.}
\gfmicro{The energy cost of identifying the largest element in a \SI{64}{\byte} cache line is \SI{0.0016}{\nano\joule}~\cite{han2016eie}. That represents an increase in 0.084\% in the energy of an \gls{LLC} eviction~\cite{damov,muralimanohar2007optimizing,tsai:micro:2018:ams}, which \emph{needs} to happen prior to \gls{PuD} execution regardless.}

\subsection{\uprog Select Unit}
\label{sec:implementation:control-unit}

\revdel{\gf{The \uprogunit aims to identify and select the appropriate bit-precision and \uprog for a \gls{PuD} operation. It consists of a 
\li~\bitprec, which evaluates the target bit-precision based on the input operands of the target \gls{PuD} operation and their associated maximum values, and
\lii~buffer space to store the selected \uprog.}}

\paratitle{Calculating Bit-Precision} \gf{The \uprogunit needs to address two scenarios when calculating the bit-precision for \gls{PuD} operations: \emph{vector-to-vector} \gls{PuD} operations, and \emph{vector-to-scalar} reduction \gls{PuD} operations.
In \omcrii{\emph{vector-to-vector}}, the target \gls{PuD} operation implements a parallel \emph{map} operation, in which inputs and outputs are data vectors. For such operations, the bit-precision can be computed \emph{a priori}, using the maximum values the \dynengine provides, \emph{even} in the presence of chains of \gls{PuD} operations. In such a case, the \gfcrii{\emph{Bit-Precision Calculation Engine}} updates the \emph{Object Tracker} with the maximum possible output value for \emph{each} \gls{PuD} in the chain. 
For example, assume a kernel that executes \texttt{D[i]=(A[i]+B[i])$\times$C[i]} as follows:   }   

\begin{center}
\vspace{-10pt}
\tempcommand{.8}
  \resizebox{0.6\columnwidth}{!}{
\begin{tabular}{ll}
\texttt{bbop\_add(tmp, A, B, 8k, 8, 1)};      & // tmp $\leftarrow$ A + B      \\
\texttt{bbop\_mul(D, tmp, C, 8k, 8, 1)};   & // D $\leftarrow$ tmp $\times$ C
\end{tabular}
}
\vspace{-10pt}
\end{center}

\noindent \gf{Assume that the maximum value of \texttt{A}, \texttt{B}, and \texttt{C} are $3$, $6$, and $2$, respectively. In this case, the \uprogunit
\li~computes the bit-precision for the addition operation as $\ceil*{\log_2 (3+6)} = 4\ bits$;
\lii~updates the \emph{Object Tracker} entry of \texttt{tmp} with the maximum value of the addition operation (i.e., $9$);
\liii~computes the bit-precision for the multiplication operation as $\ceil*{\log_2 (9\times2)} = 5\ bits$ \sg{using an $n$-bit scalar ALU};
\liv~updates the \emph{Object Tracker} entry of \texttt{D} with the maximum value of the multiplication (i.e., $18$).}

\gf{In \emph{vector-to-scalar} reduction, the \gls{PuD} operation implements a parallel \emph{reduction} operation, where the inputs are vectors and the output is a scalar value. In this case, the bit-precision \emph{cannot} be computed with \emph{only} the maximum input operands without causing \emph{overprovision\omcrii{ing}}, since in a reduction, each element contributes to the bit-precision of the scalar output. Therefore, for \emph{vector-to-scalar} reduction \gls{PuD} operations, the \uprogunit needs to 
\li~fetch from DRAM the row containing the carry-out bits produced during partial steps\footnote{\gf{\prop implements \gls{PuD} reduction operations using \emph{reduction trees}~\cite{mimdramextended}. Thus, a partial step refers to a level of the reduction tree.}} of the \gls{PuD} reduction;
\lii~evaluate \gfasplos{if} a partial step generated an overflow (i.e., check if any carry-out bit is `\texttt{1}'); and
\liii~increment the bit-precision for the next partial step \omcrii{if overflow is detected}.\revdel{ We design the \uprogunit to fully overlap the latency of bit-precision computation with \uprog execution.} }

\paratitle{Hardware Design} \gf{The \uprogunit comprises of simple hardware units:
\li~an \emph{$n$-bit ALU} to compute the target bit-precision, 
\lii~a \emph{Fetch Unit} to generate load instructions for carry re-evaluation, and
\liii~a \emph{\uprog Buffer} to store the currently running \uprog.}\revdel{\footnote{\label{ft.re.2}\revE{We do \emph{not} leverage \gls{PuD} operations to compute the target bit-precision for two reasons. 
First, the bit-precision information is required by \prop \emph{Control Unit} (placed within the memory controller) \emph{prior} to issuing \gls{PuD} operations for the target computation. 
Second, \prop calculates the target bit-precision for a given \gls{PuD} operation by identifying the largest data elements within the memory objects involved in the \gls{PuD} operation. This involves computing the target operation using at most two data elements. Offloading such computation to \gls{PuD} would incur wasted resources from the massively parallel \gls{PuD} system, high latency, and high energy cost.}}}

\subsection{\gfisca{\omcriv{Other Considerations}}}
\label{sec:implementation:alltogether}

\paratitle{\gfisca{Data Format Conversion}}
\gfisca{\revdel{Concretely, \prop might require 
\li~distributing the input operands stored vertically in one DRAM subarray to multiple DRAM subarrays (to implement the \gls{OBPS} data mapping) or
\lii~convert input operands from their two's complement representation to the RBR. 
\prop realizes both data format conversions using \gls{PuD} primitive operations.}
To distribute the bits of input operands to different DRAM subarrays, 
\prop issues \texttt{RBM} commands from the DRAM row storing bit $i$ in the source DRAM subarray to the target DRAM row in the destination DRAM subarray $i$, i.e., $row[dst]_{subarray_i} \leftarrow RBM (row[src]_i); i \in [1, bits]$. 
To convert the data stored in two's complement to its equivalent \gls{RBR} (see~\cref{sec_motivation}), \prop performs in-DRAM bitwise operations, as Table~\ref{table:data_conversion} describes.}

\begin{table}[ht]
\caption{Conversion from two's complement to RBR.}
   \centering
   \footnotesize
   \tempcommand{0.95}
   \resizebox{0.7\columnwidth}{!}{
\begin{tabular}{r||ccc}
\toprule
\textbf{Input X $\rightarrow$}                       & \multicolumn{1}{c|}{2}       & \multicolumn{1}{c|}{-1}      & -7      \\ 
\textbf{two's complement $\rightarrow$}                & \multicolumn{1}{c|}{0 0 1 0} & \multicolumn{1}{c|}{1 1 1 1} & 1 0 0 1 \\ \midrule \midrule
\textbf{Steps to convert two's to RBR $\downarrow$}    & \multicolumn{3}{c}{\textbf{Output $\downarrow$}}                                  \\ \hline \hline
Extract most-significant bit (MSB)      & \multicolumn{1}{c|}{0}       & \multicolumn{1}{c|}{1}       & 1       \\ 
\gfcrii{buffer1}: broadcast MSB to all subarrays & \multicolumn{1}{c|}{0 0 0 0} & \multicolumn{1}{c|}{1 1 1 1} & 1 1 1 1 \\ 
\gfcrii{buffer2}: \texttt{NOT}(\gfcrii{buffer1})                   & \multicolumn{1}{c|}{1 1 1 1} & \multicolumn{1}{c|}{0 0 0 0} & 0 0 0 0 \\ 
~X + 1                                  & \multicolumn{1}{c|}{1 1 1 0} & \multicolumn{1}{c|}{0 0 0 1} & 0 1 1 1 \\ \hline 
$X-$ = \gfcrii{buffer1} \& (~X + 1)                 & \multicolumn{1}{c|}{0 0 0 0} & \multicolumn{1}{c|}{0 0 0 1} & 0 1 1 1 \\ 
$X+$ = \gfcrii{buffer2} \& (X)                      & \multicolumn{1}{c|}{0 0 1 0} & \multicolumn{1}{c|}{0 0 0 0} & 0 0 0 0 \\ 
\bottomrule
\end{tabular}
}
\label{table:data_conversion}
\end{table}

\paratitle{\gfisca{System Integration}} \gfisca{
\prop leverages the \emph{same} system integration solutions as in prior \gls{PuD} systems~\cite{hajinazarsimdram,mimdramextended}, including: 
\li~ISA extensions included to the host CPU ISA that the programmer utilizes to launch \emph{bbop} instructions;
\lii~a hardware control unit, alongside the memory controller, to control the execution of \uprogs; and
\liii~a hardware transposition unit, placed between the \gls{LLC} and the memory controller, to transpose data from the native horizontal data layout to the \gls{PuD}-friendly vertical data layout.
\revdel{As in~\cite{hajinazarsimdram,mimdramextended}, we assume that the operating system (OS) can provide support for data allocation and data mapping of operands in the DRAM bank dedicated for \gls{PuD} computing and the \gls{PuD} substrate operates directly on physical addresses.} 
\revdel{We acknowledge that such system integration assumptions are potential limitations of SIMDRAM and \prop (and several other \gls{PuD} architectures~\cite{ferreira2021pluto, ferreira2022pluto, li2017drisa, deng2018dracc}). 
We leave a more robust system integration implementation to future work.}
}

\paratitle{Limitation of \gls{SLP}} \gfisca{\gls{SLP} is limited by the $t_{FAW}$ DRAM timing constant~\cite{jedec2012ddr3, jedec2012jedec, jedec2017jedec}, which corresponds to the time window during which at most four \texttt{ACT} commands can be issued per DRAM rank. 
This constraint protects against the deterioration of the DRAM reference voltage. 
DRAM manufacturers have been able to
relax $t_{FAW}$ substantially in commodity DRAM chips~\cite{micron2013tfaw}, 
as well as to perform a targeted reduction of this parameter specifically for \gls{PIM} architectures where it becomes
a performance bottleneck~\cite{he2020newton,lee2021hardware}. These advances suggest that
\omcrii{$t_{FAW}$ likely does} \emph{not} limit \prop' scalability in commodity DRAM chips.}

\paratitle{\omcrii{Proteus for Floating-Point Operations}} \prop' dynamic bit-precision computation can be employed in two different stages of floating-point arithmetic for \gls{PuD} operations, during exponent and mantissa computation, \omcrii{using} three steps. \Copy{R1.1D}{\asplosrev{\hl{First, the \mbox{\dynengine} identifies the sign, exponent, and mantissa bits in floating-point numbers stored in the evicted cache lines of \mbox{\gls{PuD}} memory objects.
Second, the \mbox{\dynengine} updates the \emph{Object Tracker} with the maximum exponent and the mantissa values for the \mbox{\gls{PuD}} memory object, in case the \omcrii{identified} maximum values are smaller than the values stored in the current evicted cache line. This process is analogous to the execution flow described in \mbox{\cref{sec:overview:execution}} for integer operands, with the addition of the fields for the maximum exponent and maximum mantissa values in the \emph{Object Tracker}.
Third, \mbox{\prop} performs the target in-memory floating-point computation by issuing~\mbox{\cite{dualitycache}}
\li~bit-serial subtraction (addition) \mbox{\gls{PuD}} operations to calculate the resulting exponents, followed by 
\lii~bit-serial addition (multiplication) \mbox{\gls{PuD}} operations to calculate the resulting mantissas for a vector addition (multiplication) instruction~\mbox{\cite{dualitycache}}. Note that the bit-serial \mbox{\gls{PuD}} operations involved in \li--\lii~are vector-to-vector \mbox{\gls{PuD}} operations. As such, \mbox{\prop} can leverage the maximum exponent and mantissa values}\omrev{\hl{ stored in the \emph{Object Tracker}}}\hl{ to set the required bit-precision for such bit-serial operations, following the same approach described in~\mbox{\cref{sec:implementation:control-unit}}.}}}

\revdel{\Copy{R1.1E}{\asplosrev{\hl{We identify another opportunity to exploit narrow values during mantissa computation by identifying \emph{repeating patterns} in the mantissa to}\omrev{\hl{ configure}}\hl{ the bit-precision for the target operation. For example, when representing \texttt{+1.3} in floating-point (sign = `\texttt{0}', exponent = `\texttt{01111111}', mantissa = `\texttt{01001100110011001100110}'), the mantissa bits `\texttt{0011}' repeats five times. 
Based on that, we can perform the bit-serial mantissa computation by
\li~issuing \mbox{\gls{PuD}} operations (e.g., adding or multiplying the mantissa) for the first iteration of the repeating pattern `\texttt{0011}' and 
\lii~copying (using RowClone~\mbox{\cite{seshadri2013rowclone}}) the remaining bits of the output mantissa depending on the identified repeating pattern (i.e., four times in our example). 
However, to implement such}\omrev{\hl{ a}}\hl{ scheme, \mbox{\prop} would need to identify variable-length patterns across the mantissa bits for all data elements involved in the computation,}\omrev{\hl{ which would incur}}\hl{ non-trivial hardware complexity. Thus, we leave the realization of}\omrev{\hl{ this optimization}}\hl{ approach as future work.}}} 
}

\section{Methodology}
\label{sec:methodology}

\label{r3.1}\Copy{R3.1}{We implement \prop using an in-house cycle-\asplosrev{\hl{level}} simulator \gfcrii{(which we open-source at~\cite{proteusgit})} and compare it to a real multicore CPU (Intel Comet Lake~\cite{intelskylake}),
a real high-end GPU (NVIDIA A100 \asplosrev{\hl{using CUDA and tensor cores}}~\cite{a100}), and a \gfasplos{simulated} state-of-the-art \gls{PuD} framework  (SIMDRAM~\cite{hajinazarsimdram}). 
In our evaluations, the CPU code uses AVX-512 instructions~\cite{firasta2008intel}.
\asplosrev{\hl{Our simulator \omcrii{is} rigorously validated against SIMDRAM~\mbox{\cite{hajinazarsimdram}} and MIMDRAM~\mbox{\cite{mimdramextended}}'s gem5~\mbox{\cite{gem5}} implementation~\mbox{\cite{mimdramgit}}. The simulator 
\li~is cycle-level accurate}\omrev{\hl{ with regard to}}\hl{ DRAM commands
 and 
\lii~accounts for the data movement cost of cache line eviction on a per-cycle basis.}}
Our simulation accounts for the additional latency imposed by SALP~\cite{kim2012case} on \texttt{ACT} commands, i.e., the extra circuitry required to support SALP incurs an extra latency of \SI{0.028}{\nano\second} to an \texttt{ACT}~\cite{hassan2022case}, which is less than  0.11\% extra latency \omcrii{of} an \texttt{AAP}. 
To verify the functional correctness of our applications, our simulation infrastructure \omcrii{performs} functional verification over application's data when performing \gls{PuD} operations.
%\revdel{ Thus, after executing an application in a \gls{PuD} configuration, we compare the produced output with and without dynamic bit-precision and \gls{PuD} offloading, looking for exact data matches.} 
We did \emph{not} observe any difference from the \asplosrev{\hl{golden}} outputs.}
\gfcriv{We open-source our simulation infrastructure at \url{https://github.com/CMU-SAFARI/Proteus}.}

Table~\ref{table_parameters} shows the system parameters we use in our evaluations.
\revdel{\changeA{A1}\revA{\label{ra.1}We evaluate SIMDRAM and \prop execution time by isolating the main kernel that SIMDRAM/\prop executes (i.e., the offloaded \gls{PuD} instructions) and evaluating its performance. 
We estimate the end-to-end speedup SIMDRAM/\prop provides for each application by applying Amdahl's law~\cite{amdahl1967validity}. 
Thus, SIMDRAM and \prop' end-to-end speedup is given by: $((1 - kernel\_time) + \frac{kernel\_time}{PUD\_impro})^{-1}$; where $PUD\_impro$ is the speedup SIMDRAM/\prop provides for the offloaded kernel compared to the CPU execution, and $kernel\_time$ is the percentage of the total execution time the offloaded kernel represents when executing the application on the baseline multi-core CPU.}}
\gfisca{To measure CPU energy consumption, we use Intel RAPL~\cite{hahnel2012measuring}. We capture GPU kernel execution time that excludes data initialization/transfer time. We use the \texttt{nvml} API~\cite{NVIDIAMa14} to measure GPU energy consumption.
We use CACTI \gfcrii{7.0}~\cite{cacti} to evaluate \prop and SIMDRAM energy consumption, where we take into account that each additional simultaneous row activation increases energy consumption by 22\%~\cite{seshadri2017ambit, hajinazarsimdram}.
\gfisca{We evaluate two SIMDRAM configurations:
\li~SIMDRAM with \gls{SLP}~\omcrii{\cite{kim2012case}} \gfasplos{and static bit-precision} (\emph{SIMDRAM\gfasplos{-SP}}), and 
\lii~SIMDRAM with \gls{SLP} and \prop' \emph{Dynamic Bit-Precision Engine} (\emph{SIMDRAM\gfasplos{-DP}})\gfmicro{. \gfasplos{In both configurations, the system implements only the 16 \uprogs proposed in SIMDRAM (i.e., there is \emph{no} \uproglib enabled). }
We evaluate four \prop configurations:}
\li~\prop \emph{LT\gfasplos{-SP}} and
\lii~\prop \emph{EN\gfasplos{-SP}}, where \prop \revCommon{selects} the \emph{lowest latency} \omcrii{(LT)} and \emph{lowest energy} \omcrii{(EN)} consuming \uprog, respectively, \agymicro{using the statically profiled bit-precision from Figure~\ref{fig:narrow_values}};
\agymicro{\liii~\prop \emph{LT\gfasplos{-DP}} and \liv~\prop \emph{EN\gfasplos{-}DP}, where \prop executes the \emph{lowest latency} and \emph{lowest energy} consuming \uprog with dynamically chosen bit-precision.}\revdel{\agymicro{\footnote{\agymicro{For statically profiled bit-precision, we \emph{must} round the bit-precision up to the nearest power-of-two\revdel{, e.g., a 13-bit operation will be implemented using 16-bit data, since high-level programming language (e.g., C/C++) are \emph{bounded} by 2's complement data representation formats}.}}}}
We use 64 subarrays in \emph{\omcrii{only} one} DRAM bank for \omcrii{our} \gls{PuD} \omcrii{evaluations}.}}\footnote{\label{ft.rc.1}\revC{\gfmicro{The column/address (C/A) bus allows the simultaneously activation of up to 84 DRAM subarrays ($\frac{t_{RAS}}{t_{CK}}$ = $\frac{32~ns}{0.38~ns}$ = 84).}}}

%Our simulation accounts for the additional latency imposes by \prop's mat isolation transistors and row decoder latches upon DRAM operations (i.e., less than 0.5\% extra latency for an \texttt{ACT}).

\begin{table}[!ht]
   \caption{Evaluated system configurations.}
   \centering
   \footnotesize
   \tempcommand{1}
  \renewcommand{\arraystretch}{0.9}
   \resizebox{0.7\columnwidth}{!}{
   \begin{tabular}{@{} c l @{}}
   \toprule
   \multirow{5}{*}{\shortstack{\textbf{Intel}\\ \textbf{Comet Lake CPU~\cite{intelcometlake}} \\ \omcrii{\textbf{(Real System)}}}} & x86~\cite{guide2016intel}, 16~cores, 8-wide, out-of-order, 3.8~GHz;  \\
                                                                           & \emph{L1 Data + Inst. Private Cache:} 256~kB, 8-way, 64~B line; \\
                                                                           & \emph{L2 Private Cache:} 2~MB, 4-way, 64~B line; \\
                                                                           & \emph{L3 Shared Cache:} 16~MB, 16-way, 64~B line; \\
                                                                           & \emph{Main Memory:} 64~GB DDR4-2133, 4~channels, 4~ranks \\
   \midrule
      \multirow{3}{*}{\shortstack{\textbf{NVIDIA}\\ \textbf{A100 GPU~\mbox{\cite{a100}}} \\ \omcrii{\textbf{(Real System)}}}} &  7~nm technology node; 826~mm$^2$ die area~\cite{a100}; 6912 CUDA cores;\\ 
                                                                            & \asplosrev{\hl{432 tensor cores}}, 108 streaming multiprocessors, 1.4~GHz base clock; \\
                                                                            & \emph{L2 Cache:} 40~MB L2 Cache; \emph{Main Memory:} 40~GB HBM2~\mbox{\cite{HBM,lee2016simultaneous}} \\
   \midrule

   \multirow{9}{*}{\shortstack{\textbf{SIMDRAM~\cite{hajinazarsimdram}}\\ \textbf{\& \prop} \\ \omcrii{\textbf{(Simulated)}}}} &  gem5-based in-house simulator~\omcrii{\cite{proteusgit, mimdramgit}};  x86~\cite{guide2016intel};  \\ 
                                                            & 1 \gfmicro{out-of-order core @ 4~GHz (\emph{only} for instruction offloading});\\
                                                                             & \emph{L1 Data + Inst. Cache:} 32~kB, 8-way, 64~B line;\\
                                                                             & \emph{L2 Cache:} 256~kB, 4-way, 64~B line; \\
                                                                             & \emph{Memory Controller:}  8~kB row size, FR-FCFS~\cite{mutlu2007stall,zuravleff1997controller}\\
                                                                             & \emph{Main Memory:}  \gfmicro{DDR5-5200}~\omcrii{\cite{jedec2020jesd795}}, 1~channel, 1~rank, 16~banks \\ 
                                      
   \bottomrule
   \end{tabular}
   }
   \label{table_parameters}
\end{table}

\label{r3.2A}\Copy{R3.2A}{\paratitle{Real-World Applications} We select \gfisca{twelve} workloads from four popular benchmark suites in our real-workload analysis (\gfisca{as Table~\ref{table:workload:properties} describes)}\revdel{, including 
\li~525.x264\_r (\texttt{x264}) from SPEC 2017~\cite{spec2017};
\lii~\texttt{pca} from Phoenix~\cite{yoo_iiswc2009};
\liii~\texttt{2mm}, 
\texttt{3mm}, 
convolution (\texttt{cov}),
doitgen (\texttt{dg}), 
fdt\gfmicro{d}-apml (\texttt{fdt\gfmicro{d}}),
gemm  (\texttt{gmm}), and
gramschmidt (\texttt{gs}) from Polybench~\cite{pouchet2012polybench};
and
\liv~heartwall (\texttt{hw}), kmeans (\texttt{km}), and backprop (\texttt{bp}) from Rodinia~\cite{che_iiswc2009}}.
%\gf{Since our base \gls{PuD}} substrate (SIMDRAM) does \emph{not} support floating-point, we manually modify the selected floating-point-heavy auto-vectorized loops to operate on fixed-point data arrays.\footnote{We only modify the three applications from the Rodinia benchmark suite to use fixed-point operations. The applications from Polybench can be configured to use integers; the auto-vectorized loops in 525.x264\_r use \texttt{uint8\_t}; pca uses integers. Prior work~\cite{fujiki2018memory} also employs fixed-point for the same three Rodinia applications.  }
We manually modified each workload to 
\li~identify loops that can benefit from \gls{PuD} computation, i.e., loops that are memory-bound and that can leverage \gls{SIMD} parallelism and
\lii~use the appropriate \emph{bbop} instructions.
\asplosrev{\hl{To identify loops that }\omrev{\hl{can leverage}}\hl{ \mbox{\gls{SIMD}} parallelism, we \omcriv{use the} \gfcrii{MIMDRAM compiler~\omcriv{\cite{mimdramgit}} for identification and generation} \omcriv{of \gls{PuD} instructions}, which uses LLVM's loop auto-vectorization engine~\mbox{\cite{sarda2015llvm, lopes2014getting,writingpass,lattner2008llvm}} as a profiling tool that outputs \mbox{\gls{SIMD}}-safe loops in an application. 
We use the clang compiler~\mbox{\cite{lattner2008llvm}} to compile each application while enabling the loop auto-vectorization engine and its loop vectorization report (i.e., \texttt{-O3 -Rpass-analysis=loop-vectorize -Rpass=loop-vectorize}).
We observe that applications with regular and wide data parallelism (e.g., applications operating over large dense vectors) are better suited for SIMD-based \mbox{\gls{PuD}} systems. 
We select applications from various domains, including linear algebra and stencil computing (i.e., 2mm, 3mm, doitgen, fdtd-apml, gemm, gramschmidt from Polybench~\mbox{\cite{pouchet2012polybench}}), machine learning (i.e., pca from Phoenix~\mbox{\cite{yoo_iiswc2009}}, covariance from Polybench~\mbox{\cite{pouchet2012polybench}}, kmeans and backprop from Rodinia~\mbox{\cite{che_iiswc2009}}), and image/video processing (i.e., heartwall from Rodinia~\mbox{\cite{che_iiswc2009}} and 525.x264\_r from SPEC 2017~\mbox{\cite{spec2017}}).}}}\revdel{\footnote{\label{r3.2B}\Copy{R3.2B}{\asplosrev{\hl{Note that \mbox{\gls{PuD}} architectures are not yet general-purpose solutions that are }\omrev{\hl{(easily)}}\hl{ well-suited for a }\omrev{\hl{very wide variety}}\hl{ of workloads. The authors of~\mbox{\cite{mimdramextended}} have conducted an extensive workload characterization analysis of 117 different applications from seven benchmark suites to identify kernels that are suited to \mbox{\gls{PuD}} architectures, with results on the twelve same applications we utilize in our paper.}}}}}\revdel{\footnote{Several prior works~\cite{damov,devic2022pim,dualitycache,fujiki2018memory,vadivel2020tdo,iskandar2023ndp,pattnaik2016scheduling} have previously shown that our selected 12 can benefit from different types of \gls{PIM} architectures.}}
\gf{Since our \omcrii{baseline} \gls{PuD} substrate (SIMDRAM) does \emph{not} support floating-point, we manually modify the selected floating-point-heavy \gls{PuD}-friendly loops to operate on fixed-point data arrays.\footnote{We only modify the three applications from Rodinia to use fixed-point, \sgi{as done by prior works~\cite{fujiki2018memory,yazdanbakhsh2016axbench,ho2017efficient}}. \sgi{Polybench} can be configured to use integers; the \gls{PuD}-friendly loops in \texttt{x264} and \texttt{pca} use integers.%
\sgdel{Prior works~\cite{fujiki2018memory,yazdanbakhsh2016axbench,ho2017efficient} also employ fixed-point for the same three Rodinia applications.}} We do \emph{not} observe an output quality degradation when employing fixed-point.}
%\revB{\label{rb.1}\changeB{B1}\prop can be modified to support floating-point arithmetic by 
%\li~leveraging a \gls{PuD} substrate that natively supports floating-point operations (e.g.,~\cite{leitersdorf2023aritpim}) and \lii~slightly modifying the \dynengine to identify narrow values in floating-point data (as done by several prior works~\cite{onur2009exploiting,ergin2004register,lipasti2004physical,valero2015enhancing}).}
{We use the largest input dataset available \gfcrii{\omcriv{for} each benchmark}.}

\label{r3.0}
\begin{table}[ht]
   \caption{Evaluated applications. \gfcriv{We measure peak GPU utilization and total memory footprint on a real system.}}
   \tempcommand{0.9}
   \resizebox{\columnwidth}{!}{%
   \Copy{R3.0}{
    \begin{tabular}{|c|c||c|c|c|c|}
\hline
\textbf{\begin{tabular}[c]{@{}c@{}}Benchmark\\  Suite\end{tabular}} & \textbf{\begin{tabular}[c]{@{}c@{}}Application\\ (Short Name)\end{tabular}} & \textbf{\begin{tabular}[c]{@{}c@{}}\revA{Peak GPU} \\ \revA{Util. (\%)}\end{tabular}} & \textbf{\begin{tabular}[c]{@{}c@{}} \asplosrev{\hl{Total Mem.}} \\ \asplosrev{\hl{Footprint (GB)}} \end{tabular}} & \textbf{\begin{tabular}[c]{@{}c@{}}Bit-Precision\\  \{min, max\}\end{tabular}} & \textbf{\begin{tabular}[c]{@{}c@{}}PUD \\ Instrs.\sgi{$^\dag$}\end{tabular}} \\ \hline \hline
\begin{tabular}[c]{@{}c@{}}Phoenix~\cite{yoo_iiswc2009}\end{tabular} & pca (\texttt{pca}) & \revA{--} & \asplosrev{\hl{1.91}} & \{8, 8\} & D, S, M, R \\ \hline
\multirow{7}{*}{\begin{tabular}[c]{@{}c@{}}Polybench\\ \cite{pouchet2012polybench}\end{tabular}} & 2mm (\texttt{2mm}) & \begin{tabular}[c]{@{}c@{}} \revA{98} \end{tabular} & \asplosrev{\hl{4.77}} & \{13, 25\} & M, R \\ \cline{2-6} 
 & 3mm (\texttt{3mm}) & \begin{tabular}[c]{@{}c@{}}\revA{100}\end{tabular} & \asplosrev{\hl{26.7}} & \{12, 12\} & M, R \\ \cline{2-6} 
 & covariance (\texttt{cov}) & \begin{tabular}[c]{@{}c@{}}\revA{100}\end{tabular} & \asplosrev{\hl{7.63}} & \{23, 23\} & D, S, R \\ \cline{2-6} 
 & doitgen (\texttt{dg}) & \begin{tabular}[c]{@{}c@{}}\revA{92}\end{tabular} & \asplosrev{\hl{33.08}} & \{10, 11\} & M, C, R \\ \cline{2-6} 
 & fdt\gfmicro{d}-apml (\texttt{fdt\gfmicro{d}}) & \begin{tabular}[c]{@{}c@{}}\revA{--}\end{tabular} & \asplosrev{\hl{36.01}}  & \{11, 13\} & D, M, S, A  \\ \cline{2-6} 
 & gemm (\texttt{gmm}) & \begin{tabular}[c]{@{}c@{}}\revA{98}\end{tabular} & \asplosrev{\hl{22.89}} & \{12, 24\} & M, R  \\ \cline{2-6} 
 & gramschmidt (\texttt{gs}) & \begin{tabular}[c]{@{}c@{}}\revA{66}\end{tabular} & \asplosrev{\hl{22.89}} & \{12, 13\} & M, D, R  \\ \hline
\multirow{3}{*}{\begin{tabular}[c]{@{}c@{}}Rodinia\\ \cite{che_iiswc2009}\end{tabular}} & backprop (\texttt{\revCommon{bp}}) & \revA{--}  & \asplosrev{\hl{22.50}} & \{13, 13\} & M, R  \\ \cline{2-6} 
 & heartwall (\texttt{hw}) &  \revA{48} & \asplosrev{\hl{0.03}} & \{17, 17\} & M, R  \\ \cline{2-6} 
 & kmeans (\texttt{km}) & \revA{36} & \asplosrev{\hl{1.23}} & \{17, 17\} & S, M, R  \\ \hline
\begin{tabular}[c]{@{}c@{}}SPEC 2017\\\cite{spec2017}\end{tabular} & 525.x264\_r (\texttt{x264}) & \begin{tabular}[c]{@{}c@{}}\revA{--}\end{tabular} & \asplosrev{\hl{0.15}} & \{1, 8\} & A, R  \\ \hline
\end{tabular}%
}
}
\resizebox{\columnwidth}{!}{$^\dag$ D = division, S = subtraction, M = multiplication, A = addition, R = reduction, C = copy}
    \label{table:workload:properties}
\end{table}

\section{Evaluation}
\label{sec:eval}

\subsection{Real-World Application Analysis} 
\label{sec:eval:real}

\paratitle{\gfisca{Performance}} \gf{Figure~\ref{fig:real_workloads} shows the CPU, GPU, SIMDRAM, \gf{and \prop}\revdel{SIMDRAM with dynamic bit-precision enabled (\emph{SIMDRAM w/ DP}), \prop \gfisca{latency-optimized (\prop \emph{LT}), and
\prop energy-optimized (\prop \emph{EN})}} performance for \gfisca{twelve} real-world applications. 
As in prior works~\revD{\cite{li2017drisa,lee20223d,zhou2023p,ferreira2021pluto,ferreira2022pluto}}, we report area-normalized results (i.e., performance per mm$^2$) for a fair comparison.\revdel{\footnote{\label{ft.rd.8}\revD{We report area-normalized performance results for fairness: the area occupied by \gls{PuD} systems (i.e., a single DRAM bank) is much smaller than the area occupied by processor-centric systems (i.e., a GPU and a CPU). The performance per $mm^2$ metric allows us to easier compare results across very distinct architectures while providing an indication of how the proposed system would scale (performance-wise) if more resources (e.g., more DRAM banks in a DRAM chip) were used for computing.}~\prtagD{D8}}} 
We make \gfmicro{four} observations. 
First, \prop \emph{significantly} outperforms all three baseline systems.  
On average across all \gfisca{twelve} applications, \gfisca{\prop \emph{LT\gfmicro{-DP}} (\prop \emph{EN\gfmicro{-DP}})} achieves 
17$\times$ (11.2$\times$), 
7.3$\times$ (4.8$\times$), and 
10.2$\times$ (6.8$\times$) the performance per mm$^2$ of the CPU, GPU, and SIMDRAM, respectively. 
Second, \gfisca{we observe that equipping SIMDRAM with \prop' \emph{Dynamic Bit-Precision Engine} to leverage narrow values for \gls{PuD} execution \emph{significantly} improves overall performance. On average\revdel{across all applications}, \emph{SIMDRAM-DP} provides 6.3$\times$ the performance \omcrii{per mm$^2$} of \emph{SIMDRAM-SP}. 
Third, \prop' ability to adapt the \uprog  depending on the target bit-precision further improves overall performance by 1.6$\times$ that of \emph{SIMDRAM-DP}.} 
\agymicro{Fourth, \prop' \dynengine \gfmicro{further increases performance by} \gfmicro{46\%}, \omcrii{over} \prop with static \gfmicro{bit-precision}.} This happens because for statically profiled bit-precision, we \emph{must} round the bit-precision up to the nearest power-of-two, \gfcrii{as high-level programming languages (e.g., C/C++) are \emph{inherently} constrained by the two's complement data representation}.
\revdel{We conclude that \prop significantly improves performance of bit-serial \gls{PuD} substrates.}}

\begin{figure}[ht]
    \centering
   \includegraphics[width=0.85\linewidth]{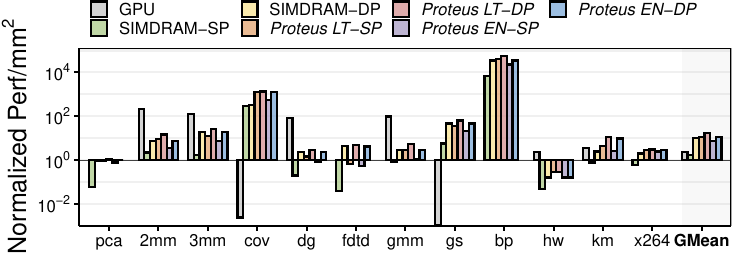}    \caption{\gfmicro{\revA{\gf{CPU-normalized performance per mm$^2$ for \gfisca{twelve} real-world applications\revdel{ in a CPU, GPU, SIMDRAM, \gfisca{SIMDRAM with dynamic bit-precision (\emph{SIMDRAM w/ DP}),} \prop latency-optimized (\prop \emph{LT}), and \prop energy-optimized (\prop \revA{\emph{EN}})}.
  Phoenix~\cite{yoo_iiswc2009} and SPEC2017~\cite{spec2017} do \emph{not} provide GPU implementations of \texttt{pca} and \texttt{x264}.}}}}
    \label{fig:real_workloads}
\end{figure}

\paratitle{\gfisca{Energy}} \gfisca{Figure~\ref{fig:real_workloads:energy} shows the \gfcri{end-to-end energy reduction the} GPU, SIMDRAM, and \prop \gfcrii{provide compared to the baseline CPU} for twelve applications. We make \gfmicro{four} observations.
First, \prop \emph{significantly} reduces energy consumption compared to all three baseline systems.  
On average across all \gfisca{twelve} applications, \gfisca{\prop \emph{EN-DP} (\prop \emph{LT-DP})} \omcrii{provides}
90.3$\times$ (27$\times$), 
21$\times$ (6.3$\times$), and 
8.1$\times$ (2.5$\times$) \omcrii{lower energy consumption than} CPU, GPU, and \emph{SIMDRAM-SP}, respectively. 
Second, \gfcrii{enabling} \prop' \emph{Dynamic Bit-Precision Engine} and \uproglib \gfcrii{allows \prop to reduce energy consumption} by \sgi{an average of} 8$\times$ and 1.02$\times$ compared to \gfcrii{\gls{PuD} substrates with statically-defined bit-precision (}\sgi{\emph{SIMDRAM-SP}) and \gfcrii{bit-serial \emph{only} arithmetic (}\emph{SIMDRAM-DP}), respectively}. 
Third, compared to \emph{SIMDRAM-DP}, \prop \emph{LT-DP} \emph{increases} energy consumption by 3.3$\times$, on average. \revC{\label{rc.6}\changeC{C6}This is because the \omcrii{highest} performance implementation of a \gls{PuD} operation often leads to an increase in the number of \aaps required for \gls{PuD} computing. 
In many cases, the energy associated with inter-subarray data copies (employed in \gls{RBR} and bit-parallel algorithms) leads to an \emph{increase} in energy consumption. Even though the inter-subarray data copy latency can be hidden by leveraging \prop' \gls{OBPS} data mapping, the extra power the DRAM subsystem requires to perform them impacts overall energy consumption.}
\agymicro{Fourth, the \dynengine \gfmicro{further reduces} \prop' energy consumption by \gfmicro{58}\%, compared to \prop with static bit-precision.}
\revdel{We conclude that \prop is an energy-efficient \gls{PuD} substrate.} }

\begin{figure}[ht]
    \centering
   \includegraphics[width=0.85\linewidth]{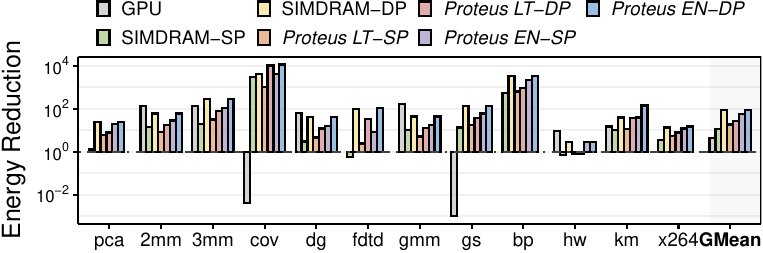}    \caption{\revA{\gf{\gfisca{\gfcrii{End-to-e}nd energy \gfcrii{reduction compared to the baseline CPU} for \gfisca{twelve} applications.\revdel{Values are normalized to the baseline CPU.}}}}}
    \label{fig:real_workloads:energy}
\end{figure}

\subsection{\gfisca{Data \gfcrii{Mapping and Representation Format} Conversion Overheads}}
\label{sec:eval:dataconverson}

\gfcrii{We evaluate the worst-case latency associated with 
\li~data \gfcrii{mapping conversion (from the conventional \gls{ABOS} data mapping to our \gls{OBPS} data mapping) and 
\lii~data representation} format conversion \gfcrii{(from \omcriv{ABOS} to \gls{RBR}) that \prop might perform during the execution of a \gls{PuD} operation.} 
Figure~\ref{fig:layout_conversion} shows \gfmicro{the} worst-case data \gfcrii{mapping and representation} format conversion latency \gfcrii{overhead} for linearly- and quadratically-scaling \uprogs.}
We make two observations.

\begin{figure}[ht]
    \centering
   \includegraphics[width=0.75\linewidth]{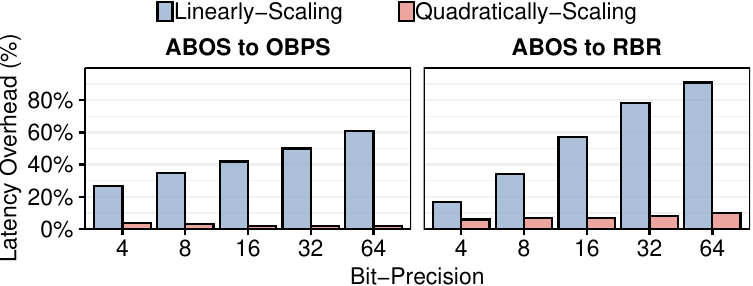}    
   \caption{\gfisca{\omcriv{Latency overheads of d}ata \gfcrii{mapping and representation} format conversion.\revdel{ \emph{ABOS} stands for all-bits in one-subarray; \emph{OBPS} stands for one-bit per-subarray; \emph{RBR} stands for redundant binary representation.}}}
    \label{fig:layout_conversion}
\end{figure}

First, data \gfcrii{mapping and representation} format conversion can \emph{significantly} impact linearly-scaling \uprogs, \gfcrii{causing} up to 60\% and 91\% \gfcrii{latency} overhead when converting from \gls{ABOS} to \gls{OBPS} and from \gls{ABOS} to \gls{RBR}, respectively. 
Second, in contrast, the impact of data \gfcrii{mapping and representation} format conversion in quadratically\omcrii{-}scaling \uprogs is low (i.e., less than 10\% latency overhead). 
This difference is because the number of in-DRAM operations required to perform data \gfcrii{mapping and representation} format conversion increases \emph{linearly} with the bit-precision (\cref{sec:implementation:alltogether}).
In most cases, the data \gfcrii{mapping and representation} format conversion is a one-time overhead that an application pays when executing a series of \gls{PuD} operations.  
On average, across our 12 applications, data \gfcrii{mapping and representation} format conversion accounts for 7.2\% of the total execution time.

\subsection{\asplosrev{Performance of Floating-Point Operations}}
\label{sec:eval:fp}

\label{r1.2A}\Copy{R1.2A}{\asplosrev{\hl{
We evaluate \prop' throughput for floating-point arithmetic operations.
Since our underlying \mbox{\gls{PuD}} architecture, i.e., SIMDRAM, does \emph{not} natively support floating-point operations, we demonstrate how \mbox{\prop} can be leveraged in a different \mbox{\gls{PuD}} architecture that can be modified to support floating-point data formats, i.e., DRISA~\mbox{\cite{li2017drisa}}. 
The main limitation of SIMDRAM when supporting floating-point computation is the lack of interconnects across DRAM columns, which is required for exponent normalization during floating-point addition. 
Since the DRISA architecture includes a shifting network within a DRAM subarray, }\omrev{\hl{it}}\hl{ can perform the required exponent normalization.}}}
\label{r1.2B}\Copy{R1.2B}{\asplosrev{\hl{We perform a synthetic throughput analysis of in-DRAM vector addition/subtraction and multiplication/division operations.}}}\footnote{\label{r1.2C}\Copy{R1.2C}{\asplosrev{\hl{We evaluate synthetic workloads instead of our twelve real-world applications since there is \emph{no} publicly available tool-chain to map real-world applications to }\omrev{\hl{the}}\hl{ baseline DRISA~\mbox{\cite{li2017drisa}} architecture. A similar approach is \omcrii{followed} in~\mbox{\cite{hajinazarsimdram}}.}}}} \label{r1.2D}\Copy{R1.2D}{\asplosrev{\hl{We use 64M-element input arrays, with randomly initialized single-precision floating-point data. We }\omrev{\hl{evaluate}}\hl{ two system configurations:
\li~DRISA 3T1C~\mbox{\cite{li2017drisa}} architecture, which performs \emph{in-situ} \texttt{NOR} }\omrev{\hl{computation}}\hl{; and
\lii~DRISA 3T1C architecture coupled with \mbox{\prop}. 
We make two observations.
First, we observe that the DRISA 3T1C architecture coupled with \mbox{\prop} achieves 1.17$\times$}\omrev{\hl{/1.15$\times$}}\hl{ and 1.38$\times$}\omrev{\hl{/1.37$\times$}}\hl{ the arithmetic throughput of the baseline DRISA 3T1C for floating-point addition/subtraction and multiplication/division operations, respectively.
Second, \prop' throughput gains are more prominent for  multiplication/division operations, since it can \omcrii{reduce} costly in-memory multiplication/division operations during mantissa computation by leveraging narrow mantissa values. 
We conclude that \prop' key ideas apply to different underlying in-DRAM processing techniques and data \omcrii{types}.}}}

\subsection{\asplosrev{\prop vs.\ Tensor Cores \omrev{in} GPU\omrev{s}}}
\label{sec:eval:tensor}

We compare the performance and energy efficiency of our real-world applications \gfcriv{that perform \gls{GEMM} operations} while running on the tensor cores in the NVIDIA A100 GPU and \mbox{\prop} for narrow data precision input operands (i.e. 4-bit and 8-bit integers).
To do so, we 
\li~identify the subset of our real-world applications that mainly perform \mbox{{GEMM}} operations and therefore are suitable for the A100's tensor core engines; and
\lii~re-implement such workloads using optimized instructions (from NVIDIA's CUTLASS~\mbox{\cite{cutlass}}) to perform tensor \mbox{{GEMM}} operations on the A100 GPU tensor cores. 
Re-implementing the GPU workloads is necessary since GPU tensor core instructions are \emph{not} automatically produced via the standard CUDA code our workloads use and there is \emph{no} reference implementation available from the original benchmark suites targeting tensor core GPUs.
\revdel{We re-implement three workloads from our real-world applications, i.e., \texttt{2mm}, \texttt{3mm}, and \texttt{gmm}, for tensor core execution.
To ensure that our re-implemented workloads efficiently utilize the tensor cores, we leverage NVIDIA's open-source CUTLASS library~\mbox{\cite{cutlass}}, which provides CUDA C++ template abstractions for high-performance \mbox{{GEMM}} operations and C++ APIs for non-standardized data types (4-bit integers, in our case). During data initialization, we ensure that the input data fits into 4-bit or 8-bit data types, depending on the evaluated configuration. As in~\mbox{\cref{sec:eval:real}}, we capture the execution time of the GPU kernel excluding data initialization/transfer time, and we use the \texttt{nvml} API~\mbox{\cite{NVIDIAMa14}} to measure GPU energy consumption.} We employ \omcrii{A100's} \omcriv{all} 432 tensor cores during GPU execution.

\label{r2.2}\Copy{R2.2}{\asplosrev{\hl{Figure~\mbox{\ref{fig:real_workloads:tensorgpu}} shows the tensor cores, SIMDRAM, and \mbox{\prop} performance per mm$^2$ (Figure~\mbox{\ref{fig:real_workloads:tensorgpu}}\gfcrii{, top}) and energy efficiency (i.e., performance per Watt in Figure~\mbox{\ref{fig:real_workloads:tensorgpu}}\gfcrii{, bottom}) for three GEMM-heavy real-world applications using 8-bit (\texttt{int8}) and 4-bit (\texttt{int4}) data types. Values are normalized to \omcrii{those obtained on real} GPU tensor cores. We make two observations.
First, \mbox{\prop} significantly improves performance }\omrev{\hl{per mm$^2$}}\hl{ and energy efficiency compared to both \omcrii{tensor cores and SIMDRAM} across all applications and data types. 
}\omrev{\hl{On average across the three applications, \mbox{\prop} \omcrii{provides} 
\li~20$\times$/43$\times$ and 8$\times$/21$\times$ the performance per mm$^2$ and 
\lii~484$\times$/767$\times$ and 9.8$\times$/25$\times$ the performance per Watt of the tensor cores and SIMDRAM, respectively, using \texttt{int8}/\texttt{int4} data types.}}\hl{  
\mbox{\prop} and SIMDRAM are capable of outperforming the tensor cores of the A100 GPU for narrow data precisions since }\omrev{\hl{both}} the throughput and the energy efficiency of bit-serial \mbox{\gls{PuD}} architectures \emph{increase} quadratically for multiplication operations as the bit-precision \emph{decreases}~\mbox{\cite{hajinazarsimdram}}. 
\revdel{As a comparison point, both SIMDRAM and \prop fail to outperform the baseline CUDA cores in the A100 GPU for the same three workloads when computing over a higher dynamic range (see \cref{sec:eval:area}). However, such performance and energy efficiency gaps shift in favor of SIMDRAM and \prop, particularly when we move from 8 to 4 bit input  operands.}  
Second, we observe that by employing dynamic bit-precision and adaptive arithmetic computation, \mbox{\prop} further improves the performance and energy gains that SIMDRAM provides compared to the A100 GPU's tensor cores, even improving performance compared to the \omrev{\hl{tensor cores}}\hl{ in cases where SIMDRAM fails to do so (i.e., for \texttt{gmm}). 
%Therefore, we conclude that bit-serial \mbox{\gls{PuD}} architectures are \emph{particularly} efficient for narrow data precision operations.
}}}

\begin{figure}[ht]
    \centering
    \Copy{R2.3}{
   \includegraphics[width=0.95\linewidth]{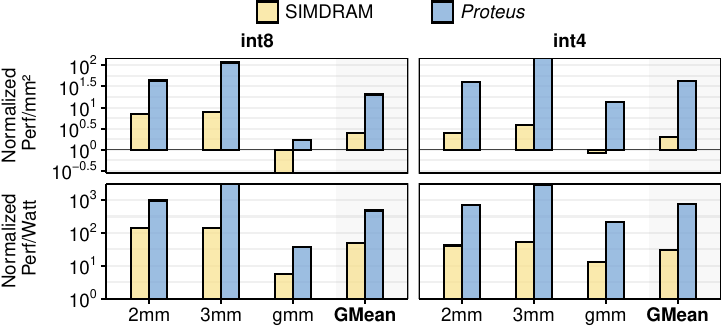}    
   \caption{\gfcrii{\hl{Performance per mm$^2$~(top) and performance per Watt~(bottom) of GEMM-intensive real-world applications using \mbox{\texttt{int8}} and \mbox{\texttt{int4}}}\omrev{\hl{, normalized to the same metric measured on 432 NVIDIA A100 tensor cores.}}}}
   \label{fig:real_workloads:tensorgpu}
}
\end{figure}

\subsection{Area Analysis}
\label{sec:eval:area}

\paratitle{DRAM Chip Area \omcriv{and Storage} Overhead} \gf{We use CACTI 7.0~\cite{cacti} to evaluate the area overhead of the primary components in the \prop design using a 22~nm technology node. 
\prop does \emph{not} introduce any modifications to the DRAM array circuitry other than those proposed by 
\li~Ambit, which has an area overhead of $<$1\% in a commodity DRAM chip~\cite{seshadri2017ambit};
\lii~LISA, which has an area overhead of 0.6\% in a commodity DRAM chip~\cite{chang2016low}; and
\liii~SALP, which has an area overhead of \gfcrii{0.15}\% in a commodity DRAM chip~\cite{kim2012case}.} 
\revB{\label{rb.3}\changeB{B3}We reserve less than 1 DRAM row (i.e., \SI{6.25}{\kilo\byte} in an \SI{8}{\giga\byte}) to store our implemented \uprogs. 
\revdel{That is enough space to store all 16 SIMDRAM \uprogs, \prop' \uprogs for addition, multiplication, division, and subtraction for the different implemented algorithms and data mapping schemes.}
In total, \gfcrii{we implement} 50 \uprogs, each of which takes \SI{128}{\byte} of DRAM space. }

\paratitle{\gfmicro{CPU} Area Overhead} 
\gf{\revdel{The main components in the \prop \emph{Control Unit} are the 
\uproglib and \dynengine.} 
We size the \uproglib to contain:
\li~16 64~B \glspl{LUT}, each \gls{LUT} holding a 8-bit  \idx);
\lii~one 2~kB \emph{\uprog Scratchpad} Memory. The size of the \uproglib is enough to hold one \gls{LUT} per SIMDRAM \gls{PuD} operations and address $2^8$ different \uprog implementations. The size of the \uprog{} Scratchpad is large enough to store the \uprog{}s for all 16 SIMDRAM operations.
%(16~\uprog{}s $\times$ 128~B max per \uprog{})
We use a 128~B scratchpad for the \dynengine. 
\gfcrii{Using CACTI,} we estimate that the \prop \emph{Control Unit} area is 0.03~mm$^2$.} 
%
%\paratitle{Transposition Unit Area Overhead} \gf{The primary components in the transposition unit are 
%\li~the \emph{object tracker},
%\lii~two transposition buffers, and 
%\liii~the \dynengine.
\gfmicro{\prop' \emph{Data Transposition Unit} \asplosrev{\hl{(one per DRAM channel)}} uses}  an 8~kB fully-associative cache with a 128-bit cache line size for the \emph{Object Tracker}, and
%\sgi{enough to store 512~entries}.
%, where each entry holds the base physical address of a \gls{PuD} memory object (19~bits), the total size of the allocated data (32~bits), the size of each element in the object (6~bits), and the maximum value (64~bits). 
two 4~kB transposition buffers.
%The \dynengine contains an $n$-bit reconfigurable comparator. 
\gfcrii{Using CACTI,} we estimate the \emph{Data Transposition Unit} area is 0.06~mm$^2$. 
Considering the area of the control and transposition units, \prop has an area overhead of only 0.03\% compared to the die area of an Intel Xeon E5-2697 v3 CPU~\cite{dualitycache}.  
%\gf{We conclude that \prop has low area cost.}
%\input{mainmatter/06_proteus/sections/06_related}
\section{Summary}
\label{sec:conclusion}

\gf{We introduce \prop, an efficient \gls{PuD} framework with \gfisca{adaptive data precision and dynamic arithmetic}. \prop fully leverages the internal parallelism inside a DRAM bank to accelerate the execution of various bit-serial and bit-parallel arithmetic operations and dynamically decides the best-performing bit-precision, data representation format, and algorithmic implementation of a \gls{PuD} operation. We experimentally demonstrate that \prop provides significant benefits over state-of-the-art CPU, GPU, and \gls{PuD} systems. 
%We hope that future work builds on \prop to ease further the adoption of \gls{PuD} architectures.
}

\chapter[DaPPA: A Data-Parallel Programming Framework for Processing-in-Memory Architectures]{DaPPA: A Data-Parallel Programming Framework for Processing-in-Memory Architectures}
\label{chap:dappa}

% Mechanism Name
\renewcommand{\prop}{{DaPPA}\xspace}

% Listing style
\definecolor{backcolour}{rgb}{0.95,0.95,0.92}
\definecolor{mygreen}{rgb}{0,0.6,0}
\definecolor{mygray}{rgb}{0.5,0.5,0.5}
\definecolor{mymauve}{rgb}{0.58,0,0.82}
\definecolor{mblue}{rgb}{0.27,0.33,0.53}

\lstdefinestyle{myC}{
  backgroundcolor=\color{backcolour},  
  basicstyle=\ttfamily\footnotesize,      
  breakatwhitespace=false,  
  breaklines=true,       
  captionpos=b,                   
  commentstyle=\color{mygreen},    
  deletekeywords={...},           
  escapechar=\%,
  xleftmargin=0pt,
  xrightmargin=0pt,
  aboveskip=\medskipamount,
  belowskip=\medskipamount,
  extendedchars=true,            
  keepspaces=true,               
  keywordstyle=\color{blue},      
  language=C++,              
  morekeywords={bbop_trsp_init, bbop_add, bbop_sub, bbop_greater, bbop_if_else, bbop_trsp_cpy, malloc, *,...},          
  numbers=left,                   
  numbersep=1pt,                   
  numberstyle=\tiny\color{mygray}, 
  rulecolor=\color{black},     
  showspaces=false,              
  showstringspaces=false,        
  showtabs=false,                 
  stepnumber=1,                    
  stringstyle=\color{mymauve},     % string literal style
  tabsize=2,	                   % sets default tabsize to 2 spaces
  title=\lstname                   % show the filename of files 
}

\section{Motivation \& Goal}
\label{sec:background}

\revdel{{The UPMEM-based PIM systems utilize a single-program multiple-data (SPMD) execution model~\cite{flynn1966very}, where multiple \emph{tasklets} (i.e., hardware threads) run the same code but operate on different pieces of data, allowing them to execute different control-flow paths at runtime. 
The number of hardware threads in a DPU is limited to 24, and the number of tasklets per DPU is determined by the programmer at compile time. 
Tasklets within the same DPU have the ability to share data and synchronize through mechanisms such as  mutexes, barriers, handshakes, and semaphores. However, tasklets in different DPUs do \emph{not} have any direct communication channel or memory sharing capability, hence they cannot directly communicate or synchronize.}
PIM-enabled memory is integrated into the host system as a \emph{loosely-coupled} accelerator, where the host CPU is responsible to offload the target kernel to the DPU cores, and move the appropriate input data from the host main memory to the PIM-enabled memory, similar to current GPU programming. }

\gft{3}{Programming general-purpose \gls{PnM} systems (such as the UPMEM \gls{PIM} system~\cite{upmem, upmem-guide, upmem2018}) require the programmer to follow three main steps.} 
 
\paratitle{Step 1: Distribute \& Transfer Input Data Across DPUs}  In the first step, the programmer distributes the input data across the multiple \gls{PIM} cores in the PIM-enabled memory, and subsequently across the many threads within a single \gls{PIM} core, by explicitly moving the input data from the host main memory to the PIM-enabled memory (\circled{1} in Figure~\ref{fig:architecture}). 
To do so, particularly in the UPMEM PIM system, the programmer makes use of one of the three CPU--DPU data transfer primitives the UPMEM SDK~\cite{upmem} provides, which allow for:
\li~\emph{serial} CPU--DPU data transfer, where a \emph{single} portion of data is moved from main memory to a \emph{single} DPU (i.e., one MRAM bank); 
\lii~\emph{parallel} CPU--DPU data transfer, where \emph{multiple} portions of data from main memory are distributed across \emph{multiple} DPUs (i.e., across many MRAM banks); and
\liii~\emph{broadcast} CPU--DPU data transfer, where a \emph{single} portion of data  from main memory is moved and replicated across \emph{multiple} DPUs (i.e., multiple MRAM banks).

\paratitle{Step 2: Handle Caching in Local Scratchpad Memory} In the second step, after data is copied to the \gls{PIM} memory, the \gls{PIM} cores can start their computations. 
For a \gls{PIM} core to process data, the programmer must carefully (and manually) orchestrate internal data movement between local DRAM banks and scratchpad memory, as well as  communication and synchronization between \gls{PIM} cores. 
In the UPMEM PIM system, the programmer needs to orchestrate
\li~data movement between the DRAM bank (MRAM) and the local scratchpad memory (WRAM) using the DPU engine \gls{DMA} (\circled{2} in Figure~\ref{fig:architecture}), and
\lii~inter-DPU  communication \& tasklet synchronization. 
For MRAM--WRAM data movement orchestration, the programmer must respect restrictive memory alignment requirements (i.e., both source addresses in WRAM and MRAM \emph{must} be 8-byte aligned) and set the appropriate transfer size (which can vary from 8 to 2048 bytes).
For inter-DPU communication, the programmer needs to identify programming patterns that require data transfers across DPUs (e.g., merging of partial results to obtain a final output) and explicitly perform DPU--CPU and CPU--DPU data transfers (since there is \emph{no} direct communication across DPUs in the PIM-enabled memory) accordingly. 
For tasklet synchronization, the programmer needs to decide which synchronization primitive (e.g., mutex, handshake, barrier, or semaphore) is the most appropriate for the given task.

\paratitle{Step 3: Consolidating Results} In the third step, 
once the main kernel finishes its execution within the \gls{PIM} cores, partial output values will be stored across the many DRAM banks in the PIM-enabled memory. 
For such output values to be visible to the host application, the programmer needs to perform DPU--CPU data transfers (\circled{3} in Figure~\ref{fig:architecture}), so that data is moved from the PIM-enabled memory to the main memory. After this transfer is complete, the host CPU might still have to perform some post-processing, depending on the target application (e.g., the host CPU might combine the partial results that each DPU produces).

\paratitle{Problem \& Goal} Even though \gft{3}{the programming model of general-purpose \gls{PnM} systems} resembles that of widely employed architectures, such as GPUs, it requires the programmer to 
\li~have prior knowledge of the underlying \gls{PIM} hardware and
\lii~manage data movement at a fine-grained granularity \emph{manually}. 
Concretely, programming a PIM-enabled system requires the programmer to perform 
\li~efficient workload partitioning across the many \gft{3}{\gls{PIM} cores} and \gft{3}{PIM threads} within the system;
\lii~manual transfer of data between standard main memory and \gls{PIM} banks, while ensuring that both host CPU and \gls{PIM} cores have access to accurate and up-to-date copies of data; and
\liii~orchestrate data movement between \gls{PIM} banks local scratchpad memory.
Such limitations can difficult the adoption of PIM architectures in general-purpose systems. 
Therefore, our \textbf{goal} in this work is to ease programmability for general-purpose \gls{PnM} systems, allowing a programmer to write efficient PIM-friendly code \emph{without} the need to manage hardware resources \emph{explicitly}.

% \subsection{Limitation of UPMEM Programming Model}

% {The use of the C programming language for programming DPUs offers a low learning curve for programmers, but it also presents various challenges. Firstly, when programming thousands of DPUs with the capability of running up to 24 tasklets, efficient workload partitioning and orchestration are crucial and must be meticulously planned. This can be achieved through the use of tasklet IDs assigned to each tasklet. Secondly, the transfer of data between the standard main memory and MRAM banks must be explicitly managed by the programmer, and it is their responsibility to maintain data coherence between the CPU and DPUs, ensuring that both have access to accurate and up-to-date copies of data. Lastly, it should be noted that the DPU architecture does not incorporate cache memories, meaning that the movement of data between the MRAM banks and WRAM must be explicitly managed by the programmer.}

\section{\prop Overview}
\label{sec:overview}

To ease the programmability of PIM architectures, we propose \prop (\underline{da}ta-\underline{p}arallel \underline{p}rocessing-in-memory \underline{a}rchitecture), a programming framework that can, for a given application, \emph{automatically} distribute input and gather output data, handle memory management, and parallelize work across the \gls{PIM} cores. 
The \emph{key idea} behind \prop is to remove the responsibility of managing hardware resources from the programmer by providing an intuitive data-parallel pattern-based programming interface~\cite{cole1989algorithmic,cole2004bringing} that abstracts the hardware components of the \gls{PnM} system. 
Using this key idea, \prop transforms a data-parallel pattern-based application code into the appropriate PIM-target code, including the required APIs for data management and code partition, which can then be compiled into a PIM-based binary \emph{transparently} from the programmer. 
While generating PIM-target code, \prop implements several code optimizations to improve performance.

Figure~\ref{fig:framework_overview} shows an overview of our \prop programming framework. \prop takes as input C/C++ code, which describes the target computation using a collection of data-parallel patterns and \prop's programming interface, and generates as output the requested computation. \prop consists of three main components:
\li~\prop's data-parallel pattern APIs,
\lii~\prop's dataflow programming interface, and
\liii~\prop's dynamic template-based compilation.

\begin{figure*}
    \centering
    \includegraphics[width=0.98\textwidth]{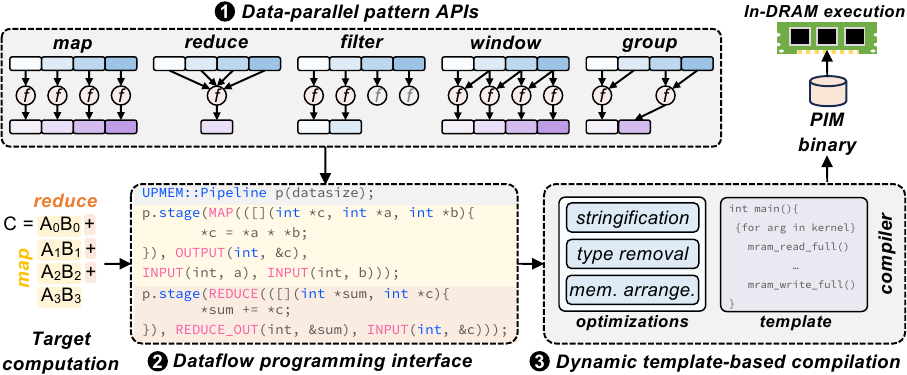}
    \caption{Overview of the \prop programming framework.}
    \label{fig:framework_overview}
\end{figure*}

\paratitle{Data-Parallel Pattern APIs} \prop's data-parallel pattern APIs (\circled{1} in Figure~\ref{fig:framework_overview}) are a collection of pre-defined functions that implement high-level data-parallel pattern primitives. 
Each primitive allows the user to express how data is transformed during computation. \prop supports five primary data-parallel pattern primitives, including:
\li~\texttt{map}, which applies a function $f$ to each individual input element $i$, producing unique output elements $y_i = f(x_i)$;
\lii~\texttt{reduce}, which reduces input vectors to a scalar;
\liii~\texttt{filter}, which selects input elements based on a predicate;
\liv~\texttt{window}, which \emph{maps} and output element as the \emph{reduction} of $W$ \emph{overlapping} input elements; 
\lv~\texttt{group}, which \emph{maps} and output element as the \emph{reduction} of $G$ \emph{non-overlapping} input elements.  The user can combine these five data-parallel primitives to describe complex data transformations in an application. 
%\prop is responsible for translating and parallelizing each data-parallel primitive to efficient CPU and UPMEM code. 

\paratitle{Dataflow Programming Interface} \prop exposes a dataflow programming interface to the user (\circled{2} in Figure~\ref{fig:framework_overview}). In this programming interface, the main component is the \texttt{Pipeline} class, which represents a sequence of data-parallel patterns that will be executed on the PIM cores. A given \texttt{Pipeline} has one or more \texttt{stage}s. Each \texttt{stage} utilizes a given data-parallel pattern primitive to transform input operands following a user-defined computation sequence. \texttt{Stage}s are executed in order, in a pipeline fashion.

\paratitle{Dynamic Template-Based Compilation} \prop uses a dynamic template-based compilation (\circled{3} in Figure~\ref{fig:framework_overview}) to generate the PIM code in two main steps.
In the first step, \prop creates an initial PIM code based on a skeleton of a PIM application. 
In the second step, \prop uses a series of transformations to 
\li~extract the required information that will be fed to the PIM code skeleton of the user program; 
\lii~calculate the appropriate offsets used to manage data across PIM banks and local \gls{PIM} scratchpad memory; and
\liii~divide computation between CPU and PIM cores. 

\paratitle{Putting All Together} Using \prop's data-parallel pattern APIs, dataflow programming interface, and dynamic template-based compilation, the user can quickly implement and deploy applications to the \gls{PnM} system without any knowledge of the underlying architecture. Figure~\ref{fig:framework_overview} showcases an example of implementing a simple vector dot-product application using \prop. 
In this example, the user defines a \texttt{Pipeline} with two \texttt{stage}s: a \texttt{map} stage and a \texttt{reduce} stage. 
\prop generates the appropriate binary for the \gls{PnM} system, executes the target computation on the PIM cores, and copies the final output from the PIM cores to the CPU.

\definecolor{mygreenl}{rgb}{0,0.6,0}
\definecolor{codebg}{rgb}{1.0,1.0,1.0}
\definecolor{standardkw}{rgb}{0.1,0.1,0.8}  % Blue
\definecolor{customkw}{rgb}{0.6,0.2,0.2}    % Dark Red
\definecolor{anti-flashwhite}{rgb}{0.95, 0.95, 0.96}

% Define listings style
\lstdefinestyle{cppminted}{
    language=C++,
    basicstyle=\scriptsize\ttfamily,
    backgroundcolor=\color{anti-flashwhite},
    frame=lines,
    framexleftmargin=0mm,
    framexrightmargin=0mm,
    rulecolor=\color{black},
    framerule=0.03em,
    xleftmargin=1.5em,
    breaklines=true,
    postbreak=\mbox{\textcolor{gray}{$\hookrightarrow$\space}},  
    showstringspaces=false,
    numbers=left,
    numberstyle=\tiny\color{gray},
    stepnumber=1,
    commentstyle=\color{mygreenl},    
    numbersep=5pt,
    keywordstyle=\color{standardkw}\bfseries,
    morekeywords={uint32_t},
    classoffset=1,
    morekeywords={Pipeline, stage, fetch, execute},
    keywordstyle=\color{customkw}\bfseries,
    classoffset=0
}

\lstdefinestyle{cppcustom}{
    language=C++,
    basicstyle=\scriptsize\ttfamily,
    backgroundcolor=\color{anti-flashwhite},
    frame=lines,
    framexleftmargin=0em,
    framexrightmargin=0em,
    rulecolor=\color{black},
    framerule=0.03em,
    xleftmargin=0em,
    breaklines=true,
    postbreak=\mbox{\textcolor{gray}{$\hookrightarrow$\space}},  
    showstringspaces=false,
    numbers=none,
    commentstyle=\color{mygreenl},    
    keywordstyle=\color{standardkw}\bfseries,
    morekeywords={uint32_t},
    classoffset=1,
    morekeywords={Pipeline, stage, fetch, execute},
    keywordstyle=\color{customkw}\bfseries,
    classoffset=0
}

\section{\prop Implementation}
\label{sec:implementation}

% In this section, we describe \prop's implementation, including the concrete design of its
% \li~data-parallel pattern APIs,
% \lii~dataflow programming interface, and
% \liii~dynamic template-based compilation.
% %Finally, we discuss \prop's limitations.

\subsection{Data-Parallel Pattern APIs}

\prop exposes \emph{data-parallel pattern APIs} that leverages skeleton-based programming and data-parallel patterns to allow users to define their algorithms using high-level abstractions~\cite{cole1989algorithmic,cole2004bringing,mccool2012structured,mccool2010structured}. 
In skeleton-based programming, applications are composed of predefined \emph{skeletons} (e.g., \texttt{map}, \texttt{reduce}, \texttt{pipeline}, \texttt{farm}), each encapsulating a common parallel computation pattern. In this way,
programmers can focus on application logic rather than low-level concerns such as tasklet management, synchronization, or inter-DPU communication. 
Data-parallel patterns further enable concurrent execution of the same operation (or sequence of operations) across multiple data elements, thereby reducing boilerplate (i.e., repeated) code and errors associated with explicit tasklet or DPU management~\cite{rauber2013parallel,hager2010introduction}. 
This separation of \emph{what} (algorithmic intent) from \emph{how} (hardware-specific optimizations) eases code reuse, scalability, and maintainability for CPUs~\cite{ernsting2017data,ernsting2012algorithmic,aldinucci2017fastflow}, GPUs~\cite{enmyren2010skepu,ernsting2017data,ernsting2012algorithmic}, special-purpose accelerators~\cite{samadi2014paraprox}, and distributed systems~\cite{ernsting2012algorithmic,aldinucci2013targeting}. 
Numerous skeleton-based frameworks (e.g., SkePU~\cite{enmyren2010skepu}, Skandium~\cite{leyton2010skandium}, and Muesli~\cite{ciechanowicz2009munster,ernsting2012algorithmic,ernsting2017data}) as well as libraries supporting data-parallel constructs (e.g., CUDA~\cite{cheng2014professional}, OpenCL~\cite{munshi2009opencl}, Intel Threading Building Blocks~\cite{pheatt2008intel}, and Apache Spark~\cite{zaharia2010spark,zaharia2016apache}) exemplify the breadth and efficacy of these high-level parallel programming approaches.
Therefore, we leverage such a programming paradigm to aid in programmability for general-purpose \gls{PnM} systems.

%We are using the established Map-Filter-Reduce patterns as a basis for our framework, as they are well understood and simple to implement. Our parallel patterns assume that the data we are operating on can be expressed as a one-dimensional vector. This is so we can understand how to partition the data. Other possible data shapes, such as matrices, are generally not supported directly. So processing a matrix, for example, would require transforming the matrix into a one-dimensional vector of elements and an algorithm that can process such a vector. For all patterns, it holds that re-ordering of elements is not possible, random element access is not possible, and multiple inputs/outputs are allowed as long as their sizes are compatible. Those limitation follow directly from the design of the parallel patterns we have chosen and while some of them might be overcome in the future, it is currently not possible. Non-vector arguments, such as scalar parameters, can also be supplied as arguments and will be broadcast to all DPUs. 

Currently, \prop supports five primary data-parallel patterns (which Figure~\ref{fig:framework_overview} illustrates), i.e., \texttt{map}, \texttt{reduce}, \texttt{filter}, \texttt{window}, and \texttt{group}.  Each data-parallel pattern takes as input one or more one-dimensional (1D) vectors, and produces as output a single 1D vector or a scalar value. 
Non-vector arguments, such as scalar parameters, can also be supplied as arguments and are broadcast across all PIM cores involved in the computation. \prop implements each one of the five data-parallel patterns as follows:
\begin{itemize}[noitemsep,topsep=0pt,parsep=0pt,partopsep=0pt,labelindent=0pt,itemindent=0pt,leftmargin=*]
    \item \textbf{map}: the \texttt{map} data-parallel pattern takes as input a 1D vector $x$ of size $N$, a \emph{pure} function $f$, and produces as output a 1D vector $y$ of size $N$, where $y_i = f(x_i)$. A \emph{pure} function consistently produces the same output for a given input (i.e., the pure function produces \emph{deterministic} outputs) and does \emph{not} induce side effects (i.e.,  the invocation of $f$ does \emph{not} modify any external state, such as global variables, files, or shared memory, and does \emph{not} depend on any non-local state that may vary over time). 
    As a result, no synchronization between PIM threads is required, and data sharing is unnecessary, making the \texttt{map} data-parallel pattern highly suitable for data-parallel execution on \gls{PnM} architectures. In our implementation of the \texttt{map} data-parallel pattern, each PIM thread \emph{independently} executes an instance of the function $f$, generating its corresponding output $f(x_i)$.

    \item \textbf{\texttt{reduce}}: The \texttt{reduce} data-parallel pattern takes as input a 1D vector \(x\) of size \(N\), a reduction function \(f\), and produces as output a single scalar value \(r\). In its simplest form, the result \(r\) is computed by repeatedly applying \(f\) over all elements of \(x\), for example: $ r = f\bigl(x_1, f\bigl(x_2, \dots f\bigl(x_{N-1}, x_N\bigr)\dots\bigr)\bigr)$. 
    For efficient parallelization, \(f\) is commonly required to be \emph{associative}, meaning that $ f(a, f(b, c)) \;=\; f\bigl(f(a, b), c\bigr) $, for all valid operands \(a\), \(b\), and \(c\). Associativity ensures that partial computations of \(f\) (i.e., partial ``\emph{reductions}'') can be combined in arbitrary groupings and orders without affecting the final result. 
    In case $f$ is pure \emph{and} associative, the \texttt{reduce} data-parallel pattern allows for intermediate values to be processed in parallel and merged incrementally, reducing the need for extensive synchronization. 
    In our implementation of the \texttt{reduce} data-parallel pattern, the input vector $x$ is equally distributed across PIM cores, and then multiple PIM threads within a PIM core compute partial reductions on each distinct subranges of the input vector. 
    These partial results are then combined (in the host CPU), using the same function $f$, in a tree-based hierarchy until a single final result \(r\) remains. 
    In case $f$ is pure and associative, the order in which PIM threads combine intermediate results does \emph{not} affect correctness, thereby enabling efficient parallel execution.
    
    \item \textbf{\texttt{filter}}: The \texttt{filter} data-parallel pattern takes as input a 1D vector \(x\) of size \(N\) and a \emph{pure} predicate function \(f\), and produces as output a new 1D vector \(y\). The predicate \(f\) tests each element of \(x\) for a condition (returning \texttt{true} or \texttt{false}). The resulting vector \(y\) contains exactly those elements of \(x\) for which \(f(x_i)\) is \texttt{true}, while preserving their original order. Formally, $ y = \bigl[x_i \;\big|\; f(x_i) = \texttt{true},\; 1 \le i \le N \bigr].$ A \emph{pure} predicate function consistently produces the same result for a given input (i.e., deterministic output) and does \emph{not} induce side effects. Purity ensures that each evaluation of \(f(x_i)\) can be performed \emph{independently} and in parallel, without synchronization or data sharing. Once the \texttt{true}/\texttt{false} decisions are computed, the \texttt{filter} data-parallel pattern gathers the qualifying elements to form the final output vector \(y\).

    %If the filter function returns true, the input element that is currently being processed is being kept, if it returns false, it is being discarded. The condition whether element $n$ is kept or discarded may only depend on the value of element $n$. If multiple inputs are present, the output will be equal to the first input specified if the element is kept.

    \item \textbf{\texttt{window}}: the \texttt{window} data-parallel pattern generalizes the \texttt{map} pattern to computations where each output element depends on a contiguous block (or ``\emph{window}'') of the input. Specifically, the \texttt{window} data-parallel pattern takes as input: \li~a 1D vector \(x\) of size \(N\), \lii~a \emph{window size} \(W\), and
    \liii~a \emph{pure} function \(f\) that operates on sub-vectors of length \(W\). It produces as output a 1D vector \(y\) of size \(M\), where \(M\) depends on how windows are defined and handled at the boundaries (e.g., \(M\) might be \(N - W + 1\) if every complete window produces exactly one output). Formally, for each valid index \(i\), $y_i \;=\; f\bigl(x_i,\, x_{i+1},\, \dots,\, x_{i+W-1}\bigr).$ Since \(f\) is assumed to be \emph{pure}, each sub-vector \(\bigl(x_i, \dots, x_{i+W-1}\bigr)\) can be processed \emph{independently}. The main distinction from \texttt{map} arises because consecutive outputs in a \texttt{window} may depend on overlapping input segments. Users often provide additional (or padding) elements for the end of the input to ensure that the last positions can form a complete window of size \(W\). This data-parallel pattern is widely used in signal processing, sliding-window algorithms, and stencil computations, and it can  be parallelized effectively on general-purpose \gls{PnM} systems by distributing different sub-vectors to different PIM cores.

    %Same as map, except that the output at position $n$ may now depend on the inputs at position $n$ to $n + W$ with $W$ being the window size. The user will also have to provide some additional elements to be used as input overlap at the very end of the data.

    \item \textbf{\texttt{group}}: the \texttt{group} data-parallel pattern partitions a 1D input vector \(x\) of size \(N\) into contiguous, \emph{non-overlapping} sub-vectors (or ``\emph{groups}'') of size \(G\). A \emph{pure} function \(f\) is then applied to each sub-vector to produce one output element per group. Formally, assuming \(N\) is divisible by \(G\), the vector \(x\) is segmented into \(\tfrac{N}{G}\) sub-vectors: $ \bigl(x_1, \dots, x_G\bigr), \quad \bigl(x_{G+1}, \dots, x_{2G}\bigr), \quad \dots, \quad \bigl(x_{N-G+1}, \dots, x_N\bigr),$~and each group is independently processed by $y_i \;=\; f\Bigl(x_{(i-1)G+1},\, \dots,\, x_{iG}\Bigr)  \quad\text{for}\quad i = 1, 2, \dots, \tfrac{N}{G}.$ Unlike the \texttt{window}, whose sub-vectors can overlap, the \texttt{group} data-parallel pattern advances its input index by \(G\) elements each step, thereby ensuring these sub-vectors are disjoint. Purity allows each group to be processed in \emph{isolation}, making the \texttt{group} data-parallel pattern highly amenable to data-parallel execution. In some applications, \texttt{group} can also be viewed as a special case of a reduction over fixed-size chunks: each chunk is ``\emph{reduced}'' into a single value by the function \(f\), without overlapping the inputs of neighboring chunks.

     %Same as window, except that the elements within a group don't overlap, so the input index moves on by $G$ instead of 1 each iteration, where $G$ is the group size. Can also be viewed as a reduction of $G$ elements without overlapping inputs.
\end{itemize}

\prop also allows the user to combine some of the five primary data-parallel patterns to implement more complex execution patterns. In our current implementation, \prop enables the combination of \texttt{window}, \texttt{filter}, and \texttt{group} data-parallel patterns into four new implementations: \texttt{window+group}, \texttt{window+filter}, \texttt{group+filter}, and \texttt{window+group+filter}. We describe each implementation below:

\begin{itemize}[noitemsep,topsep=0pt,parsep=0pt,partopsep=0pt,labelindent=0pt,itemindent=0pt,leftmargin=*]
    \item \textbf{\texttt{window+group}}: the \texttt{window+group} data-parallel pattern combines the \texttt{window} and \texttt{group} data-parallel patterns. Like \texttt{group}, the input vector is divided into contiguous, non-overlapping sub-vectors of size \(G\). However, for each sub-vector, the computation will depend not only on its own \(G\) elements but also on up to an additional \(W\) elements (forming a ``\emph{window}''). Formally, whereas the \texttt{group} data-parallel pattern would compute: $ y_n \;=\; f\bigl(x_{(n-1)G+1}, \dots, x_{nG}\bigr),$ the \texttt{window+group} data-parallel pattern extends the domain of \(f\) to: $y_n \;=\; f\bigl(x_{(n-1)G+1}, \dots, x_{nG + W}\bigr),$ with \(W\) specifying how far beyond the current group the function may read.  This expanded scope is useful for computations where nearby elements (beyond the immediate group boundary) influence the result.  As with \texttt{window} and \texttt{group}, the \texttt{window+group} data-parallel pattern assumes a \emph{pure} function \(f\), ensuring that each sub-vector plus its ``\emph{window}'' can be processed in parallel.

    \item \textbf{\texttt{window+filter}}: the \texttt{window+filter} data-parallel pattern combines the behavior of \texttt{window} (sliding or overlapping sub-vectors) with \texttt{filter} (selecting outputs based on a predicate). Like \texttt{window}, it processes an input vector \(x\) in overlapping segments (``\emph{windows}'') of size \(W\). However, rather than producing an output for every window, it applies a \emph{pure} predicate function \(f\) to each window and includes that window's contribution in the output only if \(f\) returns \texttt{true}. Formally, for each valid index \(i\), consider the window $w_i = \bigl(x_i,\, x_{i+1}, \dots, x_{i+W-1}\bigr).$ The pattern outputs $w_i \;\;\text{if}\;\; f(w_i) = \texttt{true},$ and omits it otherwise. As with both \texttt{window} and \texttt{filter}, the use of a pure predicate function \(f\). This combined pattern is particularly useful when overlapping context is necessary for deciding which segments of the data should be preserved.

    \item \textbf{\texttt{group+filter}}: the \texttt{group+filter} data-parallel pattern partitions an input vector \(x\) into contiguous, non-overlapping sub-vectors (``\emph{groups}'') of size \(G\) and then applies a \emph{pure} predicate function \(f\) to decide whether each group is retained. Formally, for \(n = 1, 2, \dots, \tfrac{N}{G}\), define the \(n\)-th group as $g_n = \bigl(x_{(n-1)G+1}, \dots, x_{nG}\bigr).$ The output includes \(g_n\) \emph{only if} \(f(g_n) = \texttt{true}\). This selective mechanism extends the \texttt{group} data-parallel pattern, maintaining data-parallel independence across groups while filtering out those that do \emph{not} satisfy the predicate.

    \item \textbf{\texttt{window+group+filter}}: the \texttt{window+group+filter} pattern extends the \texttt{window+group} pattern by selectively retaining only those results that satisfy a \emph{pure} predicate. As in \texttt{window+group}, the input vector is divided into contiguous sub-vectors of size \(G\), with each sub-vector allowed to read up to \(W\) additional elements beyond its boundary. A \emph{pure} function \(f\) then produces an output from each extended sub-vector. Formally, for the \(n\)-th sub-vector, we consider $\bigl(x_{(n-1)G+1}, \dots, x_{nG+W}\bigr),$ and compute $y_n \;=\; f\bigl(x_{(n-1)G+1}, \dots, x_{nG+W}\bigr).$ A separate \emph{pure} predicate \(p\) then determines whether \(y_n\) is kept in the output: $\text{include } y_n \;\;\text{if and only if}\;\; p(y_n) \;=\; \texttt{true}.$ Because \(f\) and \(p\) are both pure, each sub-vector can be processed independently, preserving the data-parallel benefits of the underlying \texttt{window+group} pattern while allowing for filtering of results. 
\end{itemize}

\subsection{Dataflow Programming Interface}

\prop allows the programmer to use the data-parallel pattern APIs to implement a given task or application. To do so, \prop exposes to the user a \emph{dataflow programming interface}, which Figure~\ref{fig:dataflow} illustrates. 
The \emph{key idea} behind \prop's dataflow programming interface is to allow the programmer to \emph{implicitly} describe the data movement between data-dependent sub-tasks in the target task/application. 
At a high-level, the \emph{dataflow programming interface} has three main components:
\li~a \texttt{Pipeline} (\circled{1} in Figure~\ref{fig:dataflow}), which defines a collection of transformations over the input data;
\lii~one or more \texttt{stage}s (\circled{2}), which composes the \texttt{Pipeline} while each \texttt{stage} represents an unique data-parallel pattern from \prop's data-parallel pattern APIs;
\liii~implicit dataflow (\circled{3}), where data moves sequentially, in a pipeline-fashion,  across each \texttt{stage}.

\begin{figure}[ht]
    \centering
    \includegraphics[width=\linewidth]{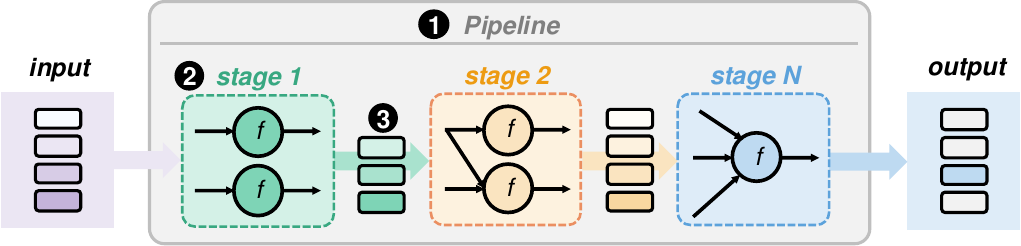}
    \caption{\prop's dataflow programming interface.}
    \label{fig:dataflow}
\end{figure}

Concretely, the dataflow programming interface exposes to the programmer one main C++ class, the \texttt{Pipeline} class, which represents a sequence of data-parallel patterns that will be executed on PIM cores. 
The \texttt{Pipeline} class has a set data vector size to guarantee length compatibility between all inputs, which can only be reduced or divided by \texttt{filter} and \texttt{group} data-parallel patterns. 
{A given \texttt{Pipeline} has one or more \texttt{stage}s, and works in three main steps.
First, the input data is transformed within each \texttt{stage} using a \emph{stage-specific function}, based on a given data-parallel pattern, producing an output data. 
The output produced by a \texttt{stage} $i$ can be used as input data for any subsequent stage $i+1$. 
Second, after the user specifies all needed data transformations that the PIM cores will execute across the different  \texttt{stage}s, they can specify the final output data that will be fetched from the PIM cores to the CPU. 
Third, the user initiates the execution of the \texttt{stage}s specified for a given \texttt{Pipeline} across the PIM cores. Note that, while defining each \texttt{stage} and its equivalent stage-specific function, the user does \emph{not} need to define \emph{any} PIM-specific command, including those related to data orchestration and parallelism distribution. 
This happens because \prop abstracts the underlying hardware characteristics from the user, allowing the programmer to focus only on implementing the functionality of the target application.  
}

\subsubsection{The \texttt{Pipeline} Class: Implementation} 

The primary interface \prop exposes to the user is the \texttt{Pipeline} class. We explain the main components of the  \texttt{Pipeline} class using a vector dot product example in Listing~\ref{listing:vecdot}. 
The \texttt{Pipeline} class has five main methods, which we describe next.

%\begin{enumerate}
\paratitle{Class Constructor} Creates a new \texttt{Pipeline} object that will be used for any subsequent in-memory operations. The class construction takes an integer \texttt{length} as parameter, which determines the length of the input and output vectors that will be processed by the instantiated \texttt{Pipeline}.
%\vspace{-10pt}
\begin{figure}[ht]
\linespread{1.0}\selectfont
%\centering
\begin{lstlisting}[style=cppcustom]
Pipeline::Pipeline(length);
\end{lstlisting}
\end{figure}
\vspace{-15pt}

\paratitle{Stage Creation} The \texttt{stage} method adds a new \texttt{stage} to the \texttt{Pipeline}. 
It takes as input:
\li~a macro string \texttt{str} that represents the stage data-parallel pattern;
\lii~a function pointer \texttt{func}, which defines the computation that will be executed by a PIM thread within a PIM core;
\liii~a tuple \texttt{argsTuple} composed of \texttt{<data type, array pointer>}, which lists the input/output vectors for the given \texttt{stage};
\liv~an extra output array \texttt{output}, which is used in case the produced output array needs to be replicated; and
\lv~the number of elements that will be used for a window (\texttt{overlap})  or a group (\texttt{groupSize)}.
For the tuple \texttt{argsTuple}, the user needs to define the \emph{type} (i.e., \texttt{ArgTypes}) of each \emph{array pointer}, which can be either an input array, an output array, an array used as \emph{both} input and output, a scalar parameters, a scalar reduction output, or a combination function (used to combine partial results across PIM cores). 
The \texttt{stage} method returns \emph{true} in case the \texttt{stage} was successfully added to the \texttt{Pipeline}.
%\vspace{-10pt}
\begin{figure}[ht]
\linespread{1.0}\selectfont
\begin{lstlisting}[style=cppcustom]
// Map / Window / Group
template<typename... ArgTypes>
bool stage(const std::string &str, std::function<void(ArgTypes*...)> func, std::tuple<ArgTyped<ArgTypes>...> argsTuple, uint32_t overlap, uint32_t groupSize);

// Filter (Input/Output)
template<typename FilterType, typename... ArgTypes>
bool stage(const std::string &str, std::function<bool(FilterType*, ArgTypes*...)> func, ArgTyped<FilterType> output, std::tuple<ArgTyped<FilterType>, ArgTyped<ArgTypes>...> argsTuple, uint32_t overlap, uint32_t groupSize);

// Filter (InOut)
template<typename FilterType, typename... ArgTypes>
bool stage(const std::string &str, std::function<bool(FilterType*, ArgTypes*...)> func, ArgTyped<FilterType> inout, std::tuple<ArgTyped<ArgTypes>...> argsTuple, uint32_t overlap, uint32_t groupSize);

// Reduce
template<typename... ArgTypes>
bool stage(const std::string &str, std::function<void(ArgTypes*...)> func, std::tuple<ArgTyped<ArgTypes>...> argsTuple);
\end{lstlisting}
\end{figure}
%\vspace{-10pt}

\paratitle{Fetch Output from PIM Memory} The \texttt{fetch} method is used to \emph{explicitly} mark an output vector (or scalar value) that needs to be copied from the PIM memory to the CPU main memory after the \texttt{Pipeline} is executed. 
This means that the marked vector will \textit{not} be treated as intermediate data (the default behavior for output vectors and scalar values within \texttt{Pipeline}).
\vspace{-10pt}
\begin{figure}[ht]
\linespread{1.0}\selectfont
\begin{lstlisting}[style=cppcustom]
Pipeline::fetch(vector);
\end{lstlisting}
\end{figure}
\vspace{-10pt}

\paratitle{Start PIM Execution} The \texttt{execute} method processes all \texttt{stage}s that have been added to the \texttt{Pipeline}, and once each \texttt{stage} finishes its execution, writes all fetched outputs (from PIM memory) to their respective vectors or scalar values (to the CPU main memory).
%\vspace{-10pt}
\begin{figure}[ht!]
\linespread{1.0}\selectfont
\begin{lstlisting}[style=cppcustom]
Pipeline::execute();
\end{lstlisting}
\end{figure}
\vspace{-10pt}

\paratitle{Get Result Length} The \texttt{getLength} method  retrieves the resulting length of an output \texttt{vector} after the PIM execution has finished. This is only needed if the \texttt{Pipeline} includes a \texttt{filter} data-parallel pattern, as the resulting length of all other data-parallel patterns can be known or calculated \emph{a priori}.
%\vspace{-10pt}
\begin{figure}[ht!]
\linespread{1.0}\selectfont
\begin{lstlisting}[style=cppcustom]
Pipeline::getLength(vector);
\end{lstlisting}
\end{figure}
    
\subsubsection{Implementing a \texttt{Pipeline}} The user follows 5 main steps to implement an application using \prop's dataflow programming interface. 
We reference the simple vector dot-product example in Listing~\ref{listing:vecdot} to illustrate each step. 
First, the user needs to allocate the input arrays that will be processed by the \texttt{Pipeline} (line 2--6).
\prop accepts as input arrays either \texttt{std::vector}s or a raw pointer. 
Second, the user creates a \texttt{Pipeline} object (line 8). 
This object will be used to construct the dot-product algorithm. The user needs to specify the length of the input and output data vectors in this step. 
A key requirement for a given \texttt{Pipeline} is that the vector length should be the same across all \texttt{stages}. In case different vector lengths are required, the user can combine several different-length \texttt{Pipeline}s.
Third, the user adds the \texttt{stage}s to the \texttt{Pipeline} (lines 9--21), where each \texttt{stage} represents a part of dot-product algorithm. 
Each \texttt{stage} contains the kernel that a PIM thread will execute, and a set of inputs and outputs. 
In this example, the user uses a \texttt{map} data-parallel pattern to implement a vector multiplication in arrays $a$ and $b$ (line 10), and a \texttt{reduce} data-parallel pattern to implement a scalar reduction of vector $c$ (line 18). 
An important point is that the user does \emph{not} need to fetch to the CPU memory the vector $c$ during the \texttt{Pipeline} execution, since it is used as the intermediate result between the \texttt{map} and \texttt{reduce} stages. 
As long as an array is \textit{only} used for intermediate data, it does \emph{not} need to be allocated in the host CPU main memory, allowing \prop to save host CPU memory space. 
Fourth, the user marks all of the outputs that need to be fetched from the PIM cores after computation (line 23). 
As described above, this is done so that \prop can avoid fetching intermediate results from PIM cores. 
Fifth, the user triggers the execution of the \texttt{Pipeline} (line 24). 
Only at this point \prop will allocate PIM cores and run each \texttt{stage} in the \texttt{Pipeline}. 
After the execution is finished, the fetched result data (i.e., $sum$) can be accessed by the host CPU.

\begin{figure}[ht]
\setcaptiontype{lstlisting}
\linespread{1.0}\selectfont
\centering
\begin{lstlisting}[style=cppminted]
uint32_t dataLength = 1048576;
std::vector<uint32_t> a(dataLength);
std::vector<uint32_t> b(dataLength);
uint32_t c; // Not fetched, no buffer needed
uint32_t sum = 0;
// Fill a and b with data here

Upmem::Pipeline p(dataLength);
p.stage(MAP(([](uint32_t *c, uint32_t *a, uint32_t *b){
    *c = *a * *b;
}),
    OUTPUT(uint32_t, &c),
    INPUT(uint32_t, a),
    INPUT(uint32_t, b)));

// REDUCE_OUT because it's a scalar and not a vector
p.stage(REDUCE(([](uint32_t *sum, uint32_t *c){
    *sum += *c;
}),
    REDUCE_OUT(uint32_t, &sum),
    INPUT(uint32_t, &c)));

p.fetch(&sum);
p.execute();
\end{lstlisting}
\caption{\prop's implementation of a vector dot-product using its dataflow programming interface.}
\label{listing:vecdot} 
\end{figure}

\subsection{Dynamic Template-Based Compilation}

After the user implements the target computation using the data-parallel pattern APIs and the dataflow programming interface, \prop generates the appropriate PIM binary using a \emph{dynamic template-based compilation}. 
{To so so, \prop uses two main components. 
First, it makes use of \emph{template-based coding}, where the PIM code is created based on an initial skeleton of a PIM application. 
Second, \prop \emph{dynamically} generates and compiles the target PIM code by applying three key optimizations, i.e., stringification, type removal, and memory arrangement, through four code transformations that 
\li~extract the required information from the user's implemented \texttt{Pipeline} to feed the PIM code skeleton; 
\lii~calculate the appropriate offsets used to manage data across PIM banks and local PIM scratchpad memory; and
\liii~divide computation between CPU and PIM cores. 
Figure~\ref{fig:host_workflow} gives an overview of \prop's code transformations targeting the UPMEM PIM system. 
It takes as user input a C++ code (implementing a given \texttt{Pipeline}) and generates as output a UPMEM binary.
This process is dynamically repeated per \texttt{Pipeline}.

\begin{figure}[ht]
    \centering
    \includegraphics[width=\linewidth]{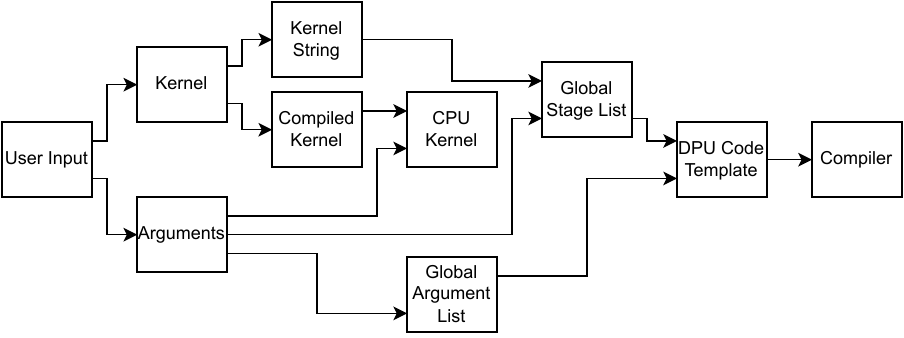}
    \caption{\prop's template-based compilation flow.}
    \label{fig:host_workflow}
\end{figure}

Next, we give an overview of each transformation that \prop performs to generate PIM code.}
In our current implementation, we make use of the Inja~\cite{inja} template engine to manage and populate the UPMEM code skeleton.

\paratitle{First Transformation} {In the first transformation, \prop extracts information regarding the function kernels (i.e., \texttt{func}) specified within \texttt{stage}s across a \texttt{Pipeline}, and its associated input/output arguments (stringification).}
{Kernel information is added to a \emph{global stage list}, while input/output arguments are added to a \emph{global argument list}.}
After this, most of the argument type information is stripped, apart from the type size and its string representation (type removal).
In this  way, we can reduce the amount of templating required, which reduces overall complexity and compilation time. {Once all \texttt{stage}s have been processed and added to the \emph{global state list}, \prop moves to the second transformation.}

\paratitle{Second Transformation} {In the second transformation,} \prop finalizes some {parameter} calculations 
{that depends on having a global view of the application, and consequently,} 
complete information about all \texttt{stage}s. 
This includes memory parameters, such as how to arrange the data in MRAM and WRAM (memory arrangement). 
{Thus, the second transformation is crucial to orchestrate data movement in the DPU system. 
In order to manage MRAM space, the second transformation sets parameters related to MRAM data allocation depending on the \texttt{stage}s and their arguments, while respecting the following approach.}
When the data is copied to the DPUs, \prop splits its total size \emph{evenly} across all DPUs while respecting the 8-byte alignment requirement. 
Within a DPU, {\prop further divides the data evenly between all} tasklets so that they can independently perform their calculations.
\prop also directly arranges the data in MRAM by calculating offsets from the MRAM base address (see Section~\ref{sec:managingdpumem}). 
The data for all \texttt{stage}s are copied simultaneously before starting the DPU process. 
If the data does \emph{not} fit in MRAM, \prop performs multiple execution rounds until the entire dataset is processed. 
Any intermediate results that have to propagate through the \texttt{Pipeline} (i.e., outputs of one \texttt{stage} that become input of subsequent \texttt{stage}s) also have their dedicated MRAM space.

{Likewise, the second transformation also generates the appropriate parameters to manage WRAM using the following approach.} 
Once the DPU starts execution, it {needs to} first initialize its memory space and set up the WRAM. 
It then iterates over each \texttt{stage}, in order. 
Each \texttt{stage} loops over WRAM blocks {(i.e.,} data segments that are small enough to fit into WRAM{)}. 
{In the UPMEM system, the kernel needs to} transfer the data from MRAM to the WRAM {prior to execution,} by passing pointers to the appropriate WRAM location as its arguments. 
{ After the computation is finished, it copies the results back to MRAM. 
Hence, the second transformation sets 
\li~the required pointers by casting them to} the correct type first {using the type information present in each \texttt{stage} argument{; and 
\lii~WRAM block index for a given WRAM access.}}
%, however, which is why the user needs to provide the type information for each argument when adding a new stage.

% Possible replacement for the second half of the paragraph above:
%
% In the UPMEM system, the DPU needs to first transfer the data from MRAM to the WRAM cache prior to execution, which is done by code described in the DPU code template. Once the data resides in the cache, a pointer for each kernel argument pointing to a specific cache location is created with the appropriate type and supplied to the user provided kernel function for the stage that is currently being executed. After a full cache block has been processed, the results will be copied back from WRAM to MRAM and processing of the next block begins.

\paratitle{Third Transformation} {In the third transformation, \prop deals with the case where the \texttt{Pipeline} data size does \emph{not} respect the 8-byte alignment requirement of the UPMEM system. 
In this case, \prop decides to execute the remaining part of the computation at the CPU. 
The third transformation is responsible to identify such computation and to generate CPU code to execute it. 
To do so, \prop creates} a separate {CPU} thread {that} will process some elements on the CPU while the DPU is processing {the portion of the data that is 8-byte aligned}. The amount of data processed on the CPU can be manually increased so that the CPU {does \emph{not}}  have to idle while the DPUs are processing the data, which can lead to performance improvements.

\paratitle{Fourth Transformation} {In the fourth transformation, \prop deals with the need of post-processing required for some \texttt{stage}s, in particular these that implement a \texttt{filter} or \texttt{reduce} data-parallel pattern.} 
For a \texttt{filter}, the results of each tasklet have to be compressed to make the data contiguous again. 
Since parallel data transfers usually achieves higher throughput in the UPMEM system~\cite{gomez2021benchmarking}, we have to copy the same amount of data from every DPU, even if the number of elements passing the filters of each DPU might differ. 
This will leave holes in the data, which we must remove in the CPU. 
Similarly, a \texttt{reduce} also needs further processing by combining the partial results of each DPU in the host CPU.

After {the four transformations are performed}, all of the {parameters} and kernels are inserted into the {UPMEM code} skeleton, and the UPMEM code is compiled.

\subsubsection{Managing DPU Memory}
\label{sec:managingdpumem}

{In order to manage the DPU memory, \prop tries to allocate as much data as possible across the total memory space of the DPUs.} In case the data cannot all fit into the combined MRAM of the DPUs, we need to perform multiple execution rounds to process all of the data. 
For this, we transfer as much data to the DPUs as possible, run the entire \texttt{Pipeline} on that part of the data, and then repeat this process until we {have} processed all of the data.
To this end, \prop performs different element count calculations, which we describe next. 

\paratitle{\underline{Element Count Calculations: General Case}}~\prop must determine how many elements of the input/output vectors, passed as arguments to function kernels, can be processed simultaneously across \texttt{stage}s of a \texttt{Pipeline}.  We describe how \prop handles this process for the general case, i.e., where the target data-parallel patterns lead to \emph{homogeneous} WRAM and MRAM accesses. 

\paratitle{1. Calculating WRAM Parameters}  
We first compute the WRAM cache element count, which defines how many elements a \texttt{stage}'s WRAM cache can hold per argument. This count is uniform across all arguments within a \texttt{stage} unless a \texttt{group} data-parallel pattern partitions it. The WRAM cache element count also determines how many function kernel invocations can occur before requiring a data reload.  
To maximize WRAM utilization, we compute the count per \texttt{stage}, considering its specific number of arguments. The process begins by summing the element type sizes of all arguments in a \texttt{stage} and dividing the total available WRAM space by this sum to obtain a preliminary count. However, since MRAM--WRAM transfers require 8-byte alignment, adjustments are necessary. We iterate over the arguments, computing the total required space, including padding. If the padded size exceeds WRAM capacity, we decrement the element count until alignment constraints are met. Finally, we determine the cache offsets for each argument by sequentially placing them in WRAM.  

\paratitle{2. Calculating MRAM Parameters}  
Next, we determine the maximum number of elements a DPU can hold in MRAM for the entire \texttt{Pipeline}. Unlike WRAM, where each \texttt{stage} is considered independently, MRAM capacity must accommodate all arguments across all \texttt{stage}s simultaneously. The calculation follows the same method as WRAM, adhering to 8-byte alignment constraints but encompassing all arguments rather than a single \texttt{stage}.  

\paratitle{3. Calculating Leftover Parameters}  
Finally, we determine the number of elements processed on the CPU. Due to UPMEM’s 8-byte alignment requirements, some data may not fit into DPUs without violating alignment rules. These excess elements are handled by the CPU. The number of DPU-processed elements follows from the elements-per-round computation:  $\text{CPU elements} = \text{total\_length} - (\text{elements\_per\_round} \times \text{nr\_rounds})$.

\paratitle{\underline{Element Count Calculations: Special Cases}}  
\prop accounts for special cases when managing DPU memory, depending on the \texttt{stage}'s target parallel pattern. We describe these cases below.  

\paratitle{\texttt{window} Data-Parallel Pattern}~The \texttt{window} data-parallel pattern requires overlapping input data between DPUs, as each DPU processes elements while looking ahead at data handled by the next DPU. At the end of the dataset, however, no additional elements remain for lookahead. To address this, one approach is to reduce the output length by the overlap size, but this complicates subsequent \texttt{stage}s that rely on a consistent output size and may \emph{not} align with user expectations. Instead, \prop allows users to provide a small vector of overlap data, which is appended to the original input to maintain the expected output size. Users preferring the first approach can manually discard excess results.  

\paratitle{\texttt{filter} Data-Parallel Pattern} Implementing the \texttt{filter} data-parallel pattern on the DPU is straightforward: the kernel function processes each element, appending those that satisfy the filter condition to the output vector. However, managing WRAM--MRAM transfers introduces complexity. There are two possible strategies for output transfers: 
\li~transferring \emph{all} available output elements \emph{after} processing a full input WRAM block, or 
\lii~waiting until a full output WRAM block is accumulated \emph{before} transferring. 
The latter is more efficient due to larger batch transfers but requires frequent checks for a full WRAM block, which is costly. To avoid this overhead, \prop implements the first approach.  A challenge in this method is ensuring 8-byte alignment, as the number of retained elements is unpredictable. To address this, we round up the transfer size to the nearest valid alignment, ensuring that no overwrites occur. Since the maximum possible output length (where all elements are retained) is always aligned by design, this guarantees safe memory handling.  During transfers, the last 8-byte section of each output WRAM block contains both valid data and padding. Since subsequent transfers cannot create gaps, the base MRAM destination must remain unchanged. To handle this, we copy the last 8 bytes of the previous WRAM block to the start of the next one, adjusting the offset so that new elements seamlessly append to the existing data.

% \begin{figure}[H]
%     \centering
%     \includegraphics[width=\linewidth]{figures/cache_filter.pdf}
%     \caption{Cache filter offset working principle}
%     \label{fig:cache_filter}
% \end{figure}

% \subsubsection{CPU Workload}

% A small section of data that doesn't fit the alignment requirements is processed by the CPU. For this we need to prepare the data first. If data is only used as an intermediate result, it doesn't necessarily require a buffer backing it from the user side as the results never leave the DPU. This of course leaves the CPU without a buffer to work with, so to handle the CPU part, a small buffer is allocated to store such intermediate results. For any inputs or fetched outputs the user provided buffers are used directly. If we have any overlap from a window function, we also need to allocate a buffer for the data. We need a buffer larger than the one the user provided in order to append the end-of-data overlap that was specified by the user.

\subsection{Handling Invalid \texttt{Pipeline} Implementations}

Certain \texttt{stage} combinations cannot be processed within a single \texttt{Pipeline} as described so far. Specifically, the outputs of \texttt{filter} and \texttt{reduce} cannot be directly used by subsequent \texttt{stage}s except for additional filtering or reduction.  
For \texttt{filter}, a \texttt{map} operation cannot process its output since each PIM core lacks knowledge of how many elements preceding PIM cores retained. Without this information, a PIM core cannot determine the correct position of its filtered results in the global output vector, making it impossible to align multiple inputs in a \texttt{map} operation when one of them is a \texttt{filter} output. Similarly, for \texttt{reduce}, each PIM core only has a partial result, meaning further \texttt{stage}s would receive incomplete or incorrect data.  

To resolve this, filtered or reduced results \emph{must} be copied back to the CPU and combined before further processing. This effectively splits execution into two separate \texttt{Pipeline}s: one handling operations up to the \texttt{filter} or \texttt{reduce} stages, and another processing the aggregated results.  

To assist in such cases, \prop provides the \texttt{PipelineFull} class, which automatically detects invalid stage combinations and partitions execution into multiple sub-pipelines. This ensures that all stage configurations can be handled correctly. The class is separate from \texttt{Pipeline} to highlight its potential performance impact, encouraging users to make an informed choice. Its interface remains identical to that of a regular \texttt{Pipeline}.  

\section{Methodology}
\label{sec:methodology}

\subsection{Implementation} As a case study, {we implement \prop targeting a UPMEM PIM system that includes a 2-socket Intel Xeon Silver
4110 CPU at \SI{2.10}{\giga\hertz} (host CPU), standard main memory (DDR4-2400) of \SI{128}{\giga\byte}, and
20 UPMEM PIM DIMMs with \SI{160}{\giga\byte} PIM-capable memory and 2,560 DPUs~\cite{upmem}.} We implemented \prop in C++, and compiled it as a shared library, therefore it can be distributed independently from the user application, if needed. 
To use \prop, we link our target application implementation to the shared library. 
%Currently, \prop is only supported on Linux platforms.

\subsection{Workloads} 

We evaluate \prop using six PIM-friendly applications from the PrIM benchmark suite~\cite{gomez2021benchmarking,gomez2022benchmarking}: vector addition (\texttt{VA}), select (\texttt{SEL}), unique (\texttt{UNI}), reduction (\texttt{RED}), general matrix-vector multiplication (\texttt{GEMV}), and image histogram small (\texttt{HST-S}). 
We use the hand-tuned implementations of such workloads from the PrIM benchmark suite as our baseline. Below, we describe the data-parallel patterns we use to implement each one of the six PrIM workloads. If not otherwise specified, we use 1M 32-bit integer elements per UPMEM core as input dataset for each workload.

\paratitle{Vector Addition (\texttt{VA})} Element-wise addition of two arrays. We implement \texttt{VA} using a \texttt{map} data-parallel pattern. 
%We evaluate the runtime of \texttt{VA} for 1M 32-bit integer elements per UPMEM core.

\paratitle{Select (\texttt{SEL})} Selects elements from the input vector based on a predicate. We implement \texttt{SEL} using a \texttt{filter} data-parallel pattern.  
%We evaluate the runtime of \texttt{SEL} for 1M 32-bit integer elements per UPMEM core.

\paratitle{Unique (\texttt{UNI})} Removes duplicates from a sorted vector. We implement \texttt{UNI} using a \texttt{window+filter} data-parallel pattern, with a window of two. Thus, our implementation accesses two consecutive elements, keeping an element if it differs from the next one.  
%We evaluate the runtime of \texttt{UNI} for 1M 32-bit integer elements per UPMEM core.

\paratitle{Reduction (\texttt{RED})} Adds all elements of an input array into a scalar value. We implement \texttt{RED} using a \texttt{reduce} data-parallel pattern. 
%We evaluate the runtime of \texttt{RED} for 1M 32-bit integer elements per UPMEM core.

\paratitle{General Matrix-Vector Multiplication (\texttt{GEMV})} Standard matrix-vector multiplication. We implement \texttt{GEMV} using a \texttt{group} data-parallel pattern by 
\li~setting the group size equal to the vector size, 
\lii~treating the vector as a scalar, and 
\liii~manually iterating over rows to perform multiplication and summation. While this approach requires the user to explicitly write the loop, it is necessary because \prop does \emph{not} inherently recognize non-vector data structures such as matrices. Additionally, this method requires the entire vector to fit within WRAM. This constraint is also present in the PrIM benchmark. We evaluate the runtime of \texttt{GEVM} for matrices with 4096 rows and 256 columns per UPMEM core.

\paratitle{Image Histogram Small (\texttt{HST-S})} Counts the number of times each value appears in a vector input. We implement \texttt{HST-S} using a \texttt{reduce} data-parallel pattern. In our \texttt{HST-S} implementation, the reduction variable is a vector, whose size depends on the amount of unique values that appear in the input vector. We evaluate the runtime of \texttt{HST-S} for 1M 32-bit integer elements per UPMEM core, with the number of histogram
bins set to 256~\cite{gomez2021benchmarking,gomez2022benchmarking}.

\section{Evaluation}
\label{sec:eval}

We first describe how \prop improves programming productivity. 
Second, we provide a performance analysis of \prop compared to hand-tuned reference implementations of our six evaluated workloads. 
Third, we discuss \prop's overheads. 

\subsection{Productivity Improvement}

We evaluate how \prop can improve programming productivity compared to the hand-tune PrIM workloads, and a prior framework for UPMEM programmability called SimplePIM~\cite{chen2023simplepim}. Similar to \prop, SimplePIM provides a series of high-level APIs that allow the user to abstract low-level implementation details when writing code for the UPMEM system. However, different than \prop, in SimplePIM, the user is still responsible to \emph{explicitly} handle CPU--DPU and DPU--DPU data communication. 
We use lines-of-code (LOC) as a productivity metric~\cite{boehm1984software,jones1985programming,demarco1986controlling}. To do so, we manually count the lines of \emph{effective} UPMEM-programming related code for each one of the six workloads, which excludes lines of code for data loading from a file to the host main memory, host memory allocation, variable definition, and time
measurements.

Table~\ref{tab:loc} summarizes the lines of effective code saved by using \prop for the six workloads compared to the hand-tuned PrIM implementation and SimplePIM. Note that there is no available reference implementation of SimplePIM for three of our workloads (i.e., \texttt{SEL}, \texttt{UNI}, and \texttt{GEMV}). 
We make two observations from the table.
First, compared to the hand-tuned PrIM implementations, \prop \emph{significantly} reduces LOC. On average across all six workloads, \prop reduces LOC by 94\% compared to the PrIM implementations. This is possible since \prop \emph{significantly} raises the abstraction level when implementing UPMEM workloads, allowing the user to mainly focus the target kernel function. 
Second, compared to SimplePIM, \prop further improves productivity by 59\%. This is due to the fact that in \prop, the user does \emph{not} need to write code for data communication or metadata bookkeeping, as in SimplePIM. 
We conclude that \prop is an efficient framework to ease UPMEM programmability.

\begin{table}[ht]
    \centering
    \caption{Lines-of-code (LOC) comparison.}
    \label{tab:loc}
    \resizebox{0.7\linewidth}{!}{
\begin{tabular}{@{}l||lll|l|l@{}}
\toprule
\textbf{Workload} & \textbf{\begin{tabular}[c]{@{}l@{}}LOC \\ PrIM~\cite{gomez2021benchmarkingcut}\end{tabular}} & \textbf{\begin{tabular}[c]{@{}l@{}}LOC \\ SimplePIM~\cite{chen2023simplepim}\end{tabular}} & \textbf{\begin{tabular}[c]{@{}l@{}}LOC \\ \prop\end{tabular}} & \textbf{\begin{tabular}[c]{@{}l@{}}LOC Red.\\ (vs. PrIM)\end{tabular}} & \textbf{\begin{tabular}[c]{@{}l@{}}LOC Red. \\ (vs. SimplePIM)\end{tabular}} \\ \midrule
\texttt{VA} & 78 & 14 & 6 & 92\% & 57\% \\
\texttt{SEL} & 120 & - & 6 & 95\% & - \\
\texttt{UNI} & 155 & - & 6 & 96\% & - \\
\texttt{RED} & 123 & 14 & 6 & 95\% & 57\% \\
\texttt{GEMV} & 180 & - & 9 & 95\% & - \\
\texttt{HST-S} & 113 & 21 & 8 & 93\% & 62\% \\ \midrule
\textbf{GMean} & \textbf{124} & \textbf{16} & \textbf{7} & \textbf{94\%} & \textbf{59\%} \\ \bottomrule
\end{tabular}
}
\end{table}

\subsection{Performance Analysis}

We compare \prop's performance for our six workloads in comparison to their PrIM hand-tuned implementations. We conduct our analysis in a real UPMEM-based system (described in Section~\ref{sec:methodology}). 
In our analysis, we configure the DPUs to use 11 tasklets (for both \prop and PrIM workloads), which is the minimum number of tasklets in order to fill the DPU instruction pipeline~\cite{gomez2021benchmarking,gomez2022benchmarking}. 
We report the average execution time across 10 execution runs.

\paratitle{End-to-End Execution Time} Figure \ref{fig:performance} shows the average end-to-end execution time to execute each one of our six workloads using PrIM and \prop implementations. In this analysis, we measure both the time it takes for CPU--DPU/DPU--CPU data transfer, inter-DPU data movement, and DPU kernel execution time. 
We make four observations from the figure. 
First, as expected, the CPU--DPU and DPU--CPU transfer times comprise the vast majority of the processing time for both PrIM and \prop implementations. 
Second, for four workloads (i.e., \texttt{VA}, \texttt{RED}, \texttt{GEMV}, and \texttt{HST-S}), \prop achieves performance on par with the hand-tuned PrIM implementations. 
Third, for two workloads (i.e., \texttt{SEL} and \texttt{UNI}), \prop performs \emph{significantly} better than PrIM implementations.  
On average across these two workloads, \prop achieves 10$\times$ the performance of PrIM implementations. 
This is due to the fact that PrIM implementations for such workloads use slower serial data transfers instead of faster parallel data transfer when copying data back from the DPUs to the CPU main memory. 
As these workloads have undefined output sizes (since they include filtering operations), PrIM transfers the output data back from each DPU \emph{individually}, one after another, after communicating the correct result size to the CPU. 
In contrast, \prop copies back from DPUs to the CPU \emph{all} the output data, and only compresses the data to remove the `holes' after we transferred it back to the CPU, hence improving transfer time.
Fourth, on average across all six workloads, \prop achieves 2.1$\times$ the end-to-end performance of the hand-tuned PrIM implementations. 
We conclude that \prop \emph{consistently} achieves performance comparable to the PrIM hand-tune reference implementations. 

\begin{figure}[ht]
    \centering
    \includegraphics[width=0.85\linewidth]{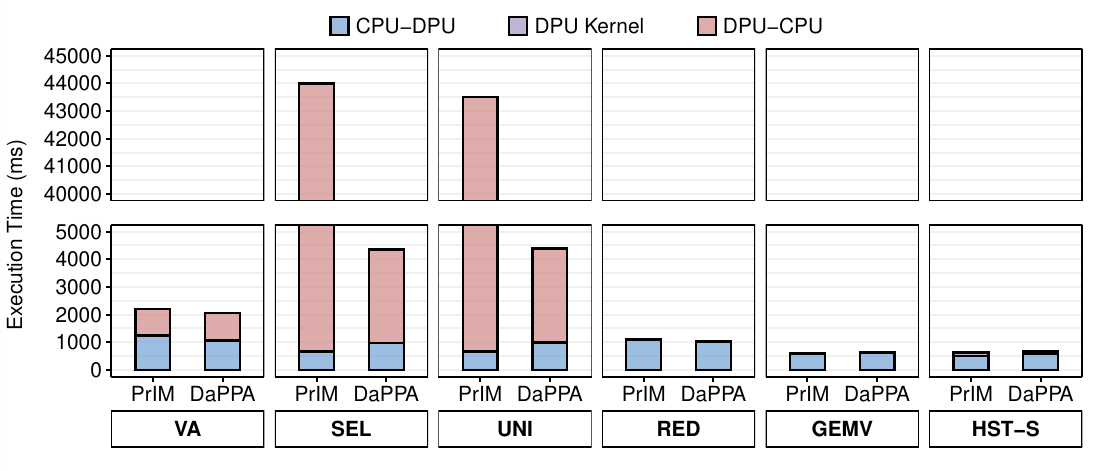}
    \caption{End-to-end execution time for six workloads using PrIM~\cite{gomez2022benchmarking} hand-tuned and \prop implementations.}
    \label{fig:performance}
\end{figure}

\paratitle{DPU Kernel Execution Time}  Figure~\ref{fig:kernel_time} shows the DPU kernel execution time (i.e., the execution time \emph{excluding} memory transfer times) for our six workloads using both PrIM and \prop implementations.
We make two observations from the figure. 
First, for most workloads, \prop either achieves on par performance or outperforms their PrIM implementations. On average across all workloads, \prop achieves 1.4$\times$ the performance of the PrIM implementations (and up to 3.5$\times$).
Second, for some workloads (i.e., \texttt{SEL} and \texttt{HST-S}), there is a large performance gap between \prop and PrIM implementations, in favor of \prop.
This is because different amounts of work are performed in the DPU in these cases: the PrIM implementations performs part of the compression/combination after the filter/reduction already in the DPU, which counts towards the DPU time, whereas \prop perform such operations within the host CPU, which counts towards the memory transfer time presented in Figure~\ref{fig:performance}.

\begin{figure}[H]
    \centering
    \includegraphics[width=0.85\linewidth]{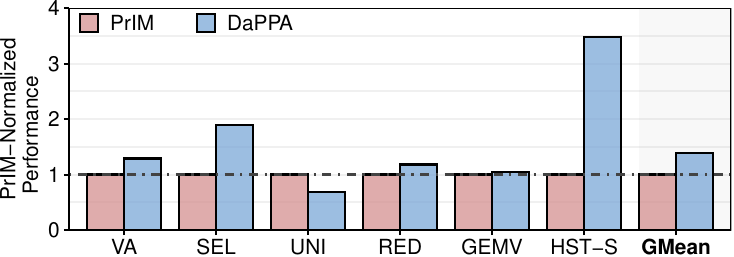}
    \caption{DPU kernel performance comparison for six workloads using PrIM~\cite{gomez2022benchmarking} and \prop implementations.}
    \label{fig:kernel_time}
\end{figure}

%In summary, the primary objective of \prop is to improve UPMEM code development time while maintaining competitive performance, rather than \emph{solely} optimizing for performance. 

%Given that data transfers constitute the dominant performance bottleneck in many UPMEM workloads, the framework's competitiveness is expected to hold in general, as reducing transfer overhead remains a challenging task, even with manual optimizations.

\subsection{\prop Execution Time Overheads}

\prop induces some additional execution overheads due to its dynamic template-based compilation. We measured such overheads for our six workload implementations, and we observe that \prop execution time overheads are of 
\li~\SI{1}{\milli\second} for substituting values into the DPU code skeleton;
\lii~\SI{150}{\milli\second} for compiling the DPU binary (recall that in \prop, the DPU binary has to be compile at runtime for each \texttt{Pipeline});
\liii~\SIrange{1}{150}{\milli\second} for other various operations (e.g., element count calculations). 
Such execution overheads are negligible in comparison to the time the UPMEM SDK takes to allocate DPUs (\SI{1200}{\milli\second}) and end-to-end DPU execution time (Figure~\ref{fig:performance}).

\section{Summary}
\label{sec:conclusion}

We introduce \prop (\underline{da}ta-\underline{p}arallel~\underline{p}rocessing-in-memory \underline{a}rchitecture), a data-parallel programming framework that simplifies programming in general-purpose \gls{PnM} systems. 
By abstracting hardware complexities through high-level data-parallel patterns, \prop enables efficient PIM execution without manual optimizations. 
\prop is comprised of three main components, a set of data-parallel pattern APIs, a dataflow programming interface, and a dynamic template-based compilation scheme.
Our evaluation shows that \prop improves performance while reducing programming complexity for different PIM workloads. 
\prop bridges the gap between ease of use and efficiency in PIM architectures, fostering broader adoption for general-purpose \gls{PnM} systems.

% % \glsresetall{}
% \ifcameraready
%     \setcounter{version}{99}
% \else
%     \setcounter{version}{9}
% \fi

\chapter{Conclusions and Future Directions}
\label{chap:conc}

In summary, the goal of this dissertation is to 
provide tools and system support for \gls{PIM} architectures (with a focus on DRAM-based solutions), to ease the adoption of such architectures in current and future systems.  
To achieve this goal, we conduct a set of research projects based on the thesis statement
that ``\textit{We can effectively exploit the inherent parallelism of \gls{PIM} architectures and facilitate their adoption across a broad spectrum of workloads through end-to-end design of hardware and software support for PIM\omt{3}{,} including benchmark suites \omt{3}{and workload analysis methodologies}, compiler/programming frameworks, and data-aware runtime systems\omt{3}{,} \omt{3}{thereby enabling large (e.g., factors or orders of magnitude) improvements in performance and energy efficiency.}}'' 
To this end, we combine workload analyses, hardware/software co-designed mechanisms, and software-level frameworks. 

First, we build a comprehensive understanding of data movement bottlenecks in modern systems. 
To this end, we conduct the first large-scale characterization of data movement across 77K functions from 345 applications, identifying key sources of data movement and evaluating the effectiveness of compute-centric (e.g., caching, prefetching) and memory-centric (e.g., \gls{PIM}) mitigation techniques (Chapter~\ref{chap:damov}). 
We develop the DAMOV benchmark suite, selecting 144 diverse functions that capture different types of data movement across various domains. 
Using DAMOV, we conduct four case studies on PIM systems, analyzing load balance, inter-PIM communication, core model impacts, and the utility of simple PIM instructions. 
We conclude that DAMOV enables rigorous study of data movement and provides a valuable tool for developing and evaluating PIM architectures. 
We open-source DAMOV and the complete source code for our new characterization methodology at~\url{https://github.com/CMU-SAFARI/DAMOV}.

Second, we enable efficient and flexible \gls{PIM} execution by overcoming the limitations of \gls{PuD} with MIMDRAM, a hardware/software co-designed system (Chapter~\ref{chap:mimdram}). 
MIMDRAM leverages fine-grained DRAM access to support SIMD/MIMD execution at a DRAM subarray, improving utilization and programmability. 
It introduces lightweight hardware extensions and memory controller modifications to enable independent execution of \gls{PuD} instructions and in-memory vector-to-scalar reduction operations. 
On the software side, MIMDRAM implements compiler passes to identify \gls{PuD}-friendly code regions, determine appropriate SIMD granularity, and schedule operations across DRAM mats. Our evaluation on twelve applications and 495 workload mixes shows that MIMDRAM provides 34$\times$ performance and 14.3$\times$ energy efficiency improvements over state-of-the-art \gls{PuD}, while incurring small area cost. 
We open-source MIMDRAM workloads and evaluation framework at \url{https://github.com/CMU-SAFARI/MIMDRAM}.

Third, we introduce \emph{Proteus}, a hardware framework that improves \gls{PuD} execution efficiency by dynamically adapting to workload characteristics (Chapter~\ref{chap:proteus}). 
\emph{Proteus} addresses the limitations of static data representation and throughput-only execution in bit-serial \gls{PuD} architectures by exploiting narrow data values and DRAM internal parallelism. 
It automatically selects the best data formats, bit-precision, and arithmetic algorithms (bit-serial or bit-parallel) for high-performance \gls{PuD} execution. Compared to CPU, GPU, and SIMDRAM, \emph{Proteus} achieves up to 90.3$\times$ energy savings and 17$\times$  higher performance per area across twelve applications, while adding only 1.6\% DRAM and 0.03\% CPU area overhead. \emph{Proteus} demonstrates the importance of data-aware execution in \gls{PIM} systems.
We open-source \emph{Proteus} workloads and evaluation framework at \url{https://github.com/CMU-SAFARI/Proteus}.

Fourth, we ease PIM programmability with DaPPA, a software framework for general-purpose PNM systems (Chapter~\ref{chap:dappa}). 
DaPPA abstracts low-level hardware management using high-level data-parallel patterns and a dataflow programming interface. 
It automates data movement, allocation, and workload distribution, allowing efficient programming without requiring detailed hardware knowledge. DaPPA comprises pattern APIs, a dataflow model, and a dynamic compilation backend. Evaluated on six PrIM workloads~\cite{gomezluna2021repo}, DaPPA improves performance by 2.1$\times$ and reduces code complexity by 94\% compared to manual implementations. Our results highlight DaPPA's potential in making PIM programming more accessible and productive.

\section{Future Research Directions}
\label{sec:conc_future}

Although this dissertation focuses on enabling system support for \gls{PIM} systems, proposing several new ideas for efficient and programmer-oblivious \gls{PIM} execution, we believe that this work is applicable in a more general sense and opens up new research directions. 
This section reviews promising directions for future work. 
%We first discuss \emph{short-term research directions} that directly stem from open-research questions that follow directly from the four works presented in this dissertation.
%Second, we discuss \emph{long-term research directions} that should be addressed to further enhance system support for future \gls{PIM} architectures. 

%\subsection{Short-Term Research Directions}

\subsection{Extending the DAMOV Methodology \& Benchmark Suite}

\paratitle{Evaluating Data Movement Bottlenecks in Other Processor-Centric Systems} First, our DAMOV methodology can be extended to evaluate data movement bottlenecks in processor-centric systems beyond general-purpose CPUs. 
Modern computing platforms increasingly incorporate heterogeneous components such as GPUs, custom accelerators, and FPGAs, each with distinct memory subsystem characteristics and bottlenecks. Applying DAMOV methodology to these platforms would require adapting its instrumentation and profiling tools to capture low-level data movement behavior in these architectures. 
For example, GPUs exhibit high main memory throughput but suffer from memory coalescing and bank conflicts, while FPGAs and accelerators often rely on statically defined memory hierarchies. Characterizing how data movement bottlenecks manifest across these systems can provide deeper insight into their performance limitations and guide the development of architecture-specific data movement mitigation mechanisms, including platform-tailored \gls{PIM} solutions.

\paratitle{Evaluating the Impact of Data Bottlenecks Caused by Storage \& Network}
Second, the DAMOV methodology can be extended to account for data movement bottlenecks arising from the storage and networking subsystems, which are increasingly critical in data-intensive and distributed applications. 
Today's systems rely heavily on high-speed storage devices (e.g., NVMe SSDs) and fast network interconnects (e.g., RDMA) that move data across devices and distributed nodes. 
Incorporating these components into DAMOV requires modifying the methodology to be able to gather profiling information regarding I/O and network traffic, enabling a holistic view of data movement across the entire system. 
Such an extension would facilitate the identification of data movement bottlenecks that span multiple layers of the system stack, such as data ingestion pipelines, checkpointing, or remote memory access, and guide the design of new hardware/software optimizations, including in-storage or in-network processing approaches.

\paratitle{Identifying \gls{PuM} Suitability}
Third, DAMOV can be extended to evaluate the suitability of \gls{PuM} techniques, which perform computation directly within the memory arrays.
Unlike \gls{PIM} solutions that rely on integrating compute units near memory, \gls{PuM} exploits the analog properties of DRAM to perform operations such as copy, initialization, or bitwise logic. 
Extending DAMOV to model and evaluate \gls{PuM} requires new characterization dimensions that capture the access patterns, data types, and computational primitives amenable to in-DRAM execution. 
DAMOV could be used to build a classification framework that identifies code regions likely to benefit from \gls{PuM}, complementing existing \gls{PIM}-focused analysis. By integrating \gls{PuM} feasibility studies into its benchmark suite, DAMOV can serve as a comprehensive platform to explore the full spectrum of memory-centric computation paradigms.

\subsection{Improving MIMDRAM \& \emph{Proteus} Execution Model}

\paratitle{Reducing the Cost of Inter-Lane Communication}
A key performance bottleneck in MIMDRAM and \emph{Proteus} arises from the costly inter-lane communication within DRAM, particularly during reduction operations. 
Due to the narrow interconnect between DRAM columns, transferring partial results across lanes requires multiple DRAM commands, significantly increasing latency and energy consumption. 
Future work can explore low-cost, high-bandwidth intra-subarray interconnects that enable more efficient broadcast and aggregation of DRAM columns. 
Such improvements could allow more efficient in-memory reductions execution, enabling broader applicability of \gls{PuM} for workloads with reduction-heavy computations.

\paratitle{Avoiding Data Layout Transformations}
Both MIMDRAM and \emph{Proteus} require data transposition to convert data from a row-major (horizontal) layout to a column-major (vertical) layout suitable for bit-serial operations. This data layout transformation incurs overheads in terms of performance, energy, and code complexity. One promising direction is to develop fine-grained inter-lane interconnects within the DRAM mat, reducing the need for full data transposition. Alternatively, future work could explore compiler- and OS-level data placement strategies that lay out data vertically across DRAM mats by construction, allowing in-place execution of vertical \gls{PuM} operations. 
A pipelined execution model, inspired by recent works like RACER~\cite{truong2021racer}, could further reduce the need for global transpositions by overlapping horizontal and vertical operations across DRAM mats.

\paratitle{Smarter Offloading Decisions}
The current MIMDRAM compiler lacks a cost model to decide whether offloading a computation to \gls{PuM} will be beneficial, leading to the possibility of offloading instructions that are better suited for CPU execution due to cache locality. Incorporating a hybrid compiler/runtime system that integrates compile-time static analysis with runtime profiling can improve offloading decisions. For example, the compiler could annotate code regions with potential memory access footprints, while a lightweight runtime system monitors cache behavior and dynamically chooses whether to execute the region in \gls{PuM} or CPU. Such hybrid schemes would reduce performance regressions and improve the robustness of \gls{PuM}-enabled systems across diverse workloads.

\paratitle{Generalizing the Execution Model}
MIMDRAM's compiler currently limits auto-vectorization to inner loops, restricting the opportunity to extract data parallelism from outer-loop structures. An improved programming model could adopt a SIMT-style approach, mapping outer loop iterations to independent threads running on different DRAM subarrays, while retaining SIMD execution within inner loops. This hierarchical execution model would allow MIMDRAM to exploit both coarse- and fine-grained parallelism across DRAM, increasing the utilization of available compute substrates. Compiler support for loop tiling and cross-subarray synchronization primitives will be necessary to efficiently orchestrate such SIMT-like \gls{PuM} execution.

\paratitle{Enabling Efficient In-DRAM Multiplication}
While \gls{PuM} enables efficient in-DRAM execution of bitwise and simple arithmetic operations, multiplication remains a costly primitive in current designs. Stochastic computing offers one possible approximation-based solution, but it introduces significant accuracy and convergence challenges. Future research should explore new algorithms that decompose multiplication into simpler DRAM-friendly primitives, or leverage emerging non-volatile memory technologies that support native multiply-accumulate operations. Additionally, \gls{PuM}-specific approximate multiplication techniques could be tuned based on application-level tolerance to inaccuracy, enabling trade-offs between precision and efficiency in domains such as machine learning and signal processing.

\paratitle{Combining \gls{PuM} and \gls{PnM} Execution}
MIMDRAM and \emph{Proteus} currently focus exclusively on \gls{PuM} execution and do \emph{not} explore the potential synergies with processing-near-memory (\gls{PnM}) designs. A combined \gls{PuM}--\gls{PnM} execution model could address several current limitations. For instance, expensive operations such as multiplication and reduction could be delegated to lightweight near-bank ALU engines, while inexpensive vector-wide operations are executed within the DRAM arrays. Such a hybrid model allows each operation to be executed where it is most efficient, reducing the burden on the DRAM mat and improving overall system throughput. Developing orchestration mechanisms between \gls{PuM} and \gls{PnM} execution units, as well as compiler/runtime frameworks to partition code accordingly, presents a promising avenue for future work.

\subsection{Enhancing \omt{3}{the} DaPPA Programming Framework}

\paratitle{Supporting Multi-Dimensional Data Structures}
DaPPA currently supports operations primarily over one-dimensional vectors, limiting its expressiveness for applications that naturally operate on higher-dimensional data structures such as matrices, tensors, or multidimensional grids. Many real-world applications in scientific computing, image processing, and machine learning rely on native support for multi-dimensional arrays. Future work can extend DaPPA's programming model to support multi-dimensional abstractions, allowing the programmer to express computation over multi-dimensional domains more naturally. This extension would also require enhancing the underlying memory allocation and scheduling infrastructure to support multi-dimensional data decomposition and distribution across the PIM cores efficiently.

\paratitle{Extending the Set of Data-Parallel Patterns}
DaPPA currently focuses on five core data-parallel patterns--\texttt{map}, \texttt{reduce}, \texttt{filter}, \texttt{window}, and \texttt{group}--which, while powerful, limit its applicability to a subset of data-parallel workloads. 
Many applications rely on more complex patterns, such as \texttt{stencil}, commonly used in scientific simulations, or \texttt{farm}, used in embarrassingly parallel workloads. Extending DaPPA to support a richer set of patterns would enable it to target a wider range of workloads and allow more programmers to express their computation within the framework. Incorporating new patterns would also open opportunities for pattern-specific compiler optimizations and hardware-aware scheduling strategies.

\paratitle{Raising the Level of Abstraction with Automatic Pattern Detection}
While DaPPA simplifies programming by abstracting away hardware-specific details, the programmer is still responsible for manually identifying and selecting the appropriate data-parallel pattern for each computation. 
Prior works~\cite{del2018finding,rul2010profile,molitorisz2015patty,tournavitis2010semi} show that it is possible to automatically detect data-parallel patterns in imperative code written in C/C++ through static and dynamic analysis. Integrating such pattern detection capabilities into DaPPA would raise the level of abstraction further, enabling legacy applications to be ported to PIM with minimal programmer intervention. This automation would reduce programming effort, lower the barrier to adoption, and allow more developers to leverage the performance benefits of PIM architectures.

\paratitle{Enabling Portability Across PIM Architectures}
Currently, DaPPA is tightly coupled to the UPMEM PIM architecture and does \emph{not} support other real-world PIM platforms. Supporting multiple backends would significantly broaden DaPPA’s applicability and facilitate comparative studies across diverse PIM systems. Future work can focus on decoupling DaPPA's intermediate representation from its hardware-specific backend, enabling portability to alternative PIM architectures such as HBM-based \gls{PIM} systems or hybrid PUM/PNM platforms. This backend modularity would position DaPPA as a general-purpose PIM programming framework, serving both research and industry use cases across heterogeneous computing environments.

%\subsection{Long-Term Research Directions}

\section{Concluding Remarks}

In this dissertation, we investigate the tools, system, and programming support required to enable \gls{PIM} architectures in current future systems efficiently. We first build a comprehensive understanding of data movement bottlenecks by characterizing 77K functions across 345 applications, leading to the DAMOV methodology and benchmark suite (Chapter~\ref{chap:damov}). 
We then propose two \gls{PuM} systems: 
\li~MIMDRAM (Chapter~\ref{chap:mimdram}), which enables flexible in-DRAM SIMD/MIMD execution through fine-grained DRAM access alongside a holistic system support design to automatically generate and execute \gls{PuD}-capable code in a system; and
\lii~\emph{Proteus} (Chapter~\ref{chap:proteus}), which dynamically adapts the data representation, precision, and arithmetic \gls{PuD} implementation for efficient in-DRAM computation. Finally, we introduce DaPPA (Chapter~\ref{chap:dappa}), a high-level programming framework for general-purpose \gls{PnM} systems, which simplifies programmability and improves performance of PIM code through data-parallel patterns.

% We firmly believe that it is time to design principled system architectures to solve the data movement problem of modern computing systems, which is caused by the rigid dichotomy and imbalance between the computing unit (CPUs and accelerators) and the memory/storage unit. Fundamentally solving the data movement problem requires a paradigm shift to a more data-centric computing system design, where computation happens where data resides or where data is generated (i.e., in or near memory/storage), with minimal movement of data.  
% Such a paradigm shift can greatly push the boundaries of future computing systems, leading to orders of magnitude improvements in energy and performance, while potentially also enabling new applications and computing platforms.

% Although many challenges remain to enable widespread adoption of processing-in-memory, we believe the mindset and infrastructure shift necessary to enable such a combined computation-storage paradigm remains to be the largest challenge. Overcoming this mindset and infrastructure shift can unleash a fundamentally energy-efficient, high-performance, and sustainable way of designing, using, and programming computing systems.
% We therefore believe the future of processing-in-memory is very bright and promising, yet there needs to be many exciting challenges to be solved across the computing stack to facilitate widespread and easy adoption. 

\gft{3}{As applications continue to grow in complexity and scale, data movement is rapidly emerging as a fundamental bottleneck in modern computing systems, significantly impacting performance, energy efficiency, and system reliability. 
Existing processor-centric architectures are increasingly constrained by this bottleneck, and it is becoming clear that incremental improvements to traditional processor-centric designs are insufficient. 
This dissertation advocates for a paradigm shift towards processing-in-memory (PIM), an approach that \emph{fundamentally} mitigates data movement by moving  computation to where the data resides.} 

\gft{3}{Although the core idea of PIM is not new, recent technological advances have made it feasible to manufacture and prototype PIM architectures that aim to accommodate the computational demands of data-intensive workloads. 
At the same time, fully embracing PIM requires overcoming significant system-level challenges due to the entrenched assumptions and abstractions imposed by decades of processor-centric designs. 
Through the contributions of this dissertation, which span workload characterization, hardware/software co-design, and programming support, we take meaningful steps toward enabling practical, efficient, and programmable PIM systems. 
However, the road to widespread adoption of PIM remains long. Continued effort is required across the entire computing stack to fully realize the promise of PIM and enable a future where systems are designed from the ground up with data movement as a \emph{first-class} design concern. 
We remain optimistic that such a future is within reach and that PIM will play a central role in shaping the next generation of high-performance, energy-efficient, and sustainable computing platforms.}

% % APPENDICES
\appendix
\cleardoublepage%
\chapter{Other Works of the Author}
\label{appendix:otherworks}

In this chapter, we first summarize other works led by the author but that are not included in this dissertation, and second, we briefly discuss other works co-lead by the author.

\section{Other Works Lead by the Author}

\subsection{Extending Memory Capacity in Modern Consumer Systems with Emerging Non-Volatile Memory: Experimental Analysis and Characterization Using the Intel Optane SSD}

In our IEEE Access paper~\cite{oliveira2023extending}, we provide the \emph{first} analysis of the impact of extending the main memory space of consumer devices using off-the-shelf \glspl{NVM}. 
We \emph{extensively} examine system performance and energy consumption when the NVM device is used as swap space for DRAM main memory to effectively extend the main memory capacity. Our empirical analyses lead us to several observations and insights that can be useful for the design of future systems and NVMs.

For our experimental evaluation, we equip real web-based Chromebook computers~\cite{chromebook} with the Intel Optane SSD~\cite{h10}. 
Our target workloads are interactive applications, with a focus on the Google Chrome~\cite{chrome} web browser. 
We choose such workloads for two reasons. First, in interactive applications, the system needs to respond to user inputs at a target output latency to provide a satisfactory user experience. 
Second, in Chromebooks, the Chrome browser serves as the main interface to execute services for the user. 
We compare the performance and energy consumption of interactive workloads running on our Chromebook with NVM-based swap space, where the Intel Optane SSD capacity is used as swap space to extend main memory capacity, against two state-of-the-art systems: 
\li~a baseline system with double the amount of DRAM than the system with the NVM-based swap space, which resembles current consumer devices but has high manufacturing cost due to the \gfi{large DRAM capacity and} relatively high cost-per-bit of DRAM; and 
\lii~a system where the Intel Optane SSD is naively replaced with a state-of-the-art (yet slower) off-the-shelf NAND-flash-based SSD, which we use as a swap space of equivalent size as the NVM-based swap space. The NAND-flash-based SSD provides a cheap alternative to extend the main memory space, but it can penalize system performance due to its high access latency. We use a memory capacity pressure test~\cite{memorypressure} to measure the impact of the new NVM swap space on user tasks that consist of loading, scrolling, and switching between Chrome browser tabs. 
We measure how the NVM device increases the 99th-percentile latency (i.e., tail latency) of each task and the total number of Chrome tabs that the user can open without discarding old tabs. 

To summarize, our experimental analysis reveals that extending the main memory space by using the Intel Optane SSD as NVM-based swap space for DRAM provides a cost-effective way to alleviate DRAM scalability issues. However, naively integrating the Intel Optane SSD into the system leads to several system-level overheads that can negatively impact overall performance and energy efficiency. We mitigate such overheads by examining and evaluating system optimizations driven by our analyses. 

We provide the following six key takeaways from our empirical analyses: 
\begin{enumerate}[noitemsep, leftmargin=*, topsep=0pt]
\item \textit{Effect of Intel Optane SSD as swap space.} Reducing DRAM size and extending the main memory space with the Intel Optane SSD as swap space provides benefits for the Chrome browser, since it can 
\li~increase the number of open tabs, and \lii~reduce system cost. 
However, it also leads to an increase in the number of tab switches with high latency compared to the baseline.

\item \textit{Reducing tail latency by enabling Zswap.} Zswap~\cite{zolnierkiewicz2013efficient} is a good mechanism to reduce I/O traffic introduced by the Intel Optane SSD, at the cost of a small increase in tab switch latency at large tab counts. The Zswap cache reduces system energy by 2$\times$ (compared to the Intel Optane SSDc3 without Zswap enabled), at the cost of increasing the high-latency tab switches by 4\% and reducing the number of open tabs by 12\%.

\item \textit{Effect of using different NVM devices.} A state-of-the-art NAND-flash-based SSD provides benefits over both the baseline and the Intel Optane SSD. Importantly, it enables more Chrome tabs to be open. These benefits come due to the larger effective main memory capacity provided by the state-of-the-art NAND-flash-based SSD over the baseline configuration. Unfortunately, these benefits come at the cost of higher tab switch latencies, compared to both the baseline and Optane configurations, due to the much longer device latencies of NAND flash memory. These large tab switch latencies degrade user experience. Taking both performance and user experience into account, emerging NVM-based SSDs such as the Intel Optane SSD are quite promising to employ in consumer devices, providing performance benefits without the undesirable user experience trade-offs incurred by NAND-flash-based SSDs.

\item \textit{System bottlenecks caused by NVMs.} The Linux block I/O layer~\cite{bovet2005understanding,chen2012energy} is a key system bottleneck when the Intel Optane SSD is used as swap space. We can mitigate some of the overheads caused by the block I/O layer by \li~employing an I/O scheduler that meets the requirements of the application's access pattern and 
\lii~using different I/O request completion mechanisms. 
    
\item \textit{Optimization 1: block I/O schedulers.} We can reduce tab switch latency by changing the default budget-fair queuing (BFQ)~\cite{bovet2005understanding} I/O scheduler in the system that uses the Intel Optane SSD as swap space. We reduce 95th- and 99th-percentile latencies by employing the None~\cite{bjorling2013linux} and the Kyber~\cite{blkmqKyb12} I/O schedulers, respectively, as those I/O schedulers reduce I/O scheduling overheads and fit the I/O access pattern of the Chrome web browser.
    
\item \textit{Optimization 2: interrupt- vs.\ polling-based I/O request completion.} On average, the interrupt-based I/O request completion mechanism provides the best performance for the system with the Intel Optane SSD device. However, the Hybrid I/O request completion mechanism can help reduce 99th-percentile latency for block I/O requests.
\end{enumerate}

Based on our analysis, we conclude that there is a large optimization space to be explore in order to \emph{efficiently} adopt emerging NVMs in consumer devices. For example, we believe that one of the main issues the system suffers from when executing interactive workloads is that scheduling decisions made by the OS do not consider the response time expected by the workload. Exposing such information to the OS could reduce tail latency and allow the scheduler to take action according to the needs of a particular workload (e.g., by prioritizing the workload with the shorter or more urgent response deadlines).  We leave the design, implementation, and evaluation of such ideas for future work.

\subsection{Surveys \& Summary Works on \gls{PIM}}

\paratitle{Accelerating Neural Network Inference With Processing-in-DRAM: From the Edge to the Cloud}  In our IEEE Micro paper~\cite{oliveira2022accelerating}, we conduct a comprehensive analysis of three state-of-the-art DRAM-based \gls{PIM} architectures, i.e., UPMEM PIM~\cite{upmem,upmem2018}, Mensa~\cite{boroumand2021google}, and SIMDRAM~\cite{hajinazarsimdram}, to evaluate their effectiveness in accelerating \gls{NN} inference workloads. 
We show that traditional compute-centric accelerators often suffer from high under-utilization and energy inefficiency for memory-bound \gls{NN} models. 
In contrast, the three evaluated PIM architectures can alleviate memory bottlenecks and significantly improve both performance and energy efficiency. We make three key observations.
First, the UPMEM \gls{PIM} architecture can outperform high-end GPUs under memory oversubscription for \gls{GEMV} workloads. Second, Mensa achieves higher throughput and lower energy usage than the Google Edge TPU by specializing accelerators for different \gls{NN} layer types. 
Third, SIMDRAM can significantly accelerate binary \glspl{NN}, outperforming CPUs and GPUs while maintaining compatibility with standard DRAM infrastructure. 
We conclude that no single PIM architecture fits all NN models, highlighting the need for heterogeneous \gls{PIM} systems with unified programming models to fully harness the benefits of multiple \gls{PIM} approaches.

\paratitle{Methodologies, Workloads, and Tools for Processing-in-Memory: Enabling the Adoption of Data-Centric Architectures} In our ISVLSI paper~\cite{oliveira2022methodologies}, we contextualize our DAMOV methodology~\cite{deoliveira2021IEEE} and SIMDRAM framework~\cite{hajinazarsimdram} as means to facilitate the adoption of \gls{PIM} architectures. Together, DAMOV and SIMDRAM provide practical tools and system-level support to guide the design, programming, and integration of \gls{PIM} systems into modern computing platforms.

\paratitle{Heterogeneous Data-Centric Architectures for Modern Data-Intensive Applications: Case Studies in Machine Learning and Databases}
In our second ISVLSI paper~\cite{oliveira2022heterogeneous}, we demonstrate the performance and energy benefits that can be achieved through hardware/software co-design tailored for \gls{PIM} architectures. Using two case studies, i.e., Mensa~\cite{boroumand2021google}, a composable framework for accelerating machine learning inference on edge devices, and Polynesia~\cite{boroumand2022polynesia}, a specialized system for hybrid transactional/analytical database processing, we show that aligning architectural design with application-specific characteristics enables efficient use of limited memory-side compute resources. Both systems achieve significant improvements over state-of-the-art platforms by customizing hardware accelerators and execution strategies to the underlying application behavior, highlighting co-design as a key enabler for realizing the full potential of memory-centric computing.

\section{\omt{2}{Other Works the Author Contributed to as Co-Author}}

{Besides the works presented in this dissertation, I had the opportunity to contribute on several different areas during my doctoral studies in collaboration with researchers from ETH Zürich, Carnegie Mellon University, University of Illinois Urbana-Champaign, Galicia Supercomputing Center, University of Toronto, Barcelona Supercomputing Center, TOBB University of Economics and Technology, Intel, NVIDIA, and Google}. In this chapter, I acknowledge these works in six categories. 

\paratitle{New \gls{PIM} Architectures}
In collaboration with Amirali Boroumand, we design:
\li~Mensa~\cite{boroumand2021google}, a new acceleration framework for edge \gls{NN} inference models that incorporates multiple heterogeneous edge \gls{ML} accelerators (including both on-chip and near-data accelerators), each of which caters to the characteristics of a particular subset of NN models and layers;
\lii~Polynesia~\cite{boroumand2022polynesia}, a hardware–software co-designed
system for in-memory hybrid transactional and
analytical processing (HTAP) databases that avoids the large throughput losses of traditional HTAP systems.

In collaboration with Gagandeep Singh, we design NAPEL~\cite{singh2019napel}, a high-level performance and energy estimation framework for \gls{PIM} architectures.
In collaboration with Nika Mansouri Ghiasi, we design ALP~\cite{ghiasi2022alp}, a new programmer-transparent technique to leverage the performance benefits of \gls{PIM} by alleviating the performance impact of inter-segment data movement between host and memory and enabling efficient partitioning of applications between host and \gls{PIM} cores.
In collaboration with João Dinis Ferreira, we design pLUTo~\cite{ferreira2022pluto}, a DRAM-based \gls{PuM} architecture that leverages the high storage density of DRAM to enable the massively parallel storing and querying of lookup tables.
In collaboration with Jisung Park, we design Flash-Cosmos~\cite{flashcosmos}, a new in-flash processing technique that significantly increases the performance and energy
efficiency of bulk bitwise operations while providing high reliability.
In collaboration with Alain Denzler, we design Casper~\cite{denzler2021casper}, a near-cache accelerator consisting of specialized stencil computation units connected to the
\gls{LLC} of a traditional CPU.

\paratitle{Understanding \& Exploiting Real-World \gls{PIM} Architectures}
In collaboration with Juan Gómez-Luna, we
\li~conduct an experimental characterization of the UPMEM-based PIM system using microbenchmarks to assess various architecture
limits such as compute throughput and memory bandwidth, yielding new insights, and present
PrIM (Processing-In-Memory benchmarks), a benchmark suite of 16 workloads from different application domains designed for the UPMEM-based PIM system~\cite{gomezluna2021benchmarking}; and 
\lii~study the potential of general-purpose PIM
architectures to accelerate ML training~\cite{gomez2023evaluating,gomezluna2022isvlsi}.

In collaboration with Maurus Item, we design TransPimLib~\cite{item2023transpimlib}, a library that provides CORDIC-based and LUT-based methods for transcendental (and other hard-to-calculate) functions in general-purpose PIM systems.

\paratitle{\gls{PIM} on \gls{COTS} DRAM Chips}
In collaboration with Ismail Emir Yuksel, we experimentally demonstrate that \gls{COTS} DRAM chips are capable of
\li~performing functionally-complete bulk-bitwise Boolean operations: \texttt{NOT}, \texttt{NAND}, and \texttt{NOR}~\cite{missingnot},
\lii~executing up to 16-input \texttt{AND}, \texttt{NAND}, \texttt{OR}, and \texttt{NOR} operations~\cite{missingnot}, and 
\liii~copying the contents of a DRAM row (concurrently) into up to 31 other DRAM rows wit $>$99.98\% reliability~\cite{yuksel2024simultaneous}.

\paratitle{Novel DRAM Architectures}
In collaboration with Ataberk Olgun, we design Sectored  DRAM~\cite{olgun2022sectored}, a new, low-overhead DRAM substrate that reduces wasted energy by enabling fine-grained DRAM data transfer and DRAM row activation.

\paratitle{DRAM Reliability \& Read Disturbance}
In collaboration with Minesh Patel, we design HARP~\cite{patel2021harp}, a new error profiling algorithm that
rapidly achieves full coverage of at-risk bits.
In collaboration with Abdullah Giray Ya\u{g}l{\i}k\c{c}{\i}, we 
\li~experimentally demonstrate on 272
real DRAM chips that lowering DRAM's wordline voltage reduces a DRAM chip's RowHammer vulnerability~\cite{yauglikcci2022understanding}; and
\lii~design Svärd~\cite{yaglikci2024spatial}, a new RowHammer mitigation mechanism that dynamically adapts the aggressiveness of existing solutions based on the row-level read disturbance profile.  

In collaboration with Ataberk Olgun, we \li~design ABACuS~\cite{olgun2024abacus}, a new low-cost hardware-counter-based RowHammer mitigation technique that performance-,
energy-, and area-efficiently scales with worsening RowHammer vulnerability; and
\lii~experimentally demonstrate for the first time that the read disturbance threshold
of a DRAM row significantly and unpredictably changes over time.

In collaboration with O{\u{g}}uzhan Canpolat, we design Chronus~\cite{canpolat2025chronus}, a new on-DRAM-die RowHammer mitigation mechanism that 
\li~updates row activation counters concurrently while serving accesses by separating counters from the data and
\lii~prevents the wave attack by dynamically controlling the number of preventive refreshes performed.

\paratitle{Virtual Memory}
In collaboration with Nastaran Hajinazar, we design VBI~\cite{hajinazar2020virtual}, a new virtual memory framework that delegates memory management duties to hardware to reduce the overheads and software complexity associated with virtual memory.

\chapter{Complete List of the Author's Contributions}
\label{appendix:complete_list}

This section lists the author's contributions to the literature in reverse chronological order under two categories:
\li~major contributions that the author led (\secref{appendix:sec:first_author_contributions}),  
and~
\lii~other contributions (\secref{appendix:sec:other_contributions}).

\section{Major Contributions Led by the Author}
\label{appendix:sec:first_author_contributions}

\begin{enumerate}
    \item Geraldo F. Oliveira, Mayank Kabra, Yuxin Guo, Kangqi Chen, A. Giray Ya\u{g}l{\i}k\c{c}{\i}, Melina Soysal, Mohammad Sadrosadati, Joaquin Olivares Bueno, Saugata Ghose, Juan Gómez-Luna, Onur Mutlu, ``\emph{Proteus: Achieving High-Performance Processing-Using-DRAM with Dynamic Bit-Precision, Adaptive Data Representation, and Flexible Arithmetic,}'' in ICS, 2025.

    \item Geraldo F. Oliveira, Ataberk Olgun, Abdullah Giray Ya\u{g}l{\i}k\c{c}{\i}, F. Nisa Bostanci, Juan Gómez-Luna, Saugata Ghose, and Onur Mutlu, ``\emph{MIMDRAM: An End-to-End Processing-Using-DRAM System for High-Throughput, Energy-Efficient and Programmer-Transparent Multiple-Instruction Multiple-Data Processing,}'' in HPCA, 2024.

    \item Geraldo F. Oliveira, Saugata Ghose, Juan Gómez-Luna, Amirali Boroumand, Alexis Savery, Sonny Rao, Salman Qazi, Gwendal Grignou, Rahul Thakur, Eric Shiu, and Onur Mutlu, ``\emph{Extending Memory Capacity in Modern Consumer Systems With Emerging Non-Volatile Memory: Experimental Analysis and Characterization Using the Intel Optane SSD.}'' IEEE Access, September 2023.

    \item Geraldo F. Oliveira, Juan Gómez-Luna, Saugata Ghose, Amirali Boroumand, and Onur Mutlu, ``\emph{Accelerating Neural Network Inference With Processing-in-DRAM: From the Edge to the Cloud.}'' IEEE Micro, 2022.

    \item Geraldo F. Oliveira, Amirali Boroumand, Saugata Ghose, Juan Gómez-Luna, Onur Mutlu, ``\emph{Heterogeneous Data-Centric Architectures for Modern Data-Intensive Applications: Case Studies in Machine Learning and Databases,}'' in IVLSI, 2022. 

    \item Geraldo F. Oliveira, Juan Gómez-Luna, Saugata Ghose, Onur Mutlu, ``\emph{Methodologies, Workloads, and Tools for Processing-in-Memory: Enabling the Adoption of Data-Centric Architectures,}'' in IVLSI, 2022. 

    \item Geraldo F. Oliveira, Nastaran Hajinazar, Sven Gregorio, Joao Dinis Ferreira, Nika Mansouri Ghiasi, Minesh Patel, Mohammed Alser, Saugata Ghose, Juan Gomez-Luna, and Onur Mutlu, ``\emph{SIMDRAM: An End-to-End Framework for Bit-Serial SIMD Computing in DRAM,}'' in ASPLOS, 2021.

    \item Geraldo F. Oliveira, Juan Gómez-Luna, Lois Orosa, Saugata Ghose, Nandita Vijaykumar, Ivan fernandez, Mohammad Sadrosadati, and Onur Mutlu, ``\emph{DAMOV: A New Methodology and Benchmark Suite for Evaluating Data Movement Bottlenecks.}'' IEEE Access, 2021.

    \item Geraldo F. Oliveira, Larissa Rozales Gonçalves, Marcelo Brandalero, Antonio Carlos S. Beck, and Luigi Carro, ``\emph{Employing Classification-based Algorithms for General-Purpose Approximate Computing,}'' in DAC, 2019.

    \item Geraldo F. Oliveira, Paulo C. Santos, Marco A. Z. Alves , and Luigi Carro, ``\emph{A Generic Processing-in-Memory Cycle Accurate Simulator under Hybrid Memory Cube Architecture,}'' in SAMOS, 2017.

    \item Geraldo F. Oliveira, Paulo C. Santos, Marco A. Z. Alves, and Luigi Carro, ``\emph{NIM: An HMC-Based Machine for Neuron Computation},'' in ARC, 2017.
\end{enumerate}

\section{Other Contributions}
\label{appendix:sec:other_contributions}

\begin{enumerate}

\item Melina Soysal, Konstantina Koliogeorgi, Can Firtina, Nika Mansouri Ghiasi, Rakesh Nadig, Haiyu Mao, Geraldo F. Oliveira, Yu Liang, Klea Zambaku, Mohammad Sadrosadati, and Onur Mutlu, ``\emph{MARS: Processing-In-Memory Acceleration of Raw Signal Genome Analysis Inside the Storage Subsystem,}'' in ICS, 2025.

\item Oguzhan Canpolat, Abdullah Giray Ya\u{g}l{\i}k\c{c}{\i}, Geraldo Francisco de Oliveira, Ataberk Olgun, Nisa Bostanci, Ismail Emir Yuksel, Haocong Luo, Oguz Ergin, and Onur Mutlu, ``\emph{Chronus: Understanding and Securing the Cutting-Edge Industry Solutions to DRAM Read Disturbance,}'' in HPCA, 2025.

\item Ataberk Olgun, Nisa Bostanci, Ismail Emir Yuksel, Abdullah Giray Ya\u{g}l{\i}k\c{c}{\i}, Geraldo Francisco de Oliveira, Haocong Luo, Oguzhan Canpolat, Minesh Patel, and Onur Mutlu,
``\emph{Variable Read Disturbance: An Experimental Analysis of Temporal Variation in DRAM Read Disturbance,}'' in HPCA, 2025.

\item Ataberk Olgun, Yahya Can Tugrul, Nisa Bostanci, Ismail Emir Yuksel, Haocong Luo, Steve Rhyner, Abdullah Giray Ya\u{g}l{\i}k\c{c}{\i}, Geraldo F. Oliveira, and Onur Mutlu, ``\emph{ABACuS: All-Bank Activation Counters for Scalable and Low Overhead RowHammer Mitigation},'' in USENIX Security, 2024.

\item Abdullah Giray Ya\u{g}l{\i}k\c{c}{\i}, Geraldo F.  Oliveira, Yahya Can Tugrul, Ismail Yuksel, Ataberk Olgun, Haocong Luo, and Onur Mutlu, ``\emph{Spatial Variation-Aware Read Disturbance Defenses: Experimental Analysis of Real DRAM Chips and Implications on Future Solutions,}'' in HPCA, 2024.

\item Ismail Emir Yuksel, Yahya Can Tugrul, Ataberk Olgun, F. Nisa Bostanci, A. Giray Ya\u{g}l{\i}k\c{c}{\i}, Geraldo F. Oliveira, Haocong Luo, Juan Gomez-Luna, Mohammad Sadrosadati, and Onur Mutlu, ``\emph{Functionally-Complete Boolean Logic in Real DRAM Chips: Experimental Characterization and Analysis,}'' in HPCA, 2024.

\item Ismail Emir Yuksel, Yahya Can Tugrul, F. Nisa Bostanci, Geraldo F. Oliveira, A. Giray Ya\u{g}l{\i}k\c{c}{\i}, Ataberk Olgun, Melina Soysal, Haocong Luo, Juan Gomez-Luna, Mohammad Sadrosadati, and Onur Mutlu, ``\emph{Simultaneous Many-Row Activation in Off-the-Shelf DRAM Chips: Experimental Characterization and Analysis,}'' in DSN, 2024.

\item Onur Mutlu, Ataberk Olgun, İsmail Emir Yüksel, and Geraldo F. Oliveira, ``\emph{Memory-Centric Computing: Recent Advances in Processing-in-DRAM},'' Invited Paper in IEDM, 2024.

\item Alain Denzler, Rahul Bera, Nastaran Hajinazar, Gagandeep Singh, Geraldo F. Oliveira, Juan Gómez-Luna, and Onur Mutlu,
``\emph{Casper: Accelerating Stencil Computation using Near-Cache Processing.}'' IEEE Access, 2023.

\item Juan Gómez Luna, Yuxin Guo, Sylvan Brocard, Julien Legriel, Remy Cimadomo, Geraldo F. Oliveira, Gagandeep Singh, and Onur Mutlu, ``\emph{Evaluating Machine Learning Workloads on Memory-Centric Computing Systems,}'' in ISPASS, 2023. 

\item Maurus Item, Juan Gómez Luna, Yuxin Guo, Geraldo F. Oliveira, Mohammad Sadrosadati, and Onur Mutlu, ``\emph{TransPimLib: Efficient Transcendental Functions for Processing-in-Memory Systems},'' in ISPASS, 2023. 

\item A. Giray Ya\u{g}l{\i}k\c{c}{\i}, Haocong Luo, Geraldo F. de Oliveira, Ataberk Olgun, Minesh Patel, Jisung Park, Hasan Hassan, Jeremie S. Kim, Lois Orosa, and Onur Mutlu,
``\emph{Understanding RowHammer Under Reduced Wordline Voltage: An Experimental Study Using Real DRAM Devices,}'' in DSN, 2022.

\item Jisung Park, Roknoddin Azizi, Geraldo F. Oliveira, Mohammad Sadrosadati, Rakesh Nadig, David Novo, Juan Gómez-Luna, Myungsuk Kim, and Onur Mutlu, ``\emph{Flash-Cosmos: In-Flash Bulk Bitwise Operations Using Inherent Computation Capability of NAND Flash Memory,}'' in MICRO, 2022.

\item João Dinis Ferreira, Gabriel Falcao, Juan Gómez-Luna, Mohammed Alser, Lois Orosa, Mohammad Sadrosadati, Jeremie S. Kim, Geraldo F. Oliveira, Taha Shahroodi, Anant Nori, and Onur Mutlu, ``\emph{pLUTo: Enabling Massively Parallel Computation in DRAM via Lookup Tables,}'' in MICRO, 2022.

\item Nika Mansouri Ghiasi, Nandita Vijaykumar, Geraldo F. Oliveira, Lois Orosa, Ivan Fernandez, Mohammad Sadrosadati, Konstantinos Kanellopoulos, Nastaran Hajinazar, Juan Gómez Luna, and Onur Mutlu, ``\emph{ALP: Alleviating CPU-Memory Data Movement Overheads in Memory-Centric Systems.}'' IEEE Transactions on Emerging Topics in Computing (TETC), 2022.

\item Juan Gomez-Luna, Izzat El Hajj, Ivan Fernandez, Christina Giannoula, Geraldo F. Oliveira, and Onur Mutlu, ``\emph{Benchmarking a New Paradigm: Experimental Analysis and Characterization of a Real Processing-in-Memory System.}'' IEEE Access, 2022.

\item Juan Gómez-Luna, Yuxin Guo, Sylvan Brocard, Julien Legriel, Remy Cimadomo, Geraldo F. Oliveira, Gagandeep Singh, Onur Mutlu, ``\emph{Machine Learning Training on a Real Processing-in-Memory System,}'' in ISVLSI, 2022. 

\item Amirali Boroumand, Saugata Ghose, Geraldo F. Oliveira, and Onur Mutlu, ``\emph{Polynesia: Enabling High-Performance and Energy-Efficient Hybrid Transactional/Analytical Databases with Hardware/Software Co-Design,}'' in ICDE, 2022.

\item Minesh Patel, Geraldo F. de Oliveira, and Onur Mutlu, ``\emph{HARP: Practically and Effectively Identifying Uncorrectable Errors in Memory Chips That Use On-Die Error-Correcting Codes,}'' in MICRO, 2021.

\item Amirali Boroumand, Saugata Ghose, Berkin Akin, Ravi Narayanaswami, Geraldo F. Oliveira, Xiaoyu Ma, Eric Shiu, and Onur Mutlu,
``\emph{Google Neural Network Models for Edge Devices: Analyzing and Mitigating Machine Learning Inference Bottlenecks,}'' in PACT, 2021.

\item Nastaran Hajinazar, Pratyush Patel, Minesh Patel, Konstantinos Kanellopoulos, Saugata Ghose, Rachata Ausavarungnirun, Geraldo F. Oliveira, Jonathan Appavoo, Vivek Seshadri, and Onur Mutlu,
``\emph{The Virtual Block Interface: A Flexible Alternative to the Conventional Virtual Memory Framework},'' in ISCA, 2020.

\item Gagandeep Singh, Juan Gomez-Luna, Giovanni Mariani, Geraldo F. Oliveira, Stefano Corda, Sander Stujik, Onur Mutlu, and Henk Corporaal, ``\emph{NAPEL: Near-Memory Computing Application Performance Prediction via Ensemble Learning},'' in DAC, 2019.

\item Michael G Jordan, Marcelo Brandalero, Guilherme M Malfatti, Geraldo F. Oliveira, Arthur F Lorenzon, Bruno C da Silva, Luigi Carro, Mateus B Rutzig, and Antonio C. S. Beck, ``\emph{Data Clustering for Efficient Approximate Computing}.'' Design Automation for Embedded Systems, 2019.

\item Marcelo Brandalero, Guilherme Meneguzzi Malfatti, Geraldo F. Oliveira, Leonardo Almeida Da Silveira, Larissa Rozales Gonçalves, Bruno Castro Da Silva, Luigi Carro, and Antonio C. S. Beck, 
``\emph{Efficient Local Memory Support for Approximate Computing,}'' in SBESC, 2018.

\item Paulo C Santos, Geraldo F. Oliveira, João P. Lima, Marco AZ Alves, Luigi Carro, and Antonio C. S. Beck, ``\emph{Processing in 3D Memories to Speed Up Operations on Complex Data Structures,}'' in DATE, 2018.

\item Paulo C Santos, Geraldo F. Oliveira, Diego G Tomé, Marco AZ Alves, Eduardo C Almeida, and Luigi Carro, ``\emph{Operand Size Reconfiguration for Big Data Processing-in-Memory,}'' in DATE, 2017.

\end{enumerate}
\chapter{Complete List of Application Functions, Representative Functions, and Evaluated Applications in DAMOV}

\section{Application Functions in the DAMOV Benchmark Suite}
\label{sec:benchlist}
\geraldorevi{\gfiii{W}e present the list of application functions in each one of the six classes of data movement bottlenecks we identify using our new methodology.}

\geraldorevi{Our benchmark suite is composed of 144 different application functions, collected from 74 different applications. These applications belong to a different set of previously published and widely used benchmark suites. In total, \gfiii{we} collect applications from 16 benchmark suites, including: BWA~\gfiii{\cite{li2009fast}}, Chai~\gfiii{\cite{gomezluna_ispass2017}}, Darknet~\gfiii{\cite{redmon_darknet2013}}, GASE~\gfiii{\cite{ahmed2016comparison}}, Hardware Effects~\gfiii{\cite{hardwareeffects}}, Hashjoin~\gfiii{\cite{balkesen_TKDE2015}}, HPCC~\gfiii{\cite{luszczek_hpcc2006}}, HPCG~\gfiii{\cite{dongarra_hpcg2015}}, Ligra~\gfiii{\cite{shun_ppopp2013}}, PARSEC~\gfiii{\cite{bienia2008parsec}}, Parboil~\gfiii{\cite{stratton2012parboil}}, PolyBench~\gfiii{\cite{pouchet2012polybench}}, Phoenix~\gfiii{\cite{yoo_iiswc2009}}, Rodinia~\gfiii{\cite{che_iiswc2009}}, SPLASH-2~\gfiii{\cite{woo_isca1995}}, STREAM~\gfiii{\cite{mccalpin_stream1995}}. The 144 application functions that are part of \bench are listed across six tables\gfiii{, each designating one of the six classes we identify in Section~\ref{sec:scalability}}:}

\begin{itemize}
    \item Table~\ref{table_1a} lists application functions \gfiv{in Class~1a, i.e.,} that are \gfiii{DRAM} bandwidth\gfiii{-}bound (characterized in Section~\ref{sec_scalability_class1a});
    
    \item Table~\ref{table_1b} lists application functions \gfiv{in Class~1b, i.e.,} that are \gfiii{DRAM} latency\gfiii{-}bound (characterized in Section~\ref{sec_scalability_class1b}); 
    
    \item Table~\ref{table_1c} lists application functions \gfiv{in Class~1c, i.e.,} that are bottlenecked by the available \gfiii{L1/L2} cache capacity (characterized in Section~\ref{sec_scalability_class1c}); 
    
    \item Table~\ref{table_2a} lists application functions \gfiv{in Class~2a, i.e.,} that are bottlenecked by \gfiii{L3} cache contention (characterized in Section~\ref{sec_scalability_class2a}); 
    
    \item Table~\ref{table_2b} lists application functions \gfiv{in Class~2b, i.e.,} that are bottlenecked by L1 cache size (characterized in Section~\ref{sec_scalability_class2b}); 
    
    \item Table~\ref{table_2c} lists application functions \gfiv{in Class~2c, i.e.,} that are compute\gfiii{-}bound (characterized in Section~\ref{sec_scalability_class2c}).
\end{itemize}

In each table we list the benchmark suite, the application name, and the function name. We also list the input size/problem size we use to evaluate each application function.% Next, we provide a short description of each application in our benchmark suite. When appropriate, we also provide a short description \gfiii{of} the different input sets used to evaluate an application.

%\vspace{-10pt}

\begin{table*}[t]
\tempcommand{1.2}
\centering
\caption{List of application functions in Class~1a.}
\label{table_1a}
\resizebox{\textwidth}{!}{%
\begin{tabular}{|c|c|c|c|c|c|}
\hline
\thead{\textbf{Class}} & \thead{\textbf{Suite}} & \thead{\textbf{Benchmark}} & \thead{\textbf{Function}} & \thead{\textbf{Input Set/ } \\ \textbf{Problem Size}} & \thead{\textbf{Representative}\\ \textbf{Function?}} \\ \hline \hline
1a & Chai~\cite{gomezluna_ispass2017} & Transpose & cpu & -m 1024 -n 524288  & No\\ \hline
1a & Chai~\cite{gomezluna_ispass2017} & Vector Pack & run\_cpu\_threads  & -m 268435456 -n 16777216 & No \\ \hline
1a & Chai~\cite{gomezluna_ispass2017} & Vector Unpack & run\_cpu\_threads  & -m 268435456 -n 16777216 & No \\ \hline
1a & Darknet~\cite{redmon_darknet2013} & Yolo & gemm  & ref & Yes \\ \hline
1a & Hardware Effects~\cite{hardwareeffects} & Bandwidth Saturation - Non Temporal & main  & ref &  No \\ \hline
1a & Hardware Effects~\cite{hardwareeffects} & Bandwidth Saturation - Temporal & main  & ref &  No \\ \hline
1a & Hashjoin~\cite{balkesen_TKDE2015} & NPO & knuth  & -r 12800000 -s 12000000 -x 12345 -y 54321 & No \\ \hline
1a & Hashjoin~\cite{balkesen_TKDE2015} & NPO & ProbeHashTable  & -r 12800000 -s 12000000 -x 12345 -y 54321 &Yes \\ \hline
1a & Hashjoin~\cite{balkesen_TKDE2015} & PRH & knuth  & -r 12800000 -s 12000000 -x 12345 -y 54321 & No \\ \hline
1a & Hashjoin~\cite{balkesen_TKDE2015} & PRH & lock  & -r 12800000 -s 12000000 -x 12345 -y 54321 & No\\ \hline
1a & Hashjoin~\cite{balkesen_TKDE2015} & PRHO & knuth  & -r 12800000 -s 12000000 -x 12345 -y 54321 & No\\ \hline
1a & Hashjoin~\cite{balkesen_TKDE2015} & PRHO & radix  & -r 12800000 -s 12000000 -x 12345 -y 54321 & No \\ \hline
1a & Hashjoin~\cite{balkesen_TKDE2015} & PRO & knuth  & -r 12800000 -s 12000000 -x 12345 -y 54321 & No \\ \hline
1a & Hashjoin~\cite{balkesen_TKDE2015} & PRO & parallel  & -r 12800000 -s 12000000 -x 12345 -y 54321 & No \\ \hline
1a & Hashjoin~\cite{balkesen_TKDE2015} & PRO & radix  & -r 12800000 -s 12000000 -x 12345 -y 54321& No \\ \hline
1a & Hashjoin~\cite{balkesen_TKDE2015} & RJ & knuth  & -r 12800000 -s 12000000 -x 12345 -y 54321 & No \\ \hline
1a & Ligra~\cite{shun_ppopp2013} & Betweenness Centrality & edgeMapSparse  & rMat & No \\ \hline
1a & Ligra~\cite{shun_ppopp2013} & Breadth-First Search & edgeMapSparse & rMat & No \\ \hline
1a & Ligra~\cite{shun_ppopp2013} & Connected Components & compute  & rMat & No \\ \hline
1a & Ligra~\cite{shun_ppopp2013} & Connected Components & compute  & USA & No \\ \hline
1a & Ligra~\cite{shun_ppopp2013} & Connected Components & edgeMapDense  & USA & No \\ \hline
1a & Ligra~\cite{shun_ppopp2013} & Connected Components & edgeMapSparse  & USA & Yes \\ \hline
1a & Ligra~\cite{shun_ppopp2013} & K-Core Decomposition & compute  & rMat & No \\ \hline
1a & Ligra~\cite{shun_ppopp2013} & K-Core Decomposition & compute & USA & No \\ \hline
1a & Ligra~\cite{shun_ppopp2013} & K-Core Decomposition & edgeMapDense & USA & No \\ \hline
1a & Ligra~\cite{shun_ppopp2013} & K-Core Decomposition & edgeMapSparse & rMat & No \\ \hline
1a & Ligra~\cite{shun_ppopp2013} & Maximal Independent Set & compute & rMat & No \\ \hline
1a & Ligra~\cite{shun_ppopp2013} & Maximal Independent Set & compute  & USA & No \\ \hline
1a & Ligra~\cite{shun_ppopp2013} & Maximal Independent Set & edgeMapDense  & USA & No \\ \hline
1a & Ligra~\cite{shun_ppopp2013} & Maximal Independent Set & edgeMapSparse  & rMat & No \\ \hline
1a & Ligra~\cite{shun_ppopp2013} & Maximal Independent Set & edgeMapSparse  & USA & No \\ \hline
1a & Ligra~\cite{shun_ppopp2013} & PageRank & compute & rMat & No \\ \hline
1a & Ligra~\cite{shun_ppopp2013} & PageRank & compute & USA & No \\ \hline
1a & Ligra~\cite{shun_ppopp2013} & PageRank & edgeMapDense  & USA & Yes \\ \hline
1a & Ligra~\cite{shun_ppopp2013} & Radii & compute  & rMat & No \\ \hline
1a & Ligra~\cite{shun_ppopp2013} & Radii & compute & USA & No \\ \hline
1a & Ligra~\cite{shun_ppopp2013} & Radii & edgeMapSparse  & USA & No \\ \hline
1a & Ligra~\cite{shun_ppopp2013} & Triangle Count & edgeMapDense  & rMat & Yes \\ \hline
1a & SPLASH-2~\cite{woo_isca1995} & Oceancp & relax  & simlarge & No \\ \hline
1a & SPLASH-2~\cite{woo_isca1995} & Oceanncp & relax  & simlarge & No \\ \hline
1a & STREAM~\cite{mccalpin_stream1995} & Add & Add  & 50000000 & Yes \\ \hline
1a & STREAM~\cite{mccalpin_stream1995} & Copy & Copy  & 50000000 & Yes \\ \hline
1a & STREAM~\cite{mccalpin_stream1995} & Scale & Scale  & 50000000 & Yes \\ \hline
1a & STREAM~\cite{mccalpin_stream1995} & Triad & Triad  &  50000000 &  Yes \\ \hline
\end{tabular}%
}
\end{table*}

\begin{table*}[t]
\tempcommand{1.2}
\centering
\caption{List of application functions in Class~1b.}
\label{table_1b}
\resizebox{\textwidth}{!}{%
\begin{tabular}{|c|c|c|c|c|c|}
\hline
\thead{\textbf{Class}} & \thead{\textbf{Suite}} & \thead{\textbf{Benchmark}} & \thead{\textbf{Function}} & \thead{\textbf{Input Set/ } \\ \textbf{Problem Size}} & \thead{\textbf{Representative}\\ \textbf{Function?}} \\ \hline \hline
1b & Chai~\cite{gomezluna_ispass2017} & Canny Edge Detection & gaussian & ref & No \\ \hline
1b & Chai~\cite{gomezluna_ispass2017} & Canny Edge Detection & supression & ref & No \\ \hline
1b & Chai~\cite{gomezluna_ispass2017} & Histogram - input partition & run\_cpu\_threads & ref & Yes \\ \hline
1b & Chai~\cite{gomezluna_ispass2017} & Select & run\_cpu\_threads & -n 67108864 & No \\ \hline
1b & GASE~\cite{ahmed2016comparison} & FastMap & 2occ4 & Wg2 & No \\ \hline
1b & GASE~\cite{ahmed2016comparison} & FastMap & occ4 & Wg2 & No \\ \hline
1b & Hashjoin~\cite{balkesen_TKDE2015} & PRH & HistogramJoin & -r 12800000 -s 12000000 -x 12345 -y 54321 & Yes \\ \hline
1b & Phoenix~\cite{yoo_iiswc2009} & Linear Regression & linear\_regression\_map & key\_file\_500MB & No \\ \hline
1b & Phoenix~\cite{yoo_iiswc2009}  & PCA & main & ref & No \\ \hline
1b & Phoenix~\cite{yoo_iiswc2009}  & String Match & string\_match\_map & key\_file\_500MB & Yes \\ \hline
1b & PolyBench~\cite{pouchet2012polybench} & linear-algebra & lu & LARGE\_DATASET & Yes \\ \hline
1b & Rodinia~\cite{che_iiswc2009} & Kmeans & euclidDist & 819200.txt & No \\ \hline
1b & Rodinia~\cite{che_iiswc2009} & Kmeans & find & 819200.txt & No \\ \hline
1b & Rodinia~\cite{che_iiswc2009} & Kmeans & main & 819200.txt & No \\ \hline
1b & Rodinia~\cite{che_iiswc2009} & Streamcluster & pengain & ref & No \\ \hline
1b & SPLASH-2~\cite{woo_isca1995} & Oceancp & slave2 & simlarge & Yes \\ \hline
\end{tabular}%
}
\end{table*}

\begin{table*}[t]
\tempcommand{1.2}
\centering
\caption{List of application functions in Class~1c.}
\label{table_1c}
\resizebox{\textwidth}{!}{%
\begin{tabular}{|c|c|c|c|c|c|}
\hline
\thead{\textbf{Class}} & \thead{\textbf{Suite}} & \thead{\textbf{Benchmark}} & \thead{\textbf{Function}} & \thead{\textbf{Input Set/ } \\ \textbf{Problem Size}} & \thead{\textbf{Representative}\\ \textbf{Function?}} \\ \hline \hline
1c & BWA~\cite{li2009fast} & Align & bwa\_aln\_core & Wg1 & No \\ \hline
1c & Chai~\cite{gomezluna_ispass2017} & Breadth-First Search & comp  & USA-road-d & No \\ \hline
1c & Chai~\cite{gomezluna_ispass2017} & Breadth-First Search & fetch  & USA-road-d & No \\ \hline
1c & Chai~\cite{gomezluna_ispass2017} & Breadth-First Search & load  & USA-road-d & No \\ \hline
1c & Chai~\cite{gomezluna_ispass2017} & Breadth-First Search & run\_cpu\_threads  & USA-road-d & No \\ \hline
1c & Chai~\cite{gomezluna_ispass2017} & Canny Edge Detection & hystresis  & ref & No \\ \hline
1c & Chai~\cite{gomezluna_ispass2017} & Canny Edge Detection & sobel  & ref & No \\ \hline
1c & Chai~\cite{gomezluna_ispass2017} & Histogram - output partition & run\_cpu\_threads  & ref & No \\ \hline
1c & Chai~\cite{gomezluna_ispass2017} & Padding & run\_cpu\_threads  & -m 10000 -n 9999 & Yes \\ \hline
1c & Chai~\cite{gomezluna_ispass2017} & Select & fetch  & -n 67108864 & No \\ \hline
1c & Chai~\cite{gomezluna_ispass2017} & Stream Compaction & run\_cpu\_threads  & ref & No \\ \hline
1c & Darknet~\cite{redmon_darknet2013} & Resnet & gemm  & ref & Yes \\ \hline
1c & Hashjoin~\cite{balkesen_TKDE2015} & NPO & lock  & -r 12800000 -s 12000000 -x 12345 -y 54321 & No \\ \hline
1c & Ligra~\cite{shun_ppopp2013} & BFS-Connected Components & edgeMapSparse  & rMat & No \\ \hline
1c & Ligra~\cite{shun_ppopp2013} & Triangle Count & compute & rMat & No \\ \hline
1c & Ligra~\cite{shun_ppopp2013} & Triangle Count & compute & USA & No \\ \hline
1c & Ligra~\cite{shun_ppopp2013} & Triangle Count & edgeMapDense  & USA & No \\ \hline
1c & PARSEC~\cite{bienia2008parsec} & Blackscholes & BlkSchlsEqEuroNoDiv  & simlarge & No \\ \hline
1c & PARSEC~\cite{bienia2008parsec} & Fluidaminate & ProcessCollision2MT  & simlarge & Yes \\ \hline
1c & PARSEC~\cite{bienia2008parsec} & Streamcluster & DistL2Float  & simlarge & No \\ \hline
1c & Rodinia~\cite{che_iiswc2009} & Myocyte & find  & 1000000 & No \\ \hline
1c & Rodinia~\cite{che_iiswc2009} & Myocyte & master  & 1000000 &  No \\ \hline
\end{tabular}%
}
\end{table*}

\begin{table*}[t]
\tempcommand{1.2}
\centering
\caption{List of application functions in Class~2a.}
\label{table_2a}
\resizebox{\textwidth}{!}{%
\begin{tabular}{|c|c|c|c|c|c|}
\hline
\thead{\textbf{Class}} & \thead{\textbf{Suite}} & \thead{\textbf{Benchmark}} & \thead{\textbf{Function}} & \thead{\textbf{Input Set/ } \\ \textbf{Problem Size}} & \thead{\textbf{Representative}\\ \textbf{Function?}} \\ \hline \hline
2a & HPCC~\cite{luszczek_hpcc2006} & RandomAccess & main & ref &  No \\ \hline
2a & HPCC~\cite{luszczek_hpcc2006} & RandomAccess & update & ref &  No \\ \hline
2a & Ligra~\cite{shun_ppopp2013} & Betweenness Centrality & Compute & rMat & No \\ \hline
2a & Ligra~\cite{shun_ppopp2013} & Betweenness Centrality & Compute  & USA & No \\ \hline
2a & Ligra~\cite{shun_ppopp2013} & Betweenness Centrality & edgeMapDense  & rMat & No \\ \hline
2a & Ligra~\cite{shun_ppopp2013} & Betweenness Centrality & *edgeMapSparse  &  USA & Yes \\ \hline
2a & Ligra~\cite{shun_ppopp2013} & BFS-Connected Components & Compute  & rMat & No \\ \hline
2a & Ligra~\cite{shun_ppopp2013} & BFS-Connected Components & Compute & USA & No \\ \hline
2a & Ligra~\cite{shun_ppopp2013} & BFS-Connected Components & edgeMapSparse  & USA & Yes \\ \hline
2a & Ligra~\cite{shun_ppopp2013} & Breadth-First Search & compute  & rMat & No \\ \hline
2a & Ligra~\cite{shun_ppopp2013} & Breadth-First Search & compute & USA & No  \\ \hline
2a & Ligra~\cite{shun_ppopp2013} & Breadth-First Search & edgeMapDense  & rMat & No \\ \hline
2a & Ligra~\cite{shun_ppopp2013} & Breadth-First Search & edgeMapSparse  & USA & Yes \\ \hline
2a & Ligra~\cite{shun_ppopp2013} & Connected Components & edgeMapDense  & rMat  & No \\ \hline
2a & Ligra~\cite{shun_ppopp2013} & Maximal Independent Set & edgeMapDense & rMat & No \\ \hline
2a & Ligra~\cite{shun_ppopp2013} & PageRank & edgeMapDense(Rmat)  & rMat & No \\ \hline
2a & Phoenix~\cite{yoo_iiswc2009}  & WordCount & main & word\_100MB & No \\ \hline
2a & PolyBench~\cite{pouchet2012polybench} & linear-algebra & gramschmidt  & LARGE\_DATASET & Yes \\ \hline
2a & Rodinia~\cite{che_iiswc2009} & CFD Solver & main  & fvcorr.domn.193K  & No \\ \hline
2a & SPLASH-2~\cite{woo_isca1995} & FFT2 & Reverse  & simlarge & Yes \\ \hline
2a & SPLASH-2~\cite{woo_isca1995} & FFT2 & Transpose  & simlarge & Yes \\ \hline
2a & SPLASH-2~\cite{woo_isca1995} & Oceancp & jacobcalc  &  simlarge & No \\ \hline
2a & SPLASH-2~\cite{woo_isca1995} & Oceancp & laplaccalc  & simlarge  & No \\ \hline
2a & SPLASH-2~\cite{woo_isca1995} & Oceanncp & jacobcalc  & simlarge &  Yes \\ \hline
2a & SPLASH-2~\cite{woo_isca1995} & Oceanncp & laplaccalc  & simlarge  & Yes \\ \hline
2a & SPLASH-2~\cite{woo_isca1995} & Oceanncp & slave2  & simlarge & No \\ \hline
\end{tabular}%
}
\end{table*}

\begin{table*}[t]
\tempcommand{1.2}

\centering
\caption{List of application functions in Class~2b.}
\label{table_2b}
\resizebox{\textwidth}{!}{%
\begin{tabular}{|c|c|c|c|c|c|}
\hline
\thead{\textbf{Class}} & \thead{\textbf{Suite}} & \thead{\textbf{Benchmark}} & \thead{\textbf{Function}} & \thead{\textbf{Input Set/ } \\ \textbf{Problem Size}} & \thead{\textbf{Representative}\\ \textbf{Function?}} \\ \hline \hline
2b & Chai~\cite{gomezluna_ispass2017} & Bezier Surface & main\_thread  &  ref &  Yes \\ \hline
2b & Hardware Effects~\cite{hardwareeffects} & False Sharing - Isolated & main  & ref & No \\ \hline
2b & PolyBench~\cite{pouchet2012polybench} & convolution & convolution-2d  & LARGE\_DATASET & No \\ \hline
2b & PolyBench~\cite{pouchet2012polybench} & linear-algebra & gemver  & LARGE\_DATASET &  Yes \\ \hline
2b & SPLASH-2~\cite{woo_isca1995} & Lucb & Bmod  & simlarge &  Yes \\ \hline
2b & SPLASH-2~\cite{woo_isca1995} & Radix & slave2  & simlarge &  Yes \\ \hline
\end{tabular}%
}
\end{table*}

%\clearpage

\begin{table*}[!t]
\tempcommand{1.2}
%\centering
\caption{List of application functions in Class~2c.}
\label{table_2c}
\resizebox{\textwidth}{!}{%
\begin{tabular}{|c|c|c|c|c|c|}
\hline
\thead{\textbf{Class}} & \thead{\textbf{Suite}} & \thead{\textbf{Benchmark}} & \thead{\textbf{Function}} & \thead{\textbf{Input Set/ } \\ \textbf{Problem Size}} & \thead{\textbf{Representative}\\ \textbf{Function?}} \\ \hline \hline
2c & BWA~\cite{li2009fast} & Align & bwa\_aln\_core & Wg2 &  No \\ \hline
2c & Chai~\cite{gomezluna_ispass2017} & Transpose & run\_cpu\_threads  & -m 1024 -n 524288 &  No \\ \hline
2c & Darknet~\cite{redmon_darknet2013} & Alexnet & gemm  & ref & No\\ \hline
2c & Darknet~\cite{redmon_darknet2013} & vgg16 & gemm  & ref & No \\ \hline
2c & Hardware Effects~\cite{hardwareeffects} & False Sharing - Shared & main  & ref & No\\ \hline
2c & HPCG~\cite{dongarra_hpcg2015} & HPCG & ComputePrologation  & ref & Yes \\ \hline
2c & HPCG~\cite{dongarra_hpcg2015} & HPCG & ComputeRestriction  & ref & Yes \\ \hline
2c & HPCG~\cite{dongarra_hpcg2015} & HPCG & ComputeSPMV  & ref & Yes \\ \hline
2c & HPCG~\cite{dongarra_hpcg2015} & HPCG & ComputeSYMGS  & ref & Yes \\ \hline
2c & Ligra~\cite{shun_ppopp2013} & K-Core Decomposition & edgeMapDense & rMat &  No\\ \hline
2c & Ligra~\cite{shun_ppopp2013} & Radii & edgeMapSparse & rMat & No \\ \hline
2c & Parboil~\cite{stratton2012parboil} & Breadth-First Search & BFS\_CPU  & ref & No\\ \hline
2c & Parboil~\cite{stratton2012parboil}  & MRI-Gridding & CPU\_kernels  & ref & No\\ \hline
2c & Parboil~\cite{stratton2012parboil}  & Stencil & cpu\_stencil  & ref & No \\ \hline
2c & Parboil~\cite{stratton2012parboil}  & Two Point Angular Correlation Function & doCompute  & ref & No \\ \hline
2c & PARSEC~\cite{bienia2008parsec} & Bodytrack & FilterRow & ref & No \\ \hline
2c & PARSEC~\cite{bienia2008parsec} & Ferret & DistL2Float & ref & Yes \\ \hline
2c & Phoenix~\cite{yoo_iiswc2009}  & Kmeans & main & ref  & No \\ \hline
2c & PolyBench~\cite{pouchet2012polybench} & linear-algebra & 3mm & LARGE\_DATASET & Yes \\ \hline
2c & PolyBench~\cite{pouchet2012polybench} & linear-algebra & doitgen & LARGE\_DATASET & Yes \\ \hline
2c & PolyBench~\cite{pouchet2012polybench} & linear-algebra & gemm & LARGE\_DATASET  & Yes \\ \hline
2c & PolyBench~\cite{pouchet2012polybench} & linear-algebra & symm & LARGE\_DATASET  & Yes \\ \hline
2c & PolyBench~\cite{pouchet2012polybench} & stencil & fdtd-apml &  LARGE\_DATASET & Yes \\ \hline
2c & Rodinia~\cite{che_iiswc2009} & Back Propagation & adjustweights & 134217728 & No \\ \hline
2c & Rodinia~\cite{che_iiswc2009} & Back Propagation & layerfoward & 134217728 & No \\ \hline
2c & Rodinia~\cite{che_iiswc2009} & Breadth-First Search & main & graph1M\_6 & Yes \\ \hline
2c & Rodinia~\cite{che_iiswc2009} & Needleman-Wunsch & main & 32768 & Yes \\ \hline
2c & Rodinia~\cite{che_iiswc2009}a & Srad & FIN & ref &  No \\ \hline
2c & SPLASH-2~\cite{woo_isca1995} & Barnes & computeForces & simlarge & No \\ \hline
2c & SPLASH-2~\cite{woo_isca1995} & Barnes & gravsub & simlarge & No \\ \hline
\end{tabular}%
}
\vspace{15pt}
\end{table*}

%\cleardoublepage
%\cleardoublepage

\clearpage
\section{Representative Application Functions}

\vspace{10pt}
\begin{table}[!b]
\begin{minipage}{\textwidth}
%\vspace{-15pt}
\caption{\gfiv{44 r}epresentative \gfv{application} functions studied in detail in this work.}
\label{tab:benchmarks}
 \tempcommand{1.2}
\centering
\resizebox{0.84 \linewidth}{!}{
\begin{tabular}{llllll}
\toprule
\textbf{Suite} & \textbf{Benchmark} & \textbf{Function} & \textbf{Short Name} & \textbf{Class}  & \textbf{\% } \\
\midrule
\multirow{3}{*}{\parbox{1.5cm}{Chai \cite{gomezluna_ispass2017}}} & Bezier Surface & Bezier & CHABsBez & 2b & 100\\
                                                  & Histogram & Histogram & CHAHsti & 1b & 100\\
                                                  & Padding & Padding & CHAOpad & 1c & 75.1\\
\hline
\multirow{2}{*}{\parbox{1.7cm}{Darknet~\cite{redmon_darknet2013}}} & Resnet 152 & gemm\_nn & DRKRes & 1c & 95.2\\
                                                  & Yolo & gemm\_nn & DRKYolo & 1a & 97.1\\
\hline
\multirow{2}{*}{\parbox{1.9cm}{Hashjoin \cite{balkesen_TKDE2015}}} & NPO & ProbeHashTable & HSJNPO & 1a & 47.8 \\
                                                  & PRH & HistogramJoin & HSJPRH & 1b & 53.1\\
\hline
\multirow{4}{*}{\parbox{1.7cm}{HPCG~\cite{dongarra_hpcg2015}}} & HPCG & ComputeProlongation & HPGProl & 2c & 34.3\\
                                               & HPCG & ComputeRestriction & HPGRes & 2c & 42.1\\
                                               & HPCG & ComputeSPMV & HPGSpm & 2c & 30.5\\
                                               & HPCG & ComputeSYMGS & HPGSyms & 2c & 63.6\\
\hline
\multirow{4}{*}{Ligra \cite{shun_ppopp2013}} & Betweenness Centrality & EdgeMapSparse (USA~\cite{dimacs}) & LIGBcEms & 2a & 78.9\\
                                             & Breadth-First Search & EdgeMapSparse (USA) & LIGBfsEms & 2a & 67.0\\
                                             & BFS-Connected Components &  EdgeMapSparse (USA) & LIGBfscEms & 2a & 68.3\\
                                             & Connected Components & EdgeMapSparse (USA) & LIGCompEms & 1a & 25.6\\
                                             & PageRank & EdgeMapDense (USA~\cite{dimacs}) & LIGPrkEmd & 1a & 57.2 \\
                                             & Radii & EdgeMapSparse (USA)& LIGRadiEms & 1a & 67.0\\
                                             & Triangle & EdgeMapDense (Rmat) & LIGTriEmd & 1a & 26.7\\
\hline
\multirow{2}{*}{\parbox{1.7cm}{PARSEC~\cite{bienia2008parsec}}} & Ferret & DistL2Float & PRSFerr  & 2c & 18.6\\
                                               & Fluidaminate & ProcessCollision2MT & PRSFlu & 1c & 23.9 \\
\hline
\multirow{2}{*}{\parbox{1.7cm}{Phoenix~\cite{yoo_iiswc2009}}} & Linear Regression & linear\_regression\_map & PHELinReg & 1b& 76.2 \\
                                              & String Matching & string\_match\_map & PHEStrMat  & 1b & 38.3\\
\hline
\multirow{9}{*}{PolyBench \cite{pouchet2012polybench}} & Linear Algebra & 3 Matrix Multiplications & PLY3mm & 2c & 100.0\\
                                             & Linear Algebra  & Multi-resolution analysis kernel & PLYDoitgen & 2c & 98.3\\
                                             & Linear Algebra  &  Matrix-multiply C=alpha.A.B+beta.C & PLYgemm & 2c & 99.7\\
                                             & Linear Algebra  & Vector Mult. and Matrix Addition & PLYgemver & 2b & 44.4\\
                                             & Linear Algebra  & Gram-Schmidt decomposition & PLYGramSch & 2a & 100.0\\
                                             & Linear Algebra  & LU decomposition & PLYalu & 1b & 100.0\\
                                             & Linear Algebra & Symmetric matrix-multiply & PLYSymm & 2c & 99.9 \\
                                             & Stencil  & 2D Convolution & PLYcon2d & 2b & 100.0 \\
                                             & Stencil & 2-D Finite Different Time Domain & PLYdtd & 2c & 39.8\\
\hline
\multirow{2}{*}{\parbox{1.8cm}{Rodinia~\cite{chen2014big}}} & BFS & BFSGraph & RODBfs  & 2c & 100.0\\
                                           & Needleman-Wunsch & runTest & RODNw  & 2c & 84.9\\
\hline
\multirow{7}{*}{\parbox{1.9cm}{SPLASH-2~\cite{woo_isca1995}}} & FFT & Reverse & SPLFftRev & 2a &12.7 \\
                                              & FFT & Transpose & SPLFftTra & 2a & 8.0\\
                                              & Lucb & Bmod & SPLLucb & 2b & 77.6\\
                                              & Oceanncp & jacobcalc & SPLOcnpJac & 2a &30.7 \\
                                              & Oceanncp & laplaccalc & SPLOcnpLap & 2a & 23.4\\
                                              & Oceancp & slave2 & SPLOcpSlave & 1b & 24.4\\
                                              & Radix & slave\_sort & SPLRad & 2b & 41.1\\
\hline
\multirow{4}{*}{\parbox{1.7cm}{STREAM~\cite{mccalpin_stream1995}}} & Add & Add & STRAdd & 1a & 98.4 \\
                                                   & Copy & Copy & STRCpy & 1a & 98.3\\
                                                   & Scale & Scale & STRSca & 1a & 97.5\\
                                                   & Triad & Triad & STRTriad & 1a & 99.1 \\
\bottomrule
\end{tabular}
}
\newline
\scriptsize{$^*$ Short names are encoded as XXXYyyZzz, where XXX is the source application suite, Yyy is the application name, and Zzz is the function (if more than one per benchmark). For graph processing applications from Ligra, we test two different input graphs, so we append the graph name to the short benchmark name as well. The \% column indicates the percentage of clock cycles that the function consumes as a fraction of the execution time of the entire benchmark}.
%\vspace{-5pt}
\end{minipage}

\end{table}

\clearpage
\section{Complete List of Evaluated Applications}
\label{sec:evaluatedapp}
%We list in Table~\ref{tab:allapps} the 345 applications we evaluate during our benchmark development. 

\vspace{5pt}

\begin{table}[!b]
\begin{minipage}{0.89\textwidth}
\centering
\caption{List of \gfv{the} \gfiv{e}valuated \gfiii{345} \gfiv{a}pplications.}
\label{tab:allapps}
\tempcommand{0.905}
\resizebox{\linewidth}{!}{
\begin{tabular}{|l|l || l|l || l |l|}
%\begin{longtable}{|l|l || l|l || l |l|}
\hline
\textbf{Benchmark Suite} &	\textbf{Application} 	&	\textbf{Benchmark Suite} &	\textbf{Application} 	&	\textbf{Benchmark Suite} &	\textbf{Application} \\ \hline \hline						
ArtraCFD~\cite{mo2019mesoscale} &	ArtraCFD	&	\multirow{5}{*}{HPCG~\cite{dongarra_hpcg2015}} & 	Global Dot Product	&	\multirow{10}{*}{SD-VBS - Vision~\cite{Thomas_CortexSuite_IISWC_2014}} & 	disparity	\\	\cline{1-1}	\cline{2-2}		\cline{4-4}		\cline{6-6}
blasr~\cite{chaisson2012mapping} &	Long read aligner	&	&	Multigrid preconditione	&	&	localization	\\	\cline{1-1}	\cline{2-2}		\cline{4-4}		\cline{6-6}
BWA~\cite{li2013aligning} &	aln	&	&	Sparse Matrix Vector Multiplication (SpMV)	&	&	mser	\\		\cline{2-2}		\cline{4-4}		\cline{6-6}
&	fastmap	&	&	Symmetric Gauss-Seidel smoother (SymGS)	&	&	multi\_ncut	\\	\cline{1-1}	\cline{2-2}		\cline{4-4}		\cline{6-6}
\multirow{12}{*}{Chai~\cite{gomezluna_ispass2017}} & 	BFS	&	&	Vector Update	&	&	pca	\\		\cline{2-2}	\cline{3-3}	\cline{4-4}		\cline{6-6}
&	BS	&	\multirow{3}{*}{IMPICA Workloads~\cite{hsieh2016accelerating}} & 	btree	&	&	sift	\\		\cline{2-2}		\cline{4-4}		\cline{6-6}
&	CEDD	&	&	hashtable	&	&	stitch	\\		\cline{2-2}		\cline{4-4}		\cline{6-6}
&	HSTI	&	&	llubenchmark	&	&	svm	\\		\cline{2-2}	\cline{3-3}	\cline{4-4}		\cline{6-6}
&	HSTO	&	\multirow{2}{*}{libvpx~\cite{libvpx}} & 	VP8	&	&	texture\_synthesis	\\		\cline{2-2}		\cline{4-4}		\cline{6-6}
&	OOPPAD	&	&	VP9	&	&	tracking	\\		\cline{2-2}	\cline{3-3}	\cline{4-4}	\cline{5-5}	\cline{6-6}
&	OOPTRNS	&	\multirow{13}{*}{Ligra~\cite{shun_ppopp2013}} & 	BC	&	\multirow{2}{*}{sort-merge-joins~\cite{balkesen2013multi}} & 	m-pass	\\		\cline{2-2}		\cline{4-4}		\cline{6-6}
&	SC	&	&	BellmanFord	&	&	m-way	\\		\cline{2-2}		\cline{4-4}	\cline{5-5}	\cline{6-6}
&	SELECT	&	&	BFS	&	\multirow{29}{*}{SPEC CPU2006~\cite{spec2006}} & 	400.perlbench	\\		\cline{2-2}		\cline{4-4}		\cline{6-6}
&	TRNS	&	&	BFS-Bitvector	&	&	401.bzip2	\\		\cline{2-2}		\cline{4-4}		\cline{6-6}
&	VPACK	&	&	BFS-CC	&	&	403.gcc	\\		\cline{2-2}		\cline{4-4}		\cline{6-6}
&	VUPACK	&	&	CF	&	&	410.bwaves	\\	\cline{1-1}	\cline{2-2}		\cline{4-4}		\cline{6-6}
clstm~\cite{ghosh2016contextual} &	clstm	&	&	Components	&	&	416.gamess	\\	\cline{1-1}	\cline{2-2}		\cline{4-4}		\cline{6-6}
\multirow{9}{*}{CombBLAS~\cite{bulucc2011combinatorial}} & 	BetwCent	&	&	KCore	&	&	429.mcf	\\		\cline{2-2}		\cline{4-4}		\cline{6-6}
&	BipartiteMatchings	&	&	MIS	&	&	433.milc	\\		\cline{2-2}		\cline{4-4}		\cline{6-6}
&	CC	&	&	PageRank	&	&	434.zeusmp	\\		\cline{2-2}		\cline{4-4}		\cline{6-6}
&	DirOptBFS	&	&	PageRankDelta	&	&	435.gromacs	\\		\cline{2-2}		\cline{4-4}		\cline{6-6}
&	FilteredBFS	&	&	Radii	&	&	436.cactusADM	\\		\cline{2-2}		\cline{4-4}		\cline{6-6}
&	FilteredMIS	&	&	Triangle	&	&	437.leslie3d	\\		\cline{2-2}	\cline{3-3}	\cline{4-4}		\cline{6-6}
&	MCL3D	&	\multirow{2}{*}{Metagraph~\cite{karasikov2020metagraph}} & 	annotate	&	&	444.namd	\\		\cline{2-2}		\cline{4-4}		\cline{6-6}
&	Ordering/RCM	&	&	classify	&	&	445.gobmk	\\		\cline{2-2}	\cline{3-3}	\cline{4-4}		\cline{6-6}
&	TopDownBFS	&	\multirow{5}{*}{MKL~\cite{mkl}} & 	ASUM	&	&	447.dealII	\\	\cline{1-1}	\cline{2-2}		\cline{4-4}		\cline{6-6}
\multirow{17}{*}{CORAL~\cite{coral}} &	AMG2013	&	&	AXPY	&	&	450.soplex	\\		\cline{2-2}		\cline{4-4}		\cline{6-6}
&	CAM-SE	&	&	DOT	&	&	453.povray	\\		\cline{2-2}		\cline{4-4}		\cline{6-6}
&	Graph500	&	&	GEMM	&	&	454.calculix	\\		\cline{2-2}		\cline{4-4}		\cline{6-6}
&	HACC	&	&	GEMV	&	&	456.hmmer	\\		\cline{2-2}	\cline{3-3}	\cline{4-4}		\cline{6-6}
&	Hash	&	\multirow{11}{*}{Parboil~\cite{stratton2012parboil}} & 	mri-q	&	&	458.sjeng	\\		\cline{2-2}		\cline{4-4}		\cline{6-6}
&	homme1\_3\_6	&	&	BFS	&	&	459.GemsFDTD	\\		\cline{2-2}		\cline{4-4}		\cline{6-6}
&	Integer Sort	&	&	cutcp	&	&	462.libquantum	\\		\cline{2-2}		\cline{4-4}		\cline{6-6}
&	KMI	&	&	histo	&	&	464.h264ref	\\		\cline{2-2}		\cline{4-4}		\cline{6-6}
&	LSMS	&	&	lbm	&	&	465.tonto	\\		\cline{2-2}		\cline{4-4}		\cline{6-6}
&	LULESH	&	&	mri-gridding	&	&	470.lbm	\\		\cline{2-2}		\cline{4-4}		\cline{6-6}
&	MCB	&	&	sad	&	&	471.omnetpp	\\		\cline{2-2}		\cline{4-4}		\cline{6-6}
&	miniFE	&	&	sgemm	&	&	473.astar	\\		\cline{2-2}		\cline{4-4}		\cline{6-6}
&	Nekbone	&	&	spmv	&	&	481.wrf	\\		\cline{2-2}		\cline{4-4}		\cline{6-6}
&	QBOX	&	&	stencil	&	&	482.sphinx3	\\		\cline{2-2}		\cline{4-4}		\cline{6-6}
&	SNAP	&	&	tpacf	&	&	483.xalancbmk	\\		\cline{2-2}	\cline{3-3}	\cline{4-4}	\cline{5-5}	\cline{6-6}
&	SPECint2006"peak"	&	\multirow{12}{*}{PARSEC~\cite{bienia2008parsec}} &	blackscholes	&	\multirow{43}{*}{SPEC CPU2017~\cite{spec2017}} & 	500.perlbench\_r	\\		\cline{2-2}		\cline{4-4}		\cline{6-6}
&	UMT2013	&	&	bodytrack	&	&	502.gcc\_r	\\	\cline{1-1}	\cline{2-2}		\cline{4-4}		\cline{6-6}
\multirow{15}{*}{Darknet~\cite{redmon_darknet2013}} & 	AlexNet	&	&	canneal	&	&	503.bwaves\_r	\\		\cline{2-2}		\cline{4-4}		\cline{6-6}
&	Darknet19	&	&	dedup	&	&	505.mcf\_r	\\		\cline{2-2}		\cline{4-4}		\cline{6-6}
&	Darknet53	&	&	facesim	&	&	507.cactuBSSN\_r	\\		\cline{2-2}		\cline{4-4}		\cline{6-6}
&	Densenet 201	&	&	ferret	&	&	508.namd\_r	\\		\cline{2-2}		\cline{4-4}		\cline{6-6}
&	Extraction	&	&	fluidanimate	&	&	510.parest\_r	\\		\cline{2-2}		\cline{4-4}		\cline{6-6}
&	Resnet 101	&	&	freqmine	&	&	511.povray\_r	\\		\cline{2-2}		\cline{4-4}		\cline{6-6}
&	Resnet 152	&	&	raytrace	&	&	519.lbm\_r	\\		\cline{2-2}		\cline{4-4}		\cline{6-6}
&	Resnet 18	&	&	streamcluster	&	&	520.omnetpp\_r	\\		\cline{2-2}		\cline{4-4}		\cline{6-6}
&	Resnet 34	&	&	swaptions	&	&	521.wrf\_r	\\		\cline{2-2}		\cline{4-4}		\cline{6-6}
&	Resnet 50	&	&	vips	&	&	523.xalancbmk\_r	\\		\cline{2-2}		\cline{4-4}		\cline{6-6}
&	ResNeXt 101	&	&	x264	&	&	525.x264\_r	\\		\cline{2-2}	\cline{3-3}	\cline{4-4}		\cline{6-6}
&	ResNext 152	&	\multirow{7}{*}{Phoenix~\cite{yoo_iiswc2009}} & 	histogram	&	&	526.blender\_r	\\		\cline{2-2}		\cline{4-4}		\cline{6-6}
&	ResNeXt50	&	&	kmeans	&	&	527.cam4\_r	\\		\cline{2-2}		\cline{4-4}		\cline{6-6}
&	VGG-16	&	&	linear-regression	&	&	531.deepsjeng\_r	\\		\cline{2-2}		\cline{4-4}		\cline{6-6}
&	Yolo	&	&	matrix multiply	&	&	538.imagick\_r	\\	\cline{1-1}	\cline{2-2}		\cline{4-4}		\cline{6-6}
DBT-5~\cite{nascimento2010dbt} & 	TPC-E 	&	&	pca	&	&	541.leela\_r	\\	\cline{1-1}	\cline{2-2}		\cline{4-4}		\cline{6-6}
\multirow{10}{*}{DBx1000~\cite{yu2014staring}} & 	TPCC DL\_DETECT	&	&	string\_match	&	&	544.nab\_r	\\		\cline{2-2}		\cline{4-4}		\cline{6-6}
&	TPCC HEKATON	&	&	word\_count	&	&	548.exchange2\_r	\\		\cline{2-2}	\cline{3-3}	\cline{4-4}		\cline{6-6}
&	TPCC NO\_WAIT	&	\multirow{23}{*}{PolyBench~\cite{pouchet2012polybench}} & 	2mm	&	&	549.fotonik3d\_r	\\		\cline{2-2}		\cline{4-4}		\cline{6-6}
&	TPCC SILO	&	&	3mm	&	&	554.roms\_r	\\		\cline{2-2}		\cline{4-4}		\cline{6-6}
&	TPCC TICTOC	&	&	atax	&	&	557.xz\_r	\\		\cline{2-2}		\cline{4-4}		\cline{6-6}
&	YCSB DL\_DETECT	&	&	bicg	&	&	600.perlbench\_s	\\		\cline{2-2}		\cline{4-4}		\cline{6-6}
&	YCSB HEKATON	&	&	cholesky	&	&	602.gcc\_s	\\		\cline{2-2}		\cline{4-4}		\cline{6-6}
&	YCSB NO\_WAIT	&	&	convolution-2d	&	&	603.bwaves\_s	\\		\cline{2-2}		\cline{4-4}		\cline{6-6}
&	YCSB SILO	&	&	convolution-3d	&	&	605.mcf\_s	\\		\cline{2-2}		\cline{4-4}		\cline{6-6}
&	YCSB TICTOC	&	&	correlation	&	&	607.cactuBSSN\_s	\\	\cline{1-1}	\cline{2-2}		\cline{4-4}		\cline{6-6}
\multirow{4}{*}{DLRM~\cite{DLRM19}} & 	RM1-large~\cite{ke2019recnmp}	&	&	covariance	&	&	619.lbm\_s	\\		\cline{2-2}		\cline{4-4}		\cline{6-6}
&	RM1-small~\cite{ke2019recnmp}	&	&	doitgen	&	&	620.omnetpp\_s	\\		\cline{2-2}		\cline{4-4}		\cline{6-6}
&	RM2-large~\cite{ke2019recnmp}	&	&	durbin	&	&	621.wrf\_s	\\		\cline{2-2}		\cline{4-4}		\cline{6-6}
&	RM2-small~\cite{ke2019recnmp}	&	&	fdtd-apm	&	&	623.xalancbmk\_s	\\	\cline{1-1}	\cline{2-2}		\cline{4-4}		\cline{6-6}
\multirow{2}{*}{GASE~\cite{ahmed2016comparison}} &	FastMap	&	&	gemm	&	&	625.x264\_s	\\		\cline{2-2}		\cline{4-4}		\cline{6-6}
&	gale\_aln	&	&	gemver	&	&	627.cam4\_s	\\	\cline{1-1}	\cline{2-2}		\cline{4-4}		\cline{6-6}
\multirow{9}{*}{GraphMat~\cite{sundaram_vldbendow2015}} & 	BFS	&	&	gramschmidt	&	&	628.pop2\_s	\\		\cline{2-2}		\cline{4-4}		\cline{6-6}
&	DeltaStepping	&	&	gramschmidt	&	&	631.deepsjeng\_s	\\		\cline{2-2}		\cline{4-4}		\cline{6-6}
&	Incremental PageRank	&	&	lu	&	&	638.imagick\_s	\\		\cline{2-2}		\cline{4-4}		\cline{6-6}
&	LDA	&	&	lu	&	&	641.leela\_s	\\		\cline{2-2}		\cline{4-4}		\cline{6-6}
&	PageRank	&	&	mvt	&	&	644.nab\_s	\\		\cline{2-2}		\cline{4-4}		\cline{6-6}
&	SDG	&	&	symm	&	&	648.exchange2\_s	\\		\cline{2-2}		\cline{4-4}		\cline{6-6}
&	SSSP	&	&	syr2k	&	&	649.fotonik3d\_s	\\		\cline{2-2}		\cline{4-4}		\cline{6-6}
&	Topological Sort	&	&	syrk	&	&	654.roms\_s	\\		\cline{2-2}		\cline{4-4}		\cline{6-6}
&	Triangle Counting	&	&	trmm	&	&	657.xz\_s	\\	\cline{1-1}	\cline{2-2}	\cline{3-3}	\cline{4-4}	\cline{5-5}	\cline{6-6}
\end{tabular}%
%\end{longtable}
}

\end{minipage}
\end{table}

\begin{table*}[h]
%\begin{minipage}{\textwidth}
\centering
%\caption{}
\label{table_aps}
\resizebox{\linewidth}{!}{
\begin{tabular}{|l|l || l|l || l |l|}
%\begin{longtable}{|l|l || l|l || l |l|}
\hline
\textbf{Benchmark Suite} &	\textbf{Application} 	&	\textbf{Benchmark Suite} &	\textbf{Application} 	&	\textbf{Benchmark Suite} &	\textbf{Application}	\\ \hline \hline
\multirow{22}{*}{Hardware Effects~\cite{hardwareeffects}} & 	4k aliasing	&	resectionvolume~\cite{palomar2018high} & 	resectionvolume	&	\multirow{14}{*}{SPLASH-2~\cite{woo_isca1995}} & 	barnes	\\		\cline{2-2}	\cline{3-3}	\cline{4-4}		\cline{6-6}
&	bandwidth saturation non-temporal	&	\multirow{19}{*}{Rodinia~\cite{che_iiswc2009}} & 	b+tree	&	&	cholesky	\\		\cline{2-2}		\cline{4-4}		\cline{6-6}
&	bandwidth saturation temporal	&	&	backprop	&	&	fft	\\		\cline{2-2}		\cline{4-4}		\cline{6-6}
&	branch misprediction sort	&	&	bfs	&	&	fmm	\\		\cline{2-2}		\cline{4-4}		\cline{6-6}
&	branch misprediction unsort	&	&	cfd	&	&	lu\_cb	\\		\cline{2-2}		\cline{4-4}		\cline{6-6}
&	branch target misprediction	&	&	heartwall	&	&	lu\_ncb	\\		\cline{2-2}		\cline{4-4}		\cline{6-6}
&	cache conflicts	&	&	hotspot	&	&	ocean\_cp	\\		\cline{2-2}		\cline{4-4}		\cline{6-6}
&	cache/memory hierarchy bandwidth	&	&	hotspot3D	&	&	ocean\_ncp	\\		\cline{2-2}		\cline{4-4}		\cline{6-6}
&	data dependencies	&	&	kmeans	&	&	radiosity	\\		\cline{2-2}		\cline{4-4}		\cline{6-6}
&	denormal floating point numbers	&	&	lavaMD	&	&	radix	\\		\cline{2-2}		\cline{4-4}		\cline{6-6}
&	denormal floating point numbers flush	&	&	leukocyte	&	&	raytrace	\\		\cline{2-2}		\cline{4-4}		\cline{6-6}
&	DRAM refresh interval	&	&	lud	&	&	volrend	\\		\cline{2-2}		\cline{4-4}		\cline{6-6}
&	false sharing	&	&	mummergpu	&	&	water\_nsquared	\\		\cline{2-2}		\cline{4-4}		\cline{6-6}
&	hardware prefetching	&	&	myocyte	&	&	water\_spatial	\\		\cline{2-2}		\cline{4-4}	\cline{5-5}	\cline{6-6}
&	hardware prefetching shuffle	&	&	nn	&	\multirow{8}{*}{Tailbench~\cite{kasture2016tailbench}} & 	img-dnn	\\		\cline{2-2}		\cline{4-4}		\cline{6-6}
&	hardware store elimination	&	&	nw	&	&	masstree	\\		\cline{2-2}		\cline{4-4}		\cline{6-6}
&	memory-bound program	&	&	particlefilter	&	&	moses	\\		\cline{2-2}		\cline{4-4}		\cline{6-6}
&	misaligned accesses	&	&	pathfinder	&	&	shore	\\		\cline{2-2}		\cline{4-4}		\cline{6-6}
&	non-temporal stores	&	&	srad	&	&	silo	\\		\cline{2-2}		\cline{4-4}		\cline{6-6}
&	software prefetching	&	&	streamcluster	&	&	specjbb	\\		\cline{2-2}		\cline{4-4}		\cline{6-6}
&	store buffer capacity	&	\multirow{8}{*}{SD-VBS- Cortex~\cite{Thomas_CortexSuite_IISWC_2014}} & 	lda	&	&	sphinx	\\		\cline{2-2}	\cline{3-3}	\cline{4-4}		\cline{6-6}
&	write combining	&	&	libl	&	&	xapian	\\	\cline{1-1}	\cline{2-2}		\cline{4-4}	\cline{5-5}	\cline{6-6}
\multirow{5}{*}{Hashjoin~\cite{balkesen_TKDE2015}} & 	NPO	&	&	me	&	\multirow{11}{*}{WHISPER~\cite{nalli2017analysis}} & 	ctree	\\		\cline{2-2}		\cline{4-4}		\cline{6-6}
&	PRH	&	&	pca	&	&	echo	\\		\cline{2-2}		\cline{4-4}		\cline{6-6}
&	PRHO	&	&	rbm	&	&	exim	\\		\cline{2-2}		\cline{4-4}		\cline{6-6}
&	PRO	&	&	sphinix	&	&	hashmap	\\		\cline{2-2}		\cline{4-4}		\cline{6-6}
&	RJ	&	&	srr	&	&	memcached	\\	\cline{1-1}	\cline{2-2}		\cline{4-4}		\cline{6-6}
HPCC~\cite{luszczek_hpcc2006} &	RandomAccesses	&	&	svd	&	&	nfs	\\	\cline{1-1}	\cline{2-2}		\cline{3-4}		\cline{6-6}
\multicolumn{4}{|l||}{\multirow{7}{*}{}} 				&	&	redis	\\			\cline{6-6}
\multicolumn{4}{|l||}{}				                &	&	sql	\\						\cline{6-6}
\multicolumn{4}{|l||}{}				                &	&	tpcc	\\						\cline{6-6}
\multicolumn{4}{|l||}{}				                &	&	vacation	\\						\cline{6-6}
\multicolumn{4}{|l||}{}				                &	&	ycsb	\\					\cline{5-5}	\cline{6-6}
\multicolumn{4}{|l||}{}				                &	ZipML~\cite{kara2017fpga} & 	SGD	\\					\cline{5-5}	\cline{6-6}
\multicolumn{4}{|l||}{}				                &	Stream~\cite{mccalpin_stream1995} &	STREAM	\\	\cline{1-1}	\cline{2-2}	\cline{3-3}	\cline{4-4}	\cline{5-5}	\cline{6-6}
\end{tabular}%
%\end{longtable}
}
\bigskip
%\end{minipage}

\end{table*}
% \hfill \hfill \hfill \hfill \vfill \vfill \vfill  \vfill \vfill \vfill 

\pagebreak

%\vspace{100pt}
%\clearpage

\balance
\begin{singlespace}
\bibliographystyle{IEEEtran}
\bibliography{references}
\end{singlespace}
\bookmarksetup{startatroot}
\end{document}